%% file: LOI.tex
\documentclass[11pt,twoside,titlepage]{book}
\usepackage{fancyhdr}
\fancypagestyle{plain}{%
\fancyhf{} 
\fancyfoot[R]{\textsf{ILD - Letter of Intent\hspace{0.5cm}\thepage}}}
\usepackage{epsfig}
\usepackage{rotate}
\usepackage[figuresright]{rotating}
\usepackage{lscape}
\usepackage{hyperref}
\usepackage{url}
\usepackage{amssymb}
\usepackage{graphicx}
\usepackage{multirow}
\usepackage{capt-of}
\usepackage{wrapfig}            
\usepackage{wasysym}            
\usepackage{textcomp}
\DeclareFontFamily{OML}{eur}{\skewchar\font127} \DeclareFontShape{OML}{eur}{m}{n}{<5> <6> 
  <7> <8> <9> gen * eurm <10><10.95><12><14.4><17.28><20.74><24.88>eurm10}{} 
\DeclareSymbolFont{greek}{OML}{eur}{m}{n} 
\DeclareMathSymbol{\upmu}{\mathord}{greek}{"16}

\renewcommand{\arraystretch}{1.25}
\usepackage{mytrc_style}
\usepackage{verbatim}

\usepackage{scrtime}

\include{setup/commondefs}

\begin{document}

\pagestyle{empty}
\begin{figure}
\includegraphics[height=27.5cm,width=18cm,viewport={80 -80 574 730}]{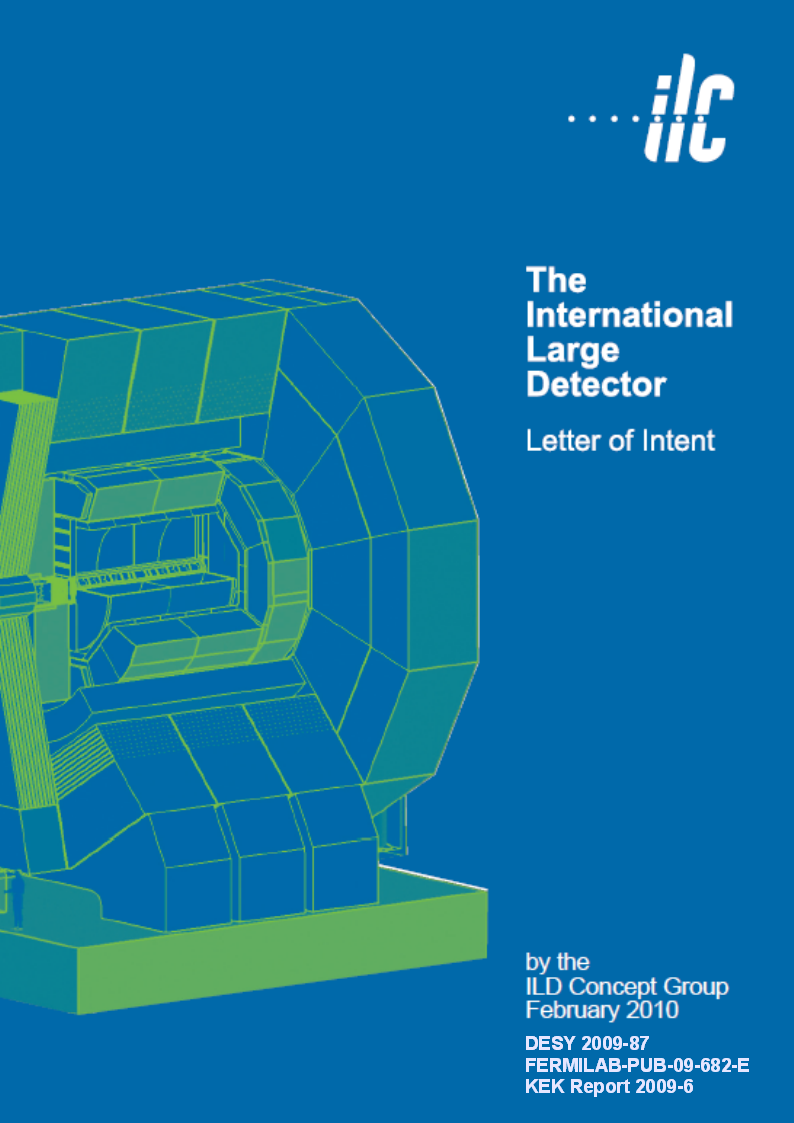}

\end{figure}
\input{setup/frontmatterfile_ild}
\newcommand{\DpName}[2]{\hbox{#1$^{\ref{#2}}$},\hfill}
\newcommand{\DpNameTwo}[3]{\hbox{#1$^{\ref{#2},\ref{#3}}$},\hfill}
\newcommand{\DpNameThree}[4]{\hbox{#1$^{\ref{#2},\ref{#3},\ref{#4}}$},\hfill}
\newcommand{\DpNameLast}[2]{\hbox{#1$^{\ref{#2}}$}\hspace{\Bigfill}}
\renewcommand{\footnoterule}{}

\setcounter{part}{1}
\setcounter{chapter}{0}
\setcounter{secnumdepth}{4}
\setcounter{tocdepth}{2}

\noindent

\cleardoublepage
\newpage
\pagestyle{fancy}
\pagenumbering{roman}
\tableofcontents
\cleardoublepage
\listoffigures
\cleardoublepage
\listoftables
\cleardoublepage
\pagenumbering{arabic}
\setcounter{page}{1}

\chapter{Introduction}
\label{introduction}
\graphicspath{{introduction/}}
\input{introduction/introduction}

\chapter{Detector Optimisation}
\label{optimization}
\graphicspath{{optimization/}}
\input{optimization/optimisation}

\chapter{Physics Performance}
\label{performance}
\graphicspath{{performance/}}
\input{performance/performance}

\chapter{The ILD Sub-Detector Systems}
\label{ILD}
\graphicspath{{ild/}}
\input{ild/ild}

\chapter{Data Acquisition and Computing}
\label{daq}
\graphicspath{{daq/}}
\input{daq/daq}

\chapter{Detector Integration\\ \hfill Machine Detector Interface}
\graphicspath{{ild/integration/}}
\label{integration}
\input{ild/integration/integration}

\graphicspath{{ild/accelerator/}}
\input{ild/accelerator/accelerator}

\chapter{Costing}
\graphicspath{{cost/}}
\label{cost}
\input{cost/cost}

\chapter{The ILD group}
\graphicspath{group/}
\label{ILDgroup}
\input{group/group}

\chapter{R\&D Plan}
\input{group/randplan}
\label{randplan}

\chapter{Conclusions}
\graphicspath{{Conclusion/}}
\label{conclusion}
\input{conclusion/conclusion}

\clearpage

\bibliographystyle{utphys_mod}
\bibliography{LOI}

\end{document}

%% file: setup/commondefs.tex
\renewcommand{\thepage}{\arabic{chapter}.\arabic{section}-\arabic{page}}
\renewcommand{\thefigure}{\arabic{chapter}.\arabic{section}-\arabic{figure}}

\newcommand\writer[1]{{}}

\newcommand{\eplus}{\mathrm{e}^+}
\newcommand{\eminus}{\mathrm{e}^-}
\newcommand{\epem}{\eplus\eminus}
\newcommand{\mpmm}{\mu^+\mu^-}
\newcommand{\tptm}{\tau^+\tau^-}
\newcommand{\lplm}{\ell^+\ell^-}
\newcommand{\eeX}{\epem X}
\newcommand{\mmX}{\mpmm X}
\newcommand{\Zzero}{\mathrm{Z}}
\newcommand{\Wboson}{\mathrm{W}}
\newcommand{\WpWm}{\Wboson^+\Wboson^-}
\newcommand{\Higgs}{\mathrm{H}}
\newcommand{\Ptau}{P_\tau}
\newcommand{\qq}{\mathrm{q}\overline{\mathrm{q}}}
\newcommand{\ttbar}{\mathrm{t}\overline{\mathrm{t}}}
\newcommand{\uubar}{\mathrm{u}\overline{\mathrm{u}}}
\newcommand{\ccbar}{\mathrm{c}\overline{\mathrm{c}}}
\newcommand{\bbbar}{\mathrm{b}\overline{\mathrm{b}}}
\newcommand{\ddbar}{\mathrm{d}\overline{\mathrm{d}}}
\newcommand{\ssbar}{\mathrm{s}\overline{\mathrm{s}}}
\newcommand{\nubar}{\overline{\nu}}
\newcommand{\GammaZ}{\Gamma_\Zzero}
\newcommand{\GammaW}{\Gamma_\Wboson}
\newcommand{\mZ}{m_\Zzero}
\newcommand{\mW}{m_\Wboson}
\newcommand{\mH}{m_\Higgs}
\newcommand{\topquark}{\mathrm{t}}
\newcommand{\mtop}{m_\topquark}
\newcommand{\Gtop}{\Gamma_\topquark}
\newcommand{\roots}{\sqrt{s}}
\newcommand{\rmsn}{\mathrm{rms}_{90}}
\newcommand{\rECAL}{R}
\newcommand{\stau}{\mbox{$\tilde{\tau}$}}

\newcommand{\mstone}  {\mbox{$ M_{\tilde{\tau}_1} $}}

%% file: setup/frontmatterfile_ild.tex
\begin{titlepage}

\begin{center}

{\sffamily\bfseries

    {\Huge I}{\huge NTERNATIONAL} 
    {\Huge L}{\huge ARGE} 
    {\Huge D}{\huge ETECTOR}
    
  \vskip 1.4cm

    {\Huge L}{\huge ETTER}
    {\Huge O}{\huge F}
    {\Huge I}{\huge NTENT}

  \vskip 1.2cm


\vskip 2cm


\vskip 2cm

{\Large The ILD concept group}

  \vskip 1.5cm

\includegraphics[height=9cm]{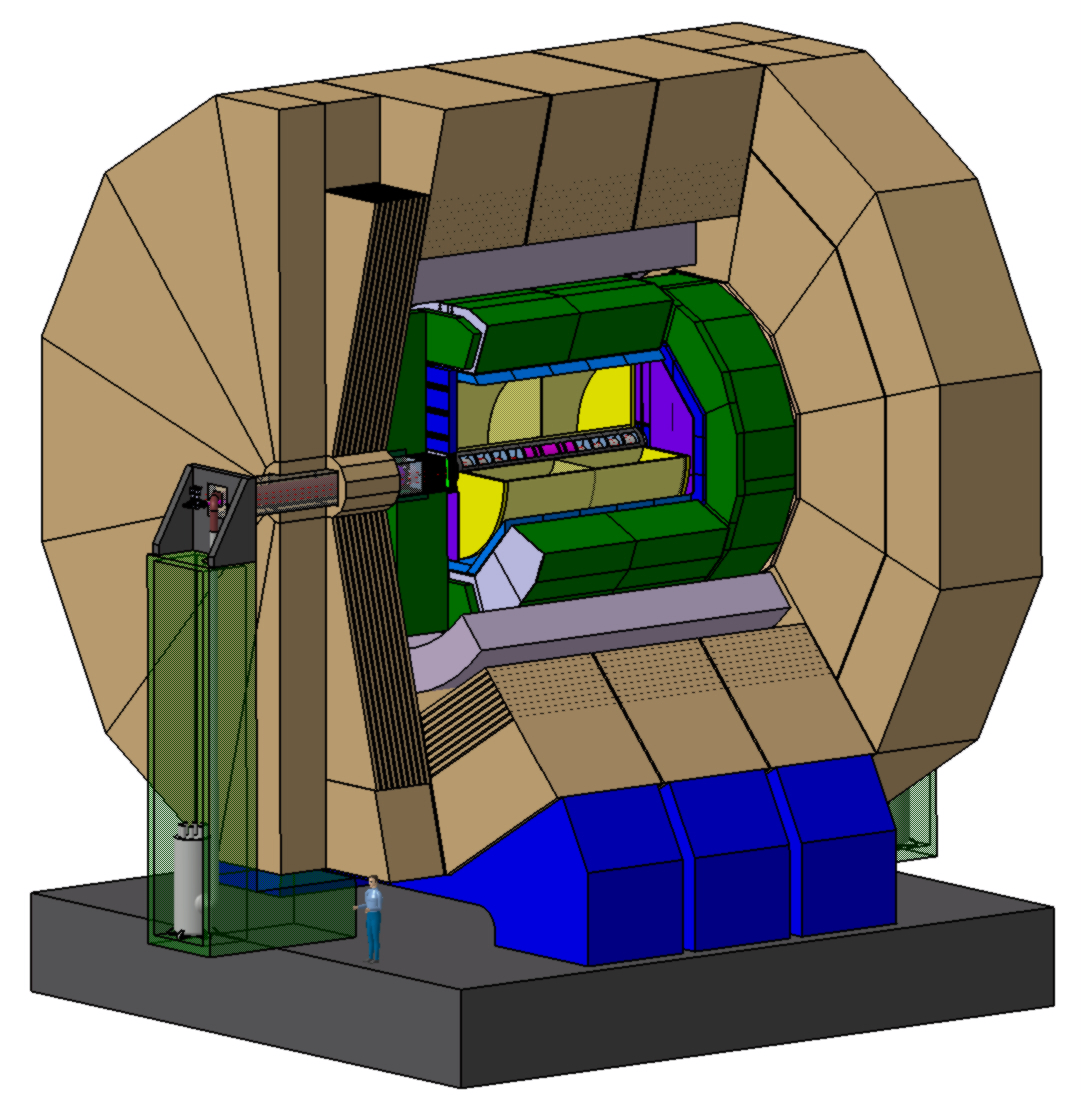}
	\vskip 1.5cm

    {February 2010}
}
\end{center}

\end{titlepage}
\newpage  ~~~~~~~
\vskip 17cm 
\hskip 5cm{\begin{minipage}{12cm}
\copyright 2010 The ILD Concept Group\\

International Large Detector - Letter of Intent\\

DESY / KEK / Fermilab\\
DESY 2009/87 - Fermilab PUB-09-682-E - KEK Report 2009-6 \\
ISSN 0418-9833\\
ISBN 978-3-935702-42-3\\

{\it http://www.ilcild.org and http://www.linearcollider.org}
\end{minipage}}

\newpage

\begin{flushright}
\section*{List of Signatories} 
\end{flushright}

\input{authors/authors.tex}\newpage

%% file: authors/authors.tex
\section*{Signatories}

\noindent{\begin{minipage}{\textwidth}
\raggedright
\mbox{Holger} {Stoeck}
\vspace{1pt}\\
{\sl \scriptsize{University of Sydney, Falkiner High Energy Physics Group, School of Physics, A28, Sydney, NSW 2006, Australia

}\normalsize}
\vspace{12pt}\end{minipage}}

\noindent{\begin{minipage}{\textwidth}
\raggedright
\mbox{Thomas} {Bergauer},
\mbox{Marko} {Dragicevic},
\mbox{Helmut} {Eberl},
\mbox{Sebastian} {Frank},
\mbox{Elena} {Ginina},
\mbox{Stephan} {Haensel},
\mbox{Hana} {Hlucha},
\mbox{Wolfgang} {Kiesenhofer},
\mbox{Manfred} {Krammer},
\mbox{Winfried~A.} {Mitaroff},
\mbox{Fabian} {Moser},
\mbox{Meinhard} {Regler},
\mbox{Manfred} {Valentan},
\mbox{Wolfgang} {Waltenberger}
\vspace{1pt}\\
{\sl \scriptsize{{\"O}sterreichische Akademie der Wissenschaften, Institut f{\"u}r Hochenergiephysik, Nikolsdorfergasse 18, A-1050 Vienna, Austria

}\normalsize}
\vspace{12pt}\end{minipage}}

\noindent{\begin{minipage}{\textwidth}
\raggedright
\mbox{Konstantin} {Afanaciev},
\mbox{Vladimir} {Drugakov},
\mbox{Igor} {Emeliantchik},
\mbox{Alexandr} {Ignatenko},
\mbox{Nikolai} {Shumeiko}
\vspace{1pt}\\
{\sl \scriptsize{National Scientific \&  Educational Centre of Particle \&  High Energy Physics (NCPHEP), Belarusian State University, M.Bogdanovich street 153, 220040 Minsk, Belarus

}\normalsize}
\vspace{12pt}\end{minipage}}

\noindent{\begin{minipage}{\textwidth}
\raggedright
\mbox{Martin} {Grunewald}
\vspace{1pt}\\
{\sl \scriptsize{University of Ghent, Department of Subatomic and Radiation Physics, Proeftuinstraat 86, 9000 Gent, Belgium

}\normalsize}
\vspace{12pt}\end{minipage}}

\noindent{\begin{minipage}{\textwidth}
\raggedright
\mbox{Alain} {Bellerive},
\mbox{Madhu~S.} {Dixit}$^{1}$
\vspace{1pt}\\
{\sl \scriptsize{Carleton University, Department of Physics, 1125 Colonel By Drive, Ottawa, Ontario, Canada K1S 5B6

}\normalsize}
\vspace{12pt}\end{minipage}}

\noindent{\begin{minipage}{\textwidth}
\raggedright
\mbox{Fran{\c c}ois} {Corriveau}
\vspace{1pt}\\
{\sl \scriptsize{McGill University, Department of Physics, Ernest Rutherford Physics Bldg., 3600 University Street, Montreal, Quebec, H3A 2T8 Canada

}\normalsize}
\vspace{12pt}\end{minipage}}

\noindent{\begin{minipage}{\textwidth}
\raggedright
\mbox{Mauricio} {Barbi}
\vspace{1pt}\\
{\sl \scriptsize{University of Regina, Department of Physics, Regina, Saskatchewan, S4S 0A2 Canada

}\normalsize}
\vspace{12pt}\end{minipage}}

\noindent{\begin{minipage}{\textwidth}
\raggedright
\mbox{Jason~M} {Abernathy},
\mbox{Dean} {Karlen}$^{1}$
\vspace{1pt}\\
{\sl \scriptsize{University of Victoria, Department of Physics and Astronomy, P.O.Box 3055 Stn Csc, Victoria, BC V8W 3P6, Canada

}\normalsize}
\vspace{12pt}\end{minipage}}

\noindent{\begin{minipage}{\textwidth}
\raggedright
\mbox{Jean-Pierre} {Martin}
\vspace{1pt}\\
{\sl \scriptsize{Universit{\'e} de Montr{\'e}al, D{\'e}partement de Physique, Groupe de Physique des Particules, C.P. 6128, Succ. Centre-ville, Montr{\'e}al, Qc H3C 3J7, Canada

}\normalsize}
\vspace{12pt}\end{minipage}}

\noindent{\begin{minipage}{\textwidth}
\raggedright
\mbox{Li} {Bo},
\mbox{Shaomin} {Chen},
\mbox{Zhi} {Deng},
\mbox{Yuanning} {Gao},
\mbox{Fanfan} {Jing},
\mbox{Yu-Ping} {Kuang},
\mbox{Yulan} {Li},
\mbox{Bo} {Li},
\mbox{Ting} {Li},
\mbox{Bo} {Liu},
\mbox{Wenbin} {Qian},
\mbox{Junping} {Tian}$^{3}$,
\mbox{Yi} {Wang},
\mbox{Zhenwei} {Yang},
\mbox{Qian} {Yue},
\mbox{Yanxi} {Zhang},
\mbox{Baojun} {Zheng},
\mbox{Liang} {Zhong},
\mbox{Xianglei} {Zhu}
\vspace{1pt}\\
{\sl \scriptsize{Center for High Energy Physics (TUHEP), Tsinghua University, Beijing, China 100084

}\normalsize}
\vspace{12pt}\end{minipage}}

\noindent{\begin{minipage}{\textwidth}
\raggedright
\mbox{Jin~Min} {Yang}
\vspace{1pt}\\
{\sl \scriptsize{Institute of Theoretical Physics, Chinese Academy of Sciences, P.O.Box 2735, Beijing, China 100080

}\normalsize}
\vspace{12pt}\end{minipage}}

\noindent{\begin{minipage}{\textwidth}
\raggedright
\mbox{Chunxu} {Yu}
\vspace{1pt}\\
{\sl \scriptsize{Nankai University, Department of Physics, Tianjin, China 300071

}\normalsize}
\vspace{12pt}\end{minipage}}

\noindent{\begin{minipage}{\textwidth}
\raggedright
\mbox{Cunfeng} {Feng},
\mbox{Xingtao} {Huang},
\mbox{Zuotang} {Liang},
\mbox{Meng} {Wang},
\mbox{Xueyao} {Zhang},
\mbox{Chengguang} {Zhu}
\vspace{1pt}\\
{\sl \scriptsize{Shandong University, 27 Shanda Nanlu, Jinan, China 250100

}\normalsize}
\vspace{12pt}\end{minipage}}

\noindent{\begin{minipage}{\textwidth}
\raggedright
\mbox{Hongfang} {Chen},
\mbox{Liang} {Han},
\mbox{Ge} {Jing},
\mbox{Wengan} {Ma},
\mbox{Ming} {Shao},
\mbox{Kezhu} {Song},
\mbox{Qun} {Wang},
\mbox{Xiaoliang} {Wang},
\mbox{Zizong} {Xu},
\mbox{Wenbiao} {Yan},
\mbox{Renyou} {Zhang},
\mbox{Ziping} {Zhang},
\mbox{Jiawei} {Zhao},
\mbox{Zhengguo} {Zhao},
\mbox{Yongzhao} {Zhou}
\vspace{1pt}\\
{\sl \scriptsize{University of Science and Technology of China, Department of Modern Physics (DMP), Jin Zhai Road 96, Hefei, China 230026

}\normalsize}
\vspace{12pt}\end{minipage}}

\noindent{\begin{minipage}{\textwidth}
\raggedright
\mbox{Zdenek} {Dolezal},
\mbox{Zbynek} {Drasal},
\mbox{Peter} {Kodys},
\mbox{Peter} {Kvasnicka},
\mbox{Jan} {Scheirich},
\mbox{Josef} {Zacek}
\vspace{1pt}\\
{\sl \scriptsize{Charles University, Institute of Particle \& Nuclear Physics, Faculty of Mathematics and Physics, V Holesovickach 2, CZ-18000 Prague 8, Czech Republic

}\normalsize}
\vspace{12pt}\end{minipage}}

\noindent{\begin{minipage}{\textwidth}
\raggedright
\mbox{Jaroslav} {Cvach},
\mbox{Michal} {Marcisovsky},
\mbox{Stanislav} {Nemecek},
\mbox{Ivo} {Polak},
\mbox{Pavel} {Ruzicka},
\mbox{Petr} {Sicho},
\mbox{Jan} {Smol{\'i}k},
\mbox{Vaclav} {Vrba},
\mbox{Jaroslav} {Zalesak}
\vspace{1pt}\\
{\sl \scriptsize{Institute of Physics, ASCR, Academy of Science of the Czech Republic, Division of Elementary Particle Physics, Na Slovance 2, CZ-18221 Prague 8, Czech Republic

}\normalsize}
\vspace{12pt}\end{minipage}}

\noindent{\begin{minipage}{\textwidth}
\raggedright
\mbox{Mogens} {Dam},
\mbox{Peter~H} {Hansen},
\mbox{Stefania} {Xella}
\vspace{1pt}\\
{\sl \scriptsize{Niels Bohr Institute (NBI), University of Copenhagen, Blegdamsvej 17, DK-2100 Copenhagen, Denmark

}\normalsize}
\vspace{12pt}\end{minipage}}

\noindent{\begin{minipage}{\textwidth}
\raggedright
\mbox{Risto} {Orava}$^{4}$
\vspace{1pt}\\
{\sl \scriptsize{University of Helsinki, Department of Physical Sciences, P.O. Box 64 (Vaino Auerin katu 11), FIN-00014, Helsinki, Finland

}\normalsize}
\vspace{12pt}\end{minipage}}

\noindent{\begin{minipage}{\textwidth}
\raggedright
\mbox{David} {Attie},
\mbox{Marc} {Besancon},
\mbox{Paul} {Colas},
\mbox{Eric} {Delagnes},
\mbox{Nicolas} {Fourches},
\mbox{Giomataris} {Ioannis},
\mbox{Francois} {Kircher},
\mbox{Pierre} {Lutz}$^{5}$,
\mbox{Christophe} {Royon},
\mbox{Maxim} {Titov}
\vspace{1pt}\\
{\sl \scriptsize{CEA Saclay, IRFU, F-91191 Gif-sur-Yvette, France

}\normalsize}
\vspace{12pt}\end{minipage}}

\noindent{\begin{minipage}{\textwidth}
\raggedright
\mbox{Jerome} {Baudot}$^{6}$,
\mbox{Auguste} {Besson}$^{6}$,
\mbox{Andrea} {Brogna}$^{6}$,
\mbox{Gilles} {Claus}$^{6}$,
\mbox{Claude} {Colledani}$^{6}$,
\mbox{Rita} {De~Masi},
\mbox{Andrei} {Dorokhov}$^{6}$,
\mbox{Guy} {Doziere}$^{6}$,
\mbox{Abdelkader} {Himmi}$^{6}$,
\mbox{Christine} {Hu-Guo}$^{6}$,
\mbox{Marc} {Imhoff}$^{6}$,
\mbox{Frederic} {Morel}$^{6}$,
\mbox{Isabelle} {Valin}$^{6}$,
\mbox{Yorgos} {Voutsinas},
\mbox{Marc} {Winter}
\vspace{1pt}\\
{\sl \scriptsize{Institut Pluridisciplinaire Hubert Curien, 23 Rue du Loess - BP28, 67037 Strasbourg Cedex 2, France

}\normalsize}
\vspace{12pt}\end{minipage}}

\noindent{\begin{minipage}{\textwidth}
\raggedright
\mbox{Helenka} {Przysiezniak}$^{2}$
\vspace{1pt}\\
{\sl \scriptsize{Laboratoire d'Annecy-le-Vieux de Physique des Particules (LAPP), Chemin du Bellevue, BP 110, F-74941 Annecy-le-Vieux Cedex, France

}\normalsize}
\vspace{12pt}\end{minipage}}

\noindent{\begin{minipage}{\textwidth}
\raggedright
\mbox{Bernard} {Bouquet},
\mbox{Stephane~L.C.} {Callier},
\mbox{Patrick} {Cornebise},
\mbox{Olivier} {Dadoun},
\mbox{Christophe} {De~La~Taille},
\mbox{Philippe} {Doublet},
\mbox{Fr{\'e}d{\'e}ric} {Dulucq},
\mbox{Michele} {Faucci~Giannelli}$^{22}$,
\mbox{Julien~L} {Fleury},
\mbox{Matthieu} {Jor{\'e}},
\mbox{Hengne} {Li},
\mbox{Gisele} {Martin-Chassard},
\mbox{Roman} {Poeschl},
\mbox{Ludovic} {Raux},
\mbox{Francois} {Richard},
\mbox{Nathalie} {Seguin-Moreau},
\mbox{Dirk} {Zerwas},
\mbox{Zhiqing} {Zhang},
\mbox{Fabian} {Zomer}
\vspace{1pt}\\
{\sl \scriptsize{Laboratoire de l'Acc\'el\'erateur Lin\'eaire (LAL), Universit\'e Paris-Sud 11, B\^atiment 200, 91898 Orsay, France

}\normalsize}
\vspace{12pt}\end{minipage}}

\noindent{\begin{minipage}{\textwidth}
\raggedright
\mbox{Cristina} {Carloganu},
\mbox{Pascal} {Gay},
\mbox{Philippe} {Gris}
\vspace{1pt}\\
{\sl \scriptsize{Laboratoire de Physique Corpusculaire de Clermont-Ferrand (LPC), Universit\'e Blaise Pascal, I.N.2.P.3./C.N.R.S., 24 avenue des Landais, 63177 Aubi\`ere Cedex, France

}\normalsize}
\vspace{12pt}\end{minipage}}

\noindent{\begin{minipage}{\textwidth}
\raggedright
\mbox{Alexandre} {Charpy},
\mbox{Catalin} {Ciobanu},
\mbox{Wilfrid} {Da~Silva},
\mbox{Guillaume} {Daubard},
\mbox{Jacques} {David},
\mbox{Christophe} {Evrard},
\mbox{Jean-Francois} {Genat}$^{24}$,
\mbox{Jean~Francois} {Huppert},
\mbox{Didier} {Imbault},
\mbox{Fr{\'e}d{\'e}ric} {Kapusta},
\mbox{Dhellot} {Marc},
\mbox{Ghislain} {Patrick},
\mbox{Thanh~Hung} {Pham},
\mbox{Philippe} {Repain},
\mbox{Aurore} {Savoy-Navarro},
\mbox{Rachid} {Sefri}
\vspace{1pt}\\
{\sl \scriptsize{Laboratoire de Physique Nucl\'eaire et des Hautes Energies (LPNHE), Universit\'e Pierre et Marie Curie, IN2P3/CNRS, Tour 43, RdC, 4, Place Jussieu, 75252, Paris-Cedex 05, France

}\normalsize}
\vspace{12pt}\end{minipage}}

\noindent{\begin{minipage}{\textwidth}
\raggedright
\mbox{Jo{\"e}l} {Bouvier},
\mbox{Daniel} {Dzahini},
\mbox{Laurent} {Gallin-Martel},
\mbox{Julien} {Giraud},
\mbox{Denis} {Grondin},
\mbox{Jean-Yves} {Hostachy},
\mbox{Sabine} {Kraml},
\mbox{Kaloyan} {Krastev},
\mbox{Eric} {Lagorio},
\mbox{Laurent} {Morin},
\mbox{Fatah-Ellah} {Rarbi},
\mbox{Christophe} {Vescovi},
\mbox{Mahfoud} {Yamouni}
\vspace{1pt}\\
{\sl \scriptsize{Laboratoire de Physique Subatomique et de Cosmologie (LPSC), Universit\'e Joseph Fourier (Grenoble 1), CNRS/IN2P3, Institut Polytechnique de Grenoble, 53 rue des Martyrs, F-38026 Grenoble Cedex, France

}\normalsize}
\vspace{12pt}\end{minipage}}

\noindent{\begin{minipage}{\textwidth}
\raggedright
\mbox{Marc} {Anduze},
\mbox{Khaled} {Belkadhi},
\mbox{Vincent} {Boudry},
\mbox{Jean-Claude} {Brient},
\mbox{Catherine} {Clerc},
\mbox{R{\'e}mi} {Cornat},
\mbox{David} {Decotigny},
\mbox{Mickael} {Frotin},
\mbox{Franck} {Gastaldi},
\mbox{Daniel~T~D} {Jeans},
\mbox{Antoine} {Mathieu},
\mbox{Paulo} {Mora~De~Freitas},
\mbox{Gabriel} {Musat},
\mbox{Marcel} {Reinhard},
\mbox{Manqi} {Ruan},
\mbox{Jean-Charles} {Vanel},
\mbox{Henri~L} {Videau}
\vspace{1pt}\\
{\sl \scriptsize{Laboratoire Leprince-Ringuet (LLR), \'Ecole polytechnique -- CNRS/IN2P3, Route de Saclay, F-91128 Palaiseau Cedex, France

}\normalsize}
\vspace{12pt}\end{minipage}}

\noindent{\begin{minipage}{\textwidth}
\raggedright
\mbox{Jean-Charles} {Fontaine}
\vspace{1pt}\\
{\sl \scriptsize{Universit\'e de Haute Alsace Mulhouse-Colmar, Groupe de Recherche en Physique des Hautes Energies (GRPHE), 61 rue Albert Camus, 68093 Mulhouse Cedex, France

}\normalsize}
\vspace{12pt}\end{minipage}}

\noindent{\begin{minipage}{\textwidth}
\raggedright
\mbox{Marc} {Bedjidian},
\mbox{Christophe} {Combaret},
\mbox{G{\'e}rald} {Grenier},
\mbox{Robert} {Kieffer},
\mbox{Imad} {Laktineh},
\mbox{Patrice} {Lebrun},
\mbox{Nick} {Lumb},
\mbox{Herv{\'e}} {Mathez},
\mbox{Kieffer} {Robert},
\mbox{Muriel} {Vander~Donckt}
\vspace{1pt}\\
{\sl \scriptsize{Universit\'e de Lyon, F-69622, Lyon, France ; Universit\'e Lyon 1, Villeurbanne ; CNRS/IN2P3, Institut de Physique Nucl\'eaire de Lyon

}\normalsize}
\vspace{12pt}\end{minipage}}

\noindent{\begin{minipage}{\textwidth}
\raggedright
\mbox{Uwe} {Renz},
\mbox{Markus} {Schumacher}
\vspace{1pt}\\
{\sl \scriptsize{Albert-Ludwigs Universit{\"a}t Freiburg, Physikalisches Institut, Hermann-Herder Str. 3, D-79104 Freiburg, Germany

}\normalsize}
\vspace{12pt}\end{minipage}}

\noindent{\begin{minipage}{\textwidth}
\raggedright
\mbox{Ringo~Sebastian} {Schmidt}$^{9}$
\vspace{1pt}\\
{\sl \scriptsize{Brandenburg University of Technology, Postfach 101344, D-03013 Cottbus, Germany

}\normalsize}
\vspace{12pt}\end{minipage}}

\noindent{\begin{minipage}{\textwidth}
\raggedright
\mbox{Hartwig} {Albrecht},
\mbox{Steve~J.} {Aplin},
\mbox{Jochen} {B{\"u}rger},
\mbox{Christoph} {Bartels}$^{12}$,
\mbox{Philip} {Bechtle},
\mbox{Jeannine} {Beck},
\mbox{Moritz} {Beckmann}$^{11}$,
\mbox{Ties} {Behnke},
\mbox{C.~Mikael~U.} {Berggren},
\mbox{Karsten} {Buesser},
\mbox{Stefano} {Caiazza}$^{12}$,
\mbox{Alan~J.} {Campbell},
\mbox{Sandra} {Christen}$^{12}$,
\mbox{D{\"o}rte} {David},
\mbox{Klaus} {Dehmelt},
\mbox{Ralf} {Diener},
\mbox{Guenter} {Eckerlin},
\mbox{Wolfgang} {Ehrenfeld},
\mbox{Eckhard} {Elsen},
\mbox{Jan} {Engels},
\mbox{Riccardo} {Fabbri},
\mbox{Manfred} {Fleischer},
\mbox{Frank} {Gaede},
\mbox{Erika} {Garutti},
\mbox{Andreas} {Gellrich},
\mbox{Ingrid-Maria} {Gregor},
\mbox{Tobias} {Haas},
\mbox{Lea} {Hallermann}$^{12}$,
\mbox{Anthony} {Hartin},
\mbox{Martin} {Harz},
\mbox{Isa} {Heinze}$^{12}$,
\mbox{Christian} {Helebrant}$^{12}$,
\mbox{Daniela} {K{\"a}fer},
\mbox{Claus} {Kleinwort},
\mbox{U.} {Koetz},
\mbox{Volker} {Korbel},
\mbox{Dirk} {Kr{\"u}cker},
\mbox{Bernward} {Krause},
\mbox{Kirsten} {Kschioneck},
\mbox{Jan} {Kuhlmann},
\mbox{Frank} {Lehner},
\mbox{Diana} {Linzmaier},
\mbox{Jenny} {List},
\mbox{Angela~Isabela} {Lucaci-Timoce},
\mbox{Benjamin} {Lutz}$^{12}$,
\mbox{Ivan} {Marchesini}$^{12}$,
\mbox{Cornelius} {Martens},
\mbox{Niels} {Meyer},
\mbox{Norbert} {Meyners},
\mbox{Joachim} {Mnich},
\mbox{Sergey} {Morozov}$^{12}$,
\mbox{Carsten} {Niebuhr},
\mbox{Alexander} {Petrov},
\mbox{Volker} {Prahl},
\mbox{Alexei} {Raspereza},
\mbox{Philipp} {Roloff}$^{12}$,
\mbox{Christoph} {Rosemann},
\mbox{K.~Peter} {Sch{\"u}ler},
\mbox{Peter} {Schade},
\mbox{Joern} {Schaffran},
\mbox{Sebastian} {Schmitt},
\mbox{Uwe} {Schneekloth},
\mbox{Felix} {Sefkow},
\mbox{Klaus} {Sinram},
\mbox{Blanka} {Sobloher}$^{12}$,
\mbox{Richard} {Stromhagen},
\mbox{Robert} {Volkenborn},
\mbox{Nanda} {Wattimena},
\mbox{Katarzyna} {Wichmann},
\mbox{Wolfram} {Zeuner}$^{20}$
\vspace{1pt}\\
{\sl \scriptsize{Deutsches Elektronen-Synchrotron DESY, A Research Centre of the Helmholtz Association, Notkestrasse 85, 22607 Hamburg, Germany (Hamburg site)

}\normalsize}
\vspace{12pt}\end{minipage}}

\noindent{\begin{minipage}{\textwidth}
\raggedright
\mbox{Matthias} {Bergholz}$^{7}$,
\mbox{Johannes} {Bluemlein},
\mbox{Maria~Elena} {Castro~Carballo},
\mbox{Hans} {Henschel},
\mbox{Hanna} {Kluge},
\mbox{Wolfgang} {Lange},
\mbox{Wolfgang} {Lohmann},
\mbox{Klaus} {Moenig},
\mbox{Martin} {Ohlerich},
\mbox{Sabine} {Riemann},
\mbox{Tord} {Riemann},
\mbox{Andr{\'e}} {Sailer},
\mbox{Andreas} {Sch{\"a}licke},
\mbox{Heinz~Juergen} {Schreiber},
\mbox{Sergej} {Schuwalow},
\mbox{Andriy} {Ushakov}
\vspace{1pt}\\
{\sl \scriptsize{Deutsches Elektronen-Synchrotron DESY, A Research Centre of the Helmholtz Association, Platanenallee 6, 15738 Zeuthen, Germany (Zeuthen site)

}\normalsize}
\vspace{12pt}\end{minipage}}

\noindent{\begin{minipage}{\textwidth}
\raggedright
\mbox{Ariane} {Frey},
\mbox{Carsten} {Hensel},
\mbox{Arnulf} {Quadt}
\vspace{1pt}\\
{\sl \scriptsize{Georg-August-Universit{\"a}t G{\"o}ttingen, II. Physikalisches Institut, Friedrich-Hund-Platz 1, 37077 G{\"o}ttingen, Germany

}\normalsize}
\vspace{12pt}\end{minipage}}

\noindent{\begin{minipage}{\textwidth}
\raggedright
\mbox{Ralph} {Dollan}
\vspace{1pt}\\
{\sl \scriptsize{Humboldt Universit{\"a}t zu Berlin, Fachbereich Physik, Institut f\"ur Elementarteilchenphysik, Newtonstr. 15, D-12489 Berlin, Germany

}\normalsize}
\vspace{12pt}\end{minipage}}

\noindent{\begin{minipage}{\textwidth}
\raggedright
\mbox{Wim} {De~Boer}$^{20}$,
\mbox{Robert} {Rossmanith}
\vspace{1pt}\\
{\sl \scriptsize{Institut f{\"u}r Experimentelle Kernphysik, KIT,Universit{\"a}t Karlsruhe (TH), Wolfgang-Gaede-Str. 1, Postfach 6980, 76128 Karlsruhe

}\normalsize}
\vspace{12pt}\end{minipage}}

\noindent{\begin{minipage}{\textwidth}
\raggedright
\mbox{Stefan} {Tapprogge}
\vspace{1pt}\\
{\sl \scriptsize{Johannes Gutenberg Universit{\"a}t Mainz, Institut f{\"u}r Physik, 55099 Mainz, Germany

}\normalsize}
\vspace{12pt}\end{minipage}}

\noindent{\begin{minipage}{\textwidth}
\raggedright
\mbox{Otmar} {Biebel},
\mbox{Ralf} {Hertenberger},
\mbox{Raimund} {Str{\"o}hmer}
\vspace{1pt}\\
{\sl \scriptsize{Ludwig-Maximilians-Universit{\"a}t M{\"u}nchen, Fakult{\"a}t f{\"u}r Physik, Am Coulombwall 1, D - 85748 Garching, Germany

}\normalsize}
\vspace{12pt}\end{minipage}}

\noindent{\begin{minipage}{\textwidth}
\raggedright
\mbox{Ladislav} {Andricek},
\mbox{Allen} {Caldwell},
\mbox{Xun} {Chen},
\mbox{Christian~M} {Kiesling},
\mbox{Shaojun} {Lu}$^{10}$,
\mbox{Andreas} {Moll}$^{10}$,
\mbox{Hans-G{\"u}nther} {Moser},
\mbox{Bob} {Olivier},
\mbox{Katja} {Seidel}$^{10}$,
\mbox{Ronald~Dean} {Settles},
\mbox{Frank} {Simon}$^{10}$,
\mbox{Christian} {Soldner},
\mbox{Lars} {Weuste}
\vspace{1pt}\\
{\sl \scriptsize{Max-Planck-Institut f{\"u}r Physik (Werner-Heisenberg-Institut), F{\"o}hringer Ring 6, 80805 M{\"u}nchen, Germany

}\normalsize}
\vspace{12pt}\end{minipage}}

\noindent{\begin{minipage}{\textwidth}
\raggedright
\mbox{Werner} {Bernreuther},
\mbox{Tatsiana} {Klimkovich},
\mbox{Hans-Ulrich} {Martyn}$^{8}$,
\mbox{Stefan} {Roth}
\vspace{1pt}\\
{\sl \scriptsize{Rheinisch-Westf{\"a}lische Technische Hochschule (RWTH), Physikalisches Institut, Physikzentrum, Sommerfeldstrasse 14, D-52056 Aachen, Germany

}\normalsize}
\vspace{12pt}\end{minipage}}

\noindent{\begin{minipage}{\textwidth}
\raggedright
\mbox{Deepak} {Kar},
\mbox{Michael} {Kobel},
\mbox{Wolfgang~F.} {Mader},
\mbox{Xavier} {Prudent},
\mbox{Rainer} {Schwierz},
\mbox{Dominik} {Stockinger},
\mbox{Arno} {Straessner}
\vspace{1pt}\\
{\sl \scriptsize{Technische Universit{\"a}t Dresden, Institut f{\"u}r Kern- und Teilchenphysik, D-01069 Dresden, Germany

}\normalsize}
\vspace{12pt}\end{minipage}}

\noindent{\begin{minipage}{\textwidth}
\raggedright
\mbox{Nils} {Feege}$^{8}$,
\mbox{Andreas} {Imhof},
\mbox{Benno} {List},
\mbox{Oliver} {Wendt}$^{8}$
\vspace{1pt}\\
{\sl \scriptsize{University of Hamburg, Physics Department, Institut f{\"u}r Experimentalphysik, Luruper Chaussee 149, 22761 Hamburg, Germany

}\normalsize}
\vspace{12pt}\end{minipage}}

\noindent{\begin{minipage}{\textwidth}
\raggedright
\mbox{Alexander} {Kaplan}$^{8}$,
\mbox{Hans-Christian} {Schultz-Coulon}
\vspace{1pt}\\
{\sl \scriptsize{University of Heidelberg, Kirchhoff Institute of Physics, Albert {\"U}berle Strasse 3-5, DE-69120 Heidelberg, Germany

}\normalsize}
\vspace{12pt}\end{minipage}}

\noindent{\begin{minipage}{\textwidth}
\raggedright
\mbox{Michal} {Czakon}$^{13}$,
\mbox{Malgorzata} {Worek}$^{13}$,
\mbox{Christian} {Zeitnitz}
\vspace{1pt}\\
{\sl \scriptsize{University of Wuppertal, Gau{\ss}stra{\ss}e 20, D-42119 Wuppertal, Germany

}\normalsize}
\vspace{12pt}\end{minipage}}

\noindent{\begin{minipage}{\textwidth}
\raggedright
\mbox{Ian} {Brock},
\mbox{Klaus} {Desch},
\mbox{Jochen} {Kaminski},
\mbox{Martin} {Killenberg},
\mbox{Thorsten} {Krautscheid},
\mbox{Adrian} {Vogel},
\mbox{Norbert} {Wermes},
\mbox{Peter} {Wienemann}
\vspace{1pt}\\
{\sl \scriptsize{Universit{\"a}t Bonn, Physikalisches Institut, Nu{\ss}allee 12, 53115 Bonn, Germany

}\normalsize}
\vspace{12pt}\end{minipage}}

\noindent{\begin{minipage}{\textwidth}
\raggedright
\mbox{Alexander} {Kaukher},
\mbox{Oliver} {Sch{\"a}fer}
\vspace{1pt}\\
{\sl \scriptsize{Universit{\"a}t Rostock, Fachbereich Physik, Universit{\"a}tsplatz 3, D-18051 Rostock, Germany

}\normalsize}
\vspace{12pt}\end{minipage}}

\noindent{\begin{minipage}{\textwidth}
\raggedright
\mbox{Peter} {Buchholz},
\mbox{Ivor} {Fleck},
\mbox{Bakul} {Gaur},
\mbox{Marcus} {Niechciol}
\vspace{1pt}\\
{\sl \scriptsize{Universit{\"a}t Siegen, Fachbereich f{\"u}r Physik, Emmy Noether Campus, Walter-Flex-Str.3, D-57068 Siegen, Germany

}\normalsize}
\vspace{12pt}\end{minipage}}

\noindent{\begin{minipage}{\textwidth}
\raggedright
\mbox{Csaba} {Hajdu},
\mbox{Dezso} {Horvath}
\vspace{1pt}\\
{\sl \scriptsize{Hungarian Academy of Sciences, KFKI Research Institute for Particle and Nuclear Physics, P.O. Box 49, H-1525 Budapest, Hungary

}\normalsize}
\vspace{12pt}\end{minipage}}

\noindent{\begin{minipage}{\textwidth}
\raggedright
\mbox{Bipul} {Bhuyan}
\vspace{1pt}\\
{\sl \scriptsize{Indian Institute of Technology, Guwahati, Guwahati, Assam 781039, India

}\normalsize}
\vspace{12pt}\end{minipage}}

\noindent{\begin{minipage}{\textwidth}
\raggedright
\mbox{Sudeb} {Bhattacharya},
\mbox{Nayana} {Majumdar},
\mbox{Supratik} {Mukhopadhyay},
\mbox{Sandip} {Sarkar}
\vspace{1pt}\\
{\sl \scriptsize{Saha Institute of Nuclear Physics, 1/AF Bidhan Nagar, Kolkata 700064, India

}\normalsize}
\vspace{12pt}\end{minipage}}

\noindent{\begin{minipage}{\textwidth}
\raggedright
\mbox{Atul} {Gurtu},
\mbox{Gobinda} {Majumder}
\vspace{1pt}\\
{\sl \scriptsize{Tata Institute of Fundamental Research, School of Natural Sciences, Homi Bhabha Rd., Mumbai 400005, India

}\normalsize}
\vspace{12pt}\end{minipage}}

\noindent{\begin{minipage}{\textwidth}
\raggedright
\mbox{B.~C.} {Choudhary}
\vspace{1pt}\\
{\sl \scriptsize{University of Delhi, Department of Physics and Astrophysics, Delhi 110007, India

}\normalsize}
\vspace{12pt}\end{minipage}}

\noindent{\begin{minipage}{\textwidth}
\raggedright
\mbox{Manas} {Maity}
\vspace{1pt}\\
{\sl \scriptsize{Visva-Bharati University, Department of Physics, Santiniketan 731235, India

}\normalsize}
\vspace{12pt}\end{minipage}}

\noindent{\begin{minipage}{\textwidth}
\raggedright
\mbox{Halina} {Abramowicz},
\mbox{Ronen} {Ingbir},
\mbox{Aharon} {Levy},
\mbox{Iftach} {Sadeh}
\vspace{1pt}\\
{\sl \scriptsize{Tel-Aviv University, School of Physics and Astronomy, Ramat Aviv, Tel Aviv 69978, Israel

}\normalsize}
\vspace{12pt}\end{minipage}}

\noindent{\begin{minipage}{\textwidth}
\raggedright
\mbox{Antonio} {Bulgheroni}
\vspace{1pt}\\
{\sl \scriptsize{Istituto Nazionale di Fisica Nucleare (INFN), Sezione di Milano, Via Celoria 16, I-20133 Milano, Italy

}\normalsize}
\vspace{12pt}\end{minipage}}

\noindent{\begin{minipage}{\textwidth}
\raggedright
\mbox{Lodovico} {Ratti},
\mbox{Valerio} {Re}
\vspace{1pt}\\
{\sl \scriptsize{Istituto Nazionale di Fisica Nucleare (INFN), Sezione di Pavia, Via Bassi 6, I-27100 Pavia, Italy

}\normalsize}
\vspace{12pt}\end{minipage}}

\noindent{\begin{minipage}{\textwidth}
\raggedright
\mbox{Simonetta} {Gentile}
\vspace{1pt}\\
{\sl \scriptsize{Istituto Nazionale di Fisica Nucleare (INFN), Sezione di Roma, c/o Dipartimento di Fisica - Universit\`a degli Studi di Roma ``La Sapienza'', P.le Aldo Moro 2, I-00185 Roma, Italy

}\normalsize}
\vspace{12pt}\end{minipage}}

\noindent{\begin{minipage}{\textwidth}
\raggedright
\mbox{Diego} {Gamba},
\mbox{Giuseppe} {Giraudo},
\mbox{Paolo} {Mereu}
\vspace{1pt}\\
{\sl \scriptsize{Istituto Nazionale di Fisica Nucleare (INFN), Sezione di Torino, c/o Universit{\'a} di Torino, facolt{\'a} di Fisica, via P Giuria 1, 10125 Torino, Italy

}\normalsize}
\vspace{12pt}\end{minipage}}

\noindent{\begin{minipage}{\textwidth}
\raggedright
\mbox{Alessandro} {Calcaterra},
\mbox{Marcello} {Piccolo}
\vspace{1pt}\\
{\sl \scriptsize{Laboratori Nazionali di Frascati, via E. Fermi, 40, C.P. 13, I-00044 Frascati, Italy

}\normalsize}
\vspace{12pt}\end{minipage}}

\noindent{\begin{minipage}{\textwidth}
\raggedright
\mbox{Massimo} {Caccia}$^{15}$,
\mbox{Chiara} {Cappellini}$^{15}$
\vspace{1pt}\\
{\sl \scriptsize{Universit{\`a} dell'Insubria in Como, Dipartimento di Scienze CC.FF.MM., via Vallegio 11, I-22100 Como, Italy

}\normalsize}
\vspace{12pt}\end{minipage}}

\noindent{\begin{minipage}{\textwidth}
\raggedright
\mbox{Takuo} {Yoshida}
\vspace{1pt}\\
{\sl \scriptsize{Fukui University, Department of Physics, 3-9-1 Bunkyo, Fukui-shi, Fukui 910-8507, Japan

}\normalsize}
\vspace{12pt}\end{minipage}}

\noindent{\begin{minipage}{\textwidth}
\raggedright
\mbox{Yasuo} {Arai},
\mbox{Hirofumi} {Fujii},
\mbox{Keisuke} {Fujii},
\mbox{Junpei} {Fujimoto},
\mbox{Yowichi} {Fujita},
\mbox{Takanori} {Hara},
\mbox{Tomiyoshi} {Haruyama},
\mbox{Takeo} {Higuchi},
\mbox{Katsumasa} {Ikematsu},
\mbox{Yukiko} {Ikemoto},
\mbox{Eiji} {Inoue},
\mbox{Hideo} {Itoh},
\mbox{Go} {Iwai},
\mbox{Nobu} {Katayama},
\mbox{Masanori} {Kawai},
\mbox{Makoto} {Kobayashi},
\mbox{Hideyo} {Kodama},
\mbox{Takashi} {Kohriki},
\mbox{Yoshinari} {Kondou},
\mbox{Akihiro} {Maki},
\mbox{Yasuhiro} {Makida},
\mbox{Takeshi} {Matsuda}$^{8}$,
\mbox{Satoshi} {Mihara},
\mbox{Akiya} {Miyamoto},
\mbox{Takeshi} {Murakami},
\mbox{Isamu} {Nakamura},
\mbox{Kazuo} {Nakayoshi},
\mbox{Shohei} {Nishida},
\mbox{Mitsuaki} {Nozaki},
\mbox{Nobuchika} {Okada},
\mbox{Tsunehiko} {Omori},
\mbox{Masatoshi} {Saito},
\mbox{Toshiya} {Sanami},
\mbox{Hiroshi} {Sendai},
\mbox{Shoichi} {Shimazaki},
\mbox{Yusuke} {Suetsugu},
\mbox{Yasuhiro} {Sugimoto},
\mbox{Kazutaka} {Sumisawa},
\mbox{Shuji} {Tanaka},
\mbox{Manobu} {Tanaka},
\mbox{Ken-Ichi} {Tanaka},
\mbox{Toshiaki} {Tauchi},
\mbox{Kazuya} {Tauchi},
\mbox{Katsuo} {Tokushuku},
\mbox{Toru} {Tsuboyama},
\mbox{Junji} {Urakawa},
\mbox{Yutaka} {Ushiroda},
\mbox{Hiroshi} {Yamaoka},
\mbox{M.} {Yamauchi},
\mbox{Yoshiji} {Yasu},
\mbox{Tamaki} {Yoshioka}
\vspace{1pt}\\
{\sl \scriptsize{High Energy Accelerator Research Organization, KEK, 1-1 Oho, Tsukuba, Ibaraki 305-0801, Japan

}\normalsize}
\vspace{12pt}\end{minipage}}

\noindent{\begin{minipage}{\textwidth}
\raggedright
\mbox{Tohru} {Takahashi}
\vspace{1pt}\\
{\sl \scriptsize{Hiroshima University, Department of Physics, 1-3-1 Kagamiyama, Higashi-Hiroshima, Hiroshima 739-8526, Japan

}\normalsize}
\vspace{12pt}\end{minipage}}

\noindent{\begin{minipage}{\textwidth}
\raggedright
\mbox{Masaki} {Asano}
\vspace{1pt}\\
{\sl \scriptsize{Institute for Cosmic Ray Research, University of Tokyo, 5-1-5 Kashiwa-no-Ha, Kashiwa, Chiba 277-8582, Japan

}\normalsize}
\vspace{12pt}\end{minipage}}

\noindent{\begin{minipage}{\textwidth}
\raggedright
\mbox{Toshiyuki} {Iwamoto},
\mbox{Yoshio} {Kamiya},
\mbox{Hiroyuki} {Matsunaga},
\mbox{Toshinori} {Mori},
\mbox{Wataru} {Ootani},
\mbox{Taikan} {Suehara},
\mbox{Tomohiko} {Tanabe},
\mbox{Satoru} {Yamashita}
\vspace{1pt}\\
{\sl \scriptsize{International Center for Elementary Particle Physics, University of Tokyo, Hongo 7-3-1, Bunkyo District, Tokyo 113-0033, Japan

}\normalsize}
\vspace{12pt}\end{minipage}}

\noindent{\begin{minipage}{\textwidth}
\raggedright
\mbox{Hirokazu} {Ikeda}
\vspace{1pt}\\
{\sl \scriptsize{Japan Aerospace Exploration Agency, Sagamihara Campus, 3-1-1 Yoshinodai, Sagamihara, Kanagawa 220-8510 , Japan

}\normalsize}
\vspace{12pt}\end{minipage}}

\noindent{\begin{minipage}{\textwidth}
\raggedright
\mbox{Yukihiro} {Kato}
\vspace{1pt}\\
{\sl \scriptsize{Kinki University, Department of Physics, 3-4-1 Kowakae, Higashi-Osaka, Osaka 577-8502, Japan

}\normalsize}
\vspace{12pt}\end{minipage}}

\noindent{\begin{minipage}{\textwidth}
\raggedright
\mbox{Akimasa} {Ishikawa},
\mbox{Kiyotomo} {Kawagoe},
\mbox{Takashi} {Matsushita},
\mbox{Hiroshi} {Takeda},
\mbox{Satoru} {Uozumi},
\mbox{Yuji} {Yamazaki}
\vspace{1pt}\\
{\sl \scriptsize{Kobe University, Department of Physics, 1-1 Rokkodai-cho, Nada-ku, Kobe, Hyogo 657-8501, Japan

}\normalsize}
\vspace{12pt}\end{minipage}}

\noindent{\begin{minipage}{\textwidth}
\raggedright
\mbox{Takashi} {Watanabe}
\vspace{1pt}\\
{\sl \scriptsize{Kogakuin University, Department of Physics, Shinjuku Campus, 1-24-2 Nishi-Shinjuku, Shinjuku-ku, Tokyo 163-8677, Japan

}\normalsize}
\vspace{12pt}\end{minipage}}

\noindent{\begin{minipage}{\textwidth}
\raggedright
\mbox{Fumiyoshi} {Kajino}
\vspace{1pt}\\
{\sl \scriptsize{Konan University, Department of Physics, Okamoto 8-9-1, Higashinada, Kobe 658-8501, Japan

}\normalsize}
\vspace{12pt}\end{minipage}}

\noindent{\begin{minipage}{\textwidth}
\raggedright
\mbox{Takahiro} {Fusayasu}
\vspace{1pt}\\
{\sl \scriptsize{Nagasaki Institute of Applied Science, 536 Abamachi, Nagasaki-Shi, Nagasaki 851-0193, Japan

}\normalsize}
\vspace{12pt}\end{minipage}}

\noindent{\begin{minipage}{\textwidth}
\raggedright
\mbox{Takashi} {Mori}
\vspace{1pt}\\
{\sl \scriptsize{Nagoya University, High Energy Physics Lab., Div. of Particle and Astrophysical Sciences, Furo-cho, Chikusa-ku, Nagoya, Aichi 464-8602, Japan

}\normalsize}
\vspace{12pt}\end{minipage}}

\noindent{\begin{minipage}{\textwidth}
\raggedright
\mbox{Takeo} {Kawasaki},
\mbox{Hitoshi} {Miyata},
\mbox{Minori} {Watanabe}
\vspace{1pt}\\
{\sl \scriptsize{Niigata University, Department of Physics, Ikarashi, Niigata 950-218, Japan

}\normalsize}
\vspace{12pt}\end{minipage}}

\noindent{\begin{minipage}{\textwidth}
\raggedright
\mbox{Hiroaki} {Ono}
\vspace{1pt}\\
{\sl \scriptsize{Nippon Dental University School of Life Dentistry at Niigata, 1-8 Hamaura-cho, Chuo-ku, Niigata 951-1500, Japan

}\normalsize}
\vspace{12pt}\end{minipage}}

\noindent{\begin{minipage}{\textwidth}
\raggedright
\mbox{Eiichi} {Nakano}
\vspace{1pt}\\
{\sl \scriptsize{Osaka City University, Department of Physics, Faculty of Science, 3-3-138 Sugimoto, Sumiyoshi-ku, Osaka 558-8585, Japan

}\normalsize}
\vspace{12pt}\end{minipage}}

\noindent{\begin{minipage}{\textwidth}
\raggedright
\mbox{Hirotoshi} {Kuroiwa},
\mbox{Kenichi} {Nakashima},
\mbox{Akira} {Sugiyama},
\mbox{Shiro} {Suzuki},
\mbox{Hiroshi} {Yamaguchi}
\vspace{1pt}\\
{\sl \scriptsize{Saga University, Department of Physics, 1 Honjo-machi, Saga-shi, Saga 840-8502, Japan

}\normalsize}
\vspace{12pt}\end{minipage}}

\noindent{\begin{minipage}{\textwidth}
\raggedright
\mbox{Yoji} {Hasegawa},
\mbox{Yasuhiro} {Ide},
\mbox{Katsushige} {Kotera},
\mbox{Miho} {Nishiyama},
\mbox{Takayuki} {Sakuma},
\mbox{Tohru} {Takeshita},
\mbox{Shunsuke} {Tozuka},
\mbox{Koji} {Yanagida}
\vspace{1pt}\\
{\sl \scriptsize{Shinshu University, 3-1-1, Asahi, Matsumoto, Nagano 390-8621, Japan

}\normalsize}
\vspace{12pt}\end{minipage}}

\noindent{\begin{minipage}{\textwidth}
\raggedright
\mbox{Masaya} {Iwabuchi},
\mbox{Ryo} {Yonamine}
\vspace{1pt}\\
{\sl \scriptsize{Sokendai, The Graduate University for Advanced Studies, Shonan Village, Hayama, Kanagawa 240-0193, Japan

}\normalsize}
\vspace{12pt}\end{minipage}}

\noindent{\begin{minipage}{\textwidth}
\raggedright
\mbox{Kazurayama} {Hironori},
\mbox{Yasuyuki} {Horii},
\mbox{Kennosuke} {Itagaki},
\mbox{Kazutoshi} {Ito},
\mbox{Yusuke} {Kamai},
\mbox{Eriko} {Kato},
\mbox{Tomonori} {Kusano},
\mbox{Tadashi} {Nagamine},
\mbox{Yoshimasa} {Ono},
\mbox{Yoshiyuki} {Onuki},
\mbox{Tomoyuki} {Sanuki},
\mbox{Rei} {Sasaki},
\mbox{Yutaro} {Sato},
\mbox{Fumihiko} {Suekane},
\mbox{Yosuke} {Takubo},
\mbox{Akira} {Yamaguchi},
\mbox{Hitoshi} {Yamamoto},
\mbox{Kohei} {Yoshida}
\vspace{1pt}\\
{\sl \scriptsize{Tohoku University, Department of Physics, Aoba District, Sendai, Miyagi 980-8578, Japan

}\normalsize}
\vspace{12pt}\end{minipage}}

\noindent{\begin{minipage}{\textwidth}
\raggedright
\mbox{Osamu} {Nitoh}
\vspace{1pt}\\
{\sl \scriptsize{Tokyo University of Agriculture Technology, Department of Applied Physics, Naka-machi, Koganei, Tokyo 183-8488, Japan

}\normalsize}
\vspace{12pt}\end{minipage}}

\noindent{\begin{minipage}{\textwidth}
\raggedright
\mbox{Toshinori} {Abe},
\mbox{Hiroki} {Kawahara},
\mbox{Sachio} {Komamiya}
\vspace{1pt}\\
{\sl \scriptsize{University of Tokyo, Department of Physics, 7-3-1 Hongo, Bunkyo District, Tokyo 113-0033, Japan

}\normalsize}
\vspace{12pt}\end{minipage}}

\noindent{\begin{minipage}{\textwidth}
\raggedright
\mbox{Toshinori} {Ikuno},
\mbox{Shinhong} {Kim},
\mbox{Yuji} {Sudo}
\vspace{1pt}\\
{\sl \scriptsize{University of Tsukuba, Institute of Physics, 1-1-1 Ten'nodai, Tsukuba, Ibaraki 305-8571, Japan

}\normalsize}
\vspace{12pt}\end{minipage}}

\noindent{\begin{minipage}{\textwidth}
\raggedright
\mbox{Donghee} {Kim},
\mbox{Guinyun} {Kim},
\mbox{Hyunok} {Kim},
\mbox{Hong~Joo} {Kim},
\mbox{Hwanbae} {Park}
\vspace{1pt}\\
{\sl \scriptsize{Center for High Energy Physics (CHEP) / Kyungpook National University, 1370 Sankyuk-dong, Buk-gu, Daegu 702-701, Korea

}\normalsize}
\vspace{12pt}\end{minipage}}

\noindent{\begin{minipage}{\textwidth}
\raggedright
\mbox{Eun-Joo} {Kim}
\vspace{1pt}\\
{\sl \scriptsize{Chonbuk National University, Division of Science Education, Jeonju 561-756, Korea (South)

}\normalsize}
\vspace{12pt}\end{minipage}}

\noindent{\begin{minipage}{\textwidth}
\raggedright
\mbox{Jik} {Lee},
\mbox{Jiwoo} {Nam},
\mbox{Shinwoo} {Nam},
\mbox{Il~Hung} {Park},
\mbox{Jongmann} {Yang}
\vspace{1pt}\\
{\sl \scriptsize{Ewha Womans University, 11-1 Daehyun-Dong, Seodaemun-Gu, Seoul, 120-750, Korea

}\normalsize}
\vspace{12pt}\end{minipage}}

\noindent{\begin{minipage}{\textwidth}
\raggedright
\mbox{Byunggu} {Cheon}
\vspace{1pt}\\
{\sl \scriptsize{Hanyang University, Department of Physics, Seoul 133-791, Korea

}\normalsize}
\vspace{12pt}\end{minipage}}

\noindent{\begin{minipage}{\textwidth}
\raggedright
\mbox{Suyong} {Choi},
\mbox{Intae} {Yu}
\vspace{1pt}\\
{\sl \scriptsize{Sungkyunkwan University (SKKU), Natural Science Campus 300, Physics Research Division, Chunchun-dong, Jangan-gu, Suwon, Kyunggi-do 440-746, Korea

}\normalsize}
\vspace{12pt}\end{minipage}}

\noindent{\begin{minipage}{\textwidth}
\raggedright
\mbox{Choong~Sun} {Kim}
\vspace{1pt}\\
{\sl \scriptsize{Yonsei University, Department of Physics, 134 Sinchon-dong, Sudaemoon-gu, Seoul 120-749, Korea

}\normalsize}
\vspace{12pt}\end{minipage}}

\noindent{\begin{minipage}{\textwidth}
\raggedright
\mbox{Nicolo} {De~Groot}$^{16}$,
\mbox{Sijbrand} {De~Jong}$^{16}$,
\mbox{Frank} {Filthaut}$^{16}$
\vspace{1pt}\\
{\sl \scriptsize{Institute for Mathematics, Astrophysics and Particle Physics (IMAPP), P.O. Box 9010, 6500 GL Nijmegen, Netherlands

}\normalsize}
\vspace{12pt}\end{minipage}}

\noindent{\begin{minipage}{\textwidth}
\raggedright
\mbox{Stan} {Bentvelsen},
\mbox{Auke} {Colijn}$^{16}$,
\mbox{Paul} {De~Jong},
\mbox{Olga} {Igonkina},
\mbox{Peter~Martin} {Kluit},
\mbox{Els} {Koffeman}$^{16}$,
\mbox{Frank} {Linde},
\mbox{Marcel} {Merk}$^{17}$,
\mbox{Antonio} {Pellegrino},
\mbox{Jan} {Timmermans}$^{8}$,
\mbox{Harry} {Van~Der~Graaf},
\mbox{Marcel} {Vreeswijk}
\vspace{1pt}\\
{\sl \scriptsize{Nikhef, National Institute for Subatomic Physics, P.O. Box 41882, 1009 DB Amsterdam, Netherlands

}\normalsize}
\vspace{12pt}\end{minipage}}

\noindent{\begin{minipage}{\textwidth}
\raggedright
\mbox{Gerhard} {Raven}$^{16}$
\vspace{1pt}\\
{\sl \scriptsize{Vrije Universiteit, Department of Physics, Faculty of Sciences, De Boelelaan 1081, 1081 HV Amsterdam, Netherlands

}\normalsize}
\vspace{12pt}\end{minipage}}

\noindent{\begin{minipage}{\textwidth}
\raggedright
\mbox{Gerald} {Eigen},
\mbox{Per} {Osland}
\vspace{1pt}\\
{\sl \scriptsize{University of Bergen, Institute of Physics, Allegaten 55, N-5007 Bergen, Norway

}\normalsize}
\vspace{12pt}\end{minipage}}

\noindent{\begin{minipage}{\textwidth}
\raggedright
\mbox{Sameen~Ahmed} {Khan}
\vspace{1pt}\\
{\sl \scriptsize{Salalah College of Technology (SCOT), Engineering Department, Post Box No. 608, Postal Code 211, Salalah, Sultanate of Oman

}\normalsize}
\vspace{12pt}\end{minipage}}

\noindent{\begin{minipage}{\textwidth}
\raggedright
\mbox{Editha~P.} {Jacosalem}
\vspace{1pt}\\
{\sl \scriptsize{MSU-Iligan Institute of Technology, Department of Physics, Andres Bonifacio Avenue, 9200 Iligan City, Phillipines

}\normalsize}
\vspace{12pt}\end{minipage}}

\noindent{\begin{minipage}{\textwidth}
\raggedright
\mbox{Marek} {Idzik},
\mbox{Danuta} {Kisielewska},
\mbox{Krzysztof} {Swientek}
\vspace{1pt}\\
{\sl \scriptsize{AGH University of Science and Technology, Akademia Gorniczo-Hutnicza im. Stanislawa Staszica w Krakowie, Al. Mickiewicza 30 PL-30-059 Cracow, Poland

}\normalsize}
\vspace{12pt}\end{minipage}}

\noindent{\begin{minipage}{\textwidth}
\raggedright
\mbox{Marek} {Adamus}
\vspace{1pt}\\
{\sl \scriptsize{Andrzej Soltan Institute for Nuclear Studies, High Energy Physics Department, P-6, Ul. Hoza 69, PL-00 681 Warsaw, Poland

}\normalsize}
\vspace{12pt}\end{minipage}}

\noindent{\begin{minipage}{\textwidth}
\raggedright
\mbox{Witold} {Daniluk},
\mbox{Eryk} {Kielar},
\mbox{Tadeusz} {Lesiak},
\mbox{Krzysztof} {Oliwa},
\mbox{Bogdan} {Pawlik},
\mbox{Wojciech} {Wierba},
\mbox{Leszek} {Zawiejski}
\vspace{1pt}\\
{\sl \scriptsize{The Henryk Niewodniczanski Institute of Nuclear Physics, Polish Academy of Sciences (IFJ PAN), ul. Radzikowskiego 152, PL-31342 Cracow, Poland

}\normalsize}
\vspace{12pt}\end{minipage}}

\noindent{\begin{minipage}{\textwidth}
\raggedright
\mbox{Pawel} {Luzniak}
\vspace{1pt}\\
{\sl \scriptsize{University of Lodz, Faculty of Physics and Applied Informatics, Pomorska 149/153, PL-90-236 Lodz, Poland

}\normalsize}
\vspace{12pt}\end{minipage}}

\noindent{\begin{minipage}{\textwidth}
\raggedright
\mbox{Jacek} {Ciborowski}$^{18}$,
\mbox{Grzegorz} {Grzelak},
\mbox{Lukasz} {Maczewski}$^{18}$,
\mbox{Piotr} {Niezurawski},
\mbox{Aleksander~Filip} {Zarnecki}
\vspace{1pt}\\
{\sl \scriptsize{University of Warsaw, Institute of Experimental Physics, Ul. Hoza 69, PL-00 681 Warsaw, Poland

}\normalsize}
\vspace{12pt}\end{minipage}}

\noindent{\begin{minipage}{\textwidth}
\raggedright
\mbox{Janusz} {Rosiek}
\vspace{1pt}\\
{\sl \scriptsize{University of Warsaw, Institute of Theoretical Physics, Ul. Hoza 69, PL-00 681 Warsaw, Poland

}\normalsize}
\vspace{12pt}\end{minipage}}

\noindent{\begin{minipage}{\textwidth}
\raggedright
\mbox{Cornelia} {Coca},
\mbox{Mihai-Octavian} {Dima},
\mbox{Laurentiu~Alexandru} {Dumitru},
\mbox{Marius~Ciprian} {Orlandea},
\mbox{Eliza} {Teodorescu}
\vspace{1pt}\\
{\sl \scriptsize{National Institute of Physics and Nuclear Engineering ``Horia Hulubei'' (IFIN-HH), Str. Atomistilor no. 407, P.O. Box MG-6, R-76900 Bucharest - Magurele, Romania

}\normalsize}
\vspace{12pt}\end{minipage}}

\noindent{\begin{minipage}{\textwidth}
\raggedright
\mbox{Aura} {Rosca}$^{19}$
\vspace{1pt}\\
{\sl \scriptsize{West University of Timisoara, Faculty of Physics, Bd. V. Parvan 4, 300223 Timisoara, Romania 

}\normalsize}
\vspace{12pt}\end{minipage}}

\noindent{\begin{minipage}{\textwidth}
\raggedright
\mbox{A.} {Bondar},
\mbox{A.F.} {Buzulutskov},
\mbox{L.I.} {Shechtman},
\mbox{Valery~I.} {Telnov}
\vspace{1pt}\\
{\sl \scriptsize{Budker Institute for Nuclear Physics (BINP), 630090 Novosibirsk, Russia

}\normalsize}
\vspace{12pt}\end{minipage}}

\noindent{\begin{minipage}{\textwidth}
\raggedright
\mbox{Marina} {Chadeeva},
\mbox{Mikhail} {Danilov},
\mbox{Vasily} {Morgunov}$^{8}$,
\mbox{Vladimir} {Rusinov},
\mbox{Evgueny~I.} {Tarkovsky}
\vspace{1pt}\\
{\sl \scriptsize{Institute of Theoretical and Experimetal Physics, B. Cheremushkinskawa, 25, RU-117259, Moscow, Russia

}\normalsize}
\vspace{12pt}\end{minipage}}

\noindent{\begin{minipage}{\textwidth}
\raggedright
\mbox{Alexander} {Olchevski}
\vspace{1pt}\\
{\sl \scriptsize{Joint Institute for Nuclear Research (JINR), Joliot-Curie 6, 141980, Dubna, Moscow Region, Russia

}\normalsize}
\vspace{12pt}\end{minipage}}

\noindent{\begin{minipage}{\textwidth}
\raggedright
\mbox{Eduard} {Boos},
\mbox{Leonid} {Gladilin},
\mbox{Mikhail~M.} {Merkin}
\vspace{1pt}\\
{\sl \scriptsize{Lomonosov Moscow State University, Skobeltsyn Institute of Nuclear Physics (MSU SINP), 1(2), Leninskie gory, GSP-1, Moscow 119991, Russia

}\normalsize}
\vspace{12pt}\end{minipage}}

\noindent{\begin{minipage}{\textwidth}
\raggedright
\mbox{Boris~A.} {Dolgoshein},
\mbox{Elena} {Popova}
\vspace{1pt}\\
{\sl \scriptsize{Moscow Engineering Physics Institute (MEPhI), Dept. of Physics, 31, Kashirskoye shosse, 115409 Moscow, Russia

}\normalsize}
\vspace{12pt}\end{minipage}}

\noindent{\begin{minipage}{\textwidth}
\raggedright
\mbox{Nicola} {D'Ascenzo}$^{8}$,
\mbox{Valery} {Galkin},
\mbox{Alexei} {Galkin},
\mbox{Dmitri} {Ossetski},
\mbox{Dmitri} {Ryzhikov},
\mbox{Valeri} {Saveliev}
\vspace{1pt}\\
{\sl \scriptsize{Obninsk State Technical University for Nuclear Engineering (IATE), Obninsk, Russia

}\normalsize}
\vspace{12pt}\end{minipage}}

\noindent{\begin{minipage}{\textwidth}
\raggedright
\mbox{Ivanka} {Bozovic-Jelisavcic},
\mbox{Stevan} {Jokic},
\mbox{Tatjana} {Jovin},
\mbox{Judita} {Mamuzic},
\mbox{Mihajlo} {Mudrinic},
\mbox{Mila} {Pandurovic},
\mbox{Ivan} {Smiljanic}
\vspace{1pt}\\
{\sl \scriptsize{VINCA Institute of Nuclear Sciences, Laboratory of Physics, PO Box 522, YU-11001 Belgrade, Serbia and Montenegro

}\normalsize}
\vspace{12pt}\end{minipage}}

\noindent{\begin{minipage}{\textwidth}
\raggedright
\mbox{Jozef} {Ferencei}
\vspace{1pt}\\
{\sl \scriptsize{Institute of Experimental Physics, Slovak Academy of Sciences, Watsonova 47, SK-04001 Kosice, Slovakia

}\normalsize}
\vspace{12pt}\end{minipage}}

\noindent{\begin{minipage}{\textwidth}
\raggedright
\mbox{Enrique} {Calvo~Alamillo},
\mbox{Mary-Cruz} {Fouz},
\mbox{Jesus} {Puerta-Pelayo}
\vspace{1pt}\\
{\sl \scriptsize{Centro de Investigaciones Energ\'eticas, Medioambientales y Tecnol\'ogicas, CIEMAT, Avenida Complutense 22, E-28040 Madrid, Spain

}\normalsize}
\vspace{12pt}\end{minipage}}

\noindent{\begin{minipage}{\textwidth}
\raggedright
\mbox{Juan~Pablo} {Balbuena},
\mbox{Daniela} {Bassignana},
\mbox{Celeste} {Fleta},
\mbox{Manuel} {Lozano},
\mbox{Giulio} {Pellegrini},
\mbox{Miguel} {Ull{\'a}n}
\vspace{1pt}\\
{\sl \scriptsize{Centro Nacional de Microelectr\'onica (CNM), Instituto de Microelectr\'onica de Barcelona (IMB), Campus UAB, 08193 Cerdanyola del Vall\`es (Bellaterra), Barcelona, Spain

}\normalsize}
\vspace{12pt}\end{minipage}}

\noindent{\begin{minipage}{\textwidth}
\raggedright
\mbox{Carmen} {Alabau~Pons},
\mbox{Markus} {Ball},
\mbox{Angeles} {Faus-Golfe},
\mbox{Juan} {Fuster},
\mbox{Carlos} {Lacasta~Ll{\'a}cer},
\mbox{Carlos} {Mari{\~n}as},
\mbox{Marcel} {Vos}
\vspace{1pt}\\
{\sl \scriptsize{Instituto de Fisica Corpuscular (IFIC), Centro Mixto CSIC-UVEG, Edificio Investigacion Paterna, Apartado 22085, 46071 Valencia, Spain

}\normalsize}
\vspace{12pt}\end{minipage}}

\noindent{\begin{minipage}{\textwidth}
\raggedright
\mbox{Jordi} {Duarte~Campderr{\'o}s},
\mbox{Marcos} {Fernandez~Garcia},
\mbox{Francisco~Javier} {Gonz{\'a}lez~S{\'a}nchez},
\mbox{Richard} {Jaramillo~Echeverr{\'i}a},
\mbox{Amparo} {Lopez~Virto},
\mbox{Celso} {Martinez~Rivero},
\mbox{David} {Moya},
\mbox{Alberto} {Ruiz-Jimeno},
\mbox{Ivan} {Vila}
\vspace{1pt}\\
{\sl \scriptsize{Instituto de Fisica de Cantabria, (IFCA, CSIC-UC), Facultad de Ciencias, Avda. Los Castros s/n, 39005 Santander, Spain

}\normalsize}
\vspace{12pt}\end{minipage}}

\noindent{\begin{minipage}{\textwidth}
\raggedright
\mbox{Bernardo} {Adeva},
\mbox{Abraham} {Gallas},
\mbox{Carmen} {Iglesias~Escudero},
\mbox{Juan~J.} {Saborido},
\mbox{Pablo} {Vazquez~Regueiro}
\vspace{1pt}\\
{\sl \scriptsize{Instituto Galego de Fisica de Altas Enerxias (IGFAE,USC) Facultad de Fisica, Campus Sur E-15782 Santiago de Compostela, Spain

}\normalsize}
\vspace{12pt}\end{minipage}}

\noindent{\begin{minipage}{\textwidth}
\raggedright
\mbox{Juan~Antonio} {Aguilar-Saavedra},
\mbox{Nuno} {Castro}
\vspace{1pt}\\
{\sl \scriptsize{Universidad de Granada, Departamento de F\'{i}sica Te\'{o}rica y del Cosmos, Campus de Fuentenueva, E-18071 Granada, Spain

}\normalsize}
\vspace{12pt}\end{minipage}}

\noindent{\begin{minipage}{\textwidth}
\raggedright
\mbox{Thorsten} {Lux},
\mbox{Cristobal} {Padilla},
\mbox{Imma} {Riu}
\vspace{1pt}\\
{\sl \scriptsize{Universitat Aut\`onoma de Barcelona, Institut de Fisica d'Altes Energies (IFAE), Campus UAB, Edifici Cn, E-08193 Bellaterra, Barcelona, Spain

}\normalsize}
\vspace{12pt}\end{minipage}}

\noindent{\begin{minipage}{\textwidth}
\raggedright
\mbox{Jordi} {Riera-Babures},
\mbox{Xavier} {Vilasis-Cardona}
\vspace{1pt}\\
{\sl \scriptsize{Universitat Ramon Llull, La Salle, C/ Quatre Camins 2, 08022 Barcelona, Spain

}\normalsize}
\vspace{12pt}\end{minipage}}

\noindent{\begin{minipage}{\textwidth}
\raggedright
\mbox{Angel} {Dieguez},
\mbox{Lluis} {Garrido~Beltran}
\vspace{1pt}\\
{\sl \scriptsize{University de Barcelona, Facultat de F\'isica, Av. Diagonal, 647, Barcelona 08028, Spain

}\normalsize}
\vspace{12pt}\end{minipage}}

\noindent{\begin{minipage}{\textwidth}
\raggedright
\mbox{Vincent} {Hedberg},
\mbox{Leif} {Jonsson},
\mbox{Bjorn} {Lundberg},
\mbox{Ulf} {Mjornmark},
\mbox{Anders} {Oskarsson},
\mbox{Lennart} {Osterman},
\mbox{Evert} {Stenlund}
\vspace{1pt}\\
{\sl \scriptsize{Lunds Universitet, Fysiska Institutionen, Avdelningen f{\"o}r Experimentell H{\"o}genergifysik, Box 118, 221 00 Lund, Sweden

}\normalsize}
\vspace{12pt}\end{minipage}}

\noindent{\begin{minipage}{\textwidth}
\raggedright
\mbox{Michael} {Campbell},
\mbox{Albert} {De~Roeck},
\mbox{Konrad} {Elsener},
\mbox{Andrea} {Gaddi},
\mbox{Hubert} {Gerwig},
\mbox{Christian} {Grefe}$^{14}$,
\mbox{Michael} {Hauschild},
\mbox{Lucie} {Linssen},
\mbox{Xavier} {Llopart~Cudie},
\mbox{Luciano} {Musa},
\mbox{Dieter} {Schlatter},
\mbox{Peter} {Speckmayer}
\vspace{1pt}\\
{\sl \scriptsize{CERN, CH-1211 Gen\`eve 23, Switzerland

}\normalsize}
\vspace{12pt}\end{minipage}}

\noindent{\begin{minipage}{\textwidth}
\raggedright
\mbox{G{\"u}nther} {Dissertori},
\mbox{Gerard} {Faber},
\mbox{Alain} {Herv{\'e}},
\mbox{Nebojsa} {Smiljkovic}
\vspace{1pt}\\
{\sl \scriptsize{ETH Z{\"u}rich, Institute for Particle Physics (IPP), Schafmattstrasse 20, CH-8093 Z{\"u}rich, Switzerland

}\normalsize}
\vspace{12pt}\end{minipage}}

\noindent{\begin{minipage}{\textwidth}
\raggedright
\mbox{Hideyuki} {Nakazawa}
\vspace{1pt}\\
{\sl \scriptsize{National Central University, High Energy Group, Department of Physics, Chung-li, Taiwan 32001

}\normalsize}
\vspace{12pt}\end{minipage}}

\noindent{\begin{minipage}{\textwidth}
\raggedright
\mbox{Paoti} {Chang},
\mbox{Wei-Shu} {Hou},
\mbox{Koji} {Ueno},
\mbox{Min-Zu} {Wang}
\vspace{1pt}\\
{\sl \scriptsize{National Taiwan University, Physics Department, Taipei, Taiwan 106

}\normalsize}
\vspace{12pt}\end{minipage}}

\noindent{\begin{minipage}{\textwidth}
\raggedright
\mbox{Gudrid~A.} {Moortgat-Pick}$^{21}$
\vspace{1pt}\\
{\sl \scriptsize{Durham University,  Department of Physics, Ogen Center for Fundamental Physics, South Rd., Durham DH1 3LE, UK

}\normalsize}
\vspace{12pt}\end{minipage}}

\noindent{\begin{minipage}{\textwidth}
\raggedright
\mbox{James} {Ballin}
\vspace{1pt}\\
{\sl \scriptsize{Imperial College, Blackett Laboratory, Department of Physics, Prince Consort Road, London, SW7 2BW, UK

}\normalsize}
\vspace{12pt}\end{minipage}}

\noindent{\begin{minipage}{\textwidth}
\raggedright
\mbox{Andre} {Sopczak}
\vspace{1pt}\\
{\sl \scriptsize{Lancaster University, Physics Department, Lancaster LA1 4YB, UK

}\normalsize}
\vspace{12pt}\end{minipage}}

\noindent{\begin{minipage}{\textwidth}
\raggedright
\mbox{Grahame} {Blair},
\mbox{Veronique} {Boisvert}
\vspace{1pt}\\
{\sl \scriptsize{Royal Holloway, University of London (RHUL), Department of Physics, Egham, Surrey TW20 0EX, UK 

}\normalsize}
\vspace{12pt}\end{minipage}}

\noindent{\begin{minipage}{\textwidth}
\raggedright
\mbox{Chris} {Damerell},
\mbox{Kristian} {Harder}
\vspace{1pt}\\
{\sl \scriptsize{STFC Rutherford Appleton Laboratory, Chilton, Didcot, Oxon OX11 0QX, UK 

}\normalsize}
\vspace{12pt}\end{minipage}}

\noindent{\begin{minipage}{\textwidth}
\raggedright
\mbox{Derek~J.} {Attree},
\mbox{Valeria} {Bartsch},
\mbox{Filimon} {Gournaris},
\mbox{Alexey} {Lyapin},
\mbox{David~J.} {Miller},
\mbox{Martin} {Postranecky},
\mbox{Matthew} {Warren},
\mbox{Matthew} {Wing}
\vspace{1pt}\\
{\sl \scriptsize{University College of London (UCL), High Energy Physics Group, Physics and Astronomy Department, Gower Street, London WC1E 6BT, UK 

}\normalsize}
\vspace{12pt}\end{minipage}}

\noindent{\begin{minipage}{\textwidth}
\raggedright
\mbox{Owen} {Miller},
\mbox{Nigel~K.} {Watson},
\mbox{John~A.} {Wilson}
\vspace{1pt}\\
{\sl \scriptsize{University of Birmingham, School of Physics and Astronomy, Particle Physics Group, Edgbaston, Birmingham B15 2TT, UK

}\normalsize}
\vspace{12pt}\end{minipage}}

\noindent{\begin{minipage}{\textwidth}
\raggedright
\mbox{Joel} {Goldstein}
\vspace{1pt}\\
{\sl \scriptsize{University of Bristol, H. H. Wills Physics Lab, Tyndall Ave., Bristol BS8 1TL, UK

}\normalsize}
\vspace{12pt}\end{minipage}}

\noindent{\begin{minipage}{\textwidth}
\raggedright
\mbox{Bart} {Hommels},
\mbox{John} {Marshall},
\mbox{Georgios} {Mavromanolakis}$^{25}$,
\mbox{Mark} {Thomson},
\mbox{David~R} {Ward}
\vspace{1pt}\\
{\sl \scriptsize{University of Cambridge, Cavendish Laboratory, J J Thomson Avenue, Cambridge CB3 0HE, UK

}\normalsize}
\vspace{12pt}\end{minipage}}

\noindent{\begin{minipage}{\textwidth}
\raggedright
\mbox{Victoria~J} {Martin},
\mbox{Hajrah} {Tabassam},
\mbox{Roberval} {Walsh}
\vspace{1pt}\\
{\sl \scriptsize{University of Edinburgh, School of Physics, James Clerk Maxwell Building, The King's Buildings, Mayfield Road, Edinburgh EH9 3JZ, UK

}\normalsize}
\vspace{12pt}\end{minipage}}

\noindent{\begin{minipage}{\textwidth}
\raggedright
\mbox{Richard} {Bates},
\mbox{Craig} {Buttar},
\mbox{Tony} {Doyle},
\mbox{Lars} {Eklund},
\mbox{Val} {O'Shea},
\mbox{Chris} {Parkes},
\mbox{Aidan} {Robson}
\vspace{1pt}\\
{\sl \scriptsize{University of Glasgow, Department of Physics \& Astronomy, University Avenue, Glasgow G12 8QQ, Scotland, UK

}\normalsize}
\vspace{12pt}\end{minipage}}

\noindent{\begin{minipage}{\textwidth}
\raggedright
\mbox{Tim} {Greenshaw}
\vspace{1pt}\\
{\sl \scriptsize{University of Liverpool, Department of Physics, Oliver Lodge Lab, Oxford St., Liverpool L69 7ZE, UK

}\normalsize}
\vspace{12pt}\end{minipage}}

\noindent{\begin{minipage}{\textwidth}
\raggedright
\mbox{David} {Bailey},
\mbox{Roger} {Barlow}$^{21}$
\vspace{1pt}\\
{\sl \scriptsize{University of Manchester, School of Physics and Astronomy, Schuster Lab, Manchester M13 9PL, UK

}\normalsize}
\vspace{12pt}\end{minipage}}

\noindent{\begin{minipage}{\textwidth}
\raggedright
\mbox{Brian} {Foster}
\vspace{1pt}\\
{\sl \scriptsize{University of Oxford, Particle Physics Department, Denys Wilkinson Bldg., Keble Road, Oxford OX1 3RH England, UK 

}\normalsize}
\vspace{12pt}\end{minipage}}

\noindent{\begin{minipage}{\textwidth}
\raggedright
\mbox{Stefano} {Moretti}$^{23}$
\vspace{1pt}\\
{\sl \scriptsize{University of Southampton, School of Physics and Astronomy, Highfield, Southampton S017 1BJ, England, UK

}\normalsize}
\vspace{12pt}\end{minipage}}

\noindent{\begin{minipage}{\textwidth}
\raggedright
\mbox{Jose} {Repond}
\vspace{1pt}\\
{\sl \scriptsize{Argonne National Laboratory (ANL), 9700 S. Cass Avenue, Argonne, IL 60439, USA

}\normalsize}
\vspace{12pt}\end{minipage}}

\noindent{\begin{minipage}{\textwidth}
\raggedright
\mbox{John} {Butler}
\vspace{1pt}\\
{\sl \scriptsize{Boston University, Department of Physics, 590 Commonwealth Avenue, Boston, MA 02215, USA

}\normalsize}
\vspace{12pt}\end{minipage}}

\noindent{\begin{minipage}{\textwidth}
\raggedright
\mbox{Lawrence} {Gibbons},
\mbox{J.~Ritchie} {Patterson},
\mbox{Daniel} {Peterson}
\vspace{1pt}\\
{\sl \scriptsize{Cornell University, Laboratory for Elementary-Particle Physics (LEPP), Ithaca, NY 14853, USA

}\normalsize}
\vspace{12pt}\end{minipage}}

\noindent{\begin{minipage}{\textwidth}
\raggedright
\mbox{Marco} {Verzocchi}
\vspace{1pt}\\
{\sl \scriptsize{Fermi National Accelerator Laboratory (FNAL), P.O.Box 500, Batavia, IL 60510-0500, USA

}\normalsize}
\vspace{12pt}\end{minipage}}

\noindent{\begin{minipage}{\textwidth}
\raggedright
\mbox{Rick~J.} {Van~Kooten}
\vspace{1pt}\\
{\sl \scriptsize{Indiana University, Department of Physics, Swain Hall West 117, 727 E. 3rd St., Bloomington, IN 47405-7105, USA

}\normalsize}
\vspace{12pt}\end{minipage}}

\noindent{\begin{minipage}{\textwidth}
\raggedright
\mbox{Z.D.} {Greenwood},
\mbox{Lee} {Sawyer},
\mbox{Markus} {Wobisch}
\vspace{1pt}\\
{\sl \scriptsize{Louisiana Tech University, Department of Physics, Ruston, LA 71272, USA

}\normalsize}
\vspace{12pt}\end{minipage}}

\noindent{\begin{minipage}{\textwidth}
\raggedright
\mbox{Dhiman} {Chakraborty},
\mbox{David} {Hedin},
\mbox{Guilherme} {Lima},
\mbox{Vishnu} {Zutshi}
\vspace{1pt}\\
{\sl \scriptsize{Northern Illinois University, Department of Physics, DeKalb, Illinois 60115-2825, USA

}\normalsize}
\vspace{12pt}\end{minipage}}

\noindent{\begin{minipage}{\textwidth}
\raggedright
\mbox{Qing} {He},
\mbox{Kirk~T} {Mcdonald}
\vspace{1pt}\\
{\sl \scriptsize{Princeton University, Department of Physics, P.O. Box 708, Princeton, NJ 08542-0708, USA

}\normalsize}
\vspace{12pt}\end{minipage}}

\noindent{\begin{minipage}{\textwidth}
\raggedright
\mbox{Bruce~A.} {Schumm}
\vspace{1pt}\\
{\sl \scriptsize{University of California Santa Cruz, Institute for Particle Physics, 1156 High Street, Santa Cruz, CA 95064, USA

}\normalsize}
\vspace{12pt}\end{minipage}}

\noindent{\begin{minipage}{\textwidth}
\raggedright
\mbox{Burak} {Bilki},
\mbox{Ed} {Norbeck},
\mbox{Yasar} {Onel}
\vspace{1pt}\\
{\sl \scriptsize{University of Iowa, Department of Physics and Astronomy, 203 Van Allen Hall, Iowa City, IA 52242-1479, USA

}\normalsize}
\vspace{12pt}\end{minipage}}

\noindent{\begin{minipage}{\textwidth}
\raggedright
\mbox{Jadranka} {Sekaric},
\mbox{Brian} {Van~Doren},
\mbox{Graham~W.} {Wilson}
\vspace{1pt}\\
{\sl \scriptsize{University of Kansas, Department of Physics and Astronomy, Malott Hall, 1251 Wescoe Hall Drive, Room 1082, Lawrence, KS 66045-7582, USA

}\normalsize}
\vspace{12pt}\end{minipage}}

\noindent{\begin{minipage}{\textwidth}
\raggedright
\mbox{Haijun} {Yang}
\vspace{1pt}\\
{\sl \scriptsize{University of Michigan, Department of Physics, 500 E. University Ave., Ann Arbor, MI 48109-1120, USA

}\normalsize}
\vspace{12pt}\end{minipage}}

\noindent{\begin{minipage}{\textwidth}
\raggedright
\mbox{Giovanni} {Bonvicini}
\vspace{1pt}\\
{\sl \scriptsize{Wayne State University, Department of Physics, Detroit, MI 48202, USA

}\normalsize}
\vspace{12pt}\end{minipage}}

\clearpage

\begin{minipage}{\textwidth}
 \setlength{\tabcolsep}{0.0cm} 
 \begin{tabular}{lp{14.9cm}} 
{ $^{1}$ } & { \scriptsize{also at TRIUMF, 4004 Wesbrook Mall, Vancouver, BC V6T 2A3, Canada}}  \\ 
 { $^{2}$ } & { \scriptsize{also at Universit{\'e} de Montr{\'e}al, D{\'e}partement de Physique, Groupe de Physique des Particules, C.P. 6128, Succ. Centre-ville, Montr{\'e}al, Qc H3C 3J7, Canada}}  \\ 
 { $^{3}$ } & { \scriptsize{also at Institute of High Energy Physics - IHEP, Chinese Academy of Sciences, P.O. Box 918, Beijing, China 100049}}  \\ 
 { $^{4}$ } & { \scriptsize{also at Helsinki Institute of Physics (HIP), P.O. Box 64, FIN-00014 University of Helsinki, Finland}}  \\ 
 { $^{5}$ } & { \scriptsize{also at Laboratoire de l'Acc\'el\'erateur Lin\'eaire (LAL), Universit\'e Paris-Sud 11, B\^atiment 200, 91898 Orsay, France}}  \\ 
 { $^{6}$ } & { \scriptsize{also at Universit\'e de Strasbourg, UFR de Sciences Physiques, 3-5 Rue de l'Universit\'e, F-67084 Strasbourg Cedex, France }}  \\ 
 { $^{7}$ } & { \scriptsize{also at Brandenburg University of Technology, Postfach 101344, D-03013 Cottbus, Germany}}  \\ 
 { $^{8}$ } & { \scriptsize{also at Deutsches Elektronen-Synchrotron DESY, A Research Centre of the Helmholtz Association, Notkestrasse 85, 22607 Hamburg, Germany (Hamburg site)}}  \\ 
 { $^{9}$ } & { \scriptsize{also at Deutsches Elektronen-Synchrotron DESY, A Research Centre of the Helmholtz Association, Platanenallee 6, 15738 Zeuthen, Germany (Zeuthen site)}}  \\ 
 { $^{10}$ } & { \scriptsize{also at Excellence Cluster Universe, Technische Universit{\"a}t M{\"u}nchen, Boltzmannstr. 2, 85748 Garching, Germany}}  \\ 
 { $^{11}$ } & { \scriptsize{also at Gottfried Wilhelm Leibniz Universit{\"a}t Hannover, Fakult{\"a}t f{\"u}r Mathematik und Physik, Appelstra\ss{}e 2, 30167 Hannover, Germany}}  \\ 
 { $^{12}$ } & { \scriptsize{also at University of Hamburg, Physics Department, Institut f{\"u}r Experimentalphysik, Luruper Chaussee 149, 22761 Hamburg, Germany}}  \\ 
 { $^{13}$ } & { \scriptsize{also at University of Wuppertal, Gau{\ss}stra{\ss}e 20, D-42119 Wuppertal, Germany}}  \\ 
 { $^{14}$ } & { \scriptsize{also at Universit{\"a}t Bonn, Physikalisches Institut, Nu{\ss}allee 12, 53115 Bonn, Germany}}  \\ 
 { $^{15}$ } & { \scriptsize{also at Istituto Nazionale di Fisica Nucleare (INFN), Sezione di Milano, Via Celoria 16, I-20133 Milano, Italy}}  \\ 
 { $^{16}$ } & { \scriptsize{also at Nikhef, National Institute for Subatomic Physics, P.O. Box 41882, 1009 DB Amsterdam, Netherlands}}  \\ 
 { $^{17}$ } & { \scriptsize{also at Vrije Universiteit, Department of Physics, Faculty of Sciences, De Boelelaan 1081, 1081 HV Amsterdam, Netherlands}}  \\ 
 { $^{18}$ } & { \scriptsize{also at University of Lodz, Faculty of Physics and Applied Informatics, Pomorska 149/153, PL-90-236 Lodz, Poland}}  \\ 
 { $^{19}$ } & { \scriptsize{also at National Institute of Physics and Nuclear Engineering ``Horia Hulubei'' (IFIN-HH), Str. Atomistilor no. 407, P.O. Box MG-6, R-76900 Bucharest - Magurele, Romania}}  \\ 
 { $^{20}$ } & { \scriptsize{also at CERN, CH-1211 Gen\`eve 23, Switzerland}}  \\ 
 { $^{21}$ } & { \scriptsize{also at Cockcroft Institute, Daresbury, Warrington WA4 4AD, UK }}  \\ 
 { $^{22}$ } & { \scriptsize{also at Royal Holloway, University of London (RHUL), Department of Physics, Egham, Surrey TW20 0EX, UK }}  \\ 
 { $^{23}$ } & { \scriptsize{also at STFC Rutherford Appleton Laboratory, Chilton, Didcot, Oxon OX11 0QX, UK }}  \\ 
 { $^{24}$ } & { \scriptsize{also at Enrico Fermi Institute, University of Chicago, 5640 S. Ellis Avenue, RI-183, Chicago, IL 60637, USA}}  \\ 
 { $^{25}$ } & { \scriptsize{also at Fermi National Accelerator Laboratory (FNAL), P.O.Box 500, Batavia, IL 60510-0500, USA}}  \\ 
 
\end{tabular}
\end{minipage}

%% file: introduction/introduction.tex
\section{ILD Philosophy}

The {\bf I}nternational {\bf L}arge {\bf D}etector (ILD) is a concept for a detector at the International Linear Collider, ILC. 
The ILC will collide electrons and positrons at energies of initially $500$~GeV, upgradeable to $1$~TeV. The ILC has an 
ambitious physics program, which will extend and complement that of the Large Hadron Collider (LHC). 
The ILC physics case has been well documented, most recently in the ILC Reference Design Report, RDR~\cite{RDR_physics}. 
A hallmark of physics at the ILC is precision. The clean initial state and the comparatively benign environment of a lepton collider 
are ideally suited to high precision measurements. To take full advantage of the physics potential 
of ILC places 
great demands on the detector performance. 
The design of ILD, 
which is based on the GLD~\cite{ref-gld} and the LDC~\cite{ref-ldc} detector concepts, 
is driven by these requirements. Excellent calorimetry and tracking are combined to 
obtain the best possible overall event reconstruction, including  the capability to reconstruct 
individual particles  within jets for particle flow calorimetry.
This requires excellent spatial resolution for all detector 
systems. A highly granular calorimeter system is combined with a central tracker which 
stresses redundancy and efficiency. 
In addition, 
efficient reconstruction of secondary vertices and excellent momentum resolution for charged particles are essential
for an ILC detector. 
The interaction region of the ILC  is designed to host two detectors, which can be 
moved into the beam position with a ``push-pull'' scheme. The mechanical design of ILD and the overall integration of 
subdetectors 
takes these operational conditions into account. 
The main features of ILD are outlined below.

The central component of the ILD tracker is a Time Projection Chamber (TPC) which provides up to 224 precise measurements along the track of a charged particle. This is supplemented by a  system of Silicon (Si) based tracking detectors, which provide additional measurement points inside and outside of the TPC, and  extend the angular coverage down to very small angles. A Si-pixel based
vertex detector (VTX) enables  long lived particles such as b- and c-hadrons to be reconstructed.
 This combination of tracking devices, which has a large degree of 
redundancy, results in high track reconstruction efficiencies, and unprecedented momentum resolution and vertex reconstruction
capabilities.
One of the most direct measures of detector performance at the ILC is the jet-energy resolution. Precise di-jet mass
reconstruction and separation of hadronically decaying $\Wboson$ and $\Zzero$ bosons are essential for many
physics channels.
The ultimate jet energy resolution is achieved when every particle in the event, charged and neutral, is measured 
with the best possible precision. Within the paradigm of particle flow calorimetry, this goal is 
achieved by reconstructing charged particles in
the tracker, photons in the electromagnetic calorimeter (ECAL), and neutral hadrons in
the ECAL and hadronic calorimeter (HCAL). The ultimate performance is reached for perfect
separation of charged-particle clusters from neutral particle clusters in the calorimeters. 
Thus, a highly granular calorimeter outside the tracker is the second key component of ILD. Sampling calorimeters 
with dense absorber material and fine grained readout are used. A tungsten absorber based electromagnetic calorimeter (ECAL)
covers the first interaction 
length, followed by a somewhat coarser steel based sampling hadronic calorimeter (HCAL). Several ECAL and HCAL
readout technologies are being pursued. 

\section{Basic Layout of ILD}
The proposed ILD concept is designed as a multi-purpose detector, which provides excellent precision in spatial and 
energy measurement over a large solid angle. It has the following components:
\begin{itemize}\addtolength{\itemsep}{-0.5\baselineskip}
\item	A multi-layer pixel-vertex detector (VTX), with three super-layers each comprising two layers. To minimise the occupancy from background hits,
the first super-layer is only half as long as the outer two. Whilst the underlying detector technology has not yet been decided, 
the VTX is optimised for excellent point resolution and minimum material thickness. A five layer geometry, VTX-SL, with the layers spaced at equal distances to the 
IP is investigated as an alternative. In either case the vertex detector has a purely barrel geometry.
\item	A system of strip and pixel detectors surrounding the VTX detector. In the barrel, two layers of Si strip detectors (SIT) are arranged to bridge the gap between the VTX and the TPC. In the forward region, a system of Si-pixel and Si-strip disks (FTD) provides low angle tracking coverage.
\item	A large volume time projection chamber (TPC) with up to 224 points per track. The TPC is optimised for excellent 3-dimensional point resolution and minimum material in the field cage and in the end-plate. It also provides d$E$/d$x$ based particle identification capabilities.
\item	A system of Si-strip detectors, one behind the end-plate of the TPC (ETD) and one in between the TPC and the ECAL (SET). 
    These provide additional high precision space points which improve the tracking measurements and provide additional
    redundancy in the regions between the main tracking volume and the calorimeters. 
\item	A highly segmented ECAL providing up to 30 samples in depth and small transverse cell size. Two technology options are considered;  Si-W and scintillator-W.
\item	A highly segmented HCAL with up to 48 longitudinal samples and small transverse cell size. Two 
options are considered, both based on a Steel-absorber structure. One option uses scintillator tiles of $3 \times 3$\,cm$^2$, 
which are read out with an analogue system. The second uses a gas-based readout which allows a $1 \times 1$\,cm$^2$ 
cell geometry with a binary or semi-digital readout of each cell. 
\item	A system of high precision, radiation hard,  calorimetric detectors in the very forward region (LumiCAL, BCAL, LHCAL). These
extend the calorimetric coverage to almost $4\pi$, measure the luminosity, and  monitor the quality of the colliding beams.
\item	A large volume superconducting coil surrounds the calorimeters, creating an axial $B$-field of nominally 3.5\,Tesla.
\item	An iron  yoke, instrumented with scintillator strips or RPCs, returns the magnetic flux of the solenoid, and
  at the same time, serves as a muon filter, muon detector and tail catcher.
\item	A sophisticated data acquisition (DAQ) system which operates without an 
    external trigger, to maximise the physics sensitivity.
\end{itemize}

\begin{figure}
	\centering
		\includegraphics[height=8cm]{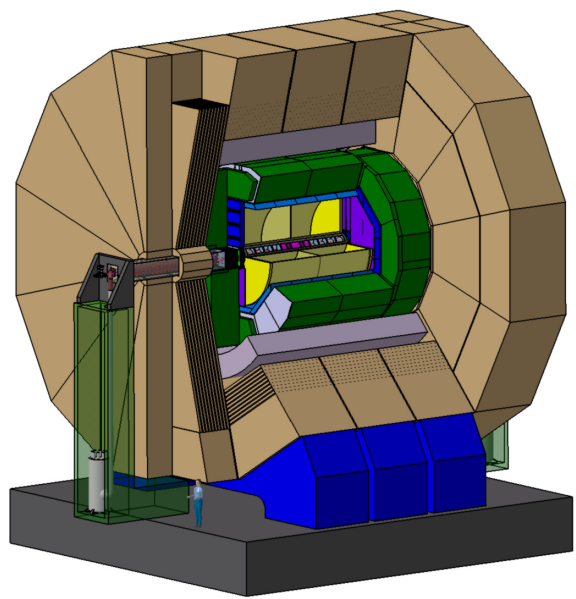}
	\caption{View of the ILD detector concept.}
	\label{fig:intro:detector}
\end{figure}

Precision physics at the ILC requires that the beam parameters are known with great accuracy. 
The beam energy and the beam polarization will be measured in small dedicated experiments, which are shared by the two detectors 
present in the interaction region. These detectors will only be covered briefly in this document, more details may be found in a 
dedicated document. The luminosity of the interaction will be measured by the luminometers integrated in ILD. 
To enable the operation of the detector in a ''push-pul'' scenario, the complete detector is mounted on a movable platform, which can move sideways out of the beam to make space for the second detector in the interaction region. The platform ensures that 
the integrity and calibration of the detector is minimally disturbed during the moving process, making the 
re-commissioning of the detector after the "push-pull" operation easier. 
The ILD detector concept is shown graphically in Figure~\ref{fig:intro:detector}.

\section{Performance Requirements}
The requirements and resulting challenges for detectors at the ILC are described in the 
ILC RDR~\cite{RDR_detector}. 
The ILC 
is designed to investigate in detail the mechanism of the electroweak symmetry breaking, and to search for and 
study new physics at energy scales up to 1\,TeV. In addition, the collider will provide a 
wealth of information on Standard Model (SM) physics, for example top physics, heavy flavour physics, and physics of the 
$\Zzero$ and $\Wboson$ bosons. 
The requirements for a detector are, therefore, that multi-jet final states, typical for many physics channels, 
can be reconstructed with high accuracy. The jet energy resolution should be sufficiently good that 
the hadronic decays of the $\Wboson$ and $\Zzero$ can be separated. This translates into a jet 
energy resolution of $\sigma_E/ E \sim 3 - 4 \%$
(equivalent to $30\%/ \sqrt {E}$ at energies below $100$\,GeV). 
This requirement is one of the most challenging for ILD and has a large 
impact on the design of the calorimeters. It also impacts  the way the tracking system is optimised. 
Nevertheless, the reconstruction of events with high precision benefits the ILD physics programme in several ways.  A more 
precise detector will result  in smaller systematic errors for many measurements, and thus 
will extend the ultimate physics reach of the ILC. In addition,  
a more precise detector implies that the luminosity delivered by the collider is 
used more efficiently, making it possible to reduce the overall running costs of the facility to reach 
a particular accuracy. 

It is difficult to anticipate the full physics programme at a new facility before the physics of the energy 
regime where it will operate is known. A detector for the ILC therefore needs to be designed 
and optimised not only in view of a limited set of benchmark reactions, but also to be as versatile and 
as flexible as possible.
Nevertheless the ILC community has defined a number of challenging benchmark physics reactions, which, to the best of our 
current knowledge, will form an important part of the physics programme at the ILC. 
The benchmark reactions stress the study of the Higgs boson and Supersymmetry (SUSY) as a model for a possible 
new physics scenario at the ILC. They also rely on excellent lepton and flavour tagging
and probe the  missing energy measurement capability of the detector. These reactions give only a
flavour of the physics reach of the ILC. 

\section{Overview of the ILD Letter of Intent}

The signatories of this Letter of Intent are an international group of physicists with strong participation from Asia, Europe and the Americas. 
With this letter of intent the undersigned express their intention to develop further the ILD concept to a point where a concrete proposal 
can be made. However, at this stage, this does not represent a firm commitment either in terms of manpower or 
resources.

This document is organised as follows: Chapter~\ref{optimization} describes the studies used to 
optimise and define the main parameters of the ILD concept; Chapter~\ref{performance} presents 
the performance of the ILD concept in terms of the low level detector response, such as momentum resolution,
jet energy resolution, and flavour-tagging performance.
It also describes a number of physics studies which demonstrate that the ILC concept is 
well optimised for physics at the ILC operating in the centre-of-mass energy range 200\,GeV to 1\,TeV;
Chapter~\ref{ILD} describes the ILD subdetector systems in the context of the ongoing R\&D programme;
Chapter~\ref{integration} describes the interface between ILD and the ILC;
Chapter~\ref{cost} discusses the current understanding of the likely cost of ILD;
and finally, Chapters~\ref{ILDgroup} and~\ref{randplan} describe the structure of the ILD
group and the necessary R\&D needed to realise this project.

%% file: optimization/optimisation.tex

\label{sec:optimization}
\label{sec:optimization-introduction}
\input{optimization/introduction}

\section{Simulation Tools and Detector Parameters}
\label{sec:optimization-simulation}
\input{optimization/simulation}

\section{Detector Optimisation for Particle Flow }
\label{sec:optimization-particleflow}
\input{optimization/particleflow}

\section{Background Considerations}
\label{sec:optimization-background}
\input{optimization/background}

\section{Detector Optimisation for Tracking}
\label{sec:optimization-tracking}
\input{optimization/tracking}

\section{Flavour Tagging}
\label{sec:optimization-flavour}
\input{optimization/flavour}

\section{Physics Performance}
\label{sec:optimization-physics}
\input{optimization/physics}

\section{Conclusions}
\label{sec:optimization-conclusions}
\input{optimization/conclusions}

\section{Choice of ILD Parameters}
\label{sec:optimization-selection}
\input{optimization/selection}

%% file: optimization/introduction.tex
The choice of the main parameters
of the ILD, such as the magnetic field, $B$, 
and overall size, is motivated by extensive simulation studies 
based on variants of the GLD~\cite{ref-gld} and LDC~\cite{ref-ldc} 
detector concepts. 
The main studies, described in the following sections, are of: 
i) the performance of particle flow calorimetry in terms of 
jet energy resolution; 
ii) the tracking performance for momentum resolution 
    and impact parameter resolution;
iii) the beam-related backgrounds and the impact of the choice of 
    $B$; 
iv) the efficiency and purity of heavy flavour tagging; 
v) and the impact on physics performance in three benchmark processes.

Ideally the overall detector cost would feed directly 
into the optimisation of the ILD detector. However, because of the 
large uncertainties in the cost of raw materials and detector 
sensors, it is felt that the approach of optimising the detector 
performance for a fixed cost is not reliable at this stage.  
Hence, whilst cost is a consideration in defining the 
parameters of the ILD concept, the main criterion is to develop a 
detector concept optimised for physics at the ILC.

%% file: optimization/simulation.tex

The optimisation of the ILD concept  was performed in parallel using 
the software tools developed by the GLD and the LDC 
groups. 
The detector models were simulated using a fairly detailed 
GEANT4~\cite{ref:geant4} simulation. A significant effort has been made 
to use a reasonable geometry for the subdetectors, including a 
description of dead regions and support structures, as
described in in Section~\ref{sec:performance-simulation}. The 
studies presented are based on full reconstruction of the simulated 
events without reference to the Monte Carlo (MC) truth information.

Six detector models were defined; 
three based on the GLD simulation (GLD, GLDPrime and GLD4LDC) 
and three based on the LDC simulation (LDC, LDCPrime, 
and LDC4GLD). 
The main parameters of the models are summarised in 
Table~\ref{tab:optdetparameter}. The models 
represent different compromises between magnetic 
field and TPC outer radius. The software frameworks 
(JSF/Jupiter/Satellites and Mokka/Marlin) used to simulate and reconstruct
the detector models are summarised below. 
The detector simulation was performed using GEANT4
(version 9.1 patch01) with the {\tt LCPhysics} physics 
list~\cite{ref:lcphys}. 

\begin{table}
\begin{center}
{\tabcolsep=0.07cm
\begin{tabular}{|l|l||c|c|c|c|c|c||c|} \hline
\multicolumn{2}{|l||}{Model Name} 
& GLD & GLD$^\prime$ & GLD4LDC & LDC4GLD & LDC$^\prime$ & LDC & ILD \\ \hline
\multicolumn{2}{|l||}{Simulator}
& \multicolumn{3}{|c| }{Jupiter} & \multicolumn{3}{|c||}{Mokka} & Mokka \\ \hline
\multicolumn{2}{|l||}{B field (T)} 
&  3.0 & 3.5 & 4.0 & 3.0 & 3.5 & 4.0 & 3.5 \\ \hline
\multicolumn{2}{|l||}{Beampipe R$_{min}$}
 & 15.0 & 14.0 & 13.0 & 15.5 & 14.0 & 13.0 & 14.5 \\ \hline
Vertex & 
 Geometry & \multicolumn{3}{|c|}{cylindrical} & \multicolumn{3}{|c||}{ladders}
 & ladders \\ \cline{2-9}
 Detector & Layers & \multicolumn{3}{|c|}{3 doublets} & \multicolumn{3}{|c||}{5}
 &3 doublets\\ \cline{2-9}
 &
R$_{min}$& 17.5 & 16.0 & 15.0 & 16.5 & 15.0 & 14.0 & 16.0 \\ \hline
Barrel&
Layers & \multicolumn{3}{|c|}{4 cylinders} & \multicolumn{3}{|c||}{2 cylinders} & 2 cylinders\\ \cline{2-9}
SIT & Radii & \multicolumn{3}{|c|}{90, 160, 230, 300} & \multicolumn{3}{|c||}{161.4, 270.1} & 165, 309\\ \hline
TPC & R$_{min}$ &  437 &  435 &  371 & \multicolumn{3}{ |c|| } {371 } & 395 \\ \cline{2-9}
drift & R$_{max}$ & 1978 & 1740 & 1520 & 1931 & 1733 & 1511 & 1739 \\ \cline{2-9} 
region &  $z_{max}$ & ~~2600~~ & ~~2350~~ & ~~2160~~ & ~~2498~~ & ~~2248~~ & ~~2186~~ & ~~2247.5~~ \\ \hline
\multicolumn{2}{|l||}{TPC pad rows} & 256 & 217 & 196 & 260 & 227 & 190 & 224 \\ \hline
ECAL & R$_{min}$ & 2100 & 1850 & 1600 & 2020 & 1825 & 1610 & 1847.4\\ \cline{2-9}
barrel & Layers & \multicolumn{3}{|c|}{33} &  \multicolumn{3}{|c||}{20(thin)+9(thick)} & 20+9\\ \cline{2-9}
 & Total $X_0$ & \multicolumn{3}{|c|}{28.4} &  \multicolumn{3}{|c||}{22.9} & 23.6 \\ \hline
\multicolumn{2}{|c||}{ECAL endcap z$_{min}$} & 2800 & 2250 & 2100 &  2700 & 2300 & 2550 & 2450 \\ \hline
HCAL & Layers & 46 & 42 & 37  & \multicolumn{3}{|c||}{48} & 48 \\ \cline{2-9}
barrel  & R$_{max}$ & 3617 & 3260 & 2857 & 3554 & 3359 & 3144 & 3330 \\ \hline
\multicolumn{2}{|l||}{$\lambda_I$ (ECAL+HCAL)} & 6.79 & 6.29 & 5.67 &  \multicolumn{3}{|c||}{6.86} & 6.86 \\ \hline 
\end{tabular}
}
\caption[Detector parameters for the optimisation studies.]{Geometrical parameters of the baseline detector models used for 
         the optimisation studies
         (GLD, GLDPrime, GLD4LDC, LDC4GLD, LDCPrime and LDC). Also shown are
     the corresponding parameters for the ILD baseline detector.  
     Unless otherwise specified, values are shown in units of mm.
\label{tab:optdetparameter}}
\end{center}
\end{table}

\subsection{GLD Software: JSF}

\label{sec:optimisation-simulation-jsf}

JSF~\cite{ref:jsf} is a ROOT~\cite{ref:ROOT} based software framework for modular applications, 
such as event generation, fast and full simulation, 
event reconstruction and data analysis. The two main components are
Jupiter~\cite{ref:jupiter} and Satellites~\cite{ref:jupiter}.
Jupiter is a GEANT4 based detector
simulation, designed to enable easy installation and modification of subdetector 
components. Satellites is a collection of event 
reconstruction modules in the JSF framework. Satellites includes 
smearing of hit points simulated by Jupiter, a ``cheated'' track finder
using MC information to associate hits to tracks, 
a Kalman Filter based track fitter, and a cheated particle 
flow algorithm (PFA).

Jupiter reads a set of detector parameters at run time from a text file,
which makes it easy to study different detector configurations.
The geometry information is saved in an output ROOT file for use by event reconstruction.
In the Jupiter detector simulation the vertex detector and 
intermediate silicon trackers are modelled as cylinders.  
The calorimeters have a 12-fold symmetry.
The electromagnetic calorimeter consists of a sandwich structure 
comprising layers of 
3\,mm of tungsten absorber, 2\,mm of scintillator, and a 1\,mm air gap.
The hadron calorimeter consists of layers comprising of 
20\,mm of iron absorber,  
5\,mm of plastic scintillator, and a 1\,mm air gap.
For the purpose of simulation, the scintillator in both the ECAL and 
HCAL is segmented into 
$1\times1$\,cm$^{2}$ readout tiles. Signals in these tiles 
can be combined at the time 
of reconstruction to simulate the strip readout structure of proposed system.
In the version of the simulation used for the studies presented here,
there is no gap between the ECAL and HCAL. 
Jupiter was executed as a module of the JSF and GEANT4 hits in each 
sensitive detector are saved in a ROOT file for subsequent study with the 
Satellites package or as an LCIO~\cite{ref:lcio}
file for reconstruction with MarlinReco~\cite{ref:MarlinReco}.  

The point resolution of the tracking chambers was implemented in the 
Satellites reconstruction. The GEANT4 hit points in the vertex detector (VTX) 
and intermediate silicon tracker (IT) were smeared with a Gaussian with 
the following resolutions. For the
VTX,  $\sigma_{r\phi}$ and $\sigma_{z}$ were taken to be 2.8$\mu$m. 
For the barrel silicon tracker, a resolution of 10\,$\mu$m was used for 
both $\sigma_{r\phi}$ and $\sigma_{z}$.
The TPC space points were smeared by Gaussian resolutions, $\sigma_{r\phi}$ and $\sigma_{z}$, given by the following physically motivated form:
\begin{eqnarray*}
\sigma^2_{r\phi}/\mu\mathrm{m}^2 &=&50^2+900^2\sin^2\phi + \left( (25^2/22)\times (4/B)^2\sin\theta\right) z;\\
\sigma^2_{z}/\mu\mathrm{m}^2 &=&400^2+80^2\times z;
\end{eqnarray*}
where $z$ is the drift length in cm,  $B$ is the magnetic field strength 
in Tesla, and $\theta$  and $\phi$ are the track angles with 
respect to the axes perpendicular to the readout plane and perpendicular 
to the pad rows, and the resolutions are given in $\mu$m.
For the calorimeter hits, no additional smearing is applied at 
reconstruction time; the simulated energy deposits in the 
scintillator tiles are used directly.

The optimisation studies in the GLD framework use the Satellites reconstruction
to investigate tracking performance and MarlinReco
for other studies. 
The interoperability 
between the two software frameworks is provided by the LCIO data format, 
{\it e.g.} after simulating the detector response with the Jupiter program, 
MarlinReco and PandoraPFA~\cite{PandoraPFA} were used for the event reconstruction.

\subsection{LDC Software: Mokka and Marlin}

\label{sec:optimisation-simulation-mokkamarlin}

The software framework developed by the LDC concept is based on the LCIO persistency format and event data model. The detailed simulation of the 
detector response is performed by the GEANT4 based
Mokka~\cite{ref:mokka} application.  The detailed subdetector geometries and 
component materials
are stored in a MySQL database. The overall detector
is then built from individual subdetectors, making it relatively straightforward 
to compare different technology choices. 
The corresponding {\tt C++} code instantiating the subdetector geometry in memory
is written such that the whole detector model can be scaled in length and radius; this
feature proved invaluable in optimising the detector geometry.
The GEAR~\cite{ref_gear} package 
provides access to these geometrical detector properties 
at the reconstruction and analysis level. The Mokka simulation of the different
subdetectors is described in more detail in Section~\ref{sec:performance-simulation}.

The Mokka generated events are processed in Marlin~\cite{Marlin}. Marlin is a 
modular {\tt C++} application framework which supports plug-in modules (called processors)
which can be loaded at runtime.
This plug-in-based design supports the distributed 
development of reconstruction algorithms and also allows comparison of
different algorithms at runtime, {\it e.g.} it is possible to run two
tracking algorithms producing parallel collections of reconstructed
tracks.

Event reconstruction is performed with the MarlinReco~\cite{ref-EUDET-Marlin} package. This consists of a set of
modules for digitisation, track finding, track fitting, 
particle flow reconstruction, and flavour tagging.
The hit smearing for the tracking detectors is implemented at the digitisation stage using the
same parameterisation as used for Satellites, except that resolutions for the
intermediate silicon tracker (SIT) are taken to be 4\,$\mu$m for $\sigma_{r\phi}$ and 
50\,$\mu$m for $\sigma_{z}$. The pattern recognition processors use Kalman Filter techniques 
and code developed for the LEP experiments. Tracks from standalone pattern recognition in the 
silicon trackers and in the TPC are combined and refitted, The resulting 
momentum resolution is discussed in Section~\ref{sec:performance-detector-tracking}.
Reconstruction of the individual particles in the event is performed with the particle flow algorithm
in the PandoraPFA~\cite{PandoraPFA}  package, currently the
best algorithm available.
The LCFIVertex~\cite{ref:lcfivertex} package provides sophisticated code for vertex finding/fitting and for the identification of heavy flavour jets 
using a neural network approach. It also provides jet charge estimation.
In addition to reconstruction algorithms, MarlinReco includes a set of analysis
tools such as algorithms for jet finding and kinematic fitting. The RAVE
toolkit~\cite{RAVE}, also available
within Marlin, provides an alternative set of vertex reconstruction based on
linear and non-linear estimators.

%% file: optimization/particleflow.tex
One of the main design considerations for a detector at the ILC is the 
ability to efficiently identify and distinguish $\Zzero\rightarrow\qq$ 
and $\Wboson\rightarrow\qq$ decays. This imposes the requirement that the
di-jet mass resolution should be comparable to the natural widths of the electroweak 
gauge bosons, $\sigma_m/m < 2.7\,\% \approx\GammaZ/\mZ\approx\GammaW/\mW$. 
In terms of jet energy resolution this requirement approximately corresponds 
to $\sigma_E/E < 3.8\,\%$. After accounting for the gauge boson widths, this
results in a $\sim2.75$ standard deviation separation of the 
$W$ and $\Zzero$ mass peaks for di-jet events. 
Most of the interesting physics at the ILC,
operating in the centre-of-mass range $\roots=0.5-1.0$\,TeV, will consist 
of final states with four or more fermions and for processes
near threshold, the gauge bosons will decay almost at rest. 
Hence the typical di-jet 
energies of interest will be in the range $80-350$\,GeV. This sets the
requirement on calorimetric performance of $\sigma_E/E \sim 30\,\%/\sqrt{E}$.
It has been demonstrated that one way of reaching this goal is
particle flow calorimetry\cite{Thomson:2007xb}. 
Whilst, the separation of $W$ and $\Zzero$ bosons defines the {\it minimum} 
requirement for the jet energy resolution,  
it should be remembered that di-jet invariant masses 
will be an important part of the event selection 
for many physics analyses; the jet energy resolution will 
affect the signal-to-background ratio in many analyses.

The ILD concept is based on the belief that particle flow 
calorimetry provides the best way of achieving the ILC jet energy
goals. Particle flow reconstruction places strong requirements on the 
subdetector technologies and the overall detector design. Particle flow calorimetry requires
efficient separation of photons and showers produced by neutral hadrons from
showers produced by the interactions of charged hadrons. 
This implies high granularity calorimetry and that both 
the ECAL and HCAL lie inside the detector solenoid. 
For high energy jets, failures in the ability 
to efficiently separate energy deposits from different particles,
the {\it confusion} term, will dominate the jet energy resolution.
The physical separation of calorimetric energy deposits from different 
particles will be greater in a large detector, scaling as the inner radius
of the ECAL, $R$, in the barrel region and the detector length, $L$, 
in the endcap region. 
There are also arguments favouring a high magnetic field, as this will tend 
to deflect charged particles away from the core of a jet. The scaling law
here is less clear. The separation between a charged particle 
and an initially collinear neutral particle will scale as $BR^2$. 
However, there is no reason to believe that this will hold for a jet of 
(non-collinear) neutral and charged particles. The true dependence 
of particle flow on the overall detector parameters ($B$ and $R$)
has to be evaluated empirically.

\subsection{Particle Flow Optimisation Methodology}

The particle flow optimisation studies for ILD use
the PandoraPFA algorithm\cite{PandoraPFA} to reconstruct events for both 
the LDC and the GLD detector models. All studies are based on full 
reconstruction of the tracking and the calorimetric information. 
The starting point for the 
optimisation studies is the LDCPrime model with a 3.5\,T magnetic field,
an ECAL inner radius of 1825\,mm and a 48 layer (6$\lambda_I$) 
HCAL. The ECAL and HCAL transverse segmentations are $5\times5$\,mm$^2$ 
and $3\times3$\,cm$^2$ respectively. The studies use variations 
of this model where
(usually) a single parameter is changed and the dependence of
jet energy resolution is determined as a function of this parameter.
For each model variation, particle flow performance was
evaluated using samples of approximately 10000 
$\Zzero\rightarrow\qq$ events (only light quarks, {\it i.e.} $q=u,d,s$)
generated with the $\Zzero$ decaying at rest (no ISR or
beamstrahlung) with 
$E_\Zzero =$ 91.2, 200, 360, and 500\,GeV. These jet energies are typical of
those expected at the ILC for $\roots = 0.5-1.0$\,TeV. For each set of events, 
the $\rmsn$ of the total reconstructed energy distribution was
determined, where $\rmsn$ is the root-mean-squared deviation from the mean
in the smallest energy range containing
90\,\% of the reconstructed events. 


\subsection{HCAL Depth}

Good particle flow calorimetry requires that both the ECAL and HCAL are within the
detector solenoid. Consequently, in addition to the cost of the HCAL, 
the HCAL thickness impacts the cost of the overall detector through the
radius of the superconducting solenoid. The thickness of the HCAL determines
the average fraction of jet energy contained within the calorimeter system.
The impact of the HCAL thickness on the particle flow performance is assessed by 
changing the number of HCAL layers in the LDCPrime model from 32 to 63.
This corresponds to a variation of $4.0-7.9$\,$\lambda_I$
($4.8-8.7$\,$\lambda_I$) in the HCAL (ECAL+HCAL).

\begin{figure}[htb]
\begin{center}
\includegraphics[width=9cm]{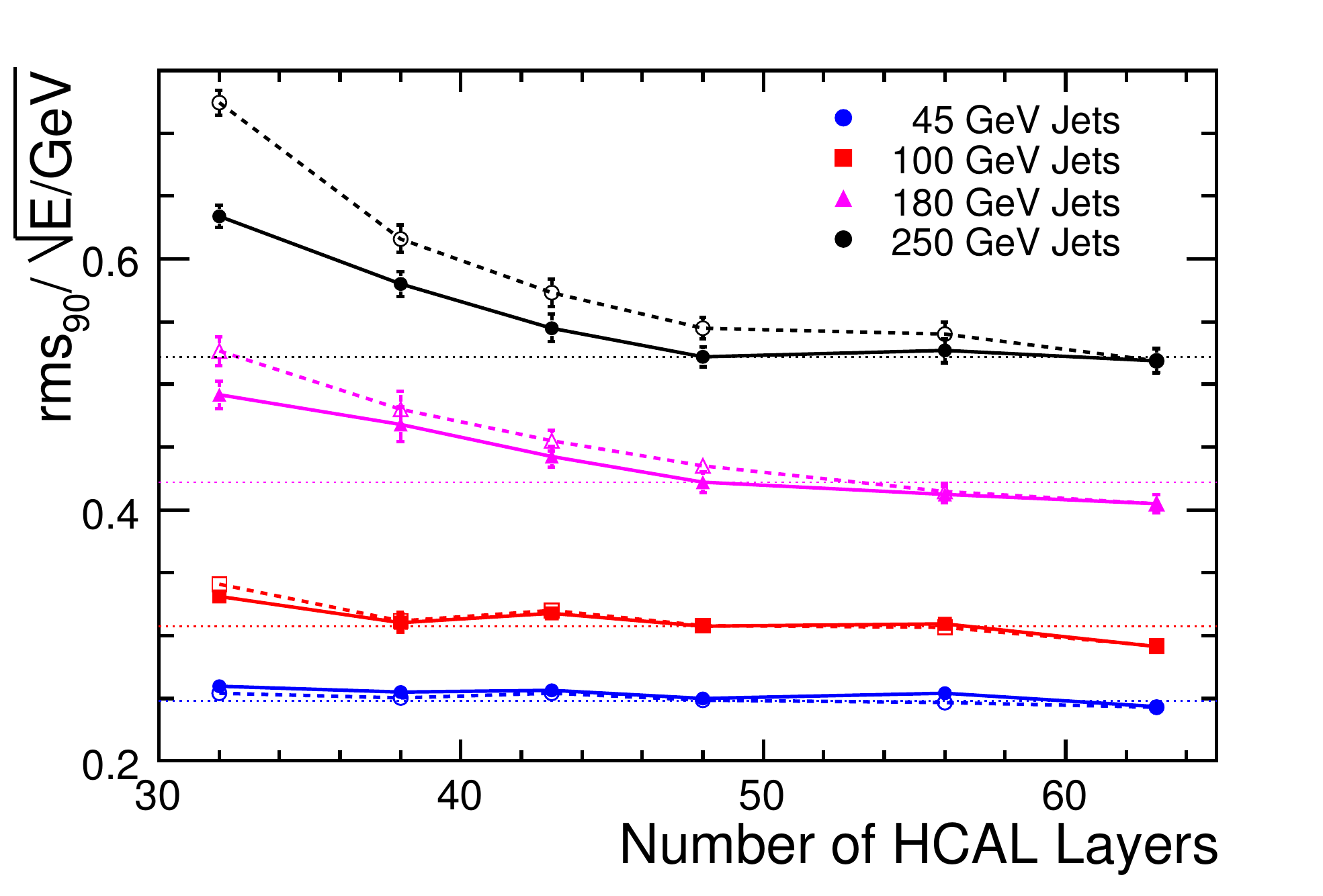}
 \caption[Jet energy resolution versus number of HCAL layers.]{Jet energy resolutions ($\rmsn$) for the LDCPrime detector model with different numbers
          of HCAL layers. Results are shown with (solid markers) and without (open markers)
          taking into account
          energy depositions in the muon detectors. All results are based on 
         $\Zzero\rightarrow\uubar,\ddbar,\ssbar$ with generated polar angle in the
         barrel region of the detector,  $|\cos\theta_{\qq}|<0.7$.
  \label{fig:pfa_hcal}}
\end{center}
\end{figure}

The study of the optimal HCAL thickness depends on the possible
use of the instrumented return yoke (the muon system) to correct for 
leakage of high energy showers out of the rear of the HCAL. The 
effectiveness of this approach is limited by the fact that,
for much of the polar angle, the muon system is behind the relatively
thick solenoid ($2\lambda_I$ in the Mokka simulation of the detector). 
Nevertheless, to assess the possible impact of using the
muon detector as a ``tail-catcher'', the energy depositions in the muon 
detectors were included in the PandoraPFA reconstruction. Whilst 
the treatment could be improved upon, it provides an estimate 
of how much of the degradation in jet energy resolution 
due to leakage can be recovered in this way. The results are 
summarised in Figure~\ref{fig:pfa_hcal} which shows the jet energy resolution 
obtained from PandoraPFA as a function of HCAL thickness.
The effect of leakage is
clearly visible, with about half of the degradation in resolution being recovered
when including the muon detector information. For jet energies of 100\,GeV or less,
leakage is not a major contributor to the jet energy resolution provided the 
HCAL is approximately $4.7\lambda_I$ thick (38 layers). However, for 
$180-250$\,GeV jets this is not sufficient; for leakage not to contribute
significantly to the jet energy resolution at $\roots=1$\,TeV, the results in 
Figure~\ref{fig:pfa_hcal} suggest that the HCAL thickness should be 
between $5.5-6.0\lambda_I$ ($43-48$ layers). To allow for uncertainties
in the simulation of the longitudinal development of hadronic showers,
and to ensure the detector is appropriate for collisions at 1\,TeV, a
48 layer HCAL was chosen for ILD. 
This was also used for the 
LDC-based models discussed below. 

\subsection{Magnetic Field versus Detector Radius}

\begin{table}[Bb]
\begin{center}
\begin{tabular}{lcc|cccc}  
\multicolumn{3}{c|}{Model}     & \multicolumn{4}{c}{$\sigma_E/E$ [\%] versus $E_\mathrm{jet}$}\\   
Name & $B$/T & $\rECAL$/m & 45\,GeV  & 100\,GeV  & 180\,GeV  & 250\,GeV  \\ \hline
SiD-like & 5.0 &  1.25       & $4.19\pm0.06$ & $3.72\pm0.06$ & $3.70\pm0.07$  & $3.94\pm0.10$    \\ 
Small    &4.5 &  1.42       & $3.90\pm0.08$ & $3.34\pm0.07$ & $3.54\pm0.06$  & $3.75\pm0.08$    \\
LDC      &4.0 &  1.60       & $3.82\pm0.06$ & $3.14\pm0.06$ & $3.26\pm0.08$  & $3.37\pm0.07$    \\
LDCPrime &3.5 &  1.82       & $3.70\pm0.06$ & $3.07\pm0.05$ & $3.15\pm0.07$  & $3.30\pm0.06$    \\
LDC4GLD  &3.0 &  2.02       & $3.60\pm0.05$ & $2.97\pm0.05$ & $3.16\pm0.06$  & $3.32\pm0.06$    \\
\end{tabular}
\caption[Jet energy resolution.]{Jet energy resolutions ($\rmsn$) for different detector parameters.
         All results are based on 
         $\Zzero\rightarrow\uubar,\ddbar,\ssbar$ events using scaled versions of the Mokka LDCPrime 
         detector  model. The results are quoted for the barrel region of the detector         $|\cos\theta_{\qq}|<0.7$. }
\label{tab:pfa_models}
\end{center}
\end{table}

The dependence of particle flow performance on $B$ and $R$ is studied
in the region of parameter space close to the LDCPrime model. 
The LDCPrime model assumes a magnetic field of
3.5\,T and an ECAL inner radius of 1820\,mm. A number of variations 
on these parameters were studied: 
i) variations of both $B$ and $\rECAL$ with
four sets of parameters considered, 
``LDC-like'' ($B$=4.0\,T, $\rECAL=1600$\,mm), 
``GLD-like'' ($B$=3.0\,T, $\rECAL=2020$\,mm), 
``Small'' ($B$=4.5\,T, $\rECAL=1420$\,mm), and
``SiD-like'' ($B$=5.0\,T, $\rECAL=1280$\,mm);
i) variations in the ECAL inner 
radius from $1280-2020$\,mm
with $B=3.5$\,T; and iii) variations in $B$ from $2.5-4.5$\,T with 
$\rECAL=1825$\,mm. In total thirteen sets of parameters 
were considered spanning a wide range of
$B$ and $\rECAL$. In each case particle flow performance was
evaluated for 45, 100, 180, and 250\,GeV jets.
Table~\ref{tab:pfa_models} compares the jet energy resolutions for LDC, LDCPrime
and LDC4GLD models. The differences between these models
is small, $\sim5\,\%$,. This is not surprising; the parameters of the 
LDC and GLD concepts on which these models are based
were chosen such that the smaller detector radius is compensated
by a higher $B$. For 
the two smaller, higher $B$ models listed in Table~\ref{tab:pfa_models} 
degradations in performance are observed.

\begin{figure}[bt]
\begin{center}
\includegraphics[width=7.0cm]{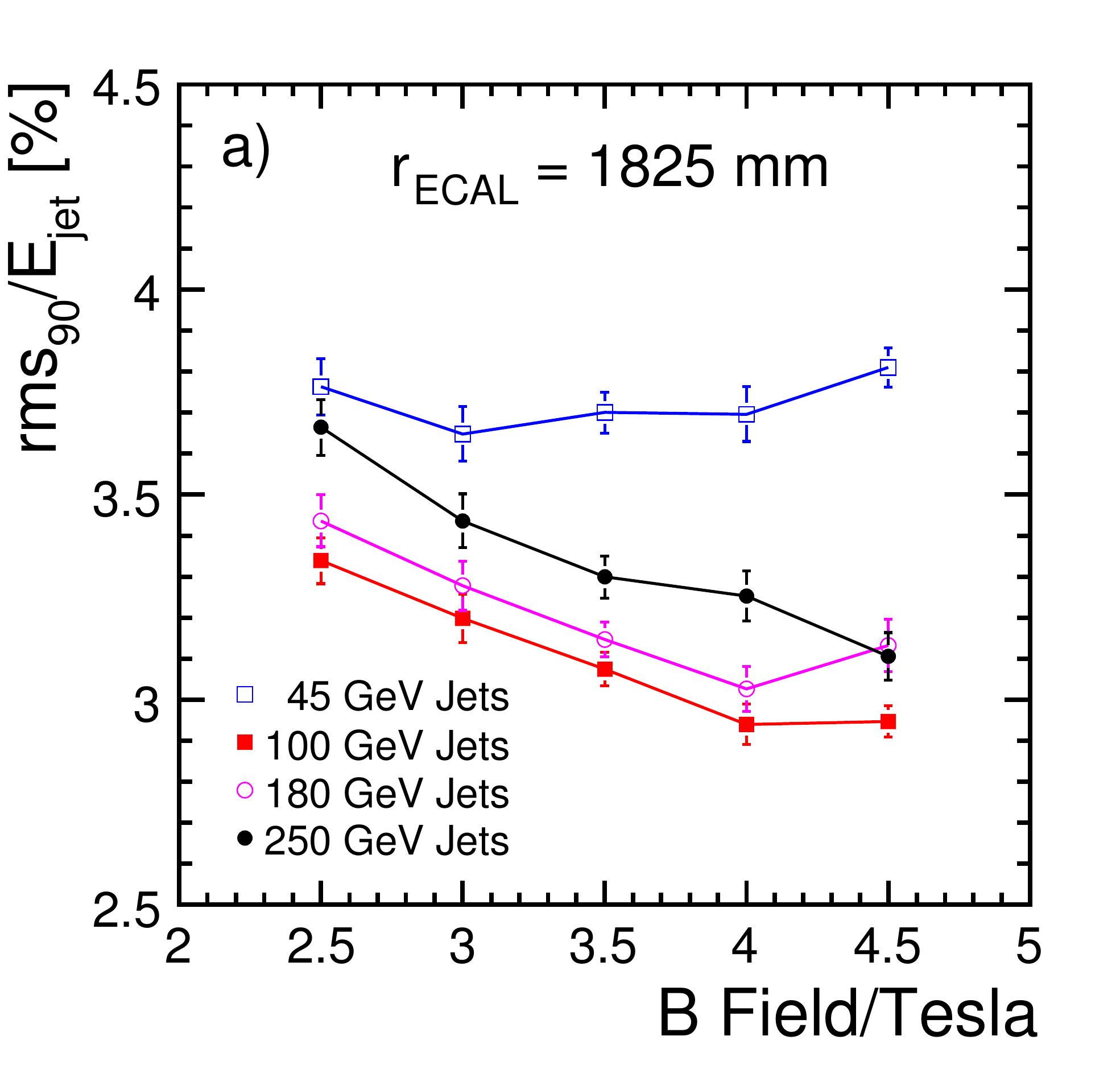}
\includegraphics[width=7.0cm]{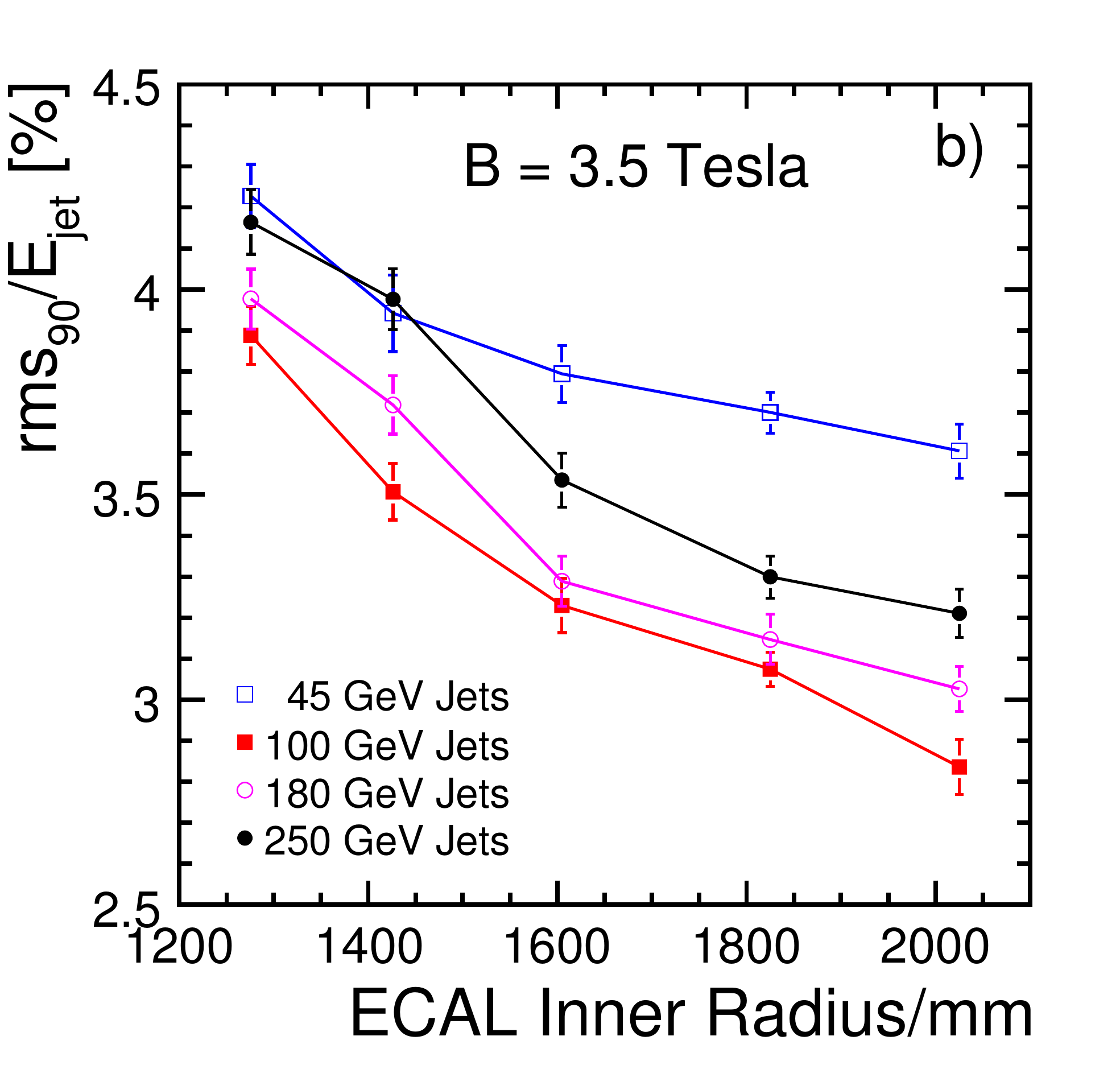}
 \caption[Dependence of jet energy resolution on B-field and ECAL inner radius.]{ a) the dependence of the jet energy resolution ($\rmsn$) on 
             the magnetic field for a fixed ECAL inner radius 
             ($B$=3.5\,T corresponds to the LDCPrime model). b) 
               the dependence of the jet energy resolution ($\rmsn$) on the ECAL
               inner radius a fixed value of the magnetic field 
             ($\rECAL$=1825\,mm corresponds to the LDCPrime model). 
 \label{fig:pfa_b_versus_r}}
\end{center}
\end{figure}

Figure~\ref{fig:pfa_b_versus_r} shows the dependence of
the jet energy resolution 
($\rmsn/\mathrm{E}_{\mathrm{jet}}$) on: a) magnetic field (fixed $\rECAL$)
and b) ECAL inner radius (fixed $B$) for four different jet energies. For 45\,GeV jets, the
dependence of the jet energy resolution on $B$ and $\rECAL$ is 
weak; for these energies the intrinsic calorimetric
energy resolutions, rather than the confusion term dominates. For
higher jet energies, where the confusion term dominates, 
the jet energy resolution shows a stronger dependence on $\rECAL$ 
than $B$.  

The jet energy resolutions listed in Table~\ref{tab:pfa_models} and those
shown in Figure~\ref{fig:pfa_b_versus_r} are reasonably well described by the 
function:
\begin{eqnarray*}
     \frac{\sigma_E}{E} & = & \frac{21}{\sqrt{E/\mathrm{GeV}}} 
           \oplus 0.7 
           \oplus 0.004 E 
           \oplus 2.1 
                   \left(\frac{\rECAL}{1825\,\mathrm{mm}}\right)^{-1.0}
                   \left(\frac{B}{3.5\,\mathrm{T}}\right)^{-0.3}
                   \left(\frac{E}{100\,\mathrm{GeV}}\right)^{0.3} \, \%.
\end{eqnarray*}
This is the quadrature sum of four terms: 
i) the estimated contribution to the jet energy resolution from the intrinsic calorimetric resolution; 
ii) the contribution from imperfect track reconstruction, estimated by comparing the jet energy resolutions with 
    those using tracks obtained from the MC information;
iii) leakage, estimated by comparing the jet energy resolutions with those for an 8\,$\lambda_I$ 
 HCAL; and 
iv) the contribution from confusion obtained empirically from a fit to the data of Table~\ref{tab:pfa_models} 
    and Figure~\ref{fig:pfa_b_versus_r}. 
In fitting the confusion term, a power-law $\kappa B^\alpha \rECAL^\beta E^\gamma$ 
provides a reasonable parameterisation of the data
\footnote{The majority of the data
points lie within 2.5$\sigma$ of the parameterisation, the only exception being the
45\,GeV and 100\,GeV jet energy resolutions for the ``SiD-like'' detector where 
the fit underestimates
the resolution.}. 
From the perspective of the optimisation of the detector, these studies show that 
for the particle flow calorimetry using the PandoraPFA algorithm, that the confusion
term scales as approximately $B^{-0.3}R^{-1}$. For particle flow
performance (with the PandoraPFA algorithm) the detector radius is more important 
than the magnetic field. This forms part of the motivation for the choice of a large detector radius
for the ILD conceptual design. 
Table~\ref{tab:pfa_comparison}
lists the relative values of $B^{0.3}R$ and relative jet energy resolutions 
from the parameterisation above for the LDC, LDCPrime and 
LDC4GLD detector models. The main conclusion of this study is that,
in terms of particle flow performance, the differences between
the LDC, LDCPrime, and LDC4GLD detector models are at the level of
$\pm5$\,\%, with the larger models being slightly prefered.     
\begin{table}[htb]
\begin{center}
\begin{tabular}{lcc|c|cccc}  
\multicolumn{3}{c|}{Model}     & $B^{-0.3}R^{-1}$   & \multicolumn{4}{c}{Relative $\sigma_E/E$ versus $E_{\mathrm{jet}}$} \\   
Name & $B$/T & $\rECAL$/m & (relative)   & 45\,GeV & 100\,GeV & 180\,GeV  & 250\,GeV   \\ \hline
LDC & 4.0 &  1.60        & 1.08 & 1.02 & 1.04 & 1.05 & 1.06   \\
LDC4GLD & 3.0 &  2.02        & 0.95 & 0.99 & 0.97 & 0.96 & 0.96   \\ \hline
\end{tabular}
\caption[Relative jet energy resolutions.]{Expected jet energy resolutions ($\rmsn$) of the LDC and LDC4GLD detector models
         relative to the LDCPrime resolution.}
\label{tab:pfa_comparison}
\end{center}
\end{table}

\subsection{Detector Aspect Ratio}

\begin{figure}[bht]
\begin{center}
\includegraphics[width=7.0cm]{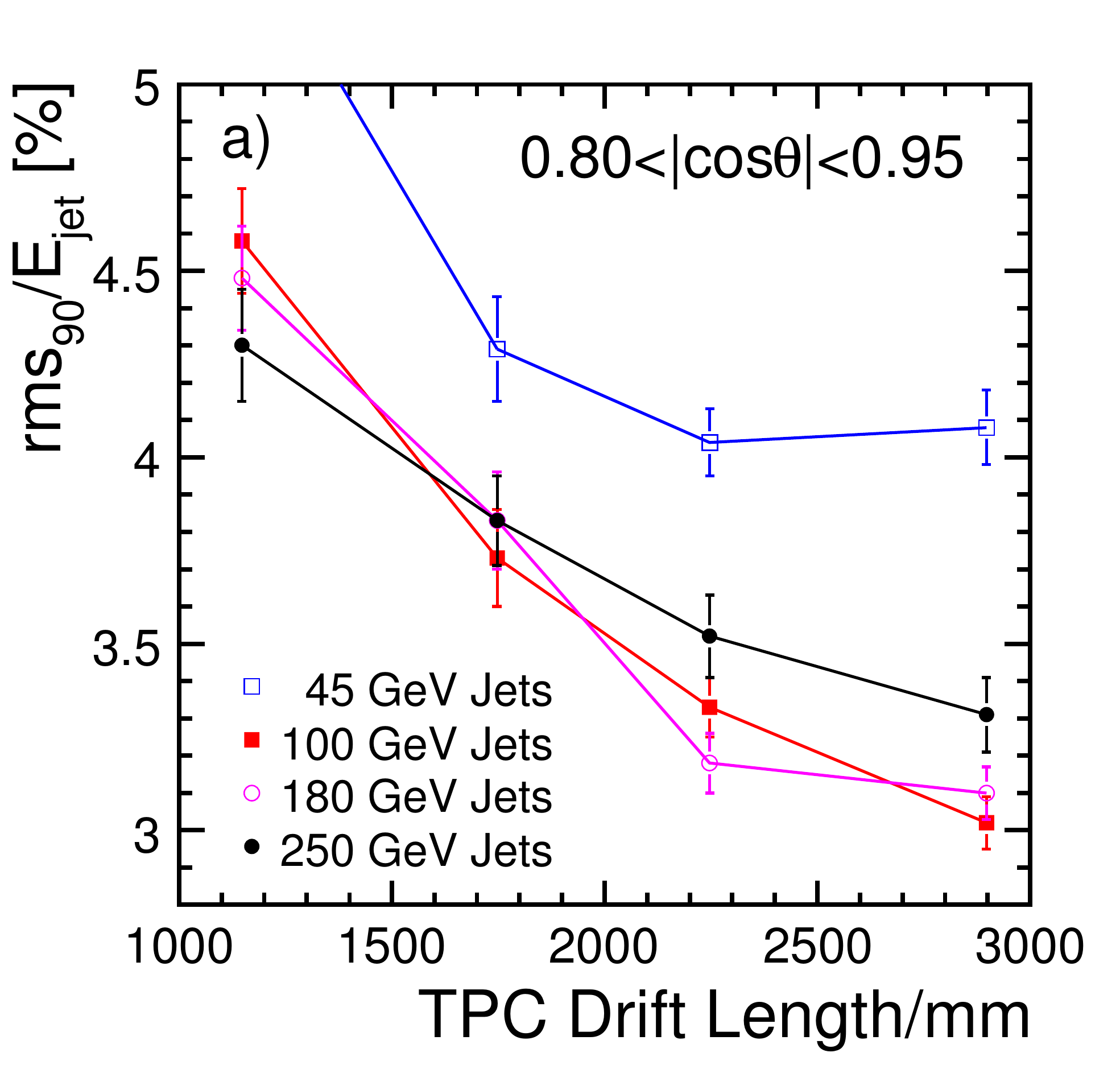}
\includegraphics[width=7.0cm]{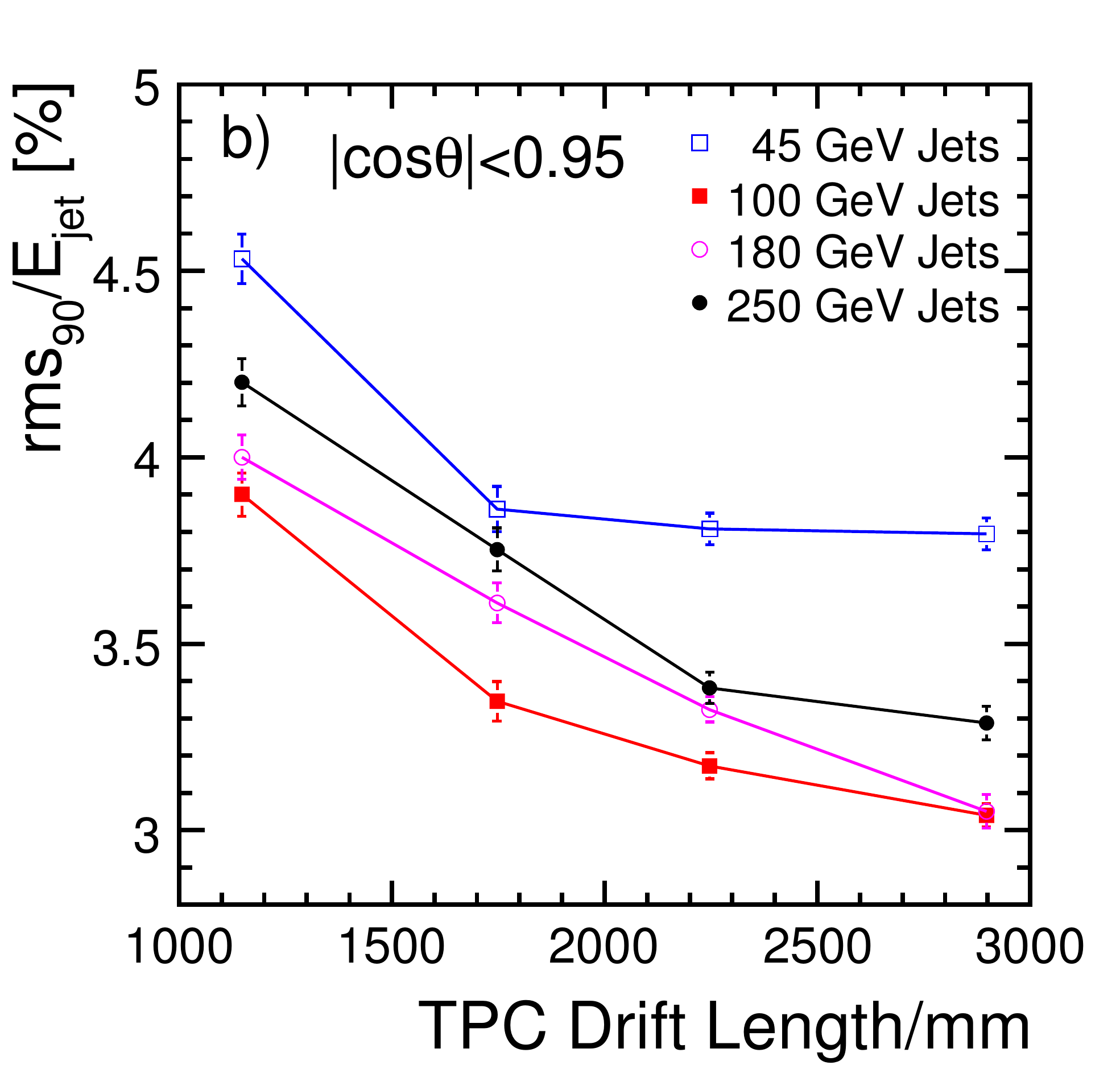}
 \caption[Jet energy resolution in the endcap region.]{a) the dependence of the jet energy resolution in the
             ``endcap'' region ($0.80<|\cos\theta_{\qq}|<0.95$) as a
             function of the TPC drift length in the LDCPrime model. 
               b) the dependence of the jet energy resolution in the
             region ($|\cos\theta_{\qq}|<0.95$) as a
             function of the TPC drift length in the LDCPrime model.
 \label{fig:pfa_aspect}}
\end{center}
\end{figure}

Although the cost of ILD 
will depend less strongly on length than on radius, 
it is, nevertheless, an important parameter
in the detector optimisation. From the perspective of particle flow, the 
main effect will be on the performance of forward jets. For forward
tracks, the importance of the $B$-field will be further diminished, and
one might expect the confusion term to scale as $L^{-1}$, 
where $L$ is the $z$-position of the endcap ECAL. 
Figure~\ref{fig:pfa_aspect}a shows the particle
flow performance for jets in the endcap region ($0.80<|\cos\theta_{\qq}|<0.95$).
For particle flow reconstruction of forward jets
it is beneficial to have the ECAL endcaps further from
the interaction region. To maintain good jet energy resolution 
in the forward region of the detector the TPC drift length needs
to be $\gtrsim 2000$\,mm. 

Figure~\ref{fig:pfa_aspect}b 
shows the length dependence of the average jet energy resolution
for jets with $|\cos\theta_{\qq}|<0.95$. 
When considering all jets, the benefits to particle flow performance 
in going beyond a TPC drift length of 2200\,mm are relatively small. 
From this study a TPC drift length of 2200\,mm looks reasonable; the 
benefits of increasing the detector length are unlikely to justify the
additional costs.

\subsection{ECAL and HCAL Granularity}

\label{sec:optimisation-granularity}
 
\begin{figure}[bB]
\begin{center}
\includegraphics[width=7.0cm]{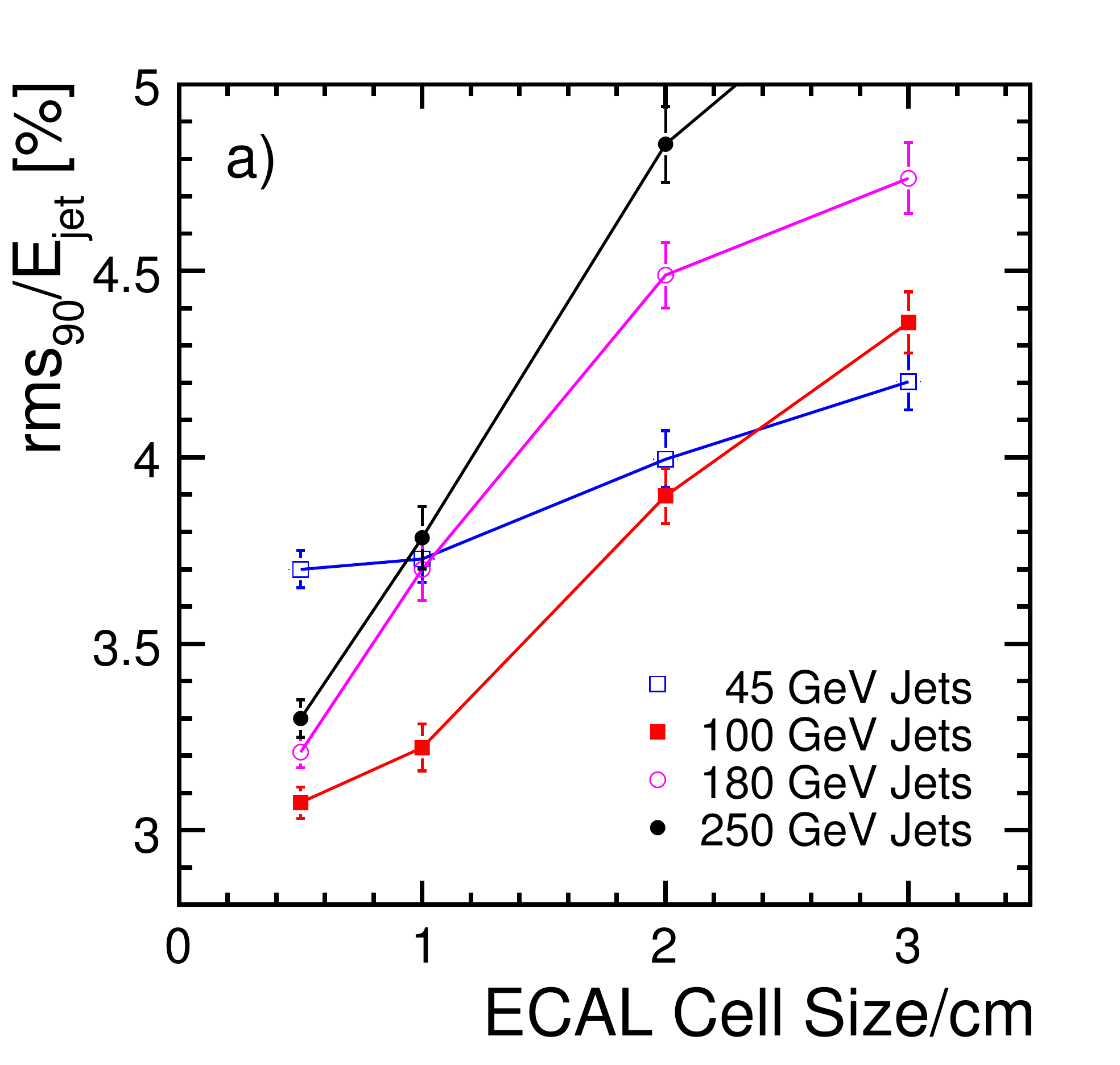}
\includegraphics[width=7.0cm]{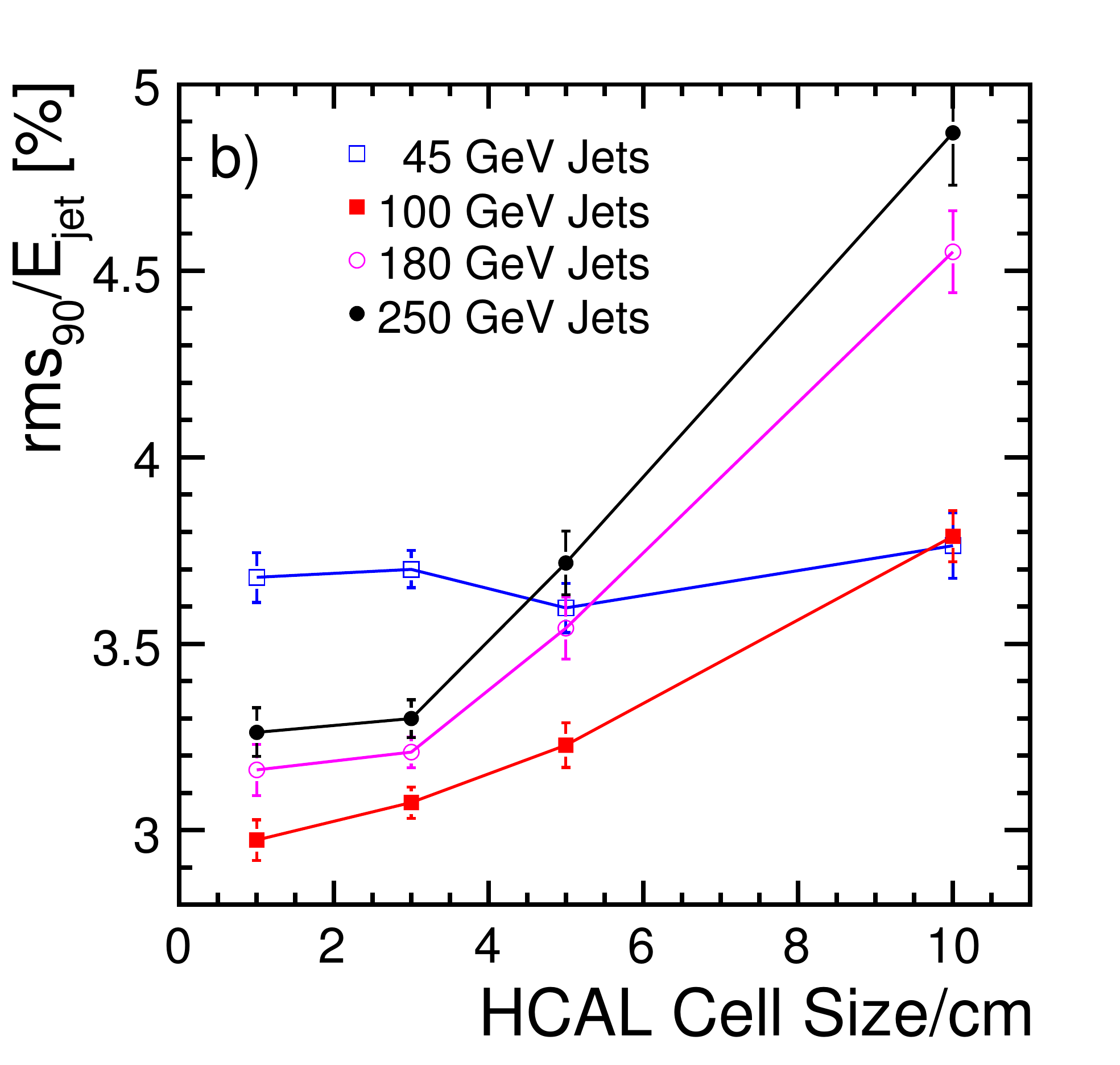}
 \caption[Jet energy resolution vs ECAL and HCAL segmentation.]{a) the dependence of the jet energy resolution ($\rmsn$) on 
             the ECAL transverse segmentation (Silicon pixel size)  
             in the LDCPrime model.  b) 
               the dependence of the jet energy resolution ($\rmsn$) on the HCAL
              transverse segmentation (scintillator tile size)  
             in the LDCPrime model.  
 \label{fig:pfa_segmentation}}
\end{center}
\end{figure}

The dependence of particle flow performance on the transverse segmentation of the
ECAL was studied using versions of the LDCPrime model with
silicon pixel sizes of $5\times5$\,mm$^2$, $10\times10$\,mm$^2$,
$20\times20$\,mm$^2$, and $30\times30$\,mm$^2$. The two main clustering
parameters in the PandoraPFA algorithm were re-optimised for each ECAL 
granularity. The particle flow performance results are summarised in 
Figure~\ref{fig:pfa_segmentation}a. For 45\,GeV jets the dependence is relatively
weak since the confusion term is not the dominant contribution to the resolution.
For higher energy jets, a significant degradation in performance is observed
with increasing pixel size. Within the context of the current
reconstruction, the ECAL
transverse segmentation has to be at least as fine as $10\times10$\,mm$^2$ to
meet the ILC jet energy requirement, $\sigma_E/E<3.8\,\%$, for the
jet energies relevant at $\roots = 1$\,TeV, with $5\times5$\,mm$^2$ 
being preferred. 

A similar study was performed for the HCAL using scintillator 
tile sizes of $1\times1$\,cm$^2$, $3\times3$\,cm$^2$,
$5\times5$\,cm$^2$, and $10\times10$\,cm$^2$. 
The particle flow performance results are summarised in 
Figure~\ref{fig:pfa_segmentation}b. From this study, it is concluded
that the ILC jet energy resolution goals can be achieved with an HCAL
transverse segmentation of $5\times5$\,cm$^2$, although for higher energy
jets there is a significant gain in going to $3\times3$\,cm$^2$.
There appears to be little motivation for $1\times1$\,cm$^2$ 
over $3\times3$\,cm$^2$ tiles.

\subsection{ECAL and HCAL detector technology}

\label{sec:optimisation-particleflow-technology}

The ILD concept incorporates two different technology options for
both the ECAL and HCAL. The two ECAL technologies 
are: i) a Silicon-Tungsten (SiW) calorimeter 
where the baseline pixel size of $5\times5$\,mm$^2$; 
and ii) a scintillator-Tungsten calorimeter where the
$1\times4$\,cm$^2$ scintillator strips in successive layers
are perpendicular to each other with the aim of achieving
a $1\times1$\,cm$^2$ effective transverse granularity. The 
particle flow studies described above were obtained using the 
simulation of the SiW calorimeter. To extend these studies
to the scintillator strip option requires additional step
in the reconstruction, namely strip-based clustering. 
First studies\cite{jeans} indicate that for 100\,GeV jets
the performance of the scintillator option with
$1\times4$\,cm$^2$ strips may approach that which would
be obtained with a scintillator segmentation of
$1\times1$\,cm$^2$. However, at this stage, further work 
is  needed to understand the limitations of the strip
based ECAL for higher energy jets and whether it is
possible to extend the approach to narrower strips to 
achieve an effective $5\times5$\,mm$^2$ segmentation.
The potential advantages of even finer segmentation, {\it e.g.} as provided
by the MAPs-based ECAL, has yet to be studied in detail.

The two HCAL technologies under consideration
are: i) an analogue steel-scintillator hadron
calorimeter (AHCAL)
with a tile size of $\sim 3 \times 3$\,cm$^2$; and ii)
a semi-digital calorimeter (DHCAL), {\it e.g.} 
using RPCs, with a readout pixel size of $1\times1$\,cm$^2$
and a three level (2 bit per cell) readout.
The particle flow studies described above used the
AHCAL option. The particle flow performance of
the semi-digital option is currently being studied 
in the context of the current PandoraPFA algorithm. 
Earlier studies (with the LDC detector model 
and a previous version of the PandoraPFA algorithm)
found that the jet energy resolution for 
100\,GeV jets with a digital (single bit) readout
 was similar to
that obtained with the AHCAL option. Further 
study is needed to establish the particle flow
performance of the DHCAL option.

%% file: optimization/background.tex
\begin{figure}[bht]
\centering
\includegraphics[width=15cm]{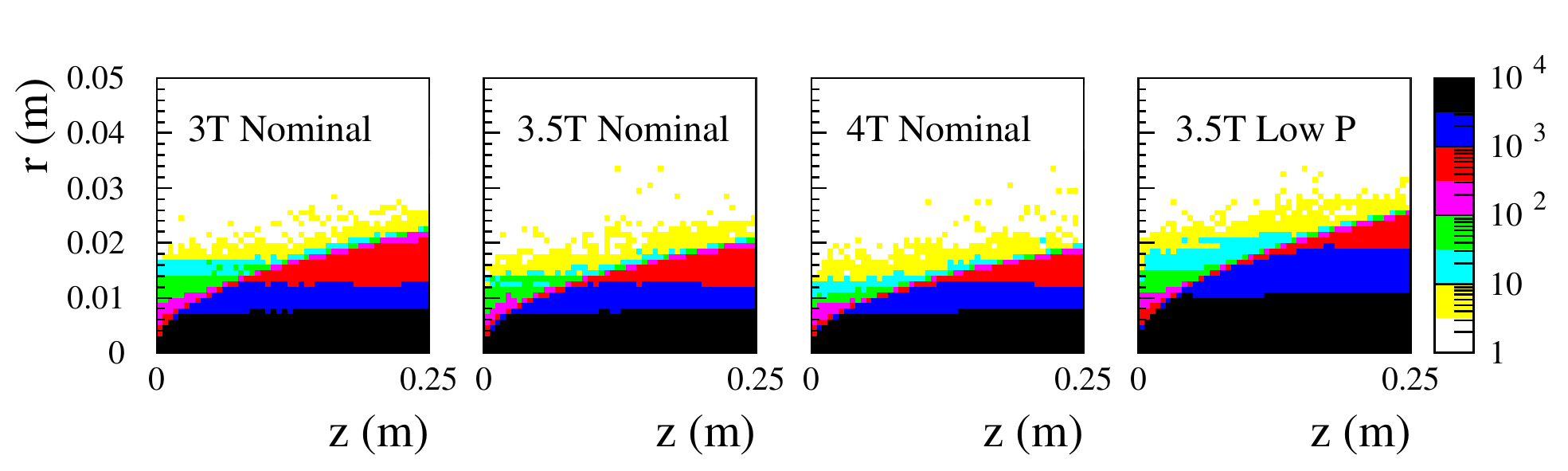}
\caption[Track density of pair background.]{Track density of the $\eplus\eminus$ pair
background ($/{\rm cm}^{-2}/{\rm BX}$) with the nominal ILC beam parameters 
at $\roots=500$\,GeV
for the detector magnetic field of 3\,T, 
3.5\,T, and 4\,T. Also shown is the background with the lowP option for
the ILC beam parameters at $\roots=500$\,GeV with $B =3.5$\,T.}
\label{fig:opt_pairbg}
\end{figure}

Beam-related backgrounds, and in particular $\eplus\eminus$
pairs created by beam-beam interactions, are an essential
input to the ILD design and optimisation. The
$\eplus\eminus$ pairs are produced at relatively
low angles to the beam direction and spiral along the
magnetic field lines parallel to the beam axis. As shown 
in Figure~\ref{fig:opt_pairbg}, the resulting pair-background 
tracks form a dense core with an approximately quadratic
envelope. The radius of the dense core for a given 
value of $z$ is roughly proportional to 
$\sqrt{B}$~\cite{Battaglia:2003kn}. The 
pair background determines the minimum radius of
the beam pipe needed 
to avoid a large source of  secondary background from 
electrons and positrons hitting
the beam pipe. In turn, the radius of the beam pipe 
determines the radius of the innermost layer of the 
vertex detector, and consequently influences the
impact parameter resolution for relatively low-momentum 
charged tracks. However, it has been
shown~\cite{ref-gld}, that if the magnetic field 
is $\gtrsim3$\,T, the required impact parameter resolution of
$5\,\mu{\rm m} \oplus 10\,\mu{\rm m}/p({\rm{GeV}})\sin^{3/2}\theta$
is achievable with a vertex detector layer thickness of 
$0.1-0.2\,\%\,X_0$/layer. 

In terms of optimisation of ILD, the 
main  effect of the pair background is to determine
the inner radius of the vertex detector, which
affects the impact parameter resolution and thus the flavour
tagging performance. However, the difference between
the radius of the core of the pair background between
a 3T and a 4T magnetic field is only 15\,\%.
In practice, the impact of the magnetic field on
the inner radius of the vertex detector is less than
this, as it is necessary to leave gaps between the dense core 
of the pair background and  
the beam pipe and between the beam pipe and the first layer of the 
vertex detector. These gaps are independent 
of the magnetic field, and when this is taken into
account, the difference of inner radius of the vertex 
detector between a B-field of 3\,T and 4\,T is only $\sim 10$\,\%.
The impact on the detector performance is discussed
in the next two sections.

Finally, it is worth noting that the inner radius of the 
vertex detector for the lowP option of the ILC machine parameters
is about 20\% larger than that for the nominal option.
Therefore, it can be concluded that the machine parameters 
have a larger impact on the inner radius of the vertex detector than the
magnetic field of the detector.

%% file: optimization/tracking.tex
The tracking system of the ILD detector concept consists
of a vertex detector (VTX) and a large volume TPC, complemented
by additional Silicon tracking layers (FTD/SIT). 
In addition, in the LDC-based models,
silicon tracking layers immediately outside the TPC are considered 
(ETD and SET). The dependence of the performance of the 
tracking system on the magnetic field and detector
size was an important consideration in optimising the 
ILD. Considerations of momentum
resolution favour a larger detector and higher 
magnetic field. As discussed above, a higher magnetic field also
allows the first layer of the vertex detector to be
closer to the interaction point (IP).
The optimisation of the tracking system is, again, a 
balance between the magnetic field and detector radius.
The parameter space spanning the LDC (smaller $R$, higher $B$-field)
and GLD (larger $R$, lower $B$) concepts is investigated
using the six detector models summarised in Table~\ref{tab:optdetparameter}.

\subsection{Momentum Resolution}

\begin{figure} 
\begin{center}
\includegraphics[width=7.0cm]{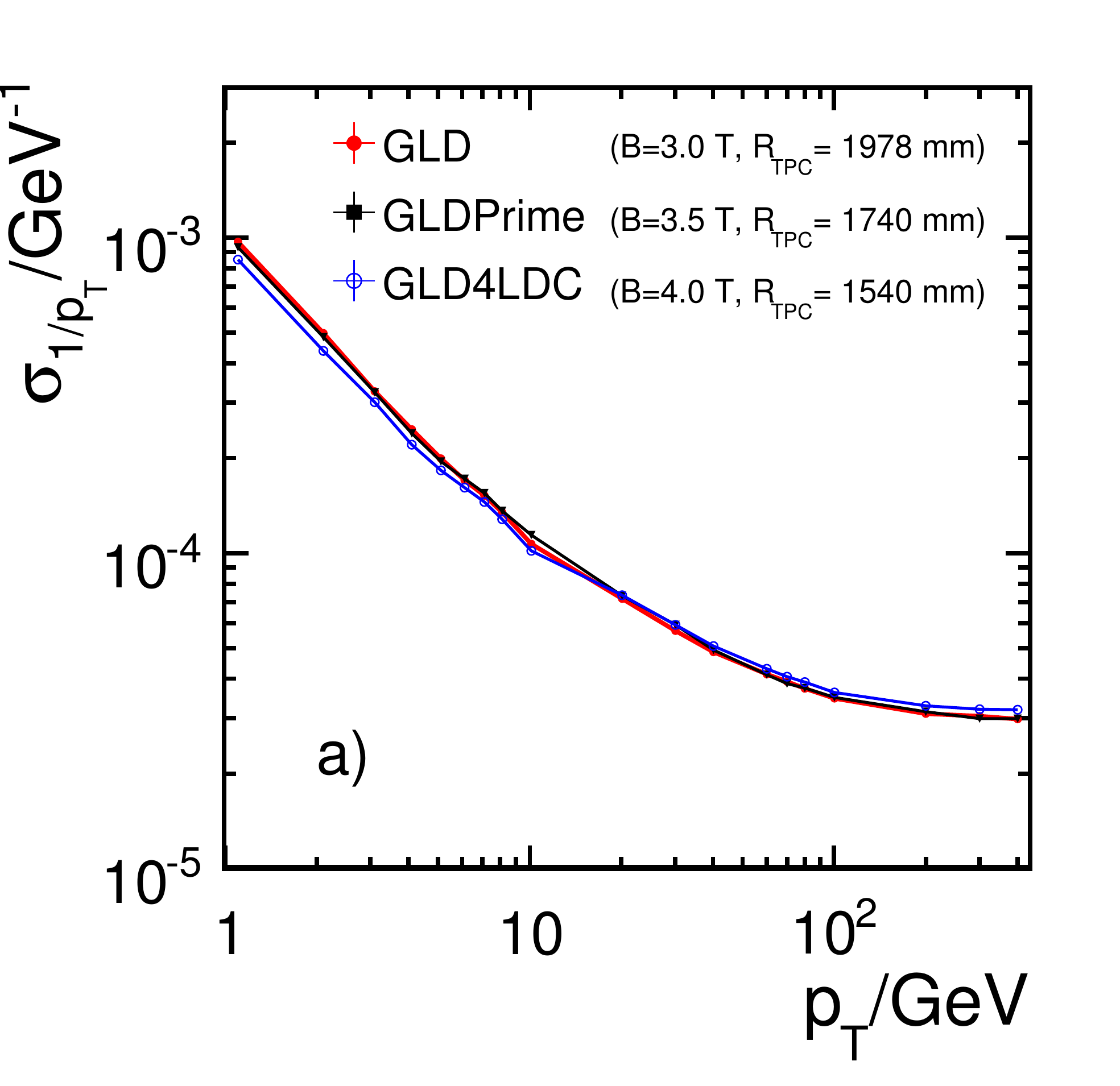}
\includegraphics[width=7.0cm]{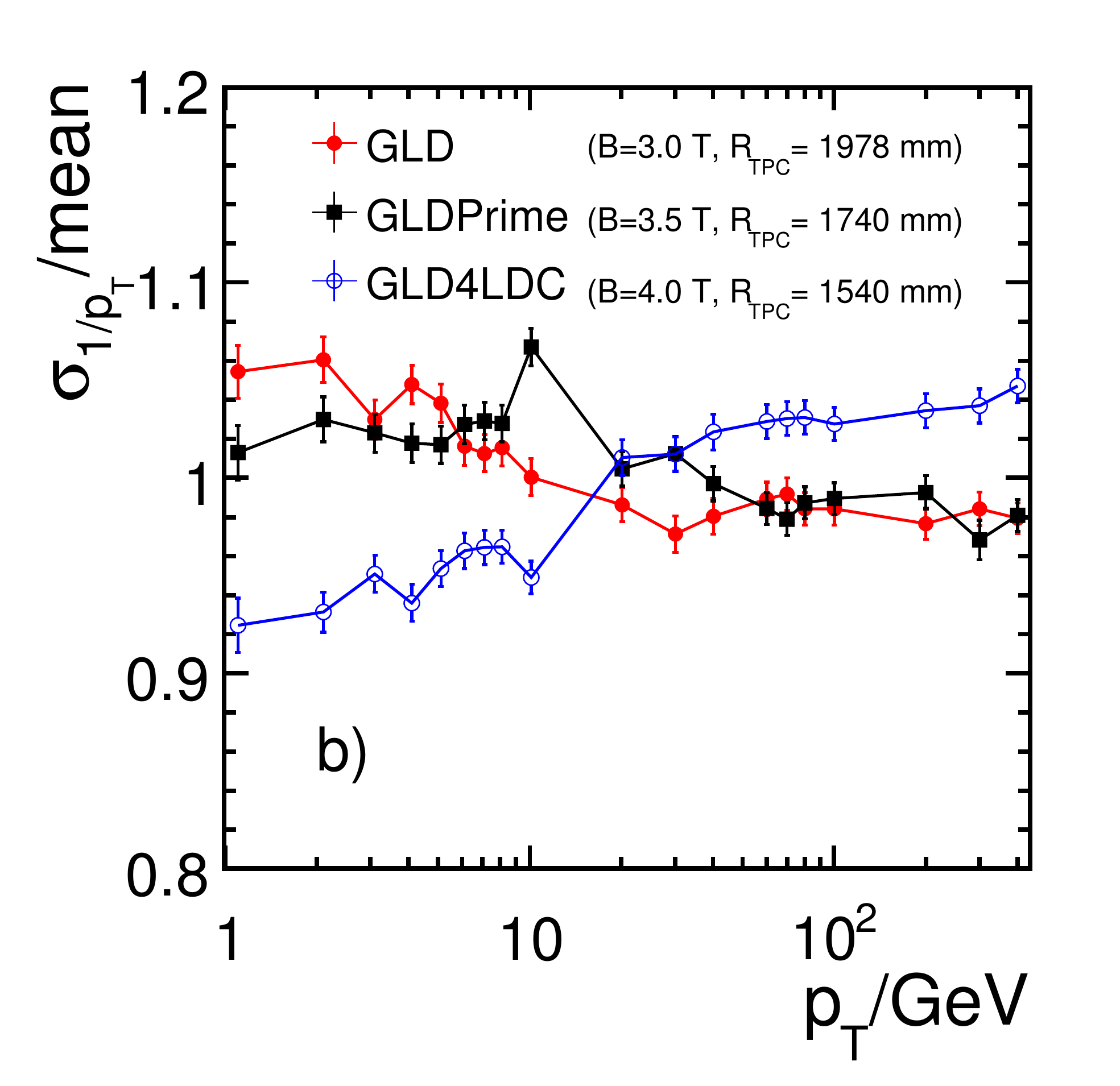}
\caption[Transverse momentum resolution versus momentum.]{(a) $\sigma_{1/p_\mathrm{T}}$ for single muon tracks at 
   $90^\circ$ to the beam axis, as a function of transverse momentum, for the
   GLD, GLDPrime, and GLD4LDC models; and (b) the ratio of $\sigma_{1/p_\mathrm{T}}$ to the 
   average of the 
   three detector models as a function of transverse momentum. To avoid the TPC central
   membrane the generated muons were displaced from the IP by a few centimetres.}
\label{fig:mreso_jupiter}
\end{center}
\end{figure}

Figure \ref{fig:mreso_jupiter}(a) shows the $1/p_\mathrm{T}$ resolution, 
as a function of $p_{\mathrm{T}}$, for single muons in 
the GLD, GLDPrime and GLD4LDC models. The results were 
obtained using the Satellites Kalman Filter 
(Section~\ref{sec:optimisation-simulation-jsf}). 
Figure \ref{fig:mreso_jupiter}(b) shows the relative $1/p_\mathrm{T}$ resolution compared
to the average of three detector models at a particular value of
$p_\mathrm{T}$, plotted as a function of $p_{\mathrm{T}}$.
Above approximately 50 GeV, the resolution obtained with the GLD4LDC model 
is $\sim$5\,\% worse than the larger detector models due to the
shorter lever arm 
of the TPC. For lower energy tracks the situation is reversed with the
higher magnetic field resulting in the resolution for 4\,T being $\sim$10\,\% 
better than for 3\,T. Similar results were obtained with the LDC-based
models using Mokka and MarlinReco (Section~\ref{sec:optimisation-simulation-mokkamarlin}).
The relative performance does not depend strongly on the angle.
The differences in resolution for the
range of $B$ and $R$ considered are small ($\lesssim10$\,\%) 
compared to those arising from different
layouts for the tracking system and the point resolutions of the components. 
For example, the use of hits in the silicon external tracker (SET) 
outside the TPC in the LDCPrime model improves the momentum resolution by 
15\,\% assuming an SET  $r\phi$ hit resolution of 4\,$\mu$m.

\subsection{Impact Parameter Resolution}

\begin{figure}
\begin{center}
\includegraphics[width=7.cm]{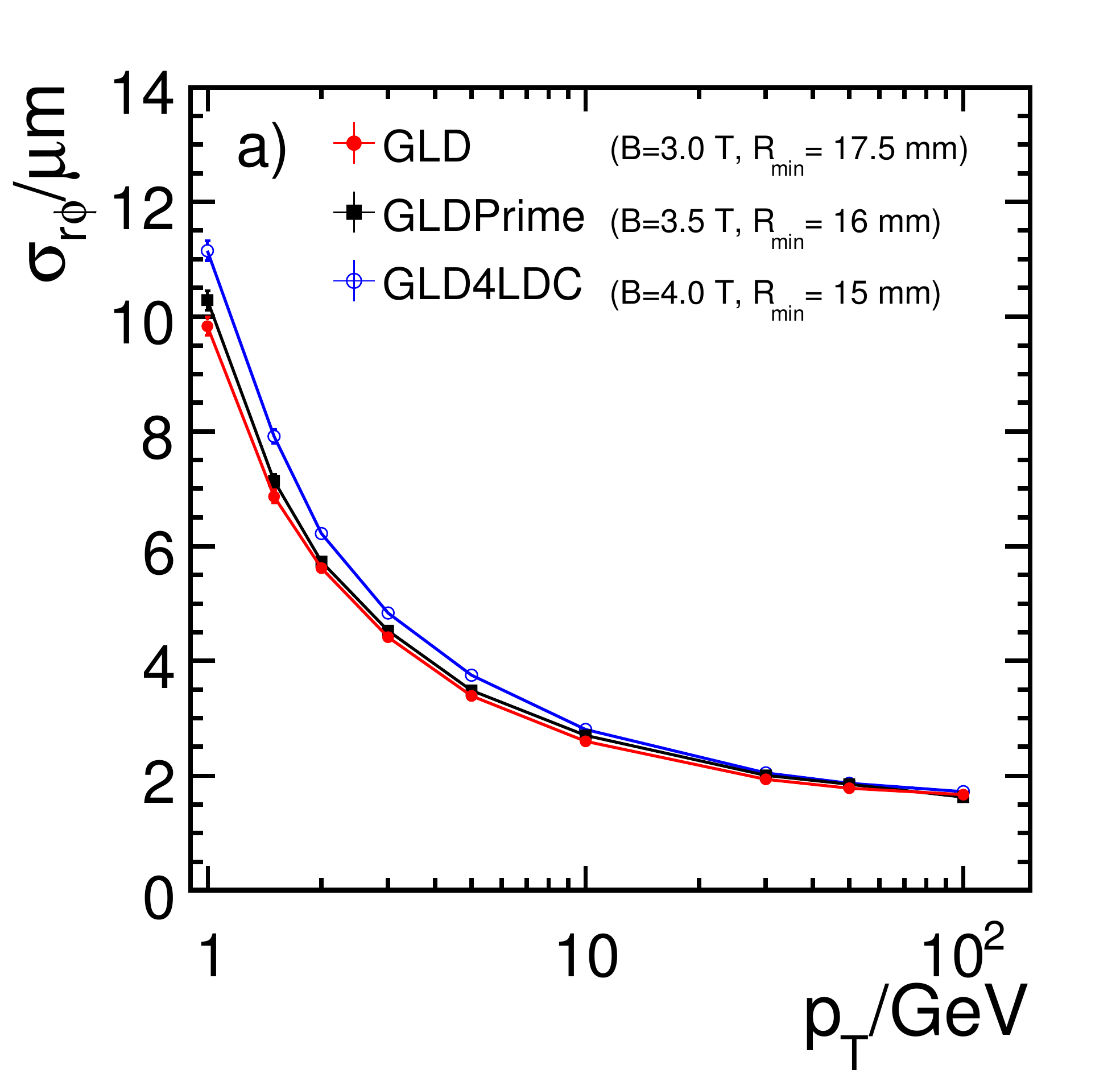}
\includegraphics[width=7.cm]{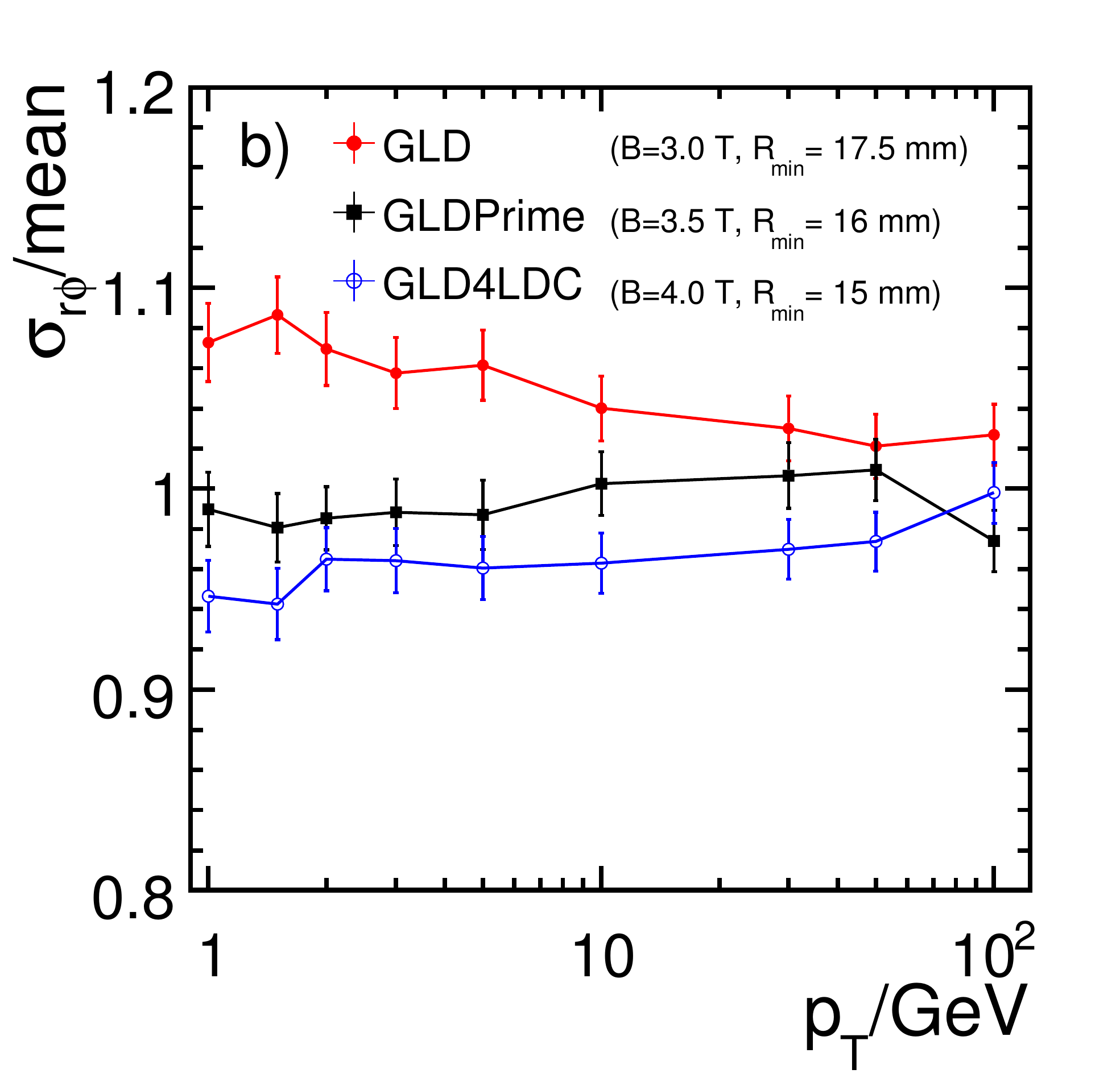}
\includegraphics[width=7.cm]{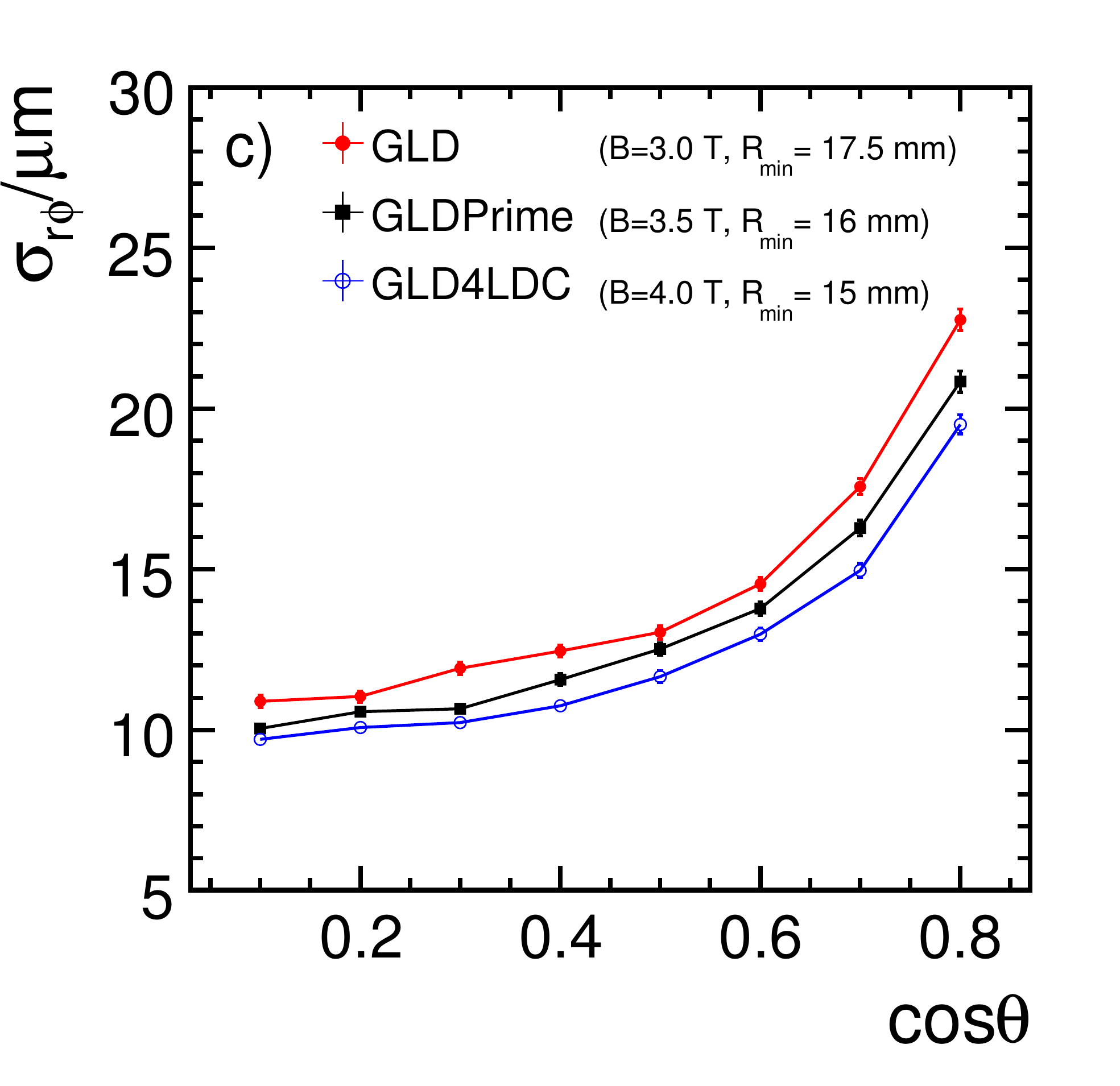}
\includegraphics[width=7.cm]{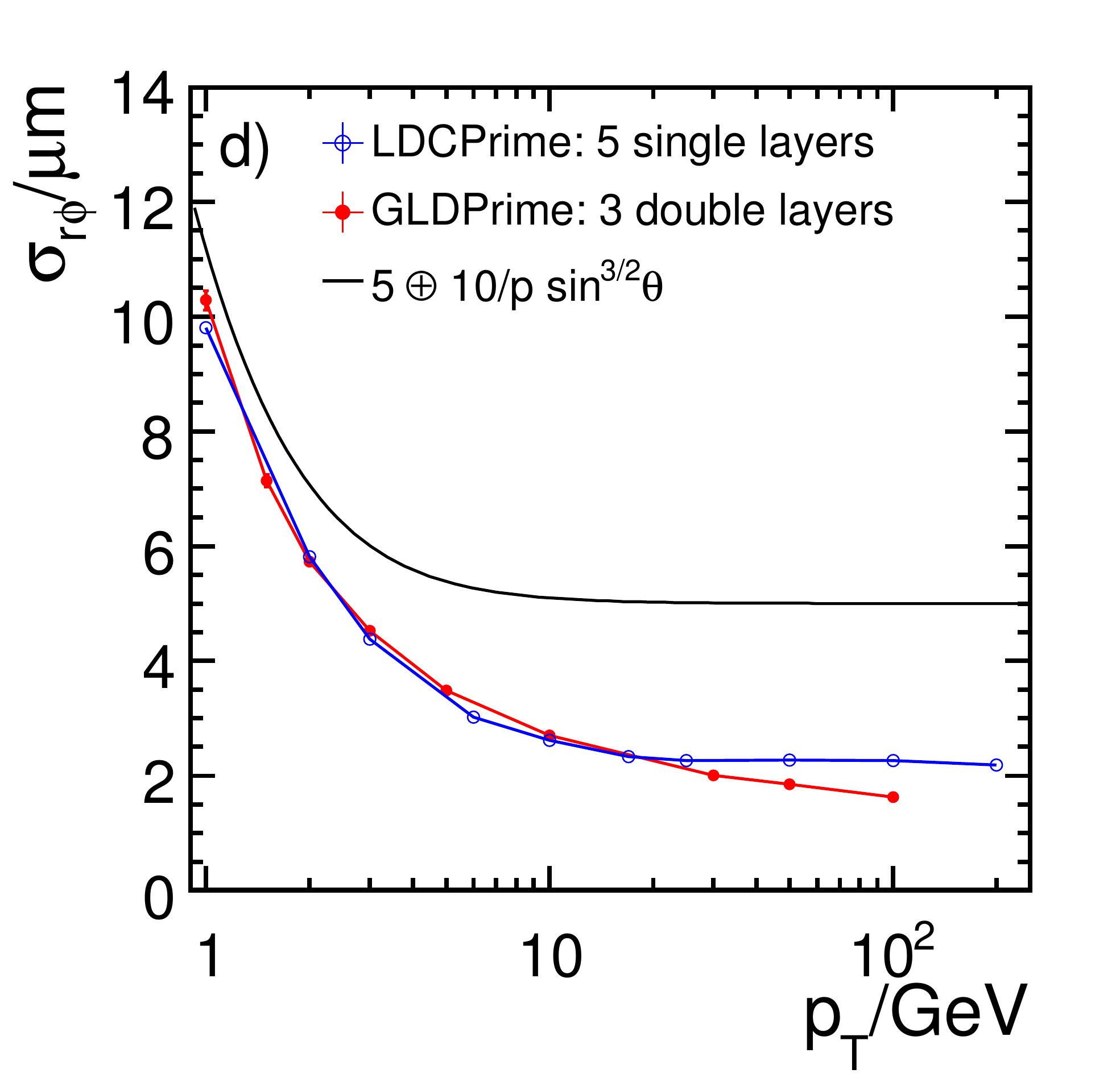}
\caption[Impact parameter resolution.]{(a) $\sigma_{r\phi}$ as a function of $p_{\mathrm{T}}$ 
for GLD, GLDPrime, and GLD4LDC, and (b) the ratio of $\sigma_{r\phi}$ to the average of the three detector models.
c) $\sigma_{r\phi}$ as a function of the track angle at the 
track energy of 1 GeV for GLD, GLDPrime, and GLD4LDC. d) 
the impact parameter resolution as a function of $p_{\mathrm{T}}$ 
for GLDPrime and LDCPrime. Also
shown is the nominal ILC goal for impact parameter resolution.
\label{fig:ipreso}}
\end{center}
\end{figure}

The impact parameter resolution, $\sigma_{r\phi}$, of the tracking system is
an important input to the heavy flavour tagging capability of the detector.
The most important detector considerations are the vertex detector
design (point resolution and material budget) and the magnetic field
which, as discussed in Section~\ref{sec:optimization-background} affects 
the minimum distance of the first layer of the vertex detector from the
interaction region, $R_\mathrm{min}$. 
Figure~\ref{fig:ipreso}a shows 
$\sigma_{r\phi}$ as a function 
of $p_{\mathrm{T}}$ for the GDC-based
detector models. The GLD4LDC model has the best resolution, 
because the higher $B$-field 
allows the innermost layer of the vertex detector to be closer to the 
interaction point (IP). However the differences between the detector
models considered, Figure~\ref{fig:ipreso}b, are relatively small $\lesssim 5-10$\,\%. 

Figure \ref{fig:ipreso}c shows the $\sigma_{r\phi}$ resolution for 1\,GeV muons
for the GLD-based detector models, plotted as a function of the track angle.
Whilst the higher magnetic fields are favoured, the differences 
between the detector models are $\lesssim 15\,\%$. For higher energy tracks,
where the effect of multiple Coulomb scattering is negligible, the differences 
between the models are even smaller. 
Although the variations in magnetic field and the corresponding inner radii 
of the vertex detector lead to relatively small differences in impact parameter
resolution, different detector layouts have a more significant impact. Figure 
\ref{fig:ipreso}d compares the 
impact parameter resolution for the GLDPrime and 
LDCPrime detector models. The GLDPrime detector 
assumes a vertex detector consisting of
six layers arranged in three closely spaced doublets 
(see Section~\ref{ild:vertex}),
whereas the LDCPrime model assumes five equally spaced layers. 
The three double layer layout results in a significantly better
impact parameter resolution for high momentum tracks 
because it gives two, rather than one, high precision 
measurements close to the IP.

\subsection{Conclusions}

For the range of $B$ and $R$ considered here,
the differences in momentum resolution are $\lesssim10$\,\%,
with higher $B$-field preferred for low $p_{\mathrm{T}}$ tracks 
and a larger $R$ preferred for high $p_{\mathrm{T}}$ tracks.
The impact parameter resolution, $\sigma_{r\phi}$, is better for
models with higher $B$ as the first layer of the vertex detector
can be placed closer to the IP. However, the differences in impact
parameter resolution obtained with a 3\,T and 4\,T magnetic field are small, 
at most 15\,\% for low momentum tracks and $\lesssim5\,\%$ for tracks above
2\,GeV. It can be concluded that for the range of $B$ and $R$ spanned
by the LDC and GLD detector concepts, the differences in 
impact parameter and momentum resolution are relatively small.
It is also concluded that the tracking resolutions depend much 
more strongly on the subdetector technologies and tracking system
layout than on the global parameters ($B$ and $R$) of the detector.

%% file: optimization/flavour.tex
Heavy flavour tagging will be an essential tool in many physics analyses 
at the ILC. The flavour tagging performance depends primarily on
the design of the vertex detector and, in particular, 
the impact parameter resolution. The flavour tagging performance
is studied using MarlinReco for the full reconstruction of the
simulated events and the sophisticated LCFIVertex package
for heavy flavour tagging ~\cite{flavourtag}.
The LCFIVertex~\cite{ref:lcfivertex} flavour tagging uses three artificial neural networks (ANNs):
 i) a {\it b}-tag to discriminate  
   $b$-quark jets from jets from charm and light quarks;
 ii) a {\it c}-tag to discriminate  $c$-quark jets from $b$ and light quark jets; and
 iii) a {\it c/b}-tag to discriminate between $c$-quark jets and $b$-quark jets.
The ANNs use different sets of discriminant variables depending on whether either
one, two, or more than two vertices are found in the jet. 
The ANN architecture is a multi-layer perceptron with $N=8$ inputs,
one hidden layer with $2N-2$ nodes, and sigmoid activation functions. 
The weights were calculated using the back propagation conjugate gradient algorithm.
Two of the most powerful inputs to the flavour tag are the joint likelihoods
(in $r-\phi$ and in $r-z$) for all tracks in the jet to have 
originated from the primary vertex. The joint 
likelihoods depend on the respective
$r-\phi$ and $r-z$ impact parameter significances of all the 
tracks in the jet. Consequently the impact parameter resolution of 
the vertex detector plays a central role in
determining the flavour tagging performance. It was demonstrated 
in the previous section that the difference in $\sigma_{r\phi}$ in going 
from 3\,T to 4\,T is rather small. Consequently, one might expect the
same to be true for flavour tagging performance.

\begin{figure}[bh]
\centering
\includegraphics[width=7.5cm]{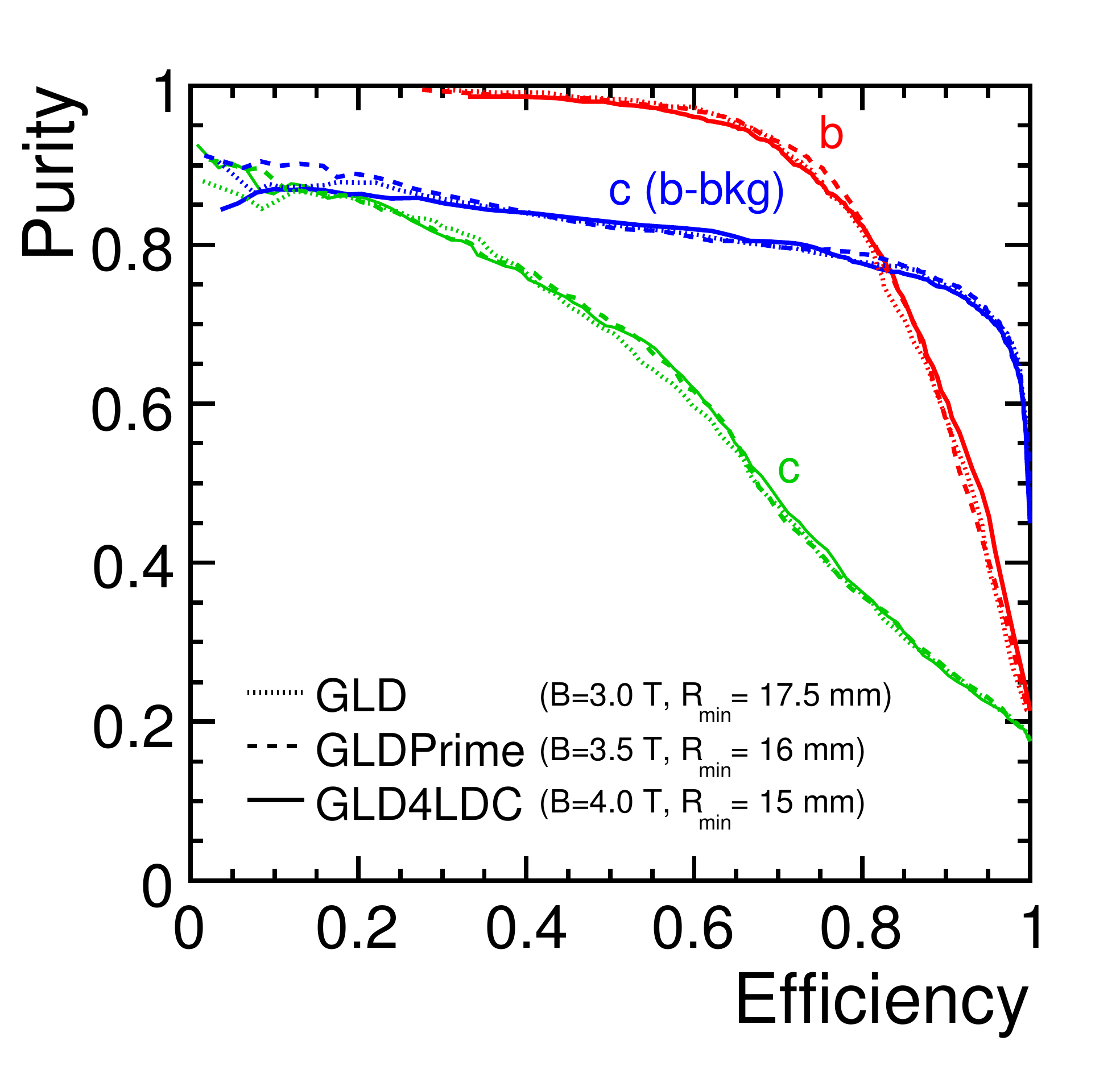}
\includegraphics[width=7.5cm]{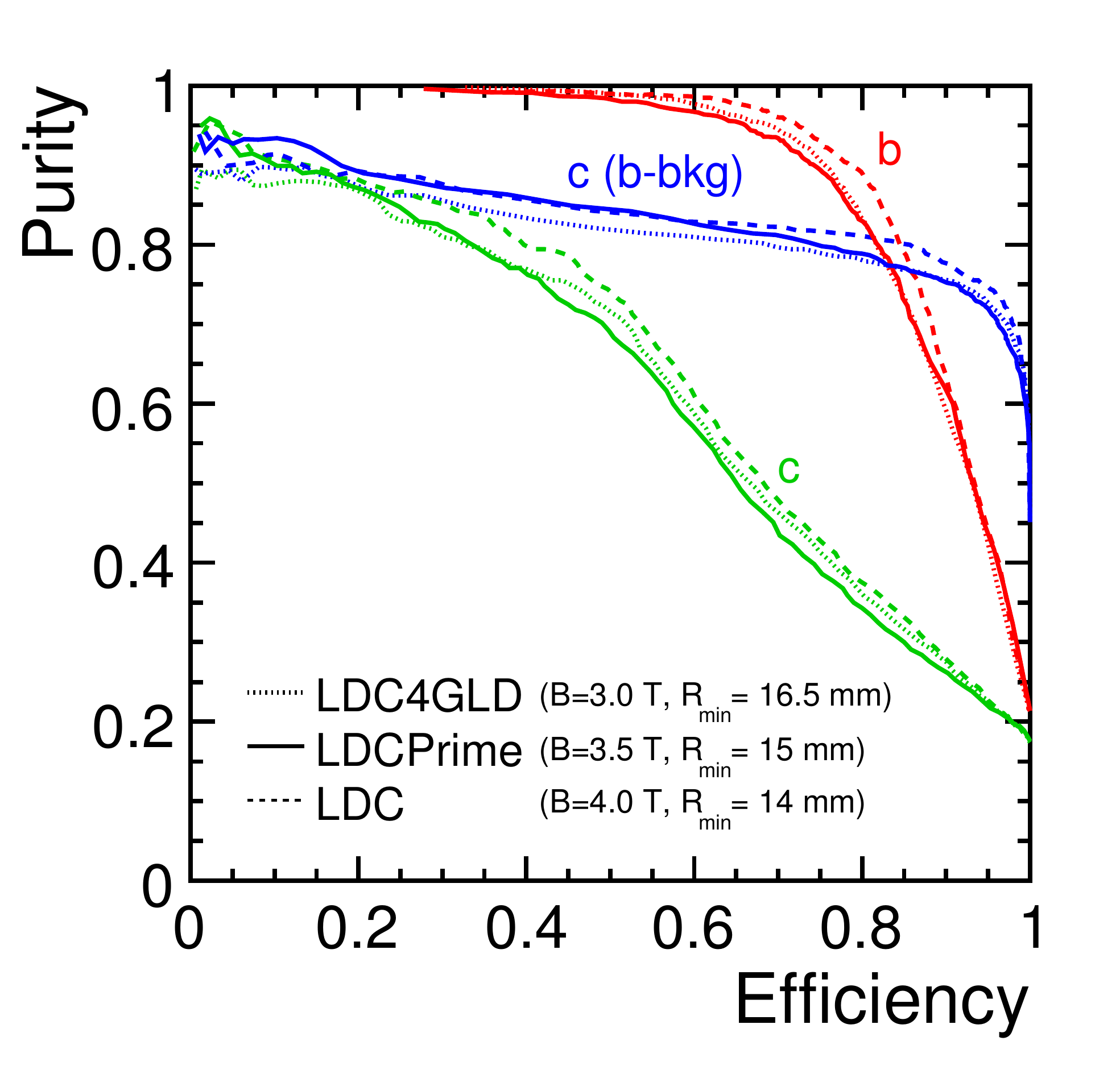}
\caption[Flavour tagging comparison.]{Flavour tagging performance for the (left) GLD-based and (right) LDC-based detector models.}
\label{fig:effpur}
\end{figure}

The dependence of flavour tagging performance on the global
detector parameters was investigated using the GLD- and LDC-based detector models.
Separate ANNs were trained for each of the models.
The samples used to evaluate the flavour tagging performance,
which  were generated with SM $\Zzero$ boson branching ratios,
were independent of those used for the training.
All the samples were generated at $\roots=m_Z$.
The results are shown in Figure \ref{fig:effpur}.
The observed 
differences in the flavour tagging performances between the GLD (LDC) models
are $\lesssim 1\,\%$ ($\lesssim 4\,\%$). 
There is a  preference for the 4.0\,T configuration, 
in particular for the $b$-tag at high efficiencies. 
However, the uncertainties on the efficiencies due to
statistics and the ANN training procedure are $\sim 2\,\%$,
and hence statistical significance of the observed differences 
are $\lesssim 2\sigma$. From this study it is concluded 
that the increased inner radius of the vertex detector when going from $B$=4\,T
to $B$=3\,T, does not have a large impact on the flavour tagging performance
of the detector.

%% file: optimization/physics.tex
The previous sections of this chapter discuss the impact of the detector 
design on the low level measurements of jet energies, track momenta, 
impact parameters
and flavour identification. Here the performance of the different detector
models in Table~\ref{tab:optdetparameter} are compared for 
three physics analyses: the
measurement of the Higgs mass, $\tau$ pair production 
and polarisation, and chargino/neutralino pair 
production.

\subsection{Higgs Recoil Mass}

\label{sec:optimization_higgs_recoil}

\begin{figure}[b]
\begin{center}
\includegraphics[width=7.5cm]{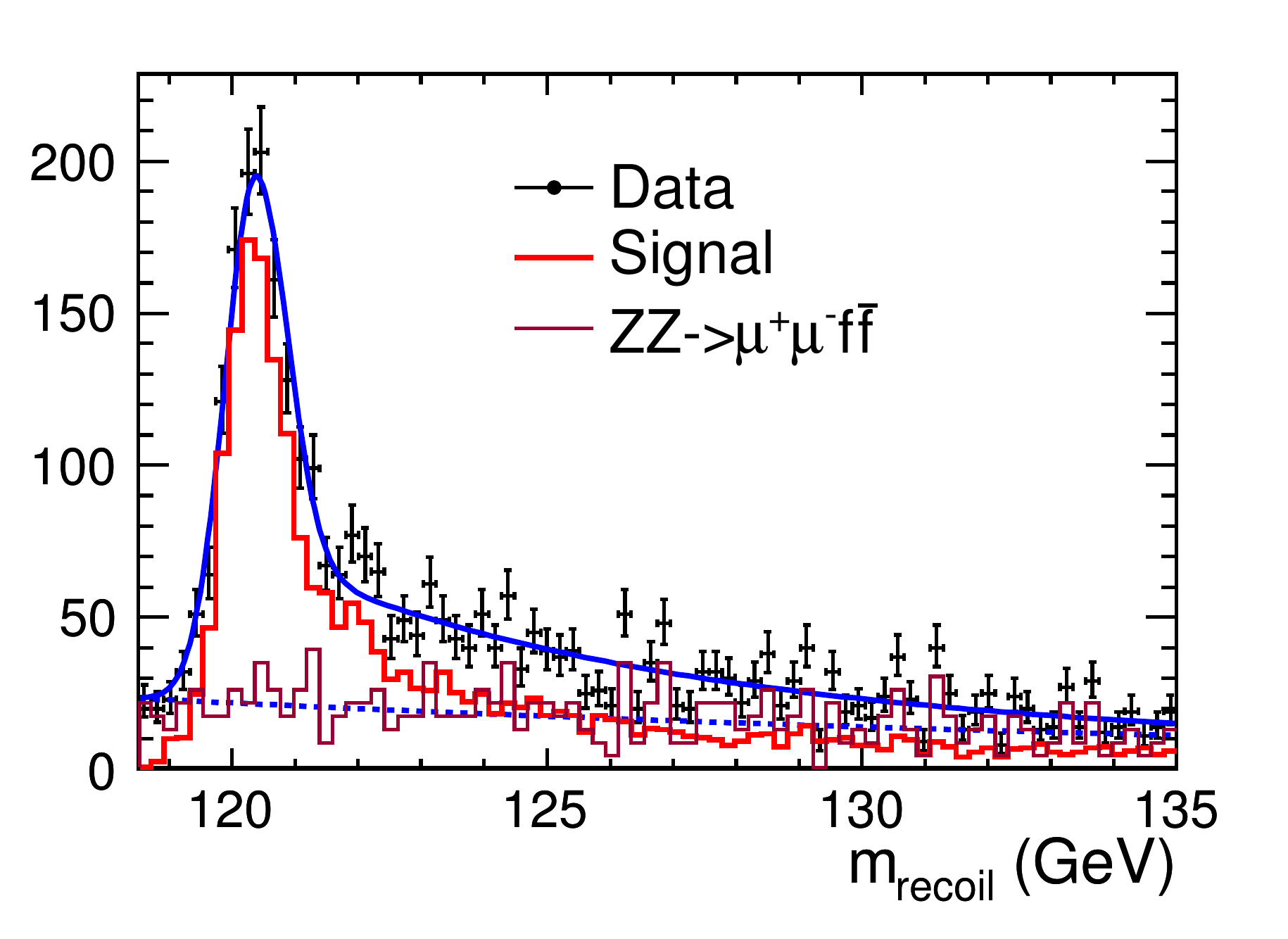}
\includegraphics[width=7.5cm]{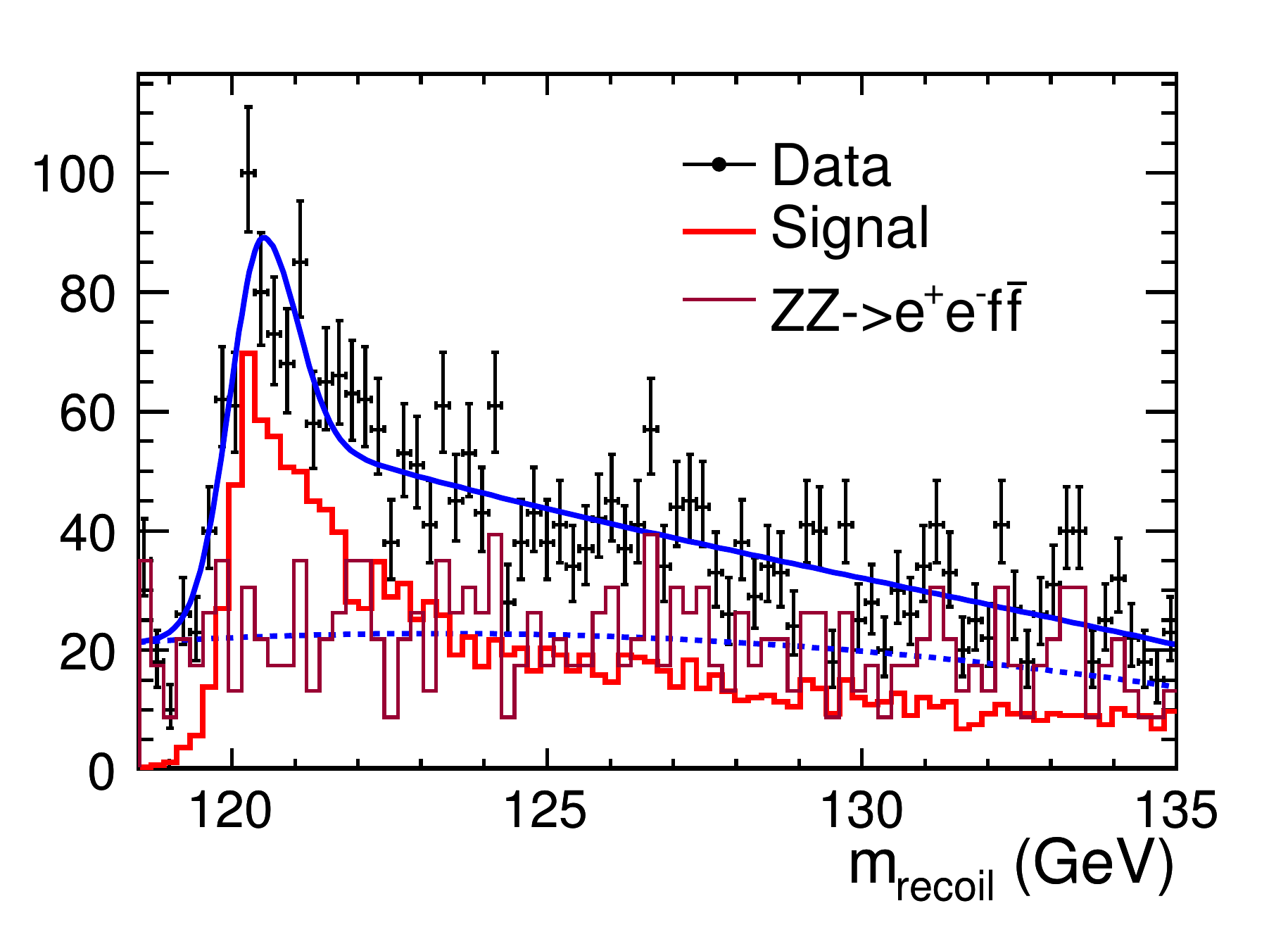}
\caption[Reconstructed Higgs recoil mass.]{Distributions of the reconstructed Higgs recoil mass
obtained with the LDCPrime model for  (a) $\Zzero\Higgs \to \mpmm X$ and (b) $\Zzero\Higgs \to \epem X$. 
 The events were generated with a top-hat beam energy distribution with a half width of 0.28\% for the electron beam and 0.18\% for the positron beam.}
\label{fig:hmass}
\end{center}
\end{figure}

One of the prime motivations for the unprecedented track momentum resolution at the
ILC is the determination of the Higgs mass from the recoil mass distribution in
$ZH \to \mpmm X$ and $ZH \to \epem X$ events. This sensitivities to
this process if the LDC, LDCPrime and LDC4GLD detector models were compared.
For this study only $\epem \to ZZ \to \epem / \mpmm Z$
 background was included. Figure \ref{fig:hmass} shows,
for the LDCPrime model, the Higgs recoil mass distribution for selected events.
To determine the Higgs mass and production cross section,
the recoil mass distributions were fitted using a Gaussian 
for the peak region with an exponential 
component for the tails~\cite{bib:higgsopt}.

The results
for the different detector models are summarised in Table \ref{tab:hrecoil}. 
These numbers should not be compared with the physics sensitivity studies
presented in Section~\ref{sec:physics-higgs-mass} as 
only $\epem \to ZZ \to \epem / \mpmm Z$ background is included and the events
were generated with a different luminosity spectrum.
\begin{table}
\centering
\begin{tabular}{|l|r|r|r|r|}
\hline
& \multicolumn{2}{c}{$ZH \to \mu^{+} \mu^{-} X$} & \multicolumn{2}{|c|}{$ZH \to e^{+} e^{-} X$} \\ \hline
& $\Delta m_{\mathrm{recoil}}$ & $\Delta \sigma$ & $\Delta m_{\mathrm{recoil}}$ & $\Delta \sigma$ \\ \hline
LDCPrime & $23\pm0.4$\,MeV & 0.28 fb & 47$\pm0.9$\,MeV & 0.49 fb \\
LDC      & $23\pm0.4$\,MeV & 0.27 fb & 47$\pm0.9$\,MeV & 0.52 fb \\ \hline
\end{tabular}
\caption[Relative Higgs mass precision.]{The measurement precision of the Higgs recoil mass ($\Delta m_{\mathrm{recoil}}$) and cross section ($\Delta \sigma$) for $\Zzero\Higgs \to \mpmm X / \epem X$. 
 The events were generated with a top-hat beam energy distribution with a half width of 0.28\% for the electron beam and 0.18\% for the positron beam.\label{tab:hrecoil}}
\end{table}

When interpreting the above results it is necessary to consider the relative importance
of momentum resolution and the beam energy spread. For the assumed beam energy spread 
(a top-hat distribution with half-widths $0.28\,\%$ and $0.18\,\%$ for the electron and 
positron beams respectively), the event-by-event 
recoil mass resolution in the peak region is $\sim400$\,MeV, which this includes
contributions from the beam energy spread and from beamstrahlung. 
From MC studies, the other major
contribution to the event-by-event recoil mass resolution arises, as expected, from 
the track momentum resolution. This is found to contribute $\sim350$\,MeV to the recoil
mass resolution. For the detector models considered in this study, the
differences in momentum resolution are $\lesssim5\,\%$, 
for the relevant momentum range. Even these small differences are diluted by the
contribution from the beam energy spread and, as verified in this study, significant
differences in the $\mH$ mass resolution are not expected.

\subsection{Tau pairs}

\begin{figure}
\begin{center}
 \includegraphics[width=7.0cm]{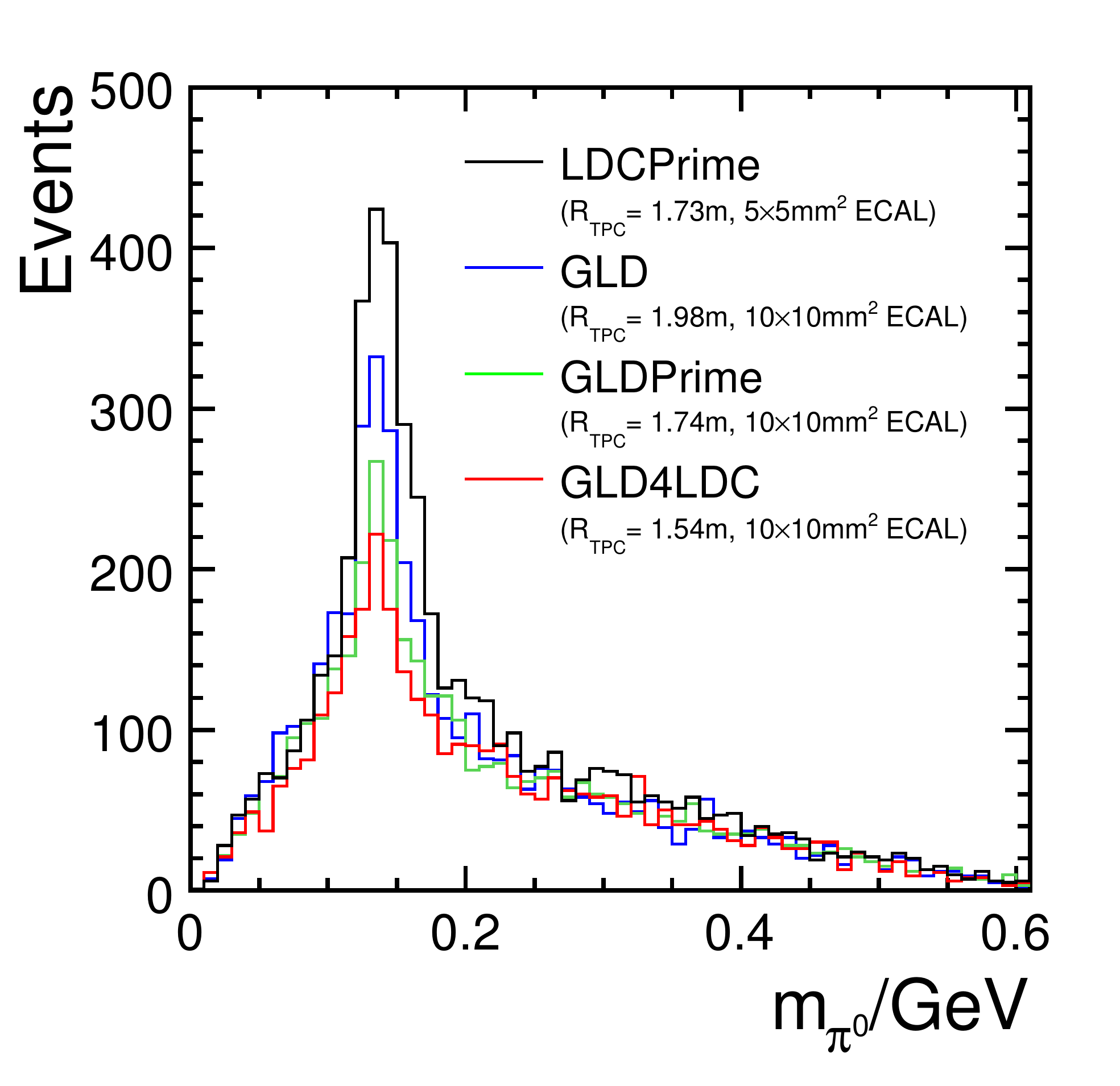}
 \includegraphics[width=7.0cm]{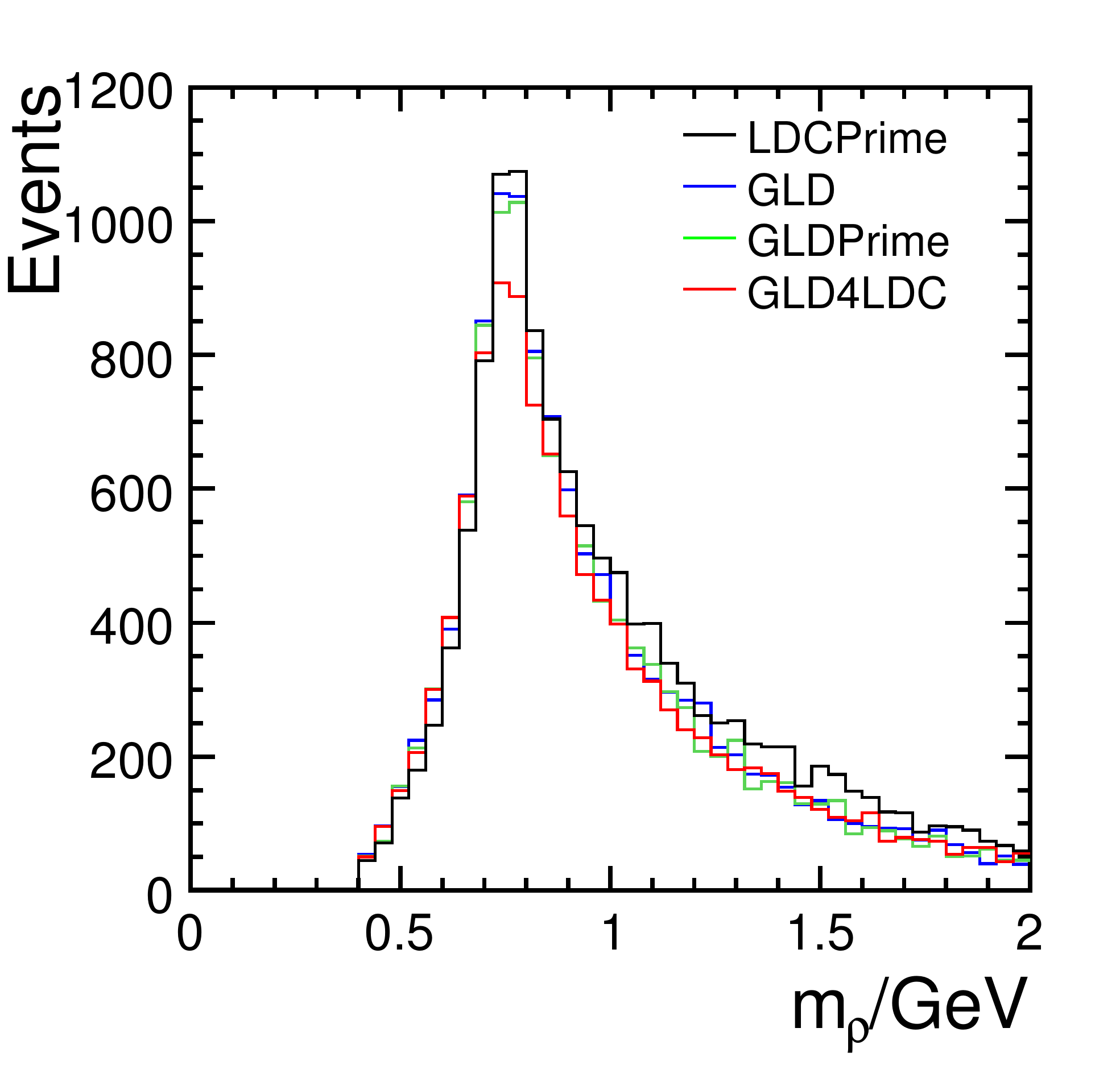}
 \caption[$\tau\rightarrow\pi\nu$ selection]{a) The reconstructed $\pi^0$ invariant mass distribution in the selected $\tau^+\tau^-$ events
             at $\roots=500$\,GeV. Only events where more than one photon is reconstructed
             are shown. b) The corresponding reconstructed $\rho^\pm\rightarrow\pi^\pm\pi^0$ mass
             distribution for decays where $\ge1$ photon cluster is reconstructed. For both plots
             the distributions include all tau decay modes.}
 \label{fig:pirhomass}
\end{center}
\end{figure}

The reconstruction of tau pair events at $\sqrt{s} = 500$\,GeV provides 
a powerful test of a number of aspects of the detector performance, {\it e.g.}
$\pi^{0}$ reconstruction and the tracking efficiency for nearby tracks
in three-prong tau decays. The performances of the  
GLD, GLDPrime, GLD4LDC and LDCPrime models are
compared for the measurement of the $\tau$ polarisation, $P_{\tau}$
which is primarily sensitive to the ability to resolve photons 
from $\pi^0$ decay from the charged hadron in $\tau\rightarrow\rho\nu$
decays.  Figure~\ref{fig:pirhomass} shows
the reconstructed $\pi^0$ and $\rho^\pm$ 
invariant mass distributions used in the tau decay selections.
The numbers of events in the $\pi^0$ mass peak reflect the efficiency
for reconstructing both photons from $\pi^0\rightarrow\gamma\gamma$ decays.
The LDCPrime detector model gives the highest $\pi^0\rightarrow\gamma\gamma$ 
reconstruction efficiency, demonstrating the advantages of smaller
ECAL pixel size ($5\times5\,$mm$^2$). For the GLD models, all 
with an ECAL pixel size of $10\times10\,$mm$^2$, the $\pi^0$ reconstruction
efficiency increases with detector radius due to the increased spatial 
separation of the two photons.

Table~\ref{tab:taupol} summarises the impact of the different
detector models on the $P_{\tau}$ measurement from
$\tau\rightarrow\pi\nu$ decays. The $\tau\rightarrow\pi\nu$ selection requires
that a tau jet consists of a single track and at most 1\,GeV of energy 
not assigned to the track. Cuts to remove $\tau\rightarrow e\nu\overline{\nu}$
and $\tau\rightarrow \mu\nu\overline{\nu}$ decays are also applied. 
The $P_{\tau}$ is determined from
the cosine of the $\pi^{\pm}$ decay angle in the $\tau$ rest-frame (which is
determined by the charged pion energy). The differences
in the different detector models are most evident in the purities of the
$\tau\rightarrow\pi\nu$ selection. The advantages of smaller ECAL pixels
(LDCPrime compared to GLDPrime) are clear and it can be seen that higher 
purities are obtained for larger detector radii. 
However, similar sensitivities to $P_\tau$ are obtained from the 
$\tau\rightarrow\pi\nu$ channel. One should not draw too strong a conclusion 
from this as the measurement of $P_\tau$
from 
$\tau\rightarrow\rho\nu$ and $\tau\rightarrow a_1\nu$ decays 
could show a stronger dependence on the detector model.

\begin{table}
\centering
\begin{tabular}{|@{\,\,}l@{\,\,}|@{\,\,}c@{\,\,}|@{\,\,}c@{\,\,}|ccc|}
\hline
Detector &ECAL/mm$^2$ & $R_\mathrm{TPC}$/m & Eff      & Purity   & $\sigma_{P_\tau}$  \\ \hline
GLD       & $10\times10$ & 1.98 & 84.5\,\% & 85.7\,\% & $0.0454\pm0.0005$   \\
GLDPrime  & $10\times10$ & 1.74 & 85.2\,\% & 83.6\,\% & $0.0452\pm0.0005$   \\
GLD4LDC   & $10\times10$ & 1.54 & 84.9\,\% & 80.8\,\% & $0.0460\pm0.0006$   \\
LDCPrime  & $5\times5$   & 1.73 & 84.1\,\% & 88.5\,\% & $0.0430\pm0.0005$   \\ \hline
\end{tabular}
\caption[$\tau^{\pm} \to \pi^{\pm} \nu$ selection.]{Summary of $\tau^{\pm} \to \pi^{\pm} \nu$ 
   selection efficiencies and purities for events generated with the GLD, GLDPrime, GLD4LDC, and LDCPrime detector
   models. The efficiencies are calculated with with respect to the
   $\tau^+\tau^-$ selection and the purities only include the background from the different tau 
   decay modes. The statistical uncertainties on the efficiencies and
   purities are all $\pm0.5\,\%$. The uncertainty on the tau polarisation measurement assumes an
   electron-positron polarisation of $(-80\,\%, +30\,\%)$ and corresponds to 80\,fb$^{-1}$ of data.
\label{tab:taupol}}
\end{table}

\subsection{Chargino and neutralino production}

\begin{figure}
\centering
\includegraphics[width=7.0cm]{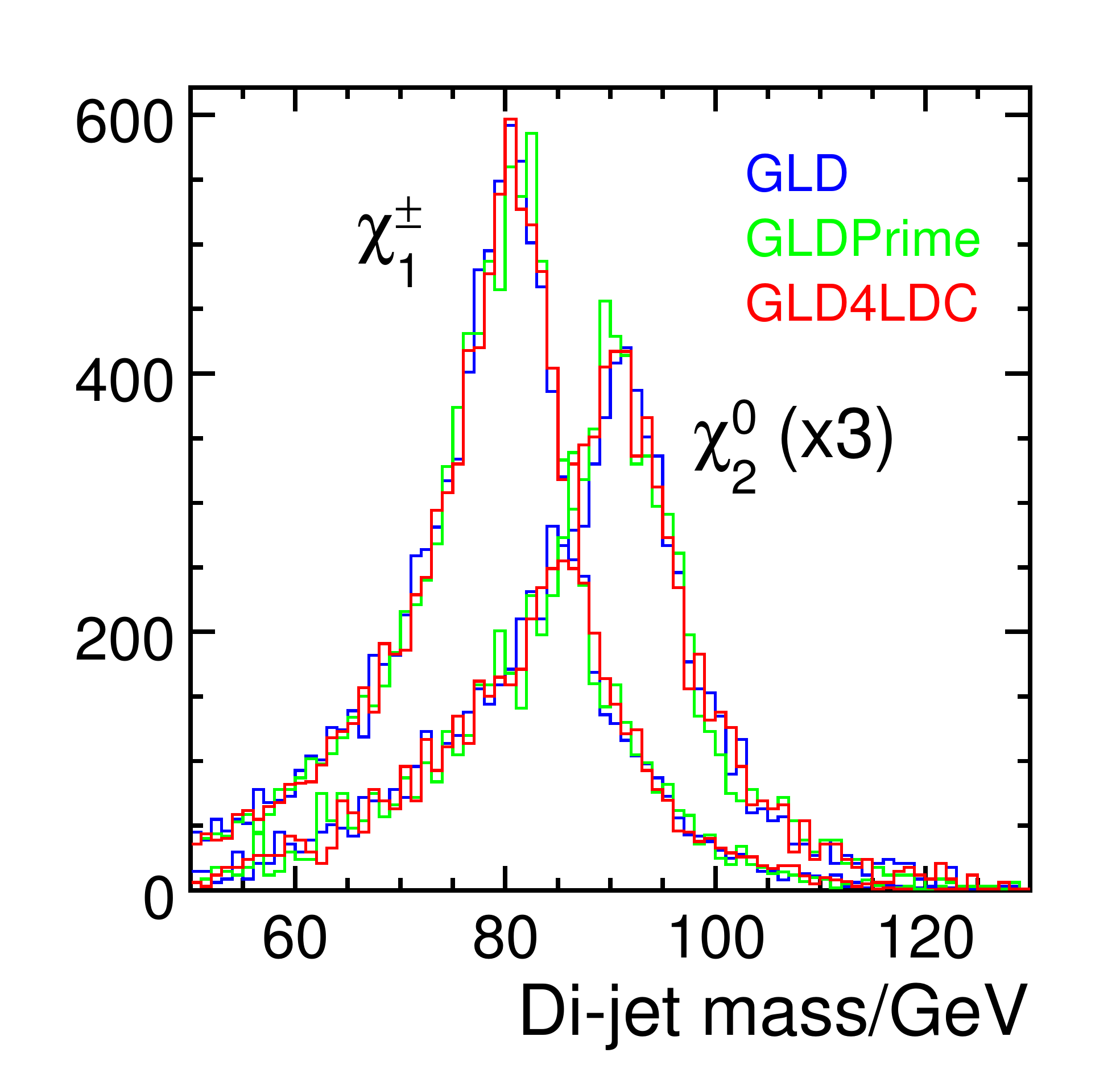}
\includegraphics[width=7.0cm]{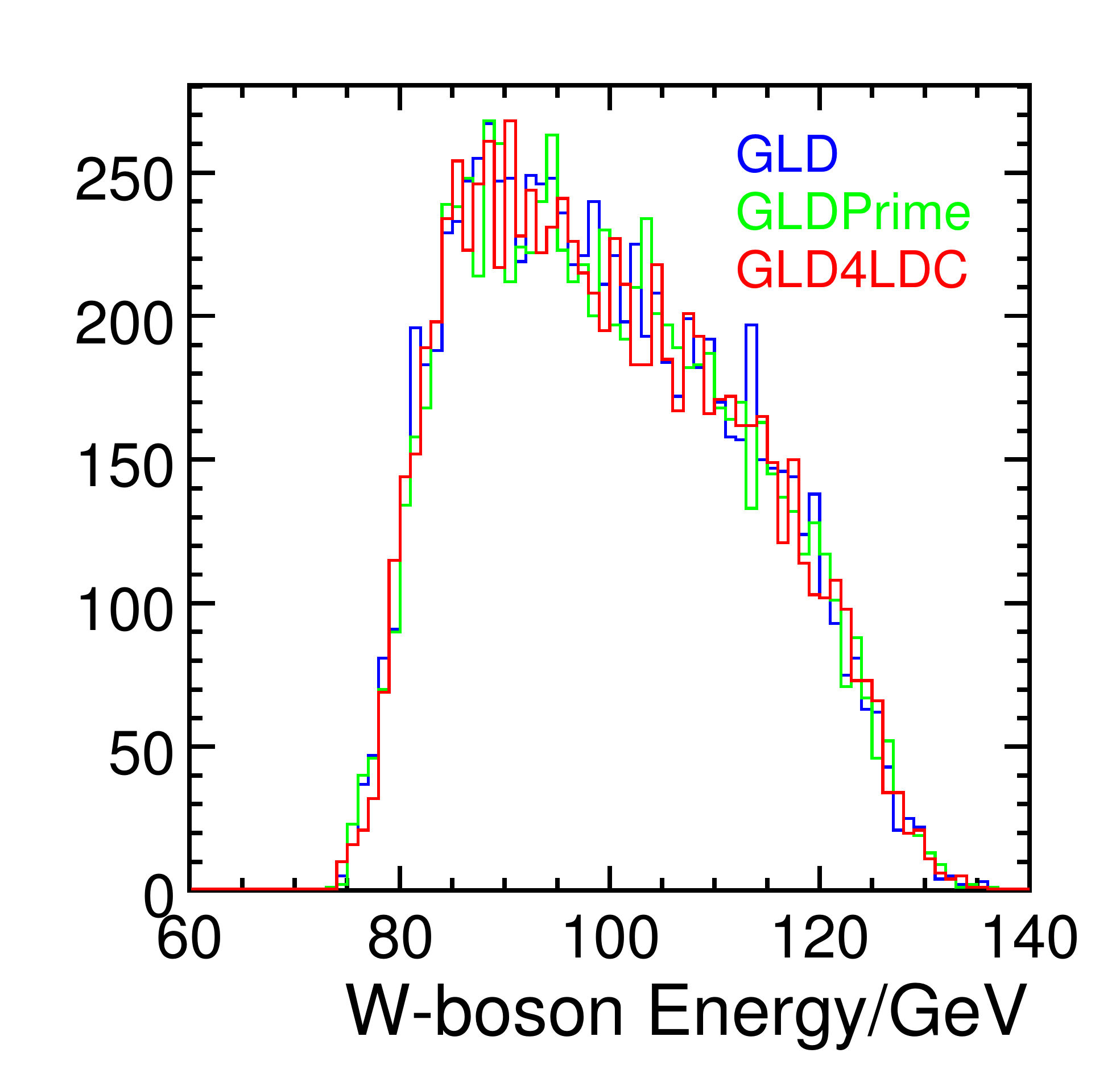}
\caption[W and $\Zzero$ mass and energy distributions.]{(a) The reconstructed masses of $W$ and $\Zzero$ bosons from the decays of $\chi_{1}^{\pm}$ and $\chi_{2}^{0}$. 
       (b) The energy distribution of the reconstructed $W$ bosons from $\chi_{1}^{\pm}$ decays.}
\label{fig:wz_mass}
\end{figure}

Chargino and neutralino pair production at $\roots=$500\,GeV
is studied in the context of SUSY point-5 benchmark 
scenario. The main signal is jets plus missing energy from
$\chi_{1}^{+} \chi_{1}^{-} \to \WpWm \chi_{1}^{0} \chi_{1}^{0}$ 
and $\chi_{2}^{0} \chi_{2}^{0} \to \Zzero\Zzero \chi_{1}^{0} \chi_{1}^{0}$.
The process  $\chi_{2}^{0} \chi_{2}^{0} \to \Zzero\Zzero \chi_{1}^{0} \chi_{1}^{0}$ is the main 
background to study $\chi_{1}^{+} \chi_{1}^{-} \to \WpWm  \chi_{1}^{0} \chi_{1}^{0}$ and {\it vice versa}.
The identification of the separate chargino and neutralino final states relies on the ability to 
distinguish $\WpWm$ from $\Zzero\Zzero$ 
and thus is sensitive to the jet energy resolution of the detector.
Figure \ref{fig:wz_mass}a shows the reconstructed invariant masses of hadronically decaying 
$W^{\pm}$ and $Z$ bosons from decays of $\chi_{1}^{\pm}$ and $\chi_{2}^{0}$, respectively. 
Neutralino and chargino event samples were   
separated based on the consistency of the reconstructed di-jet masses with the $Z$ and $W$
boson hypotheses. 
The selection efficiencies 
for $\chi_{1}^{\pm}$ and $\chi_{2}^{0}$ events for the GLD-based
detector models are summarised in Table \ref{tb:susy}. The different detector
models give statistically compatible selection efficiencies. This is consistent with the fact that
the differences in the jet energy resolutions for the three detector
models considered are at the level of $3-4$\,\% for the jet energy range $50-100$\,GeV
(Table~\ref{tab:pfa_models}).

Because $\chi_{1}^{\pm} \to W^{\pm} \chi_{1}^{0}$ and $\chi_{2}^{0} \to Z \chi_{1}^{0}$ are 
two body decays, the masses of $\chi_{1}^{\pm}$, $\chi_{2}^{0}$, and lightest SUSY particle (LSP), 
$\chi_{1}^{0}$, can be derived by using the energy distributions 
of the $W$ and $\Zzero$ bosons. The energy distributions of the reconstructed $W$ bosons are shown in 
Figure~\ref{fig:wz_mass}b. The different detector models result in very similar distributions 
and, consequently, have the same sensitivity to the $\chi_{1}^{\pm}$, $\chi_{2}^{0}$ and LSP masses.
 
\begin{table}
\centering
\begin{tabular}{|l|r|r|r|r|}
\hline
& \multicolumn{2}{|l|}{Chargino selection} & \multicolumn{2}{l|}{Neutralino selection} \\ \hline
      & Efficiency ($\chi_{1}^{\pm}$) & Efficiency ($\chi_{2}^{0}$) & Efficiency ($\chi_{1}^{\pm}$) & Efficiency ($\chi_{2}^{0}$)\\ \hline
GLD       & $47.9\pm0.3$\,\% & $1.0\pm0.1$\,\% & $11.2\pm0.5$\,\% & $33.8\pm0.6$\,\% \\
GLDPrime  & $48.4\pm0.3$\,\% & $1.0\pm0.1$\,\% & $11.4\pm0.5$\,\% & $33.3\pm0.6$\,\% \\
GLD4LDC   & $48.8\pm0.3$\,\% & $1.1\pm0.1$\,\% & $11.4\pm0.5$\,\% & $34.1\pm0.6$\,\% \\ \hline
\end{tabular}
\caption{The efficiency for $\chi_{1}^{\pm}$ and $\chi_{2}^{0}$ selection.\label{tb:susy}}
\end{table}

%% file: optimization/conclusions.tex

The studies described above informed the choice of parameters for the baseline ILD concept. The conclusions of these studies are:
\begin{itemize}\addtolength{\itemsep}{-0.5\baselineskip}
   \item {\bf B-field (vertex reconstruction):} The radius of the beam background envelope 
        scales as $B^{-0.5}$. This determines that the minimum acceptable 
        inner radius of the vertex detector goes from 
        $\sim14\,\mathrm{mm}\rightarrow16\,\mathrm{mm}$ for 
        $B=4\,\mathrm{T} \rightarrow 3\,\mathrm{T}$. The effect on impact parameter 
        resolution is $\lesssim 10\,\%$ and the resulting differences in flavour
        tagging efficiency are small ($\sim 2\,\%$). 
   \item {\bf B-field versus Radius (particle flow):} The confusion term in particle flow
        reconstruction scales as $R^{-1}$. This can be partially compensated 
        by the magnetic field, although the dependence is weak, $B^{-0.3}$.  
        For the entire range of detector parameters spanning the GLD and LDC concepts, 
        the ILC jet energy resolution requirements can be met. The differences in particle flow
        performance between the LDC and GLD parameters are small, $\lesssim6\,\%$,
        with the larger radius/lower field option being preferred. 
   \item {\bf B-field versus Radius (momentum resolution):} In terms of momentum resolution,
        the differences between the models considered are small $\lesssim10\,\%$. For
        high $p_\mathrm{T}$ tracks, larger radius/lower field detector is preferred. For low
        $p_\mathrm{T}$ tracks the opposite is true. All detector models considered here
        meet the ILC momentum resolution goals.
   \item {\bf TPC aspect ratio (particle flow):} the aspect ratio of the TPC
        ($R:z = 1:1.3$) used in the studies is close to optimum for particle flow;
        there is no significant advantage in a longer TPC and a shorter TPC would significantly degrade the performance in the forward
        region.
   \item {\bf ECAL Segmentation (particle flow):} The ECAL pixel size should be
        no greater than $10\times10$\,mm$^2$ in order to meet the ILC
        jet energy resolution goals for the jets relevant at $\roots=500$\,GeV. 
        Within the context of the current reconstruction,  
        $5\times5$\,mm$^2$ gives significant advantages over $10\times10$\,mm$^2$,
        particularly for higher energy jets.
   \item {\bf ECAL Segmentation (physics):} For the reconstruction of  
        tau decays, a $5\times5$\,mm$^2$ ECAL pixel size is 
        favoured over $10\times10$\,mm$^2$.
   \item {\bf Physics Performance:} The
        models considered give comparable physics performance.
        This is not surprising; the differences in 
        the underlying detector performance measures are small because
        the models trade-off $R$ against $B$ in such a way that {\it each}
        represents a reasonable detector choice.         
   \item {\bf HCAL Segmentation/Depth (particle flow):} For sufficient containment of 
        jets at $\roots=500$\,GeV, the HCAL should be between $5-6\lambda_I$. 
        The baseline for the
        ILD was chosen to be $6\lambda_I$ to ensure good jet containment
        for the highest energy jets and to allow for possible differences between    
        the simulation of hadronic showers and reality.
        For the current
        reconstruction,  there appears to be no significant advantage in going below
        $3\times3$\,cm$^2$. 
   \item {\bf Vertex Detector:} two detector layouts were considered: five single layers and
        six layers arranged in three doublets. Both conceptual designs  meet the
        ILC goals for impact parameter resolution, with the doublet structure giving
        an impact parameter resolution which is better, particularly
        for high momentum tracks. 
   \item {\bf SiW versus Scintillator-W ECAL:} results from studies of the strip 
        reconstruction and the
        resulting jet energy resolution of the Scintillator/Tungsten option, whilst
        promising, have yet to reach the level of sophistication where the performance
        of the strip based ECAL option can be fully evaluated. For this reason, the 
        SiW ECAL is used in the simulation of the ILD  for the physics
        studies in the next section.
   \item {\bf AHCAL versus DHCAL:} results from studies of the digital HCAL option are
        not yet at the level where its performance has been demonstrated. For this
        reason the AHCAL option with $3\times3$\,cm$^2$ tiles 
        is used in the simulation of ILD.
   \item {\bf Cost:} From the studies presented in this section it is clear any of the detector
        models listed in Table~\ref{tab:optdetparameter} are viable detectors for the
        ILC. For the same subdetector technologies, the differences in the costs for the  
        detector parameters considered are estimated to be $\sim10-20\,\%$; a large $B=3$\,T 
        detector is disfavoured on grounds of cost. However, given the large fluctuations
        in raw material costs (as seen in the last year) and the difficulty extrapolating
        detector sensor costs to the future, it is not yet possible to choose between the
        models on this basis. 
\end{itemize}

%% file: optimization/selection.tex
On the basis of the considerations above, the ILD detector parameters 
(listed Table~\ref{tab:optdetparameter}) are chosen to be close to those of 
the LDCPrime/GLDPrime models. 
The main arguments for the choices are as follows:
\begin{itemize}\addtolength{\itemsep}{-0.5\baselineskip}
   \item {\bf Choice of B-field:}  The operational magnetic field is chosen to
               be 3.5\,T, although it is assumed that the solenoid would be 
               designed for 4.0\,T to allow a safety margin in the mechanical  design. 
               This can be achieved without extrapolating significantly
               beyond the current CMS design. The arguments for a higher
               field are relatively weak: the benefits are marginal,
               and it would increase the cost of the detector. Whilst
               a lower $B$-field is not excluded, it is felt that until
               the likely ILC backgrounds and their impact on the ILD
               concept are better understood, it would be unwise to
               go to 3\,T.  
   \item {\bf Choice of Radius:} The ECAL inner radius is chosen to be 1.85\,m.
              The ILD concept is designed for particle flow 
              calorimetry and the jet energy performance is the main motivation
              for this choice.  For $B$=3.5\,T the gain in going
              to an ECAL radius of 2.0\,m is  modest and may not justify 
              the increase in cost. For a $B$=3.5\,T, the ILC jet energy
              goals suggest that the radius should be greater than $1.5-1.6$\,m.
              However, the studies presented above rely on the simulation of 
              hadronic showers. By selecting a detector radius of 1.85\,m, 
              it is likely that the ILD concept will meet the ILC jet
              energy goals, even if the current performance estimates are
              on the optimistic side.
   \item {\bf Choice of Sub-detector Technologies:} At this stage we are not
              in a position to choose the ECAL, HCAL and vertex detector
              technologies. All options are considered on an equal
              basis. Nevertheless, for the physics studies that follow it
              is necessary to define a baseline for the simulation. The
              six layer (three doublets) vertex detector layout is used
              on the basis that it gives the best impact parameter resolution.
              For the calorimetry, the SiW ECAL and the AHCAL are used in the
              simulation as they have been well studied and
              we are confident that they give the desired 
              jet energy resolution.
              The strip-based ECAL and DHCAL will be actively supported in
              simulation and software with the intention of evaluating their
              ultimate performance. 
\end{itemize}

The optimisation of the ILD parameters was performed in a 
rigorous manner using information from a number of detailed studies. 
On this basis, we are confident that the ILD concept is 
well optimised for physics at the ILC operating in the 
energy range 200\,GeV$-$1\,TeV.

%% file: performance/performance.tex
The performance of ILD is established using
a detailed GEANT4 model and full reconstruction of the simulated
events. Both detector performance measures and physics analyses are
studied. Whilst the simulation and reconstruction are not perfect,
they are at least as sophisticated as those used
in the majority of studies for previous large collider detector 
TDRs.

\section{Software for ILD Performance Studies}
\label{sec:performance-simulation}
\input{performance/simulation}

\section{ILD Detector Performance}
\label{sec:performance-detector}
\input{performance/performance-detector}

\section{Physics Performance}
\label{sec:performance-physics}
\input{performance/physics}

\section{Other Studies}
\label{sec:performance-other}
\input{performance/performance-other}

\section{Conclusions}
\label{sec:performance-conclusions}
\input{performance/conclusions}

%% file: performance/simulation.tex
\begin{figure}[t]
\begin{center}
\begin{tabular}{c c}
\includegraphics[height=6.5cm]{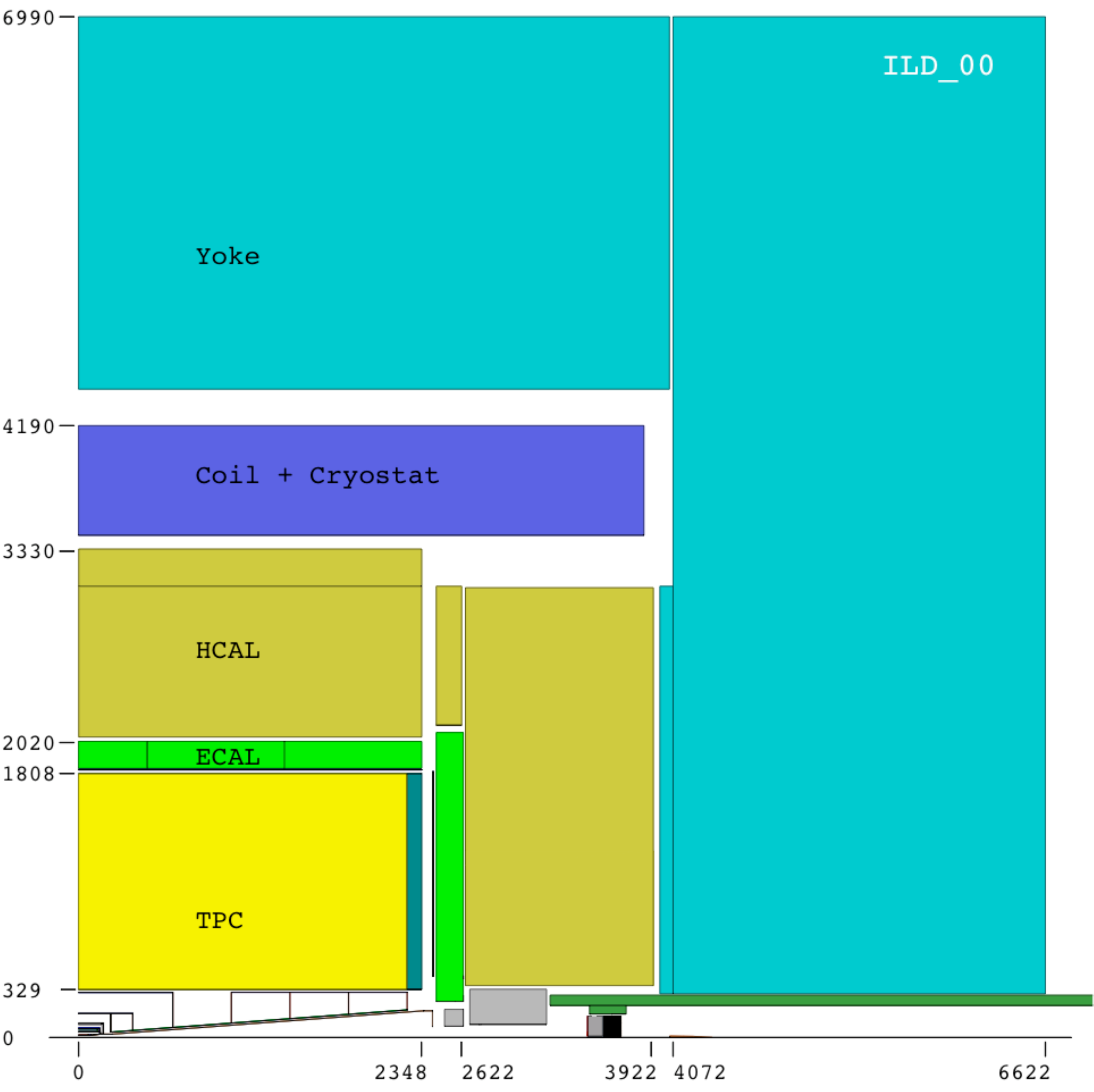}
\includegraphics[height=6.5cm]{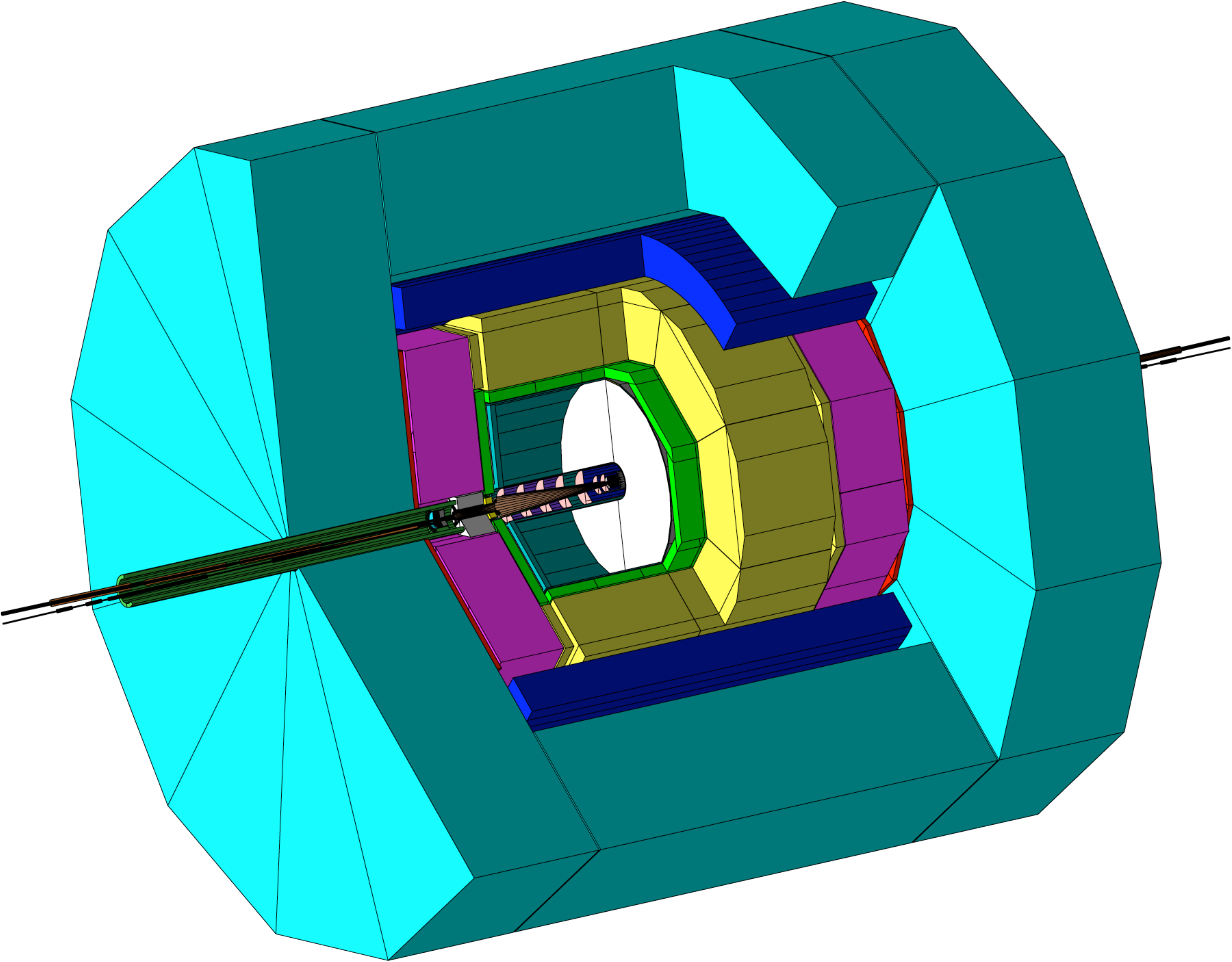}
\end{tabular}
\label{fig:performance-ild_00}
\caption[The ILD00 detector.]{The ILD\_00 detector model as implemented in Mokka. 
From the inside to the outside, the detector components are the: VTX, SIT, TPC, SET, ECAL, HCAL and 
Yoke. In the forward region the FTD, ETD, LCAL, LHCAL and BCAL are shown.}
\end{center}
\end{figure}

To demonstrate the physics capabilities of ILD,
more than 30 million Monte Carlo (MC) events have been fully simulated 
and reconstructed for the benchmark reactions \cite{ref:wws_benchmark} 
and other physics channels of interest at the ILC.
Signal samples typically correspond to an integrated luminosity of 500~fb$^{-1}$ or more.
These are combined with sufficiently large sets of SM events for
background estimation. The ``simulation reference ILD detector model'', 
ILD\_00, is 
implemented in Mokka. The silicon based tracking detectors are 
modelled with the appropriate material thicknesses and support 
structures without specifying the exact readout technology. Instead,
in the digitisation stage,  simulated hits are smeared by the 
effective point resolutions listed in 
Table~\ref{tab:pointres}. These represent the most realistic estimates from 
the relevant subdetector R\&D groups.
The SiW option with $5\times5$~mm$^2$ transverse cell size 
and the Steel-Scintillator option with 
$3\times3$\,cm$^2$ tiles are used for the ECAL and HCAL respectively. 
As discussed in 
Section~\ref{sec:optimisation-particleflow-technology}, these are the
most mature of the technology options in terms of simulation 
and reconstruction; this does not imply any pre-decision on the 
ultimate technology choice. The main parameters of the ILD\_00 model 
are listed in Table~\ref{tab:optdetparameter} and a drawing of this model 
is shown in Figure~\ref{fig:performance-ild_00}. 
Further details of the geometrical parameters can be found 
in~\cite{ref:ild_ref_model}. 

\begin{table}[htb]
\centering\small
\scriptsize
\begin{tabular}{|l|c|c||l|c|c|}
\hline         & $\sigma_{r-\phi}/{\mu \mathrm{m}}$    & $\sigma_{z}/{\mu \mathrm{m}}$   &  & $\sigma_{r-\phi}/{\mu \mathrm{m}}$    & $\sigma_{z}/{\mu \mathrm{m}}$  \\     
\hline   VTX     &    2.8     &   2.8     &   FTD      &  5.8    &  5.8    \\
\hline   SIT/SET &   7.0      &   50.0    &   ETD      &  7.0    &  7.0    \\
\hline  TPC     & \multicolumn{5}{|l|}{$ \sigma^2_{r\phi}=50^2+900^2\sin^2\phi + \left( (25^2/22)\times (4/B)^2\sin\theta\right) (z/\mathrm{cm}) \,\mu\mathrm{m}^2$ } \\
                 & \multicolumn{5}{|l|}{$ \sigma^2_{z}=400^2+80^2\times (z/\mathrm{cm}) \,\mu\mathrm{m}^2 $} \\                                               
\hline 
\end{tabular}
\caption[Simulated tracking point resolutions.]{Effective point resolutions used in the digitisation of the MC samples.\label{tab:performance_pointres} }
\label{tab:pointres}
\end{table}

Most of the subdetectors in the ILD\_00 model have been implemented including  
a significant amount of  engineering detail such as mechanical support structures, electronics 
and cabling as well as dead material and cracks.
This provides a reasonable estimate of the material budget and 
thus the effect of multiple scattering in the tracking detectors; it is 
also crucial for a realistic demonstration of particle flow performance.
In the simulation, the vertex detector has a staggered layout of 
six 50~$\mu$m thick silicon ladders and corresponding support structures.
The additional silicon tracking 
detectors, FTD/ETD (forward Si tracking) and SIT/SET (inner and outer Si tracking), 
are modelled as disks and cylinders respectively. 
The material thicknesses for these detectors give the effective radiation 
lengths listed in Table~\ref{table1}. 
The TPC model includes the material in the inner and outer field cage,
this corresponds to a total of 4.5\,\%~$X_0$ including the drift gas mixture. 
The hits from charged particles were simulated according to an end-plate 
layout with 224 rows of pads, 1\,mm wide and 6\,mm high. 
The ECAL simulation includes the alveolar structure  and also the dead regions
around the silicon pixels and between the modules. The HCAL simulation includes
steel and aluminium support structures which result in dead regions.
The energy response of the scintillator tiles was corrected according to 
Birks' law. The simulation of the forward region
includes realistic support structures and the shielding masks. 
All subdetectors are enclosed by a dodecagonal iron yoke,
instrumented in the simulation with ten layers of RPCs. 
The superconducting coil and cryostat are simulated as a 750\,mm thick aluminium cylinder,
corresponding to 1.9\,$\lambda_I$.

All events are reconstructed using 
the Kalman-Filter based track reconstruction in MarlinReco, 
the PandoraPFA particle flow algorithm and the LCFIVertex 
flavour tagging. The flavour tagging artificial neural networks (ANNs) 
have been trained using 
fully simulated and reconstructed ILD\_00 MC events.
The boost resulting from the 14\,mrad crossing angle is taken into account 
in the analyses that use BCAL hit distributions as an electron veto.

%% file: performance/performance-detector.tex
\subsection{ILD Tracking Performance}

\label{sec:performance-detector-tracking}
\input{performance/performance-tracking}

\subsection{Background Studies}
\label{sec:performance-detector-background}
\input{performance/performance-background}

\subsection{ILD Flavour Tagging Performance}
\label{sec:performance-detector-flavourtag}
\input{performance/performance-flavourtag}

\subsection{ILD Particle Flow Performance}
\label{sec:performance-detector-particleflow}
\input{performance/performance-particleflow}

%% file: performance/performance-tracking.tex
The tracking system envisaged for ILD consists of three 
subsystems each capable of standalone tracking VTX, FTD and the TPC. 
These are augmented by three auxiliary tracking systems the SIT, SET and ETD, 
which provide additional high resolution 
measurement points. The momentum resolution goal~\cite{Perf:2005TimB} is
$$\sigma_{1/{p_T}} \approx 2 \times 10^{-5}\,\rm{GeV}^{-1},$$
and that for impact parameter resolution is
$$\sigma_{r\phi} = 5\,\mu\rm{m} \oplus \frac{10}{  p(\mathrm{GeV}) \sin^{3/2}\theta }\,\mu \rm{m}.$$

\subsubsection{Coverage and Material Budget}

\begin{figure}[b]
\begin{center}
\includegraphics[width=7.0cm]{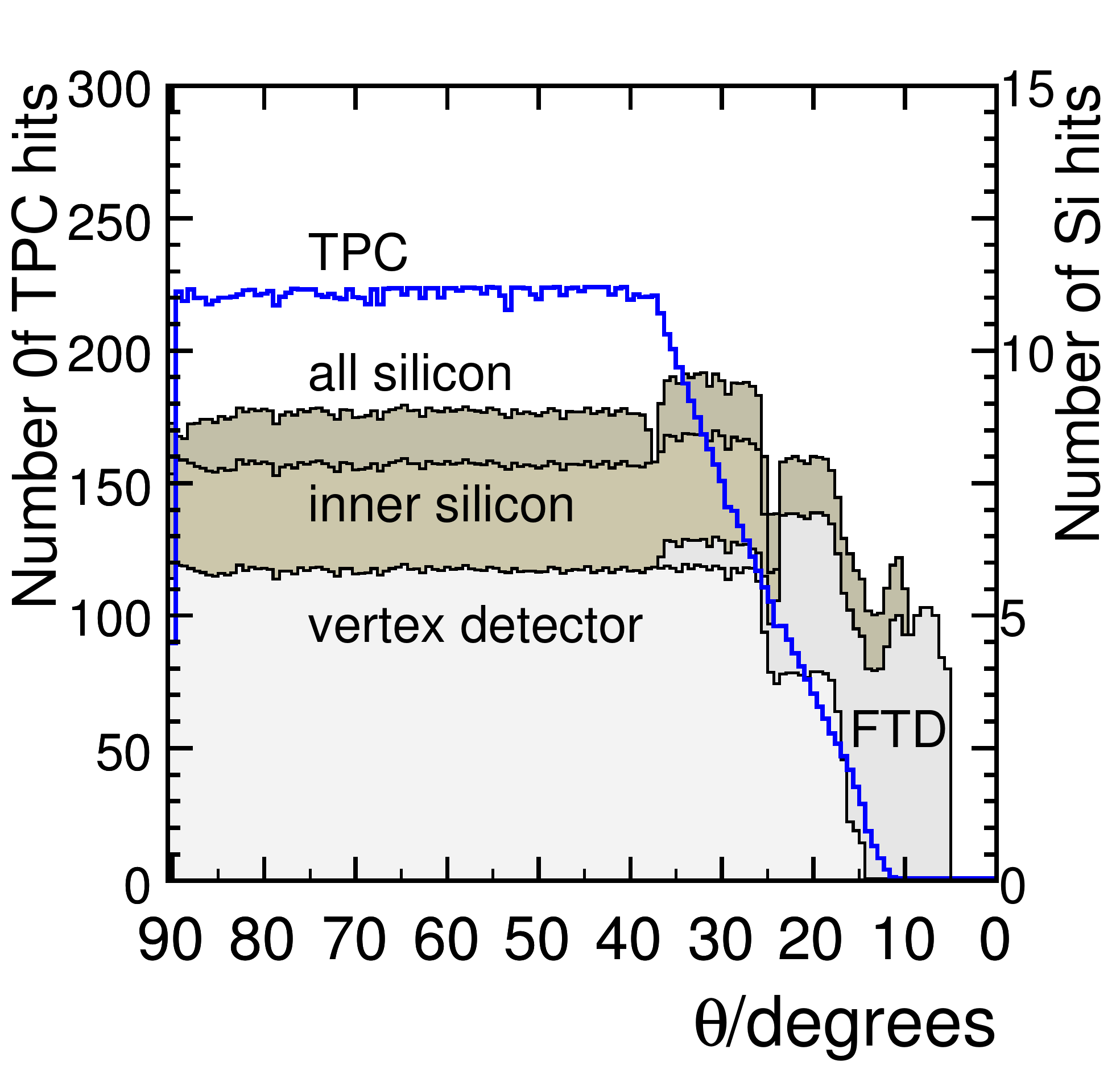}
\includegraphics[width=7.0cm]{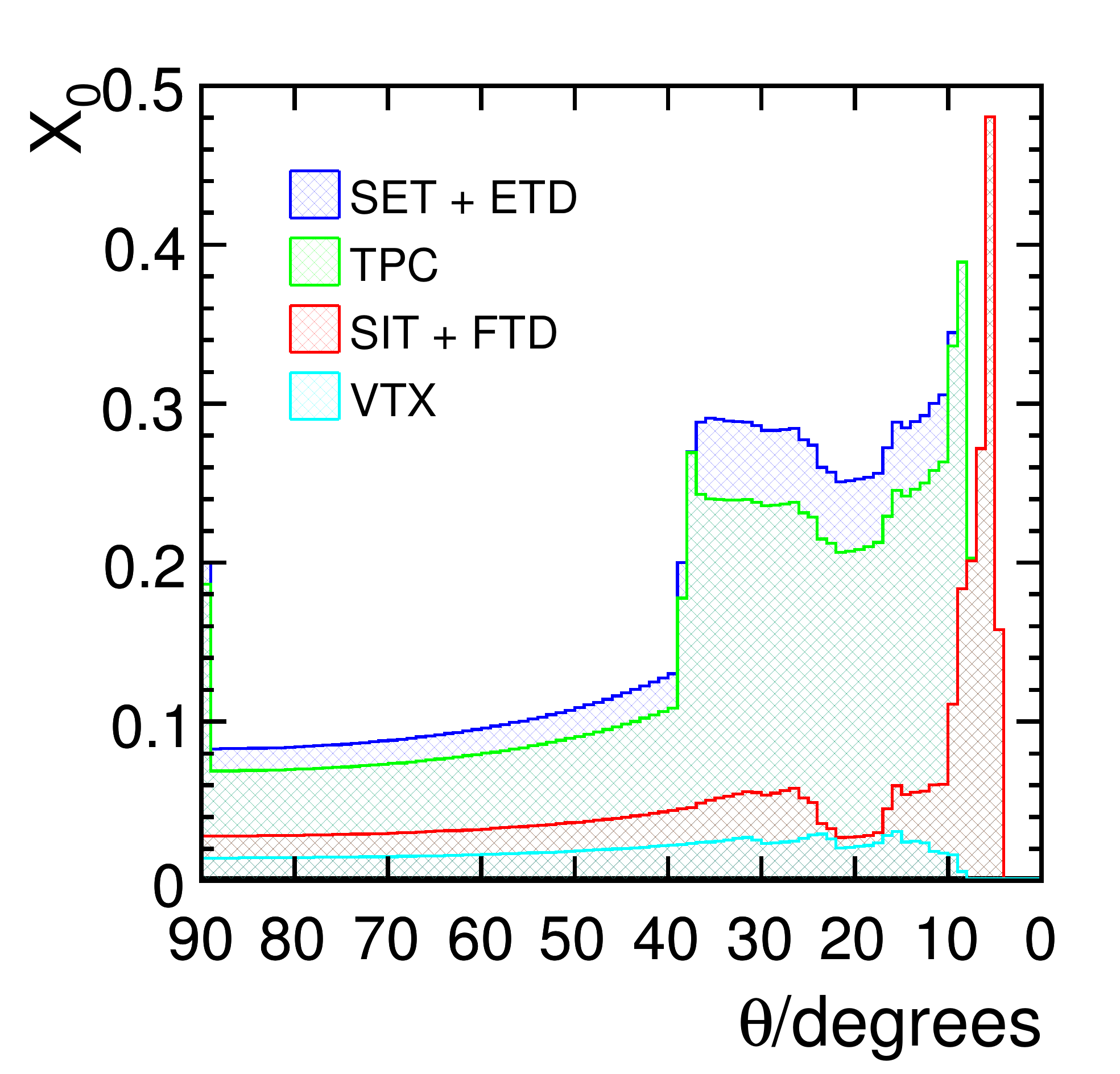}
\caption[Numbers of hits and Material budget of the ILD tracking system.]{
   a) Average number of hits for simulated charged particle tracks as a function of polar angle.  
   b) Average total radiation length of the material in the tracking detectors as a function of
      polar angle.}
\label{fig:material}
\end{center}
\end{figure}

Figure~\ref{fig:material}a shows, as a function of polar angle, $\theta$, 
the average number of reconstructed hits associated with simulated 100\,GeV muons. 
The TPC provides full coverage down to  $\theta = 37^{\circ}$. Beyond this the number 
of measurement points decreases. The last measurement point provided by the 
TPC corresponds to $\theta\approx10^{\circ}$. The central inner tracking system, 
consisting of the six layer VTX and the two layer SIT, provides eight precise measurements 
down to $\theta = 26^{\circ}$. The innermost and middle double layer of the VTX extend 
the coverage down to $\theta \sim 16^{\circ}$. The FTD provides up to a maximum of 
five measurement points for tracks at small polar angles. 
The SET and ETD provide a single high precision measurement point 
with large lever arm outside of the TPC volume down to a $\theta \sim 10^{\circ}$. 
The different tracking system contributions to the detector material budget, including 
support structures, is shown in Figure~\ref{fig:material}b.

\subsubsection{Momentum Resolution for the Overall Tracking System}

\begin{figure}[b]
\begin{center}
\includegraphics[width=7.0cm]{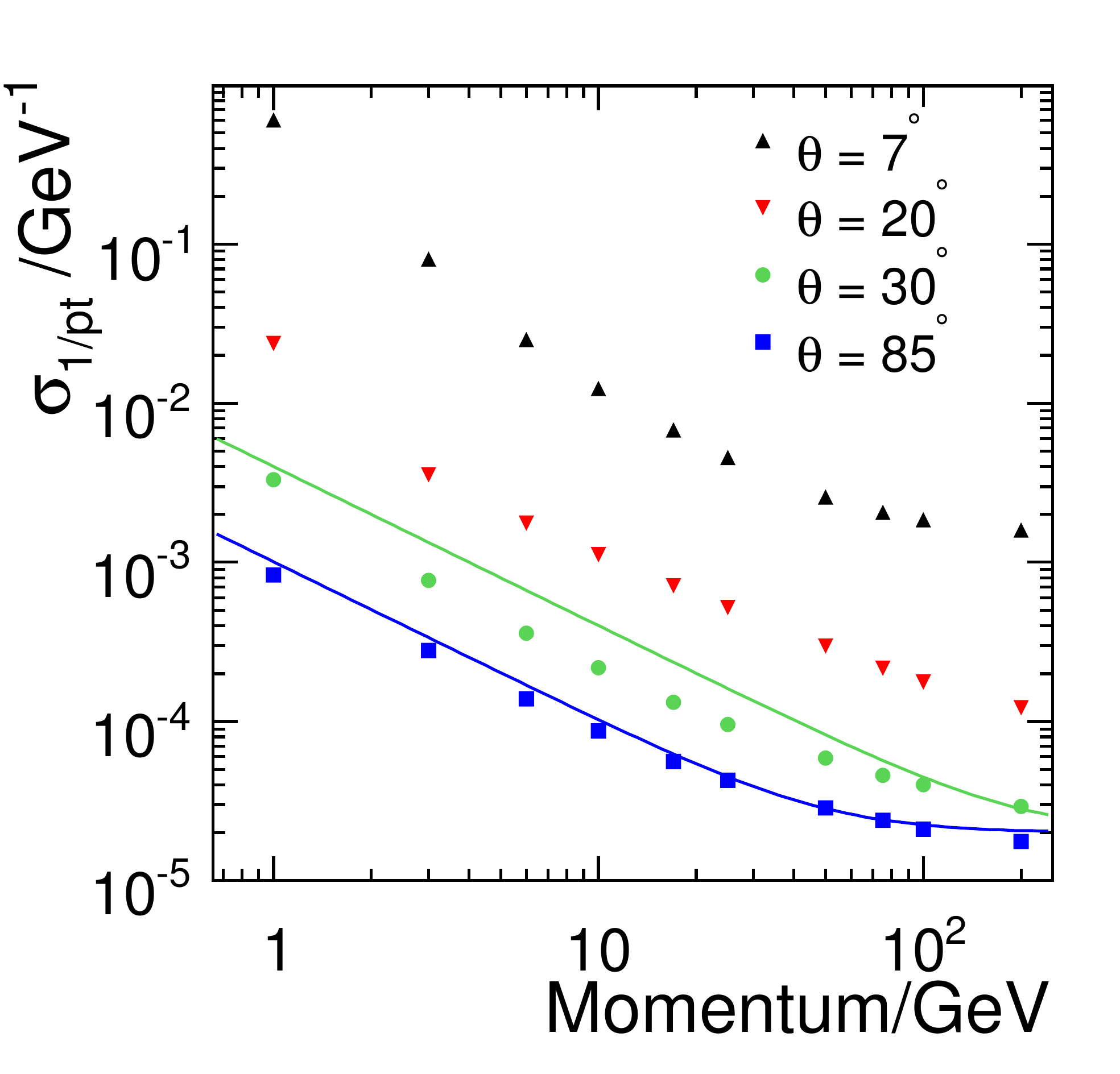}
\includegraphics[width=7.0cm]{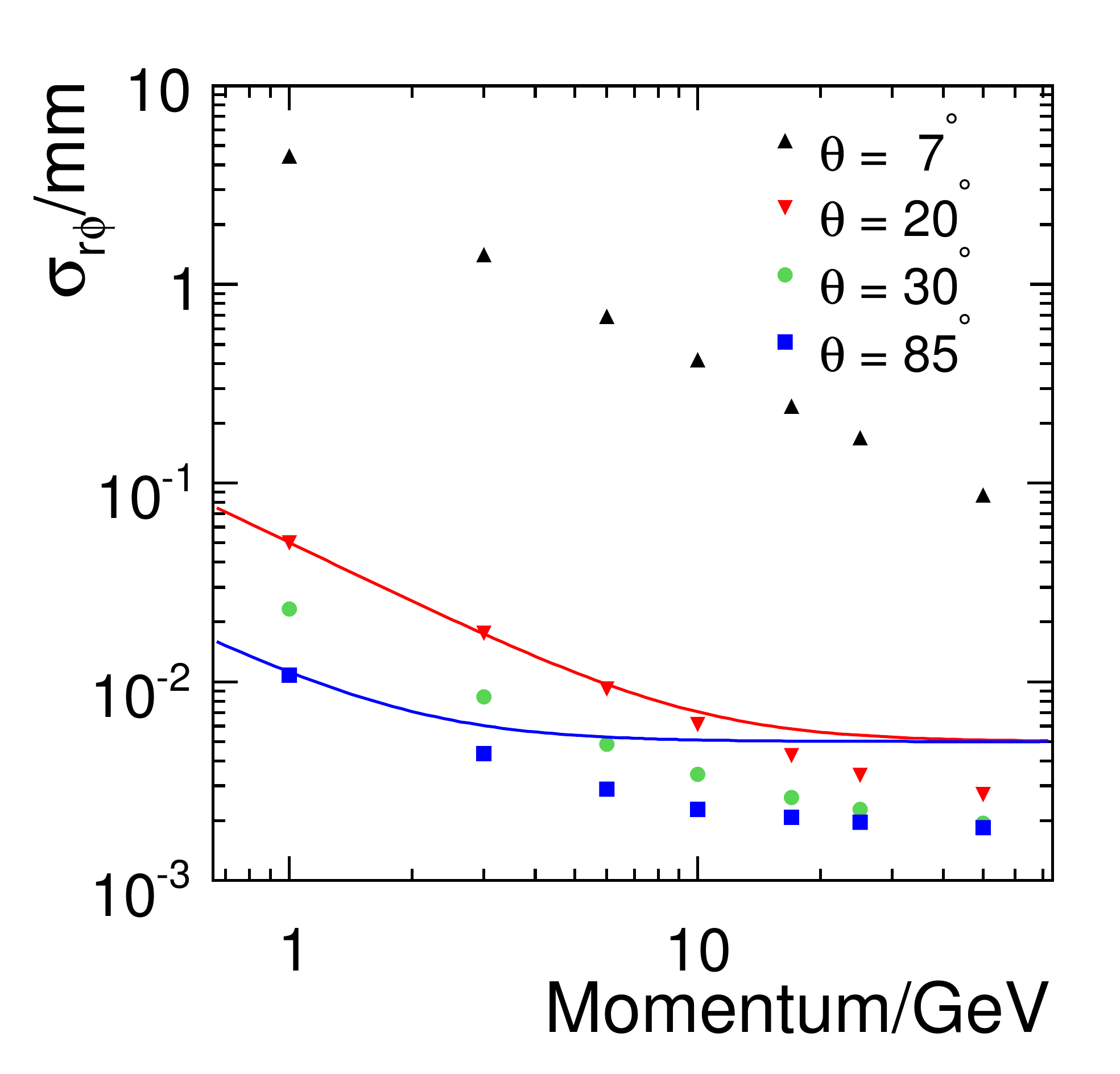}
\caption[Tracking performance.]{a) Transverse momentum resolution for muons plotted versus polar angle for four different simulated muon momenta. The lines show 
         $\sigma_{1/{p_T}} = 2\times 10^{-5} \oplus 1\times 10^{-3}/(p_{T}\sin\theta)$
         for $\theta=30^\circ$ (green) and $\theta=85^\circ$ (blue).
         b) Impact parameter resolution for muons versus
         polar angle for four different simulated muon momenta. }
\label{fig:TrackPerfMomentum}
\end{center}
\end{figure}

The momentum resolution achieved with the ILD simulation and full reconstruction
is shown in Figure~\ref{fig:TrackPerfMomentum}a. 
The study was performed using muons generated at fixed polar 
angles of $\theta = 7^{\circ},  20^{\circ},  30^{\circ}$ and $85^{\circ}$, and 
the momentum was varied over the range  $1-200$\,GeV. 
For two polar angles, this is 
compared to the expected parametric form of, 
$\sigma_{1/{p_T}} = a \oplus b/(p_{T}\sin\theta)$, with 
$a= 2\times 10^{-5}$\,GeV$^{-1}$ and $b=1\times 10^{-3}$.
As can be seen, at a polar angle of $85^{\circ}$, the required 
momentum resolution is attainable over the full momentum range from 1 GeV upwards, 
this remains true over the full length of the barrel region of the detector, where
the TPC in conjunction with the SET is able to provide the longest possible radial
lever arm for the track fit. 
For high momentum tracks, the asymptotic value of the momentum resolution 
is $\sigma_{1/p_T} = 2 \times 10^{-5}$\,GeV$^{-1}$.
 At $\theta = 30^{\circ}$, the SET no longer contributes, the effective lever-arm of the
 tracking system is  reduced by 25\,\%. Nevertheless, the momentum resolution is still 
 within the required level of performance.
 In the very forward region, the momentum resolution is inevitably worse due to 
 the relatively small angle between the $B$-field and the track momentum.

\subsubsection{Impact Parameter Resolution}

Figure~\ref{fig:TrackPerfMomentum}b shows $r\phi$ impact parameter resolution 
as a function of the track momentum.
The required performance is achieved down to a track momentum of 1\,GeV,
whilst it is  exceeded for high momentum tracks where the asymptotic resolution is 2\,$\mu$m. 
The $rz$ impact parameter resolution (not shown) is better than $\sim 10\,\mu$m down to momenta of 3\,GeV
and reaches an asymptotic value of $<5\,\mu$m for the whole barrel region. Because of the 
relatively large distance of the innermost FTD disk to the interaction point, the 
impact parameter resolution degrades for very shallow tracks, $\theta < 15^{\circ}$.
It should be noted that these studies do not account for the possible mis-alignment of the tracking systems.



\subsubsection{Tracking Efficiency}

With over 200 contiguous readout layers, pattern recognition and track reconstruction in a 
TPC is relatively straightforward, even in an environment with a large number of background hits.
In addition, the standalone tracking capability of the VTX enables the reconstruction of
low transverse momentum tracks which do not reach the TPC. 
Hermetic tracking down to
low angles is important at the ILC~\cite{TrackPerf:Foward800GeV} and
the FTD coverage enables tracks to be reconstructed to 
polar angles below $\theta = 7^{\circ}$. 

\begin{figure}
\begin{center}
\includegraphics[width=7.0cm]{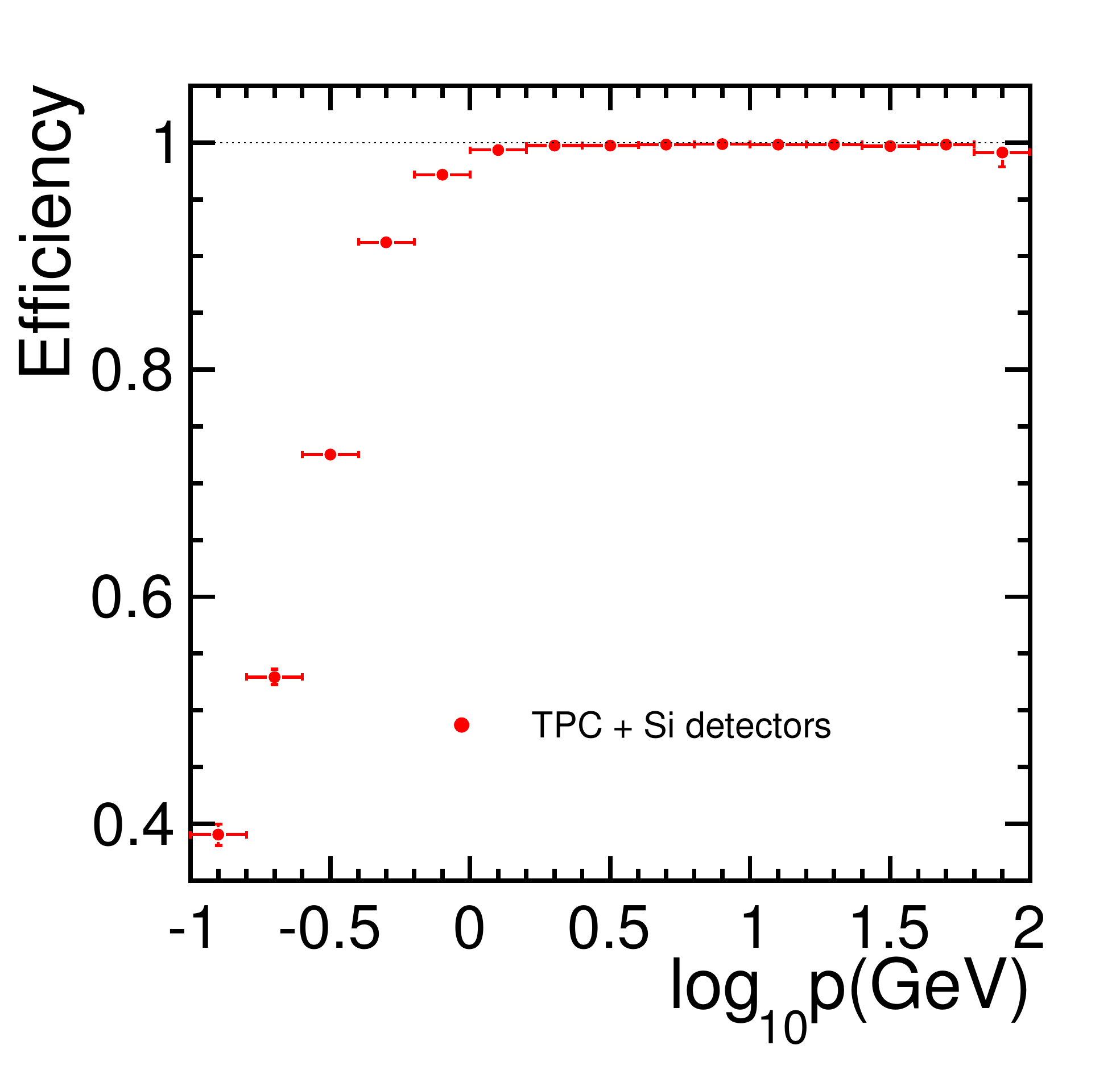}
\includegraphics[width=7.0cm]{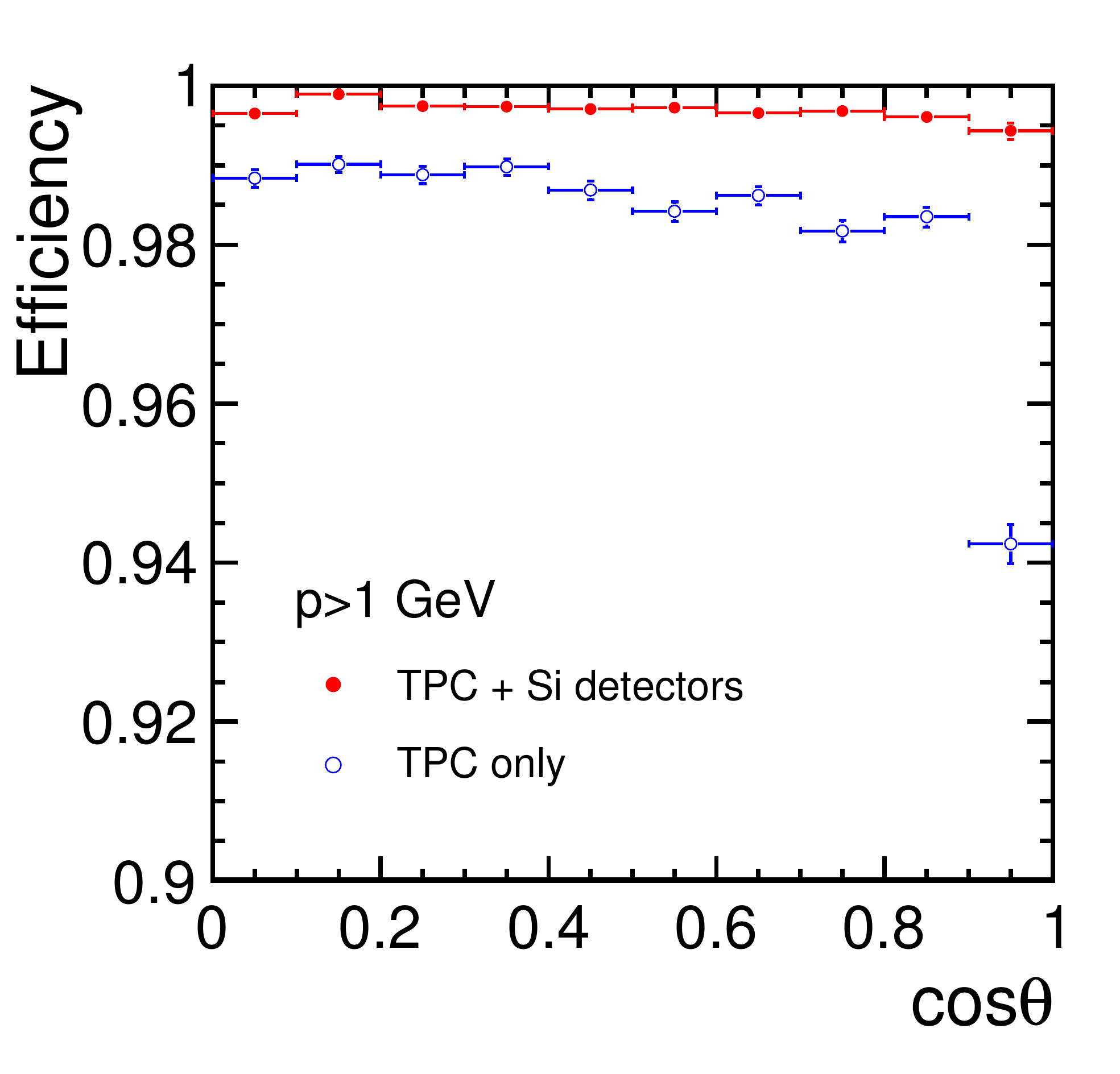}
\caption[Track finding efficiency.]{Tracking Efficiency as a function for $t\bar{t}\rightarrow$ 6 jets at 500GeV
         plotted against a) momentum and b) $\cos\theta$. Efficiencies are plotted with respect 
         to MC tracks which leave at least 4 hits in the tracking detectors including
         decays and $V^0$s.      
\label{fig:trackeff}}
\end{center}
\end{figure}

Figure~\ref{fig:trackeff} shows, as a function of momentum and polar angle, 
the track reconstruction efficiency in simulated (high multiplicity) 
$t\bar{t}\rightarrow$ 6 jet events at $\roots=$500\,GeV.
For the combined tracking system, the track reconstruction efficiency is approximately 
99.5\,\% for tracks with momenta greater than 1\,GeV across almost the entire polar angle range. 
The effects of background from the machine and from multi-peripheral 
$\gamma\gamma\rightarrow\rm{hadrons}$ events are not yet taken 
into account; dedicated studies form part of the ongoing simulation R\&D effort.
Nevertheless, a study of the TPC reconstruction efficiency as a function of the 
noise occupancy is described in Section~\ref{ild:tpc}. This demonstrates that 
there is no loss of efficiency for 1\,\% noise occupancy.

%% file: performance/performance-background.tex
The studies presented above do not include the effects of background from the machine 
and from multi-peripheral $\gamma\gamma\rightarrow\rm{hadrons}$ events. The impact
of machine background has been studied in the context of the ILD concept. These 
studies are based on the expected simulated detector hits from approximately 2000 
bunch crossings (BXs). The hits are super-imposed on simulated physics events taking
into account the 369\,ns bunch structure of the ILC and conservative estimates of the 
readout rates of the tracking detector components.

\subsubsection{Background in the TPC}

For a conservative value for the TPC gas drift velocity, 4\,cm\,$\mu$s$^{-1}$, the 
maximum TPC drift length of 2.25\,m corresponds to 150 BXs. 
Nominal background in the TPC is thus simulated as 150 BXs
appropriately shifted in $z$. 
Prior to the reconstruction, nearby 
hits are merged taking into account the expected 
$r\phi$ and $z$ extent of the charge cloud. For the TPC readout assumed
for ILD, 150 BXs of beam-related background correspond to a voxel occupancy
of approximately $0.05\,\%$ (the TPC voxel size is taken to be 1\,mm in the $\phi$ direction, 
6\,mm in $r$ and 5\,mm in $z$).

\begin{figure}[!b]
\centering
\includegraphics[height=6.6cm]{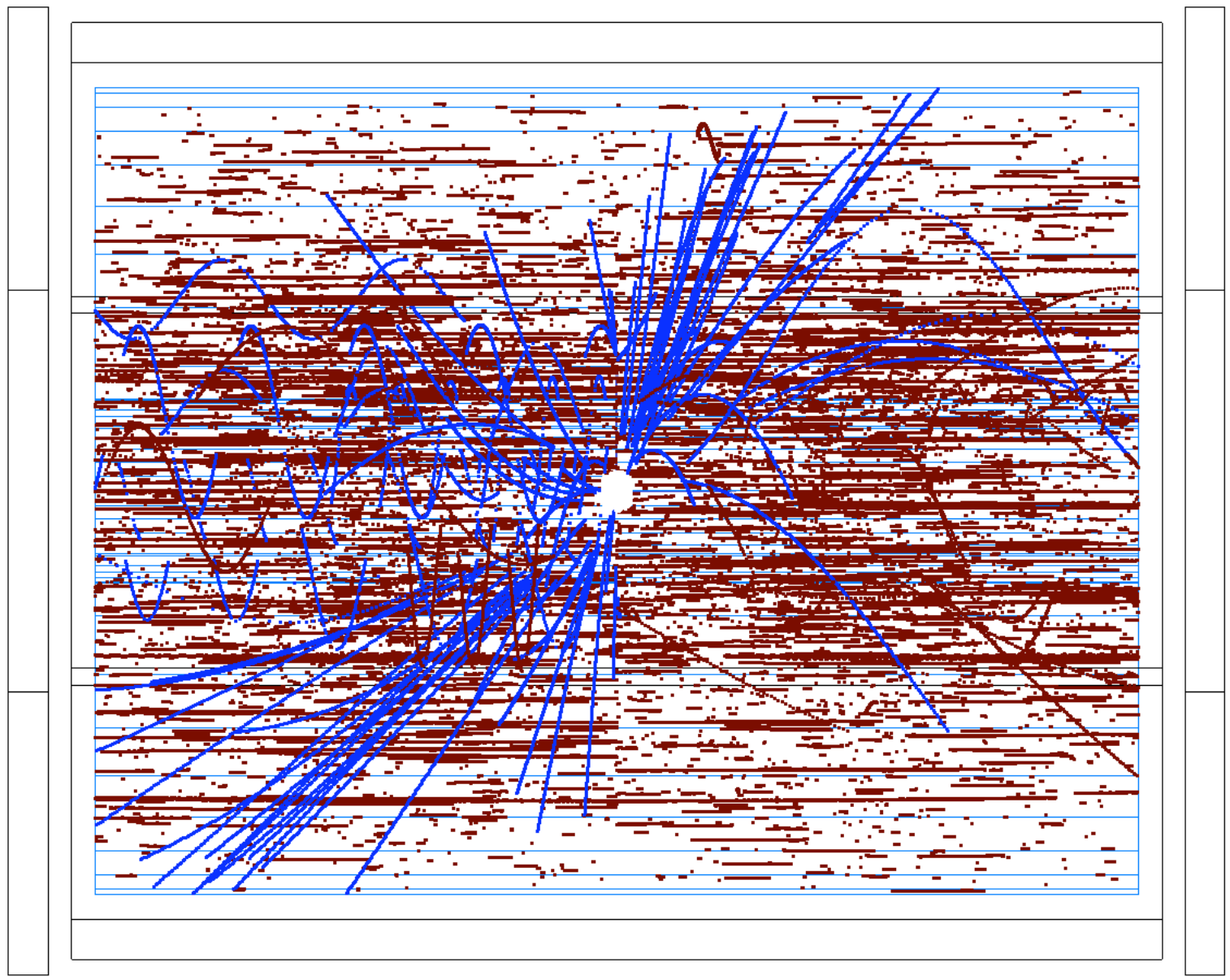}
\includegraphics[height=6.6cm]{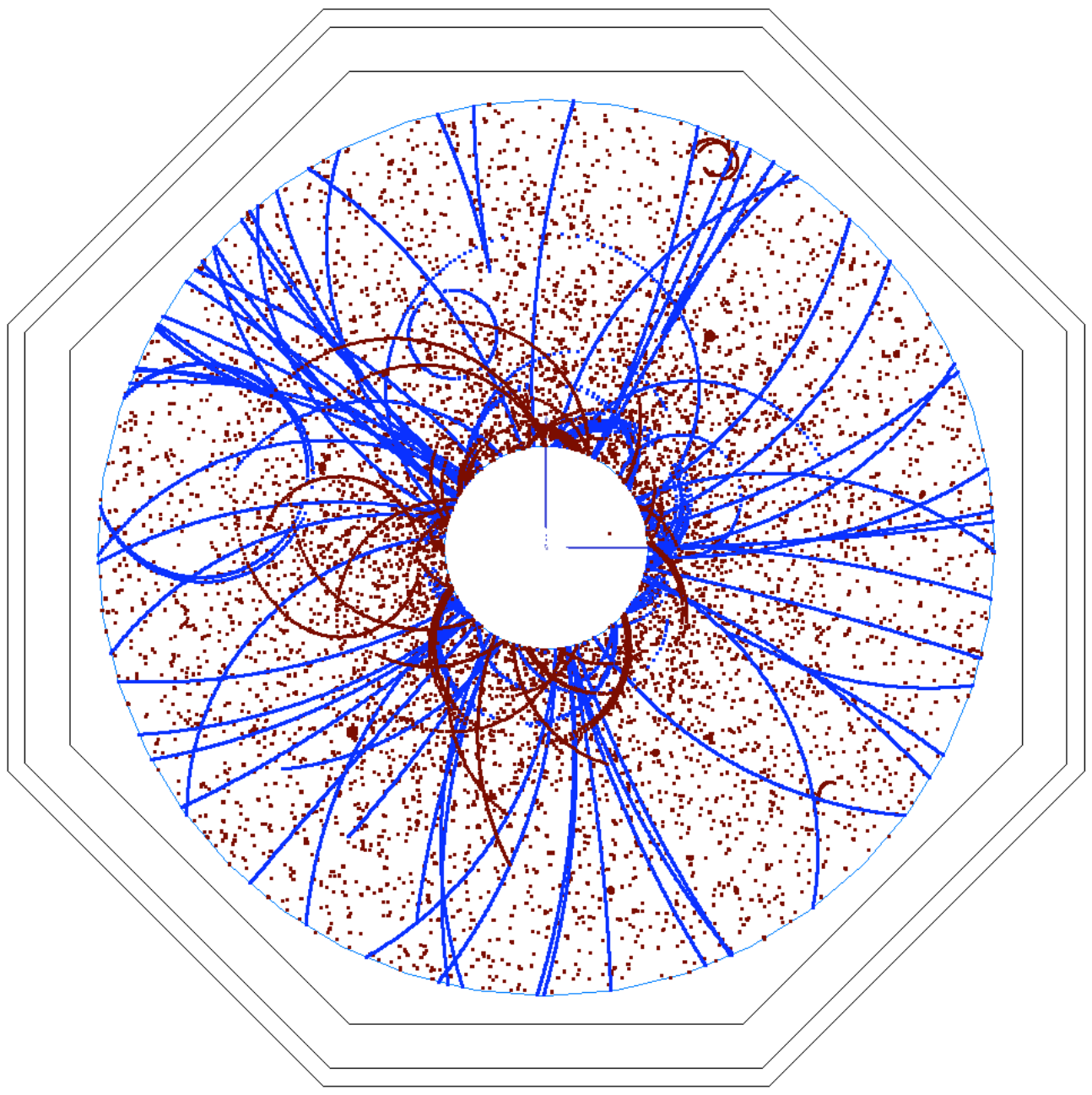}
\caption[TPC hits with 150 BXs background.]{The $rz$ and $r\phi$ views of 
   the TPC hits from a 500\,GeV $\ttbar$ event (blue) with 150 BXs of beam background (red) 
    overlayed.
\label{fig:tpcraw}}
\end{figure}

\begin{figure}[!t]
\centering
\includegraphics[height=6.6cm]{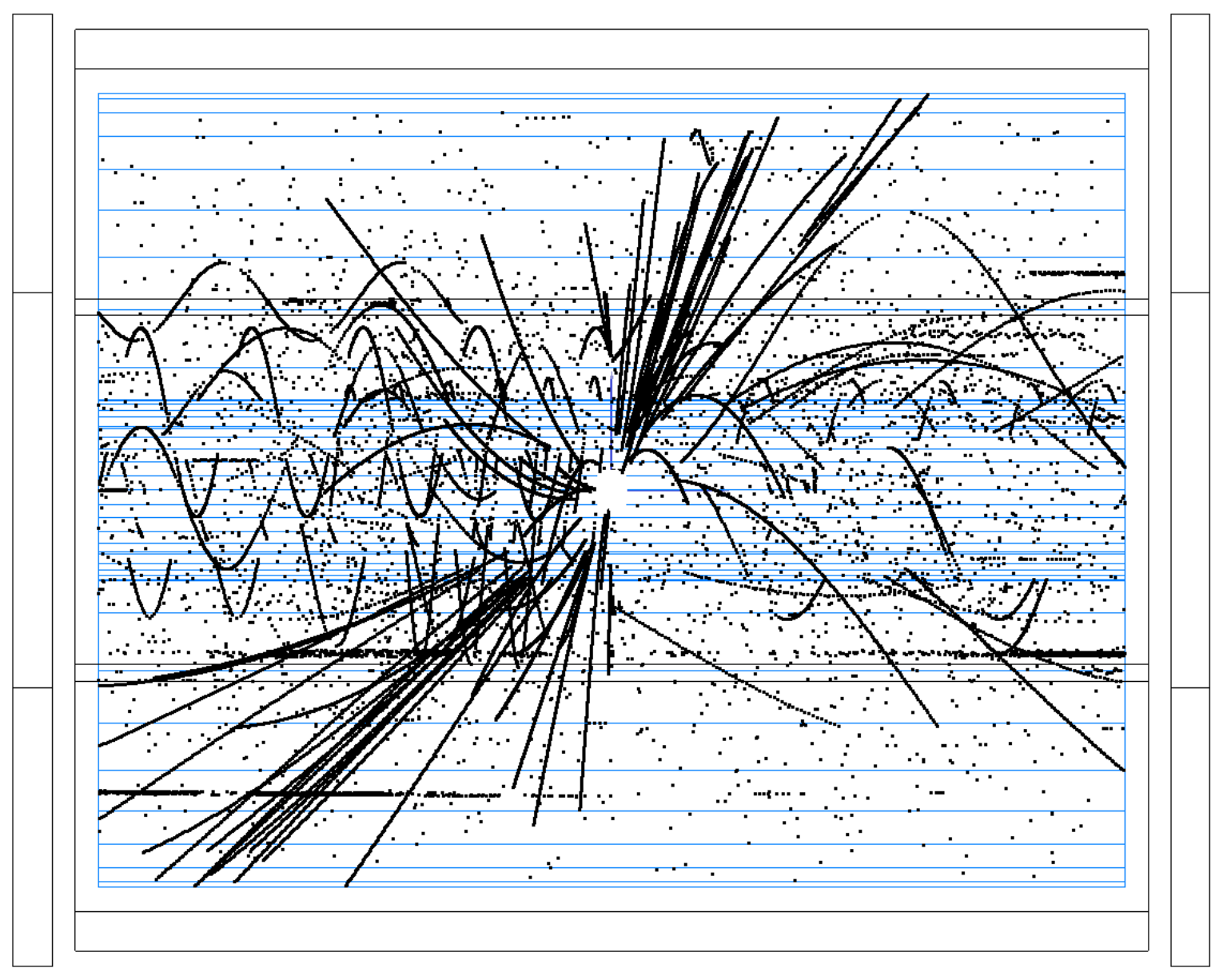}
\includegraphics[height=6.6cm]{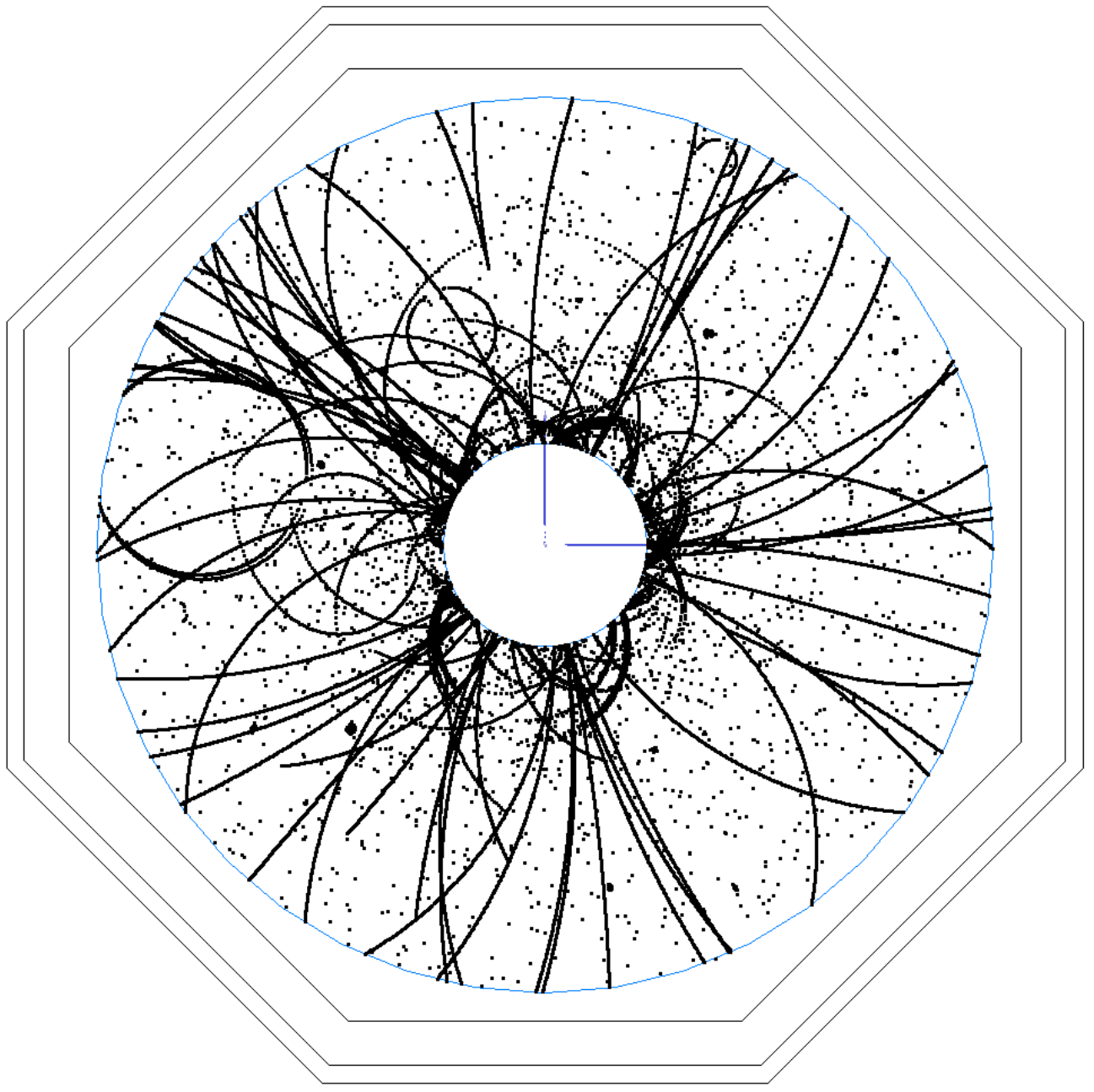}
\caption[TPC hits with background after micro-curler removal.]{The same event
   as the previous figure, with the micro-curler removal algorithm applied. This is
   the input to the TPC track finding algorithm.
\label{fig:tpcpatrec}}
\end{figure}

Figure~\ref{fig:tpcraw} shows the TPC hits for a single
$\ttbar$ event at $\sqrt{s}=500$\,GeV overlayed with 150 BXs of pair-background
hits.  On average there are 265,000 background hits in the TPC, compared to the
average number of signal hits of 23100 (8630 from charged particles with $p_T>$1\,GeV). 
Even with this level of background, 
the tracks from the $\ttbar$ event are clearly
visible in the $r\phi$ view.  
A  significant fraction of the background hits in the
TPC arise from low energy electrons/positrons from photon conversions.  These
low energy particles form small radius helices parallel to the
$z$ axis, clearly visible as lines in the $rz$ view. 
These ``micro-curlers'' deposit charge on a small number of TPC
pads over a large number of BXs. 
Specific pattern recognition software has been
written to identify and remove these hits prior to track reconstruction. (Whilst not explicitly
studied, similar cuts are expected to remove a significant fraction of
hits from beam halo muons.) 
Figure~\ref{fig:tpcpatrec} shows the TPC hits after removing hits from micro-curlers.
Whilst not perfect, the cuts remove approximately 99\,\% of the background hits
and only 3\,\% of hits from the primary interation and 
the majority of these are from low $p_T$ tracks. Less than
1\,\% of hits from tracks with $p_T>$1\,GeV originating from the $\ttbar$ event  are removed. 

\begin{figure}[!t]
\centering
\includegraphics[height=6.6cm]{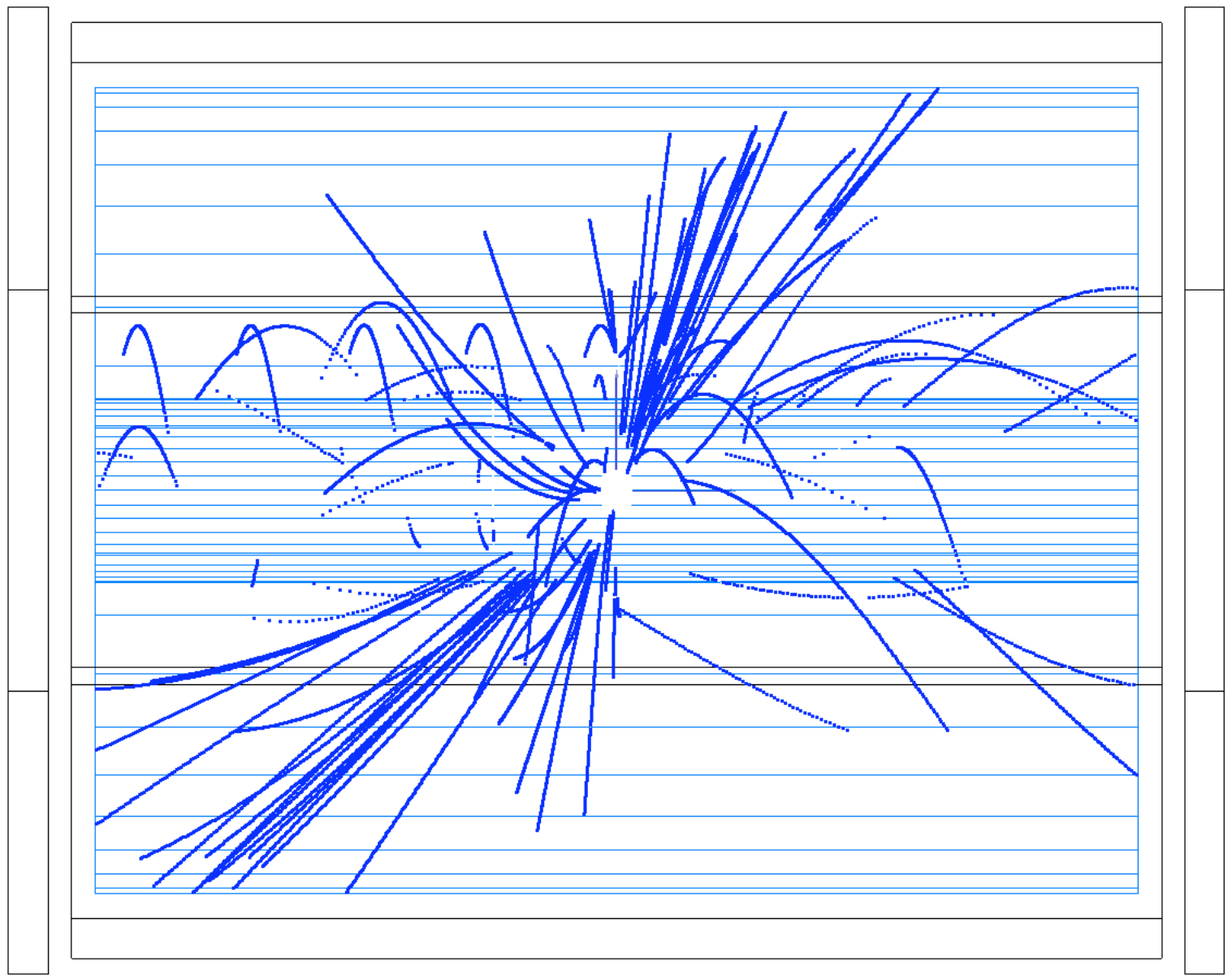}
\includegraphics[height=6.6cm]{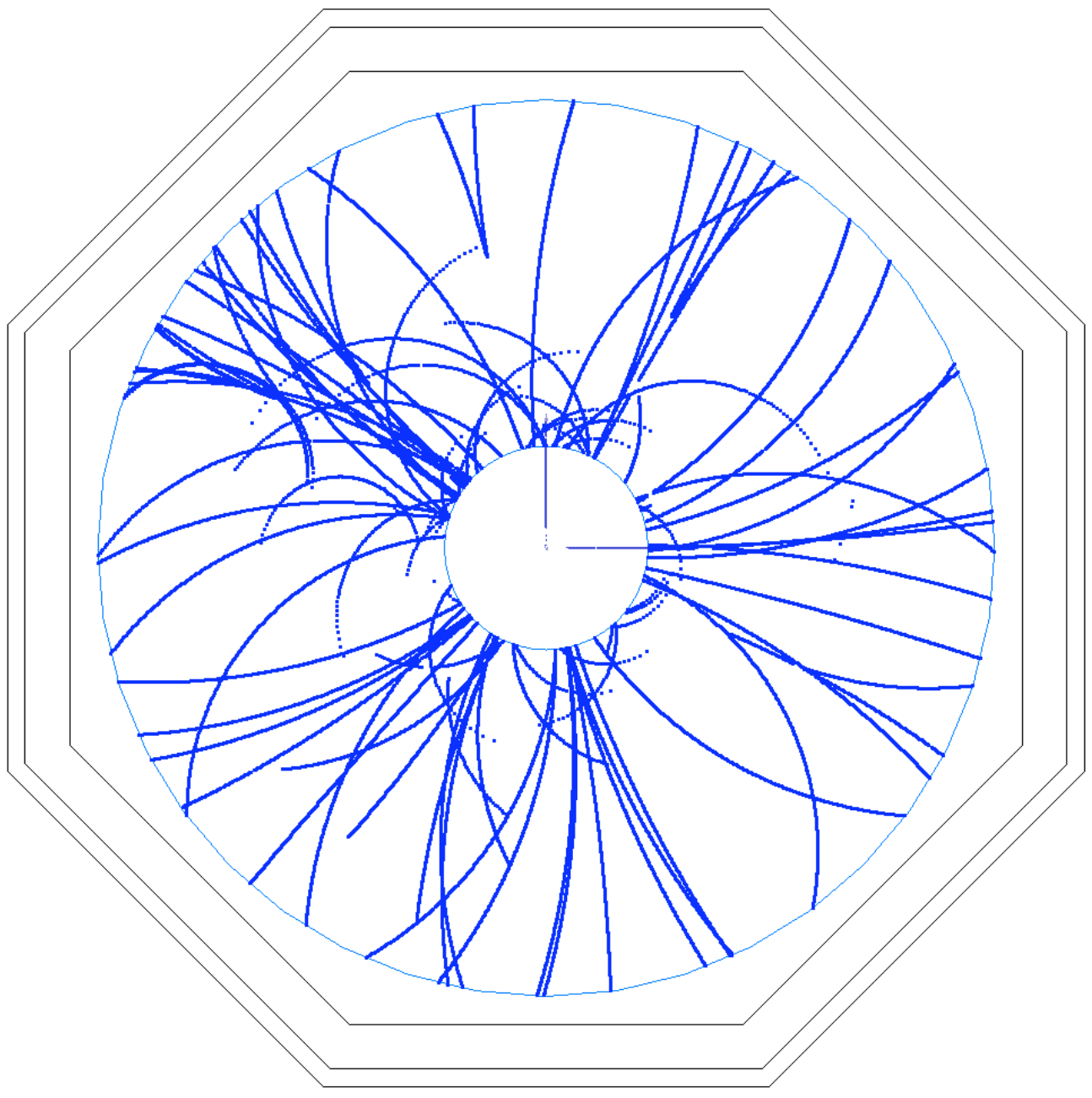}
\caption[Reconstructed TPC tracks in the presence of background.]{The same event as
          the previous plot, now showing the reconstructed TPC tracks.
\label{fig:tpctracks}}
\end{figure}

This level of background
hits proves no problem for the track-finding pattern recognition software, as can be
seen from Figure~\ref{fig:tpctracks}. Even when the background level is increased
by a factor of three over the nominal background no degradation of
TPC track finding efficiency is observed for the 100 events simulated. This study 
demonstrates the robustness of TPC tracking in the ILC background environment.

These conclusions are supported by an earlier study based on a
detector concept with $B=3.0$\,T, a TPC radius of 1.9\,m and TPC readout cells
of $3\times10$\,mm$^2$. This earlier 
study used a uniform distribution of background hits in the TPC volume, but
included a very detailed simulation of the digitised detector response and full 
pattern recognition is performed in both time and space. 
The TPC reconstruction efficiency as a function of the 
noise occupancy remains essentially unaffected for 1\,\% occupancy
(uniformly distributed through the TPC). It should be noted that this level of occupancy is 
twice the nominal occupancy at the TPC inner radius and about fifty times the typical 
total occupancy in the TPC. This earlier study is presented in more detail in Section~\ref{tpcdesign}.

\subsubsection{Background in the Vertex Detector}

\label{sec:vtxbackground}

The impact of background in the vertex detector (VTX)
depends on the assumptions made for the Silicon read-out time. If one were to
assume single BX time-stamping capability in the vertex detector, the anticipated background level is
negligible. However, it is anticipated that the readout of the Silicon pixel ladders
will integrate over many BXs.  For the studies presented here, it is assumed
that vertex 
detector readout integrates over 83 and 333 BXs for the inner two and outer four 
layers respectively. For the silicon strip-based SIT detector, single BX time-stamping
is assumed. Hence the background hits which are superimposed on the physics event
correspond to 1 BX in the SIT, 150 BXs in the TPC and 83/333 BXs in the VTX. 
It should be noted that the background studies have not yet been extended to the
FTD. 

With the above assumptions, the background in the vertex detector corresponds to  
approximately $2\times10^5$ hits per event,
with the corresponding layer occupancies listed in Table~\ref{tab:occupancy}. 
The hit occupancies account for the finite cluster size reflecting the 
fact that a single charged particle crossing a layer of the VTX
will deposit hits in multiple pixels.
The distribution of cluster sizes,
taken to be the product of the $z$ and $r\phi$ extent of the
energy deposition in the Silicon, are determined from the full simulation
of the beam related background. The mean background cluster size
is found to be 10 pixels, where a pixel is taken to be $25\times25\,\mu$m. 
\begin{table}[ht]
\begin{center}
\begin{tabular}{cccc}
  Layer        & radius/mm  & BXs    & Pixel Occupancy  \\ \hline
    0          &  16.0      &  83   &  3.33\,\% \\
    1          &  17.9      &  83   &  1.90\,\% \\
    2          &  37.0      & 333   &  0.40\,\% \\
    3          &  38.9      & 333   &  0.33\,\% \\
    4          &  58.0      & 333   &  0.08\,\% \\
    5          &  59.9      & 333   &  0.06\,\% \\
\end{tabular}
\caption[Vertex detector occupancy.]{Vertex detector occupancies for the readout times assumed in the
         background studies. The occupancies account for the finite cluster size. 
         \label{tab:occupancy}}
\end{center}
\end{table}

Pattern recognition in the environment of $10^5$ background hits is non-trivial
and required modifications to the existing Silicon track finding code. Specifically, 
tracks in the Silicon detectors are now seeded using only layers $2-5$ of the 
vertex detector and the two layers of the SIT. Seeded tracks are then 
projected inwards to pick up hits in the
inner two silicon layers (layers 0 and 1). 
Tracks with transverse momentum of $p_T<200$\,MeV are rejected. 
There are a number of questions which can be asked: i) Can the number of ``ghost'' tracks,
{\it i.e.} those formed from random combinations of hits be reduced to an acceptable
level; ii) how many genuine tracks from the pair background remain; iii)
how do any additional cuts used to reduce the beam background affect the signal; and
iv) what is the loss of efficiency due to hits from the primary interaction being merged with
the clusters of pixels from background hits. These four questions are addressed below. 

\begin{figure}[!htb]
\centering
\includegraphics[width=7.5cm]{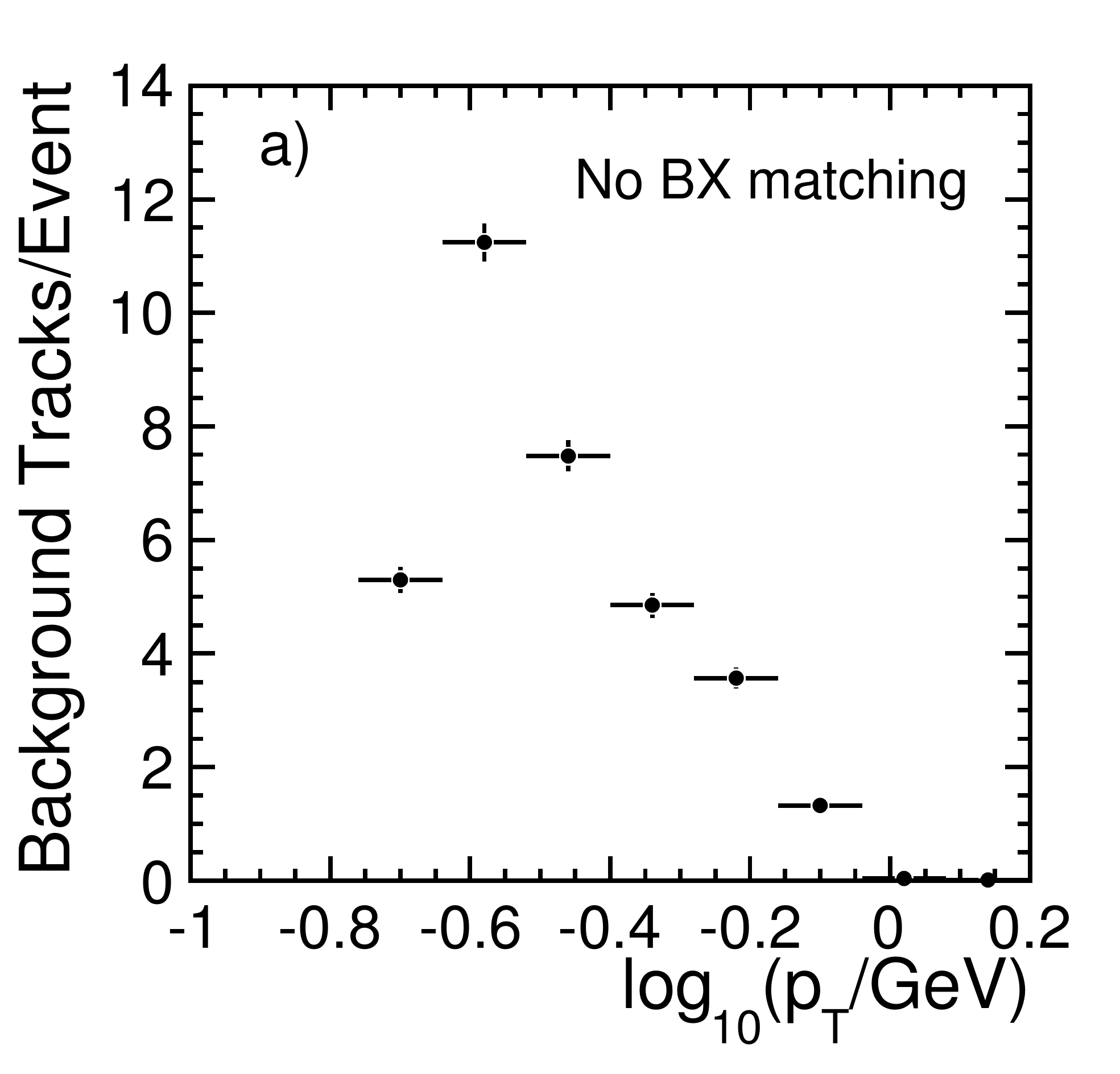}
\includegraphics[width=7.5cm]{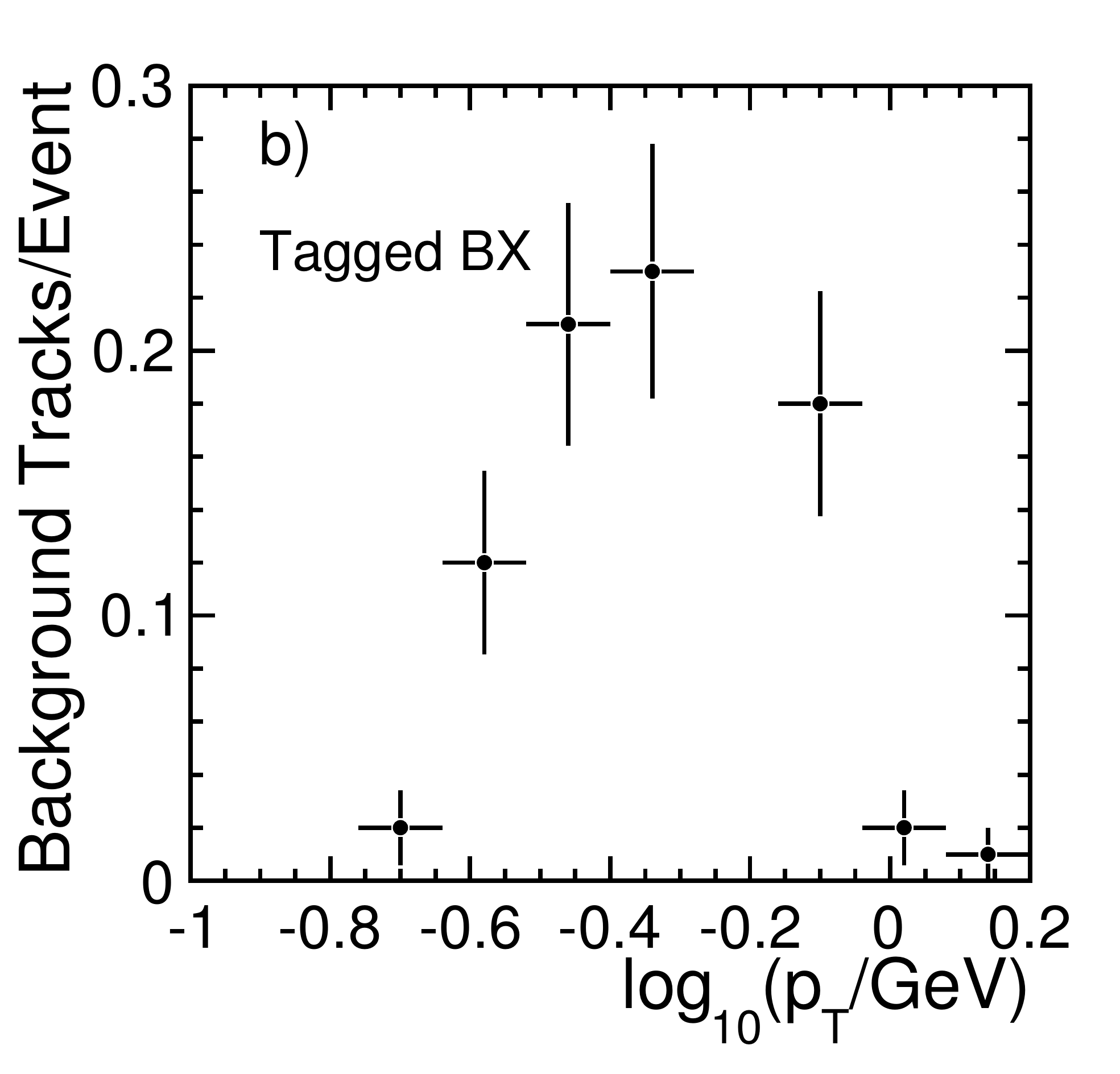}
\caption[Background tracks in the Silicon detectors.]{The $p_T$ distribution of reconstructed 
          background tracks in the Silicon detectors (VTX+SIT) a) before requiring the track to
          be in time with the physics interaction beam crossing (BX) and b) after requiring at least 1 SIT hit
          or a match to at least 10 TPC hits. A track is considered to be from background if
          more than 20\,\% of the associated hits come from the overlayed background events. 
\label{fig:ghost}}
\end{figure}
 
Figure~\ref{fig:ghost}a)  shows the   $p_T$ distribution of reconstructed tracks
in the Silicon detectors with the background overlayed. 
Due to the conservative assumptions for the VTX
readout times, an average of 34 low-$p_T$ silicon tracks are reconstructed
per physics event, reflecting the integration over $83/333$ 
BXs. The number of background tracks would be dramatically reduced by the 
requirement the reconstructed track is in the same BX as the underlying physics interaction. 
Tracks are thus required to have at least one SIT hit 
(which provides unambiguous BX identification) 
or to be associated with at least 10 TPC hits (where the matching in $z$ can be
used to identify the BX). Figure~\ref{fig:ghost} shows the 
resulting $p_T$ distribution of the remaining background tracks. 
On average 1.2 background tracks per event remain with a mean $p_T$ of 500\,MeV. 
The majority of the remaining background tracks are either from relatively high
$p_T$ electrons/positrons or from combinations of signal and background hits.
Firstly, there may be a loss of efficiency 
due to the additional requirements of associated SIT or TPC hits. 
For tracks from
the primary physics interaction ({\it i.e.} the $\ttbar$ event), the SIT/TPC 
requirements remove approximately 1\,\% of tracks with $p_T<1$\,GeV, whilst
for $p_T>1$\,GeV there is no observed loss of efficiency.

\begin{figure}
\includegraphics[width=7.5cm]{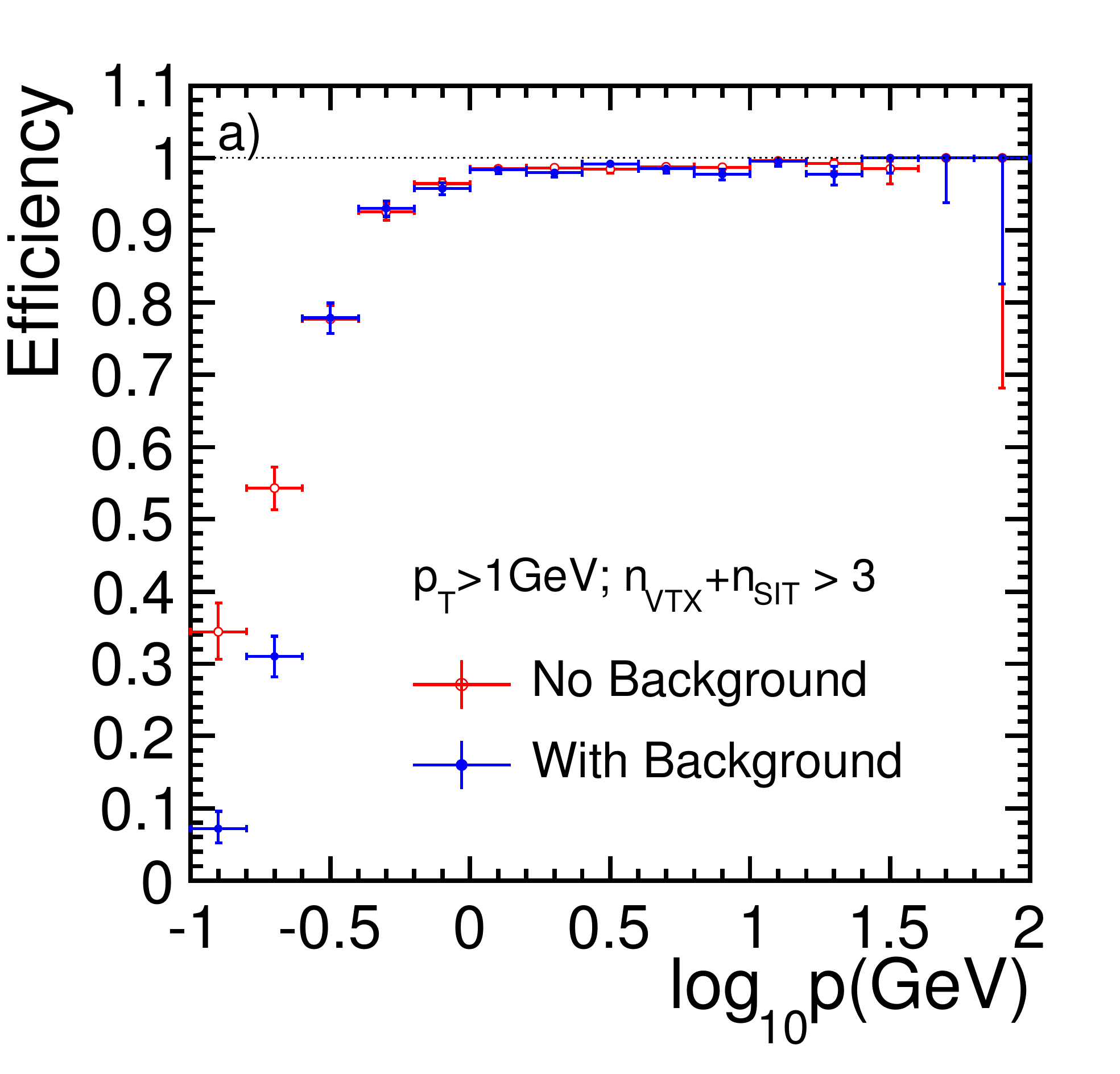}
\includegraphics[width=7.5cm]{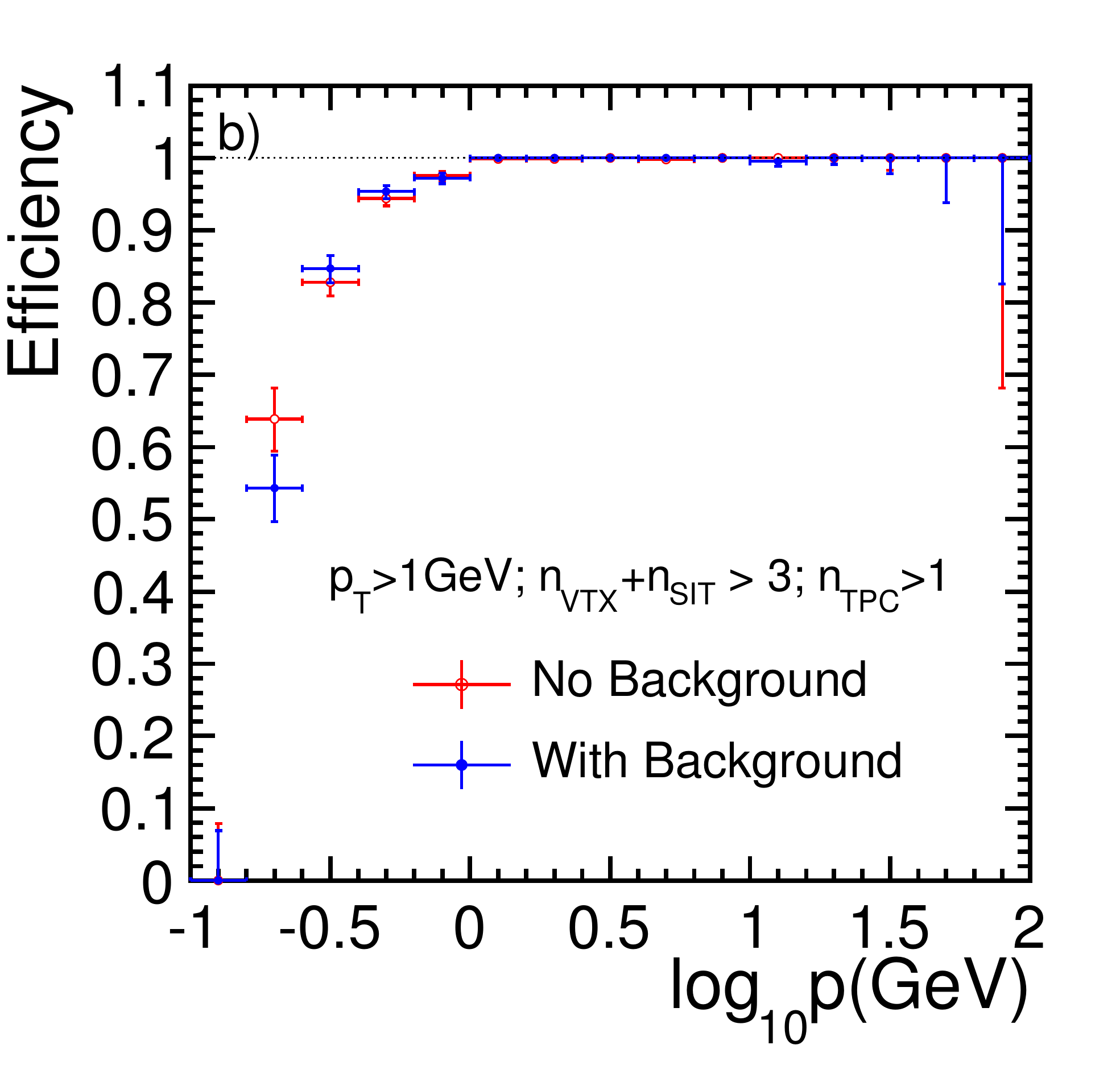}
\caption[Tracking Efficiency with Background.]{Tracking efficiency with overlayed background in the 
  vertex detector vs transverse momentum for $\ttbar\rightarrow$ 6 jets at 500GeV ($|\cos\theta|<0.8$)
  a) for all charged particles and b) for charged particles which deposit at least one hit in the TPC. 
 The tracking efficiency is shown for tracks with $p_T>1$\,GeV 
 with a total of at least 4 hits in the VTX and SIT. Tracks are considered to be well reconstructed if
 there is at least one associated SIT hit or more than 10 TPC hits and if at least
 70\,\% of the hits on the track are from the original Monte Carlo Particle.
   \label{fig:TrackPerfEfficiencyBack}}
\end{figure}

The presence of a large number background hits not only results in a small number
of background tracks, it may also lead to a degradation of the pattern recognition 
performance. In addition, if a charged particle from the interaction passes close 
to a cluster of pixels from 
the pair background, a single larger cluster will be formed. It is assumed 
such an extended cluster will not be included in the track-finding algorithm,
and hence hits close to background clusters effectively will be  lost. 
In addition to overlaying 83/333 BXs of background, the pixel occupancies 
in Table~\ref{tab:occupancy} are used to remove the appropriate number of 
hits from the primary interaction. The effect of the overlayed background and
the resulting hit inefficiencies is studied for 
simulated $\ttbar\rightarrow$ 6 jets events at $\roots=500$\,GeV.
Figure~\ref{fig:TrackPerfEfficiencyBack}a) shows
the overall track reconstruction efficiency with and without
background. The main effect of the background is
to reduce the efficiency for tracks with $p_T<300$\,MeV. For tracks
$p_T>1$\,GeV, the presence of background reduces the track finding
efficiency by less than 0.1\,\%. In the presence of background the efficiency 
for tracks with $p_T>1$\,GeV is 98.8\,\%. Care has to be taken in 
interpreting this number; the efficiencies depend on how the denominator is
defined. For example, the inefficiency 
for high $p_T$ tracks arises almost entirely from tracks which decay or
interact within the volume of the VTX/SIT. 
Figure~\ref{fig:TrackPerfEfficiencyBack}b) shows the track finding
efficiency for tracks which (in simulation) deposit energy in the
TPC gas volume. For this sample, the efficiency is greater than 
99.9\,\% for tracks with $p_T>1$\,GeV, with or without background.
It is concluded that the ILD tracking efficiency is not significantly
degraded by the nominal level of background expected at the ILC.

\subsubsection{Impact of background on physics analyses}

\label{sec:physicsbackground}

\begin{figure}[h]
\centerline{\includegraphics[width=7cm]{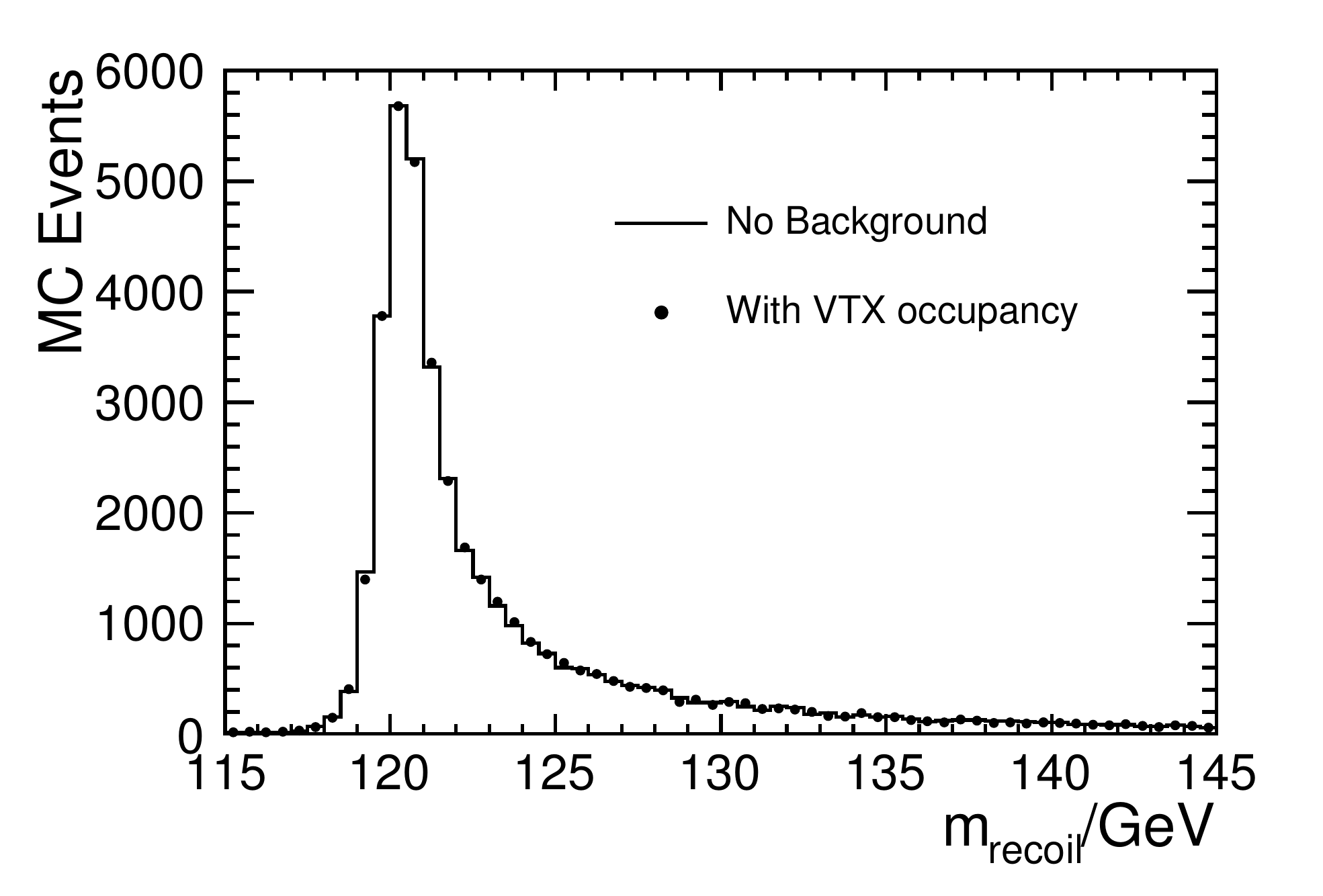}}
\caption[Higgs recoil mass distribution with background effects]{The Higgs recoil mass 
distribution in the $\mmX$ channel from 50000 generated MC 
events after selection for the case of no background and with inefficiencies
in the vertex detector due to the occupancy level expected from the pair background.
\label{fig:mhback}}
\end{figure}

To fully simulate the effect of background on a particular physics channel 
would require overlaying 1 BX in the SIT, 150 BXs in the TPC and 83/333 BXs in the 
VTX on each simulated physics event and would require vast CPU resources. 
From the studies above it is expected that the track finding inefficiencies for 
the high momentum muons
in the $\Zzero\Higgs\rightarrow\mmX$ channel will be negligibly small. In addition,
the presence of the relatively few low $p_T$ background tracks will not affect
the recoil mass distribution. The possibility that the loss of hits in the
vertex detector due to background occupancy might distort the recoil mass distribution
has been investigated (see Figure~\ref{fig:mhback}). The observed differences
are negligibly small. 

%% file: performance/performance-flavourtag.tex
Identification of $b$-quark and $c$-quark jets plays an important role within the ILC physics programme.
The vertex detector design and the impact parameter resolution are of particular importance for 
flavour tagging. 
The LCFIVertex flavour tagging (see Section~\ref{sec:optimization-flavour})
uses ANNs to 
discriminate $b$-quark jets from $c$ and light-quark jets ($b$-tag), $c$-quark jets from $b$ and 
light-quark jets ($c$-tag), and $c$-quark jets from
$b$-quark jets ($bc$-tag).

\begin{figure}[!b]
\centering
\includegraphics[width=7.5cm]{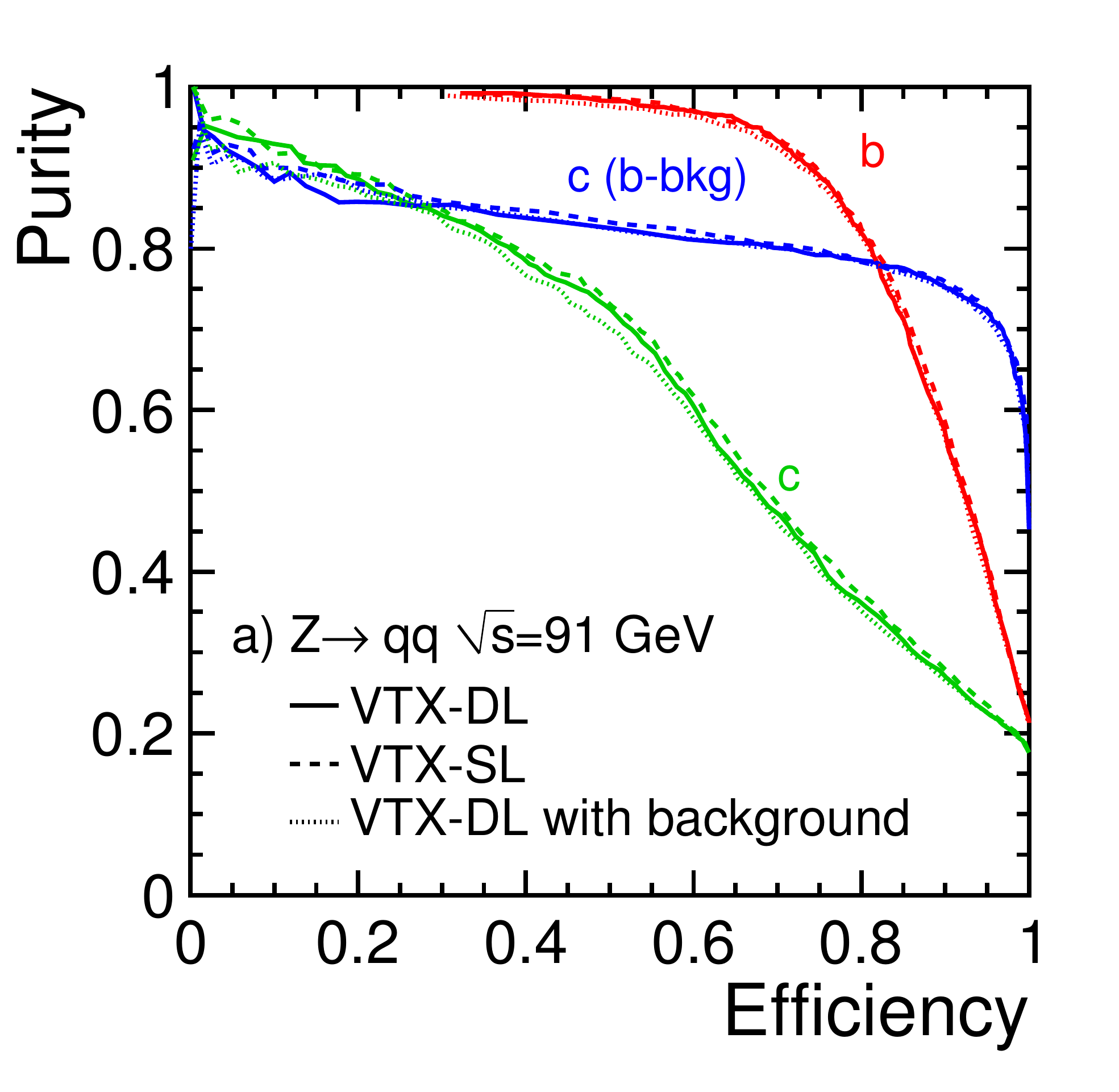}
\includegraphics[width=7.5cm]{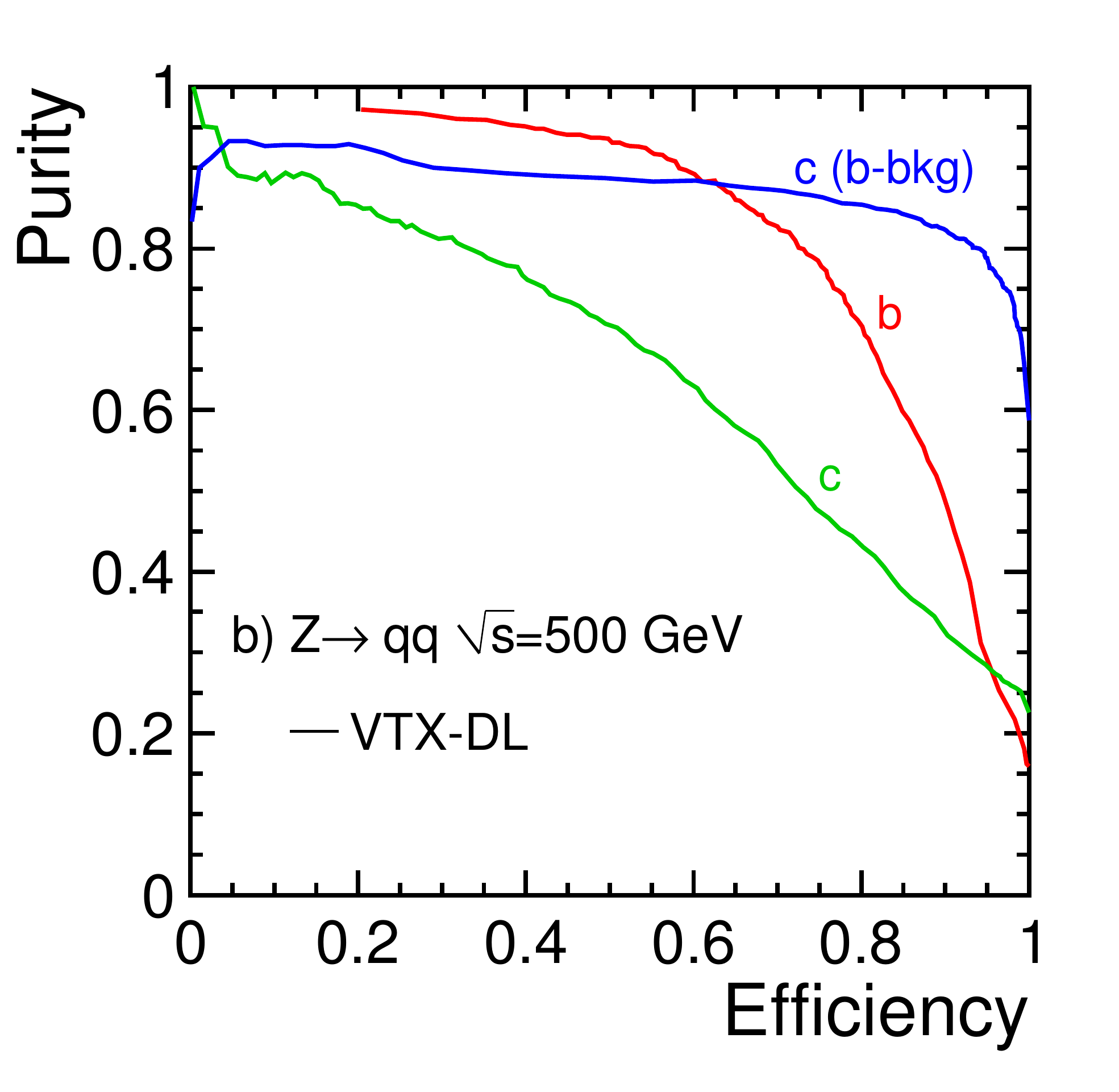}
\caption[ILD flavour tagging performance.]{a) Flavour tagging performance of the ILD detector
  for 91\,GeV  $Z\rightarrow \qq$ events 
  for both the three double-sided ladders (VTX-DL) layout and with five single-sided 
  ladder layout (VTX-SL). Also shown for the VTX-DL is the impact of background on the 
  flavour tagging performance.
  b)  Flavour tagging performance of the ILD detector
  for 500\,GeV  $Z\rightarrow \qq$ events for the VTX-DL layout.
  In all cases the acceptance corresponds to $|\cos\theta_{\mathrm{jet}}|<0.95$.
\label{fig:ild_flavourtag}}
\end{figure}

The flavour tagging performance~\cite{flavourtag} of ILD is studied for the two vertex detector geometries
considered, three double-sided ladders (VTX-DL) and five single-sided (VTX-SL) ladders.
No significant differences in the input variables for the ANNs are seen for
two geometries, and therefore the ANNs trained for the VTX-DL option were used for both
VTX configurations.  The samples used in the training consisted of 150000
$\Zzero\rightarrow\qq$, at the $\Zzero$ pole energy, equally distributed among the three decay modes
$q=b,c$ and light quarks. The test samples used to evaluate the flavour tagging performance were generated
independently and consist of $10000$ events of $Z\rightarrow \qq$ generated at both $\sqrt{s}=91$\,GeV and 
$\sqrt{s}=500$\,GeV, with the SM flavour mix of
hadronic final states. 
The ILD flavour tagging performances at 91\,GeV 
for the two vertex detector options are shown in Figure~\ref{fig:ild_flavourtag}a). The performance differences between
the two VTX geometries are small $(\lesssim 1\%)$. Uncertainites due to the statistical fluctuations
of the test sample and in those introduced in the ANN training are estimated to be $\lesssim 2\%$.
The performance for $Z\rightarrow \qq$ at $\sqrt{s}=500$\,GeV is shown in Figure~\ref{fig:ild_flavourtag}b).
It should be noted that for the 500\,GeV results the ANNs were not retrained, {\it i.e.} those 
obtained for $\sqrt{s}=91$\,GeV were used. Consequently, improvements in the performance are expected.

\subsubsection{Impact of Background}

Do to the large computational requirements in overlaying background 
hits from many BXs a parametric approach is taken to assess the impact
of background on the flavour tagging performance.
In Section~\ref{sec:vtxbackground} is was demonstrated that hits from 
pair background integrated over 83 (333) BXs for layers 0 and 1
($2-5$) of the VTX-DL doublet layout of six ladders does not result
in a significant number of background tracks and that the overall
tracking efficiency is not significantly reduced. The main impact on
flavour tagging is likely to be from the resulting hit inefficiencies
particularly in the inner layers. 
To simulate the effect of
background the pixel occupancies of Table~\ref{tab:occupancy} are
used to randomly remove Silicon hits from the events before 
track finding and flavour tagging. This results in a slight degradation 
in tracking performance in the Silicon detectors; 
the number of TPC tracks associated with a complete track in the vertex detector
(6 hits) decreases by 2\,\%. 
The resulting flavour tagging performance is shown in Figure~\ref{fig:ild_flavourtag}a). 
Although there is a suggestion
of a small degradation in the performance of the c-tag in the 
low efficiencies/high purity region, the presence of pair 
background does not significantly degrade the flavour tagging performance.

%% file: performance/performance-particleflow.tex
Many important physics channels at the ILC will consist of final states with
at least six fermions, setting a ``typical'' energy scale for ILC jets 
as approximately 85\,GeV and 170\,GeV at $\roots=500$\,GeV and $\roots=$1\,TeV respectively. 
The current performance 
of the PandoraPFA algorithm applied to ILD Monte Carlo simulated data
is summarised in Table~\ref{tab:resvsE}. The observed jet 
energy resolution ($\rmsn$) is not described by the expression 
$\sigma_E/E = \alpha/\sqrt{E/\mathrm{GeV}}$. This is not surprising, 
as the particle density increases it becomes harder to correctly
associate the calorimetric energy deposits to the particles and the confusion term increases.
The single jet energy resolution is also listed. 
The jet energy resolution ($\rmsn$) is better than 3.8\,\% for the jet energy 
range of approximately $40-400$\,GeV. The resolutions quoted in terms of
$\rmsn$ should be multiplied by a factor of approximately 1.1 to obtain 
an equivalent Gaussian analysing power\cite{PandoraPFA}.

\begin{table}[ht]
\begin{center}
\begin{tabular}{r|rrcc}
  Jet Energy        & raw rms        & $\rmsn$ &  $\rmsn/\sqrt{E_{jj}/\mathrm{GeV}}$ & $\sigma_{E_j}/E_{j}$  \\ \hline
  45 GeV            & 3.3\,GeV   &  2.4\,GeV       &  25.0\,\%   &   $(3.71\pm0.05)\,\%$  \\
  100 GeV           & 5.8\,GeV   &  4.1\,GeV       &  29.5\,\%   &   $(2.95\pm0.04)\,\%$  \\
  180 GeV           & 11.2\,GeV  &  7.5\,GeV       &  40.1\,\%   &   $(2.99\pm0.04)\,\%$  \\
  250 GeV           & 16.9\,GeV  & 11.1\,GeV       &  50.1\,\%   &   $(3.17\pm0.05)\,\%$  \\
\end{tabular}
\caption[ILD jet energy resolution.]{Jet energy resolution for $\Zzero\rightarrow$uds events with $|\cos\theta_{\qq}|<0.7$, 
            expressed as, $\rmsn$ for the di-jet energy distribution, 
             the effective
            constant $\alpha$ in $\rmsn/E = \alpha(E_{jj})/\sqrt{E_{jj}/\mathrm{GeV}}$, and the
            fractional jet energy resolution for a single jets, $\sigma_{E_j}/E_{j}$. The jet energy resolution is calculated from $\rmsn$.
\label{tab:resvsE}}
\end{center}
\end{table}

\begin{figure}[hbtp]
\begin{center}
\includegraphics[width=9cm]{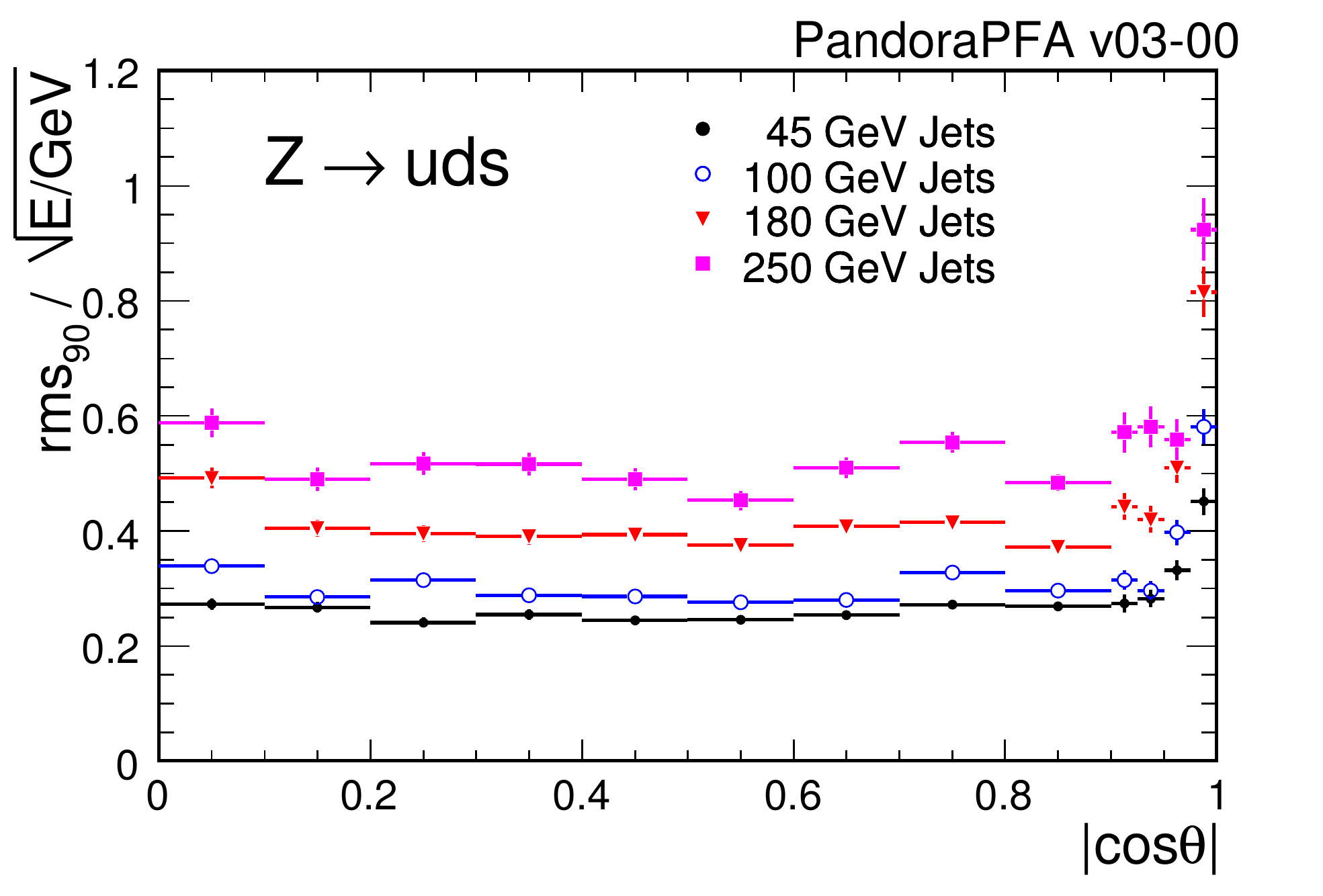}
\caption[ILD jet energy resolution versus $\cos\theta$.]{The jet energy resolution, defined as the $\alpha$ in $\sigma_E/E=\alpha\sqrt{E/\mathrm{GeV}}$, 
        plotted versus $\cos\theta_{\qq}$ for four different jet energies.
\label{fig:rmsVersusTheta}}
\end{center}
\end{figure}

Figure~\ref{fig:rmsVersusTheta} shows the jet energy resolution for $\Zzero\rightarrow$uds events 
plotted against the cosine of the polar angle of the generated $\qq$ pair,
$\cos\theta_{\qq}$, for four different values of $\roots$. Due to the
calorimetric coverage in the forward region, the jet energy resolution
remains good down to $\theta = 13^\circ$ $(\cos\theta=0.975)$.

%% file: performance/physics.tex
\input{performance/physics_introduction}

\subsection{Higgs Boson mass}
\label{sec:physics-higgs-mass}
\input{performance/physics_higgs_mass}

\subsection{Higgs Boson Branching Fractions}
\label{sec:physics-higgs-br}
\input{performance/physics_higgs_br}

\subsection{Tau-pairs}
\label{sec:physics-tau-pairs}
\input{performance/physics_tau}

\subsection{Chargino and Neutralino Production}
\label{sec:susy_point5}
\input{performance/physics_susypoint5}

\subsection{Top production}
\label{sec:physics-top}
\input{performance/physics_top}

\subsection{Strong EWSB}
\label{sec:physics-wwvv}
\input{performance/physics_wwvv}

\subsection{Lepton production in SPS1a'}
\label{sec:physics-sps1a}
\input{performance/physics_sps1a}

\subsection{Photon Final States}
\label{sec:physics-photons}
\input{performance/physics_photons}

%% file: performance/physics_introduction.tex
The ILD detector performance has been evaluated for a number of
physics processes. The analyses, described below, all use the
full simulation of  ILD and full event reconstruction.
Jet finding is performed using the Durham algorithm\cite{Durham} 
with the hadronic system being forced into the appropriate number of jets for the
event topology. The benchmark physics analyses\cite{ref:wws_benchmark} 
are studied at $\roots=250$\,GeV and $\roots = 500$\,GeV. 
Unless otherwise stated, the  results for $\roots=250$\,GeV
($\roots=500$\,GeV) correspond to an integrated luminosity of 
250\,fb$^{-1}$ (500\,fb$^{-1}$) and a beam polarisation of
$P(\eplus,\eminus) = (+30\,\%,-80\,\%)$.

%% file: performance/physics_higgs_mass.tex
The precise determination of the properties of the Higgs boson is one of the main
goals of the ILC. Of particular importance are its mass, $\mH$, 
the total production cross section, $\sigma(\epem\rightarrow\Higgs\Zzero)$,
and the Higgs branching 
ratios. Fits to current electroweak data\cite{lepwww} and direct limits from searches
at LEP and at the Tevatron favour a relatively low value for $\mH$. Studies of these measurements with ILD are described below. 
A data sample of 250\,fb$^{-1}$ at $\roots=250$\,GeV is assumed and $\mH$ is
taken to be 120\,GeV. For these values, the dominant Higgs production process is
Higgs-strahlung, $e^{+}e^{-}\rightarrow\Zzero\Higgs$. 

The Higgs boson mass can be determined  precisely  from the distribution of
the recoil mass, $m_\mathrm{recoil}$, in 
$\Zzero\Higgs\rightarrow\epem X $ and $\Zzero\Higgs\rightarrow\mpmm X $ events, where 
$X$ represents the Higgs decay products.
The recoil mass is calculated from the reconstructed four-momentum 
of the system recoiling against the $\Zzero$. 
The $\mmX$-channel yields the
most precise measurement as the $\eeX$-channel suffers from larger experimental uncertainties 
due to bremsstrahlung from the electrons and the larger background from Bhabha scattering events.
The study\cite{bib:higgsnote,bib:lithesis} is performed for two electron/positron beam polarisations:  $P(e^+,e^-) = (-30\,\%, +80\,\%)$
and $P(e^+,e^-) = (+30\,\%, -80\,\%)$.
In the simulation, Gaussian beam energy spreads 
of 0.28\,\% and 0.18\,\% are assumed for the incoming electron and positron beams
respectively. 

The first stage in the event selection is the identification of leptonically decaying $\Zzero$ bosons. 
Candidate lepton tracks are required to be well-measured, removing tracks with large uncertainties
on the reconstructed momentum. 
Lepton identification is performed using the associated calorimetric 
information resulting in an event efficiency of 95.4\,\% for identifying both in $\mmX$ events 
and 98.8\,\% for both electrons in $\eeX$ events.
Candidate $\Zzero$ decays are identified from oppositely charged pairs of identified leptons
within a mass window around $\mZ$. Background from $\Zzero\rightarrow\lplm$ is rejected
using cuts on the transverse momentum of the di-lepton system and the 
acollinearity of the two lepton tracks. Additional 
cuts reject $\Zzero\rightarrow\lplm$ events with initial and final state radiation.
The backgrounds from $e^{+}e^{-} \rightarrow \Zzero\Zzero$ and $e^{+}e^{-} \rightarrow \WpWm$ 
are reduced using a multi-variate likelihood analysis based on the acoplanarity, polar angle, 
transverse momentum and the invariant mass of the di-lepton system.

\begin{figure}[!b]
\centering
\includegraphics[width=7.5cm]{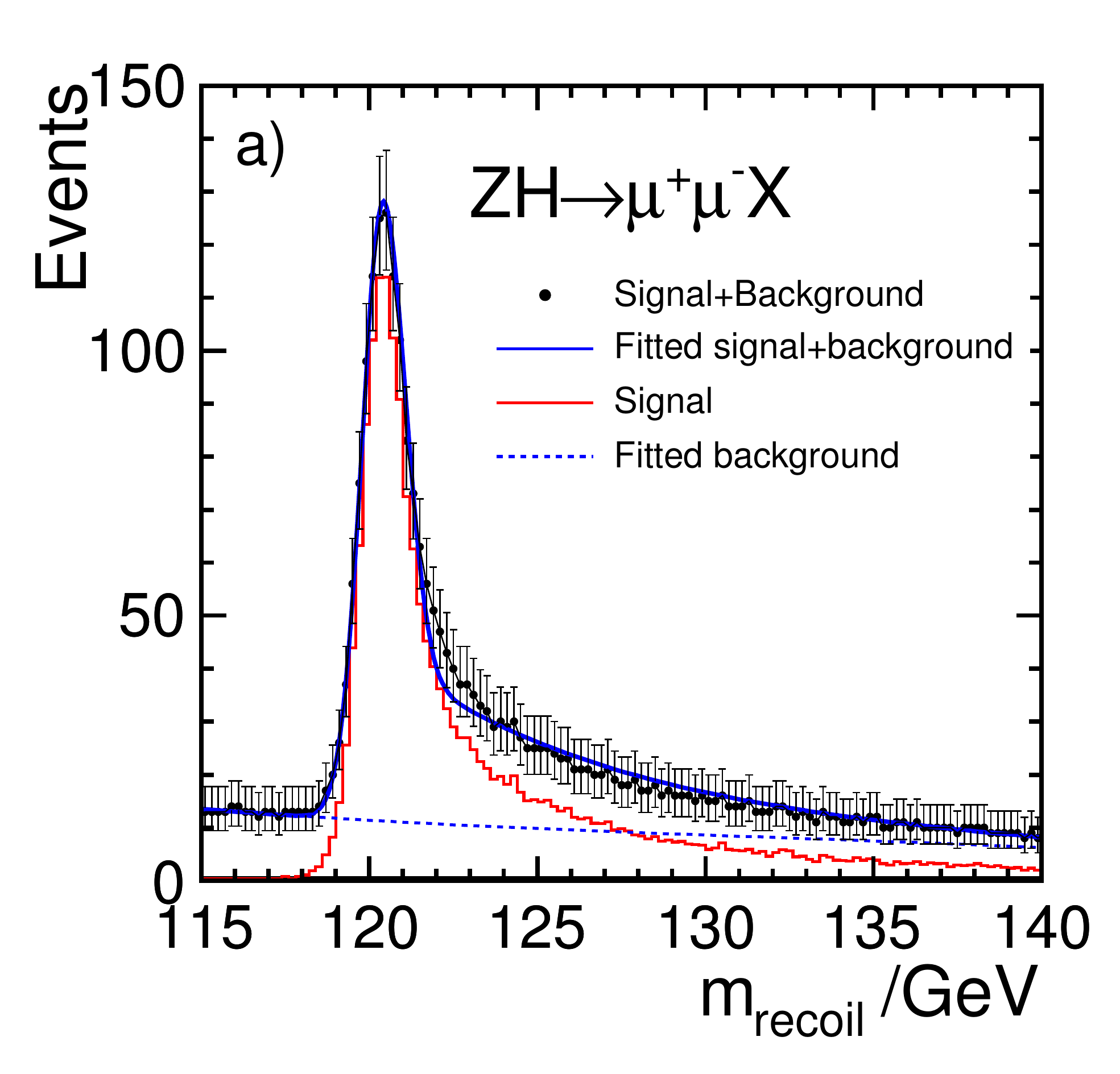}
\includegraphics[width=7.5cm]{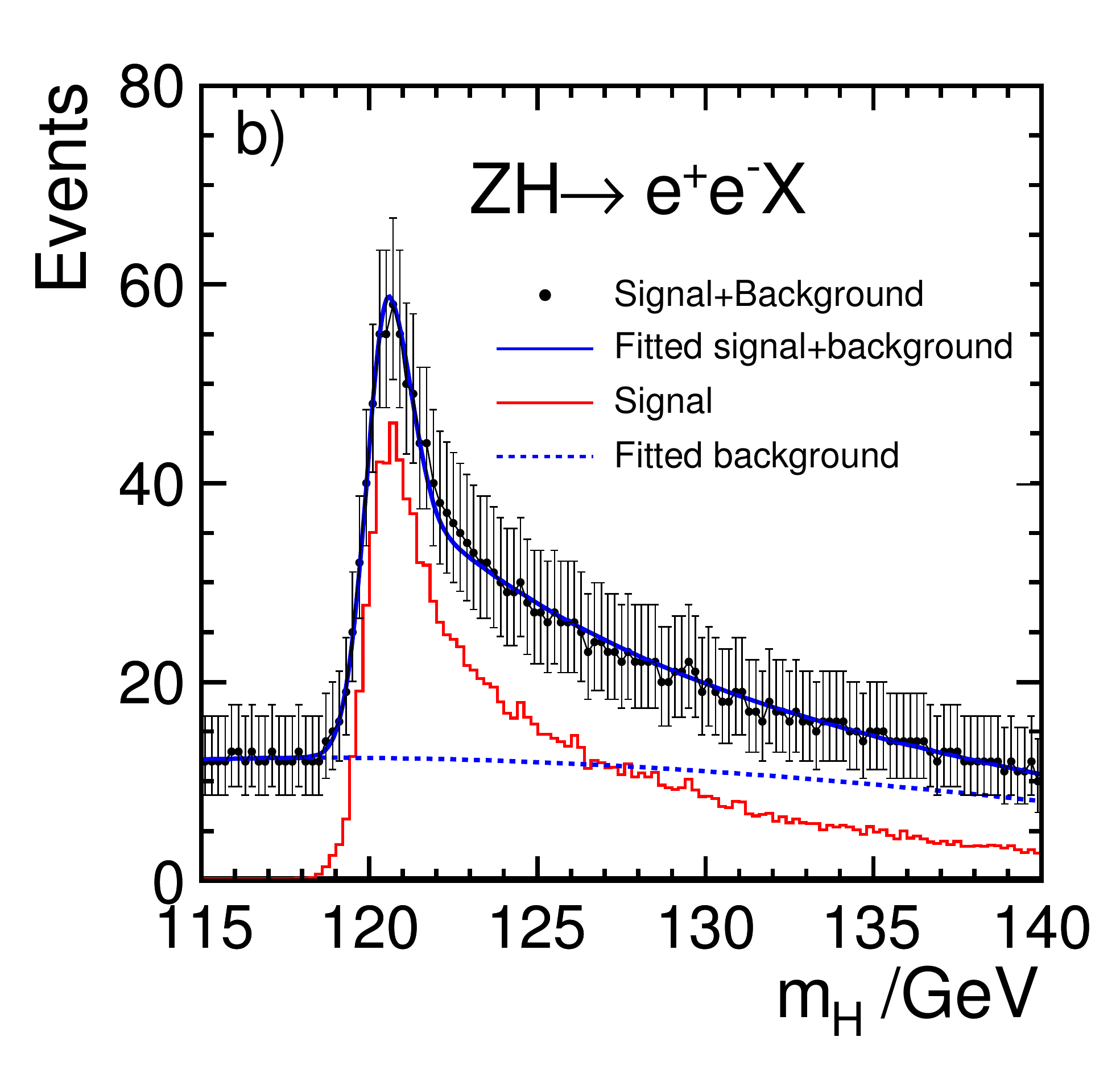}
\caption[Higgs recoil mass.]{Results of the model independent analysis of the Higgs-strahlung process 
$\epem \rightarrow \Higgs\Zzero$ in which a) $\Zzero\rightarrow\mpmm$ and b) $\Zzero\rightarrow\epem$. 
The results are shown for a beam polarisation of $P(e^+,e^-) = (+30\,\%, -80\,\%)$. 
\label{fig:fit_ppme_gpet}}
\end{figure}
The reconstructed $m_\mathrm{recoil}$ distributions are shown in Figure~\ref{fig:fit_ppme_gpet}.
The combination of signal and background is fitted using a function which assumes a 
Gaussian-like signal and that the background can be approximated by a polynomial function. 
The results of the fit for  $\mH$ and $\sigma(\epem\rightarrow\Zzero\Higgs)$ 
are listed in Table~\ref{tab:mirst}. Also shown are the results obtained when assuming the 
SM decay modes and branching fractions. In this case, labelled ``Model Dependent'',
the background is further reduced by requiring charged particle tracks in 
addition to those generated by 
the $\Zzero$ boson decay products.

\begin{table}[!h]
\centering 
\begin{footnotesize}
\begin{tabular}{|l|l|r|r|rr|}
\hline
Analysis &Polarisation ($e^-,e^+$) & Channel & $\sigma_{\mH}$ & \multicolumn{2}{c|}{Cross section}\\
\hline
\multirow{6}{*}{Model Independent}
&\multirow{3}{*}{$(+80\,\%,\, -30\,\%)$} & $\mmX$ &   40\,MeV &  $\pm0.28$\,fb & (3.6\,\%)  \\ \cline{3-6}
&                       & $\eeX$ &  88\,MeV &  $\pm0.43$\,fb &(5.1\,\%)                    \\ \cline{3-6} 
&                       & $\epem(n\gamma)X$ &  81\,MeV &  $\pm0.36$\,fb &(4.3\,\%)         \\ \cline{2-6} 
&\multirow{3}{*}{$(-80\,\%,\, +30\,\%)$} & $\mmX$ &   36\,MeV &  $\pm0.39$\,fb & (3.3\,\%) \\ \cline{3-6}
&                       & $\eeX$ &  72\,MeV &  $\pm0.61$\,fb & (4.8\,\%)                   \\ \cline{3-6}
&                       & $\epem(n\gamma)X$ &  74\,MeV &  $\pm0.47$\,fb & (4.0\,\%)        \\ \hline
\multirow{6}{*}{Model Dependent} 
&\multirow{3}{*}{$(+80\,\%,\, -30\,\%)$} & $\mmX$ &   36\,MeV &  $\pm0.26$\,fb & (3.3\,\%)   \\ \cline{3-6}
&                       & $\eeX$ &  77\,MeV &  $\pm0.38$\,fb &(4.5\,\%)                      \\ \cline{3-6} 
&                       & $\epem(n\gamma)X$ &  73\,MeV &  $\pm0.31$\,fb &(3.8\,\%)           \\ \cline{2-6} 
&\multirow{3}{*}{$(-80\,\%,\, +30\,\%)$} & $\mmX$ &   31\,MeV &  $\pm0.32$\,fb & (2.7\,\%)   \\ \cline{3-6}
&                       & $\eeX$ &  64\,MeV &  $\pm0.47$\,fb & (3.7\,\%)                     \\ \cline{3-6}
&                       & $\epem(n\gamma)X$ &  59\,MeV &  $\pm0.37$\,fb & (3.1\,\%)   \\ \hline
\end{tabular}
\end{footnotesize}
\caption[Expected statistical uncertainties on $\mH$.]{Expected statistical 
         uncertainties on $\mH$ from the recoil mass distribution in 
     Higgs-strahlung events where the $\Zzero$ decays into either $\epem$ or $\mpmm$.
           Results are listed for both the model independent and model dependent analyses.
           Also listed are the experimental uncertainties on the Higgs-strahlung 
           cross section.
           The results are given for two different beam polarisations. In the case of
           the  $\eeX$-channel results are given without ($\eeX$) and with ($\epem(n\gamma)X$) 
           the inclusion of
           identified Bremsstrahlung photons.
\label{tab:mirst}}
\end{table}

\subsubsection{Influence of Bremsstrahlung} 

\begin{figure}[!tb]
\centering
\includegraphics[width=7.5cm]{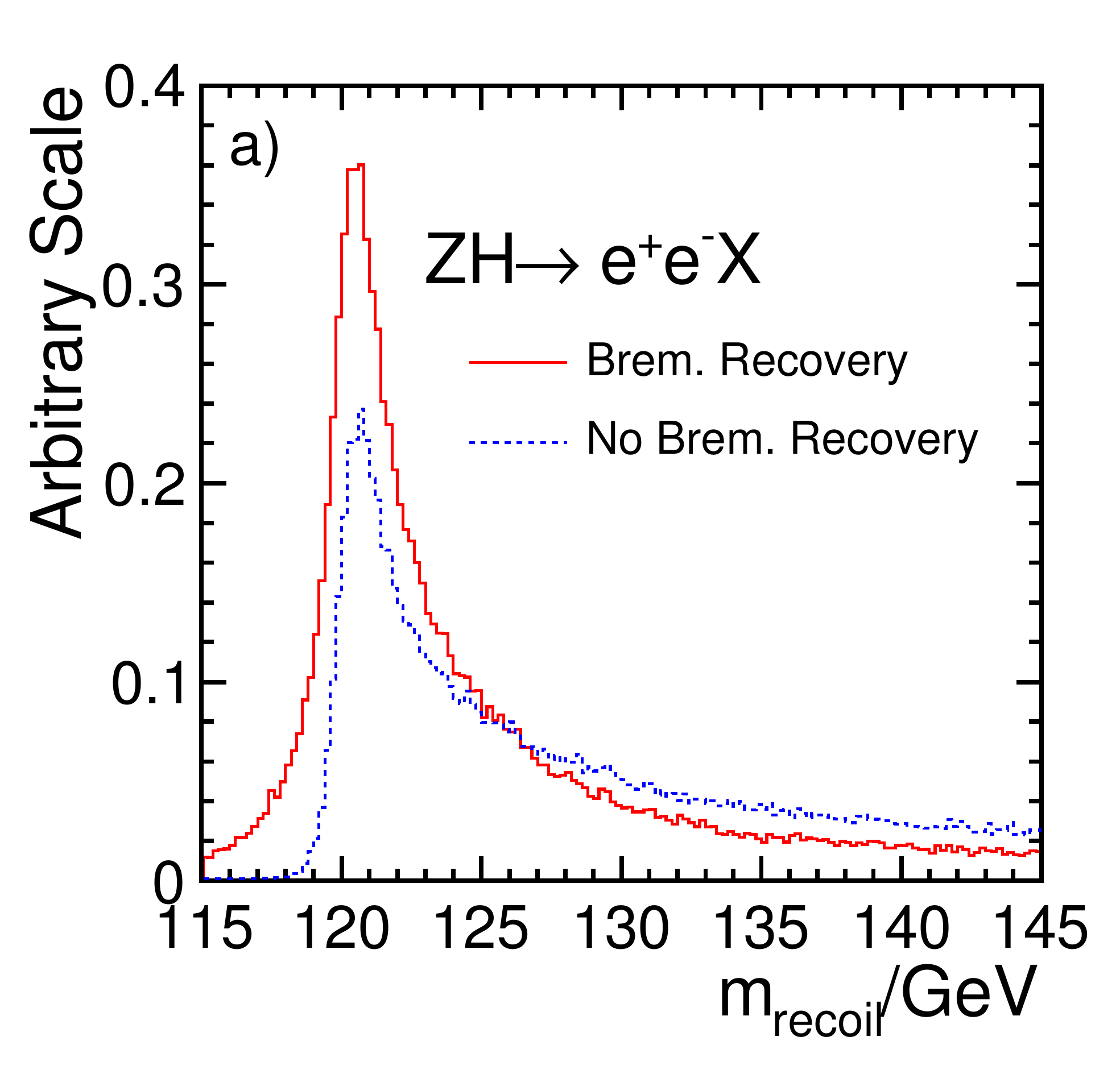}
\includegraphics[width=7.5cm]{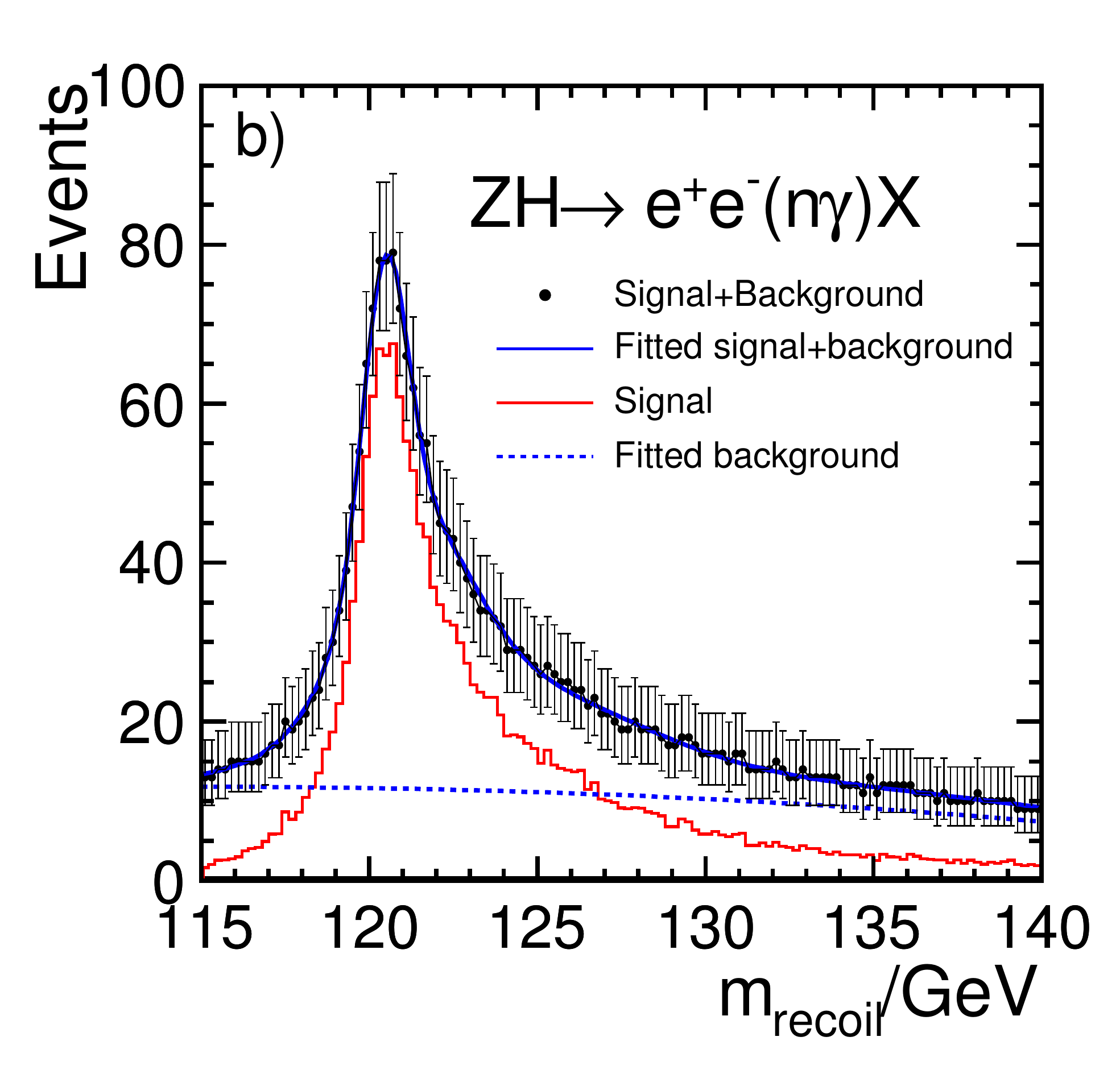}
\caption[Higgs recoil mass with Bremsstrahlung.]{The effect of including identified Bremsstrahlung
         photons in the $\eeX$ channel: a) comparison of the recoil mass distribution with and without inclusion of
         Bremsstrahlung photons and b) the fitted recoil mass distribution for the model independent
         analysis with Bremsstrahlung recovery. The plots are shown  for a beam polarisation of
         $P(e^+,e^-) = (+30\,\%, -80\,\%)$. 
\label{fig:fit_brem}}
\end{figure}

From figure~\ref{fig:fit_ppme_gpet} it is clear that Bremsstrahlung from final state
electrons and positrons significantly degrades the recoil mass resolution in the 
$\eeX$ channel. One possible strategy to mitigate this effect is to identify the
final state photons and include these in the recoil mass calculation. A dedicated
algorithm to identify Bremsstrahlung photons is used~\cite{bib:ZFinder} and
the four momenta of the $\eeX+n\gamma$ system is used in the event selection and
recoil mass calculation. Figure~\ref{fig:fit_brem}a) compares the recoil mass
distribution with and without including identified Bremsstrahlung photons.
Figure~\ref{fig:fit_brem}b) shows the recoil mass distribution for the
model independent impact analysis including Bremsstrahlung photons. To extract
the mass and cross section a modified fitting function is used.
The results of the fits ($\eeX n\gamma$) for $\mH$ and $\sigma(\epem\rightarrow\Zzero\Higgs)$ 
are listed in Table~\ref{tab:mirst}. Including Bremsstrahlung photons improves
the mass resolution by 10\,\% and the cross section resolution by 20\,\%. 
The improvement to the mass resolution is limited by the degradation in the
sharpness of the leading edge of the recoil mass distribution. It should
be noted that a more complete treatment would involve a refit of the track taking
into account the candidate Bremsstrahlung photons; at this stage no strong
conclusions should be drawn. 

\subsubsection{Influence of Beam Energy Uncertainties} 

\begin{figure}[h]
\centerline{\includegraphics[width=10cm]{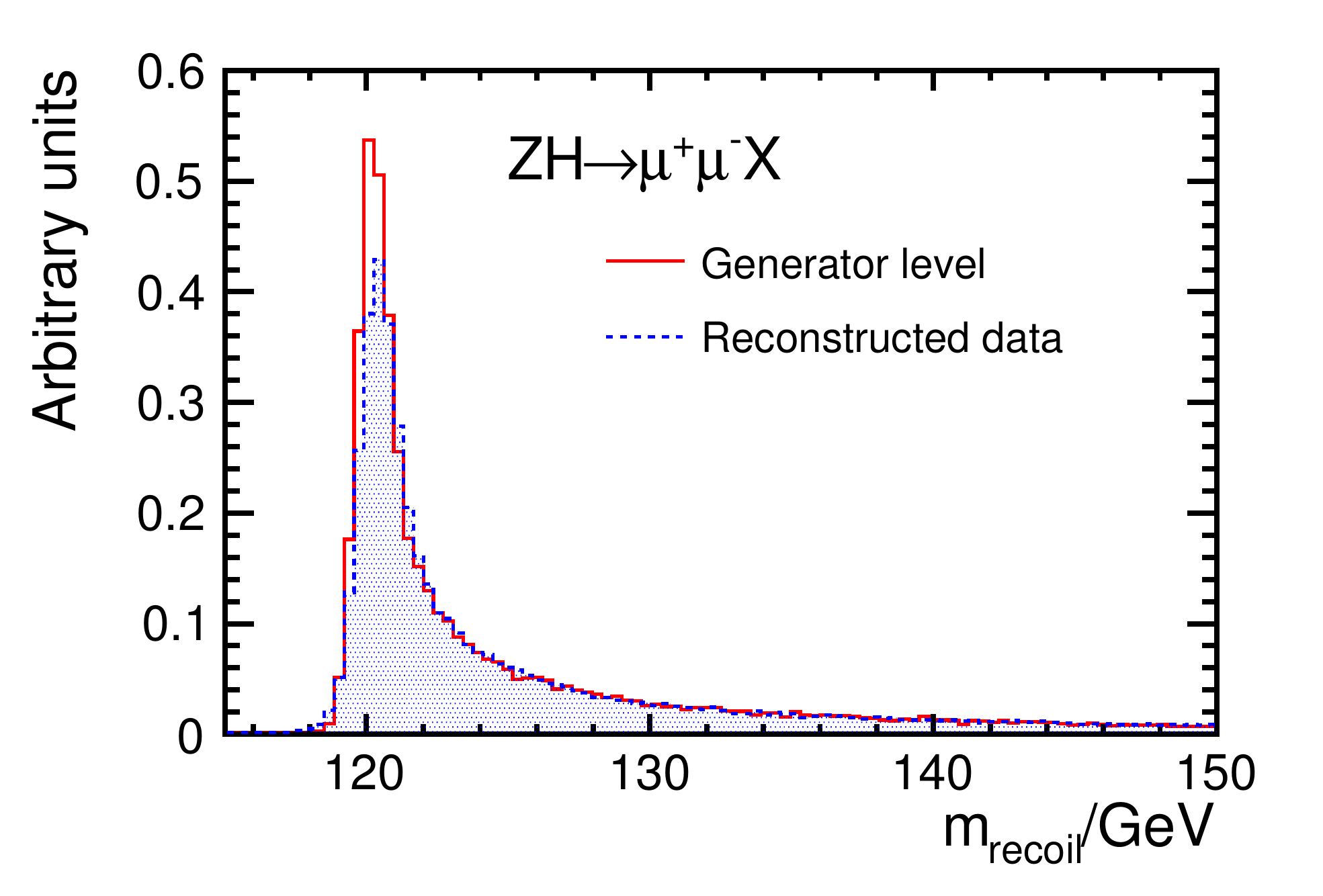}}
\caption[Generated and reconstructed Higgs recoil mass.]{The Higgs recoil mass 
distribution in the $\mmX$ channel obtained from the generator level and 
reconstructed muon pair momenta. \label{fig:mh_gen_sim}}
\end{figure}

The width of the peak of the recoil mass distribution is a
convolution of the detector response and the luminosity
spectrum of the centre-of-mass energy from the intrinsic beam 
energy spread and beamstrahlung.
For the $\mmX$ channel, the contribution
from the detector response is primarily due to the momentum resolution, whereas
for
the $\eeX$ channel bremsstrahlung dominates.
Figure~\ref{fig:mh_gen_sim}
shows the recoil mass spectrum for the $\mmX$ channel obtained from the
generated four momenta of the muon pair compared to that 
obtained from the reconstructed momenta.
The detector response leads to the broadening of
the recoil mass peak; an increase from 
560\,MeV to 650\,MeV. The contribution from momentum resolution is
therefore estimated to be 330\,MeV. For the  beam energy
spectrum used in the simulation,  the effect of 
detector resolution is not negligible, however, the dominant contribution 
to the observed width of the $\mmX$ recoil
mass distribution arises from the incoming beams rather than the response of ILD.

\subsubsection{Conclusions}

From Figure~\ref{fig:fit_ppme_gpet} and Table~\ref{tab:mirst} the following 
conclusions can be drawn: i) using the recoil mass distributions in the $\eeX$ and $\mmX$ final
states and 250\,fb$^{-1}$ of data with $P(e^+,e^-) = (+30\,\%, -80\,\%)$ beam polarisation,
$\mH$ can be determined with a statistical uncertainty of 32\,MeV independent
of its decay modes and the Higgs-strahlung cross section can be measured with a precision 
of 2.5\,\%; 
ii) the precision on $\mH$ obtained in the $\eeX$ channel is approximately a factor two worse than that obtained from
the $\mmX$ channel;
iii) with the current algorithm, the inclusion of Beamstrahlung photons in the recoil 
     mass distribution in the 
     $\eeX$ channel improves the Higgs mass resolution by approximately $10\,\%$.
iv) the ILD track resolution does not significantly degrade the $\mH$ resolution
     obtained from the $\mmX$ recoil mass distribution.

%% file: performance/physics_higgs_br.tex
The determination of the Higgs boson branching fractions is central to the ILC
physics programme. In the context of the SM, this allows a test of
the hypothesis that the strength of the Higgs coupling depends
linearly on the particle masses. The statistical uncertainties on the
branching ratios
are estimated, 
for an integrated luminosity of 250\,fb$^{-1}$ at $\roots=250$\,GeV,
based on the analysis of the  
Higgs-strahlung process $\epem\rightarrow\Zzero\Higgs$ for the three
possible $\Zzero$ decay topologies: 
$\Zzero\rightarrow\qq$, $\Zzero\rightarrow\nu\bar{\nu}$,
and $\Zzero\rightarrow\ell^+\ell^-$. Heavy flavour tagging is 
essential to the analysis; cuts on the $c$-tag and 
$b$-tag for the two jets from the candidate Higgs decay are employed. 
In addition, the $c$-tag information from two jets is combined 
into a single variable, $c$-likeness. For each topology, the uncertainty
on the exclusive cross sections are determined, 
{\it e.g.} $\sigma(\epem\rightarrow\Zzero\Higgs\to \qq\ccbar)$. 
This is combined with the 
2.5\,\% uncertainty on the total cross section, $\sigma(\epem\rightarrow\Zzero\Higgs)$, obtained from the
model Independent analysis described in the previous section,
to give the uncertainty 
on the branching ratios.

\subsubsection{\bf {\boldmath $ZH\to \ell^+\ell^-\qq$}}

\label{sec:zhllqq}

\begin{figure}
\begin{center}
\includegraphics[width=14cm]{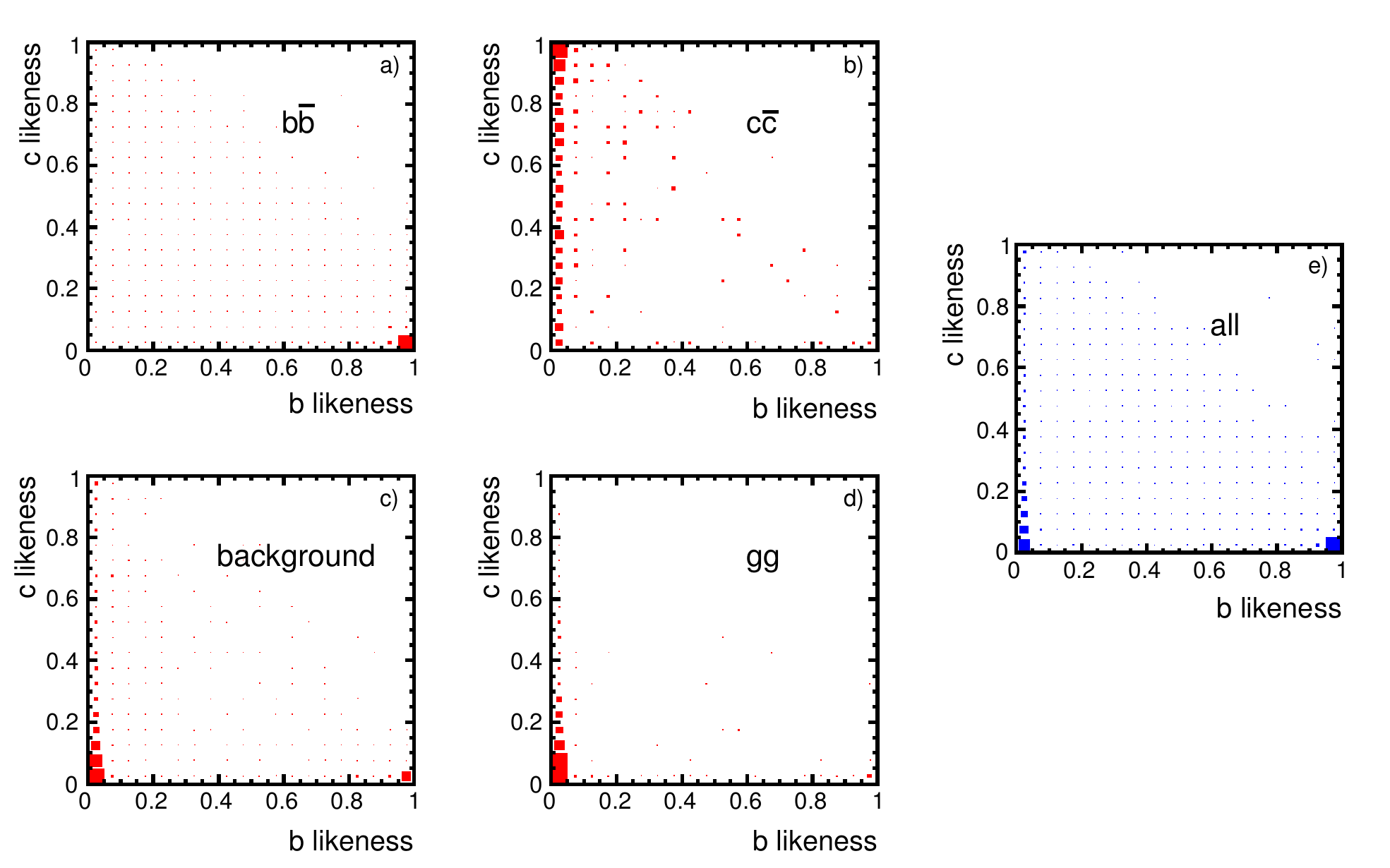}
\caption[Flavour tagging in $\Zzero\Higgs\rightarrow\ell^+\ell^-X$.]{Distributions of b- and c-likeness for exclusive samples of $H\to b\bar{b}$, $H\to c\bar{c}$, 
$H\to gg$, background and an independent combined "data" sample. \label{fig:eeh_templates}}
\end{center}
\end{figure}

Although statistically limited  compared to the other $\Zzero$ decay channels, 
$\Zzero\to\epem$ and $\Zzero\to\mpmm$ provide a clean 
signal which can be identified with high efficiency,
independent of whether the Higgs decays to $\bbbar$, $\ccbar$ or $gg$~\cite{Goldstein}. 
The dominant background is $\Zzero\Zzero$ production.
The event selection requires a 
pair of oppositely-charged electrons or muons with an invariant 
mass consistent with $\mZ$. The recoil mass is required to be
consistent with $\mH$ as is the invariant mass of the
recoiling hadronic system.  
Events in which the $\Zzero$ candidate is close to 
the beam axis are rejected to
suppress background from $\Zzero\Zzero$.
The final selection is performed by cutting on the value of a 
likelihood function formed from variables  related to the thrust, 
di-jet and di-lepton masses and angular distributions.
The hadronic system is reconstructed as two jets. To extract the
Higgs branching ratios it is not sufficient to simply apply cuts
to select, for example, $\Higgs\to\ccbar$ events since one of the
main background is from $\Higgs\to\bbbar$ for which the branching
ratio also needs to be determined. Instead, the fractions of 
$\Higgs\to \bbbar$, $\Higgs\to\ccbar$, 
$\Higgs\to gg$ and background present are determined from the 
distribution of $b$-likeness and $c$-likeness which is fitted using 
templates 
made from exclusive samples of each type as shown in Figure 
\ref{fig:eeh_templates}.
The measurement accuracy obtained  
is $(2.7\oplus2.5)$\,\% for $BR(\Higgs \to \bbbar)$, $(28\oplus2.5)$\,\% for $BR(H \to \ccbar)$ 
and $(29\oplus2.5)$\,\% for $BR(\Higgs \to gg)$.

\subsubsection{\bf {\boldmath $\Zzero\Higgs \rightarrow \nu\nubar\Higgs$}  }

The signal topology comprises two jets plus missing energy. Events are
selected based on missing mass, 
net transverse momentum, 
and net longitudinal momentum. Background containing
high momentum leptons is rejected using lepton identification cuts
and by requiring that the maximum track momentum in the event is less than
30\,GeV. The dominant remaining backgrounds are  
$\nu\bar{\nu}\qq$ and $\tau\nu_{\tau}\qq$ from 
$\Zzero\Zzero$ and $\WpWm$ respectively. These backgrounds are suppressed  
using $y_{12}$ and $y_{23}$, the $y$-cut values in the Durham jet-finding algorithm
for the transitions between one or two and two or three
reconstructed jets. The selection efficiencies for 
$\Zzero\Higgs\rightarrow\nu\nubar\ccbar$ and 
$\Zzero\Higgs\rightarrow\nu\nubar\bbbar$ are both approximately 44\,\%.
The branching ratios $BR(\Higgs \to \bbbar)$ and
$BR(\Higgs\to\ccbar)$ are determined using the $b$-, $c$-, and $bc$- flavour tags.
The reconstructed di-jet mass distribution after applying a cut on the
$c$-tag is shown in Figure~\ref{fig:hcc}b. To extract the Higgs branching ratios
the template fit of Section~\ref{sec:zhllqq} is extended to three dimensions by
including the $bc$-tag information. It is assumed that the non-Higgs background
is well understood. By fitting the signal contributions to
this distribution, the $\Higgs\rightarrow\ccbar$ and $\Higgs\rightarrow\bbbar$ cross sections
can be determined. 
The measurement accuracies for $BR(\Higgs \to \ccbar)$ and $BR(\Higgs \to \bbbar)$ are
$(13.8\oplus2.5)$ and $(1.1\oplus2.5)$\,\% repectively.

\subsubsection{\bf {\boldmath $\Zzero\Higgs\to \qq\ccbar$}}

\begin{figure}
\begin{center}
\includegraphics[width=7.0cm]{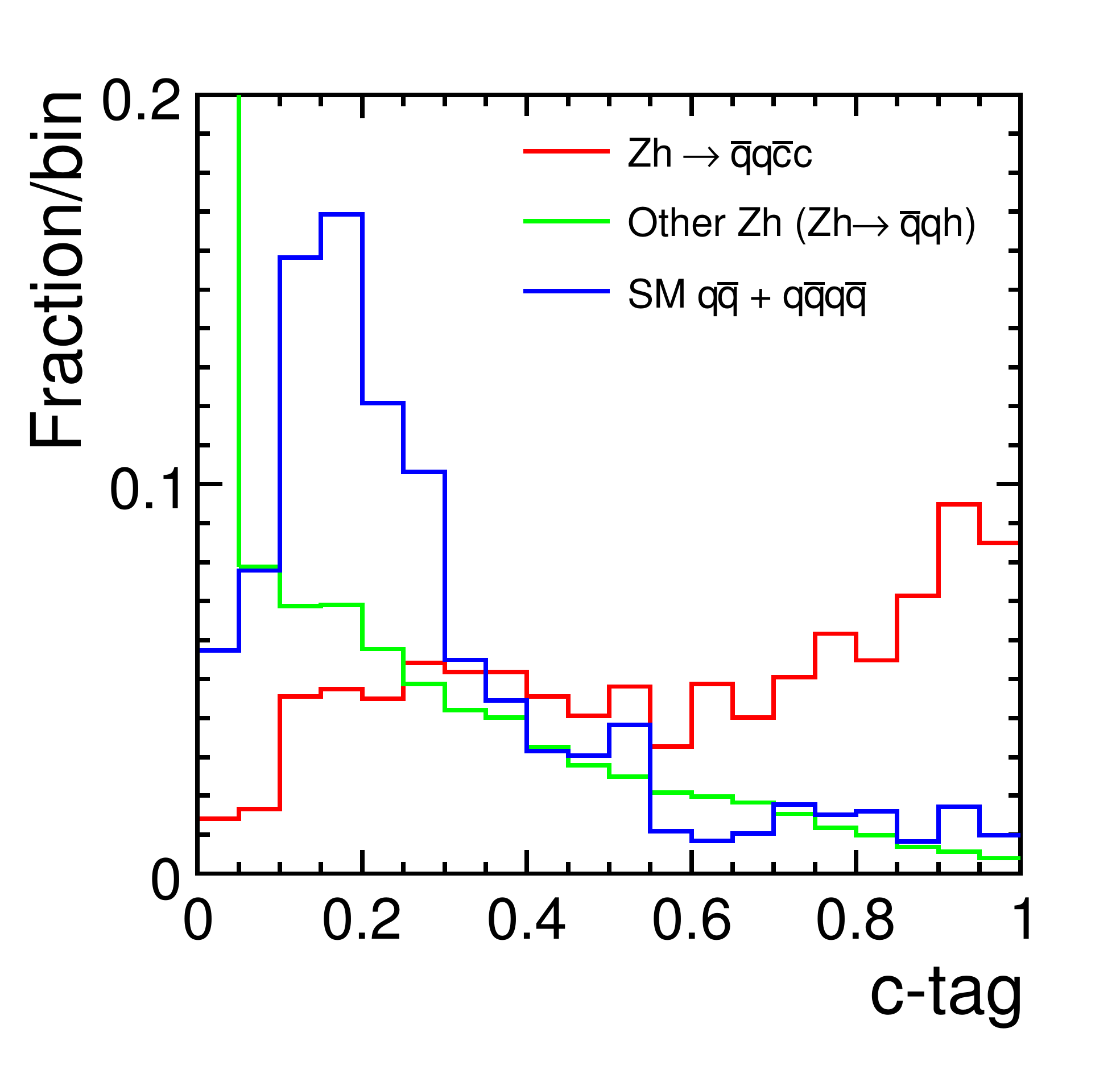}
\includegraphics[width=7.0cm]{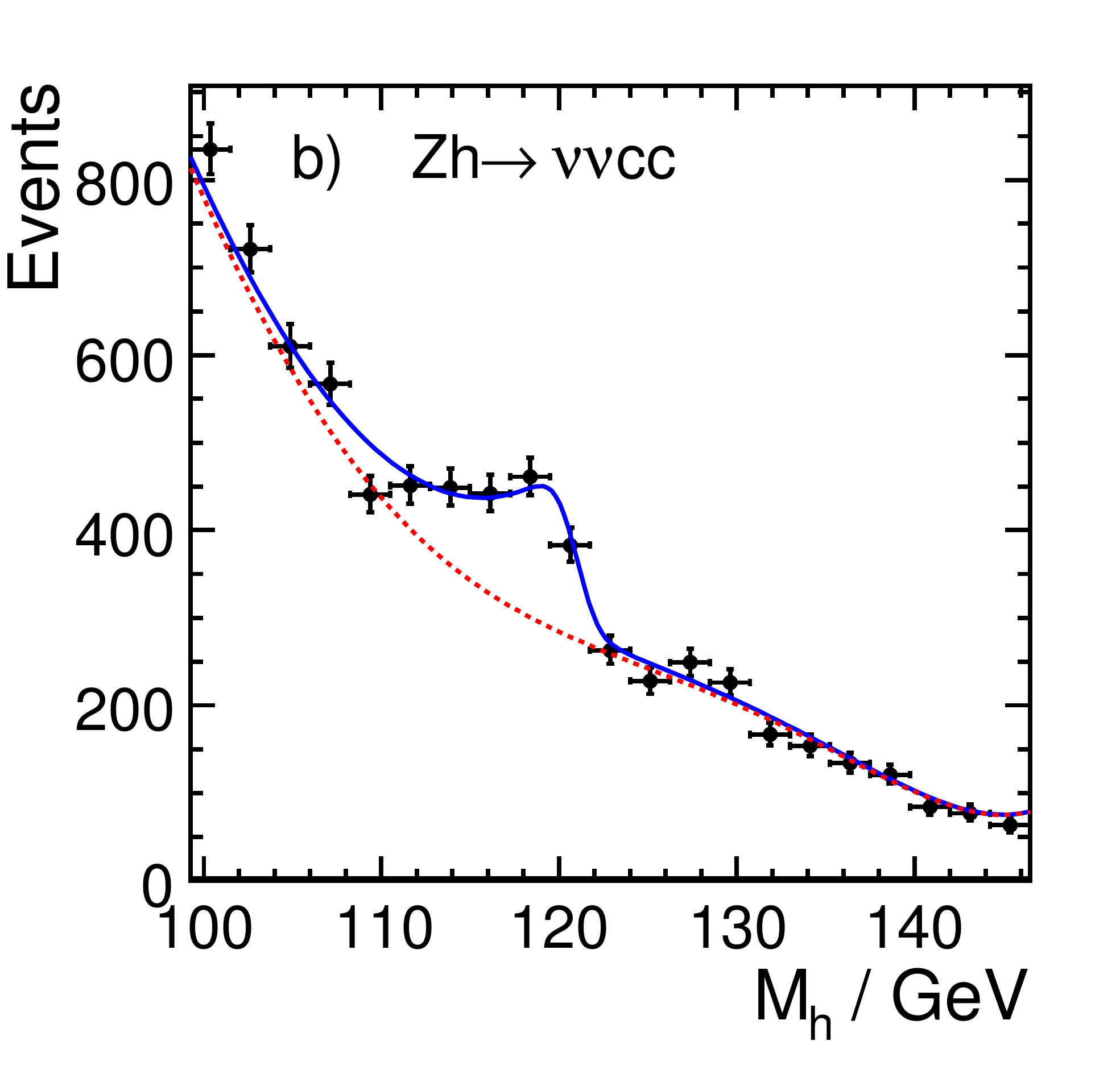}
\caption[$\Higgs\rightarrow\ccbar$ from $\Zzero\Higgs\rightarrow\qq X$
                          and $\Zzero\Higgs\rightarrow\nu\bar{\nu}X$.]{ 
       a) The $c$-tag of the two jets 
          in candidate $\Zzero\Higgs\rightarrow\qq\ccbar$ events after all other cuts
          apart from the c-tag and c-likeness cut.
        b) Distribution of the reconstructed di-jet mass for the 
      $\Zzero\Higgs \to \nu \bar{\nu} c\bar{c}$ sample prepared by $bc$-tagging. 
      \label{fig:hcc}}
\end{center}
\end{figure}

The decay topology for $\Zzero\Higgs\to \qq\ccbar$ consists of four jets,
two compatible with $\mZ$
and two compatible with $\mH$. The main backgrounds
are $\epem \rightarrow \WpWm/\Zzero\Zzero \rightarrow \qq\qq$
and four-jet events from the fragmentation of
$\epem \rightarrow \Zzero/\gamma^{\star} \rightarrow \qq$.
For the $\Zzero/\gamma^{\star} \rightarrow \qq$ 
background, four-jet events arise mainly from the
$q\bar{q}gg$ final state in which the gluon jets are generally less
energetic and are produced at relatively small angles to the quark 
jets. Consequently, cuts on event shape variables, such as the
smallest jet-jet angle, are
used to reject the $\Zzero/\gamma^{\star}$ background.
Background from $\qq\qq$ production are 
suppressed using kinematic fits.
A second fit, which imposes energy-momentum conservation 
and constrains one di-jet mass to equal $\mZ$, 
is used to reconstruct the Higgs mass, $m_\Higgs^{fit}$. 
The $\Zzero \Higgs \rightarrow \qq\ccbar$ sample is selected by requiring
$115<m_\Higgs^{fit}<125$\,GeV and using cuts on the $c$-likeness and
the $c-$tags of the two jets 
from the Higgs decay, shown in Figure~\ref{fig:hcc}.
For an integrated luminosity of 250\,fb$^{-1}$, the expected 
numbers of signal and background events after all cuts
are $37.2$ and $121.2$ respectively. This leads to a 
$(30\,\oplus 2.5)$\,\% uncertainty on 
$BR(\Higgs \rightarrow \ccbar)$ \footnote{Ongoing studies show that the expected statistical uncertainty from a more optimal analysis is more than a factor two smaller than the value quoted here.}
.  

\subsubsection{Combined Result}

The results for the Higgs branching ratios are summarised in 
Table~\ref{tab:hbr_summary}. The statistical
uncertainties are from the exclusive measurements  and 
the 2.5\,\% uncertainty on the total cross section.
After taking into account the different integrated luminosity and different centre-of-mass
energy, the combined results shown in Table~\ref{tab:hbr_summary} are broadly in 
agreement with those obtained with a fast simulation analysis performed in the context of the 
TESLA TDR~\cite{ref-TESLA_TDR}.

\begin{table}[ht]
\begin{center}
\begin{tabular}{|c|r|r|r|}\hline
Channel & $Br(\Higgs\rightarrow \bbbar)$ & $Br(\Higgs\rightarrow\ccbar)$ & 
        $Br(\Higgs\rightarrow gg)$                               \\ \hline
$\Zzero\Higgs \to \ell^+\ell^-\qq$      & $(2.7\oplus2.5)$\,\%  & $(28\oplus2.5)$\,\% & $(29\oplus2.5)$\,\% \\  
$\Zzero\Higgs \to \nu\bar{\nu}H$        & $(1.1\oplus2.5)$\,\%  & $(13.8\oplus2.5)$\,\% & $-$       \\
$\Zzero\Higgs \to \qq  \ccbar $         &         $-$           & $(30\oplus2.5)$\,\% & $-$          \\ \hline
Combined                                &  2.7\,\%  & 12\,\%  & 29\,\% \\ \hline
\end{tabular}
\caption[Higgs branching ratio measurement.]{Expected precision for the Higgs boson branching fraction
         measurements ($\roots=250$\,GeV) for the individual 
         $\Zzero$ decay channels and for the
         combined result. The expected $2.5\,\%$ uncertainty on the 
         total Higgs production cross section is added in quadrature. 
         The results are based on full simulation/reconstruction and
         assume an integrated luminosity of 250\,fb$^{-1}$. Entries marked
         $-$ indicate that results are not yet available.
	 \label{tab:hbr_summary}} 
\end{center}
\end{table}

%% file: performance/physics_tau.tex
The reconstruction of $\tau^+\tau^-$ events at $\sqrt{s} = 500$\, GeV provides a
challenging test of the detector performance in terms of separating
nearby tracks and photons. The expected statistical sensitivities for the $\tau^+\tau^-$ cross section, 
the $\tau^+\tau^-$ forward-backward asymmetry, $A_{FB}$, and the mean tau polarisation, $P_\tau$, 
are determined for and integrated luminosity of 500\,fb$^{-1}$ with beam polarisation,
$P(e^+,e^-) = (+30\,\%,-80\,\%)$.

Simulated events with less than seven tracks are clustered into candidate tau jets
each of which contains at least one charged particle. Tau-pair events are selected
by requiring
exactly two candidate tau jets with opposite charge. 
The opening angle between the two tau candidates is required to be $>178^\circ$ to 
reject events with significant ISR (including radiative return to the $\Zzero$). 
After cuts on visible energy, the polar
angles of the tau jets, and lepton identification, the purity of the $\tptm$ event 
sample is 92.4\,\%. For 500\,fb$^{-1}$ the statistical error of the cross section 
measurement, $|\cos\theta|<0.95$, corresponds to
0.29\%. The uncertainty on $A_{FB}$, determined from the numbers of $\tau^-$ in 
the forward and backward hemispheres, is $\pm0.0025$.

\begin{figure}[htb]
\begin{center}
\includegraphics[width=7.0cm]{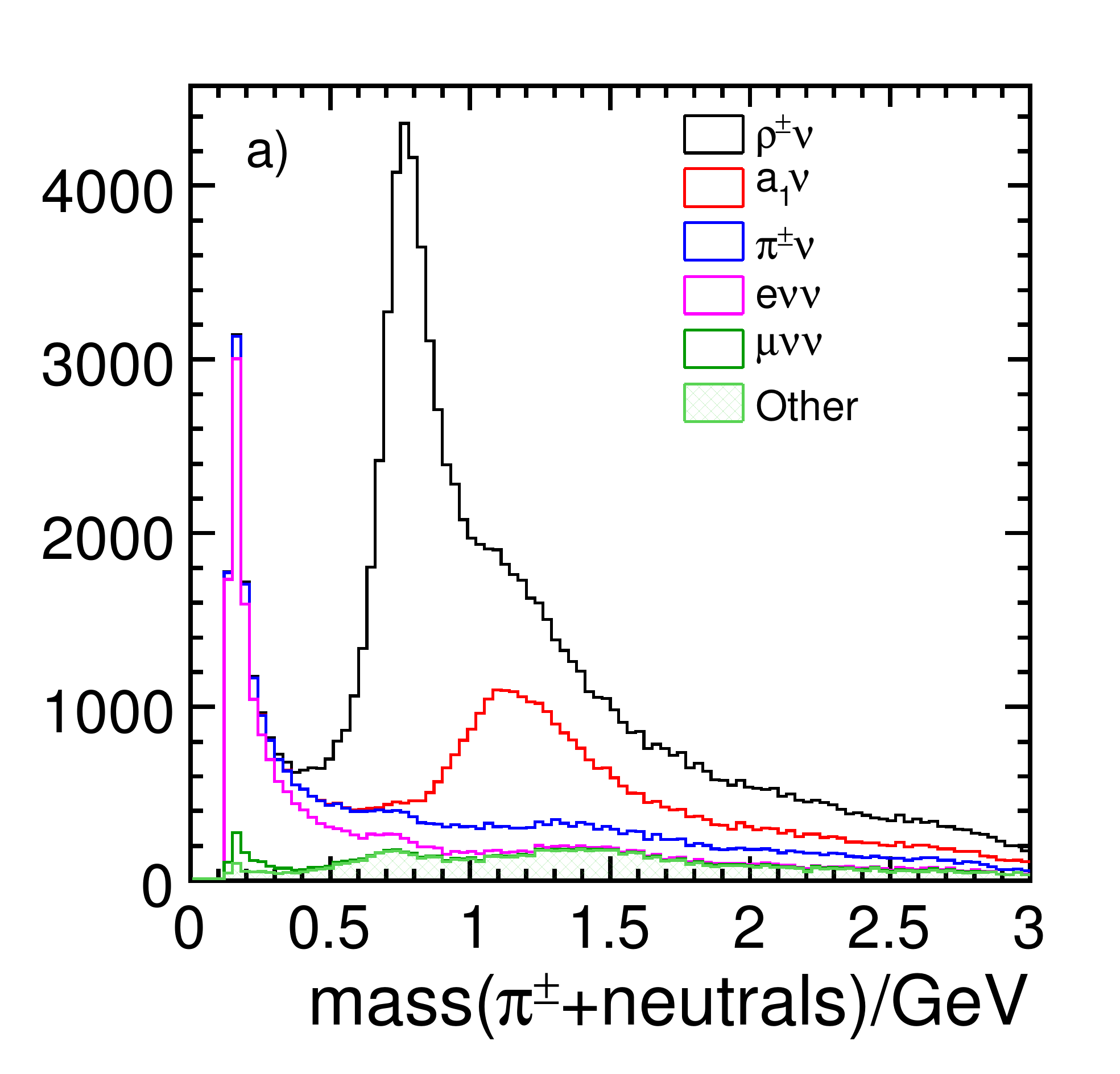}
\includegraphics[width=7.0cm]{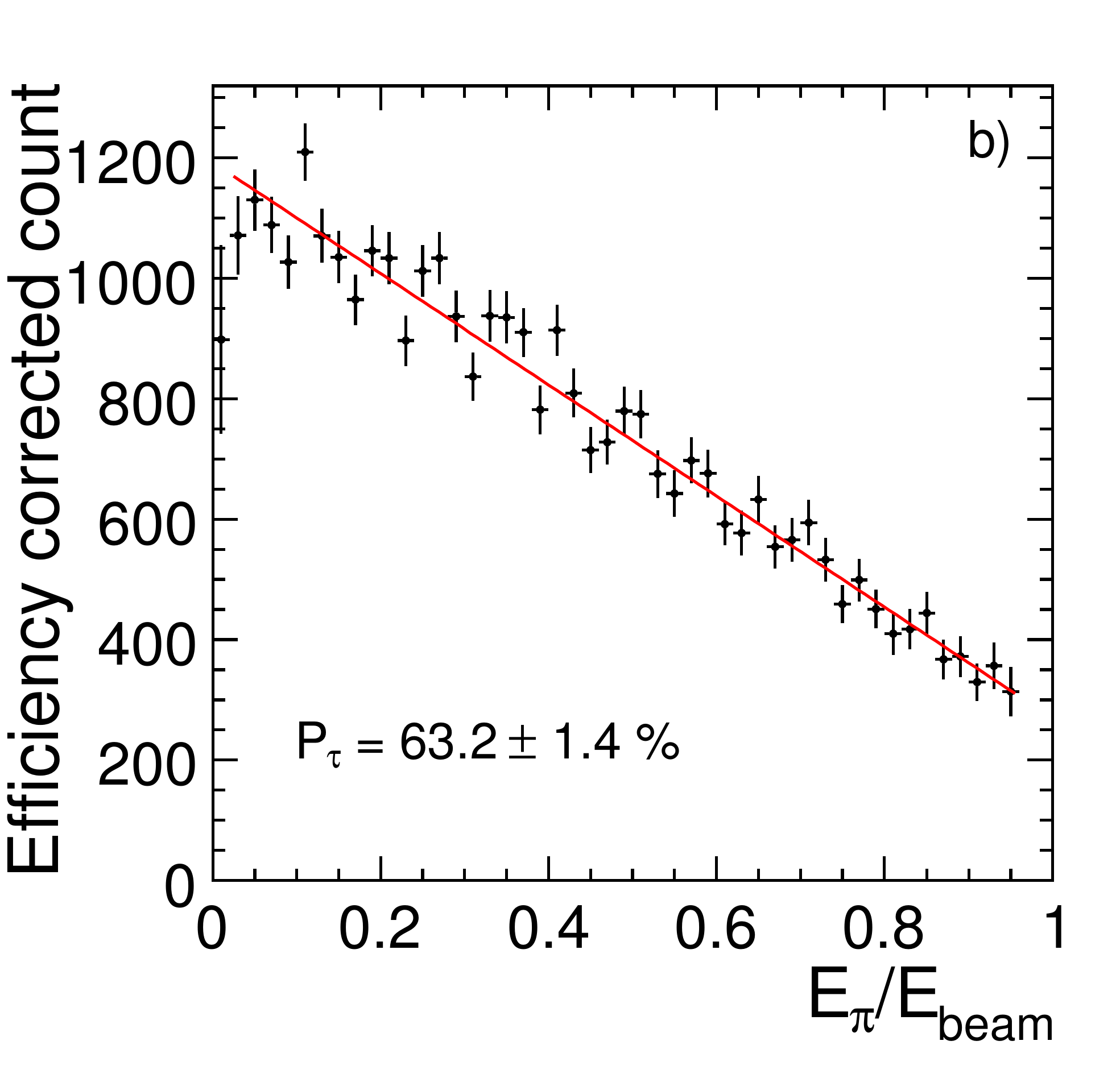}
\caption[$\tau\rightarrow\pi\nu$ selection.]{a) The invariant mass 
  distribution for selected 1-prong tau-candidates and 
  b) The efficiency corrected reconstructed pion energy 
  distribution for selected $\tau\rightarrow\pi\nu$
  candidates. \label{fig:tau-invmass}}
\end{center}
\end{figure}

The $\tau\to\pi\nu$ and $\tau\to\rho\nu$ decays have the highest sensitivity to 
$P_\tau$. The 
separation of the 1-prong decay modes relies on lepton identification and the ability to 
separate the neutral energy deposits from $\pi^0$ decays from the hadronic shower.
The invariant mass distribution for 1-prong events is shown in Figure~\ref{fig:tau-invmass}a.
A neutral network approach based on nine input variables is used to identify the
tau decays modes for each tau cone. The variables include: the total energy of the identified photons,
the invariant mass of the track and all identified photons (Figure~\ref{fig:tau-invmass}a); 
and electron and muon particle identification variables based on calorimetric information and
track momentum. Table \ref{tab:tau-modesel} shows 
the efficiency and purity achieved for the six main tau decay modes. The high granularity
and the large detector radius of ILD results in excellent separation.  

\begin{table}
\begin{center}
\begin{tabular}{|r|r|r|r|r|}
\hline
Mode 		       & Efficiency  & Purity   \\ \hline\hline
$e\nu\nu$              & 98.9\,\%    & 98.9\,\% \\ \hline
$\mu\nu\nu$            & 98.8\,\%    & 99.3\,\% \\ \hline
$\pi\nu$	       & 96.0\,\%    & 89.5\,\% \\ \hline
$\rho\nu$	       & 91.6\,\%    & 88.6\,\% \\ \hline
$a_1\nu$ (1-prong)     & 67.5\,\%    & 73.4\,\% \\ \hline
$a_1\nu$ (3-prong)     & 91.1\,\%    & 88.9\,\% \\ \hline
\end{tabular}
\caption[Efficiency and purity of tau decay mode selections.]{Purity and efficiency of the main tau 
          decay mode selections. The selection efficiency is calculated with
         respect to the sample of $\tptm$ after the requirement that the two tau candidates
         are almost back-to-back. The purity only includes the contamination from
         other $\tptm$ decays.}
\label{tab:tau-modesel}
\end{center}
\end{table}

For the beam polarisations of $P(e^+,e^-) = (+30\,\%,-80\,\%)$
and $P(e^+,e^-) = (-30\,\%,+80\,\%)$ the mean tau polarisations are
$-0.625$ and $+0.528$ respectively. For the measurement of $\Ptau$, 
only the $e^\pm\nu\nu$, $\mu^\pm\nu\nu$, $\pi^\pm\nu$, and $\rho^\pm\nu$
decay modes are used. The optimal variable approach~\cite{bib:Davier} 
is used to obtain the best sensitivity to the tau polarisation.  
In the case of the  $\pi^\pm\nu$ decay mode, the optimal observable is
the simply reconstructed $\pi^\pm$ energy divided by the beam energy, shown in
Figure~\ref{fig:tau-invmass}b. For the selected event sample and decay mode
identification the resulting statistical uncertainties on the measured 
mean tau polarisations are $\pm0.007$ and $\pm0.008$ for $P(e^+,e^-) = (+30\,\%,-80\,\%)$
and $P(e^+,e^-) = (-30\,\%,+80\,\%)$ respectively.

%% file: performance/physics_susypoint5.tex
In the SUSY ``point 5'' scenario with non-universal soft SUSY-breaking 
contributions to the Higgs masses, $\tilde{\chi}^{\pm}_1$ and 
$\tilde{\chi}^0_2$ are not only nearly mass degenerate but decay 
predominantly into $\Wboson^{\pm}\tilde{\chi}^0_1$ and 
$\Zzero\tilde{\chi}^0_1$, respectively.
This benchmark point has the following parameters: 
$M_0 = 206\,\mathrm{GeV}$, 
$M_{1/2} = 293\,\mathrm{GeV}$,  
$\tan{\beta} = 10$,
$A_0 = 0$, and
$\mu=375\,\mathrm{GeV}$ and the gaugino masses are: 
$m({\tilde{\chi}^0_1})$ = 115.7\,GeV, 
$m({\tilde{\chi}^{\pm}_1})$ = 216.5\,GeV, 
$m({\tilde{\chi}^0_2}$ = 216.7\,GeV, and
$m({\tilde{\chi}^0_3})$ = 380\,GeV.

Both
$\epem\rightarrow\tilde{\chi}^+_1\tilde{\chi}^-_1 \to qq\tilde{\chi}^0_1 qq\tilde{\chi}^0_1$ and
$\epem\rightarrow\tilde{\chi}^0_2\tilde{\chi}^0_2 \to qq\tilde{\chi}^0_1 qq\tilde{\chi}^0_1$
result in four jets and missing energy, where the di-jet masses are characteristic of the decays
of $\WpWm$ or $\Zzero\Zzero$.
Separating $\Wboson$ and $\Zzero$ decays in the fully-hadronic decay mode relies on 
good jet energy resolution. It thus provides a benchmark for particle flow based jet reconstruction.  
The analysis is complicated by the fact that the $\tilde{\chi}^0_2\tilde{\chi}^0_2$ cross section is
only 10\,\% of that for $\tilde{\chi}^+_1\tilde{\chi}^-_1$.

\begin{figure}[htb] 
\begin{center}
\includegraphics[width=6.0cm]{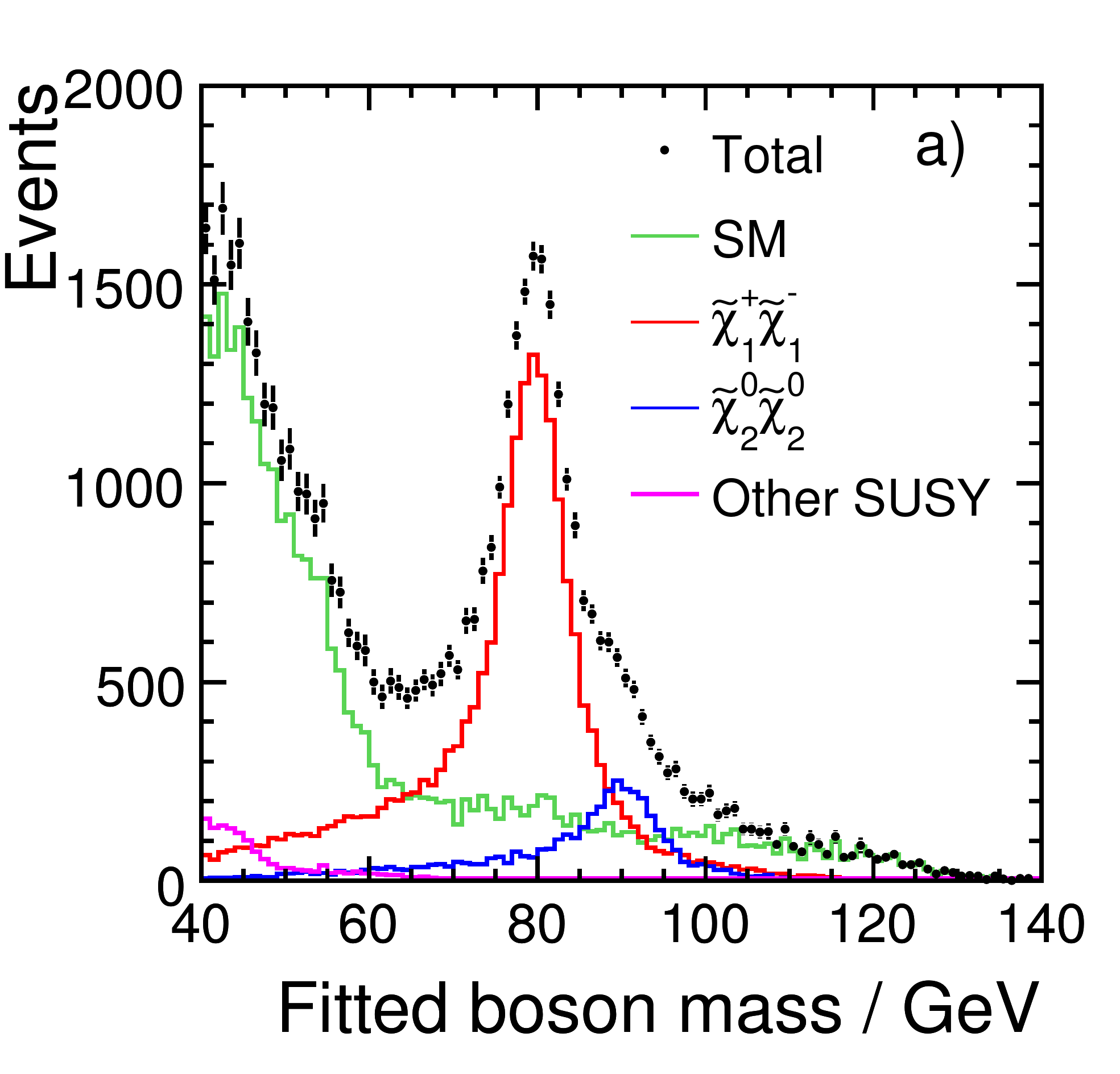}
\includegraphics[width=6.0cm]{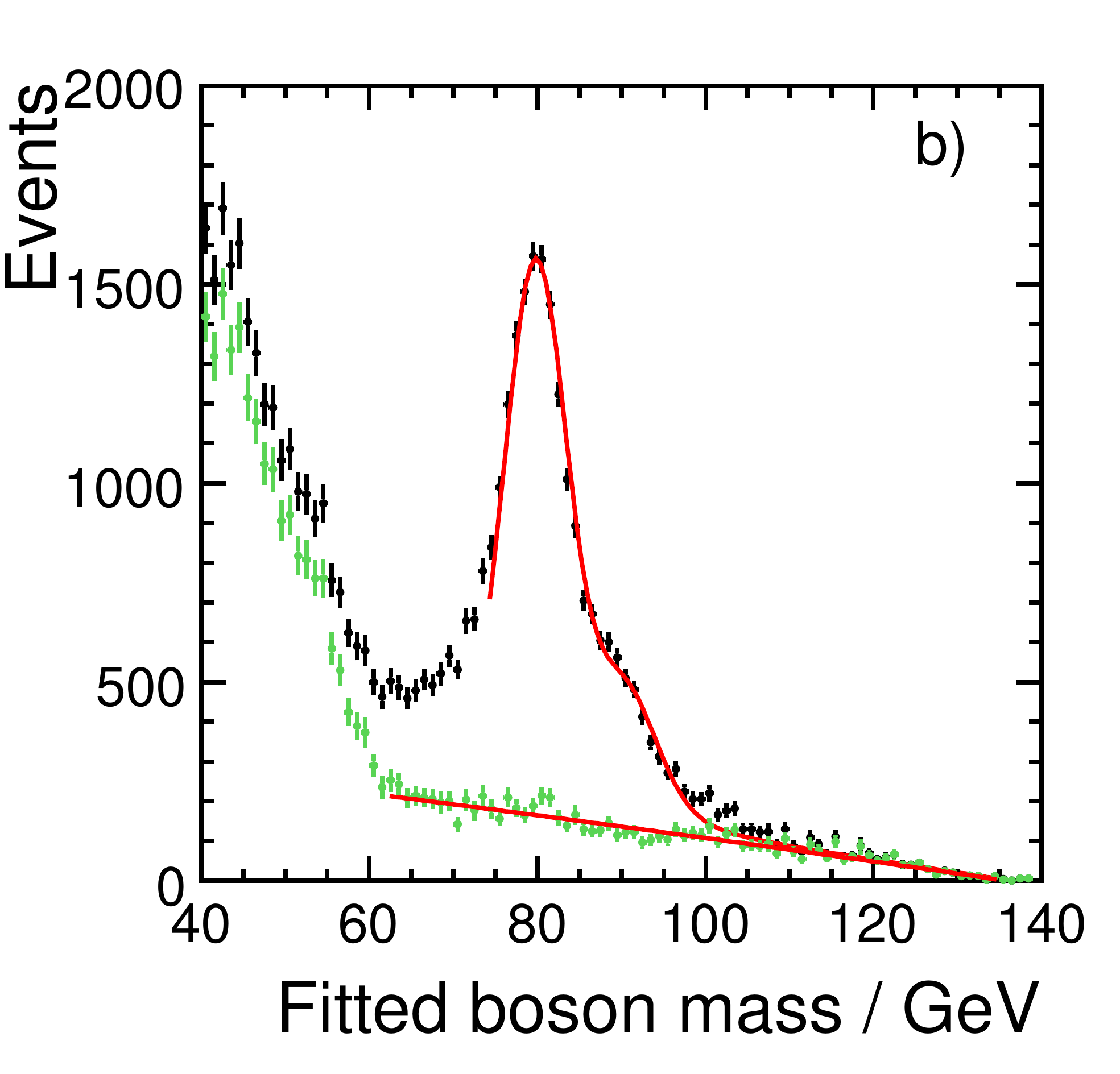}
\includegraphics[width=6.0cm]{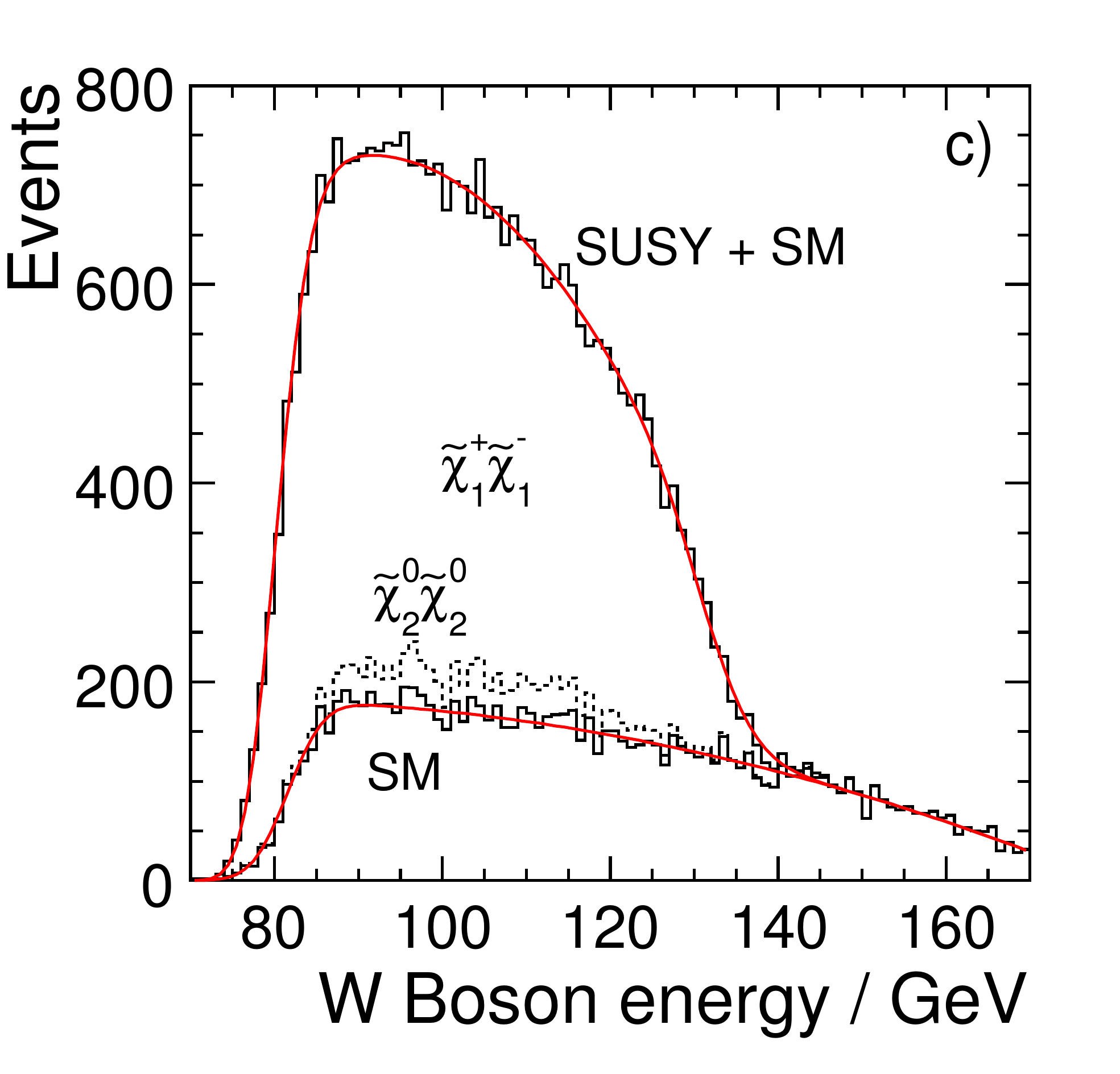}
\includegraphics[width=6.0cm]{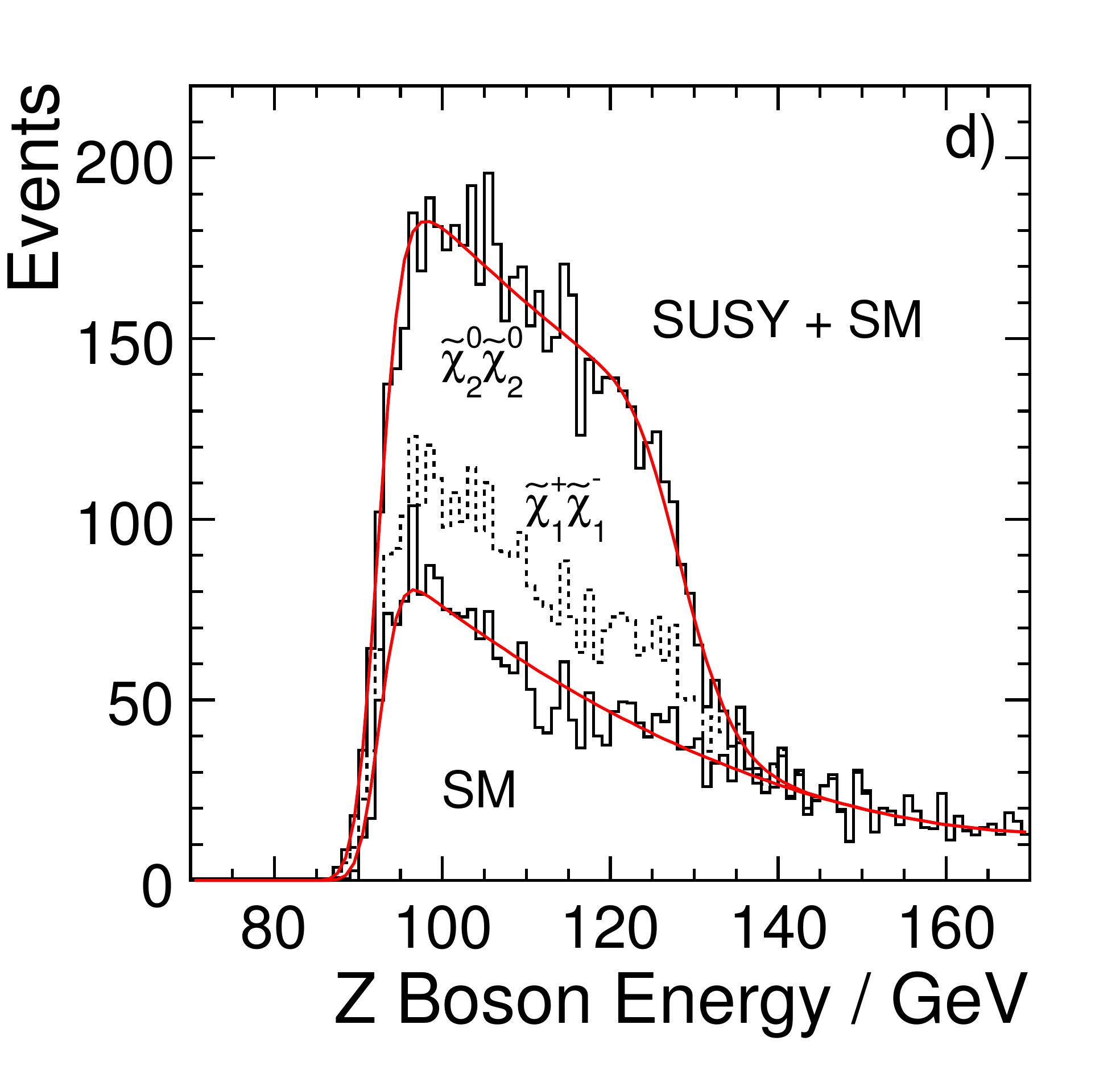}
  \caption[Chargino/neutralino selection and mass fits.]{a) Di-jet mass from the 5C kinematic fit after all selection cuts. b) 
      Fit of the background and Chargino and Neutralino contributions. The fit parameters 
      are the normalisations of the $\Wboson$ and $\Zzero$ peaks. 
      c) Energy spectra of $\Wboson$ and $\Zzero$ boson candidates after the Chargino and d) Neutralino 
         event selections, shown including fits to signal and background contributions.
      \label{fig:fitmass_allcuts}}
\end{center}
\end{figure}

The event selection starts by forcing events into four jets. A cut based preselection retains events 
consistent with a four-jet plus missing energy topology. All three possible di-jet associations 
to two bosons are considered.
A kinematic fit which constrains the two boson masses to be equal is applied;
in terms of mass resolution this is essentially equivalent 
to taking the average mass of the two di-jet systems.
Two analysis strategies are used to assess the expected uncertainty on the measured cross sections: 
{\bf i)} 
The first method aims to reduce the SM background as far as possible.
Cuts on the number of particle flow objects  in each jet, 
the direction of the missing momentum, and the missing mass are applied.
The kinematic fit is required to converge for at least one 
jet pairing. The jet pairing yielding the highest $\chi^2$ probability is used. Figure~\ref{fig:fitmass_allcuts}a shows the resulting di-jet mass distribution.
The Chargino signal has a small shoulder from the Neutralino contribution. 
The cross sections are obtained from a fit to the mass spectrum using  
a function with three components: a Breit-Wigner ($\mW, \GammaW$)
convolved with a Gaussian for the $\Wboson$-peak; 
a Breit-Wigner ($\mZ, \GammaZ$) convolved with the same Gaussian for the $\Zzero$ 
peak; and a second order polynomial.
The width of the Gaussian is fixed to 3.4~GeV reflecting the mass resolution. The 
two free parameters of the fit are the normalisations of the $\Wboson$ and $\Zzero$ peaks. 
Figure~\ref{fig:fitmass_allcuts}b shows the result of the fit. The statistical errors 
on the cross sections are $0.95\,\%$ for the Chargino signal and  $2.9\,\%$ for the Neutralino
signal.
{\bf ii)} The second approach, which does not use kinematic fitting, is to 
fit the two-dimensional distribution of the two di-jet masses in each event with MC 
templates, leaving only the normalisations of the two signal contributions free. The fit is 
performed after the preselection cuts. All three possible jet pairings are included. 
This method yields smaller statistical 
errors of $0.64\,\%$ for the Chargino and  $2.1\,\%$ for the Neutralino production
rates.

To determine the $\tilde{\chi}^{\pm}_1$ and $\tilde{\chi}^0_2$ masses, 
$\tilde{\chi}^{+}_1\tilde{\chi}^{-}_1$ and $\tilde{\chi}^0_2\tilde{\chi}^0_2$
samples are defined on the basis of the  
di-jet mass distributions (without the kinematic fit). The resulting energy spectra of 
the $\Wboson$ and $\Zzero$ candidates from the kinematic fit 
are shown in Figure~\ref{fig:fitmass_allcuts}c/d. The masses of the gauginos are
determined from the kinematic edges of the distributions located 
using an empirically determined 
fitting function for the signal and a parameterisation of the SM background.
From the fit results the upper
and lower kinematic edges  
of the $\tilde{\chi}^{\pm}_1$ sample are determined to  $\pm 0.2$\,GeV and
$\pm 0.7$\,GeV respectively. The corresponding numbers for the 
$\tilde{\chi}^{0}_2$ sample are: $\pm 0.4$\,GeV and $\pm 0.8$\,GeV.
For the SUSY point 5 parameters, the $\tilde{\chi}^{\pm}_1$ lower edge is close
to $\mW$ and, thus, does not significantly constrain the gaugino masses.
The other three kinematic edges can be used to determine the gaugino masses
with a statistical precision of  2.9\,GeV, 1.7\,GeV and 1.0\,GeV for the
$\tilde{\chi}^{\pm}_1$, $\tilde{\chi}^{0}_2$, and $\tilde{\chi}^{0}_1$ respectively.
The errors on the masses are larger than the errors on the positions of the
edges themselves. This reflects the large correlations between the extracted 
gaugino masses; the differences in masses are better determined than the sum.
If the LSP mass were known from other measurements, {\it e.g.} from the slepton sector, 
the errors on the $\tilde{\chi}^{\pm}_1$ and $\tilde{\chi}^{0}_2$ masses would
be significantly reduced.
Furthermore, the resolutions can be improved by about a factor of two using a
kinematic fit which constrains the boson masses for chargino
(neutralino) candidates not only to be equal to
each other, but also to be equal to the nominal $\Wboson$ ($\Zzero$)
mass. In this case, statistical precisions of 2.4\,GeV, 0.9\,GeV, and
0.8\,GeV are obtained for the $\tilde{\chi}^{\pm}_1$,
$\tilde{\chi}^{0}_2$, and $\tilde{\chi}^{0}_1$ respectively.

%% file: performance/physics_top.tex
Top physics will be an important part of the scientific programme at the ILC. 
In particular, the top mass, $\mtop$, and top width,
$\Gtop$, can be determined with high precision. 
The measurement of $\mtop$ and $\Gtop$ from the direct reconstruction of 
$e^{+} e^{-} \to \ttbar$ events is studied with the full ILD detector
simulation and reconstruction. Two main decay topologies are
considered: fully-hadronic, $\ttbar \to (b \qq) (\bar{b} \qq)$, and 
semi-leptonic, $\ttbar \to (b \qq) (\bar{b} \ell \nu)$.
Results are obtained for an integrated luminosity of $100 \mathrm{fb}^{-1}$ 
at $\roots=500$\,GeV, assuming unpolarised beams. 

\begin{figure}[!b]
\centering
\includegraphics[width=7.0cm]{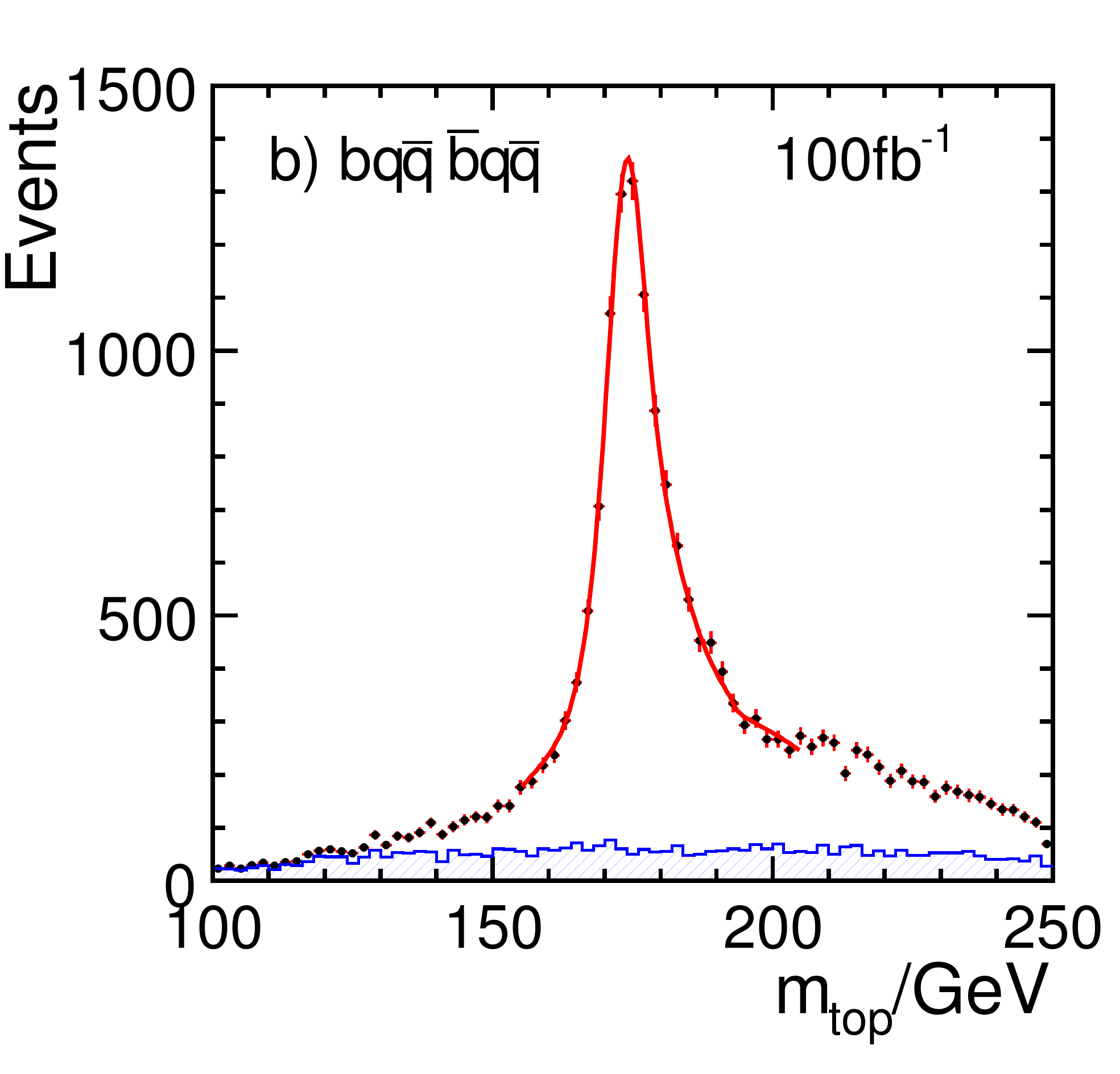}
\includegraphics[width=7.0cm]{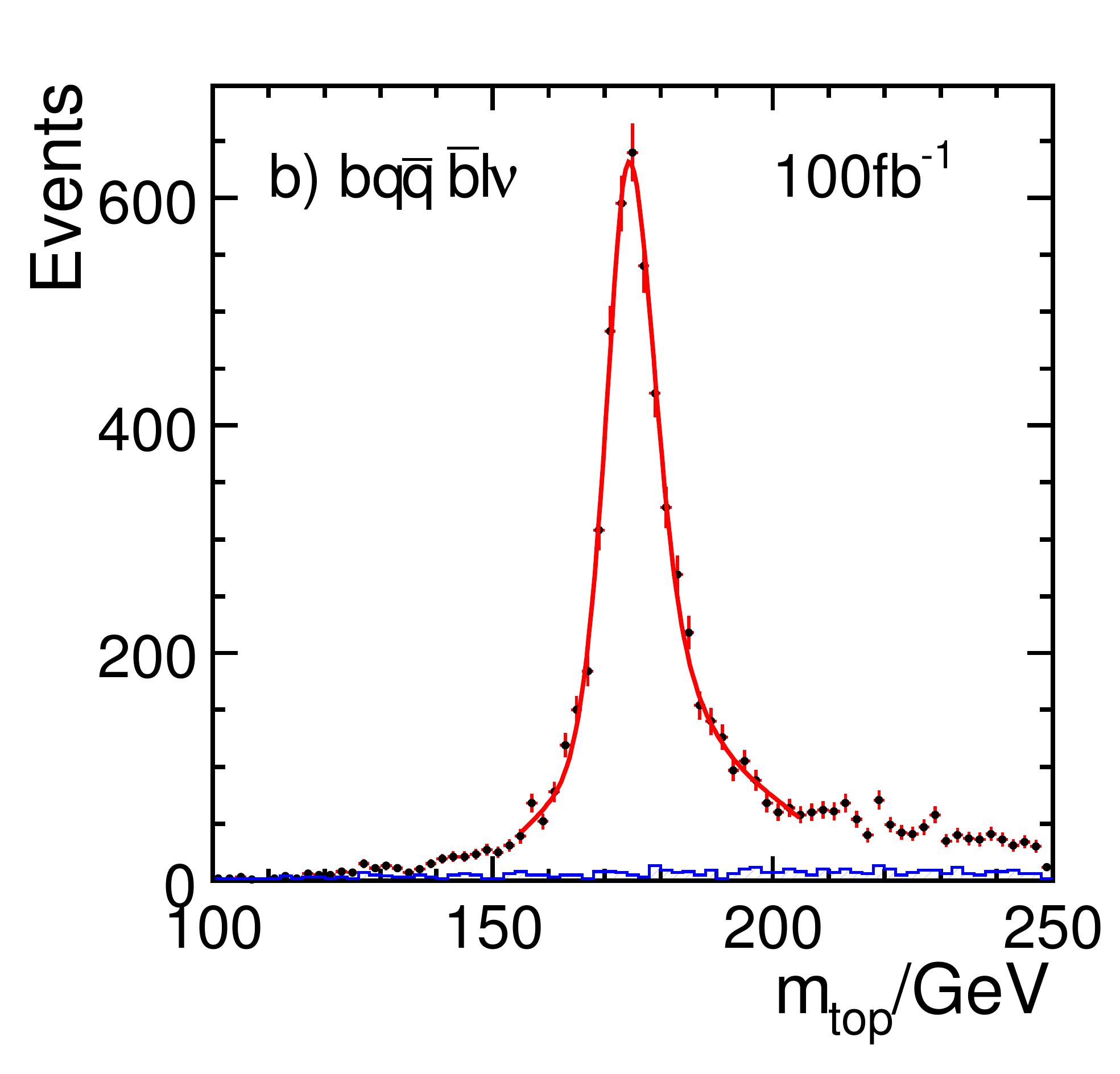}
\caption[Top quark mass.]{Distribution of the reconstructed top quark mass for a)
      the fully-hadronic $\ttbar \to (b q \bar{q}) (\bar{b} q \bar{q})$ signal sample 
      and b) the semi-leptonic $\ttbar \to (b q \bar{q}) (\bar{b} \ell \nu)$ signal sample. 
      The contribution from the non-$\ttbar$ background is indicated by the hashed distribution.
         The fits to the distributions are also shown.\label{fig:top}}
\end{figure}

Events with an isolated lepton 
are considered to be candidates for the semi-leptonic analysis, 
otherwise they are assumed to be 
candidates for the fully-hadronic analysis branch.
In the fully hadronic branch, the event is reconstructed as six jets which are 
combined to form $\Wboson$s and top quarks. 
The two b-jets originating directly from the top quark decays 
are identified using the flavour-tagging information.
The four remaining jets are considered as the decay products of the two $\Wboson$s.
The combination of the four jets into two di-jets which gives  the
smallest value of $|m_{ij}-\mW|+|m_{kl}-\mW|$ is chosen
to form the two $\Wboson$s (where $m_{ij}$ and $m_{kl}$ are the di-jet masses 
for a given jet pairing). 
Out of two possible combinations 
to pair the $\Wboson$s with the b-jets, the one which yields the smallest mass 
difference is chosen. The first step in the semi-leptonic branch is to remove 
the identified lepton and to force the remainder of the event into four jets. 
The two b-jets are identified using flavour-tagging information. 
The two remaining jets are assigned to the hadronically decaying $W$. 
The identified lepton and the neutrino are assigned to the 
leptonically decaying $\Wboson$, with the three-momentum of the neutrino
defined as the missing momentum. 
The pairing of the $\Wboson$s with the b-jets which yields the 
smallest reconstructed top mass difference is chosen.
For each analysis branch, background 
events are rejected using a multi-variate likelihood technique~\cite{BinnedLikelihood}. 
Finally, a kinematic fit~\cite{MarlinKinFit} 
is applied in order to improve the final $\mtop$ resolution. Events with a poor fit 
$\chi^2$ are rejected. The reconstructed mass distributions are shown
in Figure~\ref{fig:top}.

For an integrated luminosity of 500\,fb$^{-1}$, 
$\sigma(\epem\rightarrow\ttbar)$ can be determined with a
statistical uncertainty of 0.4\,\% using the fully-hadronic decays only. 
The invariant mass spectra are fitted with the convolution of a Breit-Wigner function 
and an asymmetric double Gaussian, the latter representing the detector resolution. 
The combinatoric background and the background from other process 
is described by a 2$^\mathrm{nd}$ order polynomial.
The fully-hadronic (semi-leptonic) analysis branch results in statistical 
uncertainties of 90\,MeV  (120\,MeV) and 60\,MeV (100\,MeV) for $\mtop$ and 
$\Gtop$ respectively.
Scaling the combined results to an integrated luminosity of 500\,fb$^{-1}$
leads to uncertainties of 
30\,MeV on $\mtop$ and 22\,MeV on $\Gtop$. The relatively small gain in 
statistical precision from a beam polarisation 
of $P(e^+,e^-) = (+30\,\%, -80\,\%)$ has not been accounted for.

\subsubsection{Top Quark Forward-Backward Asymmetry}

\begin{figure}[!b]
\centering
\includegraphics[width=10.0cm]{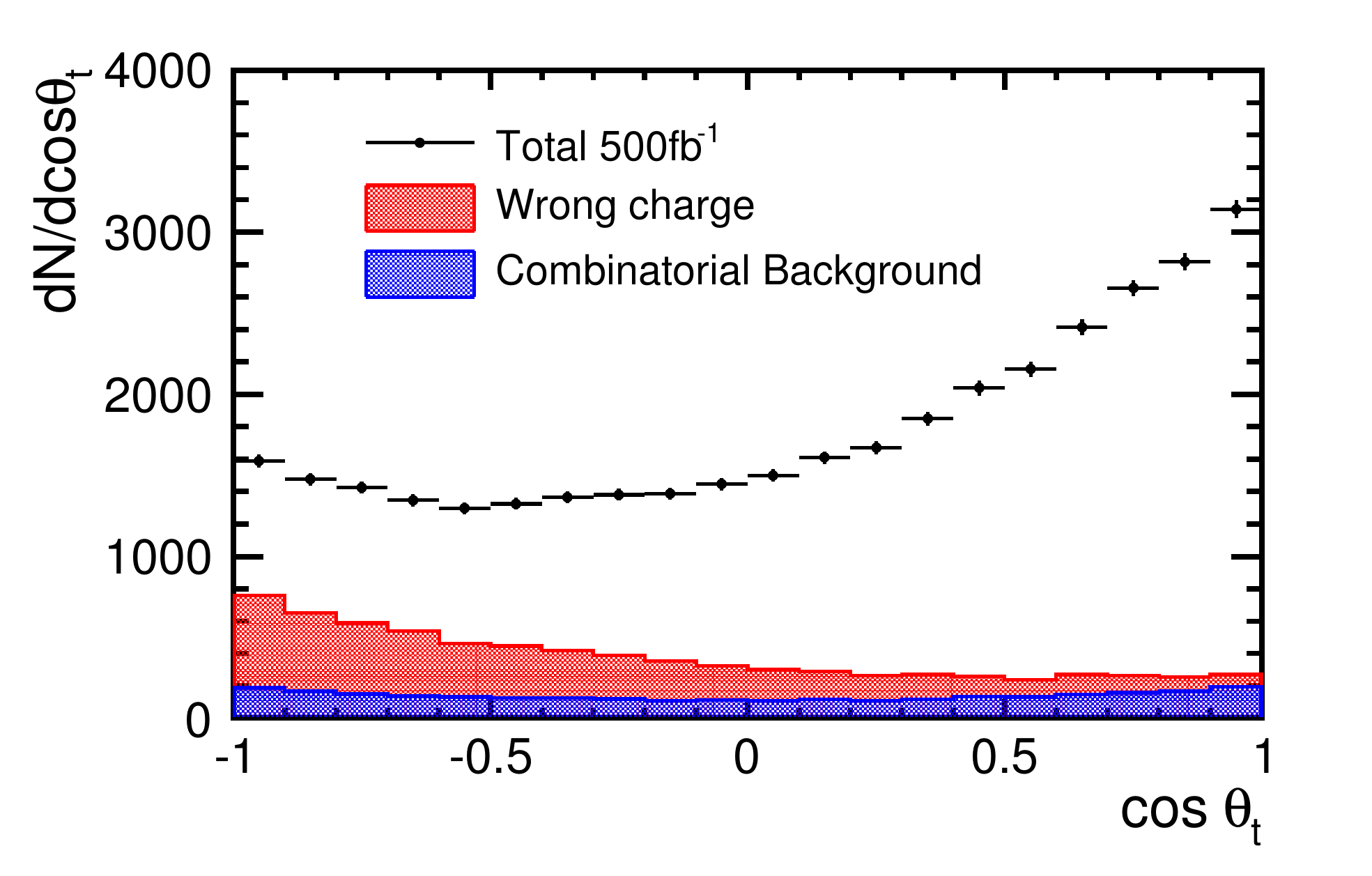}
\caption[Top quark forward-backward asymmetry.]{Distribution of the reconstructed polar angle of
         the identified top quark in fully-hadronic  $\ttbar$ events. The contributions from
         events with the wrong charge (red) and the case where the b-quark is misidentified
         are shown (blue). \label{fig:topAfb}}
\end{figure}

The top quark forward-backward asymmetry, $A_{\mathrm FB}^t$,
provides a potentially interesting
test of the SM. For the semi-leptonic the analysis is relatively straightforward
as the charge of the lepton tags the charge of the W-boson and, thus,
enables the $t$ and $\bar{t}$ to be identified. In the fully-hadronic
channel the $t$ and $\bar{t}$ can be identified by tagging the 
$b$/$\bar{b}$ from the charge of the secondary vertex from charged
$B$-hadron decays. This measurement provides a test of the vertex
reconstruction capability of ILD. This study is performed for
500\,fb$^{-1}$ with $P(e^+,e^-) = (+30\,\%, -80\,\%)$. Secondary
vertices identified by the LCFIVertex algorithm for the two identified
b-jets are used. For each of the two identified b-jets, 
the jet charge is reconstructed. Events with like 
sign combinations are rejected as are events with  two neutral
secondary vertices. In addition, the acollinearity
between the two top quark jets is required to be $<8^\circ$ to
reject events with $\sqrt{s^\prime}$ significantly less than
500\,GeV. Of the 20\,\% of fully hadronic $\ttbar$
events which pass these cuts, 79\,\% have the correctly identified
top quark charge. Figure~\ref{fig:topAfb} shows the distribution of
the cosine of the 
reconstructed polar angle of the tagged top-quark, showing
a clear forward-backward asymmetry. 
The relative numbers of events in the forward and backward hemispheres,
accounting for the charge identification/mis-identification probabilities,
 are used to determine 
\begin{eqnarray*}
    A_{\mathrm FB}^t = 0.334 \pm 0.0079.
\end{eqnarray*}

%% file: performance/physics_wwvv.tex
If strong 
electroweak symmetry breaking (EWSB) is realised in nature,
 the study of the WW-scattering processes is particularly
important. At the ILC, the $\mathrm{W}^+\mathrm{W}^-\rightarrow\mathrm{W}^+\mathrm{W}^-$
and $\mathrm{W}^+\mathrm{W}^-\rightarrow\mathrm{Z}\mathrm{Z}$ vertices can
be probed via the processes $\eplus\eminus\rightarrow\nu_{e}\overline{\nu}_{e}\qq\qq$
where the final state di-jet masses are from the decays of two W-bosons or
two Z-bosons. Separating the two processes through the reconstruction of
the di-jet masses provides a test of the jet energy resolution of the
ILD detector. 

\begin{figure}[ht]
\centering
\includegraphics[width=7.0cm]{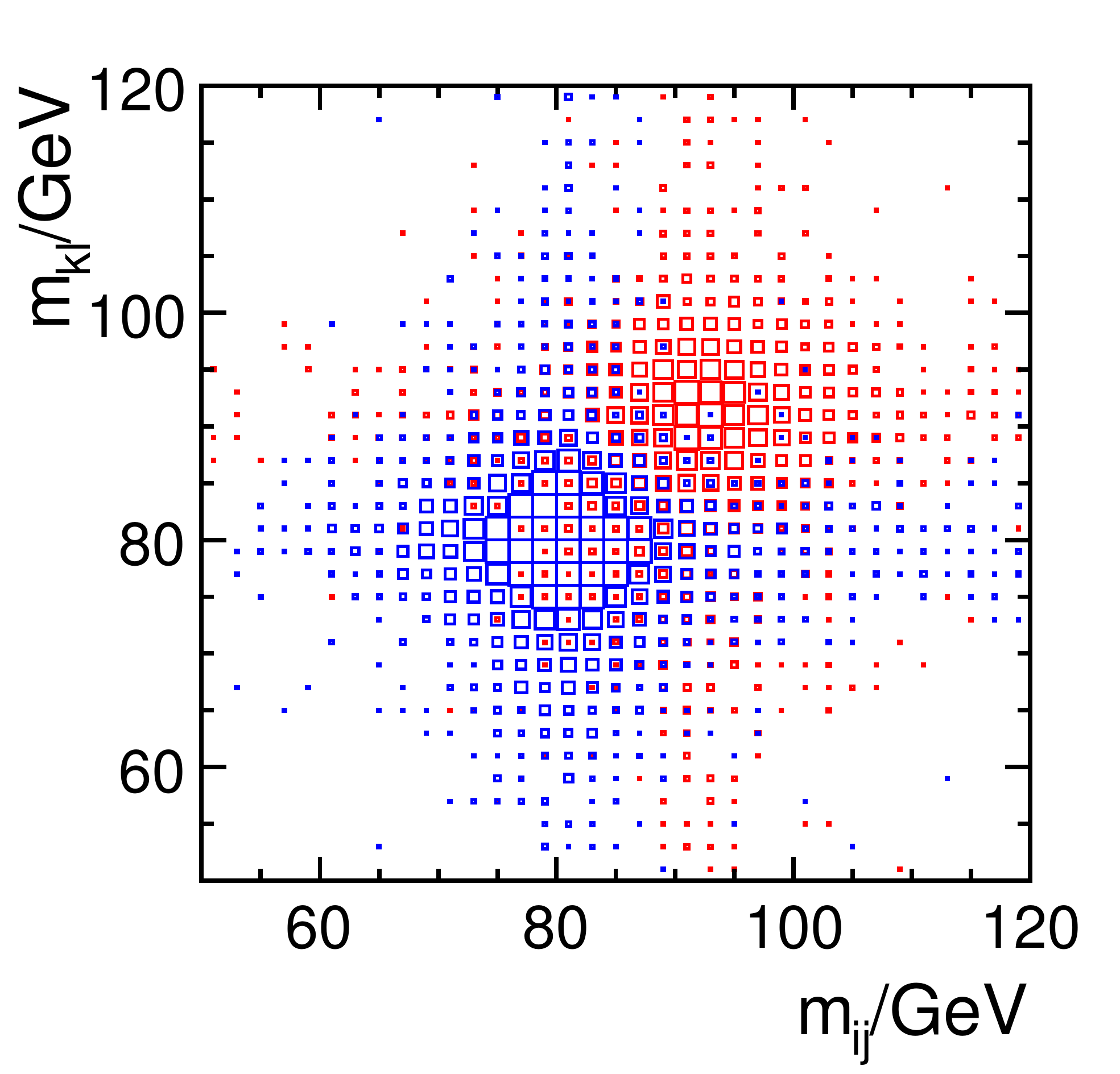}
\includegraphics[width=7.0cm]{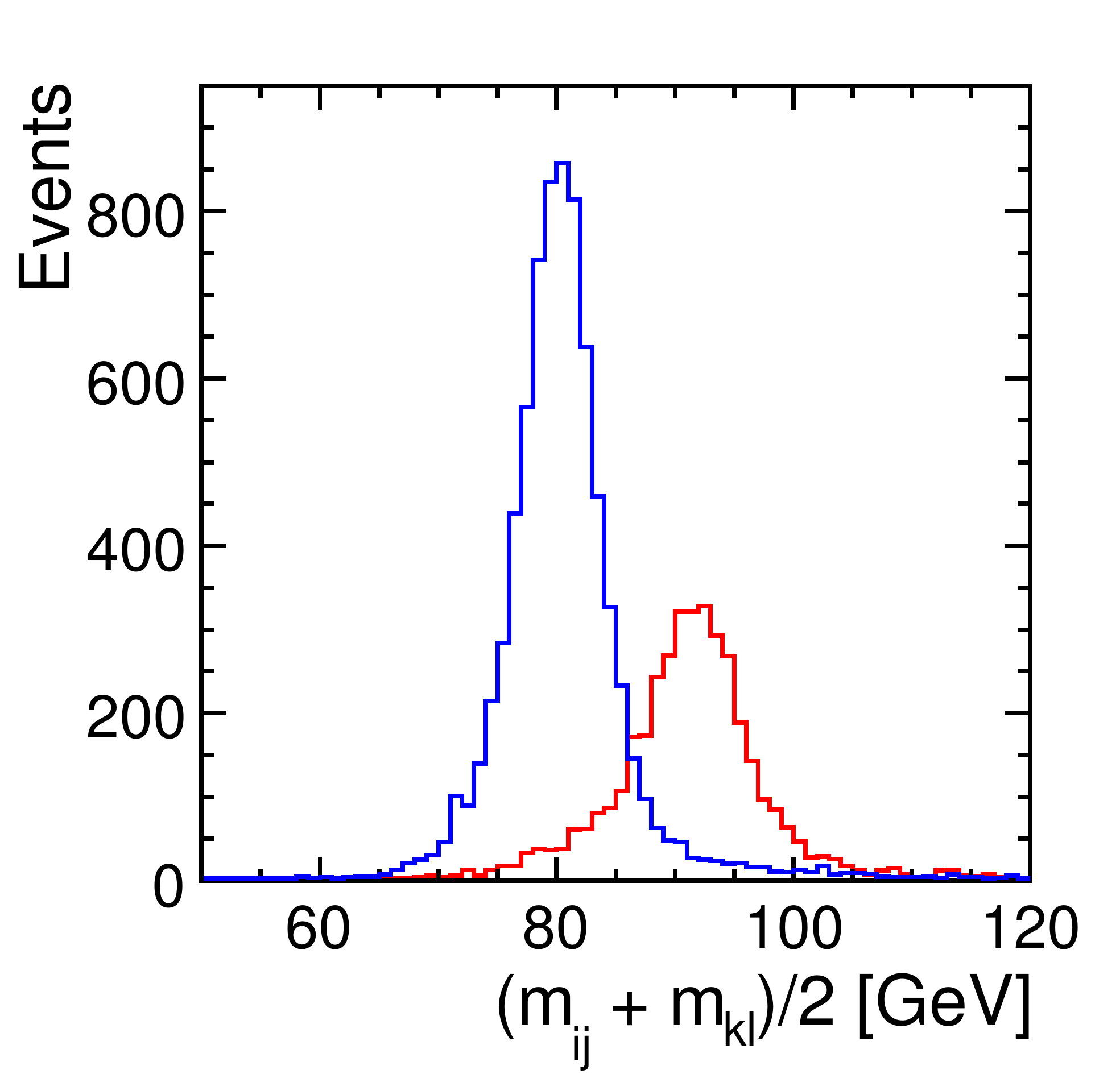}
\caption[$\nu_e\bar{\nu}_e \mathrm{WW}$
             and $\nu_e\bar{\nu}_e \mathrm{ZZ}$ mass distributions.]{a) The reconstructed di-jet mass distributions for the
             best jet-pairing in  selected 
             $\nu_e\bar{\nu}_e \mathrm{WW}$ (blue)
             and $\nu_e\bar{\nu}_e \mathrm{ZZ}$ (red) events 
             at $\sqrt{s} = 1\,TeV$.
         b) Distributions of the average reconstructed di-jet mass, 
             $(m_{ij} + m_{kl}^B)/2.0$, for the best jet-pairing 
             for $\nu_e\bar{\nu}_e \mathrm{WW}$ (blue)
             and $\nu_e\bar{\nu}_e \mathrm{ZZ}$ (red) events. \label{fig:WZ}}
\end{figure}

Strong EWSB can be described by an effective Lagrangian approach 
in which there are two anomalous quartic gauge couplings, $\alpha_4$
and $\alpha_5$~\cite{chierici} which are identically zero in the SM. 
The WW scattering events are generated at $\roots=1$\,TeV
with WHiZard~\cite{Whizard} assuming $\alpha_4=\alpha_5=0$. 
Results are obtained for an integrated luminosity of
1\,ab$^{-1}$ with $P(\eplus,\eminus) = (+0.3,-0.8)$. 
Event selection cuts, similar  
to those of~\cite{chierici, andres, bib:predrag1,bib:predrag2}, reduce the backgrounds from 
processes other than the quartic coupling 
diagrams to $\sim20$\,\% of the signal.
Of the three possible jet-pairings, the one which minimises 
$|m_{ij}-m_{W/Z}|\times|m_{kl}-m_{W/Z}|$ 
is chosen. Figure~\ref{fig:WZ} shows, for $\nu_e\bar{\nu}_e \mathrm{WW}$ and 
$\nu_e\bar{\nu}_e \mathrm{ZZ}$ events, a) 
the reconstructed di-jet mass distribution, and b) the
distribution of average reconstructed mass, 
$(m_{ij} + m_{kl})/2.0$. Clear separation between the
W and Z peaks is obtained.

The parameters $\alpha_4$ and $\alpha_5$ 
are obtained from a binned maximum likelihood fit  
to the two-dimensional distribution ($10\times10$ bins) 
of the boson polar angle in the reference frame
of boson pair and the jet polar angle in the reference frame 
of each boson, giving $-1.38 < \alpha_4 < +1.10$ and
$-0.92 < \alpha_5 <  +0.77$. These sensitivities are slightly tighter than
those from a previous fast simulation study with the TESLA detector 
concept~\cite{bib:predrag1,bib:predrag2}.

%% file: performance/physics_sps1a.tex
SUSY may provide a rich spectrum of kinematically accessible 
particles at the ILC operating at $\roots=500$\,GeV, 
for example the production of gauginos and sleptons with masses below 
$250\,\mathrm{GeV}$. The signals for new physics consist
of a complex mixture of dominant and sub-dominant processes, often with 
identical visible final states~\cite{smuon_ref1}. 
Here we consider final states consisting of missing energy and 
either two muons or two taus in mSUGRA SUSY with the SPS1a' parameter
set: $M_{0}=70\,\rm{GeV}$, $M_{1/2}=250\,\rm{GeV}$, $A_{0}=-300\,\rm{GeV}$,
sign$(\mu)=+1$, and $\tan\beta=10$. For these parameters the relevant
gaugino and slepton masses are: $m(\tilde{\chi}_{1}^{0}) = 97.7$\,GeV, $m(\tilde{\chi}_{2}^{0}) = 183.9$\,GeV, 
m($\tilde{\mu}_{R}) = 125.3$\,GeV, 
m($\tilde{\mu}_{L}) = 189.9$\,GeV and 
m($\tilde{\tau}_{1}) = 107.9$\,GeV.

\subsubsection{Muons and Missing Energy }
\label{sec:physics-smuons}

The ILC sensitivity to pair production of the lightest 
scalar muon, $\tilde{\mu}^+_{R}\tilde{\mu}^-_{R}$,
leading to a final state of two muons and missing energy has been extensively 
studied\cite{bib:TeslaTDR}.  
The study presented here concentrates on sub-dominant
di-muon plus missing energy processes which have to compete with the
large SUSY background and in particular, the challenging scenario where the
di-muon decay modes are suppressed. For these sub-dominant processes, 
the $\tilde{\chi}_{1}^{0}$, 
$\tilde{\chi}_{2}^{0}$, and $\tilde{\mu}_{L}$ masses can be 
measured from
$\tilde{\chi}^0_{2}\tilde{\chi}^0_{1}\rightarrow \mu\mu\tilde{\chi}_{1}^{0}\tilde{\chi}_{1}^{0}$ 
($\sigma = 4.1$\,fb) and
$\tilde{\mu}_{L}\tilde{\mu}_{L}\rightarrow \mu\mu\tilde{\chi}_{1}^{0}\tilde{\chi}_{1}^{0}$
($\sigma = 54$\,fb). In both cases  the signal is characterised by two energetic muons
and missing energy. Muons are identified with $95\,\%$ efficiency using 
track, HCAL and muon chamber information. Background is rejected using:
missing energy, di-muon invariant mass,
recoil mass,  transverse momentum, and the 
direction and speed of the di-muon system. 
Cuts on these variables are used to define 
$\tilde{\mu}_{L}\tilde{\mu}_{L}\rightarrow\mu\mu\tilde{\chi}_{1}^{0}\tilde{\chi}_{1}^{0}$ and
$\tilde{\chi}^0_{2}\tilde{\chi}^0_{1}\rightarrow \mu\mu\tilde{\chi}_{1}^{0}\tilde{\chi}_{1}^{0}$ event samples.

\begin{figure}[tb]
\begin{center}
\includegraphics[width=7.0cm]{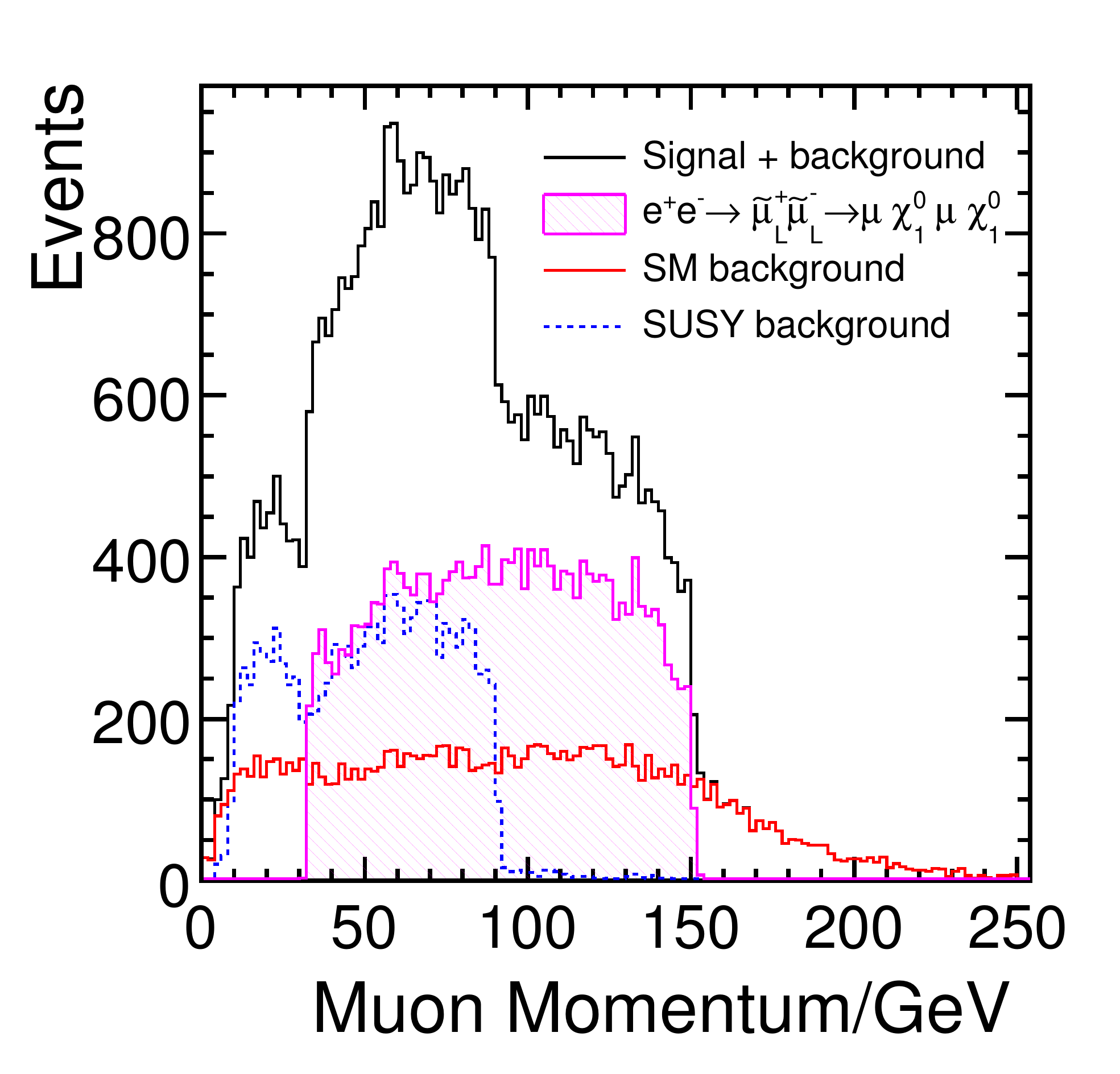}
\includegraphics[width=7.0cm]{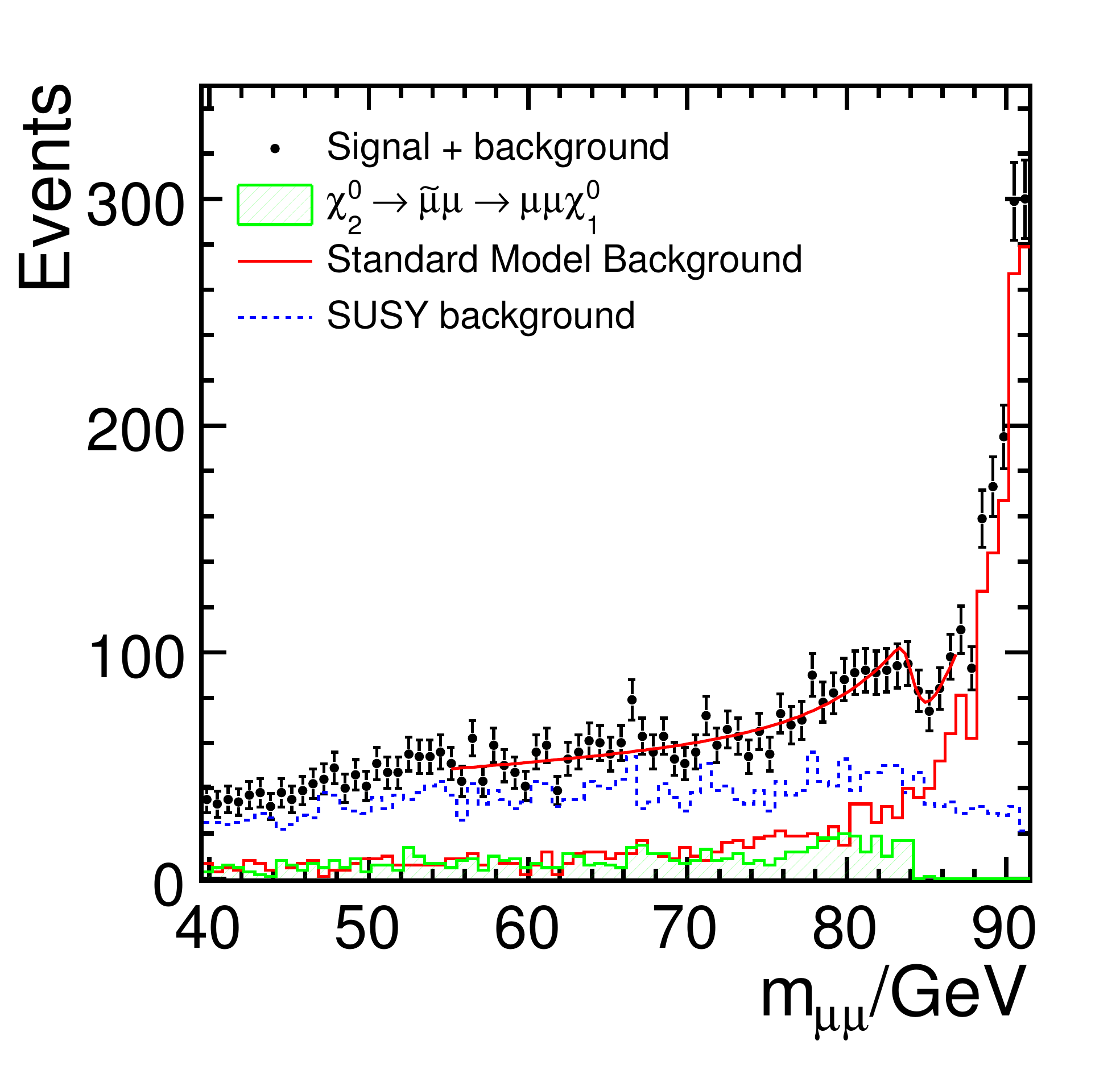}
\caption[Smuon event selection.]{(a) Distribution of the momentum of the $\mu^{+}$ in 
the laboratory frame after the selection of the $\tilde{\mu}_{L}\tilde{\mu}_{L}$ 
pair candidates. The total observed signal and the backgrounds (SUSY and SM) 
are reported. The mass of the $\tilde{\chi}_{1}^{0}$ is measured from the 
kinematic edges of the distribution of the momentum of the di-muons.
There are two entries for each event. 
(b) Fitted $m_{\mu\mu}$ spectrum. The kinematic edge corresponding to 
the $\tilde{\chi}_{2}^{0}\rightarrow \tilde{\mu}_{R}\mu$ can be extracted from the 
fit on top of the left tail of the $Z$ peak. Both plots are shown for
a beam polarisation of -80\,\%, 60\,\% and correspond to an integrated luminosity
of 500\,fb$^{-1}$.\label{fig:smu} }
\end{center}
\end{figure}

The masses of the $\tilde{\chi}_{1}^{0}$ and  $\tilde{\mu}_{L}$ are
measured from the kinematic edges of the momentum distribution of the
muons in the $\tilde{\mu}_{L}\tilde{\mu}_{L}\rightarrow
\mu\mu\tilde{\chi}_{1}^{0}\tilde{\chi}_{1}^{0}$~\cite{smuon_ref2} event selection, 
shown in Figure~\ref{fig:smu}a. The kinematic edges of the signal, 
at 32\,GeV and 151\,GeV, are fitted with a step function giving
measurements of the $\tilde{\chi}_{1}^{0}$ 
and $\tilde{\mu}_{L}$ masses 
with statistical uncertainties of $1.40\,\%$ 
and $0.27\,\%$ repectively. 
The signal cross section is determined 
with an uncertainty of $2.5\,\%$ from the number of selected events.
The sharpness of the kinematic edges, and consequently the 
mass measurements, are limited by
beamstrahlung rather than the track momentum resolution and it
is estimated that the uncertainty in the $\tilde{\chi}_{1}^{0}$
mass would be a factor two worse for the lowP option for the ILC
with the same integrated luminosity.
The $\tilde{\chi}_{2}^{0}$ mass is determined from the kinematic edge of
the di-muon mass distribution in the decay chain
$\tilde{\chi}_{2}^0\tilde{\chi}_{1}^0\rightarrow \mu\mu\tilde{\chi}_{1}^{0}\tilde{\chi}_{1}^{0}$~\cite{smuon_ref2}. 
The distribution of $m_{\mu\mu}$ after selection 
is shown in Figure~\ref{fig:smu}b.  
The kinematic edge of the signal is visible below the $Z$ peak.
A fit to the region $40\,\rm{GeV}<m_{\mu\mu}<85\,\rm{GeV}$ is used
to determine the mass of the
$\tilde{\chi}_{2}^{0}$. In this region the 
statistical significance of the excess corresponds to 9 standard deviations
and the $\tilde{\chi}_2^0$ mass resolution obtained is $1.41\,\%$. 
It should be noted that a higher positron polarisation yields 
significantly improved precision, particularly for $\tilde{\chi}^0_2\tilde{\chi}^0_1$ 
production where a positron polarisation of $60\,\%$ rather than $30\,\%$
results in 50\,\%  more signal events for a relatively small increase in
background. 

\subsubsection{Stau Production and Decay}
\label{sec:physics-stau}

For the SUSY SPS1a' parameters,
$e^+e^- \rightarrow \tilde{\tau}\tilde{\tau} \rightarrow \tilde{\chi}_1^0 \tau \tilde{\chi}_1^0 \tau$,
results in a signal of missing energy and the relatively low energy visible
decay products of the tau leptons ($E_\tau \lesssim 43$\,GeV). 
Measurements of $\tilde{\tau}\tilde{\tau}$ production requires 
precision tracking of 
relatively low momentum particles, good particle identification, 
a highly hermetic detector, and low machine background.

The stau pair event selection requires two low multiplicity tau-jets and
at least 400\,GeV of missing energy.  
The tau jets are required to have $|\cos\theta| < 0.9$ 
and to have an acoplanarity of greater than $85^\circ$.
Background is further reduced by cutting on the transverse momentum with 
respect to the transverse event thrust axis.
Given that the tau-jets are
relatively low momentum, background from multi-peripheral 
two photon processes, $\epem\rightarrow\epem X$, is particularly 
important due to the very high
cross section. This background is reduced using the beam calorimeter (BCAL) to veto
the forward going electrons/positrons. Due to the holes in the BCAL acceptance
around the 
incoming and outgoing beam pipes, the regions 
$\phi \le 110^\circ$ or $\phi \ge 250^\circ$ are not used (for details see \cite{ref:Schade}). 

\begin{figure}
\begin{center}
\includegraphics[width=7.0cm]{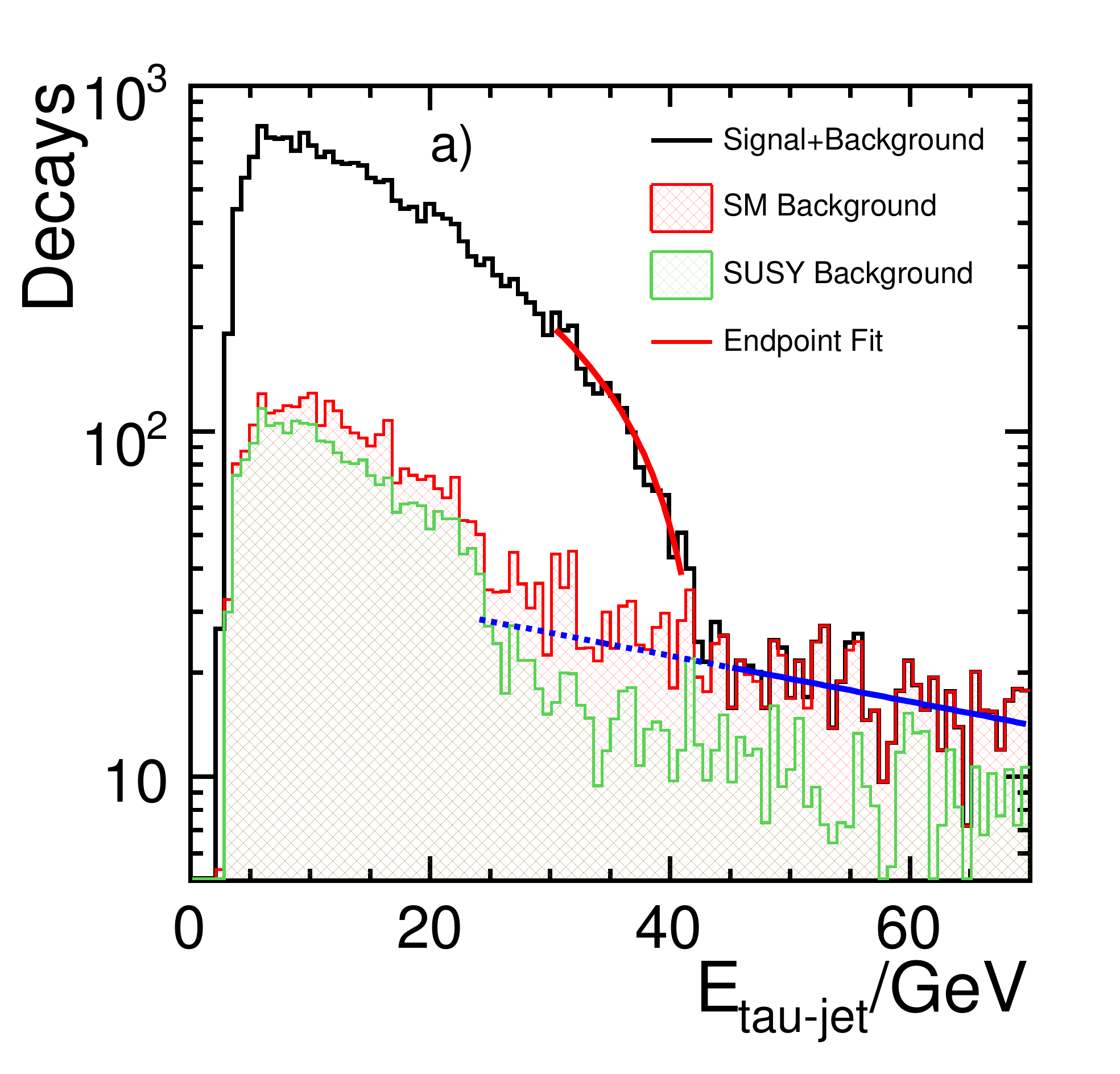}
\includegraphics[width=7.0cm]{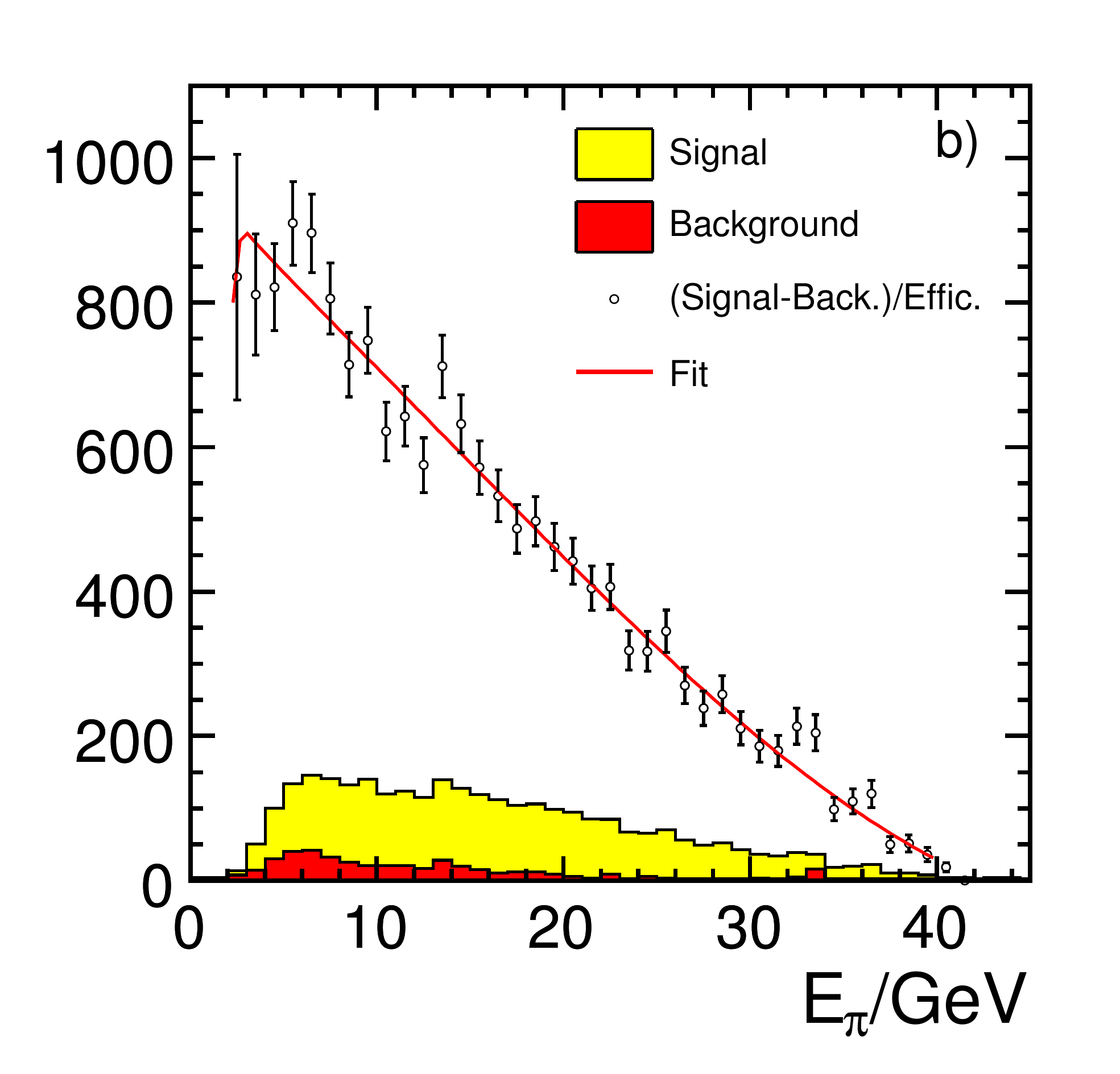}
\caption[Stau selection.]{a) Tau-jet energy spectrum, showing the fit to determine the 
            endpoint (two entries per event). b) Energy spectrum of selected tau decays and the
            fit to determine $P_\tau$. \label{fig:staus}}
\end{center}
\end{figure}

The $\stau$ mass can be extracted from the end-point of the 
tau-jet energy spectrum and knowledge of the $\tilde{\chi}_1^0$ mass, {\it e.g.}
from the study of smuon production. 
For the stau mass determination, the stau
pair selection is augmented by additional cuts on
tau-jet masses and the polar angle of the missing momentum vector.
Figure~\ref{fig:staus}a shows the distribution of the 
tau jet energy after these cuts. The
selection efficiency is 12\,\% and the sample purity is 80\,\%. The end-point 
tau energy 
is determined from a fit to the spectrum of
Figure~\ref{fig:staus}a in the region $30<E<41.5$\,GeV. The
signal, which in this region is dominated by $\tau\rightarrow\pi\nu$
decays, is described by a linear function. The resulting statistical uncertainty
on the end-point is 0.1\,GeV. 
When accounting for the uncertainty on the $\tilde{\chi}^0_1$
mass, $\sigma_{\mathrm{LSP}}$, this leads to a measurement 
precision on $\mstone$ of $0.1\,\mathrm{GeV} \oplus 1.3\sigma_{\mathrm{LSP}}$. 
Systematic uncertainties are not included.

The measurement of tau polarisation, $P_\tau$, 
in $\tilde{\tau}_1$ decays  gives direct access to the mixing of mass and interaction
eigenstates in the stau sector, and thus provides sensitivity to a number of SUSY parameters.
For SPS1a' $P_\tau = 97\,\%$. $P_\tau$ can be measured most cleanly
in $\tau \rightarrow \pi \nu$ decays where the slope of the $\pi^\pm$ energy spectrum
depends on $\Ptau$. The potential signal is large, for 
500\,fb$^{-1}$, $\sim$17500 $\pi\nu$ decays are expected.
In addition to the stau pair selection, calorimeter and 
d$E$/d$x$ information are used to identify $\tau\to\pi\nu$ decays. 
The selection 
efficiency is 13.8\,\%  and the remaining background fraction is 21\,\%.
The underlying pion spectrum is reconstructed by subtracting the background 
in a parametrised form and applying an energy dependent efficiency correction. 
The shape of the spectrum is influenced by the luminosity spectrum of the machine,
and this effect is folded into the fit function. 
Figure \ref{fig:staus} shows the reconstructed data together with the fit.
This study shows that a measurement of  $P_\tau$ with an accuracy of 
0.15 is realistic.

%% file: performance/physics_photons.tex
Physics beyond the standard model can manifest itself in final states consisting of two (or more)
photons and missing momentum. The high granularity and good photon reconstruction
capability of the ILD detector is well suited to these measurements. 

\subsubsection{Model-independent WIMP search}
\label{sec:physics-wimp}

Weakly Interacting Massive Particles (WIMPs) are possible candidates for dark matter.  If
they can pair annihilate into $\eplus\eminus$, then the reverse process can be detected 
at the ILC. In this case the two neutral (undetected) WIMPs are accompanied
by photon radiation from the incoming $\eplus\eminus$. In these $e^+e^- \rightarrow \chi\chi\gamma$ 
events, 
the photon recoil mass distribution has a characteristic onset. The location of the
onset and 
shape of the recoil mass distribution depends on the WIMP mass and spin.
Experimentally, the WIMP signal has to be resolved from the large irreducible ISR
background from $\epem\rightarrow\nu\overline{\nu}+n\gamma$. Good photon energy and angular 
resolution are required in order to detect a clear edge in the photon energy spectrum above this
large background. Assuming that the total cross-section for WIMP pair annihilation into SM fermion pairs, 
$\tilde{\chi}\tilde{\chi} \rightarrow f\bar{f} $, is known from cosmological observations, the ILC sensitivity can 
be expressed in terms of the WIMP pair branching fraction into $\epem$, $\kappa_e$. The ILD 
detector has been used to study this process~\cite{bib:wimpnote}. 
In general, $\kappa_e$ values of 0.1 are accessible for WIMP masses between 150~GeV and 200~GeV.

\subsubsection{Long-lived Neutralinos in GMSB}
\label{sec:physics-gmsb}

In the Gauge-Mediated SUSY Breaking (GMSB) scenario, $\tilde{\chi}^0_1$
decays into a Gravitino $\tilde{G}$ (the LSP) and a photon. Depending
on the SUSY parameters, the lifetime of the $\tilde{\chi}^0_1$ may be such
that it decays inside the detector tracking volume. In this case the
signal for GMSB neutralino production is a pair of photons with 
production point displaced from the origin. The signal is thus two
non-pointing photons. 
The excellent angular resolution for reconstructed photons in the ECAL
allows the $\tilde{\chi}^0_1$ lifetime to be determined from 
photon impact parameter distribution.
This process has been
studied in detail~\cite{analysis-note} with the full ILD simulation.
For neutralino lifetimes in the range $0.2-2.0$\,ns a signal to 
background ratio of approximately unity is achieved. This allows
the neutralino mass to be determined with a precision of $\pm$2\,GeV
and the lifetime can be measured to 1\,\%. 
Decay lengths
of 100\,cm can be well measured, while a decay length of 10\,cm can not 
be reconstructed with the ECAL only.

%% file: performance/performance-other.tex
The previous section described a number of physics studies related to
specific aspects of the detector performance, based on 
full simulation and reconstruction. In addition, a
number of other studies have been performed which demonstrate
the general purpose nature of the ILD concept. Three of these are described
briefly below. 

\subsection{Measurement of Beam Polarisation from WW production}
\label{sec:physics-ww}
\input{performance/physics_ww}

\subsection{Heavy Gauge Boson Production in Littlest Higgs Model}
\label{sec:physics-littlest-higgs}
\input{performance/physics_littlest_higgs}

\subsection{ZHH Production}
\label{sec:physics_ZHH}
\input{performance/physics_ZHH}

%% file: performance/physics_ww.tex
One of the unique features of the ILC is the possibility of both 
electron and positron beam polarisation~\cite{:2007sg}. The baseline 
design foresees a longitudinal electron polarisation of $80\,\%$ and a 
positron polarisation of $30\,\%$ with an option of $60\,\%$. This 
provides a tool for improving the sensitivity to new 
physics~\cite{MoortgatPick:2005cw}. For many of these applications, 
the final goal of an ultimate relative precision of $0.2\,\%$ of the 
measurement of the beam polarisations is desirable to bring the 
systematics from the uncertainty of the beam polarisation to a 
negligible level. 
 
While polarimeters~\cite{polarisation_energy} will be used 
to measure the polarisation on a bunch-by-bunch basis, 
the absolute calibration of the average luminosity-weighted 
polarisation at the interaction point can be achieved using 
$\WpWm$ production. Two methods are considered: i) the 
modified Blondel scheme~\cite{Blondel:1987wr,Moenig:2000} which uses 
the measured $\WpWm$ production cross-sections for different 
beam polarisations; and ii) the angular fit method which
uses the distribution 
of the production angle $\cos\theta_W$ of the $W^-$ with respect to 
the $e^-$ beam axis~\cite{Bechtle:2009}.

A comparison of these two methods is performed using the full simulation
of events in ILD. Semi-leptonic decays of $W$-pair events ($\qq\ell\nu$) 
are selected with an efficiency of $68.7\,\%$ and  $93\,\%$ purity.
The charge of the lepton tags the charges of two W bosons so that
the $\Wboson^-$ angular distribution can be determined. 
In  the modified Blondel scheme, the total luminosity necessary to 
reach the desired relative precision of $0.2\,\%$ is around $500\,\mathrm{fb}^{-1}$.
Using the angular fit method the same level of precision can be
achieved with an integrated luminosity  of  $250\,\mathrm{fb}^{-1}$.
The lower luminosity demand reduces the time spent on the $+\,+$ and $-\,-$ helicity 
combinations, which are less interesting from the physics point of 
view.  
To reach the desired precision for the 
measurement of the beam polarisation, ${\cal L}=250\,\mathrm{fb}^{-1}$ 
is required for the case of $60\,\%$ positron 
polarisation, while  ${\cal L}=1200\,\mathrm{fb}^{-1}$ is required
if only the baseline $30\,\%$ positron polarization is available.

%% file: performance/physics_littlest_higgs.tex
The Littlest Higgs model with T-parity (LHT) has been proposed 
as a solution to the little hierarchy problem.
Since heavy gauge bosons acquire mass terms through the breaking of the 
global symmetry, precise measurements of their masses allow a 
determination of the vacuum expectation value of the breaking ($f$).
Furthermore, since the heavy photon, $A_{\mathrm{H}}$, is a candidate for dark matter, 
the determination of its properties is important for both
particle physics and cosmology. However, at the LHC it is difficult to 
determine the properties of heavy gauge bosons 
because they have no colour charge.

The potential of an ILD-like detector concept, studied using 
the fast simulation of the GLD concept, is described in detail
in~\cite{arxiv}. Here the
processes 
$e^{+}e^{-} \to A_{\mathrm{H}} Z_{\mathrm{H}} \to A_{\mathrm{H}} A_{\mathrm{H}} H$ 
at $\sqrt{s} = 500$\,GeV and
$e^{+}e^{-} \to W_{\mathrm{H}}^{+} W_{\mathrm{H}}^{-} \to A_{\mathrm{H}} W^{+} A_{\mathrm{H}} W^{-}$
at $\sqrt{s} = 1$\,TeV, where $A_{\mathrm{H}}$, $Z_{\mathrm{H}}$, and $W_{\mathrm{H}}^{\pm}$
are the heavy gauge bosons, are studied.  The experimental signatures considered for each process are
b-jets with missing energy and four-jets with missing energy.
The masses and the vacuum expectation value, $f$, were set to 
$(M_{A_{\mathrm{H}}}, M_{Z_{\mathrm{H}}}, M_{W_{\mathrm{H}}}^{\pm}) = 
(81.9 \mathrm{~GeV}, 369 \mathrm{~GeV}, 368 \mathrm{~GeV})$, $\mH$ = 134\,GeV and $f = 580$ GeV.
It is found that the masses of $A_{\mathrm{H}}$ and $Z_{\mathrm{H}}$ can be measured 
with an accuracy of 16.2\,\% and 4.3\,\% respectively at $\sqrt{s} = 500$ GeV,  
and those of $A_{\mathrm{H}}$ and $W_{\mathrm{H}}$ can be determined 
with an accuracy of 0.2\,\% and 0.8\,\% respectively at $\sqrt{s} = 1$\,TeV~\cite{arxiv}.
In addition  $f$ can be with measured to 4.3\,\%  at $\sqrt{s} = 500$\,GeV and
0.1\,\% at $\sqrt{s} = 1$\,TeV. 
Finally it is shown that the abundance of dark matter relics can be 
determined to the 10\% and 1\% levels at $\sqrt{s} = 500$\,GeV and $\sqrt{s} = 1$\,TeV, respectively.
These accuracies are comparable to those of the current and future cosmological observations 
of  the cosmic microwave background.


%% file: performance/physics_ZHH.tex
The Higgs boson tri-linear coupling can be studied
at the ILC through the processes 
$\epem\rightarrow\nu_e\overline{\nu}_e \Higgs\Higgs$ and
$\epem\rightarrow\Zzero\Higgs\Higgs$. For $\mH=120$\,GeV, the 
cross section for the latter process is 0.18\,fb at $\sqrt{s}=500$\,GeV.
The $\qq\bbbar\bbbar$ decay mode (34\,\% of the $\Zzero\Higgs\Higgs$ decays)
has been studied using the ILD simulation and 
reconstruction~\cite{bib:giannelli}. A multi-variate selection
including the invariant masses of jet combinations and flavour tagging 
information is used. For an integrated luminosity of 500\,fb$^{-1}$, 
a precision of 90\,\% on $\sigma(\epem\rightarrow\Zzero\Higgs\Higgs)$
is achieved. It should be noted that the sensitivity does not
yet approach that of earlier fast simulation studies~\cite{bib:TeslaTDR}.
Whilst significant improvements are expected, this study demonstrates
the difficulty of this analysis; excellent particle flow and flavour
tagging performance are likely to prove essential for this important
measurement.

%% file: performance/conclusions.tex
\subsection{Detector Performance}

It has been demonstrated in Section~\ref{sec:performance-detector}
that ILD meets the requirements for an ILC detector:
\begin{itemize}\addtolength{\itemsep}{-0.3\baselineskip}
\item {\bf Track reconstruction:} The ILD tracking system provides highly efficiency track 
    reconstruction ($\sim99.5\,\%$), even in a dense multi-jet environment.
\item {\bf Momentum resolution:} 
    When hits in the TPC are combined with those in Si tracking detectors, the
    asymptotic value of the momentum resolution is
    $\sigma_{1/p_T} \approx 2\times10^{-5}$\,GeV$^{-1}$, as required.
\item {\bf Impact parameter resolution:} For either option for the VTX layout, 
    the required impact parameter resolution is achieved, with asymptotic
    values of $\sigma_{r\phi}=2$\,$\mu$m and  $\sigma_{rz}=5$\,$\mu$m.
\item {\bf Particle flow performance:} A jet energy resolution of
     $<3.8\,\%$ is achieved for jets in the energy range $40-400$\,GeV.
     For the range of energies typical of much of the ILC physics, 
     $80-200$\,GeV the jet energy resolutions is $\approx$3\,\%. 
     The performance does not depend strongly on the polar angle
     of the jet, except in the very forward region.   
\end{itemize}   
   
\subsection{Physics Performance}
 
The physics benchmark studies presented above are summarised in
Table~\ref{tab:physics_summary}. 
 However, care is needed in interpreting the results
shown. They do not represent the ultimate ILD performance as
significant improvements in the analyses are possible. 
However, the range of different measurements studied and
precision achieved demonstrate the general purpose nature
of ILD.

\begin{table}[ht]
{\tabcolsep=0.07cm
\begin{center}
\begin{tabular}{|l|c|c|c|c|}\hline
Analysis          & $\roots$ & Observable &                Precision                & Comments \\ \hline  
\multirow{3}{*}{Higgs recoil mass} &  \multirow{3}{*}{250\,GeV} & $\sigma(\epem\rightarrow\Zzero\Higgs)$ & $\pm0.30$\,fb (2.5\,\%)  & Model Independent \\
                  & & $\mH$                                          & $32$\,MeV             & Model Independent \\
                  & & $\mH$                                          & $27$\,MeV             & Model Dependent \\ \hline
\multirow{3}{*}{Higgs Decay} & \multirow{3}{*}{250\,GeV}     & $Br(\Higgs\rightarrow\bbbar)$ & 2.7\,\% & includes 2.5\,\%  \\
                              &     & $Br(\Higgs\rightarrow\ccbar)$ & $12$\,\%            & from \\
                              &     & $Br(\Higgs\rightarrow gg)$    & $29$\,\%            & $\sigma(\epem\rightarrow\Zzero\Higgs)$ \\ \hline
\multirow{3}{*}{$\tptm$}      & \multirow{3}{*}{500\,GeV}           &$\sigma(\epem\rightarrow\tptm)$& 0.29\,\% & $\theta_{\tptm}>178^\circ$\\
                              &     & $A_{FB}$                      & $\pm0.0025$ & $\theta_{\tptm}>178^\circ$\\
                              &     & $P_\tau$                      & $\pm0.007$ & exclucing $\tau\rightarrow a_1\nu$ \\ \hline
\multirow{5}{*}{Gaugino Production} &  \multirow{5}{*}{500\,GeV}    & $\sigma(\epem\rightarrow\tilde{\chi}_1^+\tilde{\chi}_1^-)$ & 0.6\,\% &  \\
   &     & $\sigma(\epem\rightarrow\tilde{\chi}_2^0\tilde{\chi}_2^0)$ & 2.1\,\% &  \\
   &     &  $m(\tilde{\chi}_1^\pm)$  & 2.4\,GeV & from kin. edges  \\
   &     &  $m(\tilde{\chi}_2^0)$    & 0.9\,GeV & from kin. edges  \\
   &     &  $m(\tilde{\chi}_1^0)$    & 0.8\,GeV & from kin. edges  \\ \hline
\multirow{4}{*}{$\epem\rightarrow\ttbar$} &  \multirow{4}{*}{500\,GeV}    & $\sigma(\epem\rightarrow\ttbar)$ & 0.4\,\% & (b$\qq$) ($\overline{\mathrm{b}}\qq$) only \\
        &                      & $m_t$       & 40\,MeV  &   fully-hadronic only \\
        &                      & $m_t$       & 30\,MeV  &    + semi-leptonic    \\
        &                      & $\Gamma_t$  & 27\,MeV  &   fully-hadronic only \\
        &                      & $\Gamma_t$  & 22\,MeV  &    + semi-leptonic    \\ 
        &                      & $A_{\mathrm FB}^t$     &   $\pm 0.0079$  &    fully-hadronic only    \\ \hline
\multirow{2}{*}{Smuons in SPS1a'} &  \multirow{2}{*}{500\,GeV} & $\sigma(\epem\rightarrow\tilde{\mu}^+_L\tilde{\mu}^-_L)$ &  $2.5$\,\%  &  \\
       &                      & $m(\tilde{\mu}_L)$          & 0.5\,GeV  &   \\ \hline
\multirow{1}{*}{Staus in SPS1a'} &  \multirow{1}{*}{500\,GeV} & $m(\tilde{\tau}_1) $       & $0.1\,\mathrm{GeV} \oplus 1.3\sigma_{\mathrm{LSP}}$ &   \\ \hline
\multirow{2}{*}{WW Scattering} &  \multirow{2}{*}{1\,TeV} & $\alpha_4$ & $-1.4<\alpha_4<1.1$  & \\
        &  & $\alpha_5$ & $-0.9<\alpha_5<+0.8$  & \\ \hline
\end{tabular}
\caption[Summary of physics sensitivities.]{A summary of the main observables presented in Section~\ref{sec:performance-physics}.
	 \label{tab:physics_summary}} 
\end{center}
}
\end{table}

%% file: ild/ild.tex
The ILD detector is strongly influenced by two basic assumptions about experimentation at a linear collider: particle flow as a way to reconstruct 
the overall event properties, and high resolution vertexing. 
Particle flow calorimetry requires a reliable and redundant tracking 
system which enables charge particle momenta to be reconstructed 
with high precision, and in particular, with very high efficiency.
ILD is built around 
a calorimeter system with very good granularity both in the transverse and in 
the longitudinal direction, and a combination of Silicon and gaseous
tracking systems. Vertexing, the other great challenge, is addressed 
by a high precision pixelated detector very close to the interaction point. 

In this section the different sub-detectors are described in more detail, 
proposed technological solutions are outlined, and necessary development 
work is highlighted, particularly where it is essential to advance the concept 
to a point where this detector could be built. 

Development of technologies for a detector at a linear collider is 
an active field, with many ideas being pursued, and great advances in 
technology are being made. ILD therefore does not at this moment 
exclude any promising technology from its consideration. Wherever possible,
ILD supports that more than one avenue is followed to eventually 
identify the best solution possible. Therefore, at this stage, all 
promising technologies are considered as possible candidates for the ILD detector. Consequently for a number of subdetectors more 
than one option are described. 

\graphicspath{{ild/vertex/}}
\input{ild/vertex/vertex}

\label{ild:vertex}

\graphicspath{{ild/silicon/}}
\input{ild/silicon/silicon}

\label{ild:silicon}

\graphicspath{{ild/tpc/}}
\input{ild/tpc/tpc}

\label{ild:tpc}

\graphicspath{{ild/calo/}}
\input{ild/calo/calo}

\label{ild:calo}

\graphicspath{{ild/fcal/}}
\input{ild/fcal/fcal}

\label{ild:fcal}

\graphicspath{{ild/coil/}}
\input{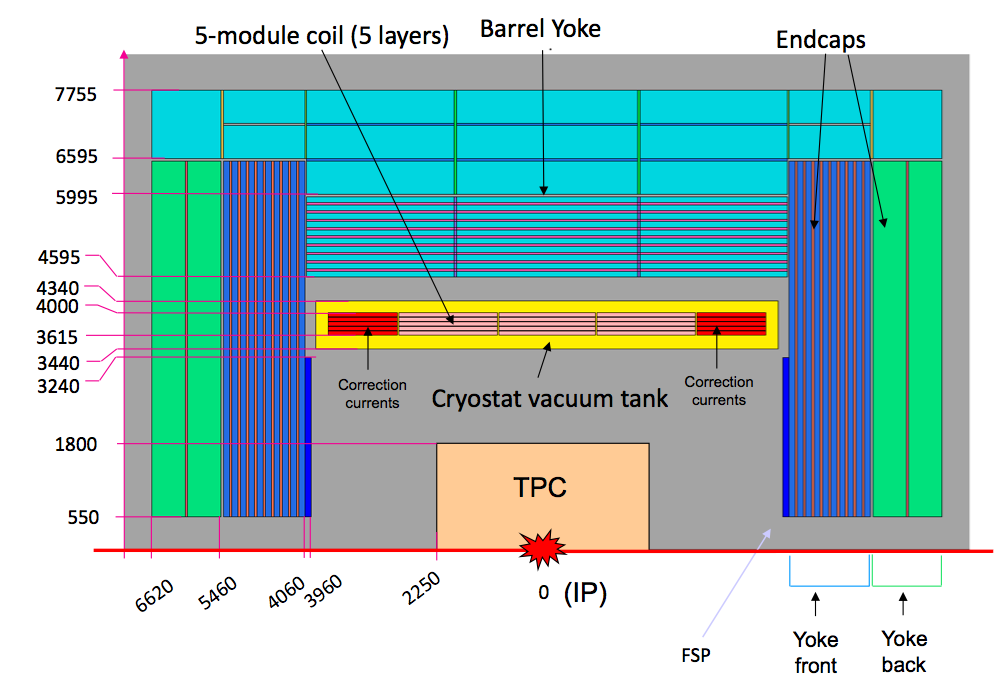}
\label{ild:coil}

\graphicspath{{ild/muon/}}
\input{ild/muon/muon}

\label{ild:muon}

\graphicspath{{ild/alignment/}}
\input{ild/alignment/alignment}
\label{ild:align}

%% file: ild/vertex/vertex.tex
\section{Vertex Detector}

 
   The Vertex Detector (VTX) is the key to achieving very high performance flavour tagging by reconstructing 
displaced vertices. It also plays an important role in the track 
reconstruction, especially for low momentum particles which don't 
reach the main tracker or barely penetrate its sensitive volume because 
of the strong magnetic field of the experiment, or due to their shallow 
production angle.

   The flavour tagging performance 
needed for physics implies that
the first measured point on a track 
should be
as close as possible to the IP. This 
creates
a major technical challenge because of the rapidly 
increasing beam-related background when approaching 
the IP. The 
sensor
technology best adapted to the high background 
environment is not yet defined. It is however clear that 
existing
technologies 
are not able to satisfy all of the
requirements defined by 
the physics goals (granularity, material budget) and those imposed by 
the running conditions near the IP ({\it e.g.} occupancy and radiation dose). 
Several alternative, innovative, pixel technologies are being considered 
and actively developed to satisfy the VTX requirements. 

   The VTX flavour tagging performance relies on a low material budget for the 
   detector sensors and the support structures. 
The VTX is also necessary in some physics studies to measure
the vertex charge (the net charge of all tracks
from the decay chain) which implies distinguishing between
the tracks from the primary vertex and the decay chain.
This is particularly challenging for low momentum tracks
in the jet.
Finally, 
secondary particle production and trajectory kinks due to 
secondary interactions with the detector material need to be 
mitigated because of their impact on the particle flow reconstruction. 
Minimising the VTX material budget 
therefore motivates
an ambitious R\&D programme.

\subsection{Physics Driven Requirements and Running Constraints}

   To identify the flavor (b or charm) of heavy-flavor jets, 
to measure the associated vertex charge, and to recognize 
tau-lepton decays, the VTX design needs to be
optimised in terms of single 
point resolution and distance between the first measured point of 
tracks and the IP. The high granularity necessary to achieve the single 
point resolution needs to be complemented with a particularly low 
material budget allowing high precision pointing with low momentum 
tracks. A high granularity is also required to separate neighboring 
tracks in a jet, a constraint which applies predominantly to the 
detector elements closest to the IP. 

  Following the usual convention, the performances of the VTX in 
terms of impact parameter resolution are summarised in a compact 
way by its well known gaussian expression:
\vspace{-3mm}
\begin{eqnarray}
\sigma_{ip} = a \oplus b/p \cdot \sin^{3/2} \theta
\end{eqnarray}
\vspace{-1mm}
The parameters $a$ and $b$ are required to be below 5 $\mu$m 
and 10 $\mu$m$ \cdot$GeV/c, respectively. Monte-Carlo studies show 
that these specifications are met with the following inputs:
\begin{itemize}\addtolength{\itemsep}{-0.3\baselineskip}
\item a single point accuracy of $\lesssim$ 3 $\mu m$, 
\item a vertex detector geometry providing a first measured point 
         of tracks at $\sim$ 15~mm from the IP.
\item a material budget between the IP and the first measured point 
         restricted to a few per mill of radiation length.
\end{itemize}
The values of $a$ and $b$ 
significantly exceed those
achieved 
so far, as illustrated by the comparison made in table \ref{VTXtab1}, 
which provides values of $a$ and $b$ obtained with vertex detectors 
operated at LEP, SLC and LHC as well as planned at RHIC.
\begin{table}[htbp!]
\begin{center}
\begin{tabular}{|l|cc|}
\hline
  Accelerator &  ${a}$ ($\mu$m) &  ${b}$ ($\mu$m$ \cdot GeV / c$)\\
\hline 
   LEP            &  25    &   70   \\  
   SLC            &  8     &   33   \\  
   LHC            &  12    &   70   \\  
   RHIC-II        &  13    &   19   \\  
{  ILD} & { $<$ 5} & { $<$ 10} \\  
\hline
\end{tabular}
\caption[Impact parameter resolution at ILD and other colliders.]{Values of the parameters $a$ and $b$ entering the expression 
of $\sigma_{ip}$ foreseen for the ILD, compared to those achieved 
with past, present or upcoming experiments at existing colliders.}
\label{VTXtab1}
\end{center}
\end{table}

  To achieve this new tagging performance standard, a 
beam pipe radius of 14~mm is envisaged, which is still compatible 
with the 
need to contain the core of the beam-related
pair background within the vacuum pipe.
The pipe is assumed to be made of machined
beryllium, 250~$\mu m$ thick, potentially covered with a 25~$\mu m$ 
thin foil of titanium against background from synchrotron radiation. Beam-related background, which 
ultimately sets the performance limits for the VTX,
is expected to be 
dominated by beamstrahlung e$^+$e$^-$ pairs. Most of these 
have low transverse momentum and remain 
trapped inside the vacuum pipe 
by
the 3.5~T solenoid field. 
Extensive Monte-Carlo simulations, based on the 
GuineaPig~\cite{guineapig} 
and CAIN generators~\cite{cain}, were performed 
to estimate the rate of e$^{\pm}$ reaching the vertex detector. The 
predicted rates amount to 5.3/4.4 $\pm$ 0.5 hits/cm$^2$ per bunch 
crossing (BX) at 15/16 mm radii, including e$^{\pm}$ backscattered 
from elements located near the outgoing beam lines \cite{VTXbg}. 
Since most of these e$^{\pm}$ have a transverse momentum close 
to the cut-off value of $\lesssim$ 10 MeV/c, they tend to penetrate 
the sensitive volume of the 
VTX sensors
at rather 
shallow angle, and tend to tend to generate pixel clusters which are elongated in 
the beam direction.
This feature may be used offline to reject a substantial fraction 
of the beamstrahlung clusters. It also impacts the radiation dose. 
The annual dose was calculated to be in the order
of 500~Gy ionising dose 
per year, with a corresponding fluence of 
$\lesssim$~10$^{11}$~n$_{eq}$/cm$^2$ at 15~mm radius.
To account for the limited accuracy of the simulated beamstrahlung 
e$^{\pm}$ rate, the latter was multiplied by a safety factor of 3 to 
derive the sensor specifications. Other backgrounds, such as photon 
and neutron gas, are expected to add marginal contributions. Overall, 
the annual radiation levels the sensors have to comply with are in  
excess of 1~kGy and of 10$^{11}$~n$_{eq}$/cm$^2$. 

  Another environmental concern entering the specifications of 
the VTX is related to the 
electrical interference associated with leakage of the 
beam-related RF from ports used for beam position monitors
and other equipment in the interaction region.
Their potential effect is 
motivating specific (delayed) signal processing read-out architectures 
of the sensors, taking advantage of the ILC beam time structure.

\subsection{Global Design Aspects}

  The VTX design is still evolving, but its prominent aspects are well 
defined. It is made of 5 or 6 cylindrical layers, 
all equipped with $\lesssim$ 50 $\mu m$ thin pixel sensors 
providing a single point resolution of 
2.8 $\mu$m all over the sensitive VTX area (see sub-section 
\ref{VTXsect4-1}). 
The innermost layer has a radius of 15-16 mm, a value for which the 
beam-related background rate is expected to still be 
acceptable.
As a consequence, the innermost layer intercepts all particles 
produced with a polar angle ($\theta$) such that $| \cos \theta |$ 
$\lesssim$ 0.97.

\begin{figure}
\begin{center}
\begin{tabular}{c c}
\includegraphics[width=0.35\textwidth]
{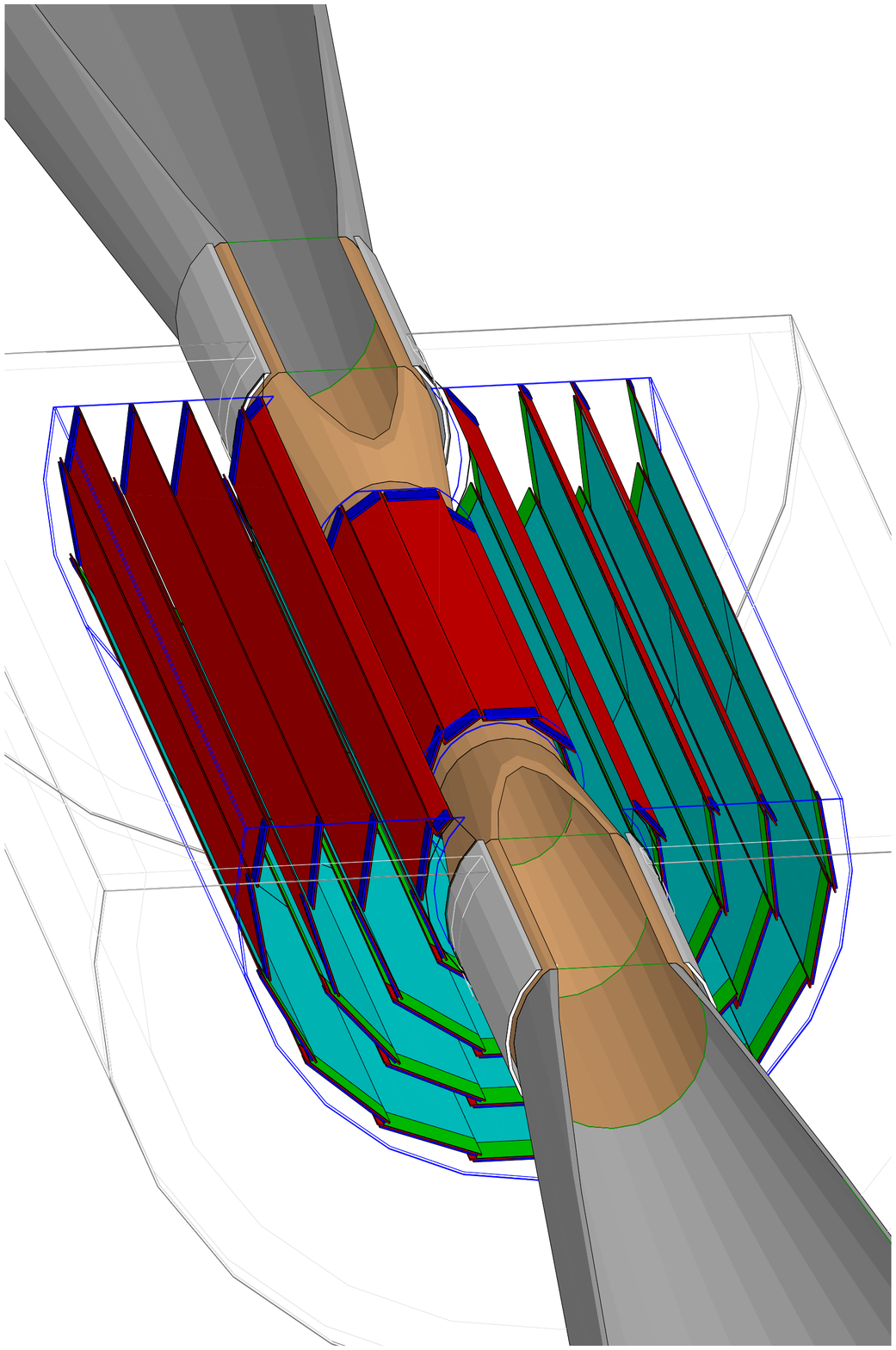}
\hspace*{2.cm}
\includegraphics[width=0.35\textwidth]
{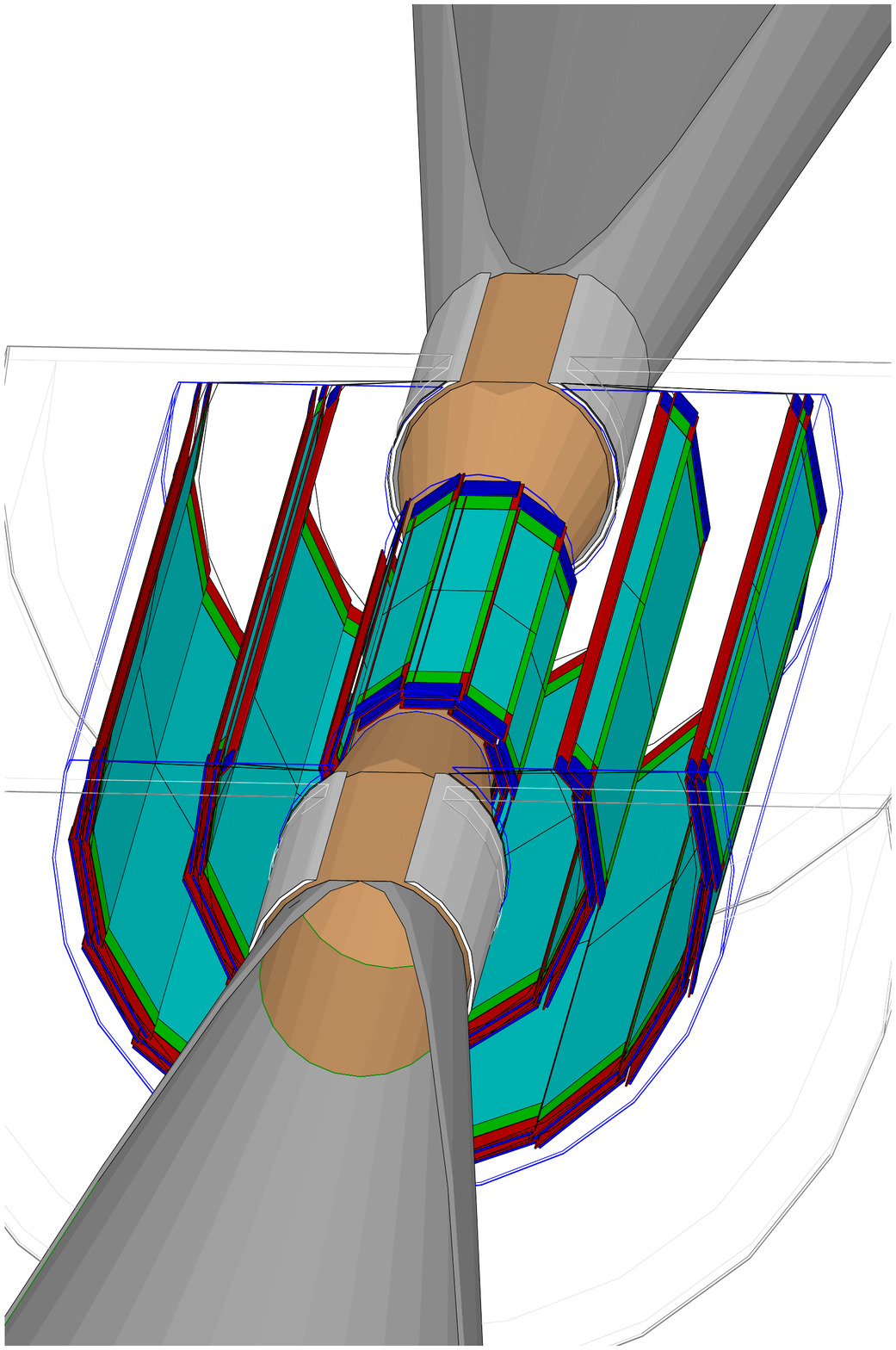}
\end{tabular}
\caption[Vertex detector geometries for the two design options.]{Vertex detector geometries of the two design options.
Left: 5 single ladders (VTX-SL). Right: 3 double ladders 
(VTX-DL).}
\label{VTXfig1}
\end{center}
\end{figure}

   Two alternative geometries are being considered, one (called 
VTX-SL) featuring 5 equidistant single layers (i.e. equipped with one 
layer of sensors only), and an alternative option (called VTX-DL) 
featuring 3 double layers (i.e. each layer being equipped with 
two, $\lesssim$~2~mm apart, arrays of sensors ). 
They are not 
associated with a specific
sensor technology. The double layer option 
allows
spatial correlations between hits generated by the same
particle in the two sensor layers equipping a ladder, even if the 
occupancy is high. It is therefore more robust against (low momentum) 
beamstrahlung background.  It is also expected to provide additional 
pointing accuracy. Moreover, it should facilitate the internal alignment,
allowing 
the use of
a large fraction of tracks traversing the overlapping 
bands of neighbouring ladders. Finally, it is expected to improve the 
modeling with tracks reconstructed at shallow angle in the very 
forward region. This geometry may however be less efficient in 
reconstructing long lived B mesons decaying outside of the beam 
pipe. It is also technically more challenging because of the 
additional difficulty to realise double ladders as compared 
to single ones. 
It may however be robust against mechanical 
distortions resulting from power pulsing the sensors inside the 
solenoid field. The two VTX geometries are displayed on 
figure~\ref{VTXfig1}. Some of their main geometrical parameters are 
listed in table~\ref{VTXtab2}.

\begin{table}[htbp!]
\begin{center}
\renewcommand{\arraystretch}{1.1}
\begin{tabular}{|l|cc|cc|cc|}
\hline
     & \multicolumn{2}{c|}{ radius [mm] } & \multicolumn{2}{c|}{ ladder length [mm]}
 & \multicolumn{2}{c|}{ read-out time [$\mu s$]} \\
  geometry          & VTX-SL   &      VTX-DL    &      VTX-SL   &      VTX-DL   &  VTX-SL    &     VTX-DL  \\
\hline 
 layer 1      &   15.0   &  16.0/18.0  &  125.0  &  125.0  &    25--50   &  25--50  \\  
 layer 2      &   26.0   &  37.0/39.0  &  250.0  &  250.0  &   50--100  &  100-200  \\  
 layer 3      &   37.0   &  58.0/60.0  &  250.0  &  250.0  &   100-200 &  100-200  \\  
 layer 4      &   48.0   &                     &  250.0  &                &   100-200 &                  \\  
 layer 5      &   60.0   &                     &  250.0  &                 &   100-200 &                 \\  
\hline
\end{tabular}
\caption[Parameters of the VTX detector options.]{Radius and ladder length for each layer of the two vertex detector 
geometries. For the double layer option (VTX-DL), the radii are provided for
each of both pixel arrays equipping a ladder.The read-out times are provided 
for each layer in the specific case of a continuous sensor read-out (see 
subsection \ref{VTXsect4}).}
\label{VTXtab2}
\end{center}
\end{table}

  The complete VTX-SL ladder thickness is equivalent to 0.11~\%~X$_0$, 
while the double ladders of VTX-DL represent 0.16~\%~X$_0$. 
These values assume 50~$\mu m$ thin silicon pixel sensors. 
The length of the innermost ladder (125~mm) 
is limited due to the radial expansion of the
pair background envelope as it diverges from the IP.
It would 
%
shrink significantly when considering 
the so-called "low-P" option
of the machine parameter.
In this case, the innermost ladders 
should be shortened to $<$ 100~mm and/or the inner radius 
should be increased in ordre to accommodate the increased 
beam-beam disruption. The loss in physics performance consecutive
to the geometrical acceptance shrinkage and 
to the potential impact parameter resolution degradation is still 
being evaluated. 
  
   Both device options are enclosed in a $\sim$~500~$\mu m$ thick 
(0.14~\%~X$_0$), 65 mm radius, cylindrical beryllium support. The 
latter is surrounded by a light foam cryostat (0.05~\%~X$_0$), 
complemented with a 0.5 mm aluminum foil (0.55~\%~X$_0$) which 
acts as a Faraday cage. The whole system, including support, 
cryostat and cage adds up to 0.74~\%~X$_0$. While the ladders of 
the three (resp. two) external layers of VTX-SL (resp. VTX-DL) 
are mounted on the beryllium support, the ladders composing 
the inner layers are supported by straight sections of the 
vacuum pipe.

   The detector alignment is expected to proceed through two 
main steps. The ladders will first be aligned inside their layer.
An overlap of $\lesssim$ 500 $\mu m$ between the sensitive 
areas of neighbouring ladders is foreseen for this purpose. Tracks 
with momentum in excess of a couple of GeV/c traversing these
overlapping bands will be used. Next the layers will be aligned 
with respect to the rest of the detector using straight tracks such as those 
of $\mu^+ \mu^-$ final states.

  The pros and cons of each design option are still being assessed. 
Moreover, the concept itself, which assumes extended cylinders, 
rather than shorter ones complemented with disks at small polar 
angle, is based on the present understanding of the minimal 
material budget which would separate the barrel from the disks. 
Depending on the evolution of technologies and materials, the 
choice between both alternatives may be reconsidered.

\subsection{Pixel Technology and System Integration Studies} 
\label{VTXsect4}

\subsubsection{R\&D on Pixels and Read-out Architectures}
\label{VTXsect4-1}

  Intensive R\&D 
has been under way for
several years, addressing the 
numerous challenging issues underlying the vertex detector 
specifications. Because of the scale of the challenge and of its
complexity, several alternative sensor technologies are being 
developed in parallel, aiming for the best suited ones. The goal of  
the development is to optimise the charge sensing system and 
the charge to electrical signal conversion, as well as the read-out, 
steering and control micro-circuits. The technologies presently 
concentrating most of the R\&D effort 
within the ILD group
are CMOS sensors~\cite{VTXmimosa1,VTXmimosa2,VTXapsel1,VTXapsel2}, 
DEPFETs~\cite{VTXreview,VTXdepfet}, 
FPCCDs~\cite{VTXfpccd1,VTXfpccd2},
and ISIS~\cite{VTXreview}.
Since recently, CMOS sensors exploiting
vertical integration technology \cite{VTX3d} are also developed.  
Alternative technological approaches mentioned in \cite{VTXreview}
may also be considered, though not currently 
developed inside the ILD
group.
The R\&D achieved so far has already demonstrated that the 
goals of 
a single point resolution of ($\lesssim$~3~$\mu m$), double 
hit separation of ($\lesssim$~40~$\mu m$) and sensor thickness 
of ($\lesssim$~50~$\mu m$) are achievable. 

   The most demanding requirement for all technologies is to 
comply with the occupancy generated by the beam related 
background in the innermost layers. Two alternative approaches 
are being investigated, one where the sensors are read out 
continuously, and one where the signal is stored during the 
whole train duration and read out during the beamless period 
separating two consecutive trains. 

  In the continuous read-out approach, most of the R\&D effort 
is invested in achieving the low noise high read-out frequency
required for the inner layers, while keeping the power consumption 
at an affordable level. Typical read-out time target values are 
summarised in table \ref{VTXtab2} for each layer. Present R\&D 
achievements indicate that the upper bounds of each time 
interval can already be considered as within reach.

  Power dissipation estimates, based on fabricated sensors and
accounting for power cycling, were performed. 
It was assumed that 
the beam time structure can be used to suppress the power 
during a large fraction of the inter-train time by about two orders 
of magnitude, estimating to about $1-2$ milliseconds before and 
after the train the transient time needed to switch on and off
all sensors in a well controlled way. In this case, Lorentz forces 
applied to the ladders are expected to remain acceptable. With
a rather conservative duty cycle of $2$\% (while the machine duty 
cycle is $0.5$\%), the average power dissipation would amount 
to a few tens of watts only (e.g. 30~W for CMOS sensors ~\cite{VTXmimosa1}). 
Such values are compatible with modest cooling, based on air 
flow, which does not require introducing additional material in 
the VTX fiducial volume.

    Power consumption may even be  mitigated more with the 
delayed read-out approach because of the very low read-out 
clock frequency it allows for, a feature which also translates 
into reduced Lorentz forces on the ladders. Moreover, if the 
signal charge is converted into an electrical signal only after 
the end of the train (e.g. like in FPCCD or ISIS devices), 
immunity against EMI can be reinforced. 

   Most pixel technologies and read-out architectures still need 
at least a couple of years until all main VTX specifications have 
been addressed. The recently considered vertical integration 
approach, which may have the highest 
potential, 
is likely to
need more time to reach maturity. It may be a technological 
solution for a second generation VTX, to be used a couple 
of years after the start of the ILC programme. It is in particular 
a 
promising solution for the machine operation near 
1 TeV, where the beam-related background may call for 
sensors substantially faster than those needed at 500~GeV.

\subsubsection{System Integration Studies}

   The R\&D on the sensors and their read-out circuits is 
complemented with studies addressing their main system 
integration issues. One of the main 
aims
of these studies is to 
tackle the design goal of 
$\lesssim$
0.1 \% X$_0$ thickness per 
layer over their active area. Attempts are made to find materials 
which combine low density and high rigidity against potential 
vibrations generated by the air cooling system and by power 
cycling (temperature gradient, Lorentz forces). The latter also
requires good thermal expansion compatibility between the 
support and the sensors. Low density materials were tested 
\cite{VTXreview}, such as silicon carbide foam, which have a 
thermal expansion coefficient close to silicon, and feature 
a density of a few per-cent only. They may actually also be 
used for the structural material of the entire VTX assembly. 

   Trials to use silicon as a support material are also made 
\cite{VTXdepfet}. They consist in using the silicon substrate 
of the sensors, excavating the silicon bulk wherever it is not 
essential for the ladder stability. The latter is provided by
''window frames'' left after the bulk excavation. This approach 
is currently followed for the upgrade of the SuperBELLE
vertex detector \cite{VTXsuperbelle}.

   Finally, a third approach consists in extrapolating from the
current state-of-the-art. It relies on the ladders equipping 
the upcoming PIXEL vertex detector of the STAR experiment 
at RHIC \cite{VTXstar}. With a total material budget of $\sim$ 0.3 \%
X$_0$, its concept may be extended to the ILD with an ultimate 
budget of $\lesssim$ 0.2 \% X$_0$. 

\subsection{Outlook}

   Definite choices concerning the sensor technology, the 
read-out architecture and the ladder design still have to wait
until full-scale fully-serviced ladders, as well as still more 
realistic simulation studies are available. For instance, a 
detailed understanding of the handling of the beam-related 
background will impact the maximal background rate acceptable, 
with direct consequences on the read-out architecture and the 
sensor technology. The validation of the VTX concept will follow, 
including the outcome of current studies of servicing issues, 
presumably around 2012.

%% file: ild/silicon/silicon.tex
\section{Silicon Tracking}

The tracking system of the ILD has been optimised to deliver outstanding resolution 
together with excellent efficiency and redundancy. The choice of ILD is a combination of 
gaseous tracking, giving a large number of hits, and the redundancy this gives, 
with a sophisticated system of silicon based tracking disks and barrels. Together the 
system achieves excellent resolution, and covers the solid angle down to the 
very forward region. 
 
An important consideration is the ability of the system to be calibrated to the 
desired precision. Here the combination of gaseous and silicon based tracking 
offers some unique advantages due to the very different nature of 
possible systematic distortions, and due to the possibility to cross-calibrate 
the different systems. For example, the Silicon tracker will help in 
monitoring possible field distortions in the TPC, as well as contributing to alignment and time 
stamping (bunch tagging). Silicon tracking is relatively easy to calibrate and as
such it is expected to provide robustness, redundancy, and ease in the calibration 
of the overall tracking system.
 
The silicon tracking system of the ILD has been developed by the SiLC collaboration. 
Detailed descriptions of the wide ranging R\&D activities conducted within the 
SiLC collaboration can be found in the latest documents and presentations 
issued by the SiLC Collaboration, and references therein~\cite{bib1, bib2, bib3, bib4, bib5, bib6}.
 
\subsection{Baseline Design of the Silicon Trackers}
Combined with the Silicon vertex detector and the central gaseous
tracker TPC, a Silicon Tracking system is proposed for the ILD. 
It is based on modern Silicon detector technology, deep sub
micron CMOS technology for the front-end (FE) electronics with a new on-detector
electronics connection and new material technology for the support
architecture. Special challenges for the ILD are a significant reduction in 
material compared to the most recent examples of large scale silicon detectors ({\it e.g.} LHC detectors), 
operating at very low power, and reaching excellent point resolution and calibration.  
The Silicon tracker is made of two sets of detectors:
\begin{itemize}\addtolength{\itemsep}{-0.6\baselineskip} 
\item The first set is located in the central barrel and is made of the 
SIT (Silicon Internal Tracker), and the 
SET (Silicon External Tracker).
Both devices have false double-sided Silicon strip detectors, together providing three precision space points. 
\item The second set is located in the forward region and is 
constituted of the FTD (Forward Tracking Detector) in the very forward region, and the 
ETD (end cap Tracking Detector), providing a space point between the TPC endplate and the calorimeter in the endcap region. 
\end{itemize}
The complete silicon tracking system is implemented in the MOKKA simulation of the ILD, including estimates 
of support structures.

\subsubsection{The Silicon Tracker in the Barrel: SET and SIT}
 
The SIT is positioned in the radial gap between the vertex detector and the
TPC. The role of the SIT is to improve the linking efficiency 
between the vertex detector and the TPC; it improves the momentum 
resolution and the reconstruction of low $p_{T}$ charged particles
and improves the reconstruction of long lived stable particles. 
 
The SET is located in the barrel part between the TPC and the central
barrel electromagnetic calorimeter (ECAL). The SET gives an entry 
point to the ECAL after the TPC end wall (3\% $X_0$). 
It acts as the third Silicon
layer in the central barrel and also improves the overall momentum
resolution. 
The SIT and SET,
in addition to improving momentum resolution (Fig. \ref{silicon:fig2}),
provide time-stamping information for separation between the bunches
and thus allowing the bunch-tagging of each event. 
These two central Silicon components may serve in
monitoring the distortion of the TPC and for the alignment
of the overall tracking.
 
\subsubsection{The Endcap and Forward Silicon Tracking: ETD and FTD}
 
The FTD is positioned in the innermost part of the tracking region, and 
covers the very forward region down to about 0.15 radians. In total seven 
disks are foreseen in this region. 
 
 
The ETD is positioned between the TPC end cap and the end cap calorimeter
system. The role of the ETD is to serve as entry point for the
calorimeter and to improve the momentum resolution for the charged
tracks with a reduced path in the TPC. Moreover it helps reducing the
effect of the material of the TPC End Plates (currently estimated to
be 15\% $X_0$). It thus might improves the matching efficiency between the
TPC tracks and the shower clusters in the EM calorimeter. It also
contributes to extending the lever arm and angular coverage of the
overall tracking system at large angle. 
Both the ETD and the FTD ensure the full tracking hermeticity.
 
 
\begin{table}[tb] 
\centering\small 
\scriptsize
\begin{tabular}{|l|l|l|l|l|l|}
\hline\hline
 \multicolumn{6}{|c|}{SIT characteristics (current baseline = false double-sided Si microstrips)} \\
\hline \multicolumn{3}{|c|}{Geometry} & \multicolumn{2}{|c|}{Characteristics} & Material \\
\hline R[mm]	& Z[mm] & cos$\theta$ & Resolution R-$\phi$[$\mu$m] & Time [ns] & RL[\%]\\
\hline 165	& 371 & 0.910 & R: $\sigma$=7.0, & 307.7 (153.8) & 0.65\\ \cline{6-6}
\cline{1-3} 309	& 645 & 0.902 & z: $\sigma$=50.0 & $\sigma$=80.0 & 0.65\\ \cline{6-6}
\cline{1-3} \cline{6-6}
\hline\hline
 \multicolumn{6}{|c|}{SET characteristics (current baseline = false double-sided Si microstrips)} \\
\hline \multicolumn{3}{|c|}{Geometry} & \multicolumn{2}{|c|}{Characteristics} & Material \\
\hline R[mm]	& Z[mm] & cos$\theta$ & Resolution R-$\phi$[$\mu$m] & Time [ns] & RL[\%]\\
\hline 1833	& 2350 & 0.789 & R: $\sigma$=7.0, & 307.7 (153.8) & 0.65\\ \cline{6-6}
\cline{1-3} 1835	& 2350 & 0.789 & z: $\sigma$=50.0 & $\sigma$=80.0 & 0.65\\ \cline{6-6}
\cline{1-3} \cline{6-6}
\hline\hline
 \multicolumn{6}{|c|}{FTD characteristics (current baseline = pixels for first 3 disks, microstrips for the other 4))} \\
\hline \multicolumn{3}{|c|}{Geometry} & \multicolumn{2}{|c|}{Characteristics} & Material \\
\hline R[mm]	& Z[mm] & cos$\theta$ & \multicolumn{2}{|c|}{Resolution R-$\phi$[$\mu$m]}  & RL[\%]\\
\hline  39-164    & 220       & 0.985-0.802 & \multicolumn{2}{|c|}{}  & 0.25\\ \cline{6-6}
\cline{1-3}   49.6-164    & 371.3	& 0.991-0.914 & \multicolumn{2}{|c|}{}  & 0.25\\ \cline{6-6}
\cline{1-3}   70.1-308    & 644.9	& 0.994-0.902 & \multicolumn{2}{|c|}{}  & 0.25\\ \cline{6-6}
\cline{1-3}   100.3-309    & 1046.1 & 0.994-0.959 & \multicolumn{2}{|c|}{$\sigma$=7.0}  & 0.65\\ \cline{6-6}
\cline{1-3}   130.4-309    & 1447.3 & 0.995-0.998 & \multicolumn{2}{|c|}{}  & 0.65\\ \cline{6-6}
\cline{1-3}   160.5-309    & 1848.5 & 0.996-0.986 & \multicolumn{2}{|c|}{}  & 0.65\\ \cline{6-6}
\cline{1-3}   190.5-309    & 2250	& 0.996-0.990 & \multicolumn{2}{|c|}{}  & 0.65\\ \cline{6-6}
\cline{1-3}  \cline{6-6}
\hline\hline
 \multicolumn{6}{|c|}{ETD characteristics (current baseline = single-sided Si micro-strips, same as SET ones)} \\
\hline \multicolumn{3}{|c|}{Geometry} & \multicolumn{2}{|c|}{Characteristics} & Material \\
\hline R[mm]	& Z[mm] & cos$\theta$ & \multicolumn{2}{|c|}{Resolution R-$\phi$[$\mu$m]}  & RL[\%]\\
\hline 419.3-1822.7     & 2426 & 0.985-0.799 & \multicolumn{2}{|c|}{x:$\sigma$=7.0}  & 0.65\\ \cline{6-6}
\cline{1-3}  419.3-1822.7     & 2428 & 0.985-0.799 & \multicolumn{2}{|c|}{y:$\sigma$=7.0}  & 0.65\\ \cline{6-6}
\cline{1-3}  419.3-1822.7     & 2430 & 0.985-0.799 & \multicolumn{2}{|c|}{z:$\sigma$=7.0}  & 0.65\\ \cline{6-6}
\cline{1-3} \cline{6-6}
\hline\hline
\end{tabular} 
\caption{The projected values of basic SIT, SET, FTD, and ETD characteristics.} 
\label{table1} 
\end{table} 
 
\subsection{Performances of the Silicon tracking system}
 
The main detector performances of the
Silicon tracking ILD system are summarized in terms of its contribution 
to i) full angular coverage, ii) momentum and impact parameter resolution, 
iii) calibration of distortions, iv) alignment, v) time stamping (bunch
tagging), and vi) redundancy and robustness of the overall tracking system.
Full simulation studies are being performed in order to best understand 
the performances of the Silicon tracking system in terms of momentum and 
spatial reconstruction and pattern reconstruction. These detailed 
simulations are completing the already available performance studies 
based on fast simulation (LiC Detector Toy Monte Carlo ``LDT'' and SGV) \cite{bib1}.
 
\subsubsection{Angular Coverage}
 
\begin{figure}
\includegraphics[height=6cm,viewport={0 0 767 621}]{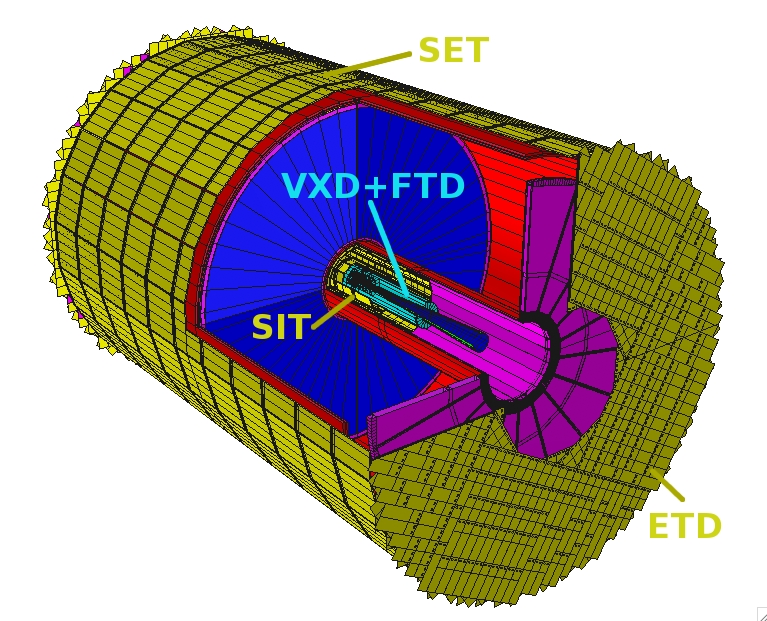}
\includegraphics[height=5.5cm]{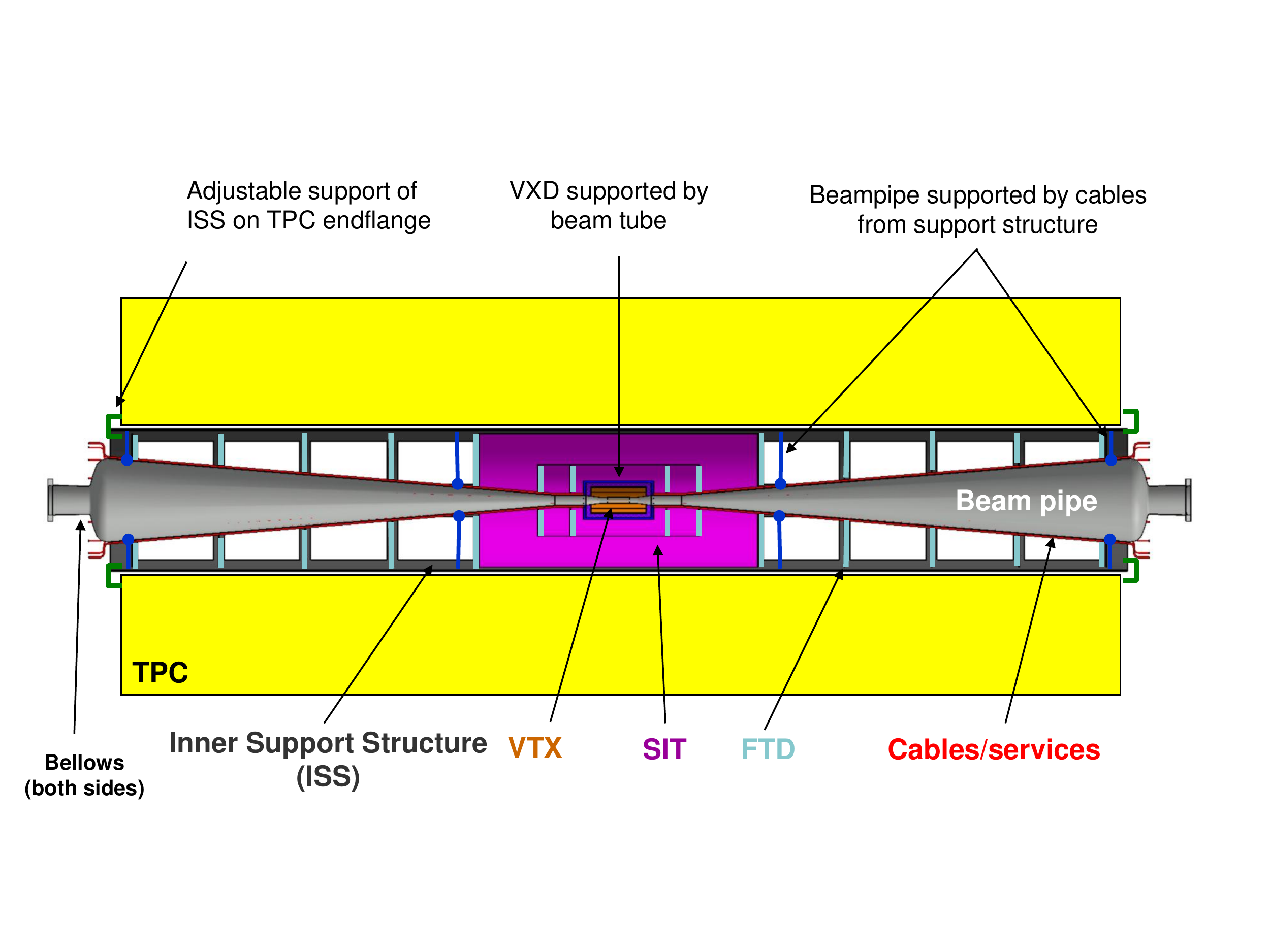}
\caption[Schematic view of the tracking system and of the SI tracking components.]
{Silicon tracking components as described in the Table \ref{table1} 
(GEANT 4-based simulation). The plot on the right shows a side-view of the inner silicon tracking system, including the support structure.} 
\label{silicon:fig1}
\end{figure}

Combining the Silicon components with the vertex 
detector and the TPC ensures efficient tracking over the full
angular coverage down to very small angles close to the beam. It helps in crucial
regions such as: i) the transition from the central barrel to the End
Cap region, ii) the end cap regions and iii) the very forward 
regions (Fig. \ref{silicon:fig1}).

\subsubsection{Momentum and Impact Parameters Performances}
 
\begin{figure}
\begin{tabular}{lc}
\centering\includegraphics[height=6.5cm,viewport={0 0 129mm 111mm},clip]{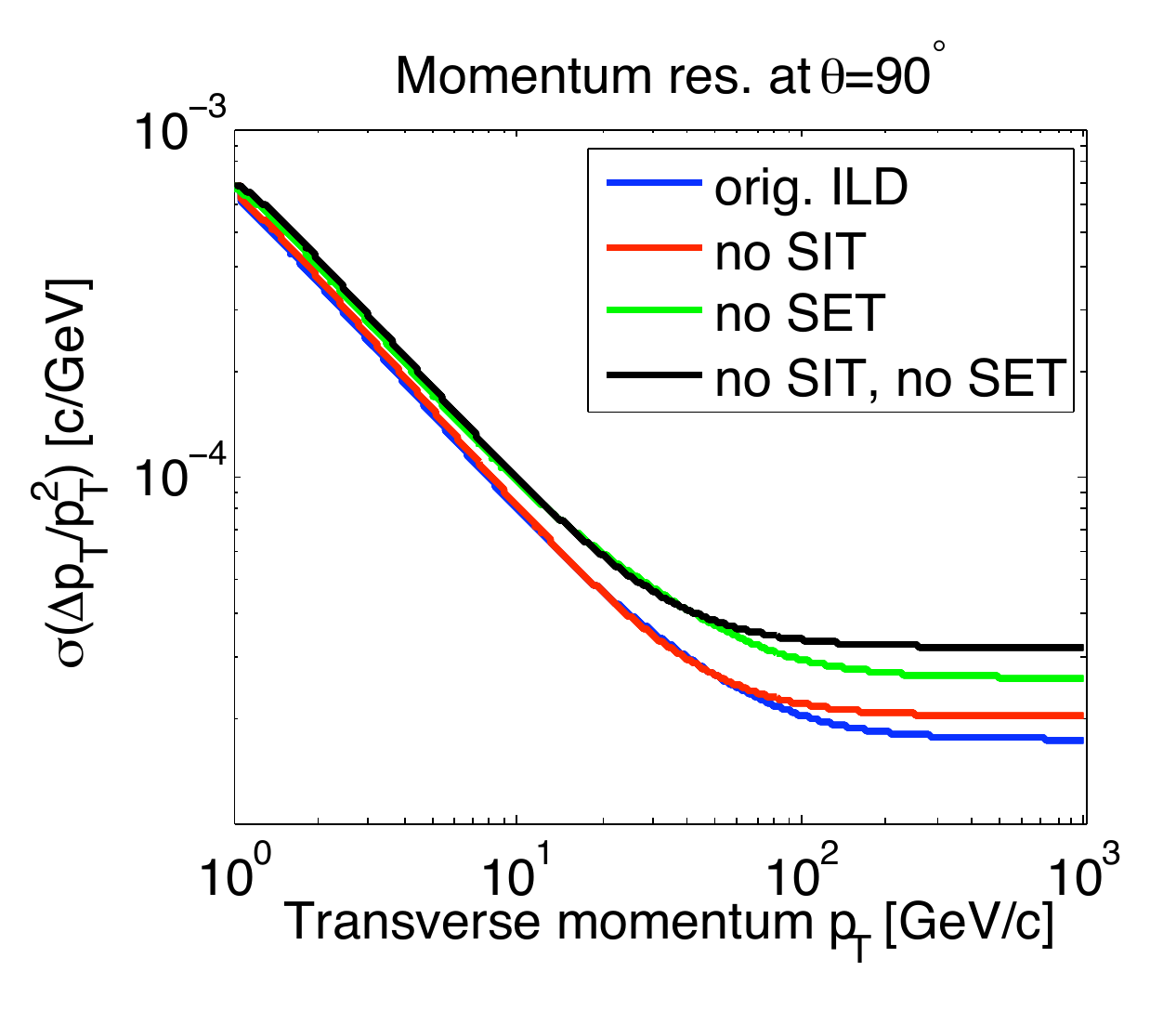}&
\centering\includegraphics[height=6.5cm,viewport={0 0 129mm 111mm},clip]{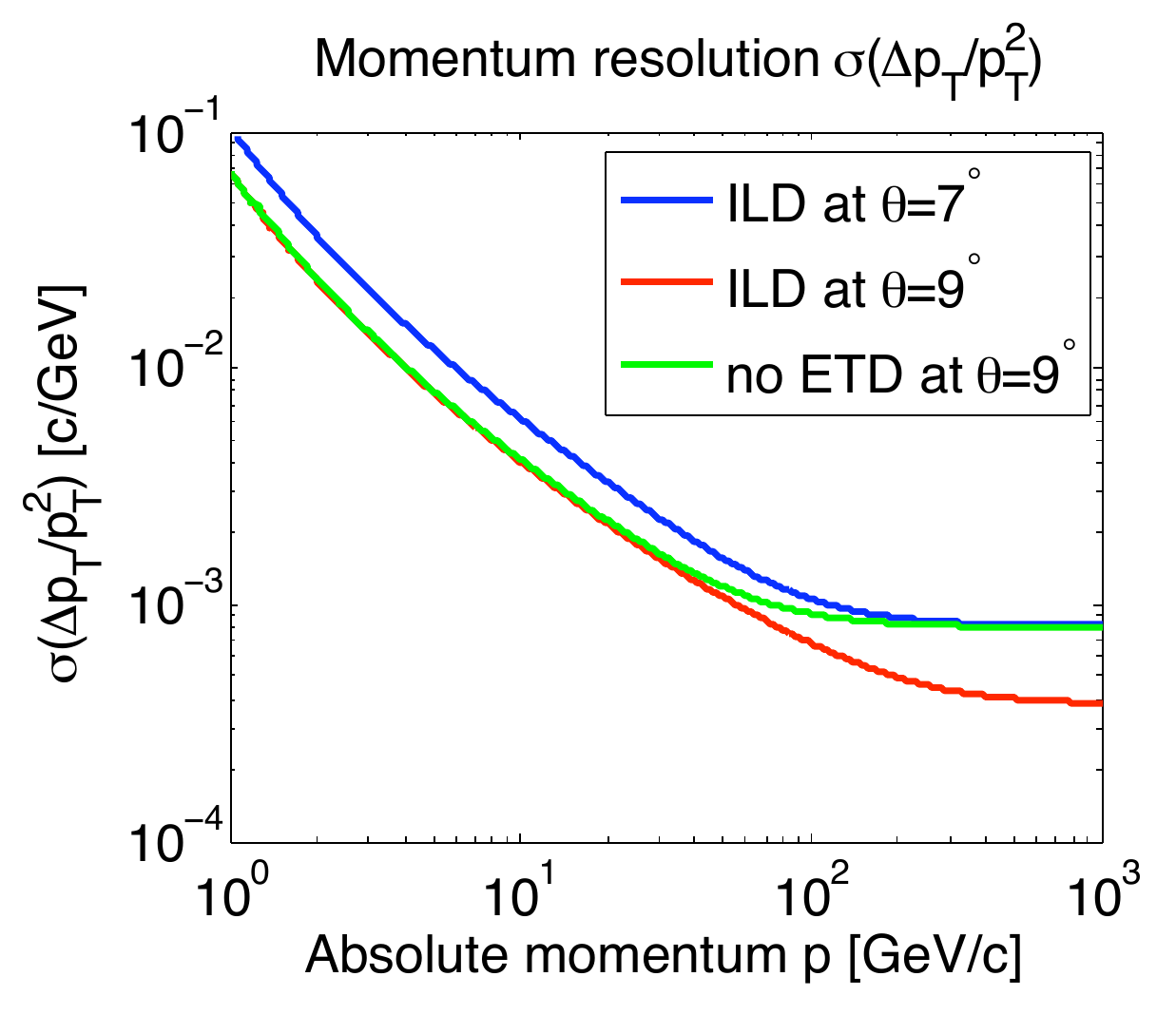} 
\end{tabular}
\caption[Transverse momentum resolution vs $p_T$.]{Left plot: barrel region, 
transverse momentum reduced resolution as function of $p_{T}$.
Right plot: forward region, transverse momentum reduced resolution as function of absolute 
momentum $p$. The different scenarios are described in the text.} 
\label{silicon:fig2}
\end{figure}

To demonstrate the contribution of the various Si tracking components to the
improvement of the overall tracking performances in terms of momentum
and impact parameter resolution, a number of studies have been performed 
based on both fast simulation and the MOKKA-Marlin and GEANT 4 simulation. 
Some of the most relevant results (evaluation using fast Monte Carlo,
muon tracking) are shown in Figure \ref{silicon:fig2}.
 
For tracks in the barrel region the present combined Silicon and 
TPC tracking system delivers an outstanding momentum resolution of 
$\sigma(\Delta p_{T} / p_{T}^{2}) < 2 \cdot 10^{-5}$~GeV as shown in Fig.~\ref{silicon:fig2}(left). The plots compare four different arrangements: 
the ILD setup as described above (blue),
a setup without the inner tracker SIT (red),
a setup without the external tracker SET (green), and
a setup without SIT and SET (black). 

Figure~\ref{silicon:fig2}(right) shows the momentum resolution 
for very forward going tracks, for three different angular ranges.  
While the addition of the silicon tracking system improves the 
momentum resolution, the impact parameter resolution remains 
virtually unchanged. 
 
Table \ref{table2} offers an alternative illustration to the improvement generated by
both SIT and SET on spatial resolution.

\begin{table}[tb] 
\centering\small 
\scriptsize
\begin{tabular}{|l|l|l|l|l|}
\hline\hline
 	& VTX ($3\times 3 \mu$m$^{2})$ & VTX ($5\times 5 \mu$m$^{2})$ & 
 VTX ($3\times 3 \mu$m$^{2})$&  
 VTX ($3\times 3 \mu$m$^{2})$  \\
	&  & & + SIT ($5\times 10 \mu$m$^{2})$ & + SET ($5\times 50 \mu$m$^{2})$  \\
\hline $\sigma$(R-$\phi$)@ R = 150cm & 1.3 cm     & 2.2 cm     & 0.6 mm    & 78 $\mu$m\\
\hline $\sigma$(z)@ R = 50cm  & 35 $\mu$m  & 60 $\mu$m  & 16 $\mu$m & 28 $\mu$m\\
\hline $\sigma$(z)@ R = 100cm & 77 $\mu$m  & 126 $\mu$m & 39 $\mu$m & 30 $\mu$m\\
\hline $\sigma$(z)@ R = 118cm & 118 $\mu$m & 192 $\mu$m & 50 $\mu$m & 39 $\mu$m\\
\hline\hline
\end{tabular} 
\caption[Extrapolation precision (R-$\phi$ and z).]{The precision of the extrapolated R-$\phi$ and z-coordinates for a 100~GeV track at 
$\theta=90^{\circ}$, at three radii in the TPC volume.} 
\label{table2} 
\end{table} 
 
\subsubsection{Distortion Monitoring and Handling}
 
The Silicon trackers are mechanically stable devices which will help 
improve the absolute alignment of the overall tracking system, and of the 
ILD as a whole. This alignment is sensitive in particular to temperature 
fluctuations, which will need to be understood to the 2 $\mu$m level.  These alignment 
systematics will be very different from the TPC ones.
 The TPC is sensitive to ambient temperature and to atmospheric pressure 
variations, to non-homogeneities in E and B fields, etc. In particular the 
E drift field in the TPC may depend on space charge transient effects due 
to variations in the machine backgrounds. 
 The SIT and SET give an independent and effective means to monitor 
accurately such effects on real data. Experience at LEP has shown that 
this capability gives an invaluable redundancy during data analysis, and a 
unique mean to disentangle and understand anomalous behaviours. It is a 
necessary complement to the unique pattern recognition capabilities of 
the TPC.
 
\subsubsection{Electronic Time Stamping}
 
Based on the performances of the front end (FE) chip currently developed,
currently a bunch crossing tagging with a precision of 160 ns can be obtained
corresponding to a shaping time around 0.5-0.7 $\mu$s and 8 sampling 
cells. This precision depends on the sampling frequency. 
A more refined estimate based on the Cleland and Stern algorithm \cite{Clelandstern}, and
function of the signal to noise, the number of samples, and the shaping
time, indicates that the currently developed framework could allow 
identifying the bunch crossing with a resolution of order of 
$20$ to $40$~ns. 

\subsection{Calibration Procedures}
 
The Front End Electronics as currently available in the
current $130$nm CMOS technology includes a full readout electronic chain
with a high level digital control of the functionality of the overall
chip. In particular a fully programmable test pulse system  is included. It will allow calibrating and monitoring of this signal processing device and play a crucial role in the 
silicon DAQ.
 
Environmental conditions around the detector due to local
temperature gradients, humidity changes, etc. will induce some
instability of the support structures comparable in size to the precision
of the detectors. Consequently, independent alignment systems monitoring these
changes will be needed. For the case of silicon trackers one can 
profit from the weak (but non-zero) absorption of infrared light
in Silicon and use laser beams as pseudo-tracks that traverse
consecutive sensors. 
 
For the SIT and FTD subdetectors, which have several Silicon layers, 
the alignment procedure is based on the use of their own tracking 
detectors as photo-sensors; the transmittance of 
Silicon to infrared beams compared to the existing AMS 
and CMS tracker systems can be improved by a further 20-30\%, 
leading to a transmittance value between 70 to 80\%.
Resolutions on the order of 2 microns can be obtained with this 
procedure~\cite{alignment}.
 
The SET and ETD are single layer detectors; the SET can be aligned
with respect to fiducial marks on the outer cage of the TPC at the
level of 100 microns by standard procedures. The monitoring of the 
SIT position could be done one order of magnitude better~\cite{integration}. 
Similar procedures can be done for the ETD. Finally, tracks will be used 
for internal alignment at the precision level of few microns, by 
using adequate $\chi ^{2}$ minimisation algorithms already employed 
in the LHC and other experiments.
 
 
\subsection{Silicon Tracker Material Budget}
 
A crucial concern of the design of the silicon tracking system has been 
to minimise the material budget. New silicon sensors and modern material technology
based on carbon fibre composite materials (CFC) provide optimal solutions for
the silicon tracking components. New front end
electronics based on DSM CMOS feature less power dissipation (see already
achieved performance of the newly developed FE chip) and allow
a direct connection onto the detector, thus removing the need for cooling. 
All these facts allow a reduction of the sensitive thickness of the sensors to at
least 250 $\mu$m, or 0.25\% X$_0$. Engineering studies have shown that a 
support structure for the envisioned silicon detectors equivalent to a 1~mm 
thick CFC layer are possible, corresponding to a contribution of $0.4\%$X$_0$ per layer. 
Together with services etc a thickness of $0.65\%$ X$_0$ per layer seems in reach
(see table~\ref{table1}).
The final goal is 0.5\% RL per layer in the innermost part of the detector, 
which will need further R\&D. A further reduction might be possible 
if new sensor and support technologies become available. 
 
\subsection{ Baseline construction and Integration of Silicon components}
 
The baseline design to construct the ILD Silicon system is an 
unified design for all the components apart from the very small FTD disks. 
The SIT, SET and ETD components will be made of Silicon strip sensors with a 
unique sensor type. The current baseline are sensors of square shape from a 
6$^{\prime\prime}$ 
wafer, $200 \mu$m thick, $50\mu$m readout pitch 
(true pitch of 25$\mu$m). The modules will be made of one up to a few sensors 
depending on the location of the module in the detector. The readout chip will be directly 
connected onto the sensor. The chip will be made 
in deep sub-micron technology (current prototypes are in 130nm technology), most 
probably in 90~nm. It is a mixed analogue-digital FE and readout chip 
with a full processing of the analogue signal, long shaping time (1$\mu$s), 
sparsification, digitisation and a high level of digital processing allowing 
full programmability of the chip.
A full prototype is presently developed \cite{FEE}. 
The power dissipation of at most 1m~W per channel is achieved and power 
cycling is included. The goal is to avoid a dedicated liquid cooling system, 
but instead to rely on a forced gas cooling as is also considered for the 
VTX and the TPC systems. Details of this however are not yet worked out. 

The integration of the silicon tracking components depends critically on the 
neigbbouring detectors. 
The SET will be made of $24 \times 2$ identical super-modules, each covering 
$2.35 \times 0.5$ m$^{2}$ supported by a light structure made from composite material. 
The support structure will be supported from the TPC end-flange, and might also 
rest at intermediate z-positions on the TPC field cage.
The two SIT layers will be made in the same way as the SET, 
namely 18 and 12 super-modules for the external and the innermost 
layers, respectively. Together with the outer four FTD disks the SIT 
layers will be supported by a CFC support structure, fixed to the 
TPC end-flange at their inner radius. 
The ETD, thanks to its XUV geometry is built in the same way as the barrel 
components and will be fixed 
to the electromagnetic end cap calorimeter. 

The support architecture of SIT, SET and ETD will be 
designed in the same way focusing on robust, very light and easy to 
mount structures. The design for the FTD disks is currently based on pixels 
(same type as the vertex detector) for the first three disks and false 
double sided strips for the other four disks. These four disks will 
be made of trapezoidal sensors and altogether 16 petals as in the 
present ATLAS forward detector. 
 
 
\subsection{R\&D needs and prospects for Silicon tracking}

The Silicon technology for large-area tracking systems will continue 
to evolve over the next years because of the stringent needs of 
the Large Collider experiments (LHC upgrades, ILC and CLIC). 
The SiLC R\&D collaboration takes an active part in these worldwide 
efforts dedicated to novel and high technology. The group will continue to develop 
novel sensors. Options include, as a first step, the edgeless planar 
micro-strip sensors followed by the 3D planar micro-strips sensors. 
The goal is to have thinner, lower voltage biased strip sensors and 
larger wafer size (8$^{\prime\prime}$). In addition, the application of pixel technology 
to at least dedicated regions of the Silicon tracking, including 3D based 
pixel technology is part of this R\&D objective. The ongoing development 
of Front End and readout electronics based on ASICs in very deep sub micron 
CMOS technology, with a high degree of processing of digital information 
on the detector, low noise, low power consumption, robustness (redundancy and 
fault tolerance), will be further pursued. New interconnection technologies 
of the ASIC directly onto the sensors by bump bonding and then by 3D 
vertical interconnect as well as new cabling techniques will be addressed. 
Challenging aspects on mechanics in order to build light, 
robust, and large area mechanical structures, with stringent mechanical 
constraints on alignment, stability (ex: push pull) and quality control 
will impact the final design and construction of these 
devices. 
A reduced material budget resulting in improved tracking performance, robustness, 
reliability, easy to build (and not expensive) are the main goals 
of this ambitious R\&D program~\cite{bib1}.

%% file: ild/tpc/tpc.tex
 
\section{The Time Projection Chamber}
\subsection{Motivation}
\label{tpcmotivation}
The subdetectors for the linear collider detector must be designed
coherently 
to cover all possible physics channels because their roles 
in reconstructing these channels are 
highly interconnected. 
Two important aspects for tracking are,
(a) precision-physics measurements require that the 
momentum of charged tracks be measured an order of magnitude 
more precisely than in previous experiments, 
and (b) high resolution measurements of the
jet-energy using the particle-flow technique require
efficient reconstruction of individual particles within dense jets.
Aspects (a) and (b) for the ILD detector 
are demonstrated in Section~\ref{sec:performance-detector-tracking} and \ref{sec:performance-detector-particleflow}.
of this document.
 
A TPC as the main tracker in a linear collider experiment offers
several advantages.
Tracks can be measured with a large number of 
three-dimensional $r\phi$,$z$ space points.
The point resolution, $\sigma_{\rm point}$, and double-hit resolution, 
which are moderate when compared to silicon detectors, are 
compensated by continuous tracking.
The TPC presents a minimum amount of material $X_0$ as required 
for the best calorimeter performance.
A low material budget also minimizes 
the effect due to 
the $\sim$10$^3$ beamstrahlung photons per bunch-crossing which 
traverse the barrel region.
Topological time-stamping 
in conjunction
with inner silicon detectors
is precise to $\sim$2~ns  
so that 
tracks from interactions at
different bunch-crossings or from cosmics can 
readily be distinguished. 
To obtain good momentum resolution 
and to suppress backgrounds, the detector will
be situated in a  
strong magnetic field of several Tesla, for which
the TPC is well suited
since the electrons drift 
parallel to $\vec{\rm B}$.  The strong B-field improves 
$\sigma_{\rm point}$ and the two-hit 
resolution by compressing 
the transverse diffusion of the drifting electrons to 
$\cal O$(1~mm)~\cite{ref-magaligruwe}.
 
Continuous tracking facilitates reconstruction of non-pointing 
tracks, e.g. from V$^0$s or certain Susy (GMSB) channels, 
which are significant for the particle-flow measurement and 
in the reconstruction of physics signatures in many 
standard-model-and-beyond scenarios. The TPC gives 
good particle identification via the specific energy 
loss dE/dx which is valuable for the majority of physics 
analyses and for electron identification. 
The TPC will be designed to be robust while easy to maintain 
so that an endcap readout module can readily be accessed 
if repair is needed.
 
A Time Projection Chamber (TPC) is chosen for the central tracker 
because of its demonstrated performance in past collider 
experiments~\cite{ref-tpcgenref}.  
The main design issues at the linear collider are covered in 
Section~\ref{tpcdesign}. 
In Section~\ref{randeffort}, 
the R\&D by the LCTPC groups\cite{ref-lcdet2007005,ref-prc08}
to determine the best state-of-the-art technology for the TPC is described.
 
\subsection{Design}
\label{tpcdesign}
There are important, and interconnected, design issues 
related to the performance, endcap, electronics, fieldcage, 
robustness in backgrounds, corrections and alignment. 
Since methods of investigating these issues have been 
established from past operational experience, 
the LCTPC groups have been actively investigating all 
aspects since 2001.
 
{\subsubsection {Performance}}
Main goals for the TPC performance
at the linear collider are given
in Table~\ref{TPC_parameters}.
Understanding the properties and 
achieving the best possible point resolution have been the object
of R\&D studies of Micro-Pattern Gas Detectors
(MPGD), MicroMegas\cite{ref-mm} and GEM\cite{ref-gem} (Section~\ref{randeffort}), 
and results from this work are reflected
in Table~\ref{TPC_parameters}. More details about the issues
are explained in the following paragraphs. 
\begin{table}[h] 
\caption[Goals for performance and design of the LCTPC]
{Goals for performance and design parameters  
for an LCTPC with standard electronics.} 
\label{TPC_parameters} 
\begin{center} 
\begin{tabular}{| l|l |} 
\hline
Size &$\phi = 3.6$m, L $= 4.3$m outside dimensions\\ 
Momentum resolution (3.5T)
&${\delta (1 / {p_t})} \sim 9 \times 10^{-5}$/GeV/c 
TPC only ($\times$ 0.4 if IP incl.)\\ 
Momentum resolution (3.5T) &${\delta (1 / {p_t})} \sim 2 \times 10^{-5}$/GeV/c 
(SET$+$TPC$+$SIT$+$VTX)\\ 
Solid angle coverage & Up to $\cos\theta \simeq 0.98$ (10 pad rows)\\ 
TPC material budget& $\sim 0.04 {\rm X}_0$ to outer fieldcage in $r$\\ 
& $\sim 0.15 {\rm X}_0$ for readout endcaps in $z$\\ 
\hline
Number of pads/timebuckets & $\sim$ 1$\times$10$^6$/1000 per endcap\\ 
Pad size/no.padrows& $\sim$ 1mm$\times$4--6mm/$\sim$200 (standard readout)\\ 
$\sigma_{\rm point}$ in $r\phi$ 
& $< 100 \mu$m (average over L$_{sensitive}$, 
modulo track $\phi$ angle)\\ 
$\sigma_{\rm point}$ in $rz$ & $\sim 0.5$~mm (modulo track $\theta$ angle)\\ 
2-hit resolution in $r\phi$ & $\sim 2$~mm (modulo track angles)\\ 
2-hit resolution in $rz$ & $\sim 6$~mm (modulo track angles)\\ 
dE/dx resolution & $\sim 5$~\%\\ 
\hline
Performance  & $>$ 97\% efficiency for TPC only ($\rm{p_t} > 1$GeV/c), and \\
                       & $>$ 99\% all tracking ($\rm{p_t} > 1$GeV/c)~\cite{ref-alexeiraspereza}\\ 
Background robustness & Full efficiency with 1\% occupancy,\\
& simulated for example in Fig.~\ref{fig:adrianfigsdan}(right)\\
Background safety factor
& Chamber will be prepared for 10 $\times$ worse backgrounds\\
& at the linear collider start-up \\
\hline
\end{tabular} 
\end{center} 
\end{table}
 
{\subsubsection {Endcap}}
The two TPC endcaps will have an area of 10~m$^2$ each. 
The readout pads, their size, geometry and 
connection to the electronics and the cooling of the 
electronics, are all highly correlated design tasks. 
The material of the endcap and its effect on ECAL for 
the particle-flow measurement in the forward direction 
must be minimized, and the goal is to keep it below 
15\%X$_0$. Designing for the finest 
possible granularity will minimize the occupancy arising 
from the TPC drifttime integrating over about 
100 bunch-crossings\cite{ref-adrianvogelthesis}. 
The sensitive volume will consist of several$\times 10^9$ 
3D-electronic standard-readout voxels (two orders of magnitude 
more than at LEP) or $10^{12}$ voxels in case of pixel readout.
Development of the layout of the endcaps, i.e. 
conceptual design, stiffness, division into sectors and 
dead space, has started, and first ideas are 
shown below (Section~\ref{randeffort}). 
 
{\subsubsection {Electronics}}
For the readout electronics, one of the important questions 
is the density of pads that can be accommodated while 
maintaining a stiff, thin, 
coolable endcap. The options being studied are 
(A)~standard readout of several million pads 
or (B)~pixel readout of a thousand times more pads using CMOS techniques. 
Table~\ref{TPC_parameters}
assumes standard readout electronics;
a similar table for pixel electronics will be made 
when the R\&D is further advanced~\cite{ref-lcdet2007005}~\cite{ref-prc08}.
A basic ingredient for the front-end electronics will be 
the use of power-pulsing which is possible due to
the bunch-train time structure and is assumed to give
a power reduction of order 100; what can be
achieved in practice is an important R\&D issue (Section~\ref{randeffort}).
 
{\noindent(A) Standard readout:}\\
Small pads, $\sim$~1mm$\times$5mm, have been found to provide good 
resolution from  
the R\&D work and
to guarantee the low occupancy in Table~\ref{TPC_parameters}.
Studies have started to establish the 
realistic density of pads that can be achieved on the endcap. 
A preliminary look at the FADC 
approach (\`a la Alice\cite{ref-alice}\cite{ref-eudet}) using 130 nm technology 
indicates that even smaller sizes 
might 
be feasible.  
An alternative to the FADC-type is the TDC 
approach (see \cite{ref-prc08}\cite{ref-eudet}) 
in which time of arrival and charge per pulse 
by time-to-charge conversion are measured.
In preparation for the possibility that the material 
budget requires larger pads, the resistive-anode 
charge-dispersion readout technique\cite{ref-madhu} 
is being studied as an option to maintain the good point 
resolution. Since this technique could compromise 
the two-track resolution, more R\&D is required.
 
{\noindent(B) CMOS pixel readout: }\\
A new concept for the combined gas amplification and readout is 
under development
~\cite{ref-prc08}\cite{ref-LC-TPC-2001}.
In this concept the ``standard'' MPGD is produced in 
wafer post-processing technology
on top of a CMOS pixel readout chip, 
thus forming a thin integrated device of 
an amplifying grid and a very high granularity endcap with
all necessary readout electronics incorporated. 
For a readout chip with $\sim 50 \mu$m pixel size,
this would result in $\sim 2\cdot10^9$
pads ($\sim 4\cdot10^4$ chips) per endcap. 
This concept offers the possibility of pad sizes small enough to observe the
individual primary electrons formed in the gas and to count the number of
ionisation clusters per unit track length, instead of measuring the integrated
charge collected.
The R\&D program (Section~\ref{randeffort}) will determine 
on what time scale this technology
will become feasible for a large TPC\cite{ref-lcdet2007005}.
 
{\subsubsection {Fieldcage}}
The design of the inner and outer fieldcages 
involves the geometry of the potential rings, the resistor chains, 
the central HV-membrane, the gas container and 
a laser system. 
These must be laid out to sustain at least 100kV at the 
HV-membrane with a minimum of material. 
The goals for the inner and outer fieldcage thicknesses are about 
1\%X$_0$ and 3\%X$_0$, respectively,
while the chamber gas adds another 1\%X$_0$.
For alignment purposes
the laser system is foreseen and may be integrated into the 
fieldcage~\cite{ref-alice}\cite{ref-star}.
The non-uniformities due to the fieldcage design
and fabrication can be minimized 
using the experience gained in past TPCs. 
 
\begin{figure}
    \begin{tabular}{cc}
    \begin{minipage}[c]{0.35\textwidth}
 		\includegraphics[height=4.5cm]{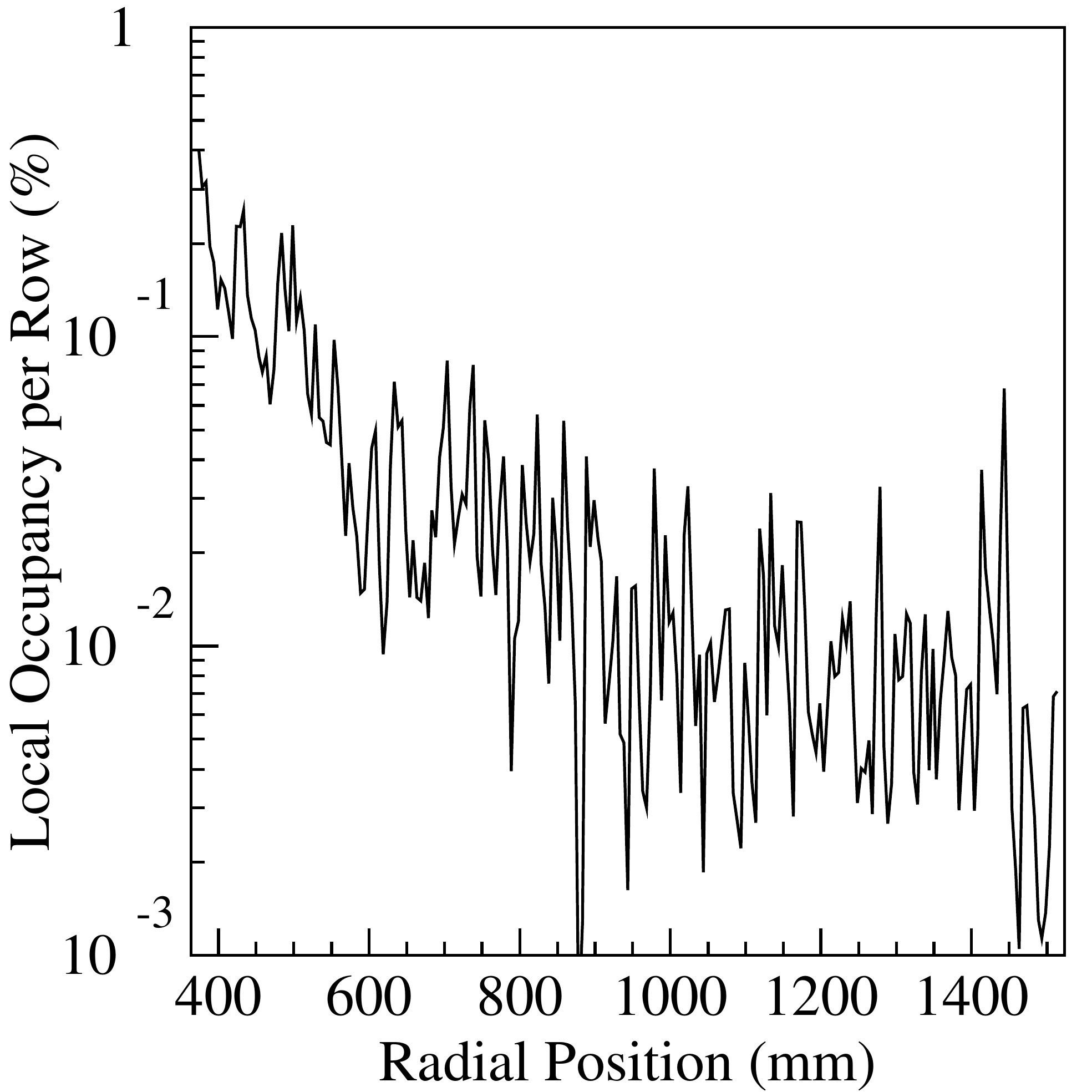}\\
 		\includegraphics[height=4.5cm]{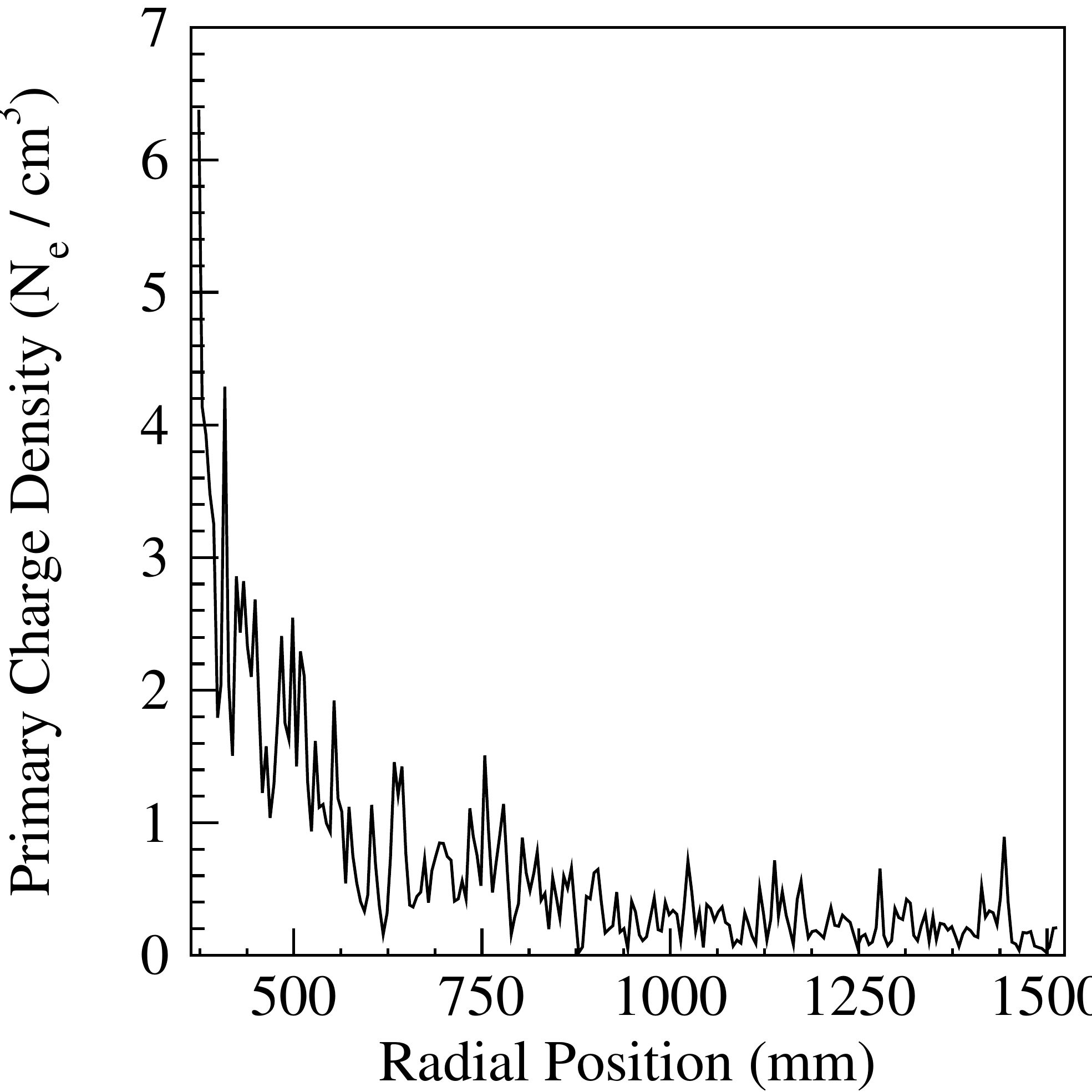}
    \end{minipage}
&
    \begin{minipage}[c]{0.75\textwidth}
 		\includegraphics[height=8cm]{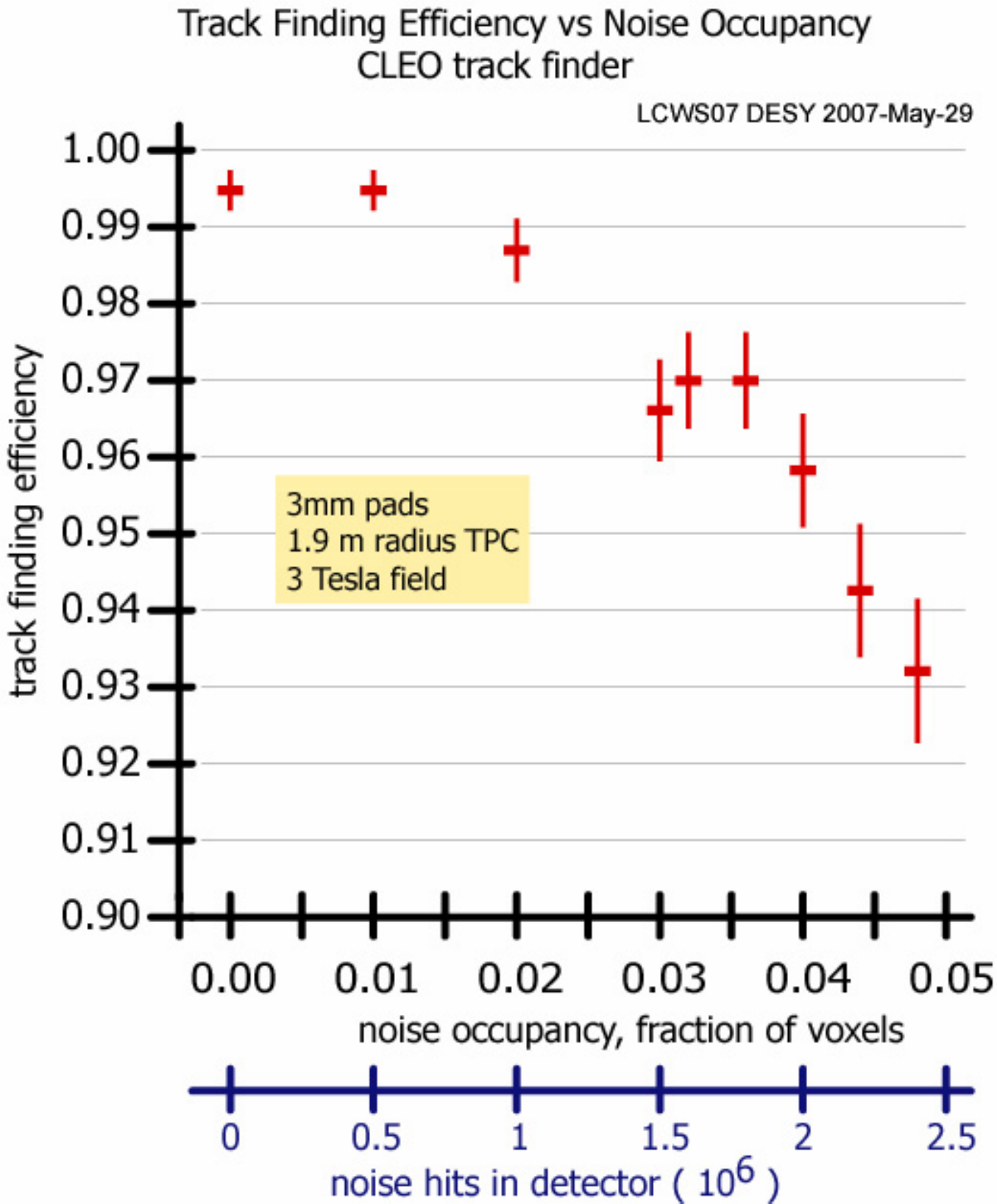}
    \end{minipage}
    \end{tabular}
  \caption[Occupancy in the TPC]{Occupancy for xyz $= 1 \times 5 \times 5 \rm{mm}^3$ voxels (left, top) 
and space charge(left, bottom)
due to the major beam-beam effects (beamstrahlung photons, electron-positron
pairs and neutrons) as simulated in \cite{ref-adrianvogelthesis}.
Study of the tracking efficiency in the presence of backgrounds (right);
this study \cite{ref-danplcws2007} assumed a conservative voxel size of
$ 3 \times 10 \times 40 \rm{mm}^3$.}
  \label{fig:adrianfigsdan}
\end{figure}

{\subsubsection{Backgrounds and robustness}}
The issues are the space-charge, covered in the next item below, 
and the track-finding efficiency in the presence 
of backgrounds which will be discussed here. 
There are 
backgrounds from the collider, 
from cosmics or other sources and 
from physics events. The main source is the collider, which 
gives rise to gammas, neutrons and charged particles 
due to $\gamma\gamma$ interactions and beam-halo muons being 
deposited in the TPC at each bunch-crossing~\cite{ref-lcdet2007005}. 
Simulations of the main sources~\cite{ref-adrianvogelthesis} arising from
beam-beam effects--gammas, pairs and neutrons--under nominal conditions 
indicate an average occupancy of 
the TPC of less than 0.1\%, 
Fig.~\ref{fig:adrianfigsdan} (left).
The TPC track finding remains robust at these occupancies; 
the continuous 3D-granularity tracking 
is inherently simple and suffers no loss in efficiency even with a uniform $1\%$ 
noise occupancy as demonstrated by the study in 
Figure~\ref{fig:adrianfigsdan}(right). Note that the latter study was
performed for a  
slightly different TPC design than the one adopted by ILD. 
The study was based on a TPC with a radius of $1.9$m, readout cells of $3 \times 10$mm$^2$ 
immersed in a 3T magnetic field. A uniform distribution of hits was assumed, and a very detailed simulation of the signal development and digitisation was performed. The 1\% uniform noise occupancy mentioned above is about twice the beam-related occupancy in Figure~\ref{fig:adrianfigsdan}(left) at the TPC inner radius and about fifty times the total occupancy in the TPC.

Since the backgrounds 
at the beginning of operation could be much larger 
until the linear collider machine is well understood, 
the LCTPC is preparing for an occupancy of 10\%.
 
{\subsubsection {Corrections for non-uniform fields}}
Both fields, (A) magnetic and (B) electric, 
can have non-uniformities which must be corrected. The
(C) chamber gas will play a crucial
role in minimizing corrections.
 
\noindent(A) {Magnetic field}\\
Non-uniformity of the magnetic field of the solenoid will be 
by design within the tolerance of 
$\int_{\ell_{\rm{drift}}} \frac {B_r}{B_z} dz < 2-10$mm 
as used for previous TPCs. This homogeneity is achieved by 
corrector windings at the ends of the solenoid. 
At the ILC, larger gradients 
will arise from the fields of the 
DID (Detector Integrated Dipole) or anti-DID, which 
are options for handling the beams inside the detector 
at an IR with $\pm$7~mrad crossing-angle. 
This issue was studied intensively and
summarized in \cite{ref-rswwlcnote}, where it is concluded 
that the TPC performance will not be degraded 
if the B-field is mapped 
to around 10$^{-4}$ relative accuracy and the 
procedures outlined 
below (under {\bf Alignment}) are followed.
These procedures will lead to an overall
systematic error due to the field components
of $\sim$ 30~$\mu$m
over the whole chamber
which has 
been shown to be sufficient~\cite{ref-rswwlcnote}
and was already achieved
by the Aleph TPC. 
Based on past experience, the field-mapping 
gear and methods will be able to accomplish 
the goal of 10$^{-4}$ for the relative accuracy. 
The B-field 
should also be monitored during running
since the currents in the DID or 
corrector windings 
may differ from the configurations mapped.
 
\noindent(B) {Electric field}\\
Three sources of space charge are
({\it i}) primary ion build-up in the drift volume,
({\it ii}) ion build-up at the readout plane and  
({\it iii}) ion backdrift, where ions created at the
readout plane could drift back into the 
TPC volume.

\begin{itemize}
\item [({\it i})] Primary ion build-up in the drift volume.
An irreducible positive-ion density due to the primary 
ionisation collected during about 1s (the time 
it takes for an ion to drift the full length of the TPC) 
will be present in the drift volume. 
The positive-ion density will be higher near the cathode, 
where the local volume integrates over backgrounds from up-to-five 
bunch trains, and using Fig.~\ref{fig:adrianfigsdan}(middle){\footnote {The 
numbers in the text derived from this figure have been 
multiplied by a safety factor of two to account for other sources of 
backgrounds.}}, 
the charge
will reach $\sim$1~fC/cm$^3$ at the inner fieldcage and   
$\sim$0.02~fC/cm$^3$ at the outer fieldcage.
The effect of the 
charge density will be established by the R\&D program, 
but the experience of the STAR TPC~\cite{ref-star}
indicates that 100~fC/cm$^3$ is tolerable\cite{ref-lcdet2007005} and
is two orders of magnitude larger
than expected for the LCTPC.\\
\item[({\it ii})] Ion build-up at the readout plane.
At the surface of the gas-amplification plane 
during an ILC bunch train of about 3000 bunch crossings
spanning 1~ms, 
there will be 
few-mm sheet layer of positive ions built up due to 
the gas amplification of the incoming charge followed 
by ion backflow.
An important property of MPGDs is that they suppress naturally 
the backflow
of ions produced in the amplification stage;
studies show that this backflow can be 
reduced to about 0.25\%~\cite{ref-lcdet2007005}.  
Using the results from Fig.~\ref{fig:adrianfigsdan} (middle), 
this layer of readout-plane ions will attain a density 
of $\cal{O}$(80) fC/cm$^3$ at the inner radius and
$\cal{O}$(2) fC/cm$^3$ at the outer radius
of the TPC.
Its effect will be simulated, but
it should affect coordinate measurement 
only by a small amount since the incoming drift electrons
experience this environment during only the last few mm of drift. 
The TPC must plan to run with the lowest possible 
gas gain, meaning of order $\sim$ 2 $\times$ 10$^3$ or less, in order to minimize this 
effect.
\item[({\it iii})] Ion backdrift and gating.
The ion buildup described in ({\it ii}) will drift 
as an ``ion sheet'' back through the TPC volume 
unless eliminated by a gating plane. In the drift volume, 
an ion sheet would be followed by sufficient drift distance 
to result in track distortions. Thus an intra-train gate 
is foreseen to guarantee a stable and robust chamber operation. 
The ILC bunch train structure requires an open-gate operation, 
without intra-train gating between bunch crossings, 
to optimally utilize the delivered luminosity. 
The gate will remain open throughout one full train and be 
closed between bunch trains. As the ion drift velocity is much 
less than that of the electrons, the gate timing allows 
collection of all of the ions. 
The added amount of material for a gating plane will be
small (e.g., $< 0.5$\%X$_0$ was the average thickness 
for the Aleph TPC gate). 
\end{itemize}

\noindent{{(C) {Chamber gas}}}\\
The choice of the gas for the LCTPC is crucial
for efficient and stable operation at the linear collider\cite{ref-magaligruwe}.
The $\sigma_{\rm point}$ resolution achievable in $r\phi$ 
is dominated by the transverse diffusion, which 
should be as small as possible; this implies that 
$\omega\tau$ for the gas should be large 
so that the transverse diffusion is compressed
by the B-field.  Large $\omega\tau$ 
means that the drifting electrons follow the
B-field, 
for which there is a program to measure well\cite{ref-rswwlcnote}, and has
the added advantage of making the chamber less
sensitive to space-charge effects
and other sources of electric field non-uniformities.
Simultaneously a sufficient number of 
ionisation electrons should be created for the position 
and dE/dx measurements. The drift velocity 
at a drift field of at most a few times 100~V/cm should be 
around 5--10~cm/$\mu$s to limit the central cathode voltage
and the event overlap.  The choice of 
operating voltage must also take into account the 
stability of the drift velocity due to fluctuations in temperature 
and pressure. 
 
{\subsubsection {Alignment}}
Achieving a momentum resolution an order of magnitude better than 
any of the collider detectors to date will be
a challenge.  The 
systematics of alignment of tracking subdetectors must be well thought 
through from the beginning to guarantee the integrity of tracking over
a radius of two meters. 
Redundant tools for solving this issue are Z-peak running, 
the laser system, the B-field map as described in \cite{ref-rswwlcnote} 
and monitored by 
a matrix of Hall-plates/NMR-probes
outside the TPC,  
and Si-layers inside the inner fieldcage and
outside the outer fieldcage.
In general based on experience at LEP\cite{ref-markronacfa8}, about 
10~pb$^{-1}$
of data at the Z peak are requested during commissioning for the
alignment of the different subdetectors, and typically 
1~pb$^{-1}$ during the
year may be needed depending on the backgound and operation of the linear
collider machine (e.g., after push-pull or beam loss).
 
The strategy learned at LEP for aligning the tracking subdetectors is
also applicable for the ILD. Needed to start with are:
a common alignment software package
for all subdetectors, the fabrication tolerances for each subdetector
$\simeq$~10--20$\mu$m internal and $\simeq$~0.1--0.2mm external (with respect to the
other subdetectors) and the B-field mapped to the requirements 
outlined in \cite{ref-rswwlcnote}. Then the steps are:
first pass through a subset of data (hadronic
tracks or $\mu$ pairs from Z-peak or from $\sqrt{s}$ running), 
each tracking detector is aligned 
internally;
second pass, the tracking subdetectors 
are lined up with respect 
to one another using a subset of data; 
finally the preceeding two steps are iterated until 
the correct momentum for $Z \rightarrow \mu\mu$ 
events is achieved.

\subsection{R\&D Effort for the LCTPC}
\label{randeffort}
All of the issues affecting the TPC performance are
being addressed by the R\&D program; a recent status report 
with extensive references to past and on-going work
is contained in \cite{ref-prc08}.
As described in the 
LCTPC-Collaboration MoA, the R\&D is proceeding in three
phases: (1) Small Prototypes (SP), (2) Large Prototypes (LP),
and (3) Design.
 
Up to now during Phase(1), 
about 6 years of MPGD experience has been gathered,
gas properties have been well measured,
the best achievable point resolution is understood,
the resistive-anode charge-dispersion technique has been 
demonstrated,
CMOS pixel RO technology has been demonstrated,
the proof of principle of TDC-based electronics has been shown and
commissioning has started for the LP. 
 
\begin{figure}
    \begin{tabular}{ll}

                \includegraphics[height=5cm]{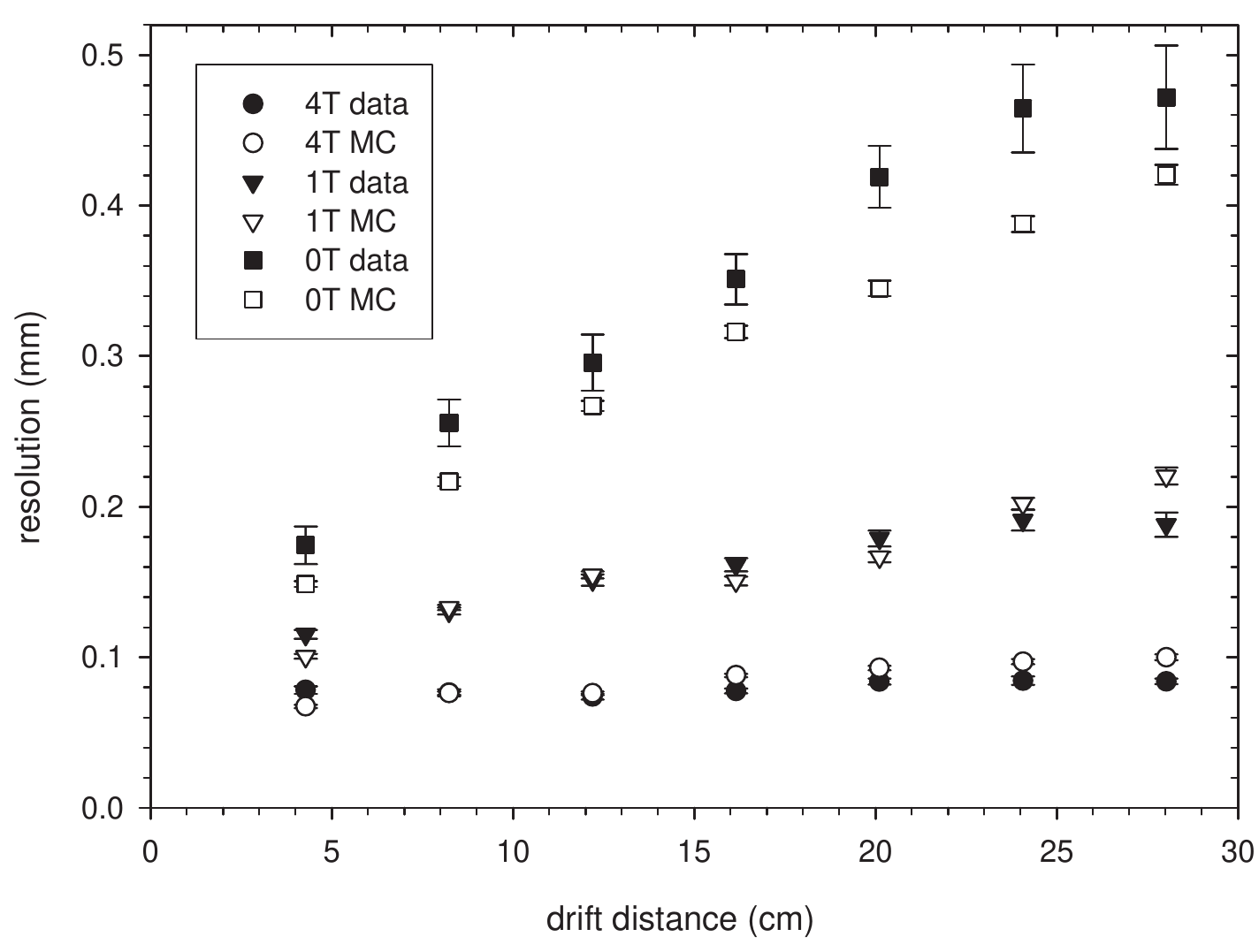}
&
		\includegraphics[height=4cm]{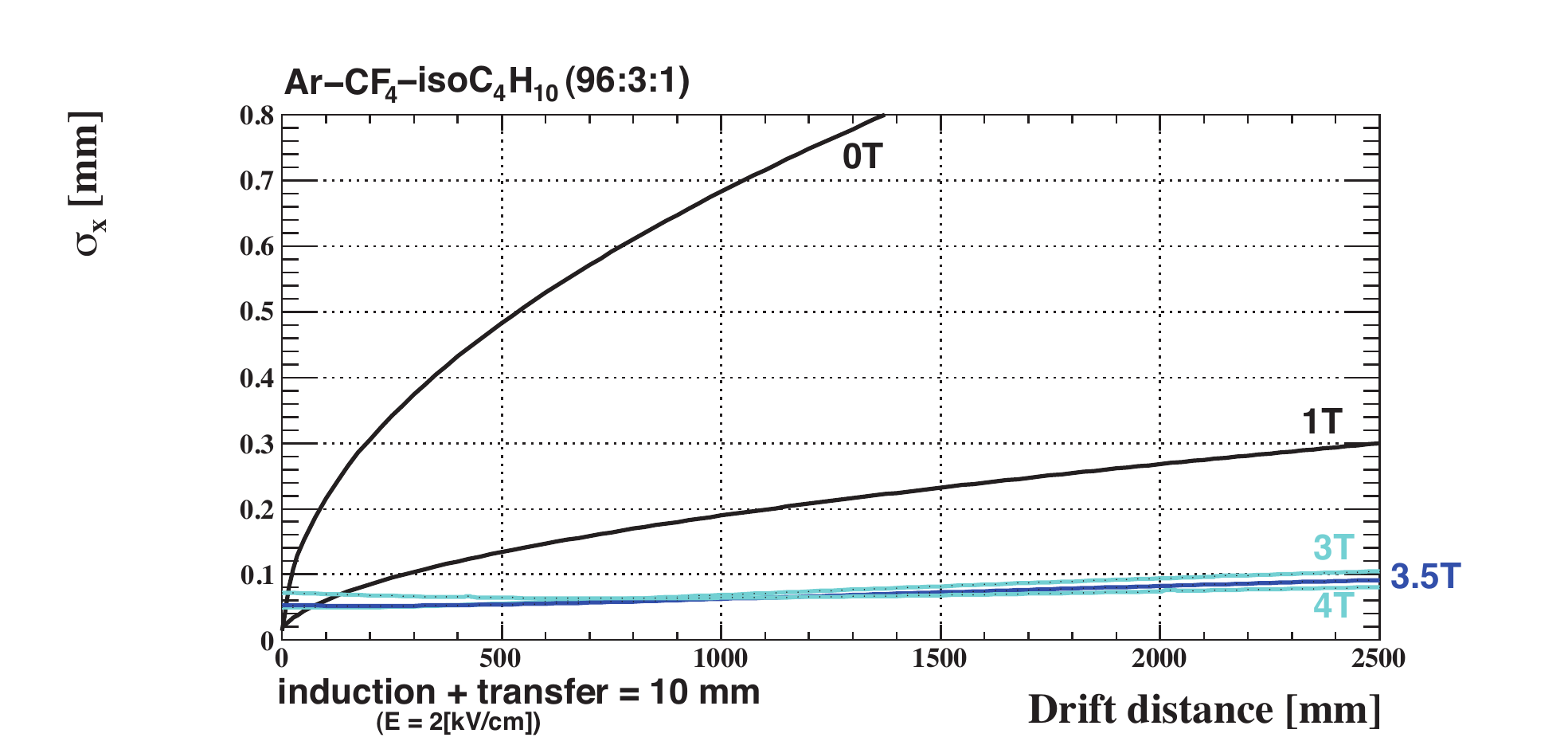}
    \end{tabular}
  \caption[R\&D Results from Large Prototype and Small Prototype Studies]
{
(left): Example of resolution results from a small prototype~\cite{ref-deannima555}
measurements with TDR gas, ArCH$_4$CO$_2$ (95-3-2); other
candidate gases are e.g. P5 and ArCF$_4$Isobutane. (Right): Theoretical resolution
for ArCF$_4$Isobutane (96-3-1) gas (right),
based on an algorithm~\cite{ref-prc08} verified during SP studies.}
  \label{fig:lpspresults}
\end{figure}

The Phase(2) LP and SP work is expected to 
take another two--three years. 
Regular bi-weekly WP phone meetings started in May 2006 where
details for the LP design were worked out and next
R\&D steps developed.
The LP commissioning is well advanced as evidenced by
Fig.~\ref{fig:lpspresults}(left),
while the fruits of the SP work resulting in 
the expected resolution are shown in
Fig.~\ref{fig:lpspresults}(center) and
Fig.~\ref{fig:lpspresults}(right).
 
The following list gives an overview 
of the currently envisioned timeline for completing the studies 
and the construction of the ILD TPC.
 
\noindent $\bullet$ 2009-12: Continue R\&D on technologies at LP, SP, pursue simulations, 
verify performance goals (details are available in \cite{ref-prc08}).\\
$\bullet$ 2009-11: Plan and do R\&D on advanced endcap; power-pulsing, electronics 
and mechanics are critical issues.\\
$\bullet$ 2011-12: Test advanced-endcap prototype at high energy and power-pulsing in
high B-field.\\
$\bullet$ 2012-18: Design and build the LCTPC.
 
Construction of endplates that satisfy the material requirements of the ILD, as well 
as the structural requirements of the TPC, will require extensive R\&D.
 
\begin{figure}
\begin{center}
    \begin{tabular}{cc}
		\includegraphics[height=4cm]{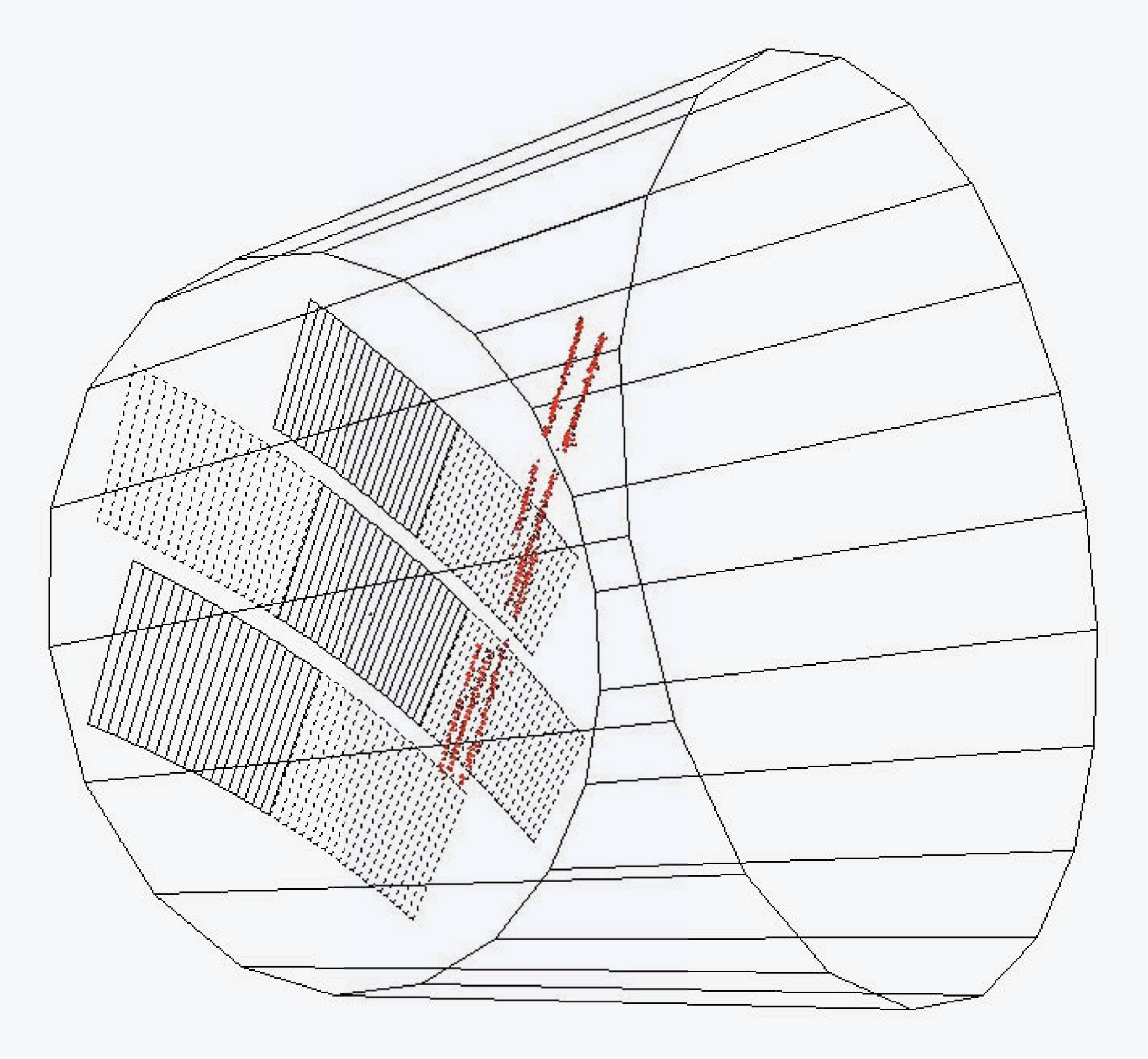} &

 		\includegraphics[height=4cm]{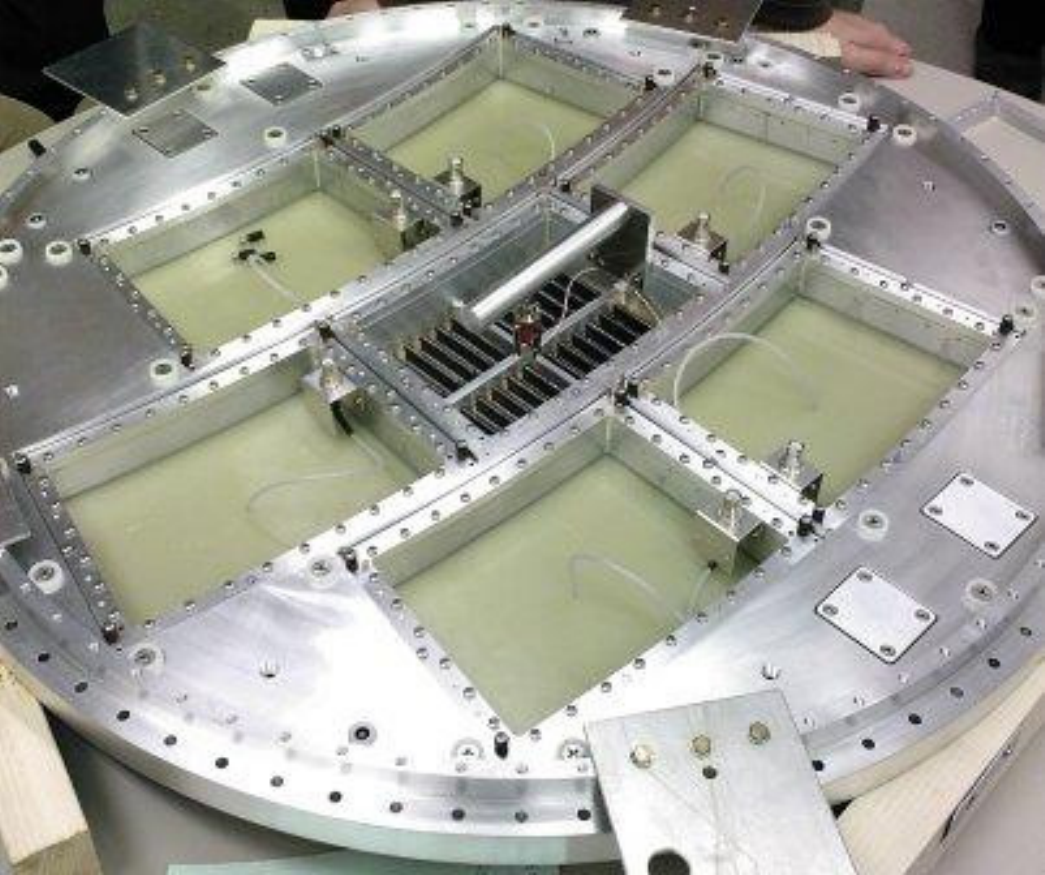}\\
 		\includegraphics[height=4cm]{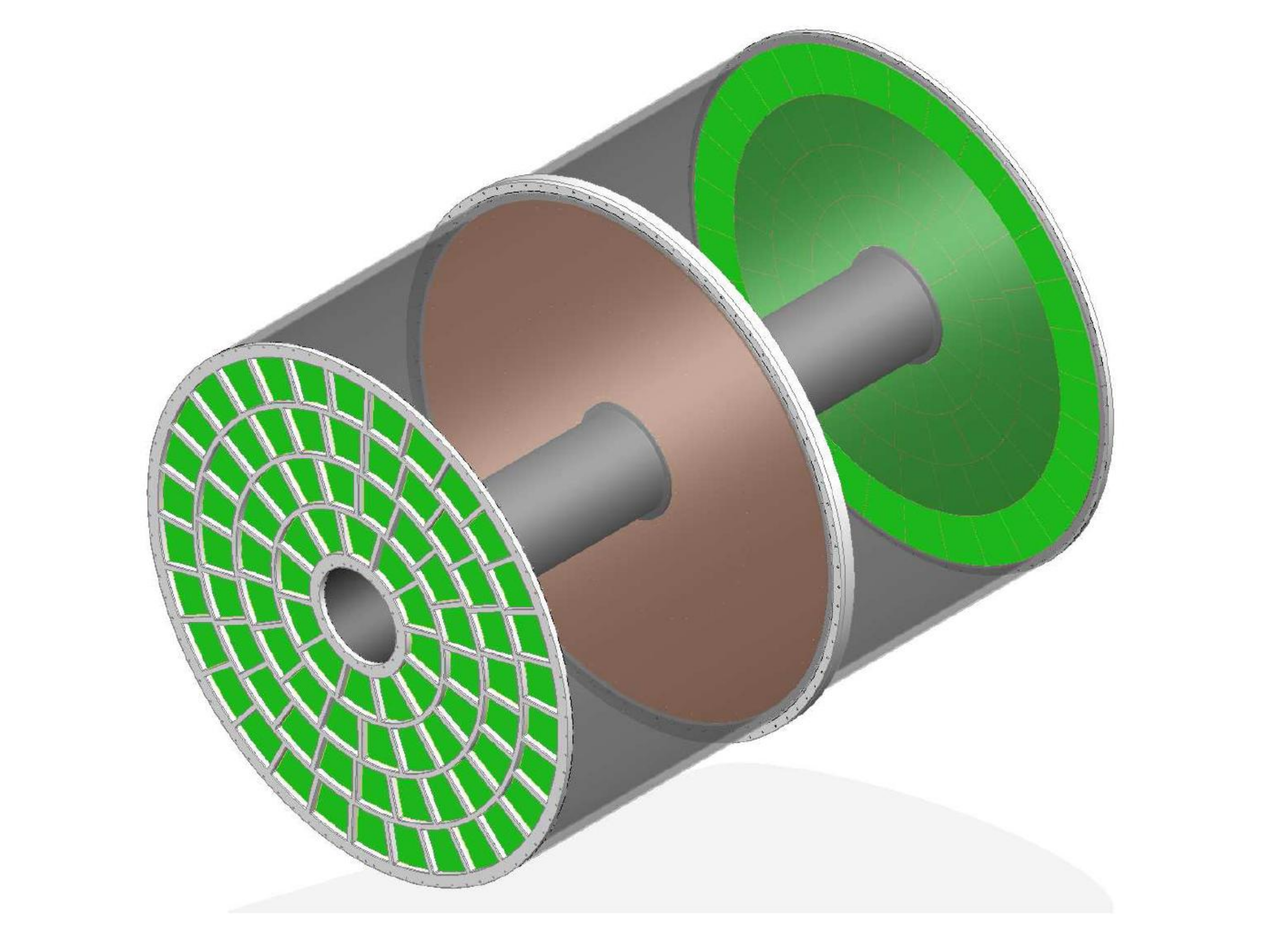}
& 		\includegraphics[height=4cm]{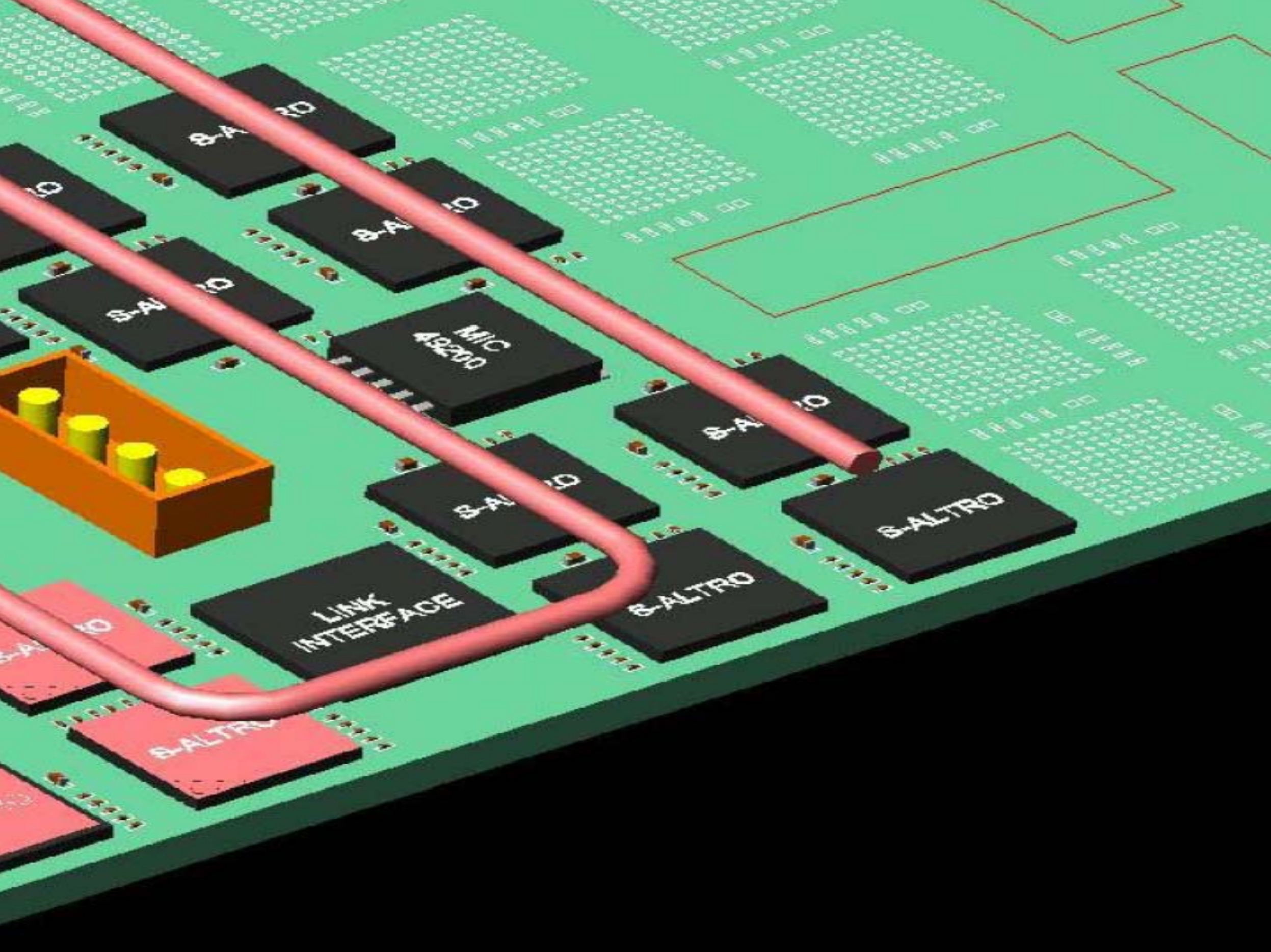}
    \end{tabular}
\end{center}
  \caption[Design ideas for a thin TPC endcap]
{(Top left): Event display from the LP beam tests. (Top right) View of the Endcap subdivision as used for the Large Prototype.
(Bottom left)Conceptual design of enplate for LCTPC.
(Bottom right) Possible layout of PCB, electronics and 
cooling for the LCTPC.}
  \label{fig:adecfigs}
\end{figure}

This work has started with first ideas having been developed
in a series of ``advanced-endcap'' meetings
during the past year. Examples are
presented in Fig.~\ref{fig:adecfigs},
and the groups
agree that there will be an evolution 
of endcaps towards a true prototype for the LCTPC.

 During the R\&D period 2009-2011, engineering studies of detailed computer 
models of advanced endplate designs will be performed.
The models will be evaluated relative to the
requirements of material limits and distribution, space limits,
rigidity in response to applied forces, manufacturing complexity and 
manufacturing precision.
Possible endcap designs fall into two general groups.
The first group is the evolution of a traditional 
machined endplate, as used in the first endcap of 
the LP, Fig.~\ref{fig:adecfigs} (left),
but with significant use of lighter materials; 
in addition, unnecessary material must be removed from the machined structure.
The second group is the simulation of true space-frame designs 
which can be constructed utilizing various 
techniques, e.g.,  fully machined, bonded 
composites and assembly of individual components.
 
During the  period 2011-2012, further study of 
designs that were successful as computer models will follow. 
Several prototypes of the advanced endplate will 
be manufactured;
both scale-models (20-50\% full size) and sections 
of the full size endplate will be used to evaluate 
the manufacturing integrity and uncover sources of 
loss of precision  or rigidity in the design.
Finite element analysis will be 
used to predict the strength of the full size endplate; this analysis 
will be calibrated by comparison with measurements on the prototypes.
 
At the beginning of the period 2012-18, the selection must be made
from the different 
technological options -- 
GEM, MicroMegas, resistive anode, pixel, electronics, endcap structure -- 
to establish a working model for the design of the LCTPC.
This design will be used for the ILD proposal 
in 2012 and include pad segmentation, electronics, mechanics, cooling 
and integration, so that performance, timeline and cost 
can be estimated reliably.
  
For the technology selection, a scenario could be 
that questions must be answered 
as to which options give the best performance
based on R\&D results from LP, SP, electronics and endcap studies.
Main performance criteria could be endcap thickness and 
$\sigma_{\rm point}$, double-hit and momentum resolution
for single tracks and for tracks in a jet environment.
Choice of criteria to use will be decided over the next two years. 
 
Finally, as to the $\sqrt{s}$ coverage, simulations in Chapter~\ref{sec:optimization} of this LOI have shown that, 
with the performance goals in
Table~\ref{TPC_parameters}, the LCTPC will give good performance
up to and well beyond 1~TeV.

%% file: ild/calo/calo.tex



\newcommand{\unit}[1]{\ensuremath{\mathrm{\,#1}}}
\renewcommand{\u}[1]{\unit{#1}}
\newcommand{\x}{\ensuremath{\times}}
\newcommand{\um}{\u{\upmu m}}
\newcommand{\uW}{\u{\upmu W}}
\def\lambdaI{\ensuremath{\mathrm{\lambda_I}}}
\def\X0{\ensuremath{\mathrm{X_0}}}
\def\RM{\ensuremath{\mathrm{R_M}}}



\section{The Calorimeter System}
\label{sec:calo}

\subsection{Introduction to calorimeters}

Tagging of electroweak gauge bosons at the ILC, based on di-jet mass
reconstruction, makes the reconstruction of multijet events a major goal for
detectors at the ILC.
The particle flow approach (see e.g. \cite{Brient:2002gh}), which consists of
individual particle reconstruction
dictates many fundamental aspects of the calorimeter design, most notably the
requirement for very fine transverse and longitudinal segmentation of the calorimeters, as studied
in Section~\ref{sec:optimization-particleflow}.  It has to be noted that a highly
granular calorimeter, optimised for PFA, leads also to a way to have a very
efficient software compensation, as it is shown in~\ref{sec:AHCAL}.
The choice of technology for the ECAL and HCAL are driven by the requirements of
pattern recognition more than the intrinsic single particle energy resolution,
although the latter is still an important consideration.
%

Several technologies for electromagnetic and hadronic calorimeters are being
pursued, with a number of prototypes in test beams.  Next generation prototypes
are being constructed with dimensions and integration issues very close to those
of final ILD detector modules. The research and development work is carried 
out in the context of the CALICE collaboration~\cite{CALICE}.

\subsection{General Layout}

The calorimeter system is divided in depth into an electromagnetic section,
optimised for the measurement of photons and electrons, and a hadronic section
dealing with the bulk of hadronic showers.  %
The two parts are installed within the coil to minimise the inactive material in
front of the calorimeters.  To follow the symmetry imposed by the beams and the
coil, the calorimeter is divided into a cylindrical barrel and two end-caps.  %

The electromagnetic calorimeter consists of tungsten absorber plates interleaved
with layers of Silicon (pads or pixels), or Scintillator detectors with very
fine segmentation of the readout.
The hadronic calorimeter is planned as a sampling calorimeter with steel absorber plates and fine grained readout. 
Two options are currently proposed.  %
The first uses scintillator cells with fine granularity and multi-bit (
analogue) readout.  The second is based on gaseous detectors and uses even finer
granularity.  Due to the large number of cells, in the second case one- or
two-bit (semi-digital) readout is sufficient.

\subsection{The Electromagnetic Calorimeter}


For the electromagnetic calorimeter
the requirements on granularity, compactness and particle separation lead to the
choice of a sampling calorimeter with tungsten (radiation length
$\X0=3.5\u{mm}$, Moli\`ere Radius ${\rm R_{M}}=9\u{mm}$ and interaction length
$\lambdaI=99\u{mm}$) as absorber material.  This allows for a compact design
with a depth of roughly 24\u{\X0} within 20\u{cm} and, compared to e.g.\ lead, a
better separation of EM showers generated by near-by particles.

To achieve an adequate energy resolution, the ECAL is longitudinally segmented
into around 30 layers, possibly with varying tungsten thicknesses.
The active layers (either silicon diodes or scintillator) are segmented into cells with a lateral size of $5-10\u{mm}$
to reach the required pattern recognition performance.

\subsubsection{Geometry and Mechanical Design}

\begin{figure}[ht]
  \centering
  \begin{tabular}[c]{cc}
  \includegraphics[width=0.45\textwidth]{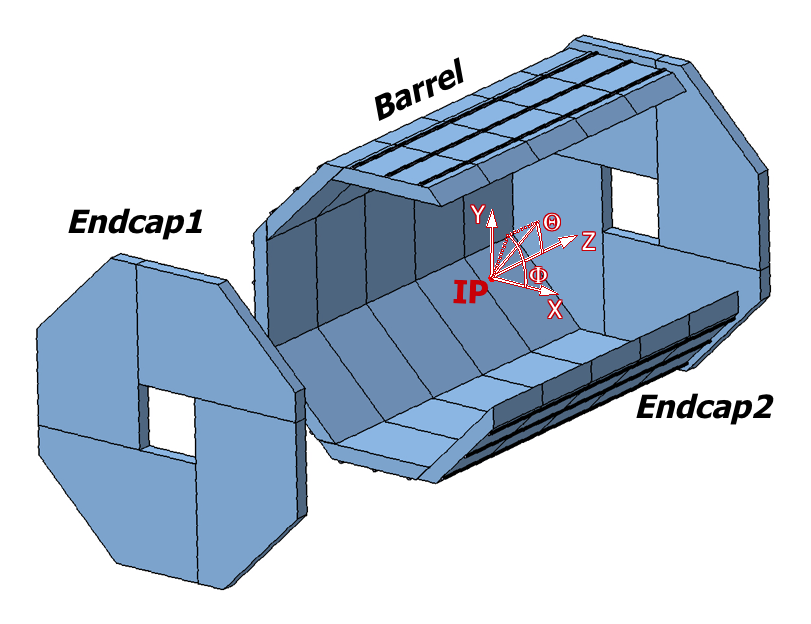} &
  \includegraphics[width=0.45\textwidth]{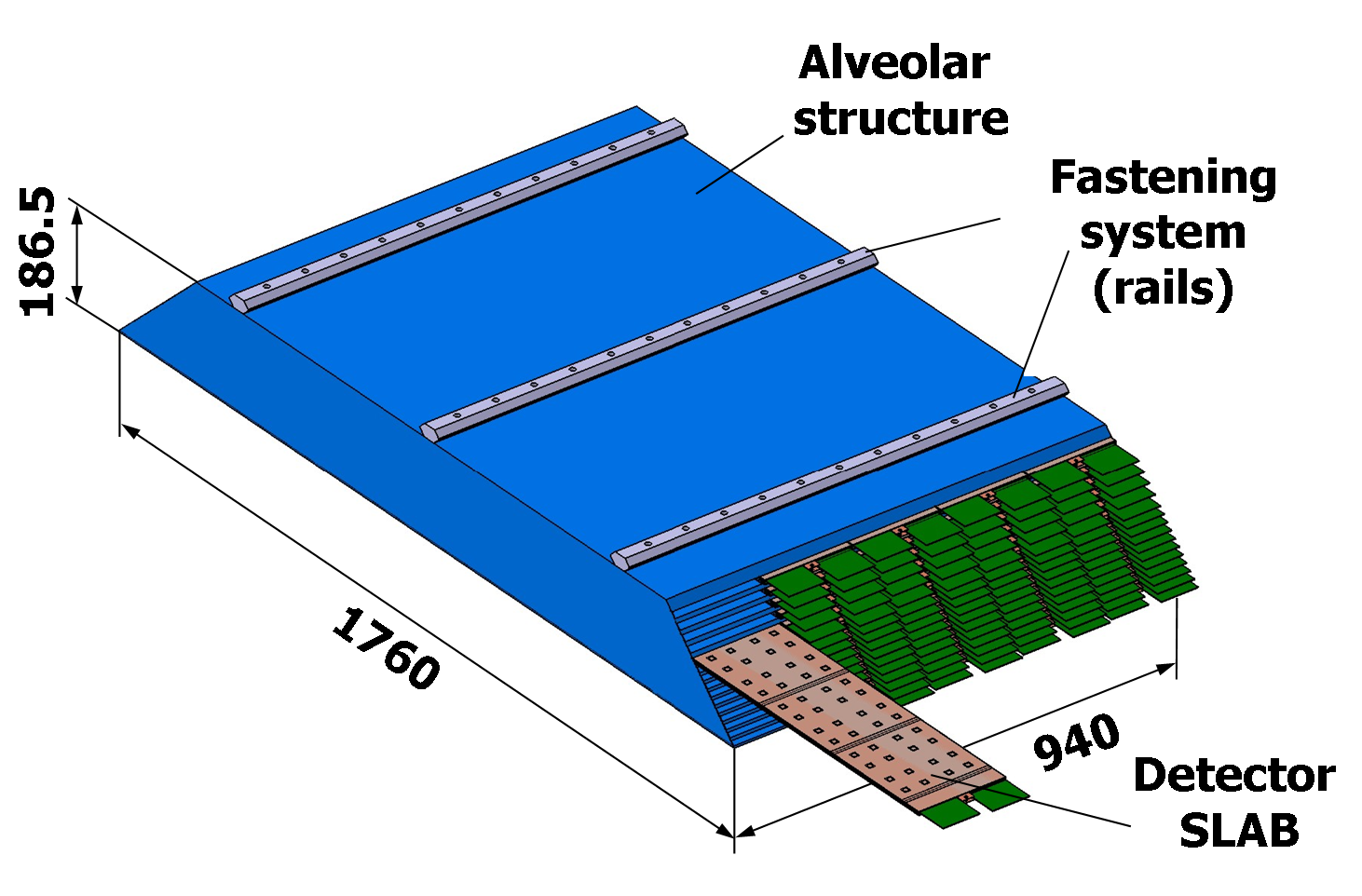}
  \end{tabular}
  \caption{Global layout of the ECAL (left) and layout of one module (right).}
  \label{fig:EcalGeom}
\end{figure}

One of the requirements for the calorimeter is to ensure the best possible
hermeticity.  %
Three regions are of particular concern for this question: the boundaries
between mechanical modules, the overlap between barrel and end-cap, and the
small angle region with the connection to the luminosity monitor.  To minimise
the number and effect of cracks in the barrel, a design with large modules is
preferred, with inter-module boundaries not pointing back to the IP.  The
cylindrical symmetry of the coil has been approximated by an eight-fold symmetry
and the modules are designed in a such a way (c.f.\ fig.~\ref{fig:EcalGeom} that
the cracks are at very large angle with respect to the radial direction.  This
octagonal shape optimises the barrel module sizes and their mechanical
properties without diverging too far from a circle.  %
One eighth of the barrel calorimeter is called a stave.  Each stave is fastened
to the HCAL front face with a precise system of rails.  Some space is left
between the ECAL and the HCAL to accommodate different services such as cooling,
electrical power and signal distribution.  %
Along the beam axis, a stave is subdivided into five modules.
The ECAL end-caps are attached to the front face of the hadronic end cap
calorimeters using a similar rail system.  %

A detailed mechanical design of the modules has been prepared, and 
is tested under real conditions in several test beam experiments. More 
details can be found in~\cite{calice1}.




\subsubsection{Optimisation}


For the final detector, a global optimisation study of the longitudinal profile
has to be performed, by varying the thickness of the Silicon and Tungsten layers
as a function of the depth, in order to minimise cost, lateral spread and
energy resolution.

The dependence of the ECAL energy resolution as a function of the longitudinal
sampling scheme has been studied in simulation~\cite{HV_Valencia}.  For a given
number of sampling layers, the energy resolution improves if the first part of
the calorimeter is more finely segmented than the latter part.
The effect of the silicon cell size on ECAL performance has been studied in
simulation, focusing on the photon reconstruction capability in di-jet events
and hadronic $\tau$ decays.  Three different cell-sizes ($5\x5\u{mm^2}$,
$10\x10\u{mm^2}$ and $20\x20\u{mm^2}$) have been investigated.  In both cases a
specialised photon reconstruction algorithm (GARLIC~\cite{GARLIC}) has
been applied.  The algorithm was separately tuned for each cell-size.

\begin{figure}[ht]
  \begin{minipage}[c]{0.4\linewidth}
    \includegraphics[width=1.2\textwidth]{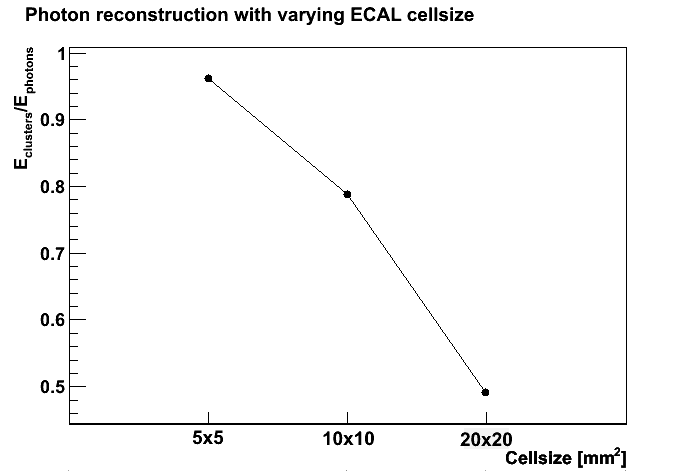}
    \caption[Reconstructed vs true energy for photons in di-jets.]{Fraction of energy identified as photon induced to true photon energy (Monte Carlo truth) in
      di-jet events at $E_\mathrm{CM} =400\u{GeV}$.}
    \label{fig:CellSizeJets}
  \end{minipage}
  \hfill
  \begin{minipage}[c]{0.45\linewidth}
    \includegraphics[width=1.1\textwidth,viewport={25 0 805 324},clip]{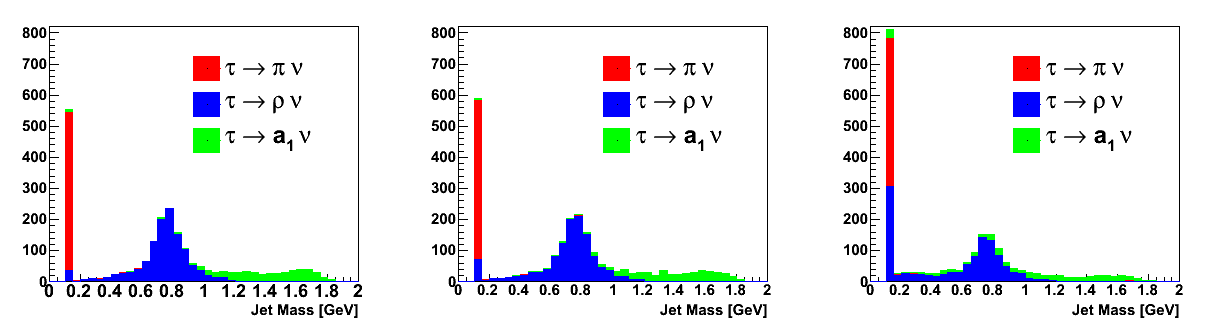}\\
        \hspace*{1.5cm} \includegraphics[width=0.55\textwidth,viewport={816 0 1235 324},clip]{CellSizeTaus.png}
    \caption[Reconstructed tau masses vs ECAL cell size.]]{Reconstructed invariant mass of hadronic $\tau$ decay products in
      $\mathrm{ZH}\rightarrow\mu \mu \tau \tau$ events for different ECAL cell
      sizes (starting at top left: $5 \times 5$mm$^2$, $10 \times 10$mm$^2$, and $20 \times 20$mm$^2$.}
    \label{fig:CellSizeTaus}
  \end{minipage}
\end{figure}

Figure~\ref{fig:CellSizeJets} shows the mean ratio of calorimetric energy
reconstructed as photons to the true photon energy, in simulated di-jet events
at $E_\mathrm{CM} = 400\u{GeV}$ for a variety of cell sizes.  A cell-size of
$5\x5\u{mm^2}$ is clearly to reconstruct the correct fraction of photon
energy inside jets. The interpretation of these result, which is based 
on a dedicated photon finding algorithm, requires care. It can 
not be applied directly to full particle flow reconstruction, 
which in general shows a weaker dependence.

Studies of $\tau$ reconstruction have been performed in ZH
($\mathrm{H}\rightarrow\tau\tau$) events at $E_\mathrm{CM} =230\u{GeV}$ with
$m_\mathrm{H} = 120\u{GeV}$.  The three decay modes $\tau \rightarrow \nu
\pi$,\, $\tau \rightarrow \nu \rho$ and $\tau \rightarrow \nu a_1$ have been
considered.  The reconstructed invariant mass of the visible $\tau$ decay
products is shown in Fig.~\ref{fig:CellSizeTaus} for the three different cell
sizes. A simple selection based on particle flow (reconstructed photons) and 
jet mass (cut at 200~MeV) allows one to reach good 
efficiency and purity, without the need for the more sophisticated analysis. 
The efficiencies and purities of the reconstruction of the various decay
channels are given in Table~\ref{tab:TauDecayPurityTable}.  Again a cell size of
$5\x5\u{mm^2}$ is favoured although the performance loss with respect to
$10\x10\u{mm^2}$ cells is smaller than in high-energy jets.

\begin{table}
  \small
  \begin{center}
    \begin{tabular}[h]{l|ccc|ccc|ccc}
      & \multicolumn{3}{c||}{$5\x5\u{mm^2}$} & \multicolumn{3}{c||}{$10\x10\u{mm^2}$} & \multicolumn{3}{c}{$20\x20\u{mm^2}$} \\
      \hline
      \hline
      & $\pi_\textrm{sim}$ & $\rho_\textrm{sim}$ & $a1_\textrm{sim}$ & $\pi_\textrm{sim}$ &
      $\rho_\textrm{sim}$ & $a1_\textrm{sim}$ & $\pi_\textrm{sim}$ & $\rho_\textrm{sim}$ & $a1_\textrm{sim} $ \\
      \hline
      $\pi_\textrm{rec}$ & 98.8 & 2.8 & 1.9 & 98.7 & 5.9 & 1.6 & 98.6 & 27.1 & 7.0 \\
      \hline
      $\rho_\textrm{rec}$ & 1.2 & 96.5 & 9.2 & 1.3 & 93.4 & 15.0 & 1.4 & 72.3 & 54.4 \\
      \hline
      $a1_\textrm{rec}$ & 0 & 0.7 & 88.9 & 0 & 0.7 & 83.4 & 0 & 0.6 & 38.6 \\
      \hline
    \end{tabular}
  \end{center}
  \caption[Photon identification in the ECAL vs cell size.]{Reconstruction efficiencies and purities of hadronic $\tau$ decays in
    $ZH\rightarrow\mu \mu \tau \tau$ events with various ECAL cell-sizes}
  \label{tab:TauDecayPurityTable}
\end{table}

\begin{figure}[ht]
  \centering
  \begin{tabular}[b]{cc}
    \includegraphics[width=0.45\textwidth]{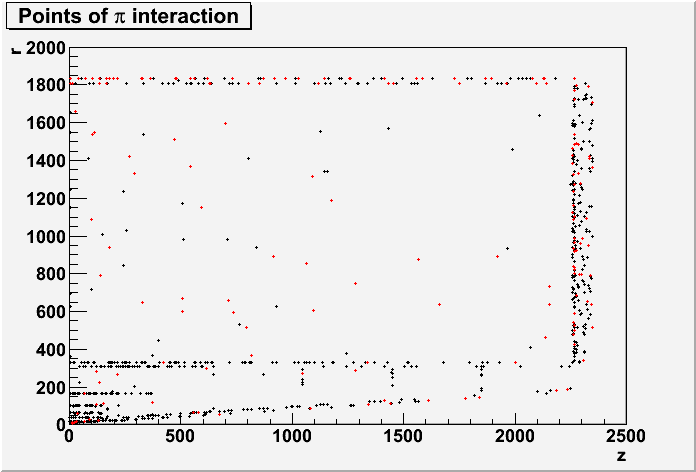} &
    \includegraphics[width=0.54\textwidth]{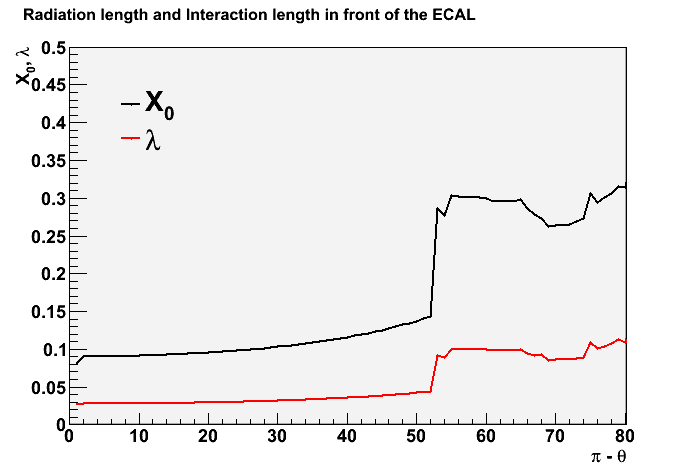}
  \end{tabular}
  \caption[Influence of material in front of the ECAL.]{ %
    Left: Interactions points of single photons in the tracker region of ILD.  
    The black points correspond to interactions that lead to the
    creation of clusters in the calorimeter found with the GARLIC photon      
    reconstruction while the red
    points correspond to interactions that did not create any clusters. %
    Right: Number of radiation and interaction length in front of the ECAL as a
    function of the polar angle.}
    \label{fig:PiInteractionMap}
\end{figure}

To study the effect of material in front of the ECAL on the particle flow
performance, 4\u{GeV} single charged pion events have been simulated.  The $\pi^0$'s
produced in interactions in the tracker region may give rise to additional
reconstructed photons in the ECAL.  The GARLIC photon identification algorithm~\cite{GARLIC}
has been applied to the single pion events.  For the approximately six percent
of pions which interact in the tracking volume, Fig.~\ref{fig:PiInteractionMap}
shows the position of the pion interaction point inside the detector for events
in which photon clusters are (red points, $55\%$) or are not (black points,
$45\%$) found.  The TPC end-plates and gas give the largest contribution to the
total number of pion interactions in front of the ECAL.  When only those
interactions which give rise to identified photon clusters are considered, the
detector components at the centre of the detector, that is, the vertex detector,
SIT, beam tube and FTD support, also give significant contributions.
Even though ILD has been designed with with minimum material in the tracker in mind, 
there is still about one pion per event which interacts in the tracker volume. This 
underlines the need for continued R\&D and continued care toward further 
material reduction in the tracker. 


\begin{table}[ht]
    \begin{center}
    \small
    \begin{tabular}{l|l|l|l}
      ~         & \% of total   & \% with &  \% of total events\\
      ~         & interactions  & clusters & with clusters \\ \hline
      VTX       & 11.9 & 64.5 & 13.9 \\ \hline
      SIT       & 11.8 & 68.7 & 14.6 \\ \hline
      Beam pipe & 10.4 & 62.9 & 11.8 \\ \hline
      FTD       & 8.9 & 66.1 & 10.6 \\ \hline
      TPC inner field cage & 5.4 & 63.8 & 6.2 \\ \hline
      TPC gas   & 17.1 & 23.0 & 7.1 \\ \hline
      TPC outer field cage & 6.5 & 50.6 & 5.9 \\ \hline
      TPC endplate  & 22.3 & 61.4 & 24.8 \\ \hline
      SET       & 3.1 & 58.0 & 3.3 \\ \hline
      ETD       & 2.8 & 35.1 & 1.8
      \end{tabular}
    \end{center}
    \caption{Interaction of pions in the different parts of the tracker region.}
\end{table}


\subsubsection{Silicon - Tungsten Electromagnetic Calorimeter}
\label{sec:SiW_ECAL}

The general requirement about compactness (small Moli\`ere radius) has led to a
sandwich calorimeter with a tungsten radiator and silicon for the sensitive
medium.
To reach an adequate energy resolution the first 12 radiation lengths are filled with 20 layers of
0.6\u{\X0} thick tungsten absorbers (2.1\u{mm}), followed by another 11 radiation lengths
made from 9 layers of tungsten 1.2\u{\X0} thick.  %
The calorimeter starts with an active layer.
For the chosen geometry the Moli\`ere radius is $19$~mm.
The choice of silicon technology for the readout layer 
permits a very high transverse granularity, now fixed at $5\x5\u{mm^2}$.

The final calorimeter will contain around $10^8$ readout cells in total.
To keep the final system as compact as possible, and reduce dead areas, 
the very
front end electronics will be embedded into the detector layers.

The challenging construction of the SiW ECAL is currently tested by a large
scale R\&D program pursued by the CALICE Collaboration.  Results from test beam
measurements demonstrating the feasibility to realise the detector have been
published in~\cite{calice1, calice_resp}.  The energy resolution has been
determined to be $(16.6\pm0.1)/\sqrt{{E(\rm GeV)}}\oplus(1.1\pm0.1)\u{\%}$ with
a MIP signal over noise ratio $S/N \approx 7.5$.  %

At present, the CALICE collaboration is preparing the construction of a
prototype module with a size and shape close to the modules envisaged for the
final calorimeter.


The detector slabs are built around an H-shaped supporting structure
incorporating a layer of tungsten absorber.  %
An active layer is placed on each side of this structure.  This active layer is
a chain of identical Active Sensor Units (ASUs), which consist of a printed 
circuit board (PCB)
integrating the Silicon sensors, Front-End electronics and electrical
infrastructure.  Each ASU can run as a standalone unit, allowing testing of each
piece before, during and after slab assembly, resulting in a high detector
yield, and thus a reduced cost.

Since the electronics are deeply embedded in the detector volume, and no space
is available for active cooling, their power consumption must be kept to a
minimum to prevent overheating.  By power-pulsing the electronics according to
the duty-cycle of the ILC machine, the consumption can be kept below 25\uW\ per
channel.


The sensors are based on high resistivity silicon (5\u{k\Omega/cm}) with 
individual pin-diodes of $5\times5$mm$^2$ size. This size 
is also feasible for the readout electronics.
%
A test batch of sensors based on 6'' wafers has been used by Hamamatsu to
produce $9\x9\u{cm^2}$ matrices.  The bonding of the sensors onto the PCB is
performed using a well controlled gluing technique.

The silicon sensors are built and integrated using well known, widely used and
well controlled technologies.  The matrix of PIN diodes is burned onto
330\u{\um} thick raw silicon wafers using standard manufacturing processes from
the microelectronics industry such as acceptor/ donator ion implantation, oxide
growth or metal deposit.  The bonding of the sensors onto the PCB is performed
using a well controlled gluing technique: standard glue (EPOTEK~410) applied by
a robotic gluing machine.
%
Prototypes of sensors have been ordered to various companies and academic
centres with two different sizes of PIN diodes. No problems due to the gluing
technique have been observed over a time span of several years.


\begin{figure}
  \centering
  \includegraphics[width=0.5\columnwidth]{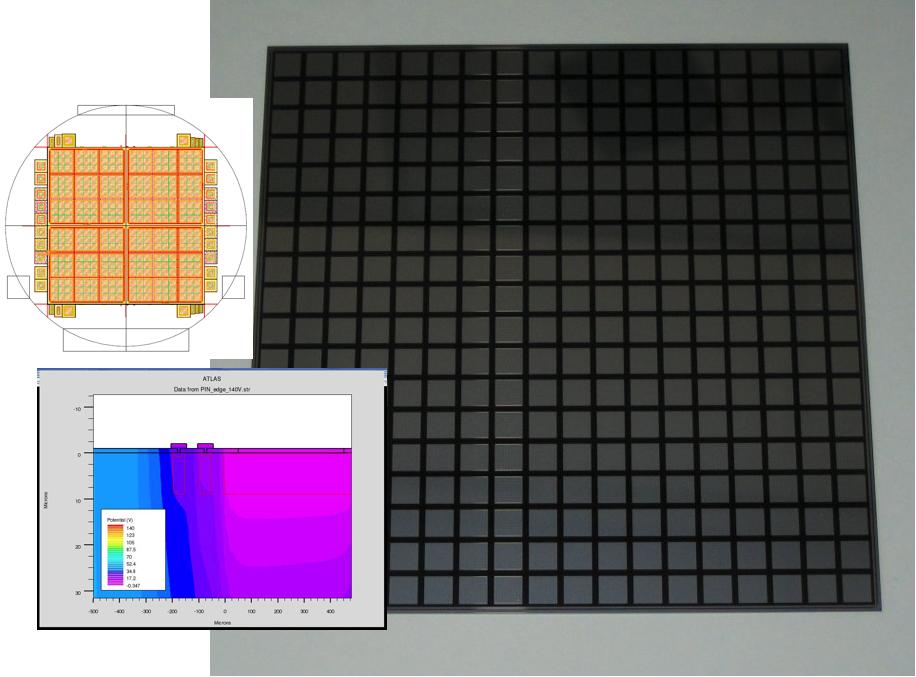}
  \caption[Si sensors for the ECAL.]{ %
    $5\x5\u{mm^2}$ pad Hamamatsu sensor (right).  %
    Layout of prototype sensors with optimised edges (upper left).  %
    In depth simulation of the potential near a guard ring (lower left). %
  }
  \label{fig:SiSensor}
  \vspace*{-10pt}
\end{figure}




The total surface of sensors for the whole ECAL is about $2500$\u{m^2}.  %
The sensors and their integration are kept as simple as possible to avoid any
dependence on a proprietary technique owned by a single manufacturer; allowing
for a variety of suppliers and manufacturers to share the production will
decrease the inherent financial risks and enable a competitive downscaling of
the costs.


\paragraph{Calibration}

The charge produced by a MIP in the silicon depends only on the silicon
thickness, and is therefore expected to be stable with time.  A single
calibration before detector assembly will therefore be sufficient.  The ASUs
will be calibrated in a muon beam before the assembly of detector slabs and
their integration into detector modules.

The VFE electronics will be calibrated by means of the VFE chips' charge
injection calibration system.

Since the tracks of muons and non-interacting charged pions in the ECAL can
easily be identified due to the ECAL's high granularity, they can be used to
monitor the calibration during the lifetime of the detector.

\subsubsection{Scintillator - Tungsten Electromagnetic Calorimeter}


The scintillator-tungsten sandwich ECAL (ScECAL) is proposed to realise a
fine-segmented calorimetry in a stable, robust and cost effective way.  %
The fine grained readout is realised by planes of $1$~cm wide and $4.5$~cm long strips, 
arranged in orthogonally in adjacent layers. 
Thanks to the strip structure, the number of necessary readout channels is
significantly reduced ($\sim 10^7$ channels) relative to the Si-W option.
Scintillator strips can be cheaply produced by the extrusion method. 
Compact photo-sensors (MPPC) and highly integrated readout electronics make dead
area in the ScECAL almost negligible.  Keeping the required granularity and
these merits, the ScECAL has good energy resolution and linearity.

\begin{figure}
  \centering
  \includegraphics[width=0.6\columnwidth]{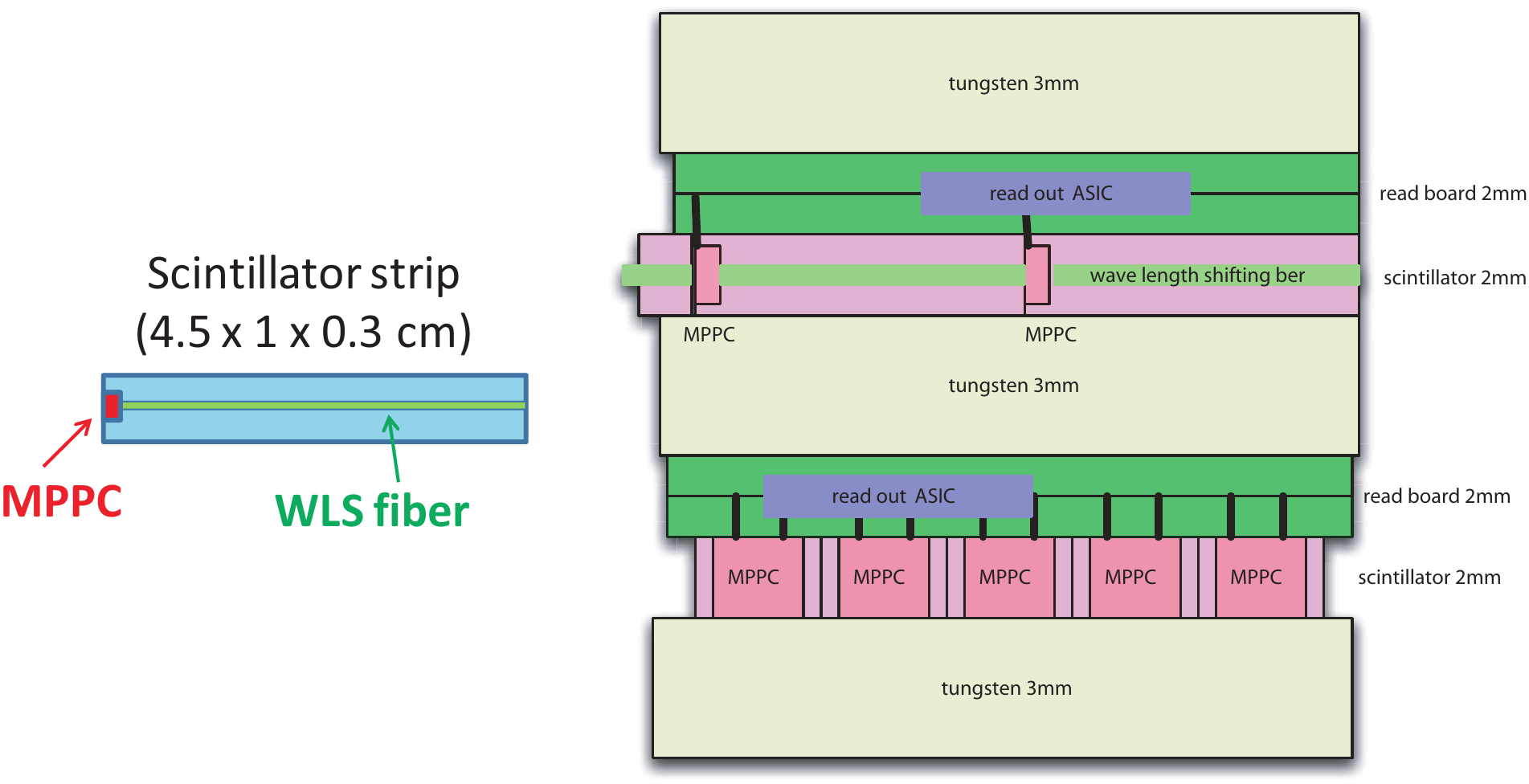}
  \caption[Layout of scitillator strips.]{Dimension of the scintillator-strip (left, view from top) and side
    view of the ScECAL layer structure (right).}
  \label{fig:ScECAL_layer}
\end{figure}

The ScECAL consists of 24 super-layers.  A schematic view of a few super-layers
of the ScECAL is shown in Figure~\ref{fig:ScECAL_layer}.  They will be mounted
in an alveolar structure similar to the case of the SiW ECAL.  %
A super-layer is made of a tungsten plate (3\u{mm} thick), scintillator strips
(2\u{mm} thick), and a readout/service layer (2\u{mm} thick).  Scintillator
strips in adjacent super-layers are arranged to be orthogonal aiming for better
effective granularity.  The thickness of a super-layer is 7\u{mm}.  The total
ScECAL thickness is 172\u{mm}, or 20.6\u{\X0} in radiation length.

\paragraph{The active layers}

The dimension of an individual scintillator strip (see
Fig.~\ref{fig:ScECAL_layer}) is $1\x4.5\x0.2\u{cm^3}$.  %
Although a strip width of 5\u{mm}, to realise an effective granularity of
$5\x5\u{mm^2}$, is thought to be feasible, further R\&D is necessary.  %
Each strip is covered by a mirror reflector film to improve collection
efficiency and uniformity of the scintillation light.  Photons from each
scintillator strip are read out via an $1\u{mm}$ diameter wavelength shifting
fibre (WLSF) embedded in a straight groove by a very compact photon sensor,
MPPC, attached at the end of the strip.

The MPPC is a version of a novel semiconductor photo-sensors
consisting of a matrix of micro APD pixels operated in Geiger mode.  Photo-detection performance and amplification power
is comparable with conventional photomultiplier tubes.  The dynamic range of an
MPPC is limited by the number of APD pixels.  A MPPC with 1600 APD pixels in an
area of 1\u{mm^2} is already commercially available.  However MPPC with
$\sim3000$ pixels should be developed to precisely measure up to $\sim
100\u{GeV}$ electromagnetic clusters.

\begin{figure}[ht]
  \begin{minipage}[t]{.47\textwidth}
    \centering
    \includegraphics[width=1.05\textwidth]{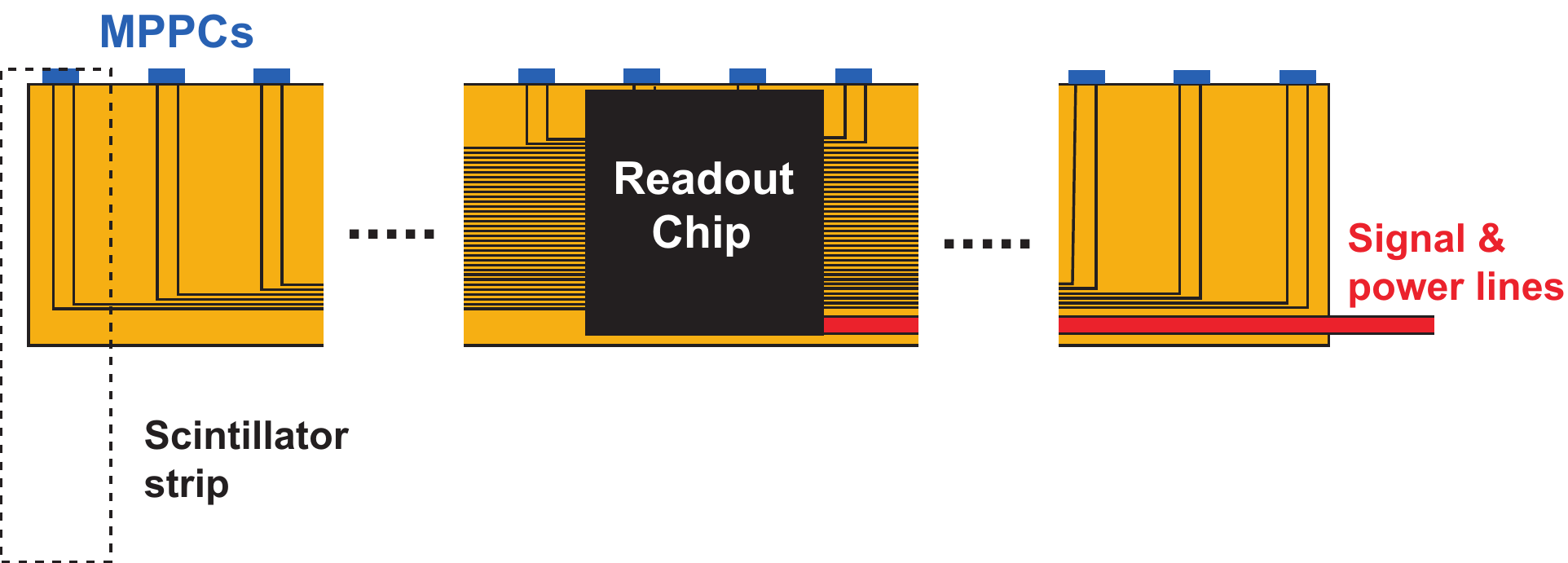}
    \caption{Layout of the MPPC, micro-strip line and readout chip on the FPC
      board.}
    \label{fig:ScECAL_FPClayout}
  \end{minipage}
  \hfill
  \begin{minipage}[t]{.47\textwidth}
    \centering
    \includegraphics[width=1.05\textwidth]{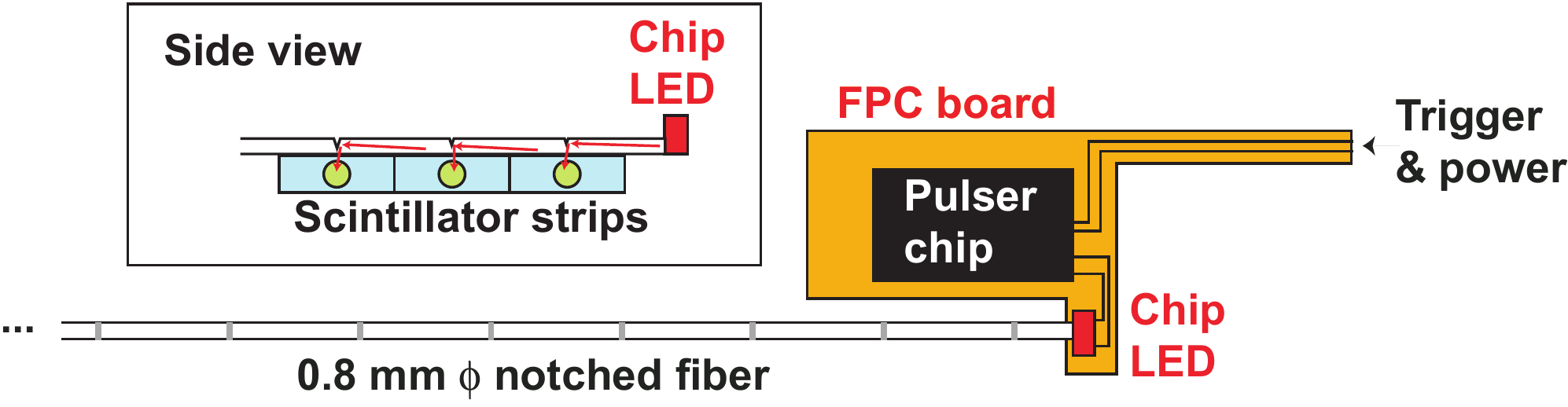}
    \caption[Calibration systen for the ScECAL.]{Schematics of the MPPC gain monitoring system with LED and notched
      fibre.  Light pulses from the LED are scattered and distributed into each
      strip.  }
    \label{fig:ScECAL_LEDsystem}
  \end{minipage}
\end{figure}

Signals from about 80 MPPC are fed into a readout chip through micro-strip lines.
They are arranged on one identical flexible readout board (FPC) (c.f.\
Fig.~\ref{fig:ScECAL_FPClayout}).  After shaping, digitisation and zero-suppression
of the analog signals on the chip, signals are taken out serially from the
detector and brought to a digitisation board by a thin FPC cable ($\sim 200\um$)
through detector gap.

\paragraph{Calibration systems}

A light distribution system has been designed to monitor possible gain drifts of MPPCs
by monitoring photo-electron peaks.  The system consists of a pulse generator, a
chip LED, and a notched fibre.  A schematic structure of the system is shown in
Figure~\ref{fig:ScECAL_LEDsystem}.  The pulse generator circuit and the chip LED
are arranged on a thin ($\sim 200\um$) FPC board.  The chip LED is directly
connected to the notched fibre to distribute lights to $\sim 80$ strips through
its notches.

Each scintillator strip can be calibrated with data by monitoring the MIP peak using
multi-hadron events at the ILC.  Monte Carlo simulation shows that more than 100 MIP
hits per strip will be obtained if running at the $\rm Z^0$.  With this method the
strips can be calibrated to better than 5\u{\%} with 1\u{fb^{-1}}
of ${\rm Z^0} \rightarrow jj$ events).

\paragraph{Status and Future R\&D plans}

The feasibility of the ScECAL has been proven by a test of a small prototype using
$1-32$\u{GeV} electron beams.  At the test clean MIP signal and electron energy
spectra are observed with negligible contamination from
electrical noise.  The energy resolution is measured to be $\sigma_E/E = 14 /
\sqrt{E} \oplus 2\u{\%}$ which is consistent with expectation from simulation.

In order to fully establish the feasibility of the ScECAL, further extensive R\&D
efforts are necessary to clarify the remaining technical issues as follows:
\begin{itemize}\addtolength{\itemsep}{-0.5\baselineskip}
\item Photon sensors: properties of the MPPC
  have to be further studied and improved.  The increase of the dynamic range is
  especially important.
\item Development of readout electronics: A highly integrated readout chip is
  needed due to the limited space in the detector.
\item Strip clustering: The strip structure is chosen in order to improve
  the effective granularity of the calorimeter.  A clustering algorithm has been
  developed which can cope with the strip structure as well as the usual
  square-tile structure.  The algorithm is being further improved, and
  performance of the strip structure must be demonstrated.
\end{itemize}

\subsubsection{Digital (MAPS) Silicon-Tungsten Electromagnetic Calorimeter}

The silicon-tungsten digital ECAL (DECAL) is an alternative to the analogue
silicon design described in Section~\ref{sec:SiW_ECAL}.
The basic principle is to replace the high resistivity pad diodes with CMOS based binary
readout pixels sufficiently small in size that, even in the core of high energy
electromagnetic showers where the density is typically equivalent to $\sim100\u{MIPs/mm^2}$, the probability of a pixel being hit by more than a single
particle will be low.  This allows the shower energy to be measured by the
number of binary pixels above threshold.  To ensure that linearity of response
is preserved even at higher energies, pixels are required to be
$O(50\x50\um^2)$, leading to $O(10^{12})$ pixels in the complete ECAL.  A very
high level of integration of the readout in the pixel is therefore mandatory. 


The active layers are based on CMOS Monolithic Active Pixel Sensors (MAPS)
which allow data
reduction and processing logic to be contained within each pixel. The target
noise level is $10^{-6}$. A new process (``INMAPS'' \cite{Ballin:2008db}, developed by the CALICE UK
groups) ensures efficient charge collection by using deep p-wells and charge collection by diffusion in the sensor. Signals (time stamp and pixel address) are stored on
the sensor during a bunch train and read out in the interval between bunch
trains. By using industry standard CMOS processes available from a large number of foundries, costs are potentially lower per unit area than analogue silicon diodes, with reduced risk to production schedules.

The performance of the DECAL has been studied using Mokka in the
context of LDC~\cite{Ballin:2007wu,Watson:2008zzd,Adloff:2007ai}, including
effects of dead area, digitisation and clustering.  A preliminary study of the
energy resolution of the DECAL to single photons, implemented by adapting only
the ECAL sensitive region in the ILD00 silicon-tungsten model in Mokka, gives
$\sigma_E/E=19.7\u{\%}/{\sqrt{E}}$. A first prototype sensor
(TPAC1.0) was designed in 0.18$\um$ process, having 28224 $(50\x50\um^2)$ pixels~\cite{Ballin:2008db}.  This
$9\x9\u{mm^2}$ sensor was fabricated and characterised, e.g.\
\cite{Stanitzki:2007zz,Crooks:2007zz,Ballin:2008zz} during 2007--8. A second revision of the sensor is 
expected for 2009. A proof-of-principle R\&D project is in progress to develop and test
  a 16 layer DECAL prototype large enough to contain electromagnetic showers~\cite{UKSPIDER} by 2012.

The DECAL option is designed to work with the same mechanical structure as the 
Si-W ECAL, thus profiting from the large R\&D done in this area. A topic for 
future R\&D
is the reduction and control of the power consumption, which at the moment is 
expected to be larger though uniformly distributed across the sensor unlike the analogue 
SiW sensor.

\subsection{The Hadronic Calorimeter}
\label{sec:HCAL}


In a particle flow calorimeter the HCAL plays a crucial role in separating and 
measuring the energy deposits of charged and neutral hadrons. Since the 
energy deposited by neutral hadrons fluctuates widely, its precise measurement 
is a key component of a well performing particle flow calorimeter. 
Consequently, the imaging capabilities of the HCAL are of prime importance and
demand high transverse and longitudinal segmentation and a design with a minimum
of uninstrumented (``dead'') regions.  However,
a very good hadronic energy resolution is also mandatory, both to assist the
topological assignment of clusters and tracks, and to optimise the precision of
the hadronic energy part characterised as neutral.
The high granularity allows the application of weighting techniques to
compensate for differences between hadronic and electromagnetic response and for
``invisible'' energy depositions (``software compensation'') and improves the
hadronic energy resolution further.

\subsubsection{Geometry and Mechanical Design}
\label{sec:HcalGeom}

The HCAL is conceived as a sampling calorimeter with steel as absorber and
scintillator tiles (analogue HCAL) or gaseous devices (digital HCAL) as active
medium.  As the HCAL must be located within the coil, the absorber has to be
non-magnetic.  Stainless Steel has been chosen both for mechanical and
calorimetric reasons.  Due to its rigidity, a self-supporting structure without
auxiliary supports (and thus dead regions) can be realised.  Moreover, in
contrast to heavier materials, iron with its moderate ratio of hadronic
interaction length ($\lambdaI = 17\u{cm}$) to electromagnetic radiation length
($\X0 = 1.8\u{cm}$) allows a fine longitudinal sampling in terms of \X0\ with a
reasonable number of layers in a given total hadronic absorption length, thus
keeping the detector volume and readout channel count small.  This fine sampling
is beneficial both for the measurement of the sizable electromagnetic energy
part in hadronic showers as for the topological resolution of shower
substructure, needed for particle separation and weighting.

\paragraph{Overall architecture}

The overall structure follows the ``short barrel'' concept, with
two large endcaps with about the same outer radius as the barrel. %
The total hadronic absorption length corresponds to a minimum of 5.5\u{\lambdaI}
in addition to the ECAL.  The endcaps are subdivided into four quadrants,
their absorber plates are oriented perpendicular to the beam line.
The mechanical engineering of the absorber structure has so far concentrated on
the barrel.
It is assumed that the solutions can be transferred to the endcaps later-on.
Compared with existing hadron calorimeters, the ILD HCAL has a rather fine
longitudinal sampling, with a correspondingly high pressure on the thickness of
the active layer gaps, but also on mechanical tolerances.  This, together with
the requirement of minimum dead zones represents a challenge to the large scale
engineering which is presently being addressed with prototypes within the
EUDET/CALICE framework.

For the barrel, two design approaches are being followed: one
with long barrel modules, subdivided only once in $z$, and with electronics and
service connections at the end faces, and a second,
with 5 rings and interfaces situated at the outer barrel perimeter.  The main
advantages of the first are the accessibility of the electronics and a maximum
filling of the detector volume limited by the coil radius, whereas the second
provides better rigidity in the transverse plane, eliminates pointing cracks and
allows for a tighter barrel end-cap transition.  %
In principle, each concept can be instrumented with both scintillator and
gaseous devices.  In practice, the detailed engineering is presently being
worked out for scintillator in the first, and for gaseous readout in the second
approach.

\paragraph{Design 1}
In the first version of the HCAL design, 
the barrel is subdivided into two sections in $z$ and eight octants in
$\varphi$, each octant has two halves which constitute the basic modules, 32 in
total.
%
\begin{figure}
\begin{tabular}{ll}
  \includegraphics[height=5.5cm]{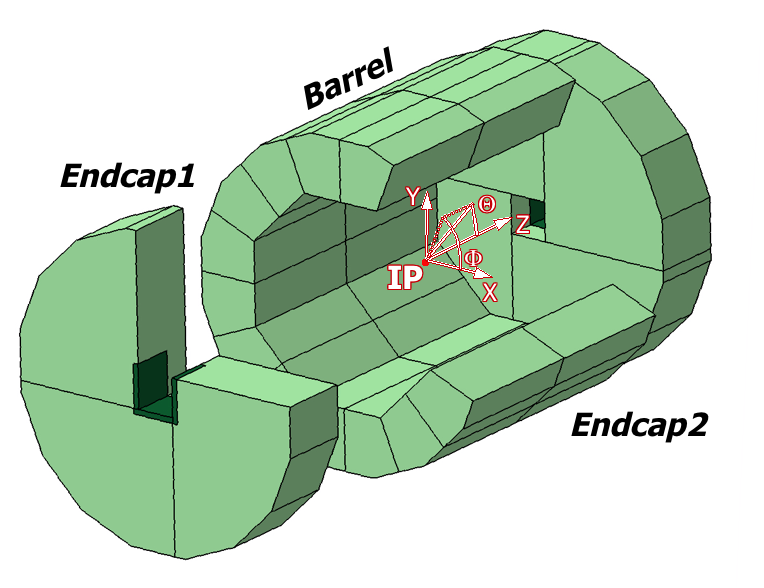}&
  \includegraphics[height=5.5cm]{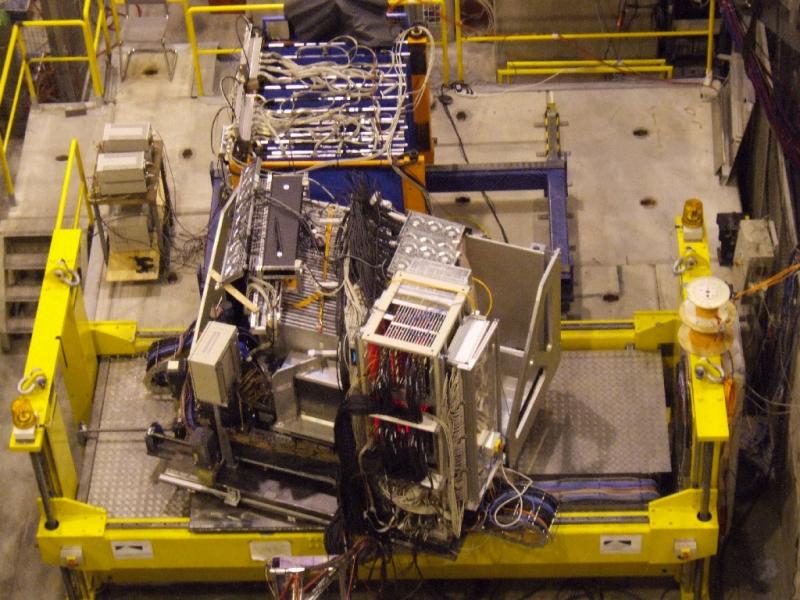}
\end{tabular}
  \caption[HCAL layout 1, view of the test beam experiment.]{Layout 1 of the HCAL (left), and view of the 
  integrated ECAL and HCAL beam test setup (right).}
  \label{fig:TeslaGeom}
\end{figure}
Each module has a weight of almost 20 tons, which is manageable with standard
installation techniques.  The modules are constructed independently of the
active layers, which can be inserted before or after installation of the modules.  %
There are 48 absorber plates, 16\u{mm} thick each, held together by 3\u{mm}
thick side panels in the $rz$ planes; no additional spacers are foreseen.
The active layers will contribute 4\u{mm} of steel to each absorption layer, and
require 5.5\u{mm} for instrumentation (3\u{mm} thick scintillator plus readout
and calibration devices). A drawing of the structure is shown in figure~\ref{fig:TeslaGeom}(left).
%
The HCAL structure has been extensively simulated using finite element methods,
including the integration of the heavy ECAL structure.  %
Maximum deformations are found to be less than 3\u{mm}, if the barrel structure
is supported by two rails in the cryostat.

Presently the boundaries between modules are pointing in $\varphi$ and in
$z$.  %
Variants with non-pointing boundaries have been validated in finite element
calculations as well, but are disfavoured to ease the mechanical
construction.  %
The pointing geometry does not degrade the performance as long as the cracks are
filled with absorber material,
and if the active instrumentation extends up to the boundary within tolerances,
which is the case in the present scintillator layer design.

\paragraph{Design 2}
This design intends to reduce cracks
both in $\varphi$ and $\theta$ and to reduce the distance between the barrel and
the endcaps.  The barrel part is made of 5 independent and self supporting
wheels along the beam axis which eliminates the $\theta$=90 degree crack.
%
\begin{figure}[ht]
  \centering
  \begin{tabular}[c]{cc}
    \includegraphics[height=5.5cm]{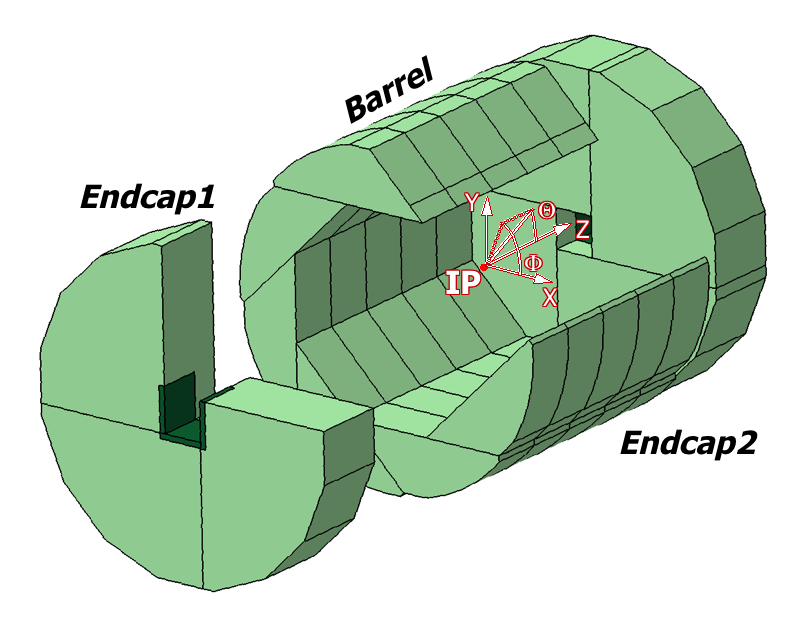} &
    \includegraphics[height=5.5cm]{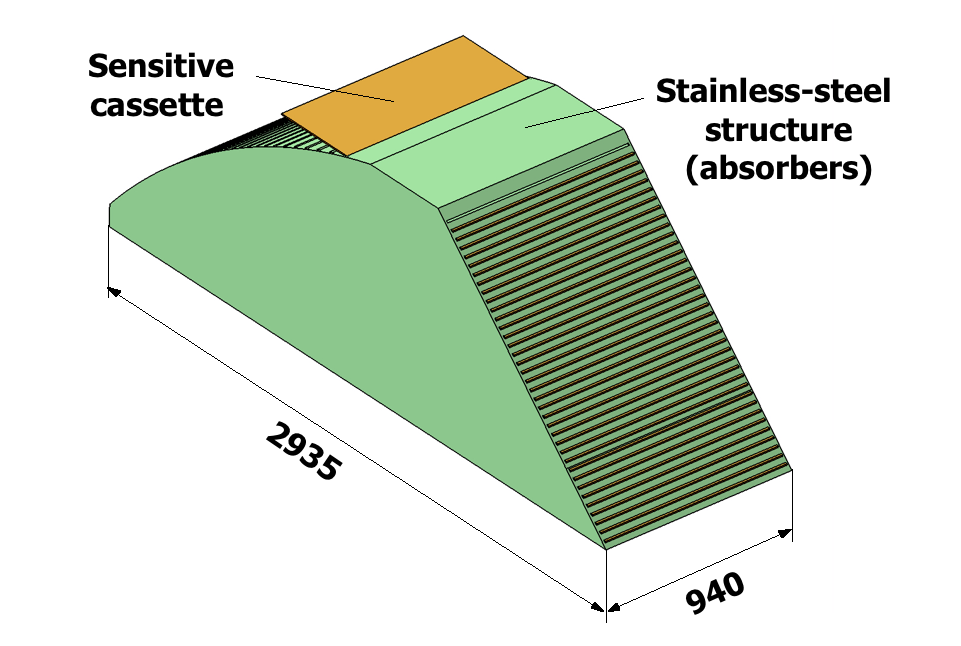}
 \end{tabular}
  \caption{Design 2 layout of the HCAL (left) and layout of one module (right).}
  \label{fig:DhcalGeom}
\end{figure}
The segmentation of each wheel in 8 identical modules is directly linked with
the segmentation of the ECAL barrel.
A module is made of 48 stainless steel absorber plates (welded with 2 transverse
10\u{mm} stainless steel plates) with independent readout cassettes inserted
between the plates.  %
They define the rigid structure on to which the corresponding ECAL modules are
mounted. A drawing of the structure is shown in figure~\ref{fig:DhcalGeom}(right).
%
The absorber plates consist of a total of 20\u{mm} stainless steel: 16\u{mm}
absorber from the welded structure and 4\u{mm} from the mechanical support of
the detector layer. %

Each wheel is independently supported by two rails on the inner wall of the
cryostat of the magnet coil.
The cables as well the cooling pipes will be routed outside the HCAL in the
space left between the outer side of the barrel HCAL and the inner side of the
cryostat.  %
%
The HCAL endcaps the same geometrical structure proposed in design~1.
The distance between the barrel and the endcaps, which have the same structure
as in design~1, is thus reduced, as only space to ensure inner detector cabling
is required.

\subsubsection{Analogue Hadronic Calorimeter}
\label{sec:AHCAL}

With the advent of novel, multi-pixel Geiger mode silicon photo-diodes,
so-called SiPMs, high granularities as required for a particle flow detector can
be realised with the well-established and robust scintillator technology at
reasonable cost.  The scintillator tiles provide both energy
and position measurement and thus allow to trade amplitude versus spatial
resolution.  The transverse segmentation suggested by simulations is about
$3\x3\u{cm^2}$ and leads to  a number of read-out
channels an order of magnitude smaller than in the digital case with $1\x1\u{cm^2}$ cells.

\paragraph{The Active Layers}

\begin{figure}[ht]
  \centering
    \includegraphics[width=0.7\textwidth]{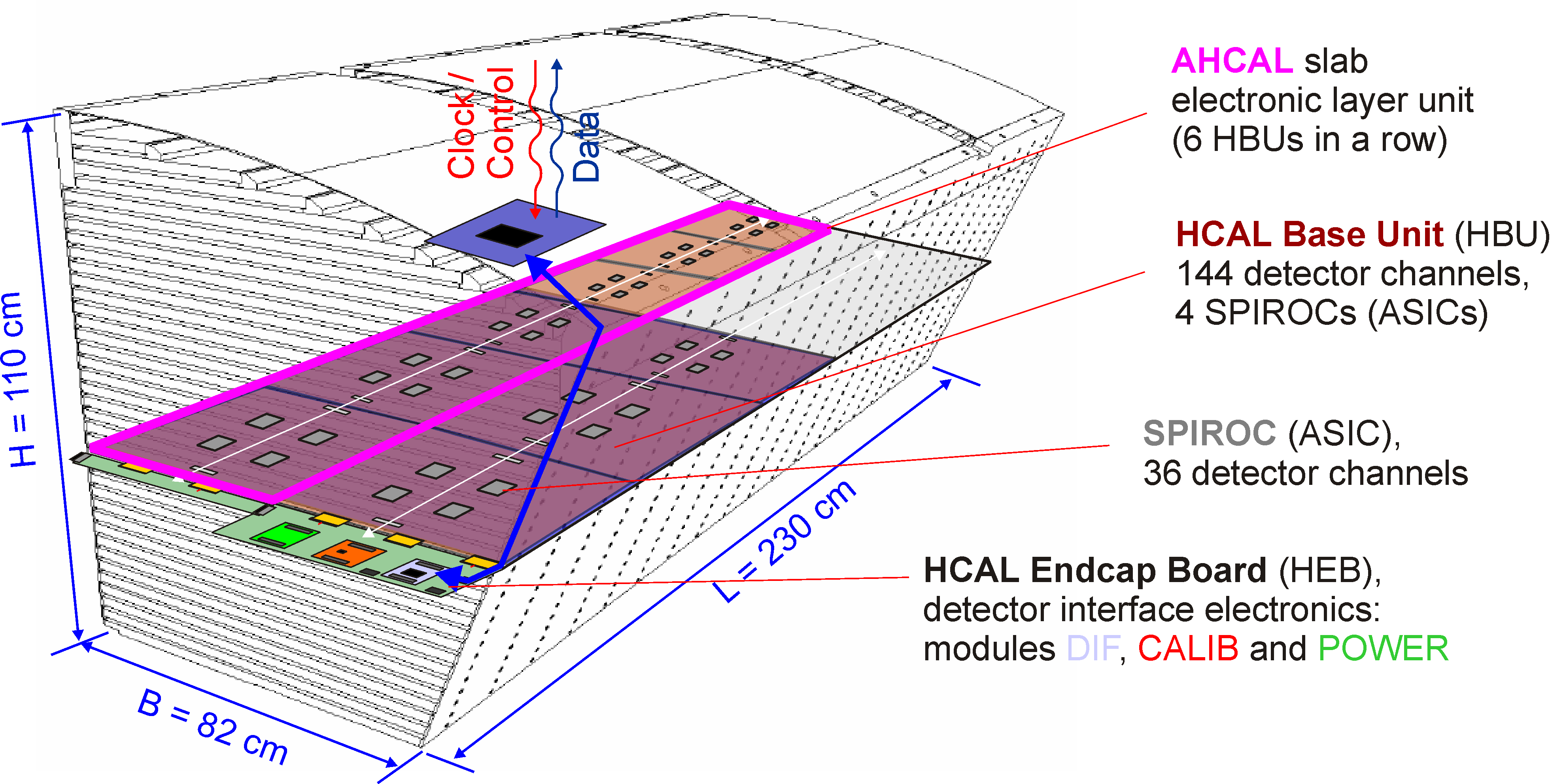}
    \includegraphics[width=0.9\textwidth]{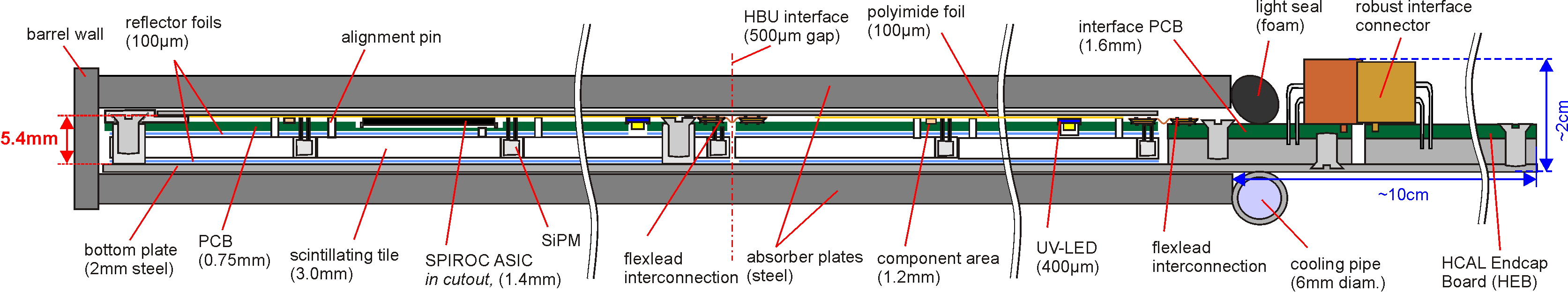}
  \caption[Details of AHCAL module design.]{Arrangement of AHCAL layers with electronic components (top),
cross section of an active layer (bottom).}
  \label{fig:AhcalInternal}
\end{figure}

The arrangement of the active layers with internal and external electronics
components is sketched in Figure~\ref{fig:AhcalInternal}.  The layer consists,
from bottom to top, of a 2\u{mm} thick steel support plate covered with
reflector foil, the scintillator tiles (3\u{mm}), the printed circuit board with
electronics components (2\u{mm}), covered with reflector foil from underneath,
and a polymide foil for insulation.
The PCB carries the SPIROC  readout ASICs (described in section~\ref{sec:VFE})and auxiliary components as well as an LED
based optical calibration system, whilst interfaces for data acquisition, clock
and control, for power distribution and for calibration system steering are
accessible at the end face.  Since the ASICs are operated in power-pulsed mode,
no cooling is needed inside the detector volume.
The PCB is subdivided into units (HCAL base units, HBUs) of smaller size,
manageable for automated mounting and soldering techniques.  The standard unit
is 12 by 12 tiles, $36\x36\u{cm^2}$, so six units are aligned along $z$ to fill
a half barrel.
In order to accommodate the variation in layer width with increasing radius, 4
different HBUs, 8 to 12 tiles wide, are needed.
At the layer edges, tiles with smaller size, e.g.\ $2\x3\u{cm^2}$, are placed
such that the uninstrumented region near the sector boundary is never larger than
5\u{mm} and
2.5\u{mm} on average.
%

The electronics at the end face will require cooling, mainly due to the use of
FPGAs in the DIF (Detector InterFace board described in~\ref{sec:DIF}).
The boards will extend 5 to 10\u{cm} in $z$, but occupy only a fraction of the
full width in $\varphi$, thus leaving space for ECAL and main tracker services
as well as for the TPC support along radial directions.  The required extra
separation between barrel and endcap is therefore much smaller.

\paragraph{Scintillators and Photo-Sensors, R\&D}

The successful operation of the 8000 channel CALICE HCAL test beam prototype
over several years has proven that the new sensor and scintillator technology is
robust and reliable.  Less than one per-mil of the SiPMs showed signs of aging
in form of increasing noise levels.  In the meantime, progress was made by
various manufacturers, e.g.\ in Russia or Japan, to provide sensors with lower
dark count rate and / or smaller inter-pixel cross-talk which allow to decrease
the noise occupancy above threshold of $10^{-3}$ in the present prototype by an
order of magnitude and thus fulfill the requirements from both physics (for
neutron hit identification) and DAQ band width.  The demands on dynamic range
are less critical than for the ECAL.

For the coupling of sensors to scintillator and PCB different approaches are
being followed, based on either wavelength-shifting WLS fibre mediated or direct
read-out with blue-sensitive photo-diodes.
The WLS option was successfully operated in the HCAL (and ECAL) testbeam
prototypes.
The production, test and integration of sensors has been industrialised further,
e.g.\ the grove for the fibre can be included in the injection moulding process
(or the hole in the extrusion process).  The positioning of the tiles must match
the precision of the PCBs, for example with
alignment pins.
Alternatively, so-called mega-tiles (plastic modules comprising several cells,
separated by groves) are also being discussed.

In the direct coupling case, the sensor is mounted in SMD style with its
sensitive surface in the PCB plane, and collects the scintillation light
directly from the tile.  The tile has to be shaped in a dedicated way to
compensate for the otherwise prohibitive light collection non-uniformities.
Verification of both concepts in beam tests are important; besides uniformity
also the stability of the light collection must be ensured.  %

Machine-related backgrounds are not a concern for the AHCAL.  %
Simulations have shown that only in the innermost regions of the end-caps,
the neutron fluence reaches levels which may degrade the visibility of single
photo-electron signals for SiPM monitoring, but not the MIP detection
capability.  One may have to revert to alternative monitoring strategies here,
or use more robust sensors which are under development.%

\paragraph{Calibration}

The calibration procedure has to relate the electronic readout signal to the
energy deposition in the cell.
For the pre-amplifiers and discriminators, charge injection is used as in the
ECAL or DHCAL case.
The gain of the photo-diodes
is monitored by means of an optical calibration system,
and adjusted via the bias voltage, by observing the spacing between single
photo-electron peaks in LED-induced pulse-height spectra.  Using test bench
measurements
this cares also for sensor efficiency variations, correlated with the gain.

We follow two approaches for the technical realisation of the LED system, one
based on a central driver located at the end faces of the modules and optical
light distribution via fibres, and one with electrical signal distribution and
surface-mounted LEDs for each tile.
%
To check for long-term effects, track segments in hadronic showers can be used
for a large fraction of the calorimeter volume.  This has been shown with test
beam data and simulated for ILD multi-jet events.  %
Also systems based on radio-active sources might need to be considered.

\paragraph{Optimisation and Performance}

The main cost- and performance driving parameters of the AHCAL are the depth and the longitudinal
and transverse segmentation.  These parameters have been varied, and their
current settings have been found, using detailed simulations and particle flow
reconstruction as described in the overall detector optimisation section \ref{sec:optimization-particleflow}.  The
simulations include a modeling of inactive regions at module boundaries which
is more conservative than the present engineering design.
%

\begin{figure}[ht]
  \centering
  \begin{tabular}[c]{cc}
    \includegraphics[width=0.49\textwidth]{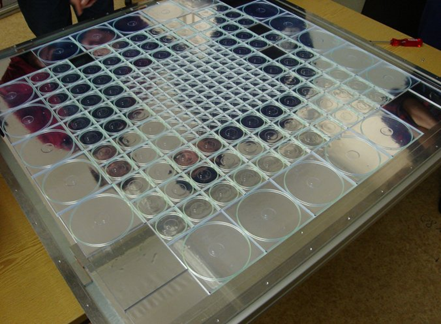} &
    \includegraphics[width=0.49\textwidth]{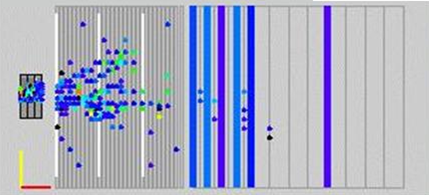}
  \end{tabular}
  \caption{AHCAL physics prototype layer (left), event display (right).}
  \label{fig:AHCALproto}
\end{figure}

The performance of a scintillator-tile HCAL with SiPM read-out and the proposed
segmentation has been demonstrated with test beam data taken with the CALICE
physics prototype.
The detector showed very good imaging capabilities which reveal the substructure
in hadronic showers, see Fig.~\ref{fig:AHCALproto}. %

\begin{figure}[ht]
  \centering
  \begin{tabular}[c]{ll}
    \includegraphics[width=0.45\textwidth]{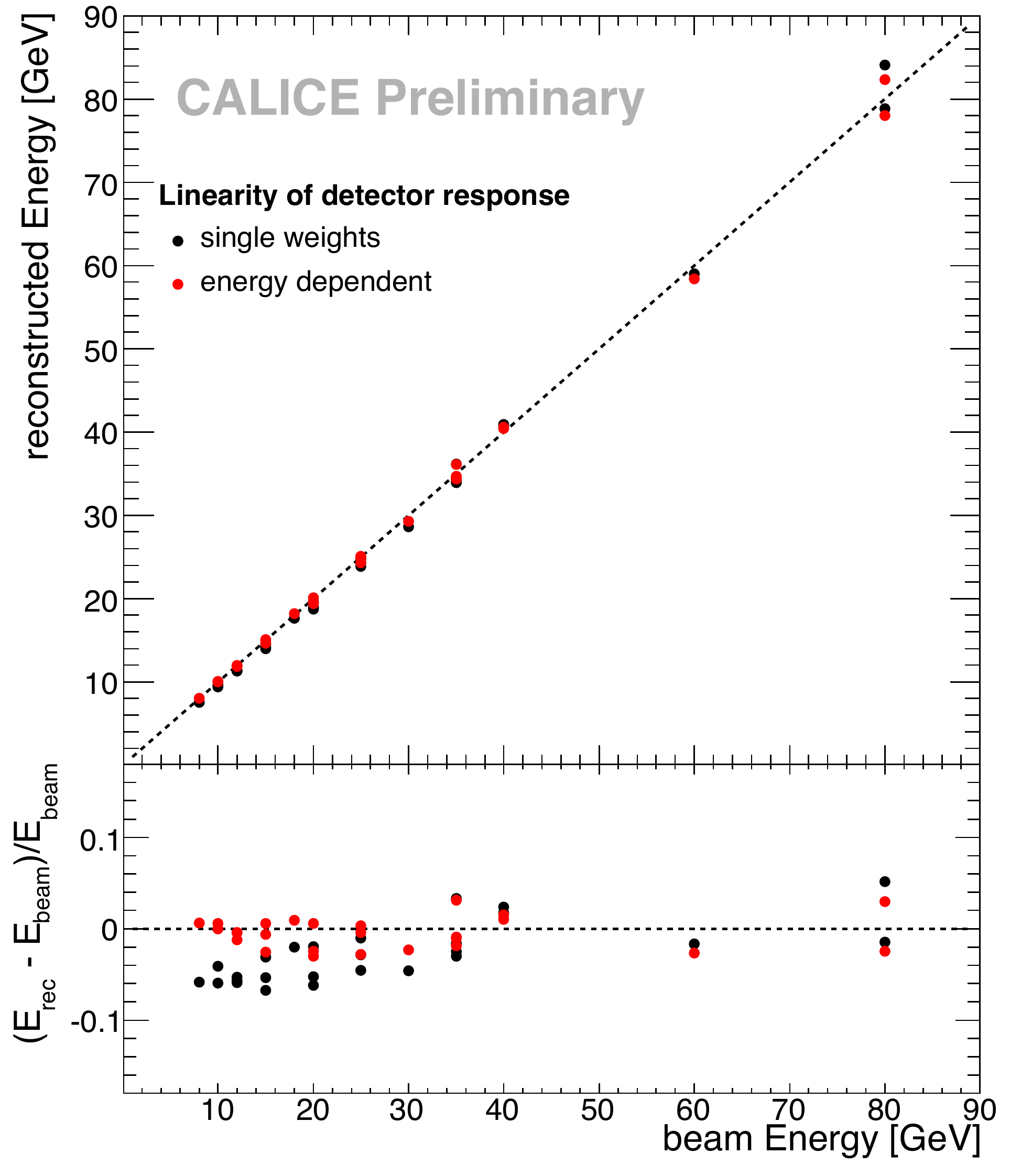} &
    \includegraphics[width=0.49\textwidth]{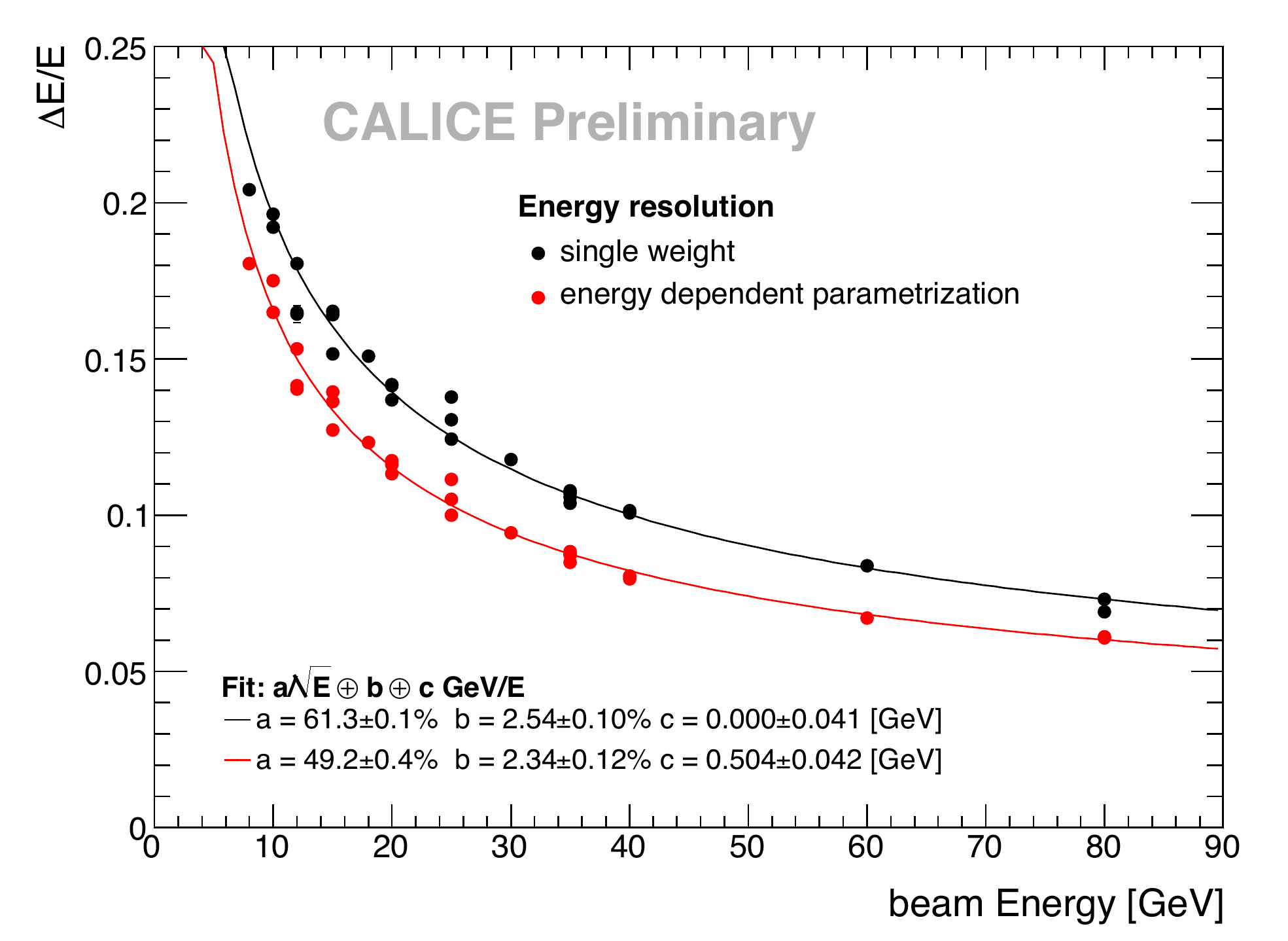}
  \end{tabular}
  \caption[Linearity and resolution for AHCAL.]{Linearity (left) and resolution (right), on electromagnetic scale and
    after weighting.}
  \label{fig:AHCALlinres}
\end{figure}

Using test bench data and in-situ measurements, temperature induced variations
and SiPM saturation effects could be corrected, and a linearity of better than
2\u{\%} for electron induced showers up to 50\u{GeV} was achieved.  %
The calorimeter is non-compensating, but the $\mathrm{e}/\pi$ ratio is not large
and the observed linearity is also good for hadronic showers, see
\hbox{Fig.~\ref{fig:AHCALlinres}}.  %
A hadronic energy resolution of $61\u{\%}/\sqrt{E}$ is obtained on the
electromagnetic scale, which can be reduced to $49\u{\%}/ \sqrt{E}$, preserving
linearity, with a simple weighting algorithm, which takes only the energy per
tile, but not yet any shower substructure into account. %

\begin{figure}[ht]
  \centering
  \includegraphics[width=0.49\textwidth]{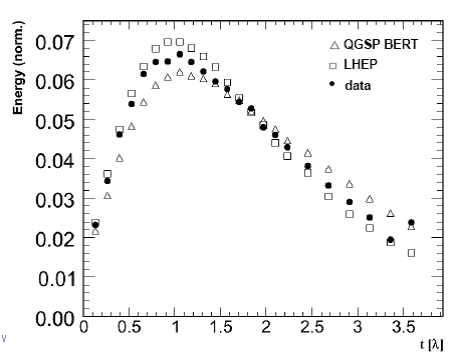} 
  \caption{Longitudinal shower profile, test beam data and simulations.}
  \label{fig:AHCALprofile}
\end{figure}

Based on experimental results from the CALICE prototype a reasonable agreement 
of the shower profiles with GEANT~4 based
simulations has been found, as shown in figure~\ref{fig:AHCALprofile}.  We found that it is essential to
model details of the detector response, such as saturation effects in the
scintillator according to Birks' law and the shaping time of the readout
electronics, in order to reach this good agreement.  These  primarily affect the
response to neutrons which would otherwise be overestimated.

\subsubsection{Semi-Digital Hadronic Gas Calorimeter}
\writer{Imad Laktineh}


The capacity to apply successfully the particle flow algorithms can be enhanced
by increasing the granularity of the different ILD sub-detectors.  In the
hadronic calorimeter this will doubtlessly help reduce the confusion between
charged and neutral hadronic particles by providing a better separation of the
associated showers.  However, the cost related to such an increase in detector
segmentation should be minimised.  To satisfy both, a gas hadronic calorimeter
with a semi-digital readout is proposed.  The study of such an HCAL has been going on
for few years in order to validate this option.

\begin{figure}
  \centering
  \includegraphics[height=3.5cm]{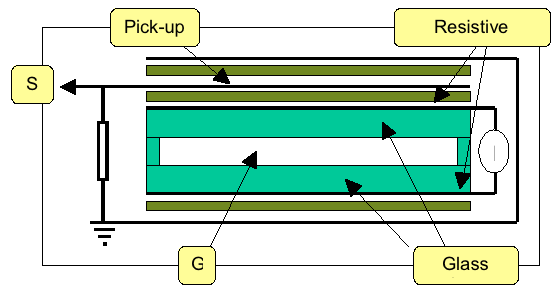}
  \caption{Single gap GRPC scheme}
  \label{fig:GRPC}
\end{figure}

The choice of gaseous detectors as the sensitive medium in the HCAL offers the
possibility to have very fine segmentation while providing high detection
efficiency.  The glass resistive plate chamber (GRPC) is one of these detectors which can
be built in large quantities at low cost.  Large GRPCs as the ones required for the
ILD HCAL can be easily produced.  This is an important advantage with respect to
other detectors since it guarantees very good homogeneity.  Several experiments
like BELLE have been using such large detectors with success for years.
However, the GRPCs to be used in the ILD HCAL need to be more elaborate.  As
the HCAL is situated inside the magnet coil, the sensitive medium thickness is an
important issue.  Very thin GRPCs are requested and 3.3\u{mm} thick GRPCs were
indeed produced and successfully tested.  In figure~\ref{fig:GRPC} a scheme of
such a single gap GRPC is shown.  %
Some key properties of these detectors are:
\begin{itemize}\addtolength{\topsep}{-0.5\baselineskip}\addtolength{\itemsep}{-0.7\baselineskip}
\item GRPC operated in avalanche mode and have been shown to show no ageing for the accumulated charge expected over the ILC running period. 
\item Test beam performed at DESY have shown that a strong magnetic field has
  negligible effect on GRPC performance.
\item GRPC detectors are insensitive to slow neutrons preventing thus an
  additional confusion.
\end{itemize}

Increasing the granularity will lead to a large number of channels. 
To limit the amount of data we propose a semi-digital readout
solution.  This simplifies the data treatment while minimising the consequences
on the energy resolution performance.  Indeed, based on several independent simulation
studies, a two-bit readout would provide better energy resolution in the
low-energy jet range (1--20\u{GeV}) and a comparable one at higher
energies when compared to an analogue readout~\cite{Matsunaga:2007zz}.

\begin{figure}[htb]
  \centering
  \begin{tabular}[b]{cc}
  \includegraphics[width=0.45\columnwidth]{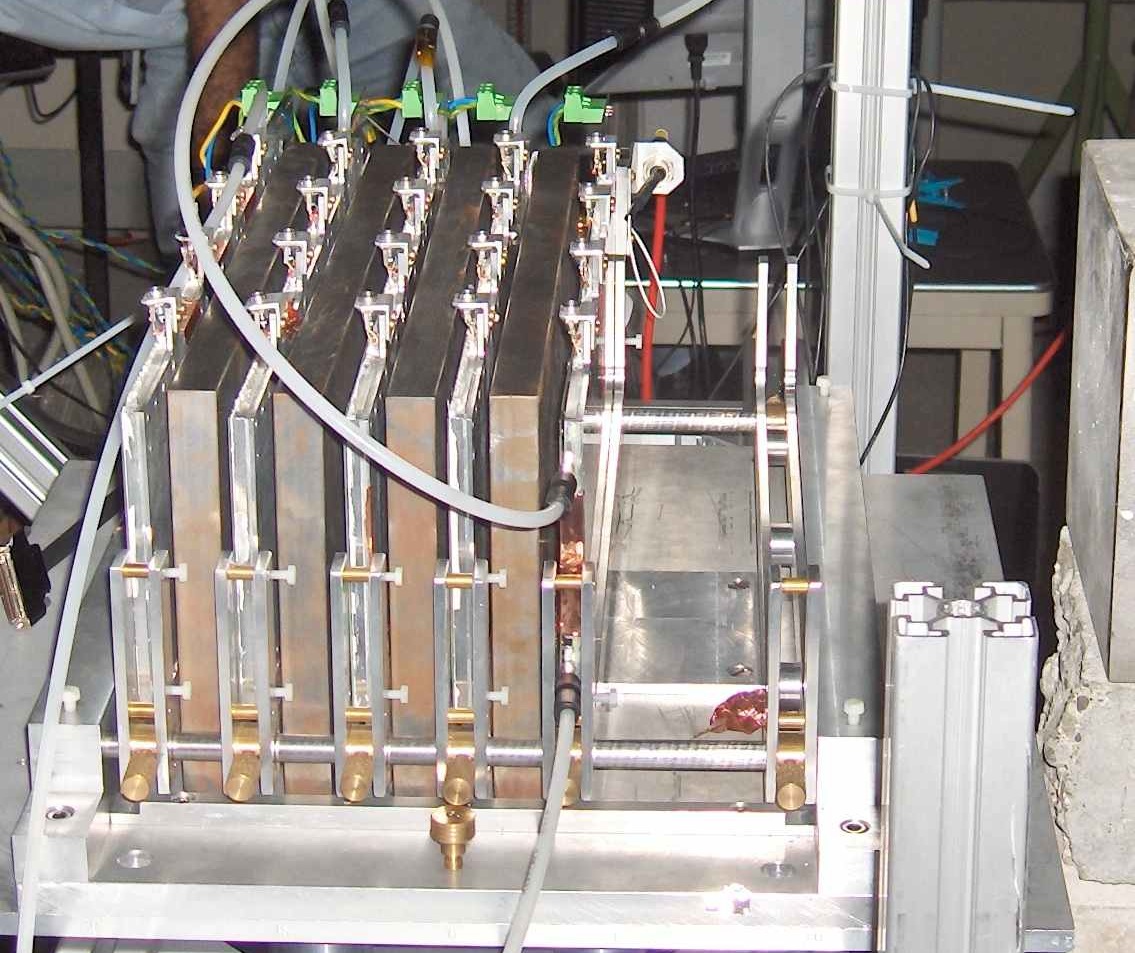} &
  \includegraphics[width=0.49\columnwidth]{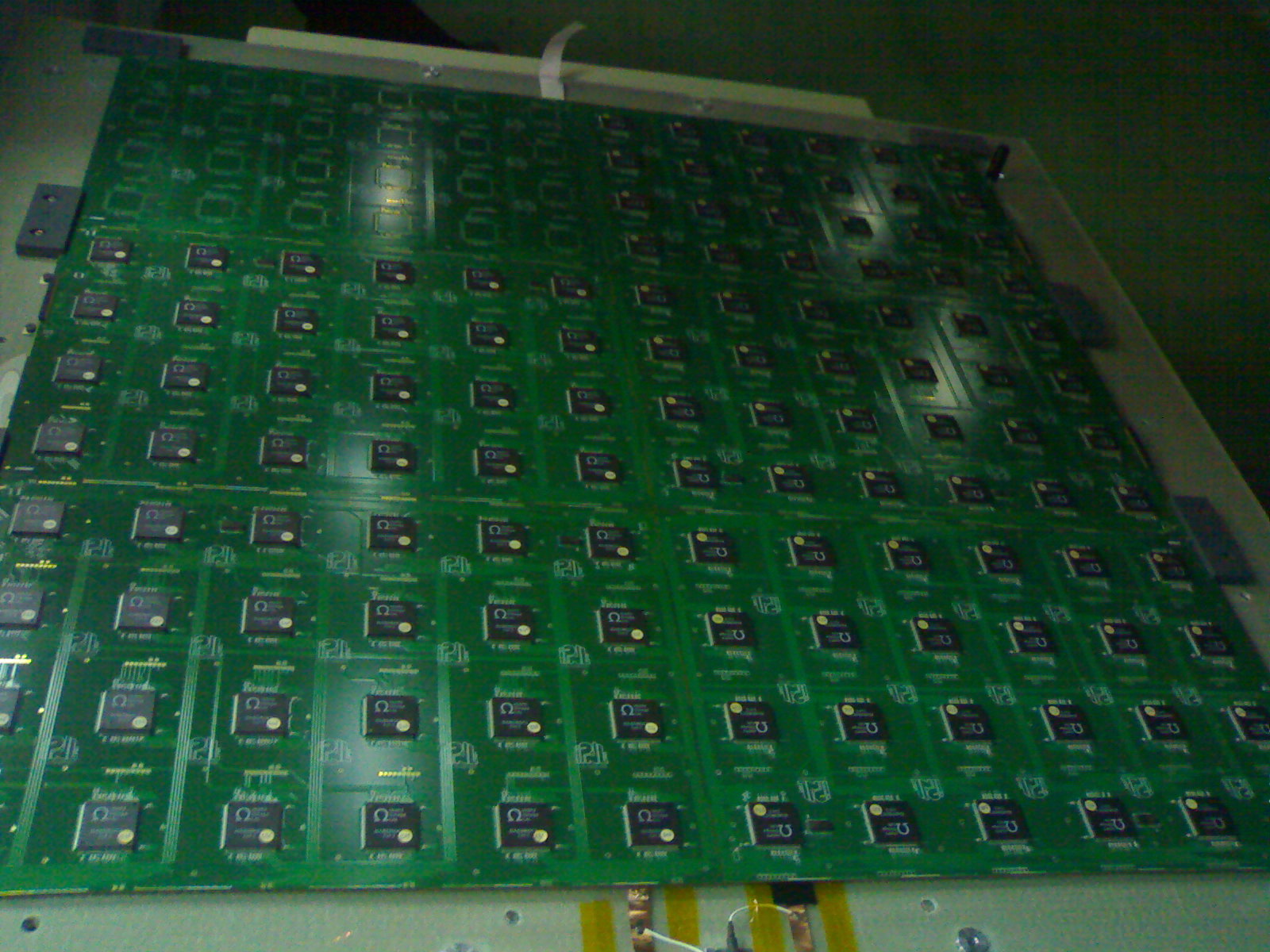}
  \end{tabular}
  \caption[Mini-DHCAL prototype.]{Mini-DHCAL prototype (left); Prototype of a large instrumented GRPC (right).}
  \label{fig:mDHCAL}
\end{figure}

Similiar to the case of the analogue HCAL the readout electronics will be 
integrated into the sensitive layer of the system, thus minimising dead 
areas. Large electronics boards are assembled together to form extra
large boards before they are attached to the GRPCs.  The board assembly will be
made possible thanks to a mechanical structure made of 4\u{mm} stainless steel plate.
In addition, to keep the HCAL as compact as possible, the fully equipped
electronic boards are designed to have less than 3\u{mm} thickness in all.
A mini hadronic calorimeter using this concept was built and
successfully tested in beam conditions at CERN in 2008 (see figure~\ref{fig:mDHCAL}).

\paragraph{The Active Layers}

R\&D activities on large GRPC detectors are being followed.  Different kinds of
spacers are tested to reduce detector noise and inefficiency while increasing
the detector robustness.  New gas distribution schemes as well as gas recycling
systems are worked out to lower gas consumption and pollution.  Although the
present GRPC detection rate of 100\u{Hz/cm^2} obtained with efficiencies greater
than 90\u{\%} is enough for the needs of ILC, a new development based on using
semi-conductive glass will lead to increase this rate. %
Multi-gap GRPCs are also investigated.  This allows reducing the spread of the
MIP charge spectrum leading to a better exploitation of the semi-digital
information.

Few large GRPCs with different options were built and are currently tested using
a 1\u{m^2} fully equipped electronics board (see figure~\ref{fig:mDHCAL}(right)).  %
This will allow to build the most appropriate GRPC detector to be used in the
ILD DHCAL.

In addition to the GRPCs activities, development on other thin and large gaseous
detectors like GEM and MICROMEGAS are also followed.

The GRPCs produces strong electric currents (a few 10\u{pC} in 10-20\u{ns}) in
the DHCAL pads.  In order to reduce cross-talk effects between the pads below
the percent level the very front end electronics is located on the other side of
the PCBs, (semi-)buried vias are used.

\paragraph{Energy Reconstruction \& Calibration}
\label{sec:DhcalCalib}

The semi-digital HCAL cell energy reconstruction can, to first order,
be estimated as $E_\mathrm{cell} = 1$, 5 and 10\u{MIP} if the charge is above
the thresholds typically placed at 0.1, 2 and 8\u{MIPs} (for the envisaged GRPC
about 0.26, 5.2 and 20.8\u{pC}).  %
Preliminary results on simulation, without algorithm optimisation, show PFA
performances comparable to the AHCAL reconstruction.

An interesting aspect of the gaseous semi-digital HCAL is the simplicity with
which the detector calibration is performed, if one is needed at all.  The sDHCAL
energy calibration requires 3 independent steps:
\begin{itemize}\addtolength{\itemsep}{-0.5\baselineskip}
\item \textit{An intercalibration of the ASIC thresholds in charge}: All 
  ASICs will have to be tested and calibrated by injecting a precisely
  controlled charge, adapted for each of the threshold, at the entrance of their
  final ASU/PCB pad.  %
  The variations can be compensated channel by channel in the ASIC by adjusting
  the channel gains (over a range of 0--2 coded on 8 bits in the current version
  of HaRDROC, described in sec~\ref{sec:VFE});
\item \textit{A calibration of the multiplicity of the RPC}: The multiplicity
  response curve of the RPC to muons as a function of high voltage applied,
  thresholds, position and gas flow and atmospheric pressure can be measured on
  a cosmic test bench or muon beam and parametrised for each type of RPC.
\item \textit{A calibration with physics}: The two first steps bring an absolute
  calibration at the level of the MIP, which can be cross-checked with cosmic
  muons or $Z \rightarrow \mu\mu$ events; the final energy scale will be a
  complex interplay in the scope of the PFA analysis between the clustering
  algorithms, jet and particle energies and types.
\end{itemize}
The definition of the calibration procedure, and an estimation of the
achievable precision, is a part of the DHCAL 1\u{m^3} programme.

\paragraph{Status and Future R\&D Plans}

A technological prototype of 1\u{m^3} HCAL based on the same principle is
currently under study.  It aims to validate at large scale the semi-digital HCAL
concept.  Questions related to the mechanical structure mentioned in the
previous section as well as the management of the limited space for services
will be addressed.  The prototype is to be built by 2010.  Combined test beams
with the different ECAL prototypes developed within the CALICE collaboration
will then be organised at FERMILAB and CERN.

\subsection{Calorimeter Readout System}

A considerable effort has been made in the framework of the CALICE
collaboration to standardise the read-out of different type of calorimeter with
embedded Very Front-End (VFE) electronics while minimising the space needed for
the configuration distribution and the data readout.  %

\subsubsection{Very Front End (VFE) ASIC description}
\label{sec:VFE}

The front-end ASICs should ensure a data format uniformity in all
the calorimeters, thus having identical back-ends to allow a
standardised detector interface board (DIF) for all detectors.

Ensuring such a compatibility between all electronics components involves a
unique read-out system based on token ring that allows a number
of ASICs to be read out by one output line, using the same protocol.
That protocol will help reducing the number of data lines outputted
from the calorimeters where the front-end ASICs are now embedded.

All the VFE will feature three operating mode : Acquisition (1\u{ms}), A/D
conversion (1\u{ms}), and data outputting during inter-bunch (199\u{ms}) using an
ultra low power protocol.  When a FE ASIC is in neither of the above modes, it is
turned to an idle mode to save up to 99.5\u{\%} of power, bringing the power
down to 10 to 25\uW\ per channel.  %

Three ASICs differing mostly on their analog front-end have been developed to
fit the different detectors.
\begin {itemize}\addtolength{\itemsep}{-0.5\baselineskip}
\item {SKIROC} ("Silicon Kalorimeter Integrated Read-Out Chip"): 64 channels
  charge preamplifier for charge measurement down to the MIP ( 3.84\u{fC}) to a
  maximum around 2500\u{MIP}. Dual gain shaping, analog memory, 12
  bit-digitisation, self-trigger capability on single MIP. 25\uW/ per channel to
  run without any active cooling ensuring therefore an extreme compactness of
  the calorimeter.
\item {SPIROC} ("Si-Pm Integrated Read-Out Chip"): auto-triggered, dual-gain voltage
  preamp, 36-channel ASIC which allows to measure for each channel the charge
  from 1 to 2000 photo-electrons with a 12 bit internal ADC and the time with a
  1\u{ns} accurate TDC.  One 8-bit 5\u{V} input DAC per channel ensures
  operation of the SiPM at its optimum bias.
\item {HARDROC}("HArdronic RPC Detector ReadOut Chip"): 24 channels semi-digital
  readout for RPCs or MicroMegas pads, allowing both good tracking and coarse
  energy measurement.  Each channel made of a variable gain low input impedance
  current preamp followed by 3 variable gain shapers and 3 low offset
  discriminators to auto-trig down to 10\u{fC} up to 10\u{pC}.  A 128 deep
  digital memory to store the encoded outputs of the discriminators as well as
  the bunch crossing identification.
\end {itemize}

Prototypes of each type have been produced in the years 2007-2008.  Boards equipped with 4
HARDROC(v1) ASICs have been designed for the DHCAL.  The
electronics readout under beam conditions has been validated.  Some key points such as the digital
daisy chaining for configuration and readout, the stability, the efficiency, and
the capability of the chip to be used without any external components have been
checked.
A small production is foreseen in Fall 2009 to equip a technological prototype
(called EUDET prototype) in 2010.

\subsubsection{Detector Interface}
\label{sec:DIF}

The ASICs are managed by specifically designed DIF (Detector InterFace) cards;
one DIF handles a full slab, whose maximum size are of $260\x141\u{cm^2}$ for
the AHCAL structure,
and $90 \x 273 \u{cm^2}$ for the DHCAL structures.
The corresponding maximum number of ASIC per slab are respectively $576$ and
$420$.  %
For an estimated occupancy per ASIC of the HCAL of 5 events / train of
2600\u{BC}, the expected data volume to be read in the inter-train is
of 336000\u{bits}; readout at a speed of 5\u{MHz} this takes
$67\u{ms}$.
During the readout phase the ASICs will be on standby except when explicitly
addressed.

\subsubsection{DAQ system}
\label{sec:calo_daq}
The data acquisition (DAQ) system is defined to start with the detector
interface boards (DIF) which service the detector slabs from the ends and which are
specific to the VFE of the subsystem.  The DIF provides a generic interface,
independent of the calorimeter type, to the DAQ system.  Because of the limited
space available for cabling and services, data are concentrated onto a single
optical communication channel with the off-detector electronics, by a link-data
aggregator (LDA) inside the detector.  The resulting data volume is mainly
determined by the zero-suppression scheme incorporated in the self-triggering
Front-End electronics.  But as the calorimeter has over 100 million readout
channels, significant demands are put on the scalability and on an attractive
price/performance ratio of the read-out electronics and the associated
data-acquisition systems.  Therefore the design should minimise the number of LDAs and
maximise the data rates on the link which are expected to be 10\u{GB} Ethernet
links.

With the data delivered over optical high-speed links, an optical switch is used
to dynamically redirect the data streams coming from the detector towards
available data receivers of the off-detector system.  The off-detector is
currently realised as a PCIexpress card hosted in a commodity PC but can be
easily implemented in a $\hbox{$\upmu$}$TCA crate for the future detector.

For the event building the machine clock will be fed into the off-detector
system to the data concentrators and the detector interfaces.  The requirements
on the clock are a low jitter and fixed latency between the machine clock and
the clock in the detector interfaces.  This part of the system needs to be
custom built to guarantee delivery times and latencies.

For all of the introduced systems (detector interface, data concentrator, off
detector system and the clock) prototypes already exist which have been built
within the EUDET~\cite{EUDET_DAQ} project.  The prototypes perform the same
tasks as in the final detector, however the prototypes are build for a proof of
principle and need to be optimised for the final detector design.

\subsection{Status and future R\&D plans}

The technology-specific R\&D will continue in CALICE.
However, in order to study the system performance of the proposed solutions, to
validate the accuracy of simulations, and to develop the reconstruction
techniques further, large test beam experiments are indispensable.  The particle
flow approach demands integrated set-ups with ECAL and HCAL together.  This
programme is pursued for the calorimeters in the framework of the CALICE
collaboration, in a cooperative way maximising the use of scarce resources with
common mechanical structures, electronics components, DAQ systems and software
frameworks.

CALICE has completed a series of full-size proof-of-principle tests with
so-called physics prototypes of both ECAL technologies and the scintillator
HCAL.  The programme is to be completed by tests with a RPC-based DHCAL at
Fermilab in 2009-10, and by a test of a MAPS-based DECAL prototype until 2012.  
Large data sets have been and will be collected and form
the basis for test and refinement of hadron shower simulations.  The emphasis in
the more realistic second generation ``technological prototypes'' is shifted
more toward a demonstration of the feasibility of a compact integrated detector
design fulfilling the ambitious demands of compactness and hermeticity.
Operational challenges not yet addressed with physics prototypes are
the power-pulsed front end electronics and the on-detector zero-suppression in
auto-triggered mode, which requires continuous and precise on-line control of
thresholds.  Therefore also technological prototypes need to undergo full-size
beam tests.  While the focus is rather on calibration and stability than on
shower physics, beam campaigns of several weeks in combined set-ups are foreseen
to start in 2010.

%% file: ild/fcal/fcal.tex
\section{Forward Detectors}

Special calorimeters are foreseen in the very forward region of the ILD
near the interaction point - LumiCal for the precise measurement
of the luminosity and BeamCal 
for the fast estimate of the luminosity~\cite{ieee1}. 
The LHCal will extend the coverage of the HCAL endcap to smaller polar angles. 
Together they will improve the hermeticity of the detector. 

A third calorimeter, GamCal, 
about 100~m downstream of the detector, will assist in beam-tuning.
Also for beam-tuning a pair monitor is foreseen, positioned just 
in front of BeamCal.

LumiCal will measure the luminosity using Bhabha 
scattering, $e^+e^- \rightarrow e^+e^-(\gamma)$ as gauge 
process .
To match the physics benchmarks, an accuracy of better than
10$^{-3}$ is needed\footnote{For the GigaZ option an accuracy
of 10$^{-4}$ is the goal~\cite{klaus}.}. Hence, LumiCal is a precision device
with challenging requirements on the mechanics 
and position control.

BeamCal is positioned
just outside the beam pipe, in front of the final focussing quadupoles. A large amount of low energy 
electron-positron pairs 
originating from beamstrahlung will deposit their energy
in BeamCal. These deposits, useful for a 
bunch-by-bunch luminosity estimate and the determination of 
beam parameters~\cite{grah1}, 
will lead, however, to a 
radiation dose of several MGy per year in the sensors
at lower polar angles.
Hence extremely radiation hard sensors are needed to instrument 
BeamCal. 

A pair monitor, consisting of a layer of pixel sensors positioned 
just in front of BeamCal, will measure the distribution of beamstrahlung pairs and
give additional information for beam parameter determination.

These detectors in the very forward region have to tackle relatively high 
occupancies, requiring special FE electronics and data transfer equipment. A small Moli\`{e}re radius is of invaluable importance for BeamCal
and LumiCal. 
It ensures an excellent
electron veto capability for BeamCal even
at small polar angles, being essential to suppress
background in new particle searches where the signatures 
are large missing energy and momentum.
In LumiCal, the precise reconstruction of electron and positron 
showers of Bhabha events
is facilitated and background processes will be rejected efficiently.

LHCal will be a hadron calorimeter extending the coverage of the 
HCAL endcaps to small polar angles. It will allow a fair hadron
shower measurement in the polar angle range of LumiCal 
and enhance the particle identification capabilities.

\subsection{The Design of the Very Forward Region} 
\begin{figure}[tb]
	\center
{\includegraphics[width=8cm]{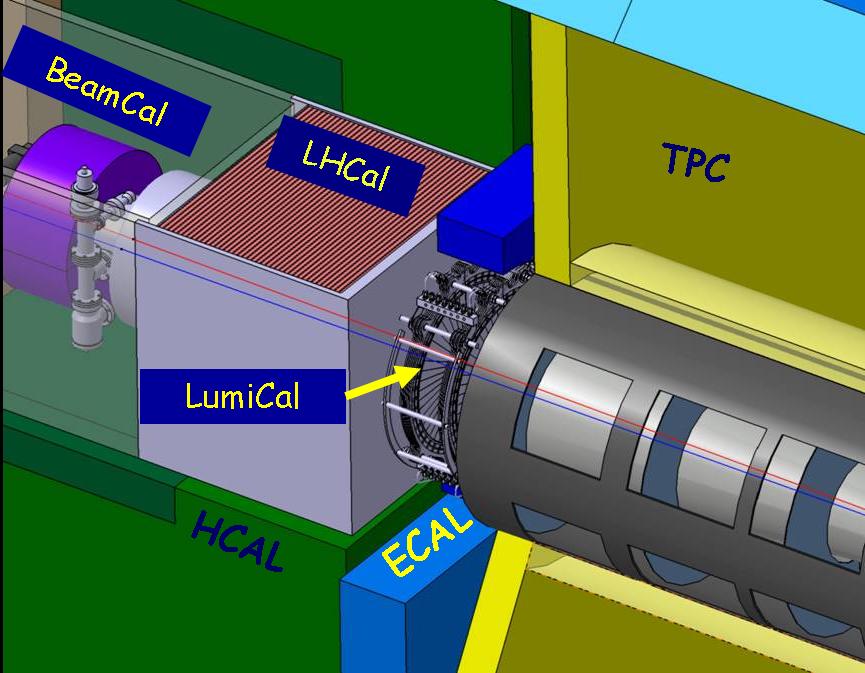}}
\caption [The very forward region of ILD.]{\label{fig:Forward_structure}
The very forward region of the ILD detector. LumiCal, BeamCal and LHCAL are carried by 
the support structure of the QD0 magnet. 
}
\end{figure} 
A sketch of the very forward region of the ILD detector is 
shown in 
Figure~\ref{fig:Forward_structure}. The design of these devices 
is complicated by the small crossing angle of the two beams.
LumiCal and BeamCal are cylindrical electromagnetic calorimeters, 
centered around the
outgoing beam. LumiCal is positioned inside and aligned with the 
forward electromagnetic calorimeter.
BeamCal is placed just in front of the final focus quadrupole.
The pair monitor will be positioned just in front of BeamCal.  

The structure of the ECAL end cap leaves a square hole at 
its centre. LumiCal needs to be positioned very precisely with a well defined fiducial zone
around the outgoing beam and is therefore
restricted to a minimal size to facility mechanical stability and
position control. The ``ECAL ring" fills the gap between the LumiCal and the ECAL. This device could be realised in the same technology as the ECAL, i.e.~a 30 layer tungsten-silicon sandwich providing 24 radiation lengths.

The 30~mm gap between the ring and the end cap is
partly filled by the electronics concentrating cards.
This gap is covered in the back by the HCAL which ensures hermeticity in that region.
The gap between the ring and the LumiCal contains the electronics of the latter and provides space for the tie-rods which suspend the magnet support structure from the coil cryostat. This gap is covered on the back by the LHCAL.
A laser position monitoring system is foreseen to monitor 
the position of LumiCal and BeamCal with respect to the beam-pipe and the 
distance between them~\cite{woitek3}. More details on the integration of the forward region are given in~\cite{ILDIntegration}.

\subsection{LumiCal} 

\begin{figure}[tb] 
\begin{minipage}{0.45\textwidth}
\begin{center}       
{\includegraphics[width=7cm,height=6cm]{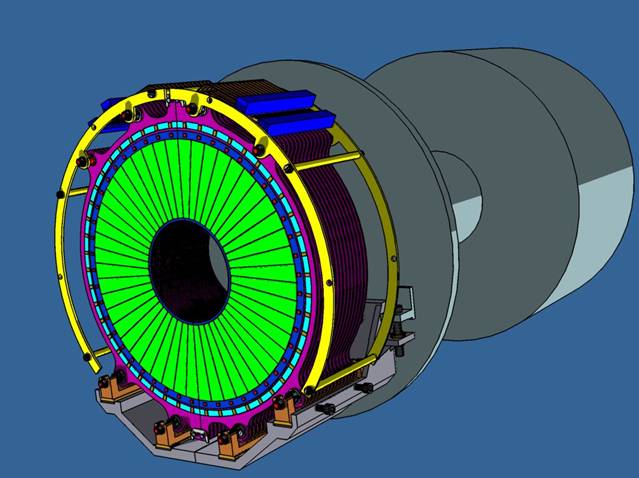}}
\caption [The Si-W LumiCal.]{\label{fig:lumical}
LumiCal designed as a silicon-tungsten sandwich calorimeter. In green the silicon sensor segments are shown  and
in yellow the mechanical frame which ensures the necessary mechanical stability.   
}
\end{center}
\end{minipage}
\begin{minipage}{0.05\textwidth}
~~
\end{minipage}
\begin{minipage}{0.45\textwidth}
\begin{center}       
{\includegraphics[width=6cm,height=6cm,angle=90]{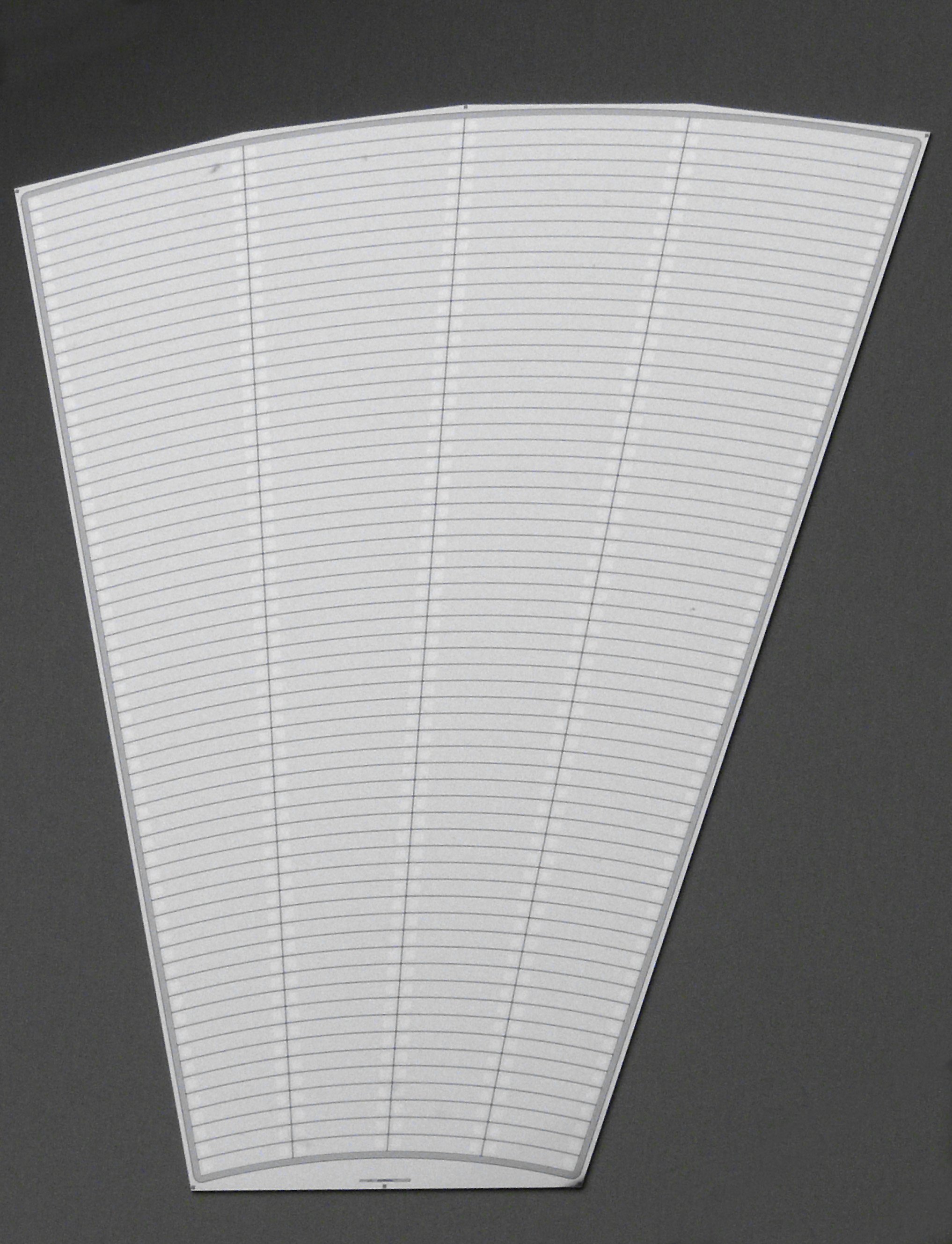}}
\caption [Prototype silicon sensor for LumiCal.]{\label{fig:Si_prototype}
A prototype of a silicon sensor for LumiCal. The sensor 
in manufactured using 6 inch wafer
of n-type silicon, the strip pitch is 1.8 mm. 
}
\end{center}
\end{minipage}
\end{figure}

Monte Carlo studies have shown that a compact silicon-tungsten 
sandwich calorimeter is a proper technology
for LumiCal~\cite{Ronen}.
In the current design~\cite{woitek1}, 
as sketched in Figure~\ref{fig:lumical}, LumiCal  
covers the polar angular range between 32 and 74~mrad.
The 30 layers of 
tungsten absorbers are interspersed with silicon sensor planes. 
The FE and ADC ASICS are positioned at the outer radius in
the space between the tungsten disks.
The small Moli\`{e}re radius and finely radially segmented 
silicon pad sensors ensure an efficient
selection of Bhabha events and a precise shower position 
measurement. The luminosity, $\cal{L}$, 
is obtained from 
$\cal{L} = \cal{N}/{\sigma}$,
where $\cal{N}$ is the number of Bhabha events counted in a 
certain polar angle range 
and $\sigma$ is the Bhabha scattering cross section in the 
same angular range
calculated from theory.
The most critical quantity to control when counting 
Bhabha scattering events 
is the inner acceptance radius 
of the calorimeter, defined as the lower cut in the polar 
angle.    
The precise determination of luminosity requires an excellent 
knowledge of the lower acceptance of the calorimeter. From Monte Carlo 
studies of the present design a tolerance of a few $\mu$m has
been estimated~\cite{achim1}.
Since there is bremsstrahlung radiation 
in Bhabha scattering, cuts on the shower energy 
will also be applied. The criteria to select good 
Bhabha events hence define requirements 
on the energy resolution and, more challenging, 
on the control of the energy 
scale of the calorimeter. The latter quantity must be known to about a few
per mille~\cite{ivan}.
Monte Carlo simulations are also used
to optimise the radial and azimuthal segmentation of silicon pad sensors 
for LumiCal~\cite{Ronen} to match the requirements on the shower measurement
performance. 
 
A first batch of prototype sensors~\cite{woitek2}, 
as shown in Figure~\ref{fig:Si_prototype},
has been delivered from Hamamatsu Corp.. 
At the first stage these sensors will be characterised and qualified, 
in a later stage, they will 
be instrumented with Front-End (FE) electronics 
for investigations in the test-beam and eventually the construction of a calorimeter prototype. 

\begin{figure}[tb] 
\begin{center}
\begin{tabular}{cc}
    {\includegraphics[width=7cm]{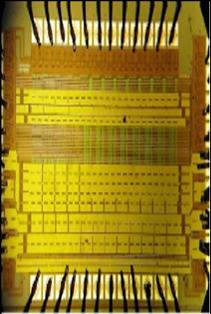}}
&
    {\includegraphics[width=7cm,height=4.5cm]{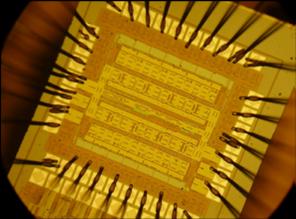}} \\
\end{tabular}
\caption [Prototypes of readout chips.]{\label{fig:ASICS}      
Prototypes of the FE (left) and ADC ASICS (right) 
prepared for systematic tests in the laboratory.} 
\end{center}
\end{figure}

Front-end and ADC ASICS are designed
with a shaping and conversion time less than 300~ns, being potentially able
to readout the calorimeter after each bunch crossing.
The range of sensor pad capacitance and the expected signal range
in electromagnetic showers originating from Bhabha 
events are taken from Monte-Carlo simulations~\cite{iftacheu1}. 
Prototypes of the FE ASICS and pipeline ADC ASICS, manufactured in 0.35~$\mu$m
AMS technology, are shown in Figure~\ref{fig:ASICS}.
The FE ASIC can be operated in low and high amplification mode. The high amplification
mode allows to measure the depositions of minimum ionising particles.
Hence muons can be used from the beam halo or from annihilations for the
calibration and sensor alignment studies. The low amplification mode 
will be used for the measurement of electromagnetic showers.   
Tests of these ASICS prototype are ongoing~\cite{FE_ASIC}. Results
on linearity, noise and cross talk measured
are matching the requirements for the performance derived from Monte-Carlo simulations.   
For 2010 multi-channel prototypes of the ASICS are planned, 
allowing to instrument
prototypes of sensor planes to investigate the performance of the 
full system in the test-beam.

\subsection{BeamCal}

\begin{figure}[tb] 
\begin{minipage}{0.45\textwidth}
\begin{center}       
{\includegraphics[width=7cm,height=6cm]{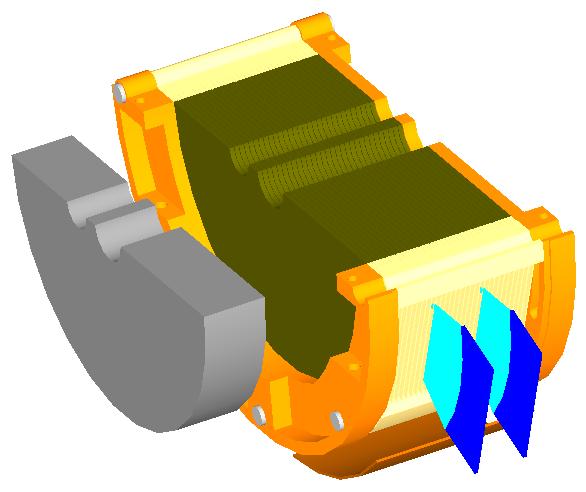}}
\caption [One half of BeamCal.]{\label{fig:beamcal}
One half of BeamCal designed as a sensor-tungsten sandwich calorimeter. 
The graphite block is shown in gray,
the 
tungsten absorber and the sensors in green and the FE electronics in blue.   
}
\end{center}
\end{minipage}
\begin{minipage}{0.05\textwidth}
~~
\end{minipage}
\begin{minipage}{0.45\textwidth}
\begin{center}       
{\includegraphics[width=7cm,height=6cm]{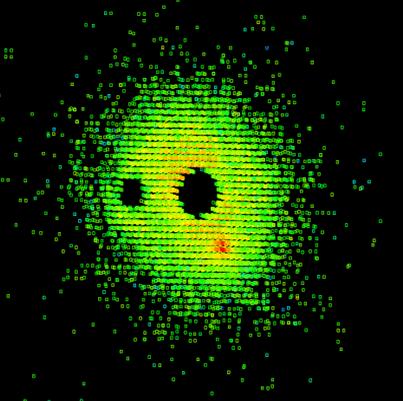}}
\caption [Distribution of energy deposited by pairs in BeamCal.]{\label{fig:beam_deposits}
The distribution of depositions of beamstrahlung pairs after one bunch
crossing on BeamCal. Superimposed is the deposition of
a single high energy electron (red spot in the bottom part). The black holes 
correspond to the beam-pipes. 
}
\end{center}
\end{minipage}
\end{figure}

BeamCal is designed as a solid state sensor-tungsten sandwich calorimeter,
as shown in  Figure~\ref{fig:beamcal}, covering 
the polar angle range between
5 and 40 mrad.
The tungsten absorber disks will be of one radiation length thickness and 
interspersed with thin sensor layers
equipped with FE electronics positioned at the outer radius. 
In front of BeamCal a 5 cm thick graphite block
is placed to absorb low energy back-scattered particles.

BeamCal will be hit after each bunch crossing 
by a large number of beamstrahlung pairs, 
as shown in Figure~\ref{fig:beam_deposits}. The energy, up to several TeV
per bunch crossing, and shape of 
these deposition
allow a bunch-by-bunch 
luminosity estimate and the determination of beam parameters~\cite{grah1}.
However, depositions of  
single high energy 
electrons must be detected on top of the
wider spread beamstrahlung.
Superimposed on the pair depositions 
in Figure~\ref{fig:beam_deposits}
is the local deposition of one high energy electron, seen as 
the red spot at the bottom.
Using an appropriate subtraction of the pair deposits
and a shower finding algorithm which
takes into account the longitudinal shower profile, 
the deposition of the high energy electron 
can be detected with high efficiency and modest energy resolution, 
sufficient to suppress the
background from two-photon processes in a search e.g. 
for supersymmetric tau-leptons~\cite{drugakov} in certain scenarios. 

\begin{figure}[tb] 
\begin{minipage}{0.45\textwidth}
\begin{center}       
{\includegraphics[width=7cm,height=5.5cm]{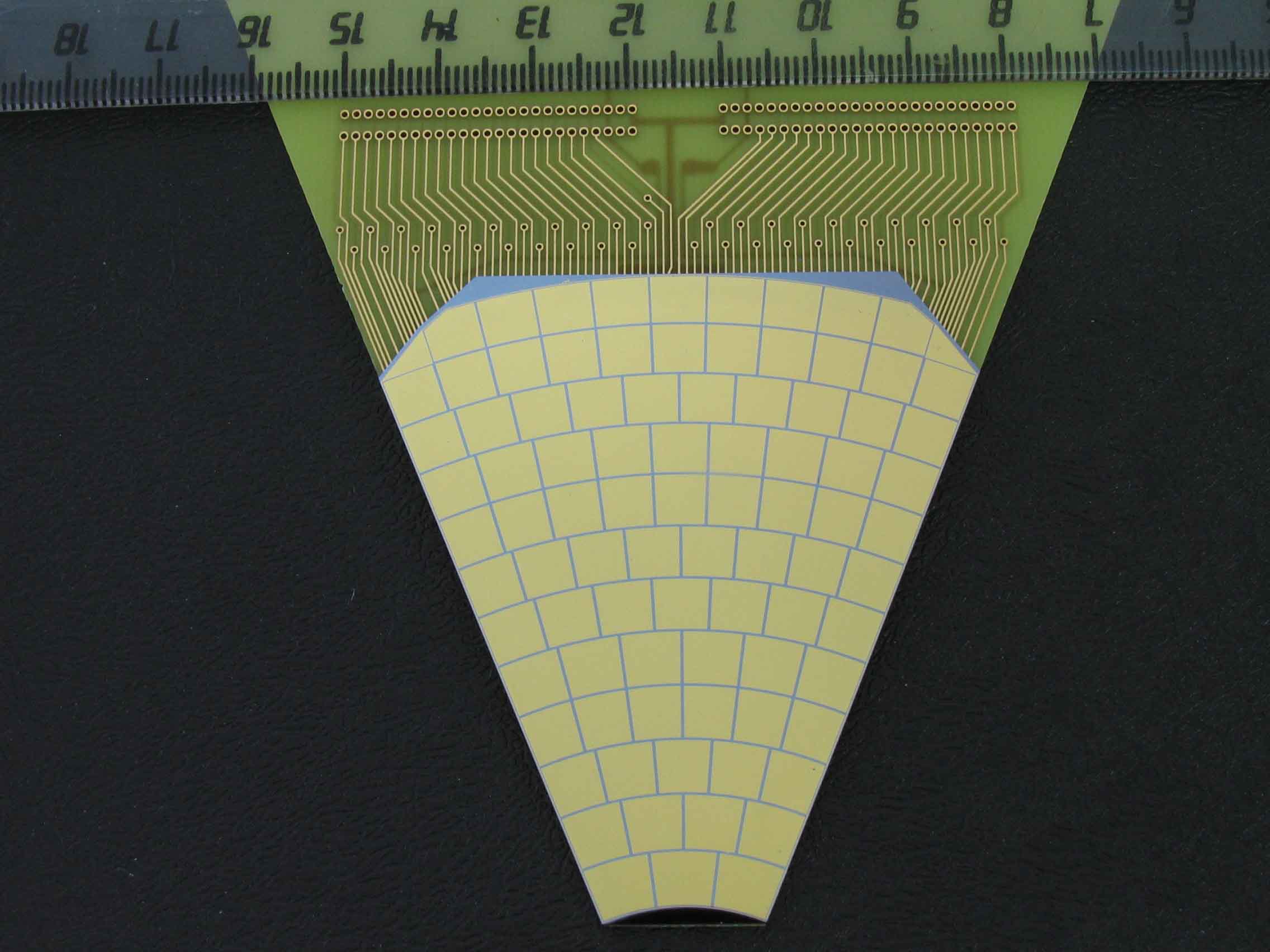}}
\caption [A GaAs prototype detector for BeamCal.]{\label{fig:GaAs}
A prototype of a GaAs sensor sector for BeamCal with pads of about 1 cm$2$ area.   
}
\end{center}
\end{minipage}
\begin{minipage}{0.05\textwidth}
~~
\end{minipage}
\begin{minipage}{0.45\textwidth}
\begin{center}       
{\includegraphics[width=8cm,height=5.5cm]{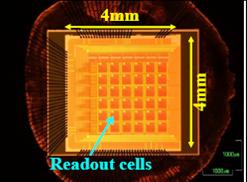}}
\caption [Pair Monitor]{\label{fig:pair_moni}
A prototype ASIC for the Pair Monitor Pixel layer. The pixel size is
400x400 $\mu$m$^2$.
}
\end{center}
\end{minipage}
\end{figure}

The challenge of BeamCal is the development of radiation hard sensors, 
surviving up to 10 MGy of dose per year.
Polycrystalline CVD diamond sensors of 
1 cm$^2$ size, and larger sectors of GaAs pad sensors
as shown in Figure~\ref{fig:GaAs} have been studied.
Polycrystalline CVD diamond sensors have been irradiated up to 
7 MGy and were found to be still operational~\cite{ieee2}. 
GaAs sensors are found to tolerate nearly 2 MGy~\cite{ieee3}.
Since
large area CVD diamond sensors are still very expensive, 
they might be used only at the innermost part of
BeamCal. At larger radii GaAs sensors seem 
to be a promising option. These studies will be 
continued in the future
for a better understanding of the damage mechanisms 
and possible improvements of the sensor materials.   

The FE ASIC development for BeamCal, 
including a fast analog summation for the beam feedback system
and an on-chip digital memory for readout in between two bunch 
trains~\cite{angel} 
is ongoing, first prototypes are expected in 2009. 

\subsection{The Pair Monitor} 

The pair monitor consists of one layer of 
silicon pixel sensors just in front of BeamCal
to
measure the distribution of the number of beamstrahlung pairs.
Monte Carlo simulation have shown 
that the pair monitor will give essential additional information
for beam tuning. For example, averaging over several 
bunch crossings, the 
beam sizes at the interaction point can 
be reconstructed with per cent precision
~\cite{sato}.
A special ASIC,
shown in Figure~\ref{fig:pair_moni},
is developed for the pair monitor.
Prototypes manufactured in 0.25 $\mu$m TSMC technology are under study. 
At a later stage, 
the pixel sensor and the ASIC are foreseen to be embedded in the same wafer.
The latter development will be done in SoI technology~\cite{takubo}.  
 
\subsection{GamCal}

GamCal is supposed to exploit the photons 
from beamstrahlung for fast beam diagnostics.
Near the nominal luminosity the energy of beamstrahlung 
photons supplements the data from BeamCal and Pair Monitor
improving the precision of beam parameter 
measurements and reducing substantially the correlations between 
several parameters~\cite{grah1}. At low luminosity the amount of depositions
on BeamCal will drop dramatically, however, GamCal will still give 
robust information for beam tuning.

To measure the beamstrahlung spectrum a small 
fraction of photons will be converted by a thin 
diamond foil or a gas-jet target about 100 m downstream of the interaction point. The created electrons or positrons
will be measured by an electromagnetic calorimeter. 
A conceptual design of GamCal exists, 
more detailed Monte Carlo studies are necessary
to fully understand the potential of GamCal for 
beam tuning and beam parameter determination.

\subsection{LHCal}

The LHCal fits in the square hole of the HCAL 
and embraces the beam tube which is centred on the outgoing beam. 
It has a thickness of four interaction lengths comprised by 40 layers of tungsten of 1~cm thickness. The sensitive medium 
could be silicon sensors similar to the ECAL ones.
LHCAL is supported by two vertical plates which are part 
of the forward structure. It would be made of two halves
separated vertically, making it easy to dismantle. 
The electonics concentrating cards would be on the top
and the bottom.



\subsection{Priority R\&D topics}  
 
The current research work covers several fields of high priority to demonstrate that the designed devices
match the requirements from physics.
These
are:
\begin{itemize}\addtolength{\itemsep}{-0.5\baselineskip} 
\item{Development of radiation hard sensors for BeamCal. 
The feasibility of BeamCal depends essentially on 
the availability large area radiation hard sensors.}
\item{Development of high quality sensors for LumiCal, 
integration of the FE electronics in a miniaturised
version and tuning of the full system to the required 
performance.}  
\item{Prototyping of a laser position monitoring system 
for LumiCal. In particular the 
control of the inner acceptance radius with $\mu$m 
accuracy is a challenge and must be demonstrated.}
\item{Development and prototyping of FE ASICs for 
BeamCal and the pair monitor. 
There are challenging requirements on the readout 
speed, the dynamic range, the buffering depth and 
the power dissipation. In addition, a system for the 
data transfer to the back-end
electronics has to be developed.}
\end{itemize}
Also of high priority, but not covered for the moment, is
the design of GamCal and an estimate of its potential 
for a fast feedback beam-tuning system.

%% file: ild/coil/coil.tex
\section{Coil and Return Yoke}
\label{sec:coil}

The basic layout of the ILD detector has always followed the strategy of tracking in a magnetic field. The ILD detector design therefore asks for a 4~T field in a large volume, with a high field homogeneity within the TPC volume and with a reduced fringe field outside the detector.

The parameters of the ILD magnet being very similar to the CMS ones (c.f.~\cite{cms_magnet}, \cite{cms_magnet2}), basic designs of both magnets are similar. An anti DiD (Dipole in Detector) is also added in the design, which allows to compensate the effect of the crossing angle for the outgoing beam (and pairs) behind the I.P.

\subsection{Physics Requirements}

The main requests from the physics for the ILD magnet are a solenoidal central field of nominal 3.5~T and maximum 4~T, in a volume of 6.9~m in diameter and a length of 7.35~m with the following requests:
\begin{itemize}
\item A high integral field homogeneity:
\[|\int_0^{2.25m}(B_r/B_z) \, dz  \, | \leq 10~mm; B_r = B_x (x/r) + B_y (y/r)\]
within the TPC volume, which is a cylinder 3.6~m in diameter and 4.5~m long. This high homogeneity requests incorporating compensation windings.
\item A fringe field in the radial direction less than 50~G at R~=~15~m to not magnetically perturb the second detector when in operation on the beam line.
\item A yoke instrumented for the detection of muons and for tail catching (see section \ref{sec:muon}).
\end{itemize}

\subsection{Magnet Design}

The magnet consists of the superconducting solenoid, including the correction coils, and of the iron yoke, one barrel yoke in three pieces and  two end-cap yokes, also in two pieces each. The anti DiD is located outside the solenoid.

Concerning the correction coils, it seemed practically simpler and less space consuming to incorporate them into the main winding, by adding extra currents in appropriate locations of the winding.

\begin{figure}[htb]
	\centering
		\includegraphics[width=14cm]{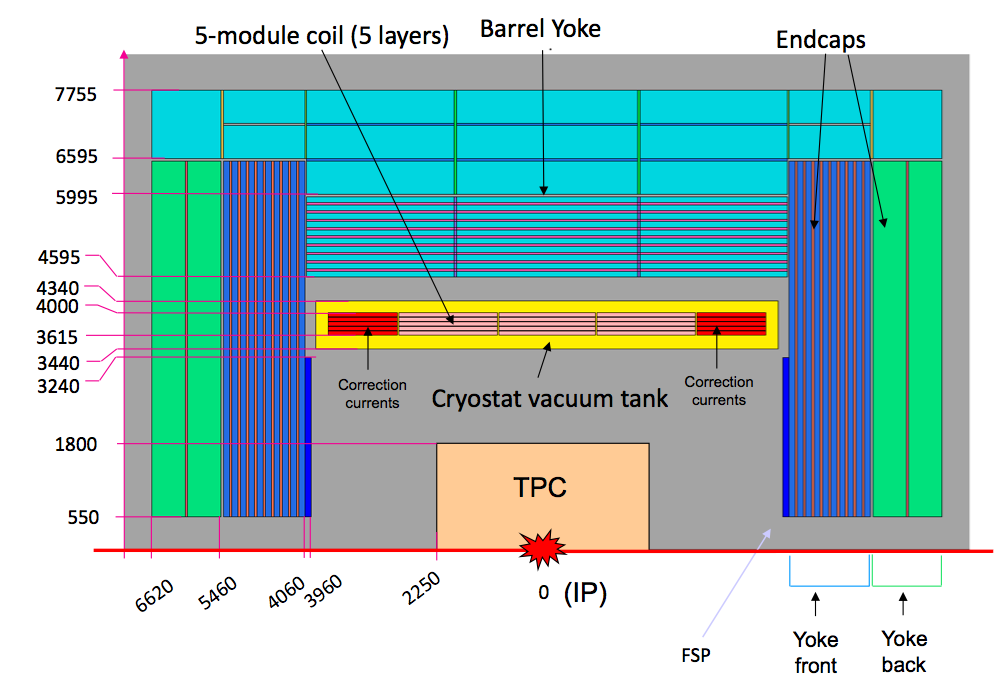}
	\caption[ILD magnet cross section.]{\label{fig:ILDcoil}Cross section of the ILD magnet.}
\end{figure}

The cross section of the ILD detector magnet is shown on Figure~\ref{fig:ILDcoil}. Its main geometrical and electrical parameters are given in Table~\ref{tab:coil:para1}.

\begin{table}[htb]
\begin{tabular}{|l|r||l|r|} \hline
Cryostat inner radius (mm)& 3440&Maximum central field (T) & 4.0 \\ \hline
Coil inner radius (mm)&3615&Maximum field on conductor (T) &5.35 \\ \hline
Coil outer radius (mm)&4065&Stored energy (GJ)& 2.0 \\ \hline
Cryostat outer radius (mm)&4340&Stored energy/ cold mass (kJ/kg)& 12.2 \\ \hline
Barrel yoke inner radius (mm)&4595&Nominal main current (kA)& 18.2 \\ \hline
Barrel yoke outer radius (mm)&7755&Nominal correction current (kA)& 15.8 \\ \hline
Coil length (mm)&7350&Ampere-turns main coil (MAt)& 1.52 \\ \hline
Cryostat length (mm)&7810&Ampere-turns correction coils (MAt)& 1.36\\ \hline
Yoke overall length (mm)&$6620*2$&&\\ \hline
\end{tabular}
\caption{\label{tab:coil:para1}Main geometrical and electrical parameters}
\end{table}


The coil is divided into five modules, electrically and mechanically connected: there are three central modules, 1.65 m long each, and two external modules, 1.2~m long each. All modules consist of a four-layer winding.

The nominal main current, 18.2~kA for a central field of 4.0~T, runs through all the turns of the solenoid. An extra correction current of about 15.8~kA is added in the turns of the four layers of the two external modules to get the integral field homogeneity.


The barrel yoke has a dodecagonal shape. It is longitudinally split into three parts. In the radial direction, the inner part of the yoke is made from 10 iron plates of 100~mm thickness, with a space of 40~mm between each to house detectors for tail catching and muon detection. Three thicker iron plates of 560~mm each with 40~mm spaces for muon detectors form the outer part of the barrel yoke. The weight of the barrel yoke is around 7000~ t.

The end-cap yokes, also of dodecagonal shape, have a similar split structure, with 10 iron plates of 100~mm thickness in the inner part, with a space of 40~mm between each to house the tail catcher and muon detectors, and two external thick plates, each 560~mm thick, to make up the total iron thickness. A 100~mm thick field shaping plate (FSP) will be added inside each end-cap to improve the field homogeneity. The weight of each end cap yoke is around 3250~t and thus the total weight of the yoke is around 13400~t.

\begin{figure}[htb]
	\centering
		\includegraphics[height=7cm]{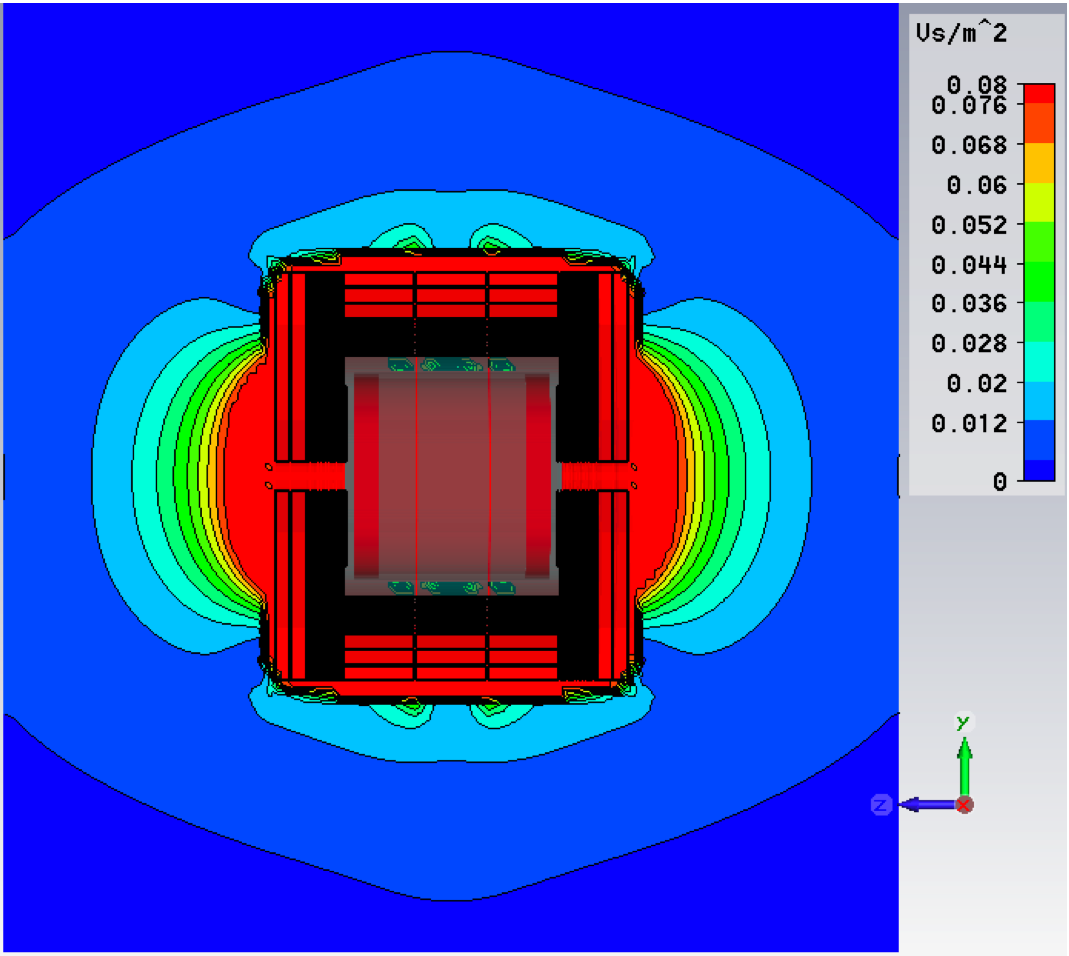}
		\includegraphics[height=7cm]{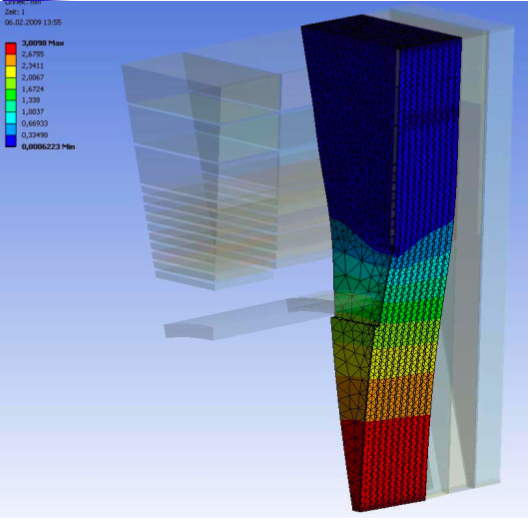}
		
	\caption[Stray fields and endcap deformation.]{\label{fig:ILDyoke}Stray fields outside the yoke (left). Deformation of an endcap segment (right).}
\end{figure}

The main design challenge of the yoke endcaps is to contain the magnetic forces themselves. A weight equivalent of $\approx$18000~t pulls at each endcap. A FEM analysis shows that if the endcaps are constructed in radially fixed segments~(c.f. figure~\ref{fig:ILDyoke}) the deformation of the endcaps due to the magnetic force could be less than 3~mm; alternative designs which lead to comparable small deformations are also under study. These deformations are far smaller than e.g. at CMS where the endcaps are deformed by $\approx$16~mm during the powering of the magnet.

\subsection{Magnetic Field}

The calculated integral field homogeneity, with the nominal values of the main and correction currents given in table~\ref{tab:coil:para1} meets the requirement (maximum value of 7~mm at 4~T). Note that the effect of the anti DiD is not taken into account in this calculation. 

With the yoke structure described, the calculated fringing field is $\approx$40 Gauss at 15~m in the radial direction and therefore fulfils the requirements (c.f. figure~\ref{fig:ILDyoke}).

\subsection{Technical Aspects}

As several technical aspects are quite similar for the ILD and CMS magnets, the experience gained during the construction of the CMS magnet will be of great help for ILD.

The conductor will consist of a superconducting cable coextruded inside a low electrical resistivity stabiliser and mechanically reinforced by adding high-strength aluminum alloy. Two different conductors will be necessary, using different superconducting cables and different ratio of mechanical reinforcement, but with the same overall dimensions.
      
The winding will be done using an inner winding technique. The magnetic forces will be contained both by the local reinforcement of the conductor and by an external cylinder. The coil will be indirectly cooled by saturated liquid helium at 4.5~K, circulating in a thermo-siphon mode.

The central barrel yoke ring will support the vacuum tank. Internal sub-detectors will be supported on rails inside the vacuum tank.

%% file: ild/muon/muon.tex
\section{Muon Detector}
\label{sec:muon}
The identification of leptons is an important part of the physics programme at the ILC. For muons above a few GeV, the 
instrumented iron return yoke is used as a high efficiency muon identifier. 
The clean environment of an electron-positron Linear Collider allows for a muon system design 
that is much simpler compared to the ones that have been developed for the hadron colliders.
There is no need to trigger on muon tracks; instead the clean nature of the events at the ILC allows the 
linking of track candidates from the inner detectors with tracks in the muon system.

In addition to its muon tagging ability the system will be instrumented to allow for a limited calorimetric 
performance. In this way it can act as a tail catcher, tagging late developing showers and thus improving the 
energy measurement. 



A muon is most easily identified by a track in a muon detector behind significant material. At the ILD, the muon system is reached by
muons with a momentum above about 3~GeV. The strong central magnetic field will keep lower energy particles from reaching the muon system. The main challenge then for these type of muons is the joining of a signal in 
the calorimeter with a track segment outside the coil. Multiple scattering in the calorimeters and the coil will 
have a large impact on this, and the efficiency of association will increase with momentum. At lower momenta, the 
signal in the calorimeter will be used to identify muons. In particular inside jets this is difficult, and more 
in - depth studies are needed within ILD to reach strong conclusions.

\subsection{Conceptual Design}
The muon system in ILD will cover a large area of several thousand square meters. 
The detectors therefore need to be reliable, easy to build, and economical. Signals
from the detectors should be large so that simple readout systems and cable routings can be 
used to the readout modules.
The detectors should have a reasonable temporal and spatial resolution. Searches
for long-lived particles and tagging of cosmics and beam halo muons requires that a 
few nsec time resolution be achievable. Since multiple 
scattering is significant, spatial resolutions in the range of cm are sufficient. Occupancies are low, so that both strip and pixel devices can be considered. The efficiency and reliability of muon identification somewhat depends on the iron longitudinal segmentation as do calorimetric performances. Mechanical construction and practical considerations indicate that plate thickness cannot be below 10~cm. For the ILD design, the total thickness needed to close the magnetic flux is $\approx$~275~cm (see \ref{sec:coil}).
It is instrumented with 10 layers of detector with 10~cm thick absorber plates in between, and a few layers at larger distance in the remainder of the yoke.


Both gas detector and extruded scintillator strips can in principle fulfill the requirements.
Plastic Streamer Tubes (PST) or Resistive Plate Chambers (RPC) are candidates for the gas detector. However, RPCs tend to be preferred over PSTs due to their reduced cost and greater flexibility in the segmentation achievable. For the RPC option one could use strips $3-4$~cm wide to obtain the desired resolution. Each gap could provide two orthogonal coordinates, with orthogonal strips on the two sides of the gas gap, while energy would be measured for non muon-hits just by hit counting. The electronics would consist of single bit information per strip; the front-end would include a variable threshold discriminator. The channel count would range around 100K. 

\begin{figure}
	\begin{center}
\includegraphics[height=8cm,viewport=0mm 0mm 200mm 168mm]{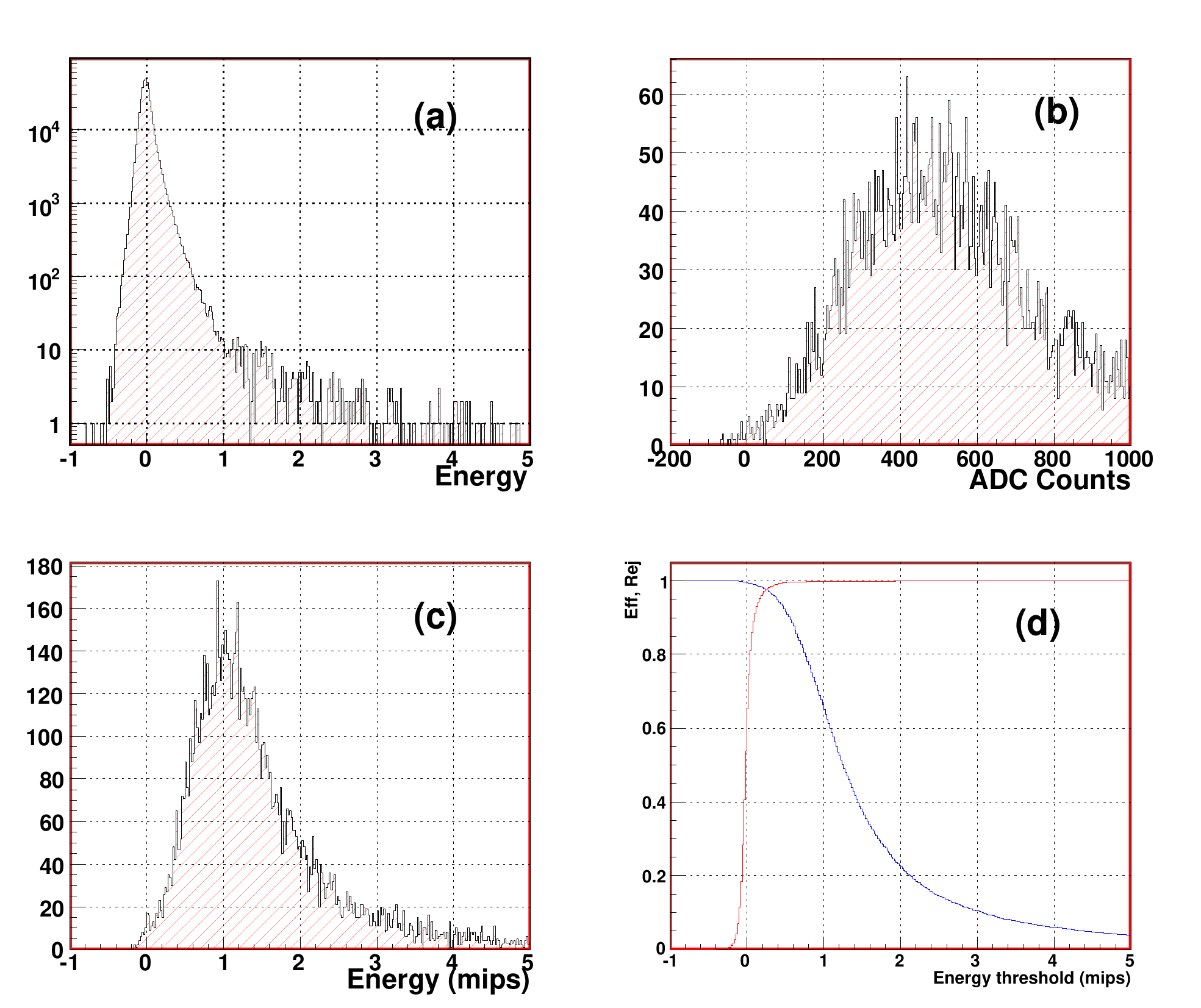}
  \caption[Strip response and efficiency.]{For a typical TCMT strip:(a) pedestal distribution (b) pedestal subtracted MIP signal from muons (c) MIP calibrated signal (d) efficiency (blue) and noise rejection (red)
  as a function of the energy threshold.}
	\label{fig:mipres}
  \end{center}
\end{figure}

An alternative solution relies on extruded plastic scintillator strips. Wavelength 
shifting fibres are embedded into the strips, and are read out at 
either end of the strips with silicon photomultipliers (SiPMs). The small size, low operating voltages and magnetic field immunity of the SiPMs implies that they can be placed inside the
detector, thus obviating the need for routing fragile fibres. Orthogonal placement of strips
can provide space-points with the required resolution.
A prototype based on this 
technology, the tail-catcher/muon tracker (TCMT), has been built and exposed to a test beam by 
the CALICE collaboration at CERN and Fermilab during the 2006-2008 period. The extended
operation of the TCMT has clearly demonstrated its excellent reliability and performance
(see Fig. \ref{fig:mipres}). It should be noted that useful synergies may exist between the muon
system and the hadron calorimeter since RPCs and scintillator are both potential technology
options for the HCAL.

\subsection{Performance}
\begin{figure}
\begin{center}
		\includegraphics[height=4cm,viewport=0 0 180mm 120mm]{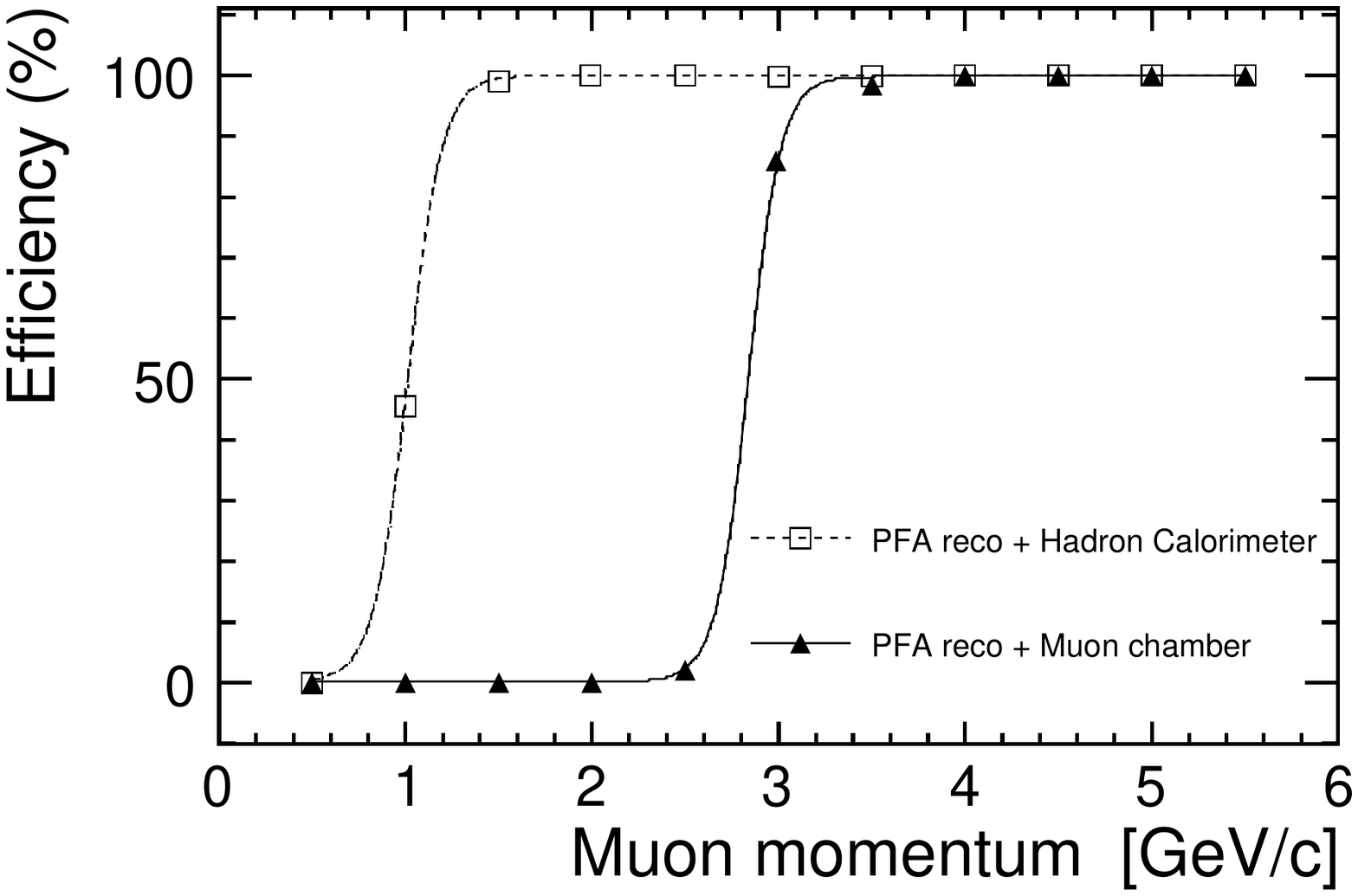}
		\includegraphics[height=7cm,viewport=0 0 135mm 200mm,angle=90]{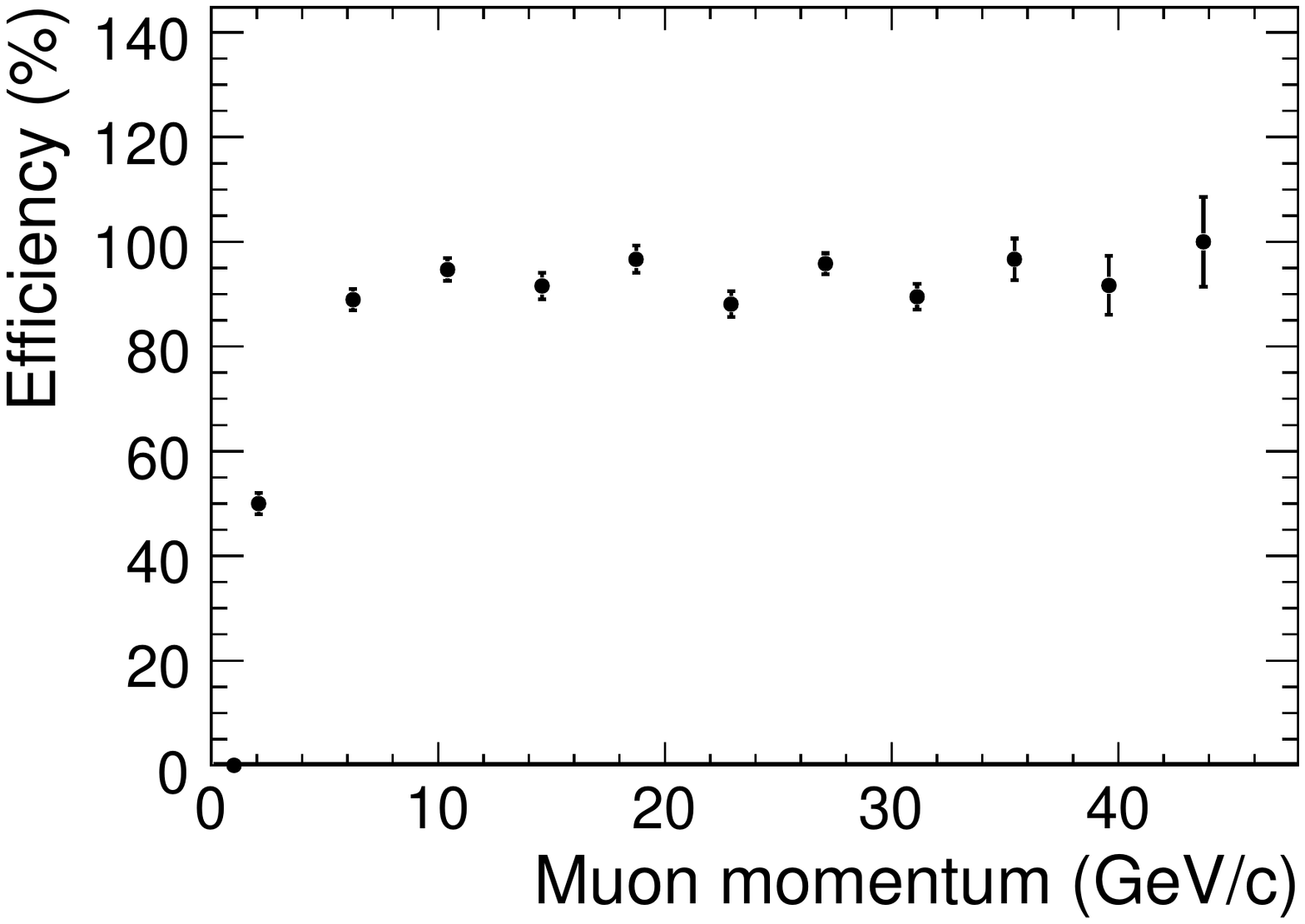}
  \caption[Muon identification efficiency.]{(top): Muon identification efficiency vs. momentum for single muon. The 
  right curve represent muon found only in the muon system, the left curves muons which were found by a combination of 
  HCAL and muon system. (bottom): Muon finding efficiency in $\bbbar$ events. }
	\label{fig:muonefficiency}
\end{center}
\end{figure}

Performances of the muon identification system described in the previous paragraph have been evaluated both with single particles and high multiplicity final states. In order to assess the linking capabilities of the proposed detector, single particle efficiency has been evaluated as shown in Figure~\ref{fig:muonefficiency}(left top). In the same figure (left bottom) the efficiency is shown for muons identified in $\bbbar$ events.  
The plots in Figure~\ref{fig:muonefficiency} show that the performance remains excellent also 
in complex high multiplicity events . 

\begin{figure}
	\centering
		\includegraphics[height=9cm,viewport=0mm 0mm 210mm 210mm]{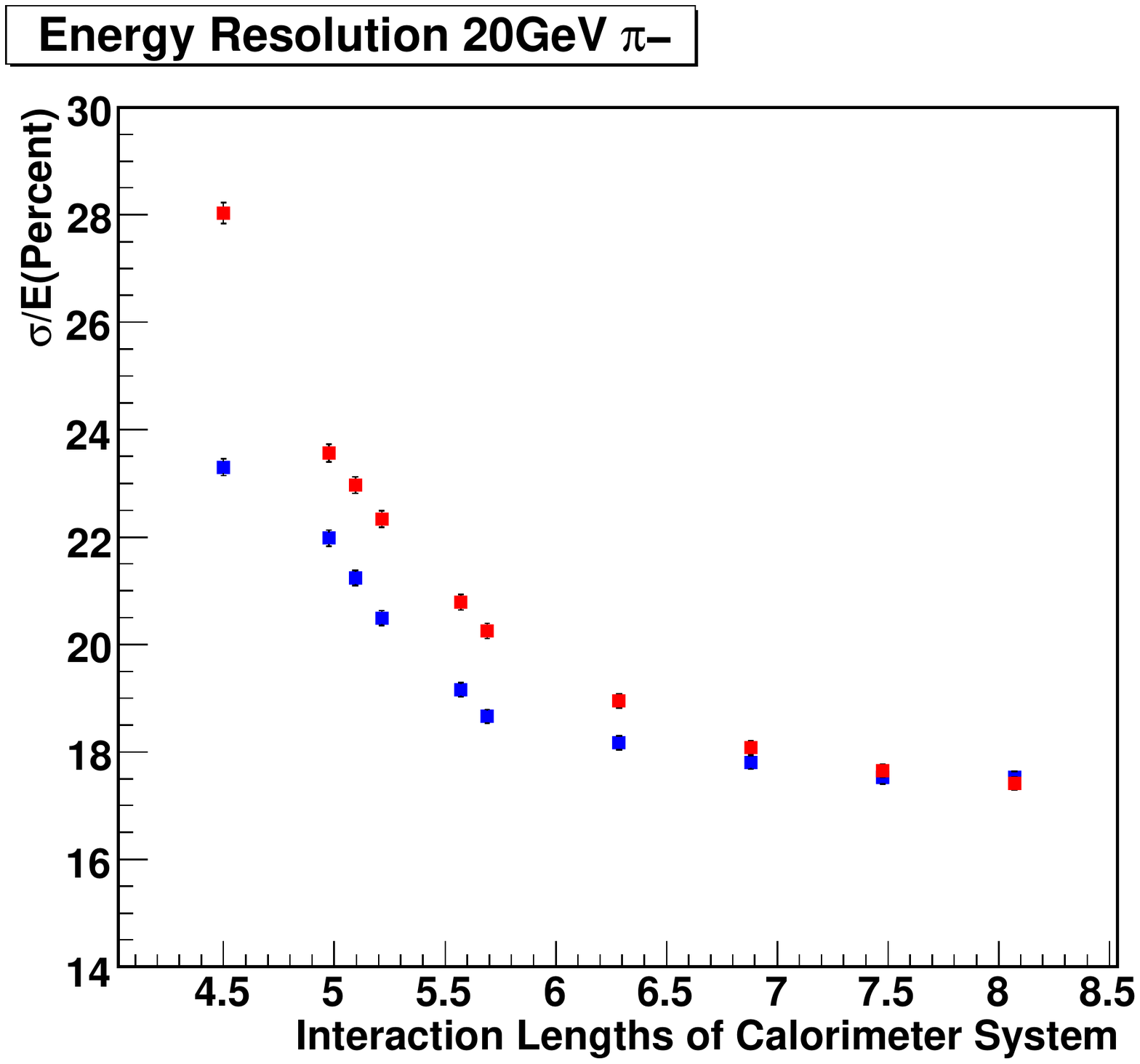}
	\caption[Correcting for energy leakage in the calorimeter.]
{Resolution improvement for 20~GeV pions by including energy beyond the coil using
CALICE testbeam data. The red points correspond to the resolution obtained for a given
thickness of the calorimeter system while the black points supplement the energy in the calorimeter
system with that from beyond the coil. The change in length of the calorimeter system, the
material contained in the coil and energy
beyond the coil are simulated by rejecting or accepting layers in the TCMT.}
	\label{fig:leak}
\end{figure}

The muon system also does have a limited calorimetric capability.
On average the energy leakage into the muon system is small but with large event-to-event fluctuations.
For K$^0_L$ the
average energy leakage is 3\% for a range of energies, but
energy leakage four or five times the average is not uncommon. The use of the muon system to estimate and
correct for the leakage is hampered by the presence of 
the coil, which introduces close to two interaction lengths of dead material between the 
last calorimeter layer and the first muon layer. Nevertheless 
a correlation between the energy of hadrons and the leakage signal 
recorded in the muon system can be observed and used to improve resolution in simulation and test beam
data (see Fig. \ref{fig:leak}).

\subsection{Outlook}
The proposed muon system for the ILD concept is well matched to the requirements as laid down in this document. Two alternative technological implementations are 
discussed, one based on gaseous detectors and the other based on plastic scintillator strips. 
The system will serve primarily as a muon identifier, but will also play an important role as a tail catcher to compensate for leakage from the calorimeter system. Continued R\&D is required to establish a detailed, realistic design and make an informed technology choice.

%% file: ild/alignment/alignment.tex
\section{Calibration and Alignment}

The ILD detector is a sophisticated precision instrument. It consists of a 
high precision tracking detector, surrounding the interaction point, followed 
by a granular calorimeter system covering nearly the entire solid angle. 

To reach the anticipated performance calibration and alignment of the sub-detectors 
and of the overall detector system are a central part of the detector design and is important for the complete life cycle 
of the experiment, from the design over the construction to the operation of 
the device. 

For the rest of this section we define calibration to be all tasks which 
deal with the internal description of the detectors.
Alignment is the relative positioning of internal parts of the subdetectors or 
of sub-detectors relative to each other.

Calibration and alignment of all sub-systems will be based on a mixture of data from dedicated calibration 
systems and from particles recorded during physics running. 

During construction tolerances must be carefully controlled, during commissioning systematic metrology of the different detector parts 
is needed. This will provide an initial alignment of the modules internally, and of the different systems relative to each other. The 
need for a high precision metrology has consequences for the mechanical design of 
the different sub-systems, which from the beginning need tolerances determined to allow 
the necessary level of mechanical precision. In particular in the tracking 
system care has to be taken to ensure that the mechanical systems are stable enough 
to allow an alignment at the level of $10 \mu$m or better. 

After installation, and throughout the lifetime of the detector, constant
measurements will be taken using dedicated hardware to monitor the 
position of the different detector components, improving the alignment from the 
initial situation. 


The final calibration of the detector will be done using particles. Tracks from the decay of 
the Z - boson will play an important role here, as they are of well known high momentum.
The best source of such tracks are from short and dedicated runs of the 
collider on the peak of the Z resonance, at 91~GeV. For the discussions in the following 
sections we assume that around $1 pb^{-1}$ of data can be collected within a few hours of running at 91~GeV. 
This will result in some 30000 Z~bosons, of which around 1000 will decay as $Z \rightarrow \mu^+ \mu^-$. 

The final high precision calibration will be derived from tracks in the data sample taken at high energies. Stiff tracks e.g. from W decays or from $q \bar {q}$ pair events will provide 
a large sample of tracks. The design of the detector has been optimised in a way to 
allow a calibration heavily relying on such data. 

The detector alignment at the ILC is a particular challenge because of the 
intended push-pull mechanism to switch between two detectors. This implies 
that each detector will move out and back in into the interaction 
region frequently, and that the overall alignment of the system should be 
re-established rapidly after a push-pull operation. As discussed in \ref{sec:mdi:pushpull} 
the switchover times between the two detectors should be of order of a few days with consequently the need to do a rapid re-establishment of calibration constants within 
a day or so. Both the detector design and the alignment concept need to 
take these requirements into account. 

Another challenge will be the need to power-pulse the detectors inside the magnetic 
coil to limit the total power consumption. Most detectors will be switched off or 
switched to reduced power in between trains of the collider. This procedure 
of power pulsing will potentially apply significant forces to the detector 
structure, during the ramp up or ramp down of the power, and stress the components 
with significant swings in temperature. Special care must be exercised
during the design of the mechanical system to ensure that the structure does 
not move during these cycles, and that the alignment does not suffer from train to train. 

In this section the calibration and alignment strategies of the overall 
detector are discussed. Calibration strategies for individual sub-detectors 
have been discussed in the relevant sections describing the 
sub-detector technologies. 

\subsection{Tracking System Calibration and Alignment}

A main purpose of the tracking system is the efficient finding of charged particles, the 
reconstruction of their momenta and their impact parameters. The anticipated precision 
for the momentum and the impact parameter are significantly above anything 
ever achieved before in a detector of this size and complexity. Using a simple 
model of the track parameters dependence on alignment tolerances, the 
following limits for the alignment of each of the tracking sub-systems have been derived: 
\begin{itemize}\addtolength{\topsep}{-0.5\baselineskip}\addtolength{\itemsep}{-0.7\baselineskip}
\item   coherent displacement of  the VTX,  2.8 $\mu$m;
\item  coherent displacement of  the SIT,  3.5  $\mu$m;
\item  coherent displacement of  the SET,  6  $\mu$m; and 
\item  coherent displacement of  the TPC,  3.6 $\mu$m.
\end{itemize}
These values must be confirmed by further studies.

An important aspect of the overall alignment of the tracking 
system is the knowledge of the 
central magnetic field. Uncertainties on the size and direction of the 
field within the tracking volume will directly impact the momentum resolution. 
Using sophisticated magnetic field probes the field will be measured to 
a precision of $dB/B<10^{-4}$. This level of precision has been 
reached in previous experiments using large volume magnetic fields \cite{ref-rswwlcnote}.

The above distortion limit defines the precision required for the magnetic field calibration. In a TPC, the drifting electrons follow the magnetic field lines; field components perpendicular to $B_z$ result in deflections of the track as measured at the readout plane. The magnet is designed to have a field uniformity of $\int {B_r(constructed)/B_z dz} = 2~\mathrm{mm}-10~\mathrm{mm}$. These deflections are largely corrected with the application of the magnetic field map. However, residual misunderstanding of the magnetic field will result in track distortions. Thus, the mapping of the magnetic field must be significantly improved beyond the initial probe precision stated above, $dB/B<10^{-4}$. Based on the limit of the internal fit sagitta above, the magnetic field map must have a precision of $\int B_r(correction)/B_z dz < 20-30  \mu$m. For the case that magnetic field distortions that are coherent along the drift length of about 2 meters, the integral is equivalent to the requirement that $dB/B<10^{-5}$. It is envisioned that 
stiff tracks as observed in either Z decays of in high energy collisions 
will provide the necessary information. 
 
\subsubsection{Silicon Tracking Alignment}
Calibration and alignment of the detectors are an important 
consideration already during the design and construction phase of the 
different silicon based sub-detectors. These aspects are discussed 
in detail in section~\ref{ild:silicon}. The alignment among different sub-detectors, 
and relative to the rest of the ILD detector, will be based on a three-fold approach: 
two laser based alignment systems will be combined with a sophisticated 
alignment strategy based on tracks. 

The extent of the challenge for alignment becomes clear if one considers the number 
of degrees of freedom which need to be determined. For the ILD silicon tracker this number is of the order of $100.000$ (calculated as six times the number of sensors). If the relative sensor positions in the module are known to the required precision (from survey data or from other hardware alignment systems) the number of degrees of freedom is reduced by a large factor (a factor five in SET that dominates the NDOF count, a factor two or three in ETD). The contribution of the outer tracking system is larger than that of the inner tracker by a factor $10$. 



If we can assume that the different support structures are basically rigid, and do 
not change dimensions internally, within the precision anticipated, we only need to worry about the overall alignment of 
the different sub-detectors relative to the rest of the detector. The goal of ILD is that 
this situation is reached for a re-alignment after a push-pull operation. In this case the 
number of degrees of freedom is greatly reduced: 
\begin{itemize}\addtolength{\topsep}{-0.5\baselineskip}\addtolength{\itemsep}{-0.7\baselineskip}
\item The outer Silicon tracker, SET, is fixed to the TPC. It is supported by a rigid carbon 
fibre structure. It probably needs to be split in the middle. In this case 12 degrees of freedom 
are needed to determine the position of the overall SET. 
\item The ETD is attached to the endcap calorimeter in one piece. In total 8 degrees of freedom 
need to be considered.
\item The SIT and the FTD are connected to a common support structure. The movement of 
this structure has 6 degrees of freedom. 
\end{itemize}
In total 26 degrees of freedom are present in this case. This reduces the initial 
problem to one which can be solved in fairly short amount of time. 

Details about the module and channel count in the silicon system are given in 
Table~\ref{tab:SILC_SI}.
{
\small
\begin{center}
\begin{table}
\begin{tabular}{l|lllll}
Component& \multicolumn{2}{c}{number of} & \# of sensors/ & \# of channels & area  \\
         & layers       &  modules    &  module        &                &  m$^2$\\
         \hline
SIT1 & layer 1 & 33 & 3 & 66000& 0.9 \\
     & layer 2 & 99 & 1 & 198000& 0.9 \\
\hline
SIT2 & layer 1 & 90 & 3 & 180000& 2.7\\
     & layer 2 & 270& 1 & 540000& 2.7 \\
\hline
SET  & layer 1 & 1260 & 5 & 2520000 & 55.2\\
     & layer 2 & 1260 & 5 & 2520000 & 55.2 \\
\hline
ETD\_F & X,U,V & 984 & & 2000000 & 30 \\
ETD\_B & X,U,V & 984 & & 2000000 & 30 \\
\hline
FTD & 7 & 350 &     &  5000000 & \\
\hline
\end{tabular}
\caption[Details of the Silicon detectors.]{Number of modules, channel count, and sensitive area for the 
different Silicon based detectors.}
\label{tab:SILC_SI}
\end{table}

\end{center}
}

\subsubsection{A laser based alignment system}
 
\begin{figure}
	\centering
		\includegraphics[height=7cm,viewport=0 0 20cm 10cm,clip]{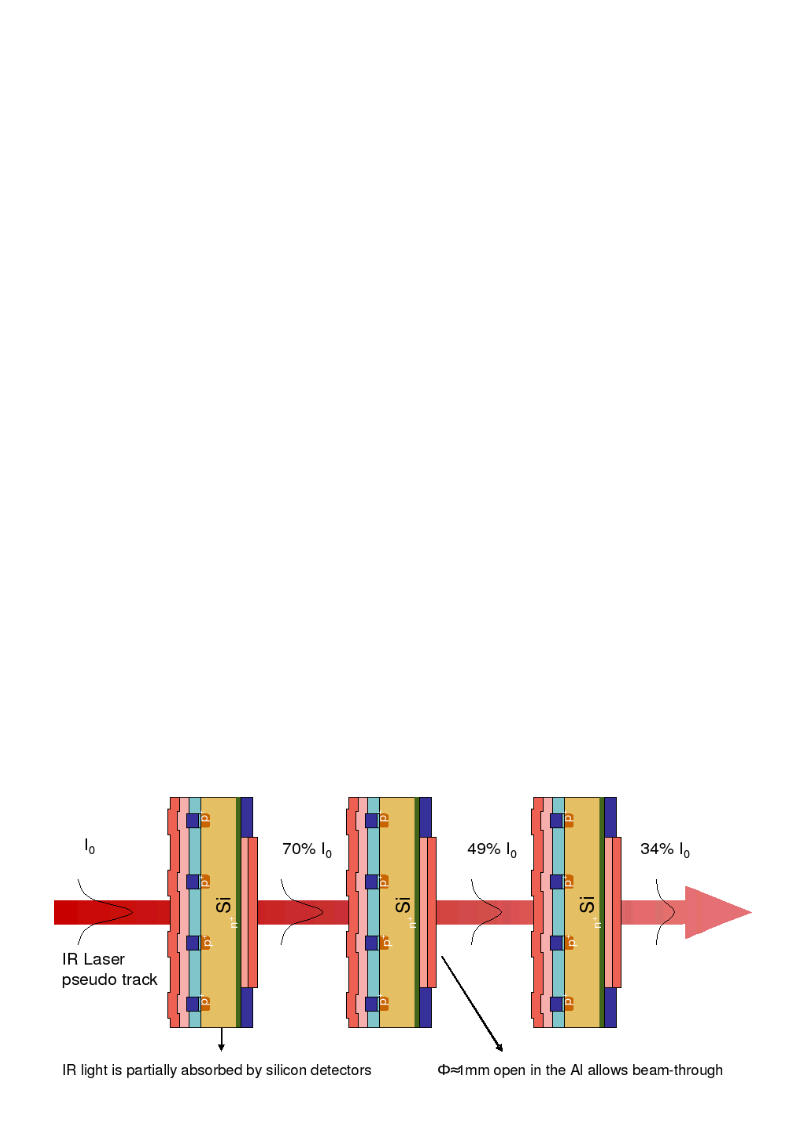}
	\caption[Infra-red laser beam traversing silicon sensors.]{Schematic view of the infra-red laser beam traversing several micro-strip sensors.}
	\label{fig:Laser}
\end{figure}

\begin{figure}
	\centering
		\includegraphics[height=7cm,viewport=0 0 20cm 12cm,clip]{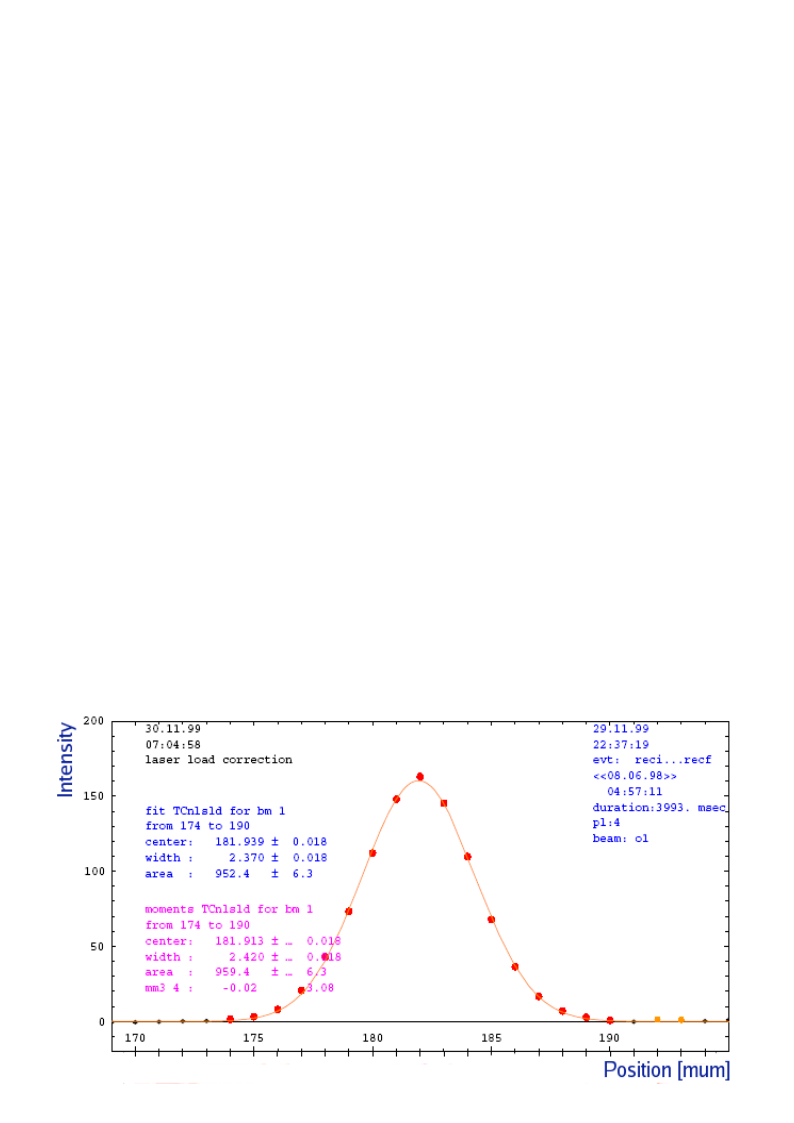}
	\caption[AMS results on the reconstruction of laser signals in silicon strip detectors.]{AMS results on the reconstruction of the laser signal on the third sensors in the stack. The data points are the result of averaging 480 readings. The position of the laser pulse can be reconstructed with a resolution of 1 micron.}
	\label{fig:AMS}
\end{figure}

A hardware position monitor system based on infra-red laser beams mimicking straight tracks will be installed in the ILD detector. 
The laser beams traverse several sensors optimized for IR laser transmittance (see Fig.~\ref{fig:Laser}). 
The signals from the laser are readout using the module sensor and front end electronics. Therefore, there is no added contribution to the error budget associated to the mechanical transfer between monitored fiducial marks and strips. The resolution expected on sensor transversal movements will depend on the number of strips illuminated by the beam and the sensor pitch. For a pitch value of $50~\mu$m with a gaussian width of the 
laser beam of $\sigma = 300 \mu$m, resolutions below $1\mu$m are achievable. A 
similar system is in use im the AMS experiment where a silicon tracker with a geometry similar to the FTD (comparable tracking volume, number of layers and cylindrical symmetry) has reached an accuracy of $2 \mu$m \cite{laser} (see Fig. ~\ref{fig:AMS}). The laser alignment system provides the possibility to precisely monitor relativly fast movements.  The current design of the ILD forward tracking disks incorporates this system. The system is able to constrain several degrees of freedom of a large fraction of the installed micro-strip sensors in the FTD and the other strip detector, and of the third tracking disk based on pixel technology. The extension to other sub-detectors is being investigated. 

Particularly 
challenging is the connection between the inner and the outer silicon detectors, because 
of the large distance between them, and the connection between the silicon tracker 
and other detector elements. Here a pixel based monitoring system (PMD), excited by an IR laser through optical fibres is under consideration. In this system special 
pixel detectors are attached to the sensors to be aligned. The pixel devices will be installed in several strategic places of the tracking system 

The IR laser systems will be complemented by a network of fiber optic sensors (Fiber Bragg System or FBS) that will monitor structural changes like deformations or relative displacements among structures, and environmental parameters as humidity and temperature. These sensors are based on Bragg gratings built into mono-mode optical fibers. In this technology, the carrier fiber is also the readout line. Compared to other traditional sensing techniques they are immune to electromagnetic interference and temperature effects. 



\subsubsection{Alignment of the Time Projection Chamber}
The large volume time projection chamber is a central piece of the ILD tracking 
system. The TPC will provide more than 200 space points along a track. The longest 
drift distance possible is around 2.3~m. The anticipated spatial resolution in the 
device is around 60-100 $\mu$m. 

In addition to the calibration issues discussed in the previous chapters, the 
TPC is particularly sensitive to the magnetic field in the detector. The survey and 
calibration of the magnetic field will be a major part of the TPC calibration. 

Mechanically the Silicon tracking detectors will all be mounted relative to the 
TPC. Most probably - though a detailed engineering design has not yet 
been done - the inner Silicon tracking system will be suspended from the 
end - plate of the TPC on either side. The external Silicon tracking in the 
barrel will be supported by the field cage, the external Silicon tracking 
behind the TPC endplate will be supported by the endplate itself, or possibly by the endcap electromagnetic calorimeter. The TPC as a whole 
will be suspended from the coil of the ILD detector. 

The PMD laser system will possibly be used to reference the TPC relative to the 
Silicon detectors, and transfer the location of the TPC to the coil. With 
this system the main degrees of rotation and shift between the TPC and the 
rest of the Silicon can be determined, and serve as starting point for the
overall determination of alignment constants. 

\subsubsubsection{Internal TPC Alignment}
The internal alignment of the TPC will be based on a well understood and measured field cage of the 
system. A construction of the field cage at the 0.1~mm level seems possible, and a survey 
of the finished field cage at the level of $30 \mu$m might be not unrealistic. 

The B-field, which is as important to the ultimate measurement precision as the mechanical components, will be mapped using probes to a level of $dB/B<10^{-4}$ as described in
~\cite{ref-rswwlcnote}. Starting from a well understood magnetic field, 
unambiguous preliminary tracks can be defined in the TPC. These will be 
used to iterativly improve the calibration of the TPC, and will eventually 
serve as starting points for tracks spanning the full tracking system.

In the TPC, the internal components, i.e. the detector readout modules, will be manufactured to tolerances of $20 \mu$m, while tolerances for placing the modules on the end-plate will be about $60 \mu$m. The internal TPC alignment process must provide the final required precision for both the mechanical alignment and the magnetic field measurement. 
Achieving these goals will require iteration. As mechanical distortions and magnetic field distortions can lead to similar track distortions, supplementary alignment systems will be used to resolve the ambiguities. 

The internal alignment in the TPC will be helped by laser systems installed on the TPC in two ways: 
using a system of mirrors straight tracks created by a laser are created inside the drift 
volume, and can be used to determine many calibration constants. In addition diffuse light 
will be shone on the cathode surface, on which an appropriate coating creates a pattern 
of charge, which can e.g. be used to calibrate field distortions and the drift velocity. 

\subsubsection{Maintaining the alignment}
Probes mounted on the fieldcage of the TPC will be used to monitor the B-field during running. Pressure and temperature will be measured continuously and will be corrected for on the fly. Cosmic-ray tracks and laser systems will be used to check for changes of the internal alignment in the TPC. Tracks from Z running will be used to extend coverage to the whole detector because lasers can only monitor a limited number of reference points, and cosmic rays give reasonable coverage of the vertical direction only. 

The PMD laser system will provide a constant stream of alignment 
data and monitor in real-time the position of the TPC relative to the 
rest of the detector. 

\subsubsection{Track Based Alignment}

While the hardware based systems are invaluable to do a fast re-calibration of the 
tracking system for the most relevant degrees of freedom, they are not suited for 
the final high precision alignment. This will need to be based on data from particle tracks.
Previous experiments have developed a sophisticated machinery and have shown that the alignment transform can be determined to a precision well below the intrinsic resolution of their detectors, assuming that a large enough sample of high quality alignment tracks 
can be collected. 
The most important consideration in selecting the track sample is to tightly constrain all degrees of freedom of the detector geometry, including those that leave the track residual distributions (nearly) unchanged. Typically, the alignment sample is composed of a mixture of collision data and tracks from other sources. High p$_T$ tracks are particularly valuable as they minimize the influence of multiple scattering. A strong constraint on the detector geometry derives from tracks that traverse overlapping detector modules (in the vertex detector and silicon tracking systems). Tracks from cosmic rays and beam halo are useful as they allow to relate different parts of the detector (upper and lower half, both end-caps). Tracks with known momentum are extremely valuable, both as a means to determine some of the weakly constrained alignment parameters and as a monitoring tool to validate the alignment. This role has traditionally been played by tracks from resonances with a well-known mass (the Z-resonance is the most popular as it provides stiff tracks, but J/$\Psi$ and Upsilon have been used as well). 

One of the main limitations of the track based system is that it will not be 
able to follow fast changes in the detector. Given the rather small production rate for $Z->\mu^+ \mu^-$ events in the ILC at 500 GeV the typical time constant to align all degrees of freedom of the detector is likely to be of the order of months . However, reduced sets of degrees of freedom, corresponding to higher-level mechanical units like ladders and rings, or even complete cylinders and disks, can be aligned with much less statistics and, hence, at a much greater frequency. In how far tracks from high energy running can be 
used is a matter of discussion, and will need further in-depth investigation.



\subsection{Calorimeter Calibration and Alignment}
A central part of the ILD detector is a highly granular calorimeter. At the moment 
a number of different technology options for the different parts are 
under consideration. In general though they all display a large number of 
channels, to obtain the excellent spatial resolution needed 
for particle flow. For any sort of stochastic calibration or alignment 
uncertainly this is an asset rather than a burden, since the precision 
with which these effects need to be known scale with $1/\sqrt{N}$, where 
$N$ is the number of channels. 
Nevertheless the detailed procedures and 
the way calibration is implemented differ significantly from 
technology to technology, and will be discussed separately below.

\subsubsection{Si-W electromagnetic calorimeter}
The information used for defining the alignment and calibration procedures of the ILD
electromagnetic calorimeter comes from two sources: a very detailed simulation of the
calorimeter and the results from a prototype exposed to beam for now four years.
Alignment:
The requirements on alignment come from the precision we can reach in measuring the
position of a shower. This is of the order of $1 \mathrm{mm}/\sqrt{E}$. Therefore an alignment precision of
$100\mu$m is the goal. The main uncertainty comes from the play of the slabs inside the
alveoli. With a survey of the module and a careful positioning we can reach $250\mu$m, we
need then an alignment with tracks in situ. A small number of electrons, thousands,
should be enough to align the calorimeter with respect to the tracking system at the
required precision.

\subsubsubsection{Energy calibration}
An early study had shown that the energy measurement in our calorimeter is robust
against dead channels. Provided they are quite randomly distributed, a fraction up to $5\%$
dead channels does not harm the resolution. The mean response is restored by
estimating the dead channels response from their neighbours and the resolution is very
marginally touched. The measurement of the channel noise provides the identification of
the dead ones.

A good energy calibration is the result of a suite of dedicated actions. First the design of
the detector is chosen to provide an intrinsic stability with variables like temperature,
humidity, radiation, voltages, etc. Second a monitoring of these variables and of the
detector response evolution with them is ensured. Third the cells are, at construction
time, inter-calibrated at an adequate level of precision. Finally the absolute calibration is
determined in test beams for few modules and globally at running time. The large number of cells is an asset, as teh calibration fluctuations decrease with $\sqrt N$ os the number of cells. 
The estimation of energy uses a
combination of two estimators, the deposited energy and the counting of cells.
The fully depleted silicon diodes offer a very stable behaviour. The tungsten plates can be
checked. A complete cosmic ray testing produces the accuracy needed by measuring the
minimum ionising particle peak. This has been done in the prototype with success. It is
estimated to be a work of 200 days. The electronics is monitored accurately by injecting
charges calibrated by a band gap device. It can be noted that the two estimators of
energy are sensitive to very different systematic effects providing a powerful global test.
After the inter-calibration, the absolute calibration is done by comparison with the
tracker or using electrons and photons kinematically constrained like Bhabha's or return
to the Z. This does not require any running at the Z peak

\subsubsection{Scintillator Tungsten Electromagnetic Calorimeter}
The scintillator ECAL consists of 1~cm $\times$ 4.5~cm scintilator strips, readout 
with Multi Pixel Photon Counters (MPPC). 
The calibration of these devices, similar to the analogue HCAL, 
needs to be done for the energy scale and limearity, and should be 
monitored against changes of environmental conditions such as temperatur etc.. 
The calibration will be based on a mixture of built-in calibration 
systems and the use of particle, either from cosmic-ray muons, or from test beams, before 
installation. 

A further calibration of the strips may be done using 
tracks during operation of the device. To reach an accuray of $5\%$ per channel, 
about 100 calibration quality tracks are needed. An ideal 
data sample for this would be an extended run at the Z-pole; to collect enough tracks 
about $100 pb^{-1}$ would be sufficient. This amount of data will only be available 
at rather infrequent intervals, since it corresponds to a few weeks of running 
on the Z pole. However once calibrated the system is intrinsically 
stable if monitored well, and no frequent channel-by-channel recalibration 
is needed. 

The photon sensor MPPC has a powerful built-in calibration capability, which 
is also used in the AHCAL. By identifying the response to single, two, three 
etc photons, a precise response and energy dependence can be established 
from data, for every single MPPC channel. THe single photon signals 
will be produced by a system of light fibres and blue LEDs, which 
illuminate each channel with a well defined intensity. This system 
will be used to monitor the time dependence of the calibration constants 
as well. 

\subsubsection{Hadron calorimeter alignment}
Depending on the choice of technology the hadronic calorimeter has cell 
sizes as small as $10 \times 10$mm$^2$, thus requiring a mechanical 
precision at the mm level for the absorber and sensor structures. 
This precision will be somewhat less stringent for the analogue hadronic 
calorimeter, which has larger cells. From the construction and 
survey the location of the cells to this accuracy will be known for 
the installed detector. The alignment required between different parts of the 
calorimeter system are also at a level of mm and will be 
established with data, based on muons from different sources. 

\subsubsubsection{Scintillator analogue hadronic calorimeter energy calibration}

The scintillator tile HCAL, segmented in 48 layers and 32 barrel, 32 end-cap modules,
has 8 million read out channels. While electromagnetic and hadronic energy scales can be
established with sample structures of the HCAL alone or in conjunction with the ECAL
exposed to beams of muons, electrons and hadrons, the inter-calibration of the detector
cells must be established with muon beams for all active layers of the detector. Based on
test beam experience, we estimate that this can be accomplished in about two months.
The calibration accuracy is maintained using LED monitoring of the photo-sensor gain,
in-situ MIP calibration based on track segments in hadron showers and classical slow-control
recording of the relevant operation parameters, temperature and bias voltage.
These methods have been successfully applied to test beam data.

\begin{figure}
	\centering
		\includegraphics[height=7cm,viewport=4 0 20cm 14cm]{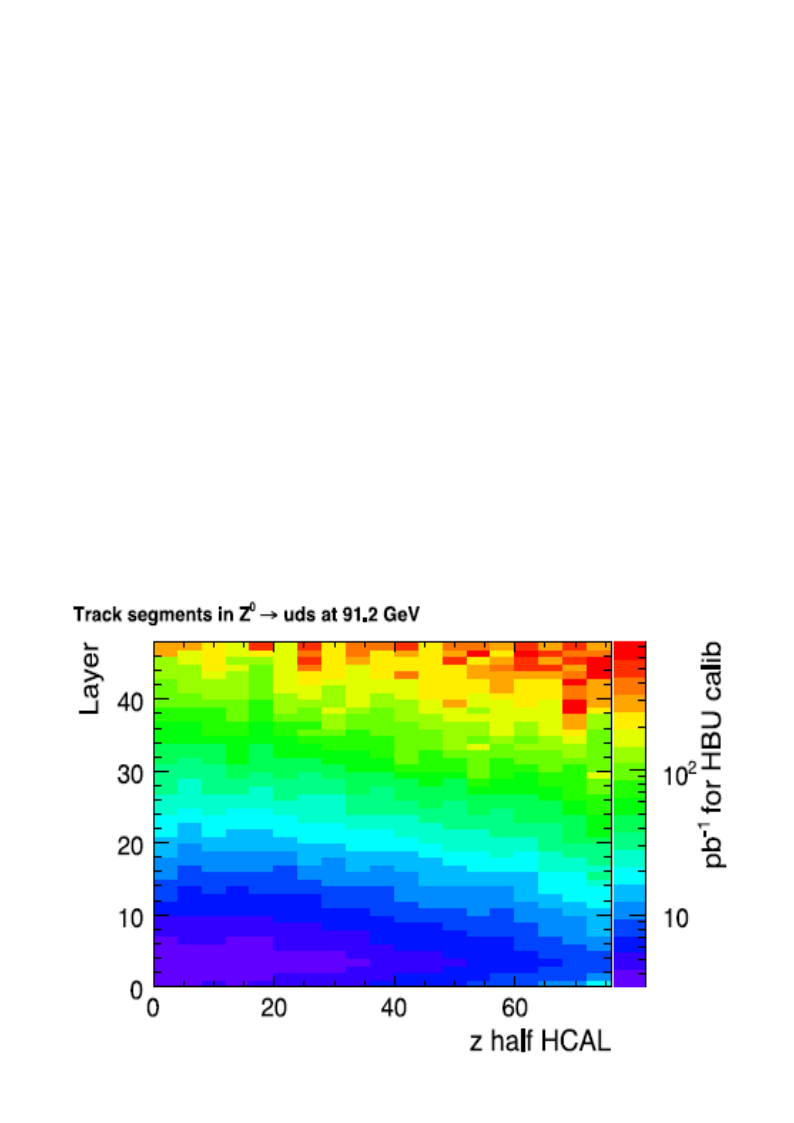}
	\caption[Required luminosity for in-situ calorimeter calibration.]{Required luminosity for an in-situ calibration of the AHCAL, as af function of position of the cell to be calibrated. The scale on the right gives the amount of integrated luminosity needed for one HCAL base unit (HBU).}
	\label{fig:AHCAL_calib_data}
\end{figure}

Simulating ILC events and using algorithms bench-marked with test beam data, we have
determined the required luminosity for in-situ MIP calibration of individual cells and of
average values for sub-sections of the detector, like module layers. A cell-by-cell in-situ
calibration is not possible with realistic running times, but it is also not necessary.
Average values for individual module layers can be obtained with a comfortable accuracy
of 3\% from a data set corresponding to $10pb^{-1}$ at the Z resonance or $20 fb^{-1}$ at 500~GeV.
For the innermost 20 layers, this accuracy is achieved with $ 1pb^{-1}$ or $2 fb^{-1}$, respectively (see Fig.~\ref{fig:AHCAL_calib_data} for more details).

Using detailed simulations of the ILD detector and reconstruction based on the Pandora
PFA, we have modeled different scenarios of statistically independent as well as coherent
mis-calibration effects, affecting the entire HCAL or parts of it. Purely statistical
variations, like those arising from calibration errors or random aging effects, do not affect
the resolution at all. However, they may degrade the in-situ MIP calibration capability.
From this, a moderate requirement of the inter-calibration stability of 10\% is derived.
Coherent effects which could for example arise from uncorrected temperature variation
induced changes of the response are potentially more harmful, as they directly show up
in the constant term, if they affect the entire detector. However, these are easy to detect,
and even a 5\% variation only mildly propagates into the jet energy resolution. Systematic
effects in sub-sections like layers are unnoticeable unless they exceed about 15\%,
comfortably in range of the in-situ calibration method accuracies.

\begin{figure}
	\centering
		\includegraphics[height=6cm,viewport=0 0 20cm 14cm,clip]{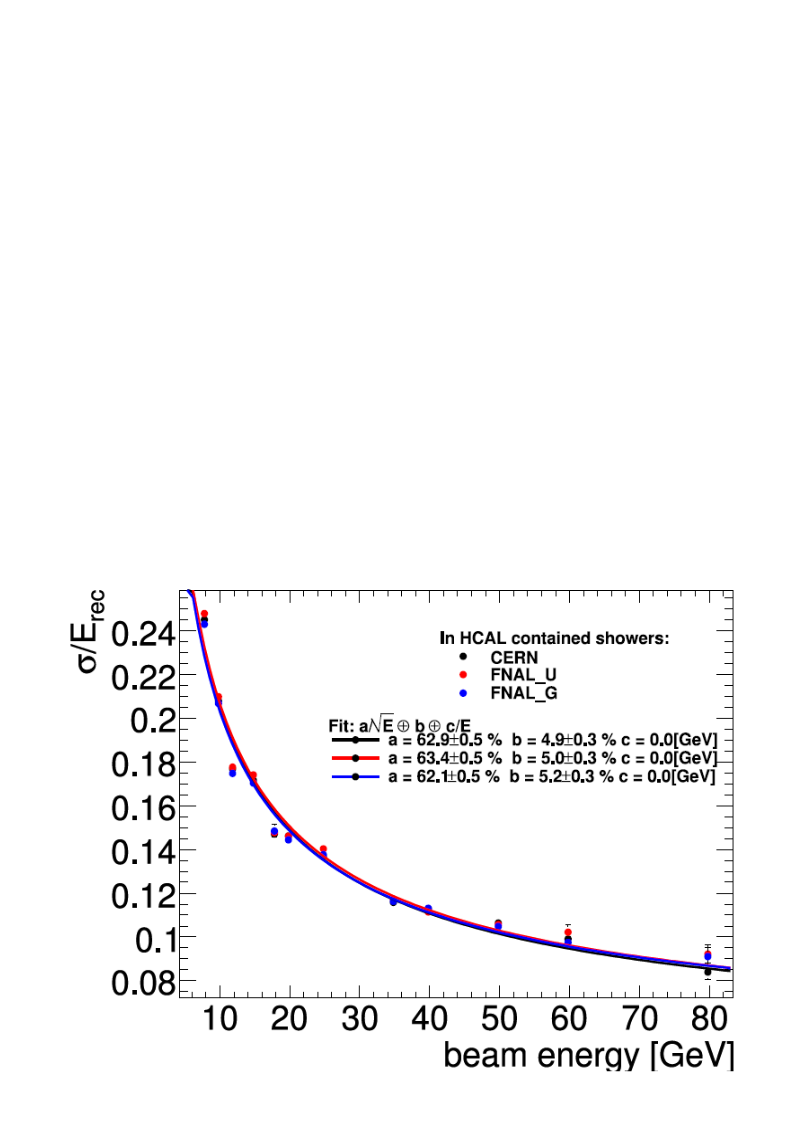}
	\caption[Comparison of measured energy resolution for different test beam locations.]{Comparison of measured energy resolution from three independent test beam experiments. The calibration constants were obtained in one experiment, and transported to the other ones. Excellent agreement is achieved, demonstrating the intrinsic stability of 
	the system.}
	\label{fig:HCAL_calibration}
\end{figure}

The concept was experimentally verified by using data from two independent test beam experiments, performed with the same module, at two different locations CERN and Fermilab. 
The module which had some 9000 channels was calibrated based on data in one location, 
and the calibration constants were then applied to the data recorded at 
the second location. The resulting comparison of the energy 
resolution of the detector is shown in Fig.~\ref{fig:HCAL_calibration}.

We convinced ourselves of the validity of these simulation based estimates by treating our
test beam experiment like a collider detector, using cell-by-cell inter-calibrations only
from data taking at a different site, under different conditions. Applying only in-situ
monitoring techniques, we re-established the scale and reproduced the resolution.
Imperfections absent in the simulation showed up, but were successfully compensated.
All in all, we conclude that the high granularity and channel count is a blessing rather
than a curse. On one hand, thanks to the law-of-large-numbers suppression of statistical
effects, the requirements on individual cell precision are very relaxed. Coherent effects,
on the other hand, can be studied with any desired combination of channels, be it layers,
longitudinal sections, electronics units or according to any other supposed hypothesis of
systematic effects. This has been proven to work with data taken at CERN and at FNAL.

\subsubsubsection{Semi-Digital hadronic calorimeter energy calibration}

The most important features of the semi-digital HCAL option which have an impact on the calibration
strategy are the large number of $\approx$ 70 million electronic channels, their semi-digital
readout which means that efficiency is the meaningful quantity rather than energy, and
the stable and homogeneous nature of the sensitive medium made of resistive plate chambers.
The electronics calibration includes a gain correction procedure, a noise level
measurement and a linearity measurement. The gain correction intends to reduce the
dispersion of the electronic channels response to a given charge. A dispersion of a few
percent for the threshold levels of the semi-digital electronics readout is currently
obtained. The procedure is to inject charges corresponding to the lowest
threshold level with different gains. The procedure is completely automated and will be
applied to all the electronics boards before installation.

We foresee to apply the same procedure in situ. Based on our experience with a fully
equipped $1$m$^3$ detector (9216 channels) we estimate that 200 minutes are needed to
calibrate all the detector channels in parallel. The frequency of the calibration is
under study. Preliminary results based on procedures applied one year apart on the same
electronics board showed small variation ($<2$\%).

To perform the detector calibration, a procedure will be used at construction time to
qualify every piece, and a global control will be made at running time. The homogeneity
of each GRPC detector will be tested by exposing all the HCAL detectors to cosmic rays
before installation. To estimate the effort involved we can say that in order to achieve an
efficiency measurement resolution better than 1\% for each square cm of all the detectors
using benches hosting five detectors, 5000 hours will be needed.

After installation, thanks to the detectors homogeneity only global efficiency of each
detector needs to be controlled. This will be done using:
\begin{itemize}\addtolength{\topsep}{-0.5\baselineskip}\addtolength{\itemsep}{-0.7\baselineskip}
\item Cosmic rays: Their number will depend on the detector depth. At sea level and taking
into account the ILC duty cycle of $5\%$ only a few hours are needed to calibrate the horizontal
detectors. More time is needed for the inclined ones.
\item Beam halo muons: At the ILC with the best currently proposed shielding scheme,
660 halo beam muons are expected per second to traverse the detector. Only a few seconds of running will be sufficient to calibrate the end-cap detectors based on these muons.
\item Tracks produced in hadronic showers: There are few of them in each hadronic shower.
\item Muons produced in data: Those produced from direct decay of Z$\rightarrow \mu^+\mu^-$ and those resulting from decays in the tau tau
and b$\bar {\mathrm {b}}$ channels become an essential source in case the GigaZ scenario is approved.
With $10^{33}cm^{-2}s^{-1}$ of instantaneous luminosity less than five hours are needed to
calibrate all the detectors.
\end{itemize}

In addition different procedures will be used to monitor the behaviour of the calorimeter,
like following the leakage current, following the ratio of pads above the different
thresholds. If necessary, we can consider injecting radioactive gas to check the
response homogeneity.

\subsection{Conclusions}
In this section we have discussed the current state of thinking about aligning the different 
detectors in ILD. We propose to base the alignment on a mixture of 
harware alignment systems,
and a sophisticated use of tracks taken from data. 
Even though many details of the alignment systems still have to be finalised, 
and a lot of technical developments are still needed, we have established
a number of key components central to the proposed procedures in 
test beam experiments. We are therefore confident that the 
system is adequate and that we realistically can 
expect to reach the anticipated precision required for physics at the ILC.

We consider the ability of 
the collider to deliver some luminosity on the peak of the Z resonance to be important and 
very beneficial for a fast calibration, and in particular a fast re-calibration 
after a push pull operation. 

%% file: daq/daq.tex

\newcommand{\fixme}[1]{\textsf{\textbf{(#1)}}} 

\label{sec:daq_intro}

As outlined in the detector outline documents \cite{ref-gld,ref-ldc,ref_SID} the data acquisition (DAQ)
system of a detector at the ILC has to fulfill the needs of a high luminosity, high precision experiment
without compromising on rare or yet unknown physics processes. 
Although the average collision rate of the order of a few kHz is small compared to the LHC,
peak rates within a bunch train will reach several MHz due to the bunched operation.
In addition the ILC physics goals require higher precision in many measurements than 
has ever been achieved in a colliding beam experiment. 
This improved
accuracy can only be achieved by a substantially bigger number of readout channels than in previous
detectors.
Taking advantage of the bunched operations mode at the ILC, event building without a
hardware trigger, followed by a software based event selection was proposed in \cite{ref-TESLA_TDR}
and has been adopted for the ILD. This will assure the needed flexibility as well as
scalability  and will be able to cope with the expected complexity of the physics and detector
data without compromising on efficiency or performance.

The very large number of readout channels for the ILD will require signal processing
and data compression already at the detector electronics level as well as high bandwidth for
the event building network to cope with the data flow.
The recently commissioned LHC experiments have up to $10^{8}$ front-end readout channels and a
maximum event building rate of 100 kHz, moving data with up to 100 GB/s \cite{ref-CMS_TRIDAS}.
The proposed ILD DAQ system will be less demanding in terms of data throughput but the number
of readout channels is likely to be a factor of 10 or more larger.
The computing requirements for the ILC event processing in terms of storage and CPU
are also going to be less demanding than those of the LHC experiments. 
The details of the DAQ and computing system depend to a large extent on the developments in microprocessors and
electronics and the final design of the different sub detector electronic components.
Therefore the DAQ and computing system presented here will have to be rather conceptual, highlighting some
key points to be addressed in the coming years.

\section{DAQ Concept}
\label{sec:daq_concept}
In contrast to past and recent colliders, such as HERA, Tevatron or LHC, which have a continuous rate
of equidistant bunch crossings the ILC has a pulsed operation mode.
The nominal parameter set \cite{RDR_machine} of the ILC with
\begin{itemize}\addtolength{\itemsep}{-0.5\baselineskip}
\item
2625 bunch crossings in a train about 1ms long,
\item
369 ns between bunch crossings inside a bunch train and
\item
a bunch train repetition rate of 5 Hz
\end{itemize}
results in a burst of collisions at a rate of 2.6MHz over 1ms 
followed by 200ms without any interaction. The overall collision rate of 
13kHz is significantly smaller then the expected event building rate
for the LHC experiments.

The burst structure of the collisions at the ILC immediately leads to the
suggested DAQ system with
\begin{itemize}\addtolength{\itemsep}{-0.5\baselineskip}
\item
dead time free pipeline of 1 ms,
\item
no hardware trigger,
\item
front-end pipeline readout within 200 ms and
\item
event selection by software.
\end{itemize}
Rapidly developing fast network infrastructures and high performance computing
technologies, as well as the higher integration and lower power consumption of electronic
components are essential ingredients for the proposed data acquisition system.
Furthermore for such large systems a restriction to standardised components is vital to
achieve maintainability at an affordable effort, requiring commodity hardware and industry
standards to be used wherever possible.

\begin{figure}
	\centering
		\includegraphics[height=9cm,viewport={0 0 243mm 158mm}]{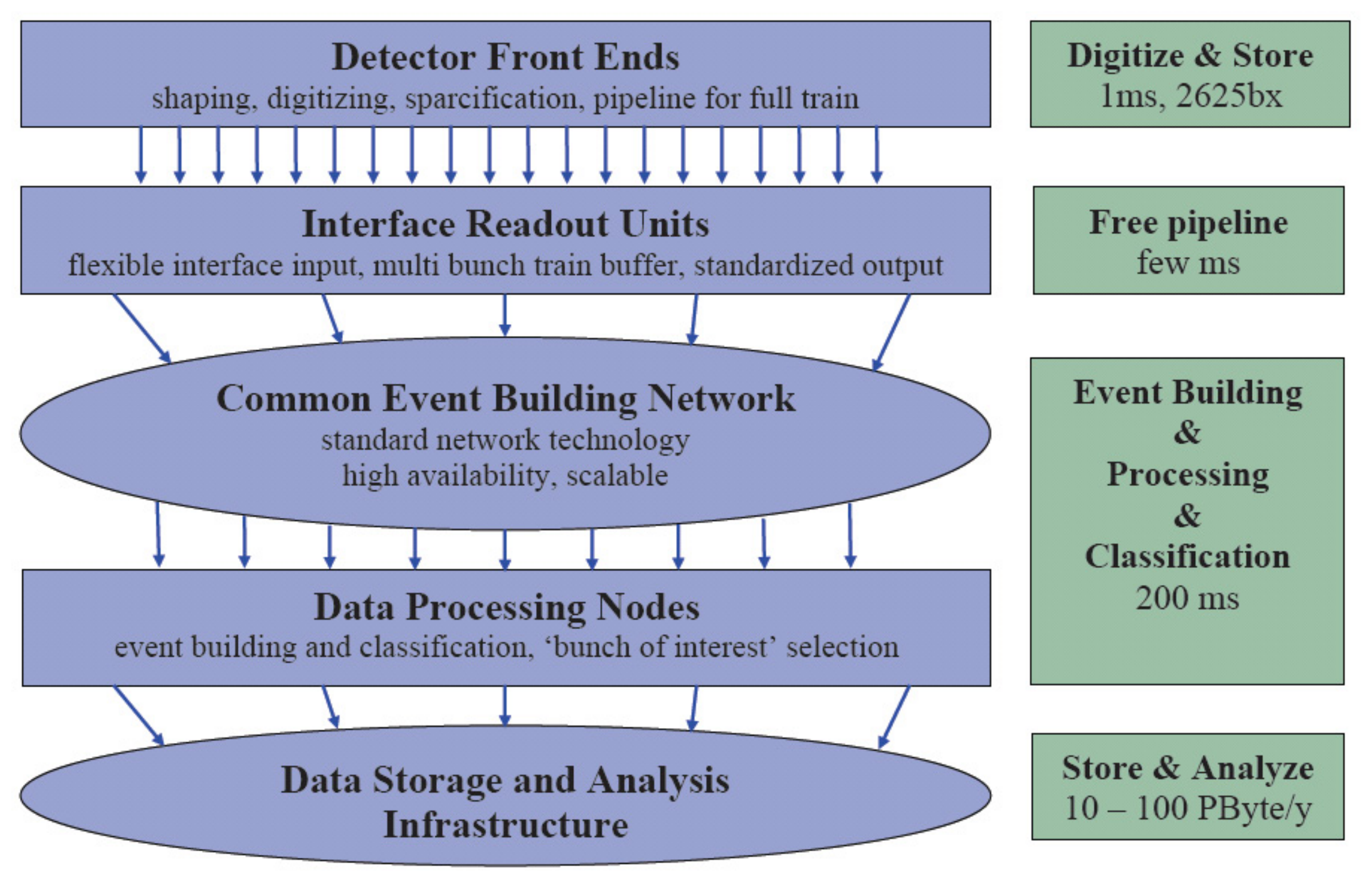}
	\label{fig:DAQ_Layout}
	\caption{General layout of the DAQ system}
\end{figure}

The general layout of the proposed DAQ system is shown in figure~\ref{fig:DAQ_Layout}
The front end electronics on the detector or sensor level has to be detector specific
and will digitize and store the data of $\approx$2600 bunch crossings. But already on
or near the detector a standardized interface with additional buffering and processing
capability will assure a common protocol for the subsequent event building which is
currently estimated to be done by a standard switched network technology like 10G Ethernet.
Event building of all data from the bunch train will be done in a single processing unit.
Hence all data of the complete train will be available for the event processing without further
data transfer which is essential since many detectors will integrate over several bunch
crossings.
The purpose of this online event processing will be mainly event classification, calibration,
alignment and data quality monitoring.
Although no event rejection is foreseen a scheme of event finders may be used to
identify 'bunches of interest' which could then be used for the physics
analysis or for fast analysis streams.

The data volume will be dominated by machine background which in turn is mainly pair production
from beam-beam interaction( see section~\ref{ild:accelerator:backgrounds}). For the nominal ILC parameter set it is expected that this background
produces per bunch crossing
$\approx 8000$ hits in the HCAL, $\approx 150$ hits in the ECAL, $\approx 400$ hits in the TPC
and $\approx 3$ hits/$cm^2$ in the inner layer of the Vertex detector. Physics events
which are less then 0.1 per bunch train will hence contribute less then 1\% of the data
recorded.
Details of the simulated background are described in Section \ref{ild:accelerator:backgrounds}.

Table~\ref{tab:DataVolume} lists the expected data volume per train for the major ILD detector components. Only values for the configuration used in the ILD detector simulation (see table~\ref{tab:optdetparameter}) are shown as an example. For different technologies and options numbers may vary (DHCAL for instance $< 20$MB).
The occupancies are quoted for the detector-spcific time intervals. Some detectors 
digitise individual bunches, others like the TPC integrate over several bunches or the 
full bunch train. 

The total data volume per trains is estimated to be $\approx 340$MByte, hence the event building
network has to cope with 1.7 GB/s. Assuming a safety factor of 10 on the background estimation
and further contributions from electronic noise hits the maximum bandwidth anticipated is less
than 20 GB/s. For the ILC Low-P parameter set backgrounds will increase by a factor 5 to 10 for
the different subdetectors, with a resulting higher bandwidth demand.
%
\begin{table}
	\centering
		\begin{tabular}{|l|r|r|r|} \hline
		Subdetector & Channels $[10^6]$ & Occupancy [\%] & Data volume $[MB]$ \\ \hline
		VTX  & 800  & 1.0 &   50 \\
		TPC     & 2    & $<0.1$ &   12 \\
		FTD     & 1  & 9  &    2 \\
		SIT     & 1    & 30  &    6 \\
		SET     & 5    &  1  &    1 \\
		ETD     & 4    & 10  &    7 \\
		ECAL    & 100  & $<0.1$ &   3  \\
		HCAL    & 8    & 1 &  130 \\
		MUON    & 0.1  & $<0.1$ & $ \leq 1 $ \\
		LCAL    & 0.2  & 70  & 4    \\
		BEAMCAL & 0.04 & 100 & 126  \\
    \hline
		TOTAL	& $\approx$920 &  & $\approx$340 \\
    \hline
		\end{tabular}
	\caption{Data Volume in MB per bunch train for the major ILD detector components}
	\label{tab:DataVolume}
\end{table}

Since machine parameters and beam conditions like beam energy or polarisation will be a vital
input for the high precision physics analyses they should be stored alongside with the data.
The time structure and data volume are similar, hence a common DAQ and data storage model
is envisaged.

\section{Front end electronics}
\label{sec:daq_fee}
In contrast to the central DAQ system the front end readout electronics for the
different subdetector prototypes has to be designed now to allow for a realistic
engineering and detector performance tests. Several approaches are underway for
the calorimeters, TPC, silicon trackers and vertex detectors. Common to all the
designs is a highly integrated front end electronics with signal shaping, amplifying,
digitizing, hit detection, data storage and highly multiplexed data transfer
to reduce the number of cables. Some designs foresee data processing like noise
detection or cluster finding already at this stage to further reduce cables.
For a detailed description see for example  Section~\ref{sec:calo_daq} on the calorimeter readout
which had been developed for the EUDET project with a unified test beam DAQ system
for various ILC Calorimeters.

For a highly granular detector like the ILD with the resulting large channel counts
both the material budget as well as the power consumption are areas of concern.
Minimizing the number of cables by data processing and multiplexing
already on the sensor level is required as well as high density electronics with
low power consumption. A common approach to reduce the power consumption is to
turn the front end electronics off in the train gaps. First systems have been
designed and build with this power pulsing capability.


\section{Detector Control and Monitoring}
\label{sec:daq_dcm}
Modern data acquisition, detector control and monitoring systems are closely coupled
to ensure good efficiency and data quality.
An overall experiment control system will keep DAQ, detector control
and monitoring synchronised, and assure proper timing, conditions and error handling.
The systems should be designed such that subdetectors can be treated independently
for commissioning or calibration runs in parallel to collision data taking.

In addition the ILD will be operated in a truly worldwide collaboration with partners
all over the world.
Similar to the global accelerator network (GAN) a  global detector network (GDN) 
is proposed to operate the ILD detector remotely from the participating institutes.
First experience was gained with the CALICE remote control room at DESY during test
beams at CERN and FNAL. In addition the experience from the CMS remote operation
centers at CERN, FNAL and DESY will be taken into account.
The design of the DAQ and control system should have remote operation features
built in from the start.

\section{Data Processing}
\label{sec:daq_dataprocessing}
\subsection{Event Building and Prompt Reconstruction}
\label{subsec:daq_eventbuilding}
Event building and prompt reconstruction will be performed on the \textit{Online Filter Farm} -- 
a sufficiently large farm of processing units near the detector, connected to the front end electronics
via the Common Event Building Network as shown in figure~\ref{fig:DAQ_Layout}. 
Every data processing unit of the Online Filter Farm will process the data of one complete bunch 
train at a time. The raw data of a complete bunch train is kept in the 
raw data file after compression. This is essential as many detectors will integrate over several bunch
crossings or even the full bunch train. The event reconstruction will be an iterative procedure where 
in a first step a preliminary reconstruction will be done on the data from every subdetector. 
Bunches of interest are then identified by exploiting correlations in time and space between the data from 
all subdetectors. After calibration and alignment finally a full event reconstruction is performed on the event data.
The reconstructed events are then written to the storage systems 
in an object oriented data format that is suited for further analysis with appropriate pointers 
into the raw data file containing the bunch train data. A first version of such an event data model 
has been developed and is in used for several test beam efforts and for the offline analyses within ILD \cite{ref:lcio}. An event filter mechanism run at prompt reconstruction 
will provide the necessary meta data for fast event selection at the physics analysis level.

\subsection{Offline computing}
\label{subsec:daq_offline}
The further offline data processing will exploit the existing Grid infrastructure for distributed computing
using a multi-tier like approach similar to what is done for the LHC-experiments \cite{ref-ATLAS-CMP_TDR,ref-CMS-CMP_TDR}. 
The offline computing tasks such as the  production of more condensed files with derived physics quantities (DST/AOD), Monte Carlo
simulations and re-processing of the data will be distributed to the various tiers of the ILC-computing Grid.
Setting up a data Grid and suitable data catalogues will allow the physicists to efficiently access the data 
needed for their analyses.

\
\section{Software}
\label{sec:daq_software}
As is fairly standard with current HEP experiments, ILD will have a modular software framework based on object oriented programming 
languages such as C++. A component based plugin system together with well defined abstract interfaces will allow the flexible
combination and exchange of algorithms with minimal configuration overhead. Using the same software components in the 
online and offline computing as much as possible will facilitate frequent data re-processing with improved calibrations and 
algorithms. A common object oriented persistency format that is used from the prompt reconstruction to the final analysis will 
allow transparent access to lower level data objects at later stages of the data processing chain. The actual raw data format 
containing the data read out per bunch train will essentially be defined by the front end electronics.
Using abstract interfaces for the object oriented event data model will provide the flexibility to change the underlying persistency
format in case the need might arise at some point in the lifetime of the experiment.
The long term archiving of the data even beyond the experiments lifetime should be taken into account when choosing the 
persistency format. 
Data for calibration, slow control and alignment will be stored in a conditions database system that provides a suitable timestamp mechanism
and versioning capabilities.

ILD already has two fully functional software frameworks (Sections \ref{sec:optimisation-simulation-jsf}, 
\ref{sec:optimisation-simulation-mokkamarlin}) which have been used for the
massive Monte Carlo production for detector optimisation. 
One of the frameworks is also used by a number of ILC testbeam experiments for detector R\&D \cite{ref-EUDET-Marlin} providing important feedback 
on the usability of the software in real world experimental conditions.

\section{Outlook and R\&D}
\label{sec:daq_randd}
Due to the timescales involved and the rapid changing computing and network market
a decision on the DAQ hardware will be done as late as possible
to profit from the developments in this area.

Nevertheless key elements of the DAQ systems have to be defined to guide the R\&D of the
subdetectors front end electronics especially when entering the technical prototype stage.
This includes standardized interfaces to the central DAQ, online calibration and alignment
strategies as well as detector control and monitoring concepts with remote operation build in.
To gain experience and prepare for the decision on how to build the central DAQ system new
developments and evolving standards, like ATCA \cite{ref_ATCA} for example, should be evaluated
in the next years.

In addition efficient event processing and event classification strategies taking
realistic background simulations into account have to be developed. The ongoing 
test beam and simulation efforts will be a first step to learn how to treat the data but a more
dedicated effort toward online processing and bunch tagging in a multi bunch environment will
be needed to ensure efficient processing.

GRID resources currently available for post-LOI studies are limited. Especially, the inter-regional 
bandwidth ( between Europe and Asia ), CPU and storage resources especially in Asia, and 
the connectivity to OSG-GRID are the area which need more improvement.

A continues effort on improving the software framework according to needs and adapting to
new computing hardware, introducing multi-threading capabilities for multi core processors
for example, is needed to keep up with the activities both in the subdetector R\&D as well
as the computing infrastructure


%% file: ild/integration/integration.tex
\section{Mechanical Concept}

\begin{figure}[tb]
	\centering \includegraphics[width=8cm]{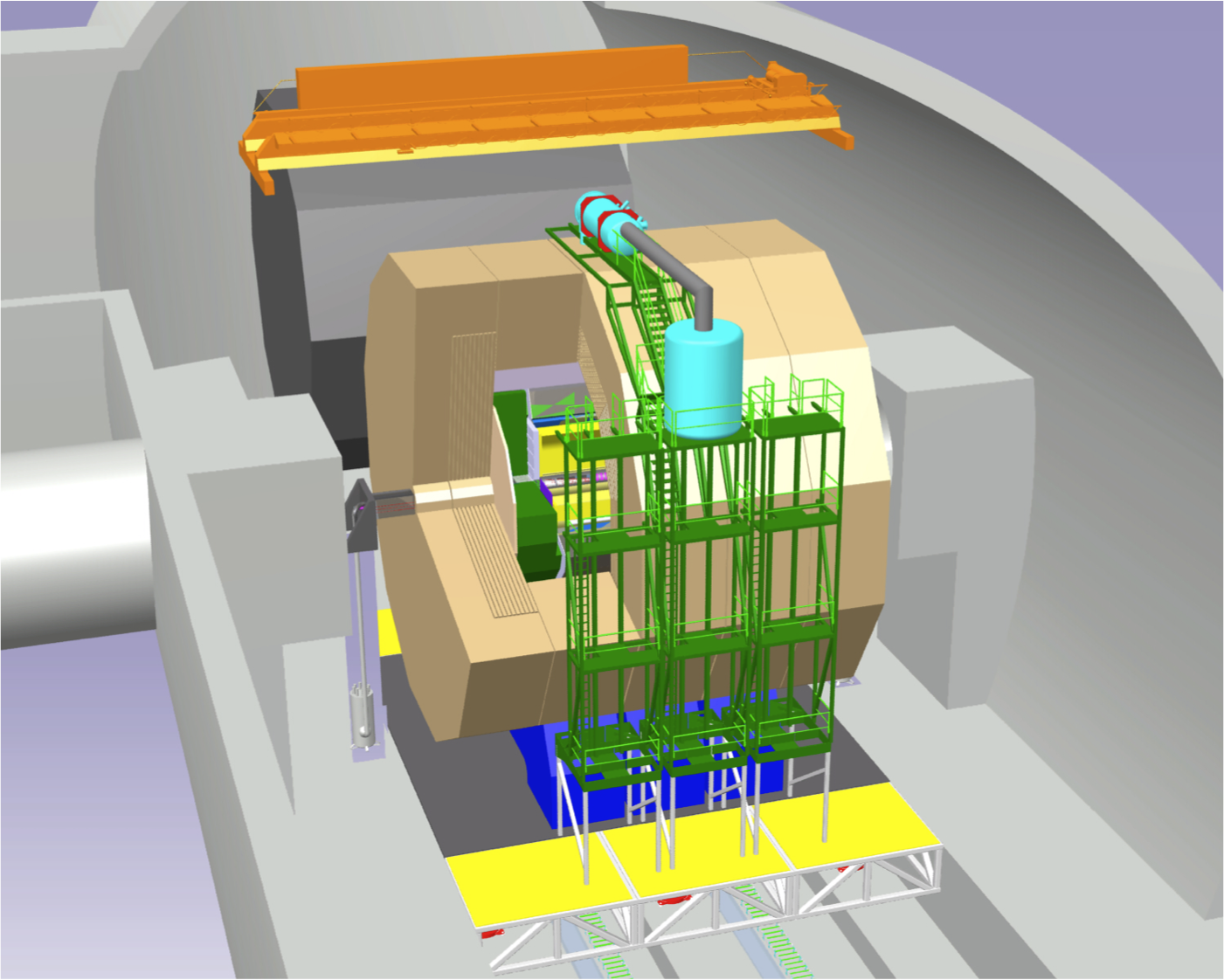}
	\caption[The ILD detector.]{\label{fig:ILD00} The ILD detector and its services.}
\end{figure}
The mechanical design of the ILD detector is shown in figures~\ref{fig:intro:detector} and \ref{fig:ILD00}. The major components are the five parts of the iron return yoke: three barrel rings and two endcaps. The central barrel ring carries the cryostat with the solenoid coil in which the barrel calorimeters are installed. The TPC and the outer silicon envelope detectors are also suspended from the cryostat using tie rods. The endcap calorimeters are supported by the endcap yoke sections which can be moved independently from the barrel sections to allow an opening of the detector at the beam line. The beam pipe, the vertex detector and the other inner silicon detectors are supported from a structure of carbon fibre reinforced plastic (CFRP), which hangs at the flanges of the TPC field cage. The whole structure can be aligned with respect to the beam axis using actuators and a laser alignment system. The QD0 magnets are mounted independent of the yoke endcaps in a support structure that carries the magnets and the forward calorimeters. This structure is supported from a pillar outside of the detector and is suspended from the solenoid cryostat using tie rods. The QD0 magnets are also monitored by a laser alignment system and can be moved using actuators.

A full 3D CAD model of the ILD detector exists and is the baseline for all engineering and technical studies presented in this LoI. The actual dimensions and masses of the ILD subcomponents can be found in~\cite{ILD_Dimensions}. The engineering model of ILD has been synchronised to the detector model used in the full detector MC studies. A detailed description of the integration of the ILD detector including a conceptual scheme for the cabling is given in~\cite{ILDIntegration}.




\section{Detector Assembly and Opening}
\label{sec:integration:opening}
The ILD detector will be assembled in large parts in a surface building above the underground experimental hall. The pre-assembled sections will then be lowered into the underground cavern using a temporary portal crane. The largest and heaviest ($\approx$3500t) part will be the central barrel ring with the solenoid coil and the barrel calorimeters installed. 
The underground assembly sequence comprises the following steps:
\begin{enumerate}\addtolength{\itemsep}{-0.5\baselineskip}
\item The first pillar for the support of the QD0 magnet is installed. This pillar needs to be movable in the garage position but will not move on the beam line. The service helium cryostat is also carried by the pillar.
\item The QD0 magnet is suspended from the pillar together with its support structure.
\item The endcap yoke with the pre-mounted endcap calorimeters are installed.
\item The first part of the barrel yoke is installed.
\item The central part of the barrel yoke carrying the coil and the barrel calorimeters is installed; the cables are routed through the slits between the central yoke ring and the other rings.
\item The TPC is inserted, the cables follow the same routes as the cables of the barrel calorimeters.
\item The inner part of the detector including the beam pipe and the inner silicon detectors is inserted into the TPC, cables are routed to the outside following the same routes as the cables of the barrel calorimeters.
\item The third part of the barrel is lowered.
\item The second pillar, QD0 (with support) and the second yoke endcap are installed.
\end{enumerate}

\begin{figure}[tb]
	\centering	 \includegraphics[height=6cm]{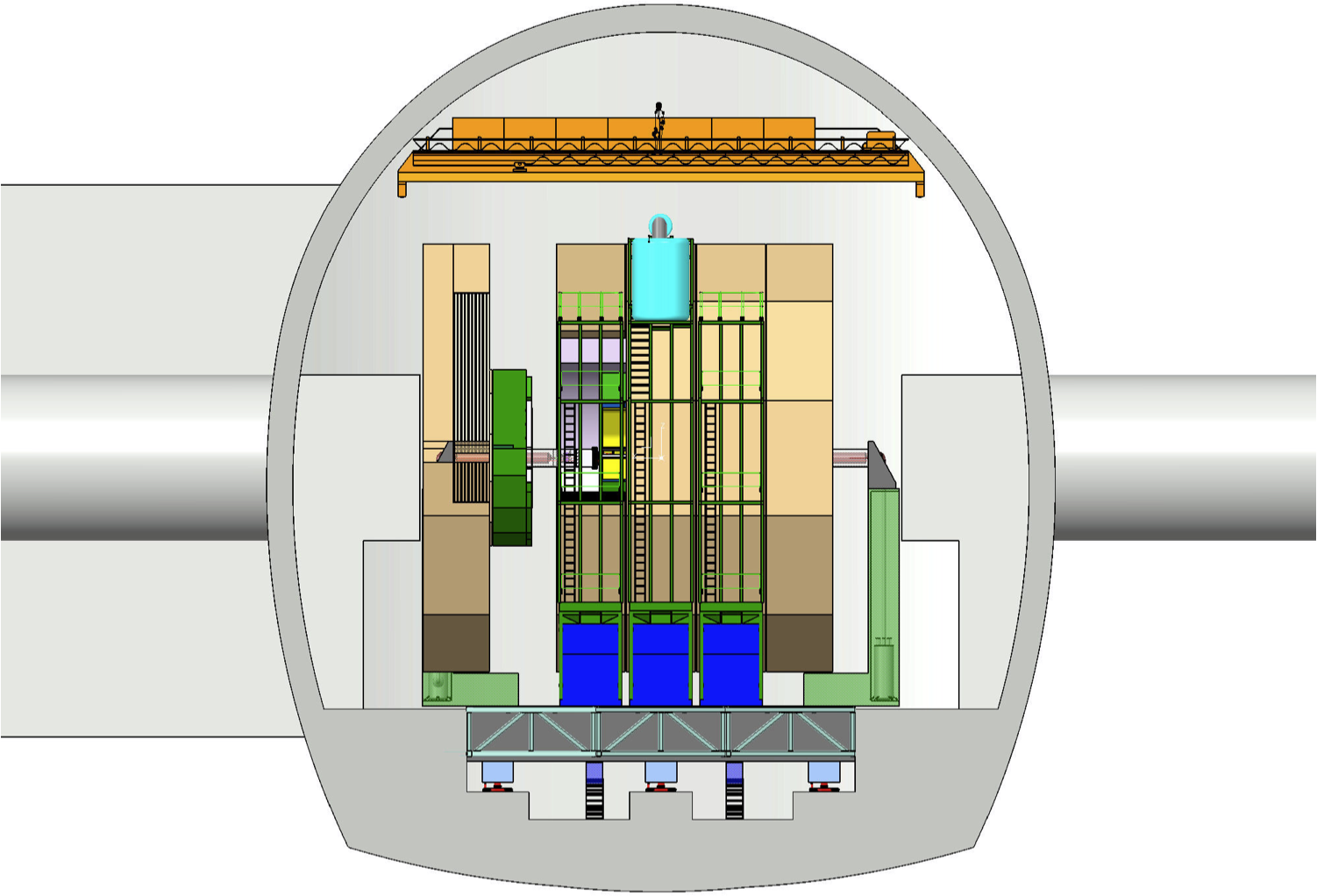}
	
	\centering \includegraphics[height=6cm]{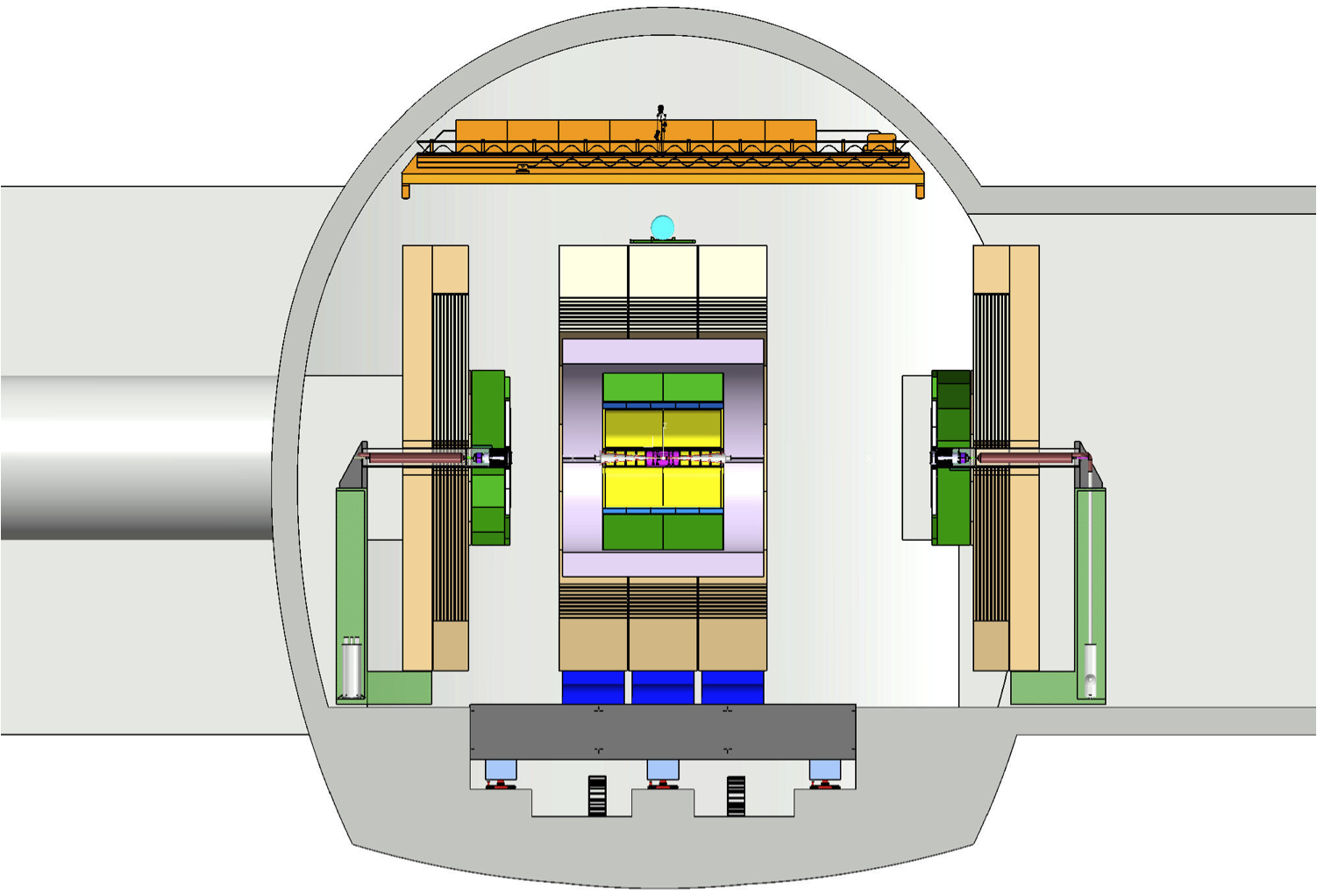}
	\caption[Detector opening procedures.]{\label{fig:ILDopening} Detector opening on the beamline (top) and in the garage position (bottom).}
	\end{figure}

The detector can be opened and maintained in the garage position using the above described procedures. The space required is available in a garage zone with side access tunnels~(c.f. figure~\ref{fig:ILDopening}). It is planned to allow the opening of the detector endcaps also at the beam line. As space is limited there due to the machine elements, only limited access can be reached. The QF1 magnets including their ancillaries extend to $\approx$9~m from the IP. Preliminary studies show that access space of $\approx$1~m can be gained between the endcap and the barrel yoke. This would allow limited access to the inner detector for short maintenance and connection/deconnection actions at the beam line (c.f. figure~\ref{fig:ILDopening}). Major interventions would be done in the garage position taking advantage of the push-pull system.

\section{Civil facilities and services}
\subsection{Detector Services}
\label{sect:integration:services}
A number of services are needed for the operation of the ILD detector. The concept of push-pull puts stringent requirements on the design, as the services need to be designed for a moving detector. Therefore they must be integrated in the design of the detector and the civil facilities from the beginning.

\subsubsection{Primary Services}
Primary services are usually provided by installations that are located on surface due to their dimensions, possible impact on the detectors (vibrations, etc.) and related risks. Examples are water chillers that provide cooling water, high to medium voltage power transformers (e.g. 18 kV/ 400 V AC tri-phase), UPS facility (Diesel generator), helium storage and compressor plant for the solenoid coil and gas and compressed air plants.

\subsubsection{Secondary Services}
\label{sect:integration:secon_services}
Secondary service plants often need to be close to the detector and should be located in the underground areas. Typical secondary services are temperature stabilised cooling water, voltage supplies for front-end electronics, AC/DC converters for the super-conducting coil and cryogenics and vacuum services. Data connections for the transmission of the detector readout also need to be included. Due to the push-pull design, these services are permanently connected and run in cable-chains towards the detector. As the flexible pipes and cables in the chains need to be kept within reasonable lengths, it would be very convenient to locate a small service cavern for the secondary services at the end of  the main underground cavern with independent ventilation and limited crane access. Electrical noise and vibrations are kept away from the vicinity of the detector, and people working in the main underground cavern would also be protected from physical noise coming from various equipment.
The main benefit of the usage of cable chains is the permanent connection of the detector to all its services and readout cables. The chains can be equipped when the detector is still being assembled on the surface, and this would greatly speed up the connection and commissioning time in the underground cavern once the detector parts are lowered. The hall floor can be kept clean and without obstructions by the use of cable chains.

\subsubsection{On-Board Services}
Some secondary services need to be carried on board with the detector if the connection through cable chains are found to be technically difficult or too expensive. As this increases the risks for the detector operation in the push-pull scenario, these on-board services should be kept to a minimum. Examples are the service cryostats for the helium supply for the solenoid and the QD0 magnets.

\subsubsection{Cryogenics}

\begin{figure}[tb]
	\centering
		\includegraphics[width=13cm]{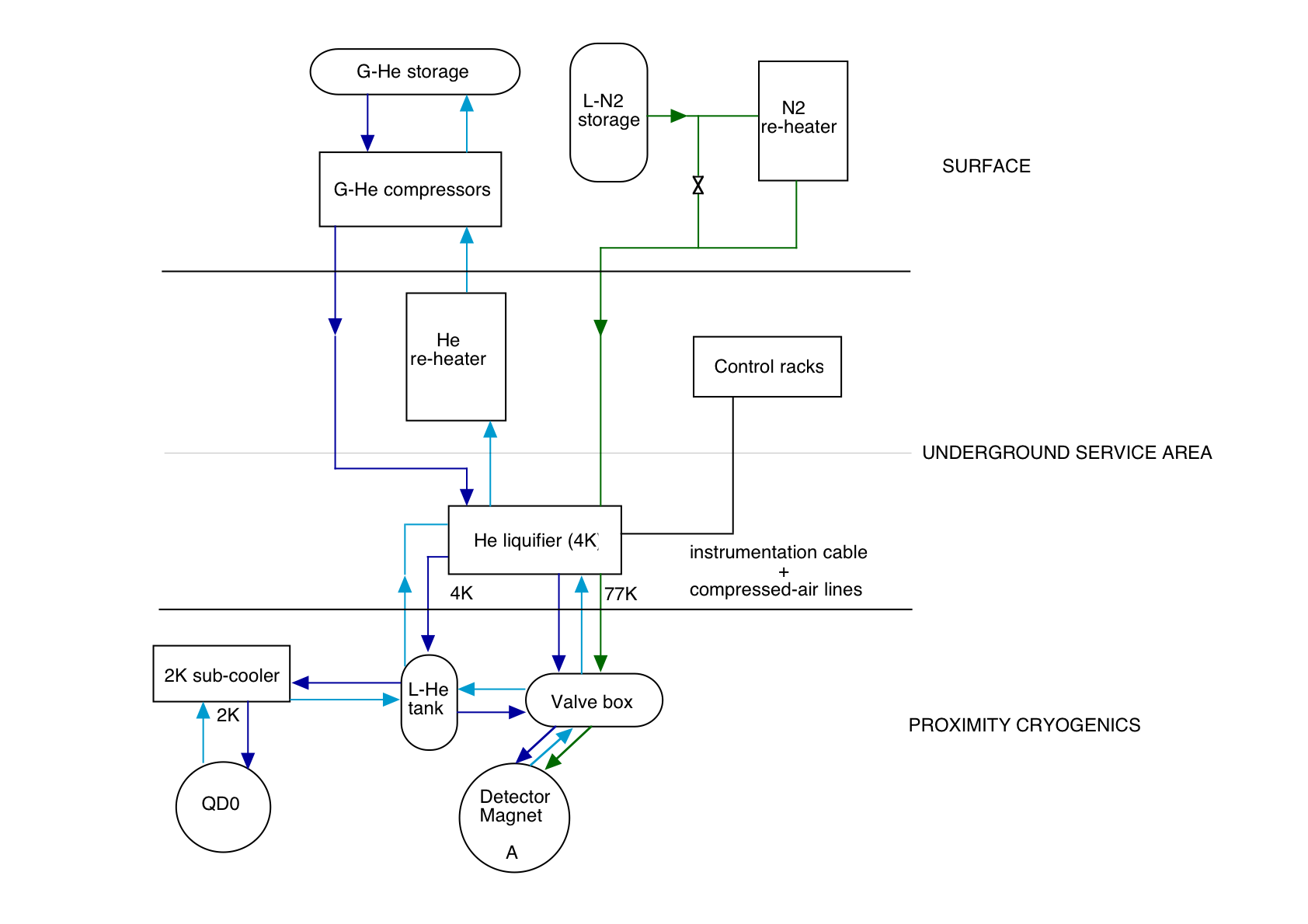}
	\caption[Cryogenics block diagram.]{\label{fig:cryogenics}Diagram of the cryogenic services for the detector.}
	
\end{figure}

Figure~\ref{fig:cryogenics} shows the block diagram of the cryogenics needed for the operation of the detector solenoid coil and the QD0 magnets. While the primary facilities like the helium storage and compressors are on the surface, the helium liquefier (4K) and the re-heater are in the underground hall. Directly on-board of the detector are the valve box, which distributes the helium to the coil, and the liquid helium tank. Also the 2K sub-cooler and the service cryostat for the QD0 magnet are moving with the detector.

\subsection{Surface Assembly Hall}
The detector will be assembled in a surface hall. The RDR baseline design of the surface hall (100 $\times$ 25m, 25~m high, 400~t crane capacity) is well suited for the assembly of the ILD detector. A portal crane with a capacity of 3500~t needs to be installed temporarily at the main shaft to lower the pre-assembled detector elements into the underground cavern. 

\subsection{Underground Experiment Hall}
\begin{figure}[tb]
	\centering
		\includegraphics[width=13cm]{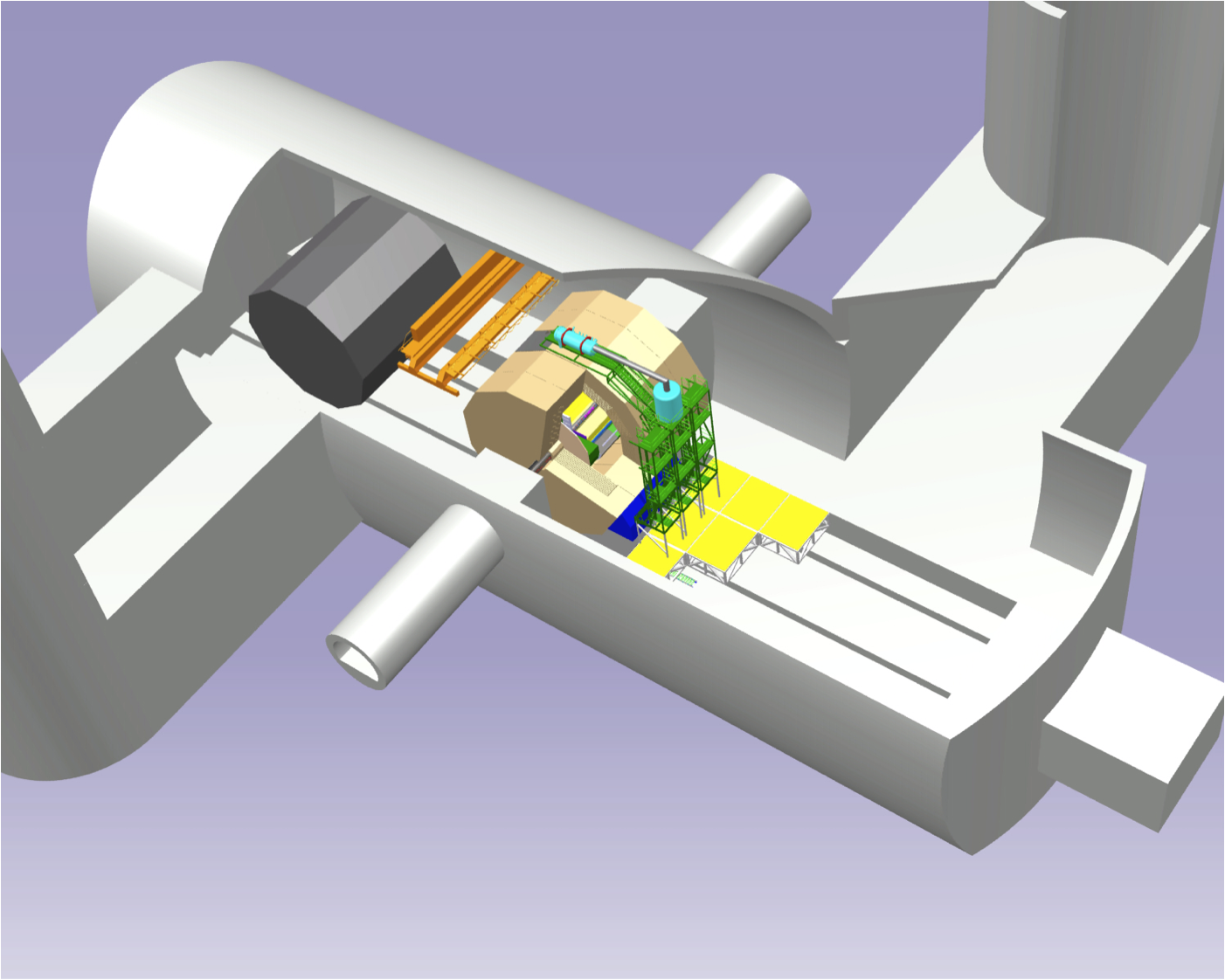}
	\caption[Design study for the underground experimental hall.]{\label{fig:ILDhall2}Design study of the underground experiment hall with ILD (left) and the second detector in push-pull configuration.}
	
\end{figure}

The underground experiment hall needs to accommodate both push-pull detectors. The design of the hall presented in the ILC RDR~\cite{RDR_machine} has not been optimised taking into account realistic assumptions for the services of the detector in the push-pull environment. Figure~\ref{fig:ILDhall2} shows a design study of the underground hall which has been optimised for ILD taking into account the following criteria:  
\begin{itemize}\addtolength{\itemsep}{-0.5\baselineskip}
\item minimum impact on civil engineering cost with respect to the RDR baseline (reduced main cavern diameter and length),
\item enhanced safety and reduced time losses by moving the shaft from the cavern ceiling to a side alcove providing also the necessary space for the full opening of the detector,
\item small service cavern at the end of the hall for the secondary services (\ref{sect:integration:secon_services}),
\item optimised for push-pull.
\end{itemize}

A hall width of 25~m is sufficient for the detector assembly and maintenance procedures as the side alcove for the vertical shaft increases the parking positions substantially. The beam height has been assumed to be 12~m from the floor of the hall. This allows for the detector to rest on a 2-m-high platform. It must be noted that if the experimental area is a deep construction in a terrain requesting a cylindrical or semi-cylindrical hall for question of rock stability the space situated below the floor level has to be filled with concrete up to the proper level. The platform and the cave-like structures containing the supporting mechanism and the cable are thus just a special part of this filling. 
As the heavy parts of the detector will be moved on air pads and on the platform, the crane capacity in the underground hall is modest. Two 40t cranes, which can be connected to form an 80t crane are largely sufficient.

\section{Push-pull Operations}
\label{sec:mdi:pushpull}
The present ILC baseline design foresees one interaction beamline that needs to be shared by two detectors in a push-pull configuration. While one detector is taking data on the beamline, the other one is parked in its garage position in the same underground hall. Following a still-to-be-defined time schedule, the detector on the beam line is moved away to its own garage position to make space for the waiting detector to collect data. This push-pull scenario has never been tested at existing accelerators and poses unprecedented engineering challenges to the detector designs.

The LoI concept groups and the ILC Beam Delivery System group have agreed on a set of minimum functional requirements~\cite{MinimumRequirements} which define the boundary conditions for the push-pull operations. Most of these requirements comprise geometrical boundaries, like the size of the underground hall or the limits of the garage position of the detectors. But also physical limits for ionising radiation and magnetic fields need to be defined to allow a friendly co-existence of two detectors in one hall. In addition direct requirements come from the machine itself. As the QD0 final focus magnets will move with the detectors, requirements on alignment tolerances and vibration limits have been defined.

The timescale for the push-pull operation needs still to be defined, but it is clear that the time for the exchange of both detectors needs to be minimised to maximise the integrated luminosity. The full push-pull procedure comprises for the outgoing detector:
\begin{itemize}\addtolength{\itemsep}{-0.5\baselineskip}
\item securing the beams,
\item powering down of the detector solenoid,
\item removing the radiation shield between detector and hall,
\item disconnecting all local supplies (in principle only the main bus-bars),
\item disconnecting the beam pipe between the QD0 and the QF1 magnets,
\item moving the detector out towards its garage position,
\item connecting back the main bus-bars in the garage position.
\end{itemize}

For the incoming detector the procedure is reversed, but additionally needs to include time for alignment and eventually calibration of the detector system at the beam line.
It is envisaged to complete the full push-pull operations on a timescale of about two days after procedures have been optimised based on experience. However, as the full understanding of the challenges requires a detailed engineering design of the hall and the definition of the procedures, a final evaluation of the push-pull operation for ILD is beyond the scope of this LoI and needs to be studied in the following Technical Design Phase. Nevertheless this section describes our conceptual understanding of the ILD operations.

\subsection{Moving the ILD Detector}
The ILD detector will be placed on a concrete platform to avoid possible damages due to non-synchronised movements or from vibrations during push-pull and also to ease internal alignment challenges. The concrete platform will have a size of approximately 15 $\times$ 20m and needs to be $\approx 2$m thick. Figure~\ref{fig:ILDhall2} shows the ILD detector on its platform on the beam line. The platform will move on the hall floor using a system of rollers suitable solution for this one-dimensional movement or air pads that may provide more easily a millimetric positioning tolerance.

\begin{figure}[tb]
	\centering
		\includegraphics[width=11cm]{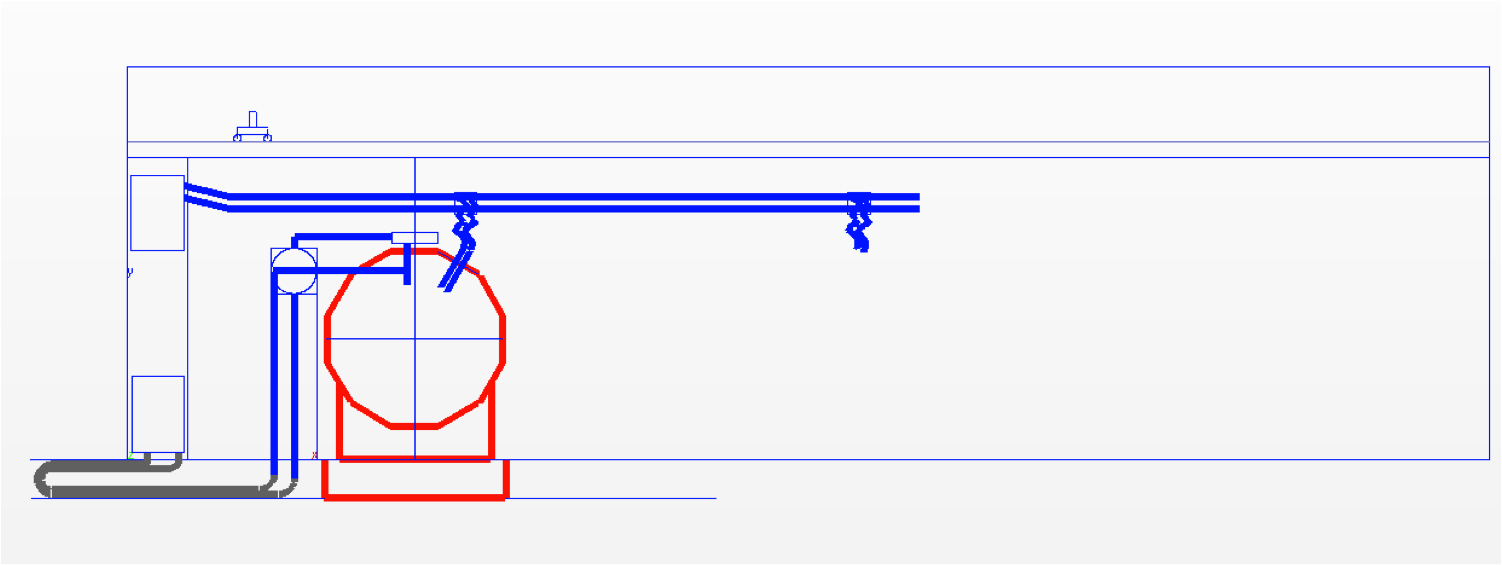}
	\centering
		\includegraphics[width=11cm]{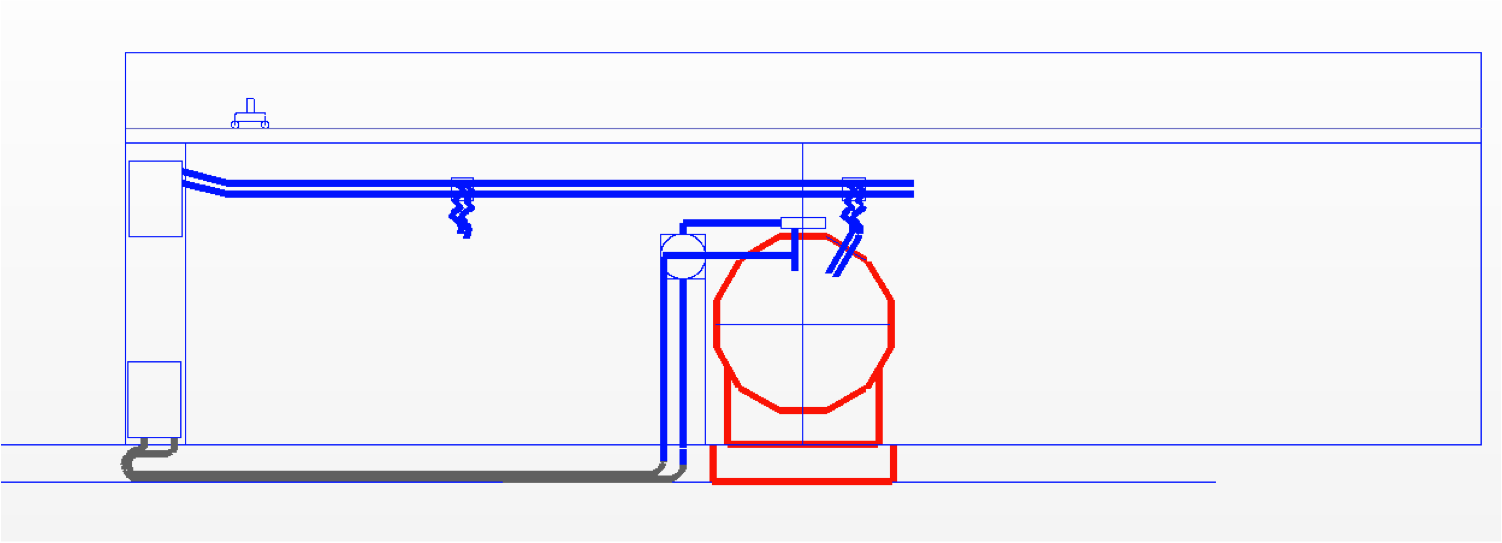}

	\caption[ILD detector in the garage position and on the beam line.]{\label{fig:push-pull}ILD detector in the garage position and on the beam line. The detector will be disconnected from the power bus bars (top, blue) during movement, while the cables and cryogenic service lines run in cable chains (bottom, grey). }
\end{figure}

All supplies for the detector will be provided by using flexible supply lines that move with the platform. This includes the cryogenic lines that supply the detector solenoid and QD0 systems with 4K helium. As the development of flexible cryo lines for 2K helium is challenging, the QD0 magnet will be connected permanently to a service cryostat that moves together with the detector on the platform. As the detector solenoid does not need to be powered during the movement, the detector can be disconnected from the power bus bars and re-connected in the parking position. Figure~\ref{fig:push-pull} shows a schematic drawing of the movement of the detector with the cable chains and the power bus bars.
The detector elements can be moved on the platform by using either a roller system, or by using air pads similar to the solution CMS has adopted. As the axial space at the beam line is limited, only the opening of the detector endcaps is foreseen to allow a limited access to the inner detectors. In the garage position more space is needed to allow major maintenance work, e.g. the removal of the TPC. More details of the opening procedures are described in section~\ref{sec:integration:opening}.

\subsection{Shielding}
The ILD detector will be self-shielding with respect to maximum credible accident beam loss scenarios. Detailed simulations show \cite{integration:self-shielding} that a proper design of the detector provides shielding which is sufficient to still allow access to the detector hall for professional workers. This is important to fulfil the minimum requirements which are needed to allow access to the other detector in its garage position during beam operations.

A movable concrete shield needs to fill the gap between the detector and the walls of the underground hall. As this shielding needs to fit both detectors, no engineering effort has been pursued so far to find a detailed design. This has been referred to the Technical Design Phase where it will be studied in collaboration with the second detector concept group. Nevertheless, these kind of shieldings have been used in other accelerator experiments (e.g. at HERA) and pose no conceptual design challenge.

\subsection{Alignment and Calibration}
The ILD detector will be moved on a platform that will be repositioned very precisely on the IP. Nevertheless, the alignment of the detector after being brought to the beam position is not trivial. The functional requirements ask for an alignment accuracy of the detector axis $\pm$1mm and 100$\mu$rad after push-pull. The requirements for the QD0 magnet are even tighter: $\pm$ 200 $\mu$m and 5 $\mu$rad.
ILD will be equipped with a laser-interferometric alignment system like MONALISA~\cite{MONALISA}. This system allows for alignment of both QD0 magnets which are carried by the detector to the ILC beam lines on each side of the hall. In addition, the detector itself can be positioned within the required tolerances using this system. The QD0 magnets will be placed on actuators that allow for an independent alignment of the magnets with respect to the detector.

After the alignment of the detector described above and the commissioning of the beam, some calibration data taken at the energy of the Z resonance would allow to check the alignment of the subdetectors. The internal alignment of the subdetectors can then be done most precisely in offline analysis. This technique was used at LEP where experience shows that about 1~pb$^{-1}$ of calibration data on the Z peak will be sufficienct after the push-pull procedure at ILC~\cite{ref-markronacfa8}.  If the ILC machine cannot easily switch between Z-peak and $\sqrt{s}$ running, then other techniques than used at LEP will be developed.


\section{R\&D Plans: Detector Integration}
\label{sec:integration:rand}\addtolength{\itemsep}{-0.5\baselineskip}
This Letter of Intent describes a conceptual design of the integration efforts for the ILD detector. Though no show-stoppers have been identified, a challenging engineering programme needs to be set up to transform this conceptual design into a technical design based on engineering studies. Main points of work will be among others:
\begin{itemize}
\item Integration of the subdetectors in close collaboration with the subdetector R\&D groups.
\item Engineering design of the yoke.
\item Opening and assembly procedures including a study on the tooling needed for it.
\item Development of an engineering solution for push-pull. This includes the very important issue of cryogenic supplies for a moving detector.
\item Find an optimised underground hall design in close collaboration with the ILC machine CF\&S group and the other detector concept groups. This includes a common shielding strategy.
\item Elaborate the alignment strategy and find technical solutions for interferometric systems ‡ la MONALISA. 
\end{itemize}

%% file: ild/accelerator/accelerator.tex

\section{The Interaction Region}

\begin{figure}[htb]
	\centering \includegraphics[width=13cm]{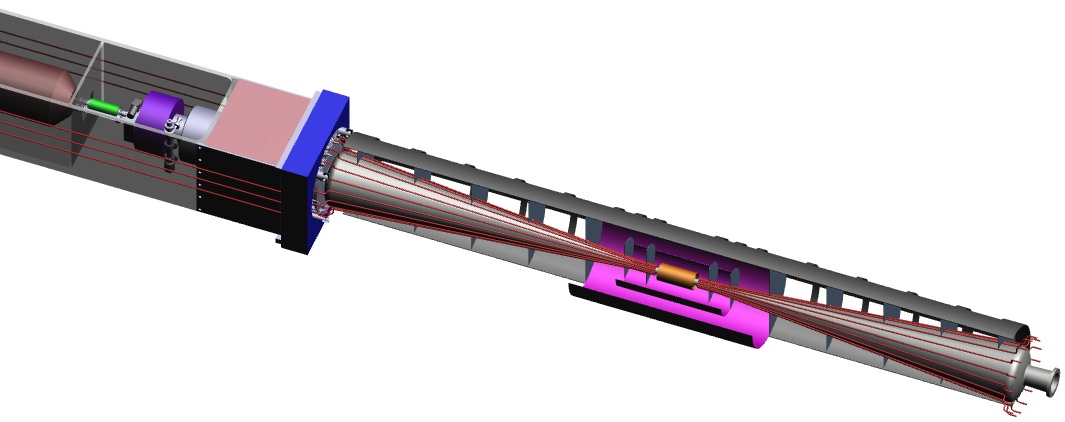}
	
	\caption[Interaction region of the ILD detector: inner detectors.]{\label{fig:ILDIR2}Interaction region of the ILD detector. Shown are the vertex detector (yellow), the SIT (pink), the ECAL plug (blue) with the LumiCal, the LHCAL (light red) and the BeamCal (violet). The routing of the cables is also shown.}
\end{figure}

The interaction region of the ILD detector comprises the beampipe, the surrounding silicon detectors, the forward calorimeters with the masking system and the QD0 magnet with its ancillaries and the support structure. Figure~\ref{fig:ILDIR2} shows a blow-up of this region.

\subsection{The Beampipe}

The design of the beam tube has to obey several constraints:
\begin{enumerate}\addtolength{\itemsep}{-0.5\baselineskip}
\item It must not interfere with the luminosity.
\item Its central part must be small enough to optimise the measurement of the impact parameter and large enough not to interfere with the background.
\item It must comply with a crossing angle of 7 mrad.
\item It must be as light as possible to reduce photon conversions and hadron interactions, withstanding nevertheless the atmospheric pressure.
\item It must not induce electromagnetic perturbations generating heat.
\item It has to be pumped down to an agreed level.
\end{enumerate}

\subsubsection{Mechanics}
\begin{figure}[htb]
	\centering \includegraphics[width=13cm]{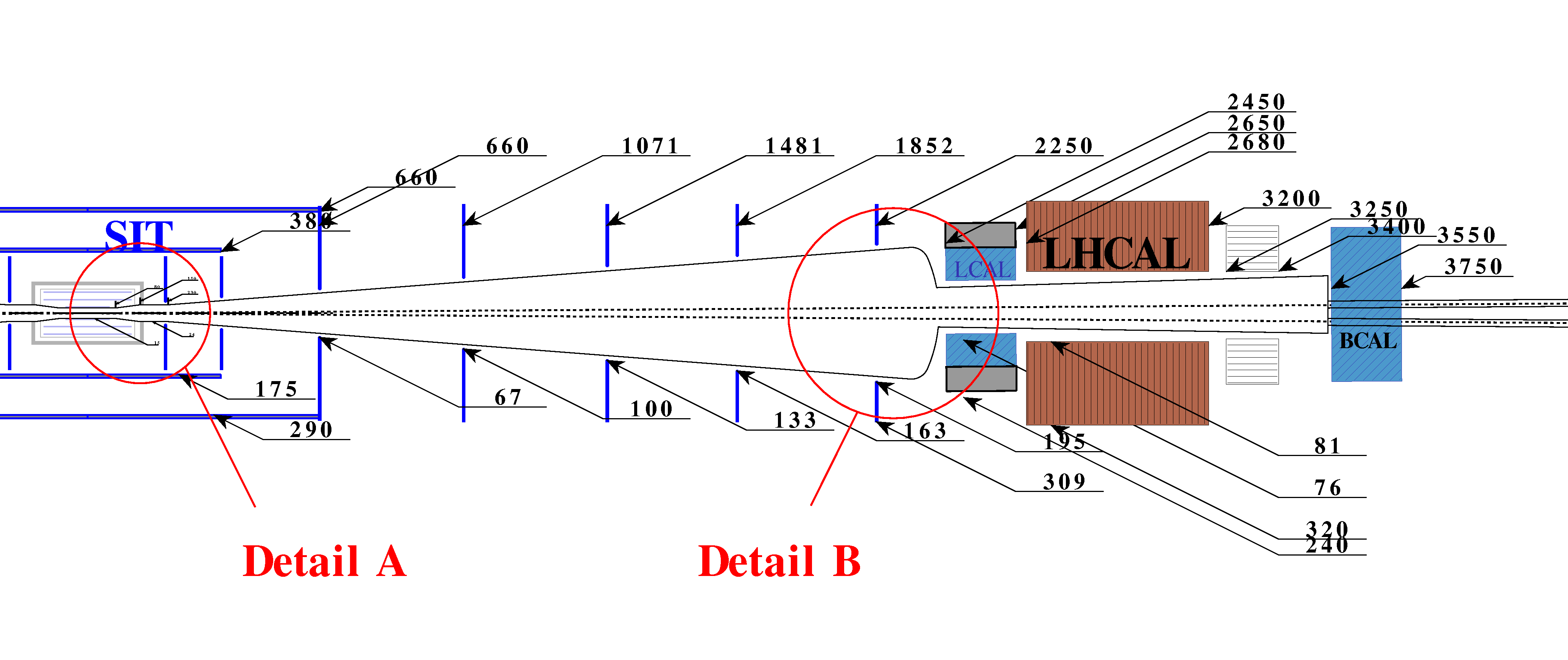}
	
	\caption[Beam pipe geometry.]{\label{fig:bpgeom}Beam pipe geometry.}
\end{figure}

The current mechanical design is shown on figure~\ref{fig:bpgeom}. The tube is conical, offering very little matter in front of the luminosity monitor LumiCal (LCAL). The tube is made of beryllium with some ring reinforcements at the level of the forward detectors. The part inside the vertex detector is a cylinder, the connection with the large cone is such that background does not interfere. This very thin tube (8~kg total) has to be supported from the inner detector structure. The mechanical behaviour has been studied in detail~\cite{videau_beampipe}. The most important constraint is related to the buckling where a safety coefficient of 6 is being considered, see figure~\ref{fig:bpvacuum}. To go beyond this first approach, one would need to work with manufacturers because the technology becomes very important. The example of LHCb tube shows that such a tube results from a strong R\&D by the manufacturer on the way to realise cones and to perform the weldings. Figure~\ref{fig:bpgeom} shows the dimensions, figure~\ref{fig:bpvacuum} shows the buckling behaviour.

\subsubsection{Vacuum}
\begin{figure}[htb]
	\includegraphics[width=7cm]{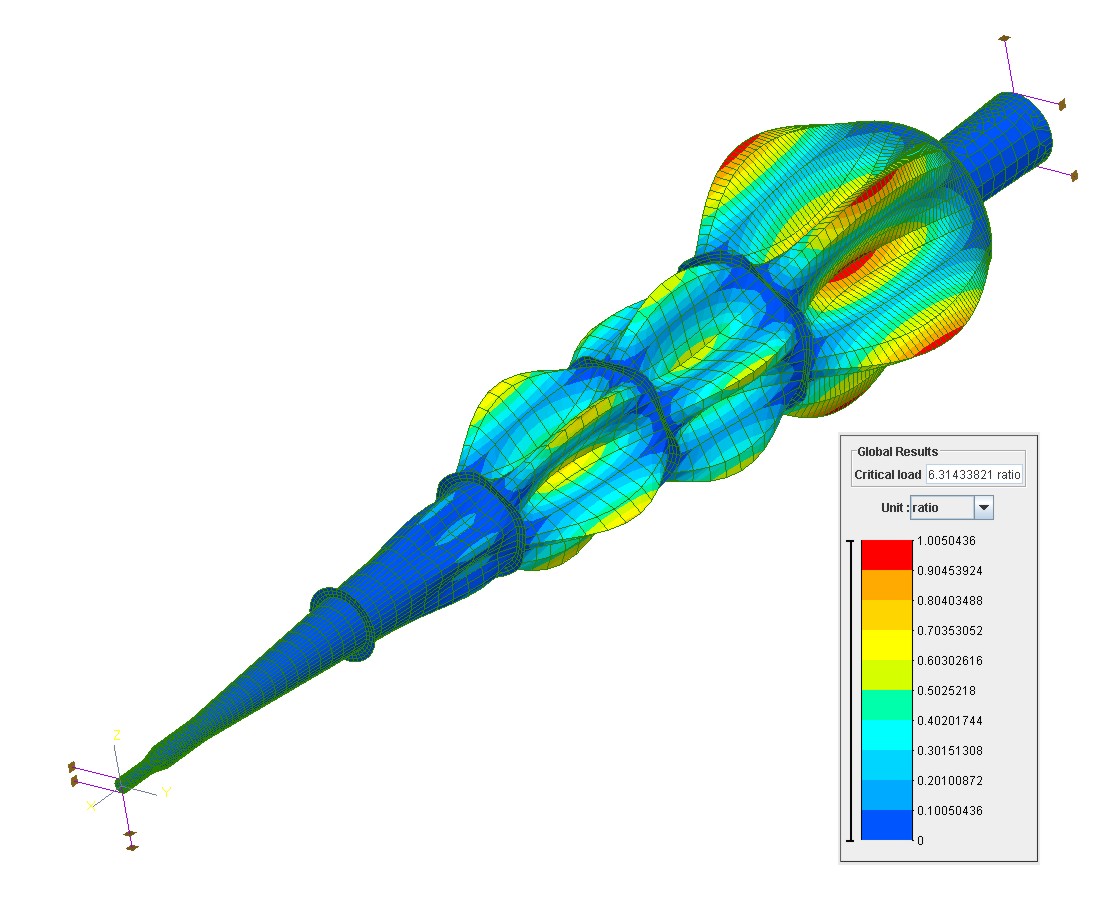}\includegraphics[width=7cm]{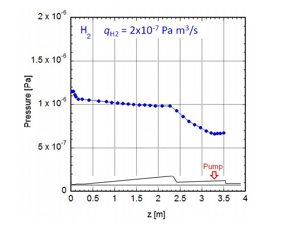}
	\caption[Beampipe buckling and vacuum profiles.]{\label{fig:bpvacuum}Beampipe buckling (left) and vacuum profile for H$_2$ (right).}
\end{figure}

The vacuum situation in the beam pipe has been studied in detail~\cite{suetsugu_beampipe}. It is assumed that the whole beam pipe would be pumped by vacuum pumps which are located in the space between the LHCAL and the BeamCal, about 3.3~m from the IP. Assuming effective pumping speeds of 0.72 (0.12) $m^3s^{-1}$ for H$_2$ (CO) yields in pressures of approximately 1$\times$10$^{-6}$ (6$\times$10$^{-7}$)~Pa. The limit on the effective pumping speed is given by the conductance of the small pipe at the back side of the conical part. The vacuum profile for H$_2$ is shown in figure~\ref{fig:bpvacuum}.

\subsubsection{Wakefield Losses}

The wakefields generated by the passing beam in the beam pipe result in parasitic losses which have been studied taking into account the ILC Low-P beam parameters~\cite{suetsugu_beampipe}. The parasitic losses are in the range of 20-24~W so that air cooling will be sufficient to remove this additional power. Higher order modes can be excited in the beam pipe, the resulting parasitic losses are small and will be dissipated in the surface region of the beam pipe. 

\subsection{Support of the Final Focus Magnets}

\begin{figure}[htb]
	\centering \includegraphics[width=10cm]{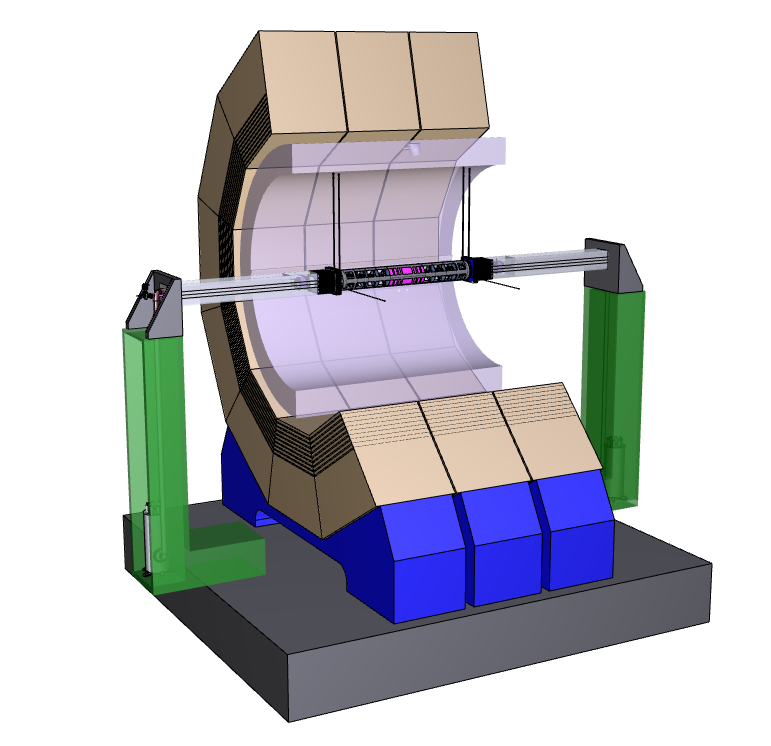}
	
	\caption[Support of the magnets in the detector.]{\label{fig:qd0support}Support of the magnets in the detector. The inner detector part and the beam pipe are suspended from the TPC end flanges, not shown in this figure.}
\end{figure}

While the QF1 magnets of the final doublet will stay fixed in their positions, the QD0 magnets need to move with the detector during push-pull operation. The magnets are installed in a support structure which is supported from pillars residing on the push-pull platform. The support structure has a square cross section and is suspended from the solenoid cryostat using carbon-fibre tie rods (c.f. figure~\ref{fig:qd0support}. This assembly allows the opening of the yoke end caps without interference with the alignment of the QD0 magnets.
The inner silicon detectors (SIT, vertex detector) and the beam pipe are supported by a CFRP structure which is suspended from the TPC end flanges. 
The QD0 magnets reside on actuators in their support structure and are monitored using an interferometric laser alignment system like MONALISA~\cite{MONALISA}. The service cryostat for the supply of the QD0 magnets on the beam line, in the garage position and during the push-pull operations are located at the bases of the pillars and move with the platform.

The stability of the QD0 support structure has been studied. The vibrations induced by ground motions are at most 2.2~nm at 8.3~Hz which is below the limits of 50~nm defined in the minimum requirements document~\cite{MinimumRequirements}.

\section{Machine-Induced Backgrounds}
\label{ild:accelerator:backgrounds}

Machine-induced backgrounds have been studied in detail for the ILD detector and its predecessors GLD and LDC~\cite{ref-gld,ref-ldc}. The main relevant background are pairs from beamstrahlung which are produced in the highly charged environment of the beam-beam interaction. The background levels found are well below the critical limit for most sub-detectors. The sub-detector most sensitive to beam-related backgrounds is the vertex detector, which features an inner radius value dictated by the maximum affordable beamstrahlung hit rate.

Table~\ref{table:pairbackground} summarises the expected background levels in the ILD subdetectors for several beam parameter sets: the nominal ILC beam parameters for 500 and 1000 GeV cms energy and the Low-P parameter set\footnote{The Low-P parameter sets might require modifications to the baseline detector design which are discussed in section~\ref{sec:lowp}. The numbers in the table are valid for the baseline detector design only.}. The numbers are the result of a study using the same full ILD detector Monte Carlo that has been used for the physics studies described in this LoI (c.f. sections~\ref{optimization} and \ref{performance}), i.e. with the nominal detector geometries and 3.5~T magnetic field. In these background  simulations, the double-layer option of the vertex detector has been chosen. It should be noted that the background numbers for the single-layer option are 15\% larger for the inner layer due to the smaller radius. The ILC parameter sets used are the ones described in the ILC Reference Design Report~\cite{RDR_machine}. It should be kept in mind that the numbers given are per bunch crossing (BX). As in the different ILC parameter sets the bunch crossing distance varies, the number of hits per subdetector readout must be scaled accordingly, if the corresponding subdetector integrates over several bunch crossings.  

\begin{table}
\footnotesize
\begin{tabular}[hbt]{|l|l|c|r|r|r|} \hline
Subdetector & Units & Layer & Nom-500 & Low-P-500& Nom-1000 \\ \hline
VTX-DL&hits/cm$^2$/BX&1& 3.214$\pm$0.601&7.065$\pm$0.818 &7.124$\pm$1.162 \\ \cline{3-6}
			 & &2&1.988$\pm$0.464 &4.314$\pm$0.604 &4.516$\pm$0.780\\ \cline{3-6}
			 & &3&0.144$\pm$0.080 &0.332$\pm$0.107 &0.340$\pm$0.152 \\ \cline{3-6}
			 & &4&0.118$\pm$0.074 &0.255$\pm$0.095 &0.248$\pm$0.101 \\ \cline{3-6}
			 & &5&0.027$\pm$0.026 &0.055$\pm$0.037 &0.046$\pm$0.036 \\ \cline{3-6}
			 & &6&0.024$\pm$0.022 &0.046$\pm$0.030 &0.049$\pm$0.044 \\ \hline
SIT&hits/cm$^2$/BX&1&0.017$\pm$0.001 &0.031$\pm$0.007 &0.032$\pm$0.012 \\ \cline{3-6}
			 & &2&0.004$\pm$0.003 &0.016$\pm$0.005 &0.008$\pm$0.002 \\ \hline
FTD&hits/cm$^2$/BX&1&0.013$\pm$0.005 &0.031$\pm$0.007 &0.019$\pm$0.006 \\ \cline{3-6}
			 & &2&0.008$\pm$0.003 &0.023$\pm$0.007 &0.013$\pm$0.005 \\ \cline{3-6}
			 & &3&0.002$\pm$0.001 &0.005$\pm$0.002 &0.003$\pm$0.001 \\ \cline{3-6}
			 & &4&0.002$\pm$0.001 &0.007$\pm$0.002 &0.004$\pm$0.001 \\ \cline{3-6}
			 & &5&0.001$\pm$0.001 &0.006$\pm$0.002 &0.002$\pm$0.001 \\ \cline{3-6}
			 & &6&0.001$\pm$0.001 &0.005$\pm$0.002 &0.002$\pm$0.001 \\ \cline{3-6}
			 & &7&0.001$\pm$0.001 &0.007$\pm$0.002 &0.001$\pm$0.001 \\ \hline
SET&hits/BX&1&5.642$\pm$2.480 &57.507$\pm$10.686 &13.022$\pm$7.338 \\ \cline{3-6}
			 & &2&5.978$\pm$2.360 &59.775$\pm$8.479 &13.711$\pm$7.606 \\ \hline
TPC&hits/BX&-&408$\pm$292 &3621$\pm$709&803$\pm$356 \\ \hline
ECAL&hits/BX&-&155$\pm$50 &1176$\pm$105&274$\pm$76 \\ \hline
HCAL&hits/BX&-&8419$\pm$649 &24222$\pm$744&19905$\pm$650 \\ \hline

\end{tabular}
\caption[Background numbers for ILD.]{\label{table:pairbackground}Pair induced backgrounds in the subdetectors for nominal (500~GeV and 1~TeV) and Low-P (500~GeV) beam parameters. The numbers for the ECAL and the HCAL are summed over barrel and endcaps. For the vertex detecor, the double-layer option has been chosen for this simulation, the numbers for the single-layer option differ. The errors represent the RMS of the hit distributions of the simulation of $\approx$ 100 bunch crossings (BX).}
\end{table}

\begin{figure}[htb]
	\centering \includegraphics[width=11cm]{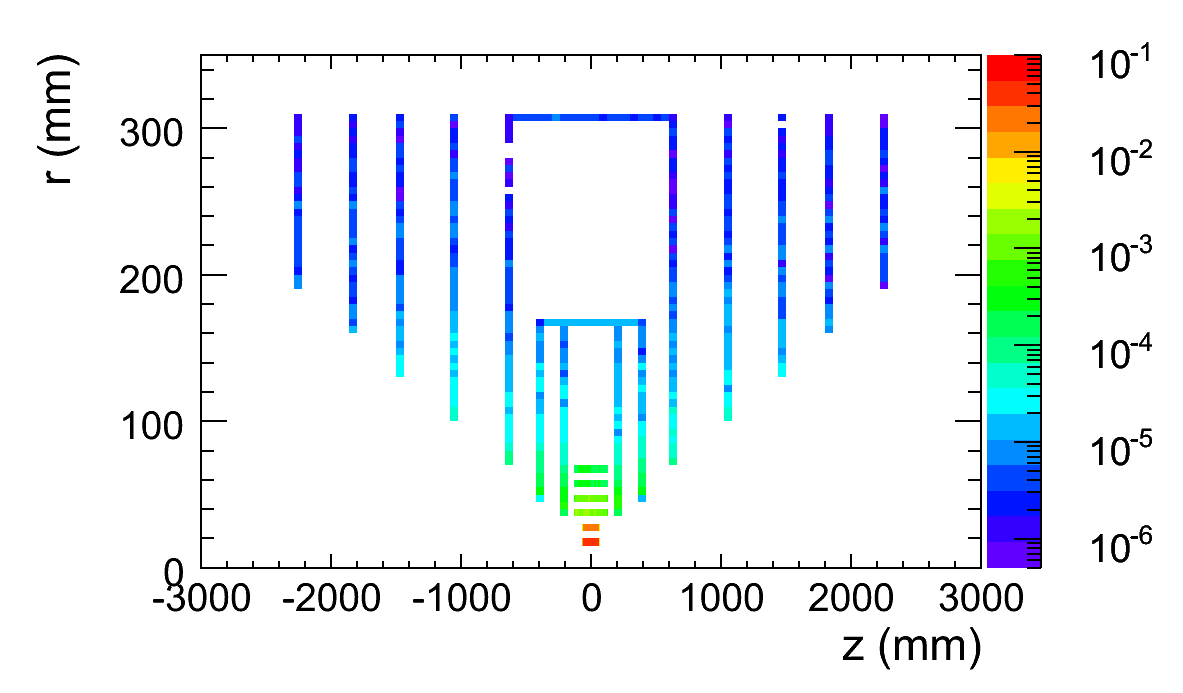}
	\caption[Background hit densities on the inner silicon detectors.]{\label{fig:background_densities}Distribution of the background hit densities on the inner silicon detectors (VTX, SIT, FTD) in units of [hits/mm$^2$/BX].}
\end{figure}

Figure~\ref{fig:background_densities} shows the distribution of the background hit densities on the inner silicon detectors (VTX-DL, SIT, FTD). The correlation of the hit densities with the distance from the interaction point can clearly be seen. The most critical point is the innermost layer of the vertex detector. 

\subsection{Background Uncertainties}

As the vertex detector is most critical with respect to beam induced backgrounds, detailed studies have been performed to understand the influence of different detector geometries and simulation parameters like the choice of range cut parameters in Geant4. The number of hits on the vertex detector change up to 30\% which gives an order of magnitude of the uncertainties for these simulation results~\cite{VTXbg}. Another study of the uncertainties of the background simulations has been done in~\cite{ref-adrianvogelthesis} where a general safety factor of 5-10 has been suggested.

\subsection{Provisions for the Low-P Beam Parameters}
\label{sec:lowp}
The Low-P beam parameter set assumes an ILC machine with less RF-power available. The number of bunches in a train is reduced while the loss in luminosity is compensated by squeezing the bunches to smaller sizes during the collisions. This results in larger losses due to beamstrahlung photons and therefore in a diluted luminosity spectrum. On the other hand the number of pairs produced in beamstrahlung gamma collisions is enhanced significantly.

Table~\ref{table:pairbackground} shows how the increased pair production results in an increase of background hits in the subdetectors. Especially the inner vertex detector layer suffers from the enhanced backgrounds\footnote{It should be remembered, however, that the numbers in the table are given per bunch crossing. As the bunch spacing in the RDR Low-P option is increased to 480.0~ns as compared to the 369.2~ns in the nominal case, the background numbers per readout need to be scaled down by $\approx$25\% for Low-P to make them comparable to the nominal case.}. Matters to remedy the situation are still under study a possible solutions is an increased radius of the inner vertex detector layer. Also the ladder of the inner layers might need to be shortened to keep their read-out electronics out of the hot cone of the beamstrahlung pairs.

\section{Measurement of Energy and Polarisation}

Traditionally the methods of measuring the beam energy and the beam polarisation are also parts of the Machine Detector Interface. As the polarimeters and the energy spectrometers are not a part of the detectors at the ILC but are common facilities in the machine, we will refer here only to the common design efforts~\cite{polarisation_energy}.

\section{R\&D Plans: Machine Detector Interface}
As in the area of detector integration~\ref{sec:integration:rand}, also in the integration with the accelerator many topics need to be studied in engineering detail level before the construction of the detector could be envisaged. Important R\&D topics are e.g.:
\begin{itemize}\addtolength{\itemsep}{-0.5\baselineskip}
\item Engineeering study of the beam pipe including its vacuum behaviour in close collaboration with possible manufacturers.
\item Support and monitoring of the QD0 magnets. Adaptions to new BDS schemes which would allow larger L* optics.
\item Detailed study of the beam induced backgrounds.
\end{itemize}

%% file: cost/cost.tex
The cost of the ILD detector as presented in this section has been evaluated 
based on a common methodology of costing, and a detailed work breakdown 
structure for each of the sub-detectors. 
Costing a detector as complex as ILD at this early stage of the project is
difficult at best, and certainly not precise. For this document, 
a rather crude evaluation has therefore been done focusing on the  cost 
driving items. 
Another important aspect was as well to understand the scaling of the 
cost of the different sub-systems with the main design parameters, to 
realize the impact of the sub-system on the overall cost, and 
to evaluate possible cost savings versus performance.
 
\section{Methodology of Costing}
 
The method used here follows the prescriptions from the GDE at the time of the RDR. 
For each important item a work breakdown structure (WBS) has been developed, and each part 
has been costed to the best of our knowledge up to a level equivalent to about 1\% 
of the total cost.  
The bases of the cost have been experience when constructing the LHC detectors and ILD prototypes, and, in 
some cases, quotes from manufacturers. An attempt has been made to use consistent cost 
numbers across different sub-detectors.  
The prices are expressed in ILC units (ILCU) which correspond in 2006 values to 1 US dollars,  117 Yen or 0.83 EUR.
 
No attempt was made to guess the impact of future escalation. 
Contingencies are currently not taken into account. 
No R\&D costs are included, except in some cases costs for industrialisation.
No maintenance and operation has been estimated.
In some instances, prices for the same item vary widely between different countries and regions. In this case
the price used is the one proposed by the group in charge of this component.
Different options have been costed independently, but the tables and numbers correspond to the detector
which has been simulated for physics studies.
 
It should be noted that the level of detail and even of understanding for the different 
sub-detectors and options may be different. This may reflect in the cost estimates
as well as in the performances.
The manpower has been estimated roughly from past experiences for the different items 
and is added globally. Some options have no estimate yet.

\section{ILD Work Breakdown Structure}
 
The actual WBS for the different sub-systems can be found in \cite{WBS} .
The list of sub-systems under consideration is given in table~\ref{costs}.
 
As a guide we tried to estimate for the different systems the following items:
\begin{itemize}\addtolength{\itemsep}{-0.5\baselineskip} 
\item the amount of material and a unit price, the manufacturing,
\item the sensors,
\item the front-end electronics including printed circuit boards,
\item the local acquisition, testing and calibration,
\item the transportation (not knowing where it is made and where is the experimental site),
\item the assembly on site, tooling,
\item the spares and miscellaneous.
\end{itemize}
 
\section{ILD Current Cost Evaluation}
 
The following material costs have been used in the estimate:
\begin{itemize}\addtolength{\itemsep}{-0.5\baselineskip} 
\item Tungsten at 120 ILCU/kg (from a quotation for 40t of pure tungsten plates 
with tolerances),
\item Stainless steel for the Hcal at 18 ILCU/kg (from Atlas), 
and for the cryostat 15 ILCU/kg for SS304,
\item Yoke steel (low carbon) at 4.1 ILCU/kg,
\item Silicon strips for tracking at 7 ILCU/$cm^2$,
\item Silicon sensors for ECal at 3 ILCU/$cm^2$,
\item SiPM(MPPC) at 1.2 ILCU per piece from industrial quotation.
\end{itemize}

The cost estimates for the different sub-systems in ILD are listed 
in table~\ref{costs}.
 In addition we estimate the cost for other options to be 41 MILCU for the DHCal
and 35.7 for the SC-ECal.
 
\begin{table}
\begin{center}
\begin{tabular}{|c|c|c|c|}\hline
 Item               & cost  &  fraction in \%& man-years\\ \hline\hline
 Magnet yoke        & 68.4  & 16.8           &\\ 
 Muon system        &  8.4  &  2.1           &100\\ 
 Magnet coil        & 47.6  & 11.7           &200\\ 
 Magnet ancillaries & 11.0  &  2.7           &\\ \hline
 AHCal              & 48.3  & 11.9           &300\\ 
 Si-ECal            &112.0  & 27.5           &300\\ 
 Silicon tracking       & 21.6  &  5.3           &200\\ 
 Vertex             &  2.9  &  0.7           &100\\ 
 TPC                & 34.3  &  8.4           &100\\ \hline 
 Forward calorimeter&  5.3  &  1.3           &\\ 
 Beam tube          &  1.6  &  0.4           &\\ 
 Integration        &  1.7  &  0.4           &\\ \hline
 Global DAQ         &  1.2  &  0.3           &\\ \hline
 Offline computing  & 30.0  &  7.4           & \\ \hline
 Transport          & 13.0  &  3.2           & \\ \hline\hline
 Total              &407.0  &100.00          &\\ \hline
\end{tabular}
\caption[Table of sub-system costs.]{ Costs in MILCU and estimate of the manpower in man-years for the technologies
retained in the simulation for physics studies. Options for two major sub-systems are included, but 
not used in the sum.}
\end{center}
\label{costs} 
\end{table}

The cost driving items are the yoke, and the calorimeters. 
The price for the yoke is as large as it is as a direct consequence of the push-pull operation. 
The restriction on the stray-field outside the detector can only be met with a rather 
thick iron yoke, inflating the price significantly. 
The cost for the integration is an estimate of the scenario described in section~\ref{integration}, 
and might vary significantly with different scenarios. It includes the extra cost for 
the large platform on which the detectors moves, as well as the extra costs of the 
cryogenics needed to allow a cold move of the detector. 
In the absence of platform and cable chain, the lower part of the experimental hall would have to be
filled with concrete. The cost for this is subtracted. Some integration tooling has been added.
The offline computing represents a significant cost. Owing to the continued large advances 
in computing technology, we have estimated this at $20\%$ of the equivalent cost for the LHC detectors.

A first estimate of the manpower needed has been done for each sub-system (see
table~\ref{costs}, last column). Detailed estimates are available only for 
the major components, the rest is estimated to be around 100 MY in total.
The average cost per MY has been taken to be 93 kILCU including overheads. This value is typical 
for the mix of qualifications needed for a sophisticated project like the ILD. 
The estimate only includes the manpower needed to build the detector, and does not 
include needs to finish the R\&D or work out a detailed design of the detector. 
The manpower is then estimated to be in total 1400 MY, or 130 MILCU. 
 
The overall ILD cost could then be 530 MILCU   + 100 / - 50 MILCU.\\
 
The study has been carried out assuming that the detector is in a push-pull configuration.
Most of the sub-system costs are only marginally affected by this assumption, with 
the exception of the integration costs, as discussed above. It has been estimated 
that without these requirements the total cost of the detector might be reduced 
by some $5-10\%$.
 
\section{Cost Scaling Laws with Detector Parameters}
The parameters which have been considered for possible scalings are the following:
\begin{itemize}\addtolength{\itemsep}{-0.5\baselineskip} 
\item the magnetic field;
\item a characteristic transverse size chosen as the inner radius of the ECAL barrel;
\item a characteristic longitudinal size chosen as the length of the ECAL barrel or TPC;
\item the number of samples for the ECAL (for a given number of radiation  lengths);
\item the number of samples for the HCAL (for a given layer interaction length or a given total 
interaction length);
\item the calorimeter cell sizes.
\end{itemize}
 
The study was done under the assumption that the technologies remain the same. 
This implies that the range for scaling is rather small, basically at the level of 
25\%. To go further would in most cases require reassessing the technology choices.
 
The nominal field is 3.5~T, but the magnet is designed to withstand 4T. 
Reducing the field below 3.5T might offer cost savings, but also 
result in a large loss of the physics potential of the detector. ILD therefore 
does not consider this option of de-scoping. There is a clear impact on the 
overall cost of the coil from the correction coils. 
Reducing the requirements on the field quality would result in cost 
savings. Whether or not this might be acceptable is currently unknown. 
 
\begin{figure}
\begin{center}
\begin{tabular}{c c c}
\includegraphics[width=7cm]{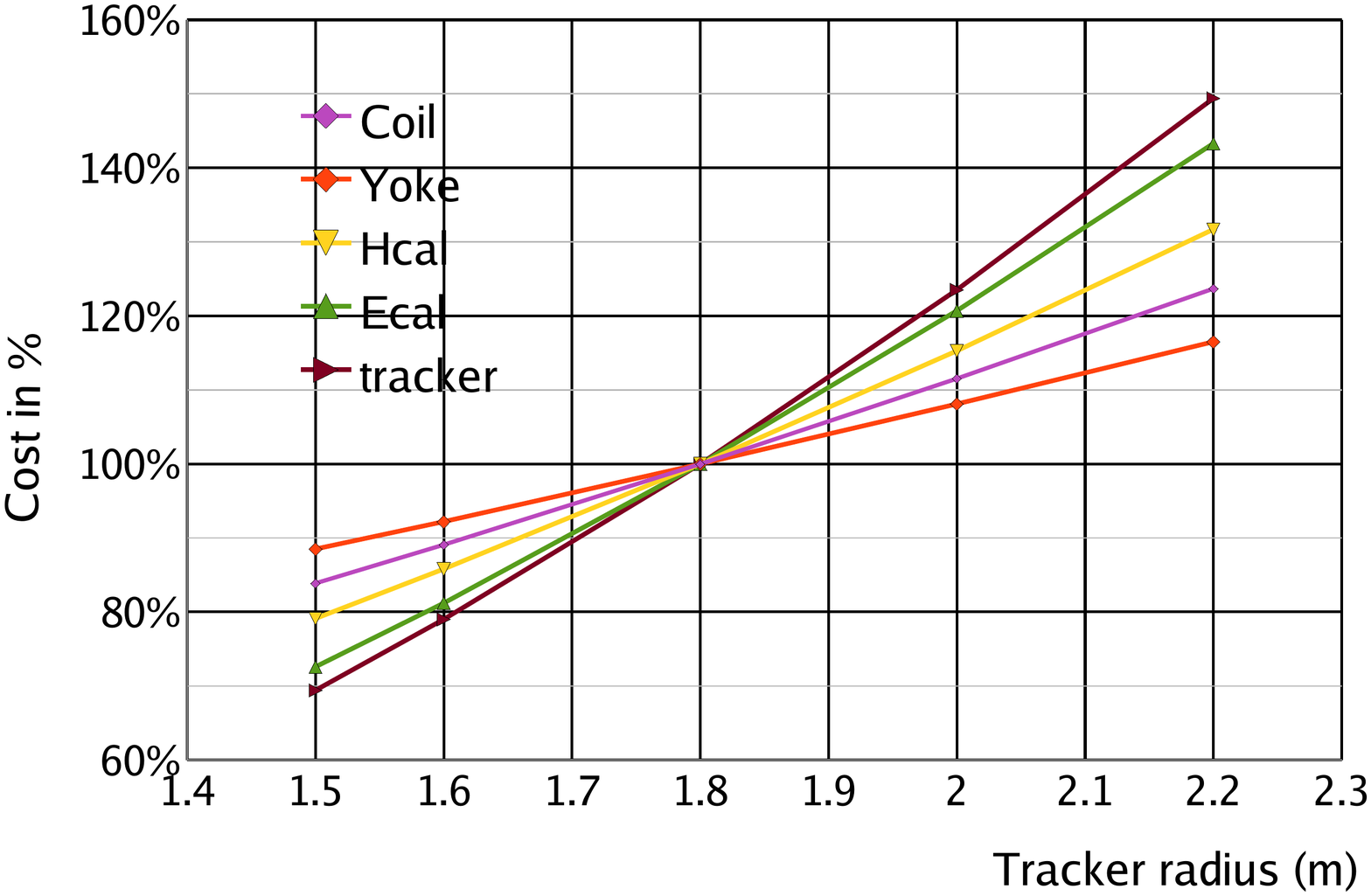}&  &
\includegraphics[width=7cm]{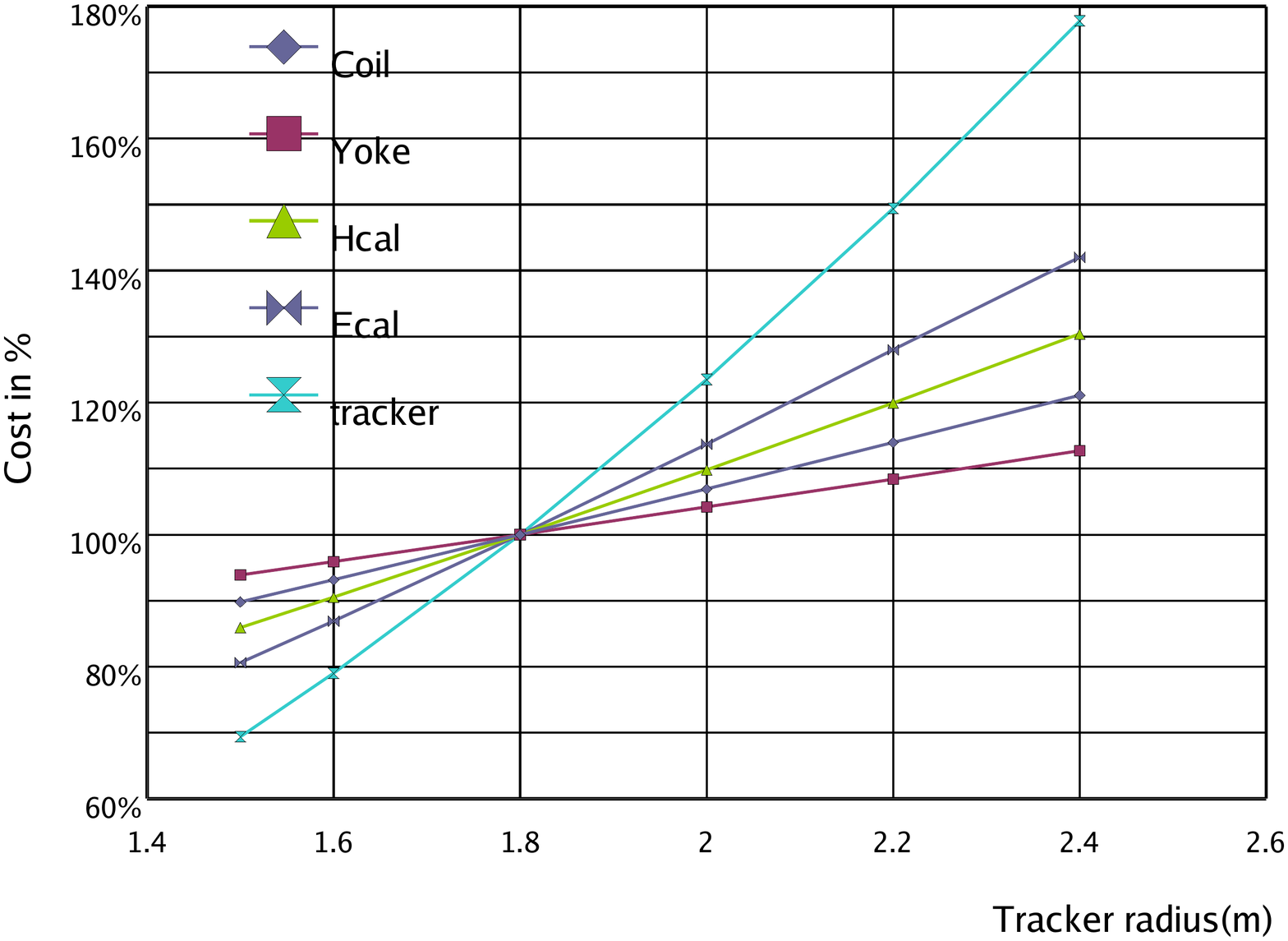}
\end{tabular}
\caption[Cost vs. transverse detector size.]{Dependence of the cost of the main items with the size, on the left for a constant angle,
on the right changing only the transverse size}
\label{Gsize}
\end{center}
\end{figure}

\begin{figure}[ht]
 \begin{minipage}[t]{.47\textwidth}
   \centering
   \includegraphics[width=1.05\textwidth]{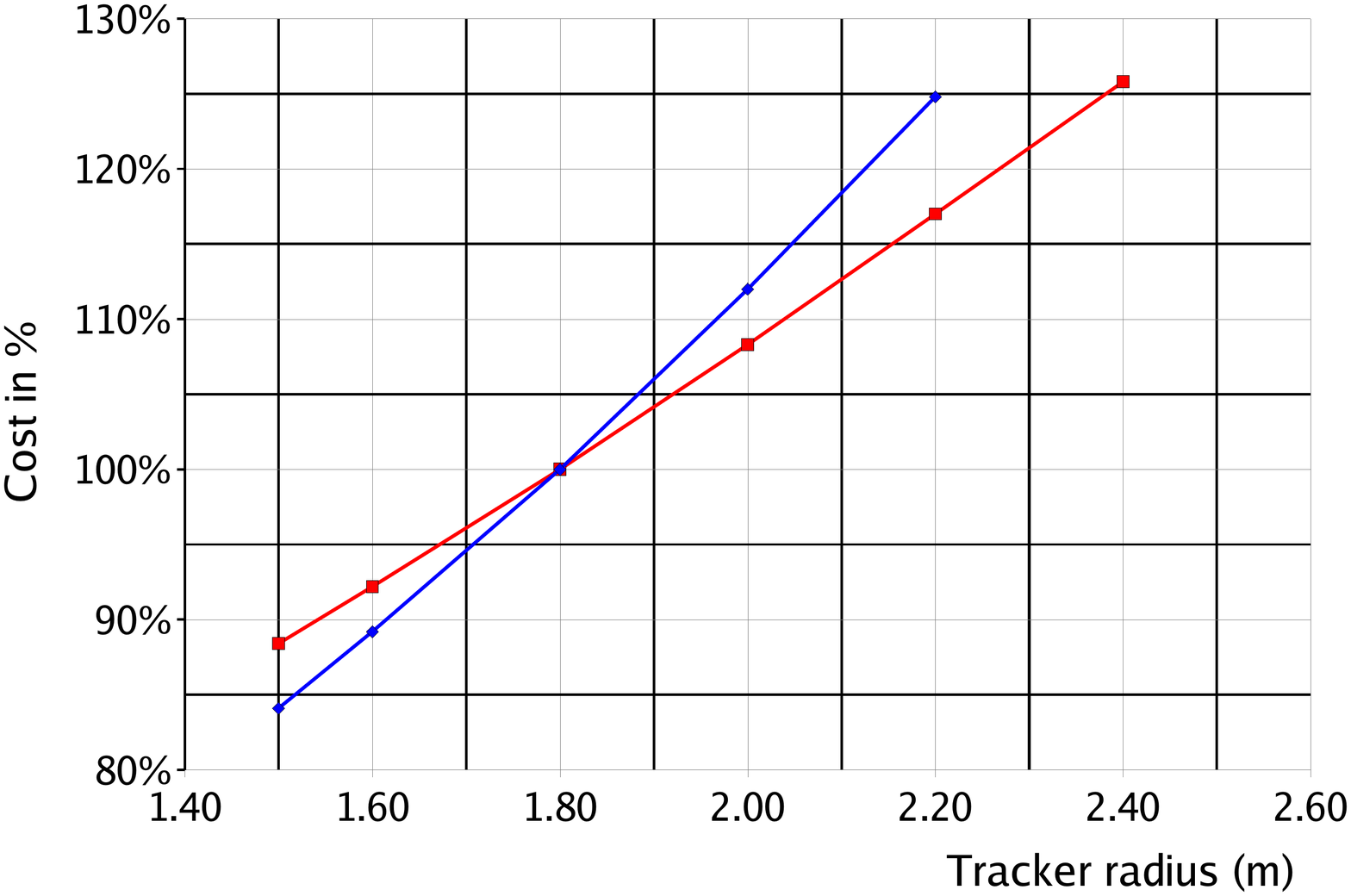}
   \caption[Cost vs. longitudinal detector size.]{Dependence of the total cost with the size of the detector, in blue when the
aspect ratio is kept, in red when the radius only changes.}
   \label{Gtotsize}
 \end{minipage}
 \hfill
 \begin{minipage}[t]{.47\textwidth}
   \centering
   \includegraphics[width=1.05\textwidth]{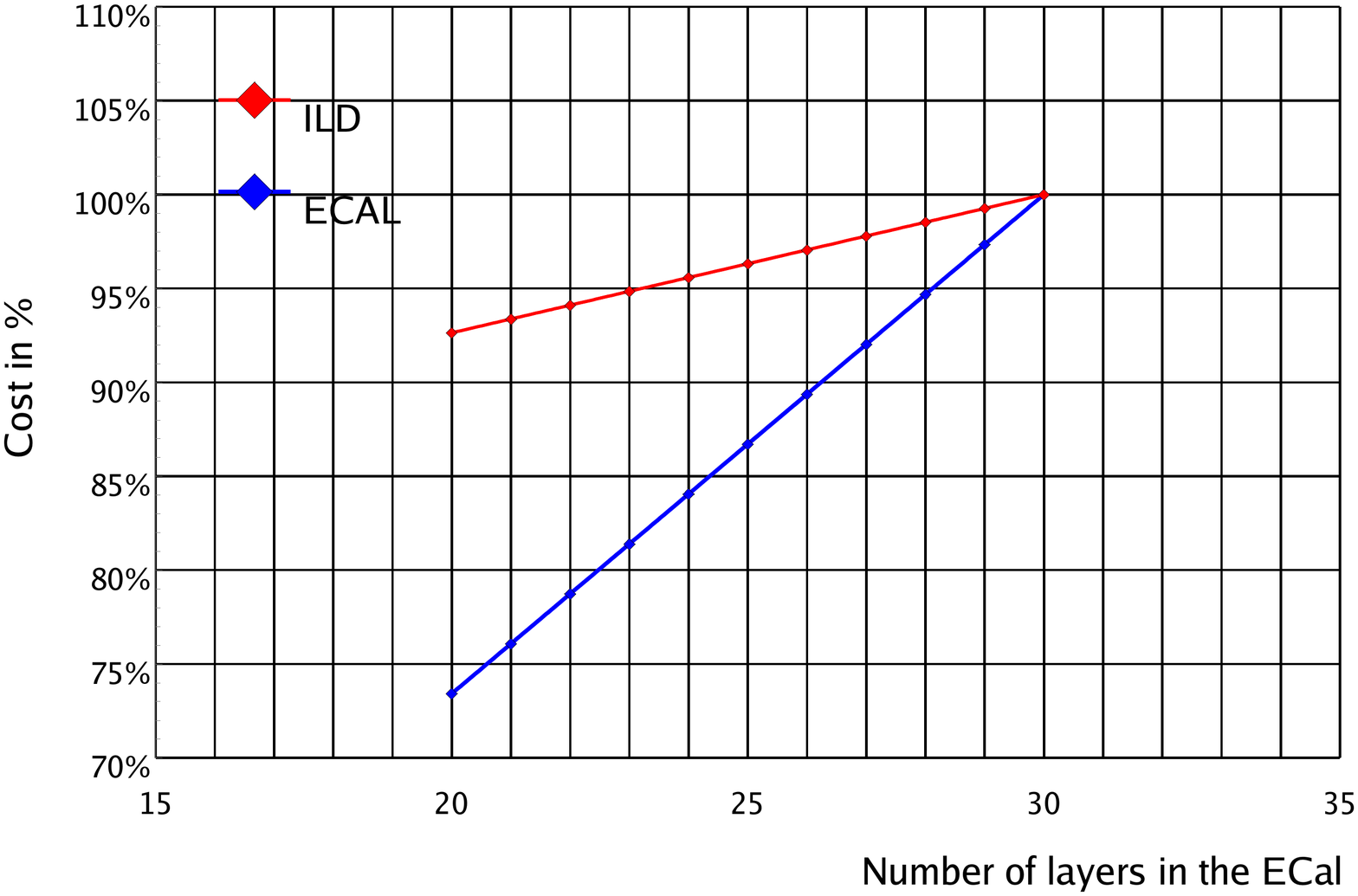}
   \caption{Dependence of the cost with the number of layers in the ECal. }
   \label{EC_layer}
 \end{minipage}
\end{figure}

The dimensions of the detector parts inside the TPC are dictated by considerations 
of background and assembly. They are not very relevant for costing. 
Therefore a characteristic transverse scaling parameter is the radius of the 
transition from the tracker into the calorimeter, a characteristic longitudinal 
scaling parameter is the length of the TPC. 
Moving these parameters impacts calorimetry, coil and yoke. 
There are two ways we can envisage to scale down the size of the detector, either by keeping the aspect ratio
constant, or by reducing only the radius.
The estimate of the cost variation with a constant aspect ratio from the reference design 
is shown in figure~\ref{Gsize} on the left. The figure on the right shows the 
impact of changing solely the TPC radius. The figure~\ref{Gtotsize} illustrates the total cost 
scaling of ILD in the same conditions.
 
The scaling of the ECal sampling  has been done under the assumption that the total number of radiation lengths in the ECAL is 
kept constant. The area of sensitive medium and the number of readout channels then scale proportional to
the number of samplings. 
On the other hand, as the total amount of radiator does not change, the thickness of the absorber plates changes and
the manufacturing of the plates varies in a non negligible way.
Reducing the number of samples will reduce the overall thickness of the ECAL even when the total 
amount of absorber material stays constant. For example, going from 30 to 20 samples will 
reduce the radial thickness by 20~mm. In the cost scaling we do not consider the impact on the surrounding detectors
(see figure~\ref{EC_layer}).
 
\begin{figure}
\begin{center}
\begin{tabular}{c c c}
\includegraphics[width=7cm]{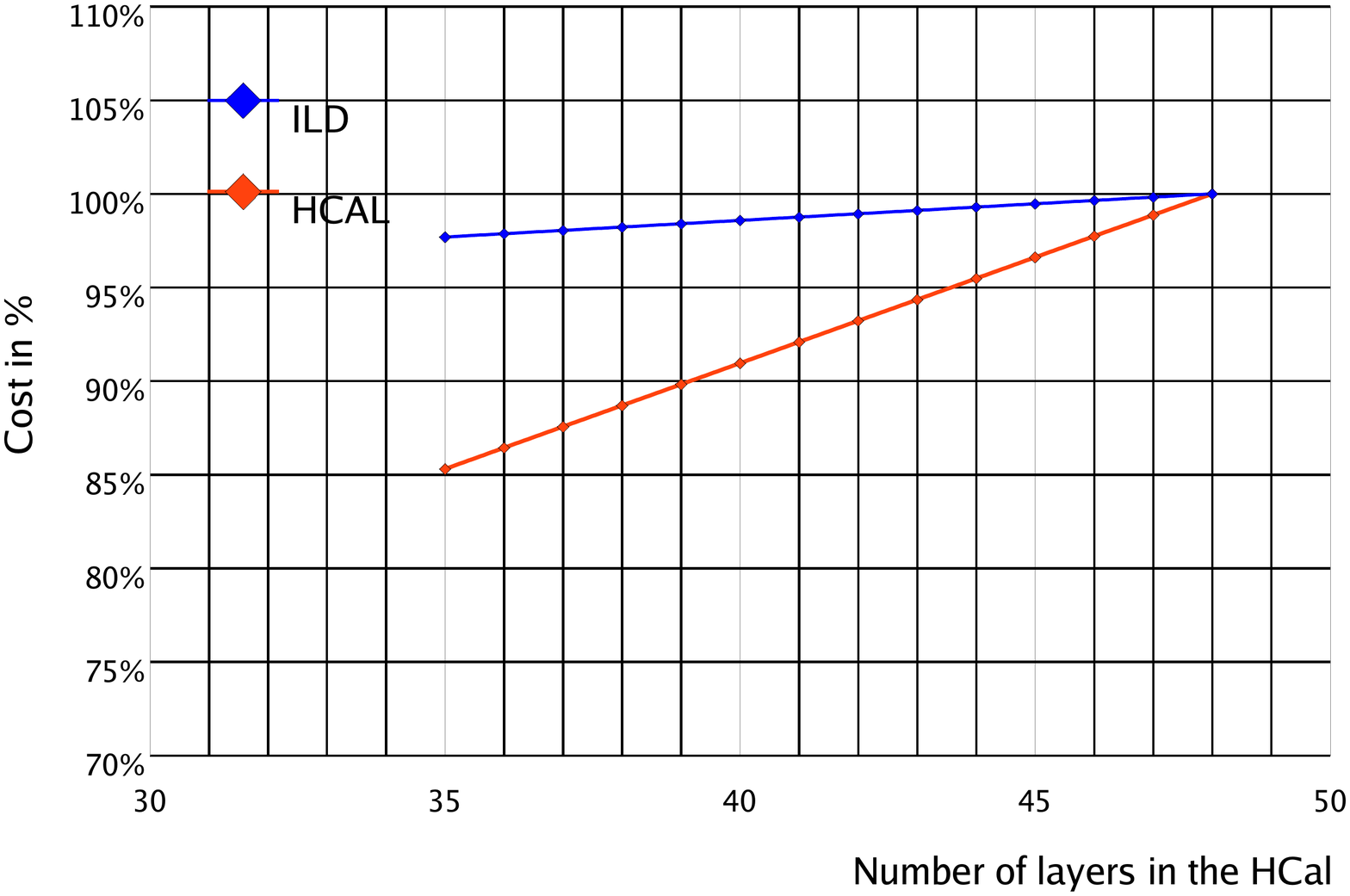}& &
\includegraphics[width=7cm]{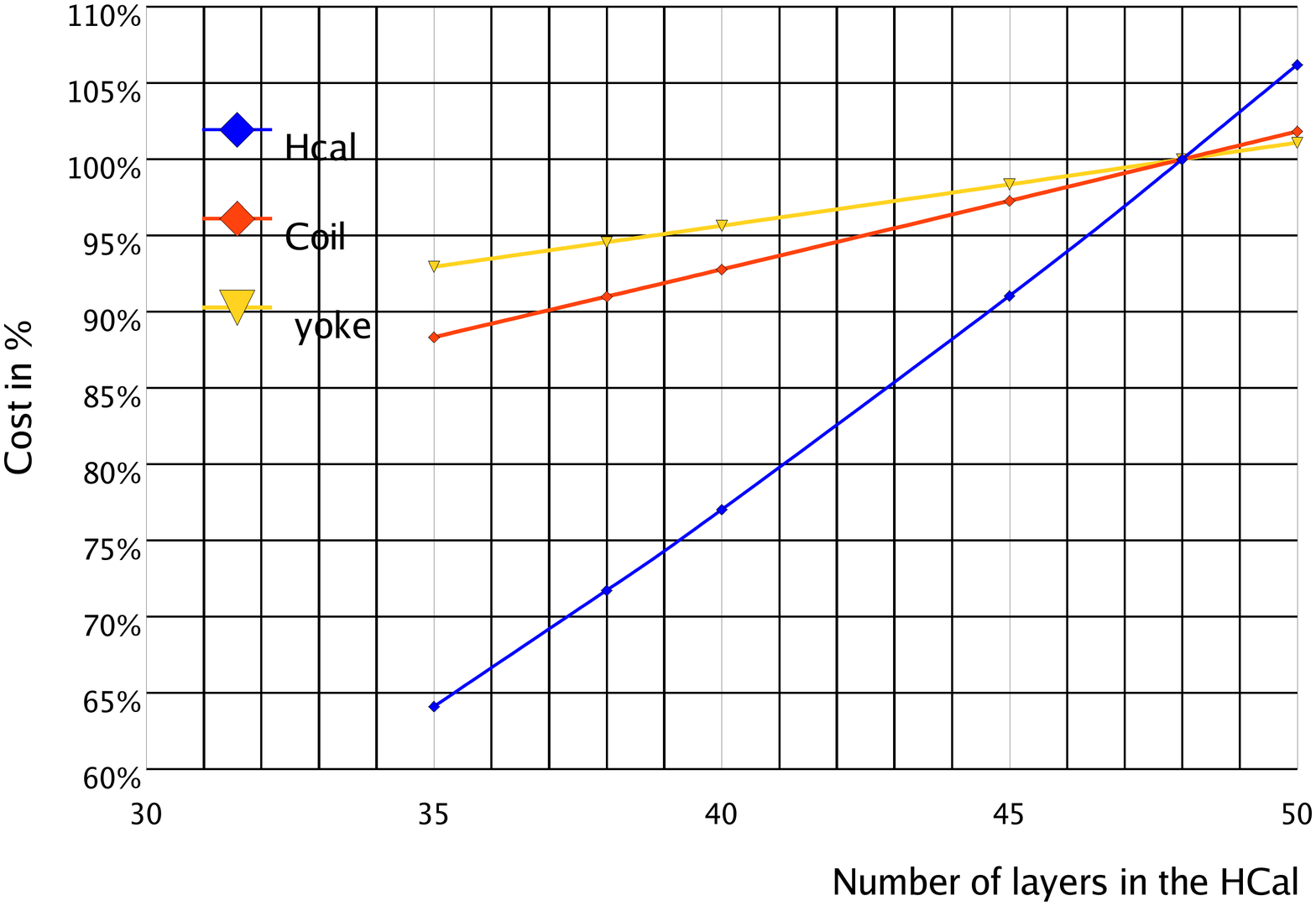}
\end{tabular}
\caption[Cost vs number of layers in the HCAL.]{Dependence of the cost with the number of layers in the HCal, 
on the left if you keep the total number of interaction lengths, on the right 
if you keep the thickness of the layers. }
\label{HC_layer}
\end{center}
\end{figure}

In the case of the HCAL we investigate two different scenarios for reducing the sampling: 
either we keep the depth in radiation length constant by increasing the layer thickness,
or we keep the layer thickness the same, changing the total number of interaction lengths. 
The same approach can be taken for the analogue and the digital HCAL option. 
This impacts the radiator, sensors
and all the surrounding subsystems. 
The global impact on the ILD cost is about 7\% when changing from 48 to 40 layers in the HCAL 
as can be seen in figure~\ref{HC_layer}.

 
The cell sizes of the electromagnetic calorimeter are already quite at a lower limit, as long as the 
 design currently under development is used. 
To go below, a new design, may be a different approach will be needed.
The MAPS ECAL which cost has not been estimated may be one of the possibilities. 
Increasing the cell sizes within the same technology will have only a minor impact on the 
cost, as the cost roughly scales with the area of silicon, not the number of readout channels. 
There is of course some effect due to a 
different cost of the printed circuit boards and other ancilliary equipment. 
We estimate that reducing the number of cells by an order of magnitude reduces the 
cost of the ECAL by less than 10\%, or 3\% of the total detector cost. 
The impact on the cost for the scintillator version may be larger but it is unlikely that scaling up
the size in this version would be considered.
 
For the hadronic calorimeter changing the cell sizes will result in a changed number of 
FE chips, calibration devices etc. We estimate that a reduction of the 
number of readout channels by an order of magnitude reduces the cost of the 
digital HCAL by about 20\%, of the analogue HCAL by about 10\%. This has to be balanced with 
a large performance loss.
 
\section{Conclusion }
The cost of the ILD detector has been estimated to be about 500 MILCU. It
includes the material and labour to build the detector, but does 
not include cost escalation and contingencies. 
The dependence of the cost on the main detector parameters has been studied, 
and effects of order $10\%$ or less per item on the total detector cost 
have been found.  To illustrate the possibilities, a cost reduction of 20\% can be reached by reducing 
the number of HCAL layers from 48 to 40, the number of ECAL layers to 20, the inner radius of the ECAL to 160~cm, and 
the length accordingly. This reduces clearly the performances of the detector (See section~\ref{optimization}). 
It should be noted that in many instances, a reduced performance of the detector 
translates into a longer running time of the accelerator until the desired physics 
measurements can be made. 
 
The quoted cost of the ILD detector is comparable to the total cost of the 
recently completed large LHC detectors. 
 
Although costs quoted in many instances are based on actual costs of 
prototypes, together with educated guesses toward mass production etc, 
there are still large uncertainties. 
A more reliable cost estimate will only be possible when a more complete and detailed 
engineering of the ILD detector will have been done.

%% file: group/group.tex
The ILD concept group was formed in 2007 by the merger of the GLD and the LDC groups. 
The ILD group has members from all three regions of the world, but is 
particularly well anchored in Europe and in Asia. More than 650 signatories from about 170 institutions support this Letter of Intent for ILD. 
Since the ILD group is not yet a collaboration, the membership of the ILD
group has not been very clearly defined. The signatories of the Letter of Intent
is the first set of names that comes close to the membership of the ILD
group. They are, however, in a state of flux. Anybody who has contributed or intends to contribute to the ILD detector concept study is welcome to
sign the LoI and can do so without any formal evaluation. On the
other hand, the management of the ILD group has been defined clearly and is in
operation with well defined membership and distribution of responsibilities.

The combined leadership of the two former groups GLD and LDC
elected a joint steering board charged to produce a single letter of intent.
The newly formed detector concept was named `ILD' which stands for
`International Large Detector'. The joint steering board consists of two
representatives from each of the three regions - Asia, Europe, and North
America. The joint steering board then elected
working group leaders, subdetector contacts, and representatives for the
research directorate.

There are four working groups -  optimisation, MDI/integration, costing, and
software - and each has two conveners. The optimisation working group is
charged with optimising the detector parameters based on simulations and to
evaluate physics performance of the resulting detector. This working group
played a key role in unifying the detector parameters of the GLD and LDC
detector concepts, and continues to be the main framework for physics
analyses. The MDI/integration working group was formed to fill the immediate
need to liaise with the accelerator activities on such issues as dimension
and shape of the experimental hall, design of push-pull operation, support
of the final quads, etc. As the name suggests, the MDI/integration working
group handles the issues of overall integration of the detector. There are
three technical coordinators which belong to the MDI/integration working
group.  We believe that system engineers with strong authority are not
needed at this stage of the detector development. The costing working group
essentially consists of two conveners only, and is charged with estimating
the cost of the whole detector and to represent the ILD group in discussions
to define common costing rules with other LoI groups in the framework of the
ILC research directorate. The responsibility of the software working group
is to unify the softwares of the two former concept groups, and manage the
development of the resulting software system.

The ILC detector R\&D groups play a critical role in the ILD detector
development. They are sometimes called `horizontal' collaborations as
opposed to LoI groups which are viewed as vertical organisations. Examples
are the CALICE collaboration for calorimeters, the SiLC collaboration for
silicon trackers, the LCTPC collaboration for TPC, and the FCAL
collaboration for the calorimeters and instrumentation in the forward
region. These groups are in principle independent of LoI groups and often
bound by a "Memorandum of Understanding" to form more formal collaborations. Subdetector contact
persons have been selected for vertexing, silicon trackers, TPC, ECAL, HCAL,
FCAL, Muon system, DAQ, and solenoid. The number of contact persons is two
per subdetector except for FCAL and muon system for which number is one.
When relevant detector R\&D collaborations exists for a given subdetector, a
person appropriate as a liaison  for that group was chosen. The subdetector
contacts are charged to act as contact points to R\&D groups and to make sure
that the relevant detector technology is applied to the ILD environment.

In addition, the ILD group nominates members of the common task groups which
serve under the ILC research director. There are two representatives of the
ILD LoI group, two for MDI, one for engineering tools, three for the R\&D
common task group, two for the physics group,  and two for the software
group. They are often the same persons as the corresponding working group
conveners or subdetector contacts, and act as a link to the research
directorate.

\begin{figure}
	\centering
		\includegraphics[width=15cm]{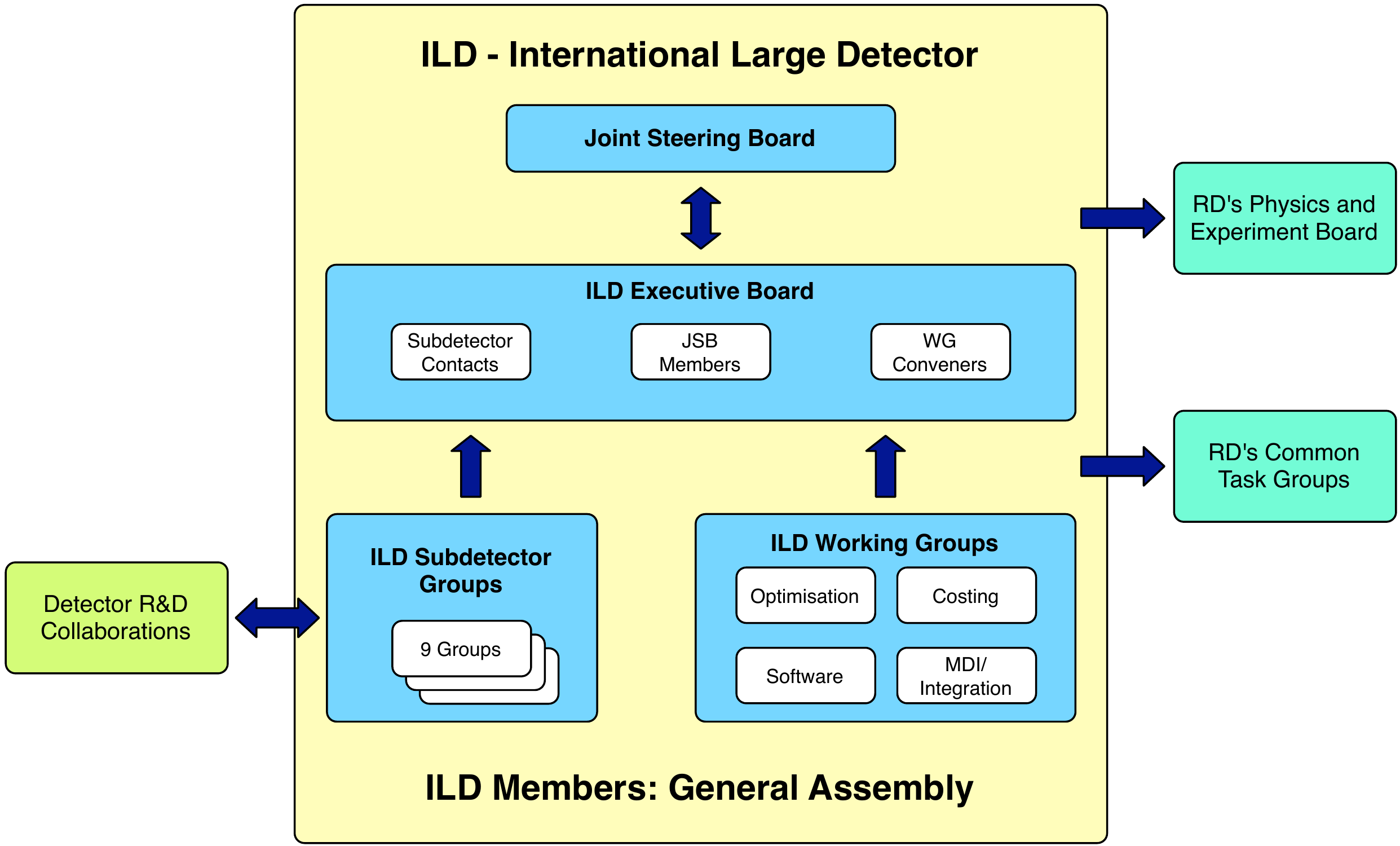}
	\caption{Structure of the ILD group.}
	\label{fig:ILD_structure}
\end{figure}

In figure~\ref{fig:ILD_structure} the organisational structure of the ILD group is 
shown. 

The joint steering board, working group conveners, subdetector contacts, and
representatives for the research directorate form an entity called the
executive board. The ILD executive board forms the core of the ILD-related
activities, and meets roughly biweekly over internet or in person. General ILD meetings lasting a half day to a full day are usually attached
to each ILC workshop. Separately we have a few dedicated ILD workshops per
year of a few days each. At these meetings and dedicated workshops we have
general ILD assembly meetings which take in opinions and comments from the wider
community.

%% file: group/randplan.tex
The ILD detector has been developed under the assumptions that particle flow is 
the most best method for event reconstruction and that an excellent 
vertex reconstruction is needed for many physics channels. To reach 
the proposed performance of ILD, significant advances in detector 
technology are needed, compared to existing detectors. Over the 
past years R\&D collaborations have formed to address the main 
issues in technological development. ILD is closely cooperating with 
these R\&D collaborations and is discussing and coordinating the 
needed work with these groups. 

Particle flow has consequences in many parts of the detector: it is essential that 
a calorimeter is built that is capable of imaging the shower, and that it is 
complemented by a very powerful and efficient tracking system. 

ILD has chosen a solution where a highly granular calorimeter is combined 
with a powerful and highly redundant tracking system, which contains 
both a large volume gaseous tracker (time projection chamber) and a
high precision Si tracker with excellent tracking and vertexing capabilities. 
The main technological innovation relevant for ILD is the granularity in 
particular of the calorimeter, and the overall very low material 
budget projected for the tracking system while reaching excellent 
spatial resolution for charged particles. 

Over the past years significant R\&D has taken place to establish the 
basic feasibility of the main technologies which have been proposed. 
Major projects have been started to study granular calorimeters, both 
electromagnets and hadronic, which are organised in the context of the 
CALICE collaboration. A novel type of time projection chamber is 
under development, based on micro-pattern gas detectors, organised 
by the LC-TPC collaboration. New low-mass systems have been developed 
for all parts of the Si tracking system, in the 
context of the SiLC, LCFI, the MAPS, the DEPFET collaboration and other groups. 
Instrumentation in the very forward direction, which is special in 
that it is the only place in the ILC detector where radiation hardness 
is required, is studied in the context of the FCAL collaboration. Large 
area muon chambers, which can also serve as tail catchers, are being 
developed, also in the context of the CALICE collaboration, and by independent efforts. 

The first round of R\&D established for most subdetectors the basic feasibility of these systems. 
Test experiments with highly granular calorimeters were successfully 
completed at both CERN and Fermilab. The concept of a micro-pattern 
TPC could be established with small prototypes. First prototypes of 
extremely low-mass Silicon-based detectors have been developed and 
tested. 

Internationally the linear collider experimental community wants to 
be able to make a reliable and well understood proposal for 
experiments at the ILC by 2012. This requires that a fundamental 
understanding of the major detector components needs to be 
achieved by this time, and that a first version of an engineering 
solution is available. The work needs to advance to a different 
level than described in this document, and goes beyond the feasibility tests of the 
last years. It has to address system integration aspects within the 
sub-detector - integrated mechnical design, realistic integration of 
readout electronics, power management, cooling etc - but also 
address questions of integrating different sub-detectors. 

Based on this, ILD considers that the experimental investigation
of particle flow has the highest priority of 
R\&D. This has a number of different aspects: 
\begin{itemize}\addtolength{\itemsep}{-0.5\baselineskip}
\item  Develop technological solutions for a granular electromagnetic and hadronic calorimeter. 
\item Develop the alternatives of Si-based and scintillator-based electromagnetic calorimeter to a point that both can be proposed, 
and that a future ILD collaboration can take a rapid decision after approval. 
\item Develop the alternatives of analogue and digital hadronic calorimetry 
to a point where a rapid decision between the options can be taken by a future ILD 
collaboration after approval. Ensure that both the analogue and the digital hadronic calorimetry are tested 
at the same level of sophistication in large scale setups. 
\item Continue to develop the simulation and software tools needed to 
study particle flow in detail, and continue to refine the particle 
flow reconstruction tools. 
\item Develop an experimental program, which can convincingly demonstrate 
the feasibility of the concept of particle flow. 
\end{itemize}

The very forward detector systems in ILD are small but very different 
from the rest of the detector. They are the only devices where 
significant radiation hardness is needed. The flux of electromagnetic 
radiation seen by these devices is even larger than what is 
anticipated at the LHC. Dedicated R\&D is therefore needed 
to develop adequate technologies for the precision 
calorimeters needed at very forward angles. 

Particle Flow relies on a very powerful tracking system. ILD is special 
in all the ILC concept groups that it is proposing a TPC as central tracker. 
This is central to the overall performance and robustness of the 
system. The experimental proof that a TPC can be built and operated 
with the required precision and stability has very high priority for ILD. 

ILD includes in its tracker a very powerful Si tracking system. 
The main challenge here is the development of a technology which 
is powerful enough and which meets the requirements in terms 
of material, power consumption and speed. Solutions seem to be in reach 
for the strip tracking system, many different options are being 
studied for the pixel-based vertex detector part of the system. ILD considers 
powerful Si tracking systems to be an essential part of the or
detector concept. ILD stresses the need to pursue a broad range of different 
Silicon detector technologies, in particular for the vertex detector, so 
that an optimal solution can be chosen as close in time as possible to the 
construction of the detector. In particular in the rapidly and quickly 
evolving field of Si detectors which are dominated by commercial developments 
a final choice of technology should be delayed to the latest possible moment. ILD 
considers it essential to follow the technological developments, to develop 
as many alternative solutions as possible to be in a position to pick 
the optimal one quickly, once needed. 

Testing sophisticated hardware components requires adequate testing facilities. 
The next generation of test beam experiments should address the interplay of different 
sub-detectors, in addition to novel technologies of the sub-detectors. The experiments 
should comprise vertexing, tracking and calorimetric components together, in 
an interchangeable way, in a sufficient magnetic field, at a hadron beam. A beam with 
an ILC-like particle bunch structure would be a big bonus. It would 
test integration aspects, aspects of a common data acquisition, and the data from such 
an experiment would be very useful in the continuing improvement of the 
understanding of the different reconstruction techniques needed for ILD.
ILD considers the creation of such an integrated test facility of 
high importance for the eventual success of the programme. 

Even though most of the work currently done is on technologies for sub-detectors, 
the overall detector integration has to be considered as well. The concept 
of push - pull, which is currently favored to save one beam line, 
will require dedicated designs for many of the sub-detectors. These 
points need to be known early on, and may require some significant 
R\&D on their own. Points of concern are sensitivity to vibrations, 
reproducibility of alignments, and the external monitoring of 
inter- detector alignments. 

The ILD group intends to continue its work toward a full technical design of the 
detector at the end of the technical design phase 2 (around 2012), in 
step with the plans of the GDE for the machine. The different R\&D plans of the 
sub-detector components are detailed in the individual sections. For the large components 
calorimetry and time projection chamber, large second generation prototypes should be 
running and delivering results by the end of 2012, such that - if the ILD group 
is transformed into a collaboration, and if the ILC project gets approval - 
a selection of technologies is possible. The steps needed to advance all 
options to this point are described in the individual sections. For sub-detectors 
like the vertex detector, intense R\&D into sensor technologies will continue. 
A decision on the technology can be taken at a later stage, but the 
integration aspect of the VTX detector will need to advanced enough in 2012, that 
a realistic design can be proposed. The same is true for the other 
smaller sub-systems discussed in the context of ILD. 



%% file: conclusion/conclusion.tex
In the summer of 2007, the GLD concept study group, whose membership was largely
based in Asia, and the
LDC concept study group, which was mostly based in Europe with a strong north american membership, joined 
forces to produce
a single Letter of Intent for a detector at the International Linear Collider, and formed the ILD concept group.
Both the GLD and LDC concepts used
the particle flow algorithm for jet reconstruction and a TPC for the
central tracker. The basic
parameters of the two concepts such as the size of the detector and
the strength of the solenoid
field, however, were quite different and had to be unified
in order to write this letter of intent for ILD.
Also, other critical details such as the interaction region design had to be
unified. This was a non-trivial task, neither politically nor sociologically.

The newly-formed concept
study group, the ILD group, created a management team and engaged in intense
studies to
define the ILD detector concept by scientifically optimising the detector
designs. The process
has worked remarkably well, and we present here the outcome of this
study as well
as the large amount of studies that preceded separately by the two older
concept groups.
The ILD detector concept is now well defined, even though some technology choices are
still open. One of the merits of unifying the detector concepts was that it
revitalised the studies on
physics performance and detector designs. We believe that
the level of sophistication of the simulation and physics analyses has
reached a high degree of sophistication for a detector group at this stage. This was achieved through collaboration and
competition, and is the result of a productive learning process.

The unification had also positive effects on the subdetector R\&D efforts. Most R\&D on
detector technologies relevant to the GLD and LDC groups is being performed
within the
framework of detector R\&D collaborations such as LCTPC, SiLC, CALICE, and
FCAL which pursue their own goals of detector technology development. Members of the detector concept groups participate in the R\&D collaborations and make sure that the detector technologies are successfully applied to the detector concept designs. By the creation of the ILD concept group, the application efforts became more focused. Currently, the ILD management includes
subdetector contacts who are also key members of the detector R\&D collaborations.
This scheme is working efficiently such that we can finish basic R\&D in
time for the Technical Design Report which is envisaged around 2012.

Overall, the ILD group structure is efficient while keeping flexibility and
openness. Even though we are still short on person power and
funding at this time, we believe that we are well
positioned to successfully complete a technical design for a detector at
the International Linear Collider. The ILD group is firmly committed to 
the ILD project.